\documentclass[11pt]{report}
\pdfoutput=1
\usepackage{amsmath}
\usepackage[a4paper,top=0.8in,left=1in,right=1in,bottom=1in]{geometry}
\usepackage{graphicx}
\usepackage{graphbox}
\usepackage{appendix}
\usepackage{tocloft}
\usepackage[square]{natbib}
\usepackage[breaklinks,colorlinks,citecolor=blue]{hyperref}

\newcommand{\vc}[1]{{\textbf{\em #1}}}
\newcommand{\vca}[1]{{\boldsymbol #1}}
\newcommand{\cD}{{\mathcal D}}
\newcommand{\cH}{{\mathcal H}}
\newcommand{\cJ}{{\mathcal J}}
\newcommand{\cL}{{\mathcal L}}
\newcommand{\cO}{{\mathcal O}}
\newcommand{\cS}{{\mathcal S}}
\newcommand{\cW}{{\mathcal W}}
\newcommand{\der}{\partial}
\DeclareMathSymbol{\mg}{\mathrel}{symbols}{"1D}
\newcommand{\ml}{\ll}
\newcommand{\be}{\begin{equation}}
\newcommand{\ee}{\end{equation}}
\newcommand{\ba}{\begin{eqnarray}}
\newcommand{\ea}{\end{eqnarray}}
\newcommand{\bea}[1]{\begin{align}#1\end{align}}
\renewcommand{\d}{\mathrm{d}}
\newcommand{\ga}{\alpha}
\newcommand{\gb}{\beta}
\newcommand{\gd}{\delta}
\renewcommand{\ge}{\epsilon}
\newcommand{\gz}{\zeta}
\newcommand{\gth}{\theta}
\newcommand{\gk}{\kappa}
\newcommand{\gs}{\sigma}
\newcommand{\gf}{\phi}
\newcommand{\gvf}{\varphi}
\newcommand{\get}{\eta}
\newcommand{\getpa}{\eta^\parallel}
\newcommand{\getpe}{\eta^\perp}
\newcommand{\gx}{\xi}
\newcommand{\gxpa}{\xi^\parallel}
\newcommand{\gxpe}{\xi^\perp}

\newcommand{\gc}{\chi}
\newcommand{\fnl}{f_\mathrm{NL}}
\newcommand{\gint}{g_\mathrm{int}}
\newcommand{\bc}{\bar{c}}
\newcommand{\bv}{\bar{v}}
\newcommand{\bA}{\bar{A}}
\newcommand{\bH}{\bar{H}}
\newcommand{\bN}{\bar{N}}
\newcommand{\tf}{\tilde{f}}
\newcommand{\tg}{\tilde{g}}
\newcommand{\tn}{\tilde{n}}
\newcommand{\tU}{\tilde{U}}
\newcommand{\tV}{\tilde{V}}
\newcommand{\tW}{\tilde{W}}
\newcommand{\lh}{\left(}
\newcommand{\rh}{\right)}
\newcommand{\Var}{\mathrm{Var}}
\newcommand{\nn}{\nonumber}
\newcommand{\siml}{\raisebox{-.1ex}{\renewcommand{\arraystretch}{0.3}
$\begin{array}{@{}c} \scriptstyle < \\ \scriptstyle \sim \end{array}$}}
\newcommand{\simg}{\raisebox{-.1ex}{\renewcommand{\arraystretch}{0.3}
$\begin{array}{@{}c} \scriptstyle > \\ \scriptstyle \sim \end{array}$}} 
\newcommand{\half}{{\textstyle\frac{1}{2}}}

\renewcommand{\ae}{\alpha_{(1)}}
\newcommand{\aee}{\alpha_{(2)}}
\newcommand{\tvarphi}{\tilde{\varphi}}
\newcommand{\tdvphi}{\dot{\tilde{\varphi}}}
\newcommand{\dvphi}{\dot{\varphi}}
\newcommand{\dphi}{\dot{\phi}}
\newcommand{\ha}{\hat{\alpha}}
\newcommand{\hr}{\hat{\rho}}

\makeatletter
\renewcommand{\@tocrmarg}{\@pnumwidth plus1fil}
\makeatother

\defcitealias{Planck:2006aa}{Planck scientific programme}
\defcitealias{planck2013-01}{Planck 2013~I}
\defcitealias{planck2013-12}{Planck 2013~XII}
\defcitealias{planck2013-24}{Planck 2013~XXIV}
\defcitealias{planckint-17}{Planck intermediate~XVII}
\defcitealias{planck2015-01}{Planck 2015~I}
\defcitealias{planck2015-09}{Planck 2015~IX}
\defcitealias{planck2015-10}{Planck 2015~X}
\defcitealias{planck2015-11}{Planck 2015~XI}
\defcitealias{planck2015-12}{Planck 2015~XII}
\defcitealias{planck2015-13}{Planck 2015~XIII}
\defcitealias{planck2015-17}{Planck 2015~XVII}
\defcitealias{planck2015-20}{Planck 2015~XX}
\defcitealias{planck2015-25}{Planck 2015~XXV}
\defcitealias{planck2018-01}{Planck 2018~I}
\defcitealias{planck2018-02}{Planck 2018~II}
\defcitealias{planck2018-03}{Planck 2018~III}
\defcitealias{planck2018-04}{Planck 2018~IV}
\defcitealias{planck2018-06}{Planck 2018~VI}
\defcitealias{planck2018-09}{Planck 2018~IX}
\defcitealias{planck2018-10}{Planck 2018~XX}

\begin{document}

\begin{titlepage}

\begin{center}  
  \includegraphics[width=5cm]{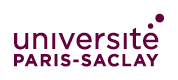}\hspace{1cm}
  \includegraphics[width=4cm]{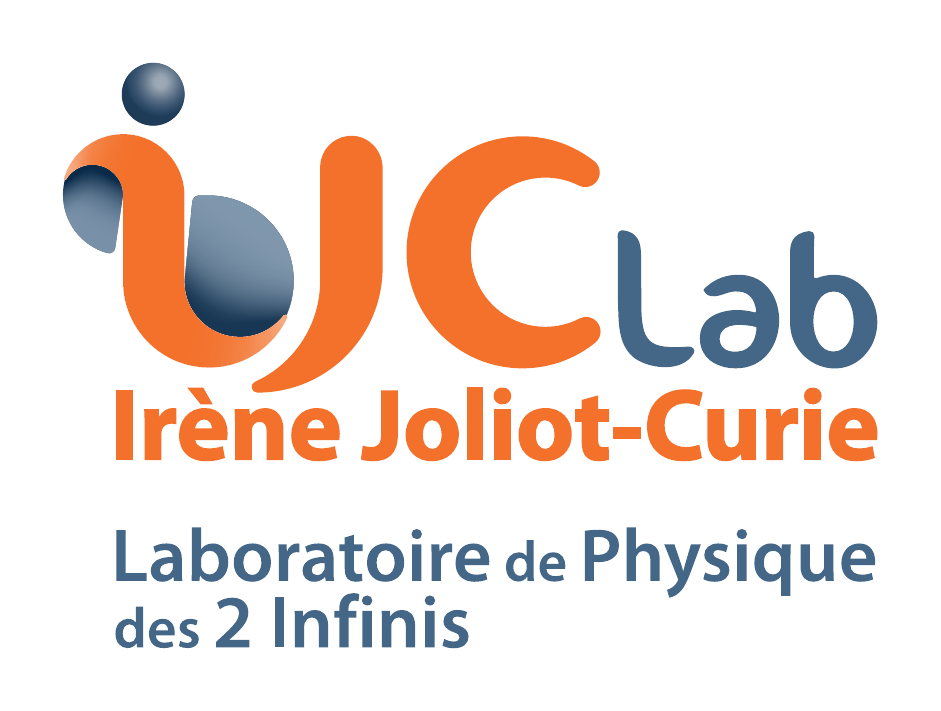}\hspace{1.3cm}
  \includegraphics[width=3cm]{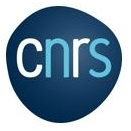}
\end{center}

\vspace{2cm}

\begin{center}
  \Huge\bf Non-Gaussianity in Cosmology:\\ from Inflation to the CMB\\
  \vspace{2cm}
  \Large Habilitation thesis {\em (Habilitation \`a Diriger des Recherches)}\\
  \vspace{0.2cm}
  \large defended in Orsay, France, on 25 November 2020, by\\
  \vspace{1.5cm}
  \LARGE\bf Bartjan van Tent
\end{center}

\end{titlepage}

\thispagestyle{empty}

\noindent
{\bf Abstract:}
A period of cosmological inflation, driven by one or more scalar
fields, is still the best candidate to solve various problems in the
standard Big Bang model. We want to find as many observational
constraints on the inflationary quantities as possible, in order to
pin down the correct inflation model, which in its turn will provide
information about the underlying high-energy theory. Fortunately the
non-Gaussianity of inflationary perturbations, as encoded in the
bispectrum (or 3-point correlator), has become an important additional
way of distinguishing between models, going beyond the linear Gaussian
perturbation quantities of the power spectrum.

This habilitation thesis provides a review of my work on both the
theoretical and the observational aspects of these non-Gaussianities.
In the first part a formalism is described, called the long-wavelength
formalism, that provides a way to compute the non-Gaussianities in
multiple-field inflation. Applications of this formalism to various
classes of models, as well as its extensions, are also treated.
In the second part an estimator is described, called the binned
bispectrum estimator, that allows the extraction of
information about non-Gaussianities from data of the cosmic microwave
background radiation (CMB). It was in particular one of the three estimators
applied to the data of the Planck satellite to provide the currently
best constraints on primordial non-Gaussianity. Various extensions of the
estimator and results obtained are also discussed.

\vspace{\fill}

{\flushleft \small \textbf{Composition of the jury :}
\bigskip

\begin{tabular}{|p{8cm}l}
\textbf{R\'eza Ansari} & President \\ 
Professor, IJCLab, Universit\'e Paris-Saclay & \\
\textbf{Ana Ach\'ucarro} & Rapportrice \\ 
Professor, Lorentz Institute of Theoretical Physics, Leiden University,
Netherlands & \\ 
\textbf{David Langlois} & Rapporteur \\ 
Directeur de Recherche CNRS, APC, & \\
Universit\'e de Paris & \\ 
\textbf{Sabino Matarrese} & Rapporteur \\ 
Professor, Department of Physics and & \\
Astronomy G.~Galilei,
University of Padua, Italy & \\ 
\textbf{Fran\c{c}ois Bouchet} & Examiner \\ 
Directeur de Recherche CNRS, IAP, & \\
Sorbonne Universit\'e & \\ 
\textbf{Martin Bucher} & Examiner \\ 
Directeur de Recherche CNRS, APC, & \\
Universit\'e de Paris & \\ 
\end{tabular}
}

\clearpage

\thispagestyle{empty}

\strut

\vspace{5cm}

{\hfill\em To my parents, in memory of my father.}

\clearpage
\tableofcontents

\chapter{Introduction}
\label{introduction}

The main subject of most of my research after my PhD thesis is
the non-Gaussianity of cosmological fluctuations. Here non-Gaussianity refers
to the distribution of the fluctuations evaluated at different spatial
coordinates in the universe (but for a given time).
When all equations are linearized, it turns out that the distribution is
necessarily Gaussian. The non-Gaussianity is caused by non-linear corrections.
Roughly speaking my
non-Gaussianity research can be divided into two different parts. The first
is the development of a formalism, {\em the long-wavelength formalism},
to compute the non-Gaussianity as produced
in inflation models, in particular multiple-field inflation models, and
the application of this formalism to study those models. The second is the
development of an estimator, {\em the binned bispectrum estimator},
to extract information about this inflationary
non-Gaussianity from cosmic microwave background (CMB) data in an optimal way,
its numerical implementation, and its application to the data of the Planck
satellite.

In this introductory chapter we will set the stage by going into some detail
about (multiple-field) inflation, linear cosmological fluctuations, and the CMB,
all of which are required as background knowledge for my research.
In chapter~\ref{NGinflsec} and associated appendices we will then describe the
first part of my post-PhD research as defined above (non-Gaussianity in
multiple-field inflation), while in chapter~\ref{NGCMBsec} and associated
appendices we will do the same for the second part (non-Gaussianity in the CMB).
Some conclusions are presented in chapter~\ref{conclusion}.

My papers on these subjects are of two different types: papers in which
the long-wavelength formalism (for the first part) and the binned bispectrum
estimator (for the second part)
are developed, and papers which contain applications or specific extensions
of the formalism/estimator. The latter type are relatively self-contained
papers (especially once the formalism/estimator is known), while the
former type corresponds to a succession of papers over many years, in which
the formalism/estimator was continuously further refined.
For the former type, just reading all the papers in order is not necessarily
the most efficient way to learn about them, as certain parts discussed in
older papers were improved upon in newer papers. Hence in this thesis
we provide a description of the formalism and estimator in their final state
and a summary of their derivation as synthesized from several papers, so that
it is in principle not necessary to look at the original papers (except for
the reader interested in all the details). On the other hand, for the
self-contained papers
of the second type (applications and specific extensions), we provide only a
brief summary in the main text of some of them, and then refer to the
appendices where these selected papers have been included almost verbatim.

\section{Inflation}
\subsection{Main ideas}

The observation by Hubble that the universe is expanding, naturally led
Lema\^{\i}tre and Gamow to propose the hot Big Bang theory, the idea that
the universe was smaller and hotter in the past.\footnote{There will be
  almost no references in this introductory chapter. In my opinion
  the basics of cosmology, single-field slow-roll inflation, and the CMB
  have now become part of the ``general knowledge of physics'' that is
  taught in undergraduate and graduate courses and can be found in any book
  on cosmology, see e.g.~\citep{Dodelson:2003ft,Peter:2013avv}.
  Just as one does not cite the original papers when using
  basic elements of general relativity or quantum mechanics, I feel there is no
  longer a need to cite the original paper by Guth when talking about
  inflation, for example. This chapter is just my way of presenting that
  general knowledge, to remind the reader who might not use it on a daily
  basis. There will only be some limited references in the parts of this chapter
  that I feel are not (yet) part of the general knowledge, like multiple-field
  inflation.} The confirmed predictions
of nucleosynthesis (a cosmic origin for the light elements by nuclear fusion
in the hot early universe) and the cosmic microwave background radiation (a
uniform radiation left over from the hot early stages of the universe) then
led to the general acceptance of this theory. However, some issues remained.
In the early 1980s it was discovered that several of these can be solved
by introducing a period of inflation.
In its simplest form inflation is a period in the very early universe
where the total energy density of the universe was dominated by the
potential energy of a scalar field, called the inflaton. This
potential energy is (almost) constant, either because the inflaton
field is trapped in a false vacuum behind a potential barrier, or
(in most models) because the potential is very flat and the field rolls
down very slowly. As a consequence this potential energy plays the
role of a cosmological constant and the universe expands
exponentially.

Such a brief period of enormous expansion in the very early universe
solves many problems of the standard Big Bang theory: why the universe
is so homogeneous on the largest scales despite the absence of causal
contact between the different regions (horizon problem), why the
universe is so flat (flatness problem), why we see no magnetic
monopoles and other topological defects left over from phase
transitions at very high energies (monopole/topological defect
problem), etc. These problems are all solved because of the very
different relation between the energy density $\rho$ and the scale factor $a$
for a cosmological constant ($\rho \propto a^0$) as compared to for matter
($\rho \propto a^{-3}$) or for radiation ($\rho \propto a^{-4}$). This
means for example that any topological defect ($\rho \propto a^{-1},a^{-2},a^{-3}$
depending on the type: domain walls, cosmic strings, or magnetic monopoles,
respectively) or the term corresponding to the curvature of the
universe ($\rho \propto a^{-2}$) loses energy density
faster than the dominant component during cosmological constant
domination, so that topological defects or spatial curvature will naturally
disappear during inflation. During radiation or matter domination it is the
opposite, which is the origin of those problems.
It turns out that all these
problems can be solved if inflation lasts for at least 60 e-folds (i.e.\ the
universe expands by at least a factor $\mathrm{e}^{60}$ during
inflation).\footnote{While 60 is
  the number generally quoted in the literature, the exact number depends on how
  much the universe expanded after inflation up until the current time, which
  in its turn depends on the energy scale of inflation and the reheating
  temperature. Given observational constraints, the value 60 is approximately
  the upper limit (of the minimum amount of inflation required), but it could
  easily be somewhat lower. For simplicity we will always use the number 60 in
  the text, but it should be understood that the actual number could be
  different.}
At least in slow-roll models of inflation this is generally not hard to achieve,
which means that as far as these issues are concerned, there is not much of
an observational constraint on inflation models.

While the energy scale of inflation is currently unknown, theoretical
arguments generally put it around the Grand Unified Theory (GUT) scale
of about $10^{15}$--$10^{16}$ GeV. These arguments are simple: in
order to solve for example the flatness problem, you want inflation to
happen as early as possible (otherwise a closed universe might already
have recollapsed before inflation could start, for example). On the
other hand, one does not want to rely on unknown quantum gravity
physics at the Planck scale ($10^{19}$~GeV). Hence the slightly lower
GUT scale is a better option, where we also have many theoretical
scalar field candidates.  In addition, observational constraints from
Planck and Bicep2-Keck on the tensor-to-scalar ratio (see
section~\ref{CMBpowspecsec}) have by now put an upper limit on the
inflationary energy scale of about $1.7\cdot 10^{16}$~GeV ($95\,\%$
CL) \citepalias{planck2018-10}. This corresponds to a tiny fraction of a second
after the Big Bang (about $10^{-37}$~s if we assume radiation domination
before inflation).

Soon after inflation was initially proposed by Guth, Linde,
Starobinsky and others in the early 1980s to solve the above problems
of the standard Big Bang theory, it was realized that inflation also
solves another important problem in cosmology: it provides a source
for the small density fluctuations that later through gravitational
collapse are the seeds of all the structure in the universe: galaxies,
clusters, etc. These small
fluctuations are observed very precisely in the cosmic microwave
background radiation, and so CMB observations allow us to put much
more stringent observational constraints on the various inflation
models than just having at least 60 e-folds of inflation.
In addition, the fact that inflation can solve all these problems,
including one for which it was not even initially designed, is
generally considered as a very strong point in favour,
and puts the bar high for any alternative theory.

The way inflation generates these fluctuations is as follows. The
small-scale quantum vacuum fluctuations that appear all the time
during inflation as in any other quantum field theory, and that would
in a Minkowski background simply disappear again as usual, now are
immediately stretched to very long wavelengths before they can
disappear again, because of the enormous expansion during
inflation. There is a characteristic length scale in an expanding
universe, which is the Hubble length $H^{-1}$, or in comoving units
$(aH)^{-1}$ ($H=\dot{a}/a$ being the Hubble parameter, where the overdot here,
unlike in the rest of this thesis, denotes a derivative with respect to cosmic
time).\footnote{In this whole thesis 
  natural units will be used, in which $c=\hbar=k_B=1$.}
During inflation this is also called the horizon, although that is
strictly speaking not correct (it would only be the event horizon in
an exact De Sitter universe). The equations for fluctuations during
inflation have a very different behaviour in what is generally called
the sub-horizon (or sub-Hubble) limit, where the fluctuations have
wavelengths much smaller than the Hubble length, than in the opposite
super-horizon limit. Where on sub-horizon scales we have
oscillations, on super-horizon scales we get a growing (or constant
with the proper choice of variables) mode and a decaying mode, and the
latter rapidly disappears. With the disappearance of the decaying
mode, the fluctuations lose their quantum nature (``squeezing'') and
effectively become classical fluctuations.

\begin{figure}
  \begin{center}
  \includegraphics[width=0.8\textwidth]{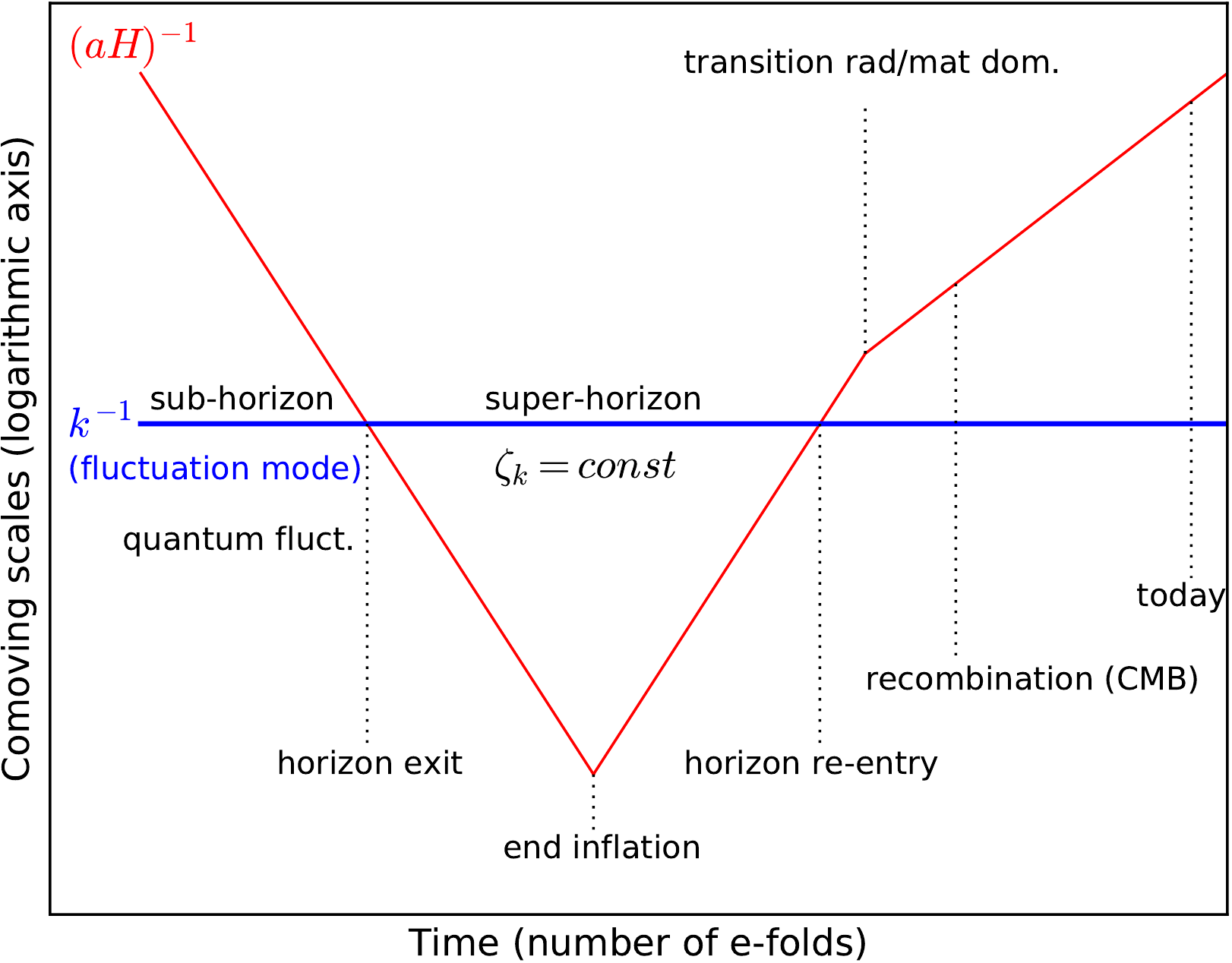}
  \end{center}
  \caption{A figure showing the horizon exit of a cosmological fluctuation
    with comoving wave number $k$ during inflation, and its subsequent horizon
    re-entry long after inflation.}
  \label{Fig:horizoncrossing}
\end{figure}

Later, long after inflation, when
these density fluctuations re-enter the horizon, they will start growing under
the influence of gravity and form the seeds of structure formation. Note that
in comoving units, where the wavelength (or wave number $k$) of a fluctuation
mode is constant, the comoving Hubble length $(aH)^{-1}$ rapidly decreases
during inflation ($\ln[(aH)^{-1}] \propto -t$, with $t$ the number of e-folds);
hence the horizon exit of the fluctuations. But after inflation, during
radiation ($\ln[(aH)^{-1}] \propto t$) and matter
($\ln[(aH)^{-1}] \propto \frac{1}{2} t$)
domination, it grows, so that the fluctuations will eventually re-enter
the horizon, see figure~\ref{Fig:horizoncrossing}. To be precise we will
now define the moment during inflation when $k=aH$ as the time of horizon exit
of that mode $k$, with $k>aH$ corresponding to the sub-horizon region and
$k<aH$ to the super-horizon region. In particular the time of horizon exit
of the modes that are observable in the CMB will play an important role
in what follows. As these are the largest wavelengths in our observable
universe, they left the horizon approximately 60 e-folds before the end
of inflation. That specific time will be denoted by $t_*$.

The proper way of treating linear fluctuations in single-field
inflation (with in particular gauge issues being non-trivial) was
worked out during the 1980s and culminated in the review paper by
Mukhanov, Feldman, and Brandenberger \citep{Mukhanov:1990me} in 1992. In the rest
of this work we will look in particular at multiple-field
inflation. Given that scalar fields are quite common in high-energy
theories, for example the superpartners of fermion fields in
supersymmetric theories or the scalar fields that naturally appear in
Kaluza-Klein compactifications of theories with extra dimensions, it
is a natural progression to extend inflation theory beyond a single
field and consider the impact of having more than one
scalar field.

\subsection{Background equations}
\label{sec:background}

Having discussed the main ideas of inflation in words, we will now give the
corresponding equations, as far as they will be required to understand
the research that is the subject of this thesis. The following subsections
are based on \citep{GNvT1,GNvT2,thesis,vT}.
An extensive list of references to earlier works on linear fluctuations
in multiple-field inflation can be found at the beginning
of section 4.1 of \citep{thesis}. With the exception of \citep{Sasaki:1995aw,
Nakamura:1996da}, those earlier works were generally limited to specific
models, usually with only two fields and standard kinetic terms.

On the matter side we assume a very general multiple-field inflation model
with an arbitrary number of scalar fields $\gf^A$ (where $A$ labels the
different fields) and a potential $W(\gf^A)$ with arbitrary
interactions. We also allow for the possibility of a non-trivial
field manifold with field metric $G_{AB}$.
The matter Lagrangean density then is
\be
\cL_\mathrm{m} = - \frac{1}{2} g^{\mu\nu} \der_\mu \gf^A G_{AB} \der_\nu \gf^B
- W(\gf^A).
\label{matterL}
\ee
On the metric side we take as background a standard flat
Friedmann-Lema\^{\i}tre-Robertson-Walker metric, describing a homogeneous
and isotropic universe in expansion with scale factor $a(t)$:
\be
\d s^2 = -N^2(t)\,\d t^2+a^2(t)\,\d \vc{x}^2.
\label{Bmetric}
\ee
As any curvature will rapidly become sub-dominant during inflation
(exactly the reason why inflation solves the flatness problem), taking a
flat metric to describe the last 70 or so e-folds of inflation that we are
interested in is a good approximation. The lapse function $N(t)$ encodes
our choice of time coordinate. The most common choices are cosmic/comoving
time ($N=1$), conformal time ($N=a$), and the number of e-folds
$t \equiv \ln a$ (with $N=1/H$).
In most of this work we will use the number of e-folds $t$ as time
coordinate, and denote derivatives with respect to this time with
overdots.\footnote{In the literature the use of the symbol $N$ for the
  number of e-folds is more common, with $t$ being reserved for cosmic time.
  As in this work cosmic time will not be used and $N$ is already used for
  the lapse function, we denote the number of e-folds by $t$.}
The Hubble parameter of the universe is denoted by $H(t)$.\footnote{In
  the case of other time coordinates, the expansion information of the
  universe is encoded in $a$ while $H$ is directly derived from it using
  $H\equiv\der_t a/(Na)$. However, when using the number of e-folds as time
  coordinate, $a$ is a trivial function, and the expansion information is
  encoded in $H$, which can in this case not be derived from~$a$.}

In terms of the number of e-folds, the background field equation for $\gf^A$
and the Friedmann equation for $H$ take the following form:
\be
\cD_t \dot{\gf}^A + (3-\ge)\dot{\gf}^A + \frac{G^{AB} W_B}{H^2} = 0, \qquad\qquad
H^2 = \frac{\gk^2 W}{3-\ge}.
\label{fieldeq}
\ee
Here $\gk^2 \equiv 8\pi G$ and the index on $W$ denotes a derivative with 
respect to the fields: $W_B = W_{,B} = \partial W / \partial \gf^B$. 
The quantity $\ge$ is a short-hand notation of which the physical interpretation
will be discussed in the next section. It is defined as
\be
\ge \equiv - \frac{\dot{H}}{H} = \frac{\gk^2}{2} \dot{\gf}^A G_{AB} \dot{\gf}^B
\equiv \frac{\gk^2}{2} \dot{\gf}^2
\label{defeps}
\ee
(where the second equality follows from the Friedmann equation for $\dot{H}$,
which we have not given explicitly here but which is easily deduced).
Here we have defined $\dot{\gf}$ as the length of the vector with components
$\dot{\gf}^A$.

In (\ref{fieldeq}) we also encounter for the first time the covariant
space-time derivative $\cD_\mu$ (of which we indicate the temporal component
as $\cD_t$):
\be 
\cD_\mu A^{A\ldots}_{B\ldots}=\der_\mu A^{A\ldots}_{B\ldots}
+ \lh \Gamma^A_{CD} A^{D\ldots}_{B\ldots} + \ldots
- \Gamma^D_{CB} A^{A\ldots}_{D\ldots} - \ldots \rh \partial_\mu \phi^C.
\label{def_cov_muder}
\ee
This is a derivative with respect to the space-time coordinates $x^\mu$, but
it is covariant with respect to the field manifold ($\Gamma^A_{BC}$ is the
Christoffel symbol defined from the field metric $G_{AB}$ in the usual way).
In fact, the quantity $A^{\ldots}_{\ldots}$ is a scalar with respect to 
space-time but can carry any number of field indices and derivatives with
respect to $\phi^A$. Explicitly for the $\cD_t\dot{\gf}^A$ in (\ref{fieldeq})
this means that it is equal to
$\ddot{\gf}^A + \Gamma^A_{BC} \dot{\gf}^B \dot{\gf}^C$.
In the case of a trivial field metric
all covariant derivatives are of course equal to normal derivatives.

\subsection{Slow-roll parameters and basis choice}
\label{sec:slowroll}

If the potential is almost flat and the field slowly rolls down,
certain terms in the equations will be small compared to others.
To quantify this we
can introduce a set of slow-roll parameters. It is important to keep
in mind that the introduction of these parameters is not yet an
approximation: the equations are still completely exact and the
slow-roll parameters can be considered as just a short-hand
notation. It only becomes an approximation (the slow-roll
approximation) if we then say that some of these
parameters are small and start neglecting certain terms. We will do
that in certain later sections, but not here.

The first slow-roll parameter is $\ge$ defined in (\ref{defeps}). It will be
small if the kinetic energy of the fields is small compared to their potential 
energy. The other slow-roll parameters are vectors in field space and can be 
defined as follows with $n\geq 2$ (in terms of the number of e-folds):
\be
\eta^{(n)A}\equiv\frac{1}{H^{n}\dot{\gf}}
\lh H \cD_t\rh^{n-1} \lh H \dot{\gf}^A \rh.
\label{defeta}
\ee
The most important ones are for $n=2$ (simply called $\eta^A$) and
$n=3$ (called $\xi^A$). For example, for $\eta^A$ the expression above becomes
\be
\eta^{A} = \frac{1}{\dot{\gf}}
\lh \cD_t \dot{\gf}^A - \ge \dot{\gf}^A \rh.
\label{defeta_detail}
\ee
The components of $\eta^A$ will be small if the corresponding field component
rolls slowly, with its acceleration being small compared to its velocity (see
below for a more precise statement).

As we have a multi-dimensional field space, we need a basis, and a very useful
orthonormal basis was first proposed in \citep{GNvT1}\footnote{A version of
  this basis in the case of two fields only and a trivial field metric was
  independently proposed in \citep{Gordon:2000hv}.} (with some minor
refinements later added in appendix A of \citep{TvT1}). 
In this basis the basis vectors are not constant, but defined with respect
to the field trajectory, which allows us to easily distinguish between
effectively single-field effects and truly multiple-field effects.
The basis is defined as follows (note that e.g.\ $\boldsymbol\eta$ is
the vector with components $\eta^A$). The first basis vector 
$\vc{e}_1$ is the unit vector in the direction of the field velocity. Next, the
direction of the basis vector $\vc{e}_2$ is given by the direction of that part
of the field acceleration that is perpendicular to $\vc{e}_1$. This
orthogonalization process is then continued with higher-order time derivatives,
until a complete basis is found.

Using the $\boldsymbol\eta^{(n)}$ defined
above we can define the basis vectors via an iterative procedure as
\be
\vc{e}_n \equiv \frac{\boldsymbol{\eta}^{(n)} - \sum_{i=1}^{n-1} 
  \eta^{(n)}_i \vc{e}_i}{\eta^{(n)}_n}
\label{basis}
\ee
for $n \geq 2$, with $\vc{e}_1 \equiv \boldsymbol{\dot{\gf}}/\dot{\gf}$
and $\eta^{(n)}_i \equiv
\vc{e}_i \cdot \boldsymbol{\eta}^{(n)}$, where the inner product is defined
using the metric $G_{AB}$. There is an arbitrariness in the
choice of sign of the basis vectors, which in the original definition was fixed
by choosing $\eta^{(n)}_n$ to be non-negative by taking the absolute value: 
\be
\eta^{(n)}_n \equiv \left | \boldsymbol{\eta}^{(n)} - \sum_{i=1}^{n-1} 
\eta^{(n)}_i \vc{e}_i \right |.
\qquad\qquad\mbox{(old definition)}
\ee
While being a perfectly valid choice analytically, this choice does mean that
certain basis vector components and slow-roll parameters make sudden sign flips
when one or more fields are oscillating, and that is hard to deal with
numerically. Hence we later \citep{TvT1} proposed a different choice for
$\eta^{(n)}_n$, which is identical except for the overall sign, and which
eliminates the sudden sign flips:
\be
\eta^{(n)}_n \equiv - \varepsilon_{A_1 \cdots A_n} e_1^{A_1} \cdots
e_{n-1}^{A_{n-1}} \eta^{(n) \, A_n},
\qquad\qquad\mbox{(new definition)}
\ee
where $\varepsilon$ is the fully antisymmetric symbol. From the fact that
$\boldsymbol{\eta} = \sum_{i=1}^n \eta^{(n)}_i \vc{e}_i$ it immediately follows
that
\be
\varepsilon_{A_1 \cdots A_n} e_1^{A_1} \cdots e_n^{A_n} = -1,
\ee
so that this choice means that the basis has a definite handedness. Note that in
the case where the fields do not oscillate, the two definitions have the same
overall sign (hence the choice of the minus sign). To have the expressions for
the time derivative of the basis vectors and of the $\eta^{(n)}_n$ unchanged, we
see that we also need the relation
\be
\varepsilon_{A_1 \cdots A_n} e_1^{A_1} \cdots e_{n-1}^{A_{n-1}} e_{n+1}^{A_n} = 0
\ee
to be satisfied. Then all results and expressions developed with this basis are
unchanged when going from the old to the new definition.

An interesting consequence of these relations, including the orthogonality 
relation
\be
\vc{e}_m \cdot \vc{e}_n = \delta_{mn},
\ee
is that for the cases of two and of three fields we have sufficient conditions
to write all basis vectors in terms of $\vc{e}_1$, without knowing anything 
about the dynamics. For two fields we have
\be
\label{e2e1}
\vc{e}_2 = (e_1^2, -e_1^1),
\ee
with $(e_1^1)^2+(e_1^2)^2=1$. The expressions for three fields can be found
in \citep{TvT1}.

Having defined this basis, we can now look at the components of the various
vectors in this basis. For example, for the $\eta^A$ slow-roll parameters we
define the parallel and perpendicular components as follows:
\be
\eta^\|\equiv \eta^{A}e_{1A},\qquad \eta^\perp \equiv \eta^{A}e_{2A}.
\end{equation}
For most vectors the use of the symbol $\perp$ only makes sense in the case
of two-field inflation, as otherwise there would be many perpendicular
directions. However, by construction $\eta^A$ only ever has components in
the $e_1$ and the $e_2$ directions, so that the symbol $\eta^\perp$ is
unambiguous.
The parameters $\eta^\|$ and $\eta^\perp$ will be small if the components
of the field acceleration parallel and perpendicular to the field
velocity, respectively, are small compared to the field
velocity.\footnote{This remark is exact when acceleration in terms of
  cosmic time is considered. When using the number of e-folds as time
  coordinate, as we do here, there is a correction term as seen in
  (\ref{defeta_detail}). However, that correction disappears for
  $\eta^\perp$.}  The parameter $\eta^\perp$ is quite fundamental to anything
concerning multiple-field inflation: as long as it is negligible we
are in an effectively single-field situation, but as soon as it
becomes significant we have truly multiple-field effects. With a trivial
field metric a non-zero $\eta^\perp$ automatically means a curved field
trajectory. However, it should be noted that in the case of a non-trivial
field metric even a straight field trajectory could have a non-zero $\eta^\perp$
due to the $\Gamma^A_{BC}$ terms in its definition.

In the context of the slow-roll approximation, $\ge, \getpa, \getpe$ are
called first-order slow-roll parameters, while the components of $\xi$ are
second-order slow-roll parameters. Now one might wonder about the fact
that we call $\getpe$ a slow-roll parameter, given that the actual 
slow-roll approximation (in the spirit of a field slowly
rolling along its trajectory) would only require $\ge$, $\getpa$
and higher-order parallel slow-roll parameters to be small, and say nothing
about the perpendicular parameters. However, in many cases where we need
to make the slow-roll approximation in order to make analytical progress,
it turns out that we also need to assume that the perpendicular parameters
are small (and even $\chi$, defined below). Hence we will call all these
parameters slow-roll parameters, and assume all of them to be small in the
slow-roll approximation (sometimes adding the word ``strong'' to be explicit).
It should also be noted that for the models that we have studied explicitly,
it is anyway not possible to have a large $\getpe$ while $\getpa$ stays small.

For later use we will define the following quantities:
\be
\tW_{A_1\ldots A_n}=
\lh\frac{\sqrt{2\ge}}{\gk}\rh^{n-2}\frac{W_{A_1\ldots A_n}}{3H^2}, \qquad\qquad
\tW_{m_1\ldots m_n}=\tW_{A_1\ldots A_n}e_{m_1}^{A_1}\cdots e_{m_n}^{A_n},
\label{defW}
\ee
where the $m$ indices denote the components of the basis. The $A_i$ indices
on $W$ denote multiple derivatives with respect to the fields $\gf^{A_i}$,
which in the case of a non-trivial field metric should be taken as
covariant derivatives (e.g.\ $W_{AB} = W_{,A;B} = W_{,AB} - \Gamma^C_{AB}W_{,C}$).
In order to 
distinguish explicit components of these two different quantities, indices like
1 and 2 will indicate components in the basis defined above (e.g.\ $\tW_{21}$),
while field indices like $\gf$ and $\gs$ (in the case of inflation with
fields $\gf$ and $\gs$) will be used to indicate components in terms of
the original fields (e.g.\ $\tW_{\gs\gs}$).
From the definition of the slow-roll parameters and using the field equation
(\ref{fieldeq}) and its derivative,
\be
\cD_t^2 \dot{\gf}^A + 3(1-\ge) \cD_t \dot{\gf}^A
-2\ge(3+\getpa) \dot{\gf}^A + \frac{G^{AB}W_{B;C}\dot{\gf}^C}{H^2} = 0,
\ee
one can show that
\begin{align}
  \getpa & =-3-3\tW_{1}, & \getpe & =-3\tW_{2},&&\nonumber\\
  \gxpa & =-3\tW_{11}+3\ge-3\getpa, & \xi_2 & =-3\tW_{21}-3\getpe
  & \xi_3 & = -3 \tW_{31}.
\label{srpareq}
\end{align}
We also introduce the parameter
\be
\chi \equiv \tW_{22}+\ge+\getpa.
\label{defchi}
\ee
Despite its similarity to the expressions for the components of
$\boldsymbol{\xi}$, the parameter
$\chi$ is a first-order slow-roll parameter and not a second-order one. The 
reason is that within the slow-roll approximation cancellations occur in
the right-hand sides of (\ref{srpareq}), making the slow-roll parameters on
the left-hand side one order smaller than the individual terms on the 
right-hand side. However, no such cancellation occurs in (\ref{defchi}).

We can compute the time derivatives of the basis vectors and find:
\be
\cD_t \vc{e}_i = \frac{\eta^{(i+1)}_{i+1}}{\eta^{(i)}_i} \, \vc{e}_{i+1}
- \frac{\eta^{(i)}_i}{\eta^{(i-1)}_{i-1}} \, \vc{e}_{i-1}
\ee
(where the second term should be omitted in the case $i=1$). Explicitly for
the first two basis vectors this means:
\be
\cD_t \vc{e}_1 = \eta^\perp \vc{e}_2,\qquad\qquad
\cD_t \vc{e}_2 = \frac{\xi_3}{\eta^\perp} \vc{e}_3 - \eta^\perp \vc{e}_1.
\ee
For the derivatives of the slow-roll parameters we get:
\be
\begin{split}
\dot{\ge}&= 2\ge(\ge+\getpa), \qquad
\dot{\eta}^{\parallel} = \gxpa+(\getpe)^{2}+(\ge-\getpa)\getpa,\qquad
\dot{\eta}^{\perp} = \xi_2 +(\ge-2\getpa)\getpe,\\
\dot{\chi}&=\ge\getpa+2\ge\chi-(\getpa)^{2}+3(\getpe)^{2}+\gxpa+\frac{2}{3}\getpe\xi_2+\tW_{221}+2\frac{\xi_3}{\getpe}\tW_{32},\\
\dot{\xi}^{\parallel}&=-3 \tW_{111}+2\getpe\xi_2+(2\ge-3)\gxpa+9\ge\getpa+3(\getpa)^{2}+3(\getpe)^{2},\\
\dot{\xi}_2&=-3 \tW_{211}-\getpe\gxpa+(2\ge-3)\xi_2+9\ge\getpe+6\getpa\getpe-3\getpe\chi + \frac{(\xi_3)^2}{\getpe},\\
\dot{\xi}_3&=-3 \tW_{311}+(2\ge-3)\xi_3 -3\getpe\tW_{32}- \frac{\xi_2\xi_3}{\getpe},
\end{split}
\label{srderivatives}
\ee
or more generally
\be
\cD_t \boldsymbol{\eta}^{(n)} = \boldsymbol{\eta}^{(n+1)}
+ ((n-1)\ge-\getpa)\boldsymbol{\eta}^{(n)}.
\ee

\subsection{Linear perturbations}

As mentioned at the beginning of the section, it is in particular the fact that
inflation also explains the origin of the cosmological fluctuations, for which
it was not explicitly constructed, that has made the theory so convincing. In
addition it is these fluctuations that allow inflation models to be
observationally constrained. So we will now add small inhomogeneous fluctuations
to the homogeneous background. As observations tell us these fluctuations were
very small initially, linearizing all equations is a good first approximation,
and the resulting equations and solutions will be discussed in this subsection
(as this is only an introduction and not the main subject of this thesis, only
a summary is presented, for details see \citep{GNvT1,GNvT2,thesis,vT} as in the
previous subsections, or \citep{Mukhanov:1990me} for the single-field case).
Of course, with CMB observations having become as precise as they are with
WMAP and Planck, we can now put constraints on the even smaller second-order
perturbations. That is the main topic of this thesis and will be treated in the
next chapters.

In this subsection only we will not use the number of e-folds as time
coordinate, but conformal time $\tau$ defined by a lapse function $N=a$. In
terms of conformal time the scale factor is an overall multiplicative factor
of the background metric (i.e.\ the metric is a conformal transformation of
the Minkowski metric, hence the name). It turns out that in particular the
sub-horizon equations will be easier to solve in terms of this time coordinate.
We denote a conformal time derivative by a prime.

The scalar fields are perturbed as follows:
\be
\gf^A_\mathrm{full}(\tau,\vc{x}) = \gf^A(\tau) + \gd\gf^A(\tau,\vc{x})
\ee
with $\gd\gf \ll \gf$. For the metric we write:
\bea{
g_{\mu\nu}^\mathrm{full}(\tau,\vc{x})
= \: & a^2 \left( \begin{array}{cc}
-1 & 0 \\ 0 & \delta_{ij} \end{array}\right)
+ a^2 \left( \begin{array}{cc}
-2\Phi & B_{,j} \\ B_{,i} & 2(E_{,ij}-\Psi\delta_{ij}) \end{array}\right)
\nonumber\\
& + a^2 \left( \begin{array}{cc}
0 & S_j \\ S_i & F_{i,j}+F_{j,i} \end{array}\right)
+ a^2 \left( \begin{array}{cc}
0 & 0 \\ 0 & h_{ij} \end{array}\right).
}
Here the first term is the background metric, the second contains the
scalar perturbations encoded in four scalar functions $\Phi, \Psi, E, B$,
the third contains the vector perturbations encoded in two divergenceless
vectors $\vc{S}, \vc{F}$, and the last term contains the tensor
perturbations given by the symmetric transverse traceless tensor $h_{ij}$.
Two of the scalar and two of the vector degrees of freedom are gauge degrees
of freedom related to the choice of coordinates. It is not hard to construct
combinations of the scalar degrees of freedom (including the scalar
field perturbation $\gd\gf^A$) that are invariant under coordinate
transformations, and similarly for the vector degrees of freedom (the tensor
$h_{ij}$ is already gauge-invariant). It turns out that working with these
gauge-invariant combinations is equivalent to choosing the longitudinal
($B=E=0$) and vector ($F_i=0$) gauges, so that is what we will do.

Up to linear order the scalar, vector and tensor perturbations decouple. We
will not consider vector perturbations in this thesis, as they are absent
if there is no vector matter source, and we will only consider scalar fields as
matter sources. Moreover, even if vector perturbations were sourced at some
point, they decay afterwards. Tensor perturbations (gravitational waves), on
the other hand, are always present, even in the absence of tensor matter
sources, as the two tensor degrees of freedom represent the two physical
degrees of freedom of the metric and are hence directly sourced by metric
quantum fluctuations. However, there is no difference between the tensor
perturbations in single-field and in multiple-field inflation, and so they
are not so interesting from our point of view, and we will rarely mention
them. Our main interest will be the scalar perturbations.

By plugging the above expressions for the field and the metric into the Einstein
equations, we obtain equations of motion for the various quantities. A first
result (that follows from the off-diagonal $ij$ part of the Einstein equation)
is that $\Phi=\Psi$. Hence there is in the end only one metric scalar degree
of freedom. And even that is not physical, but is only present because it is
sourced by one of the physical scalar degrees of freedom of the inflaton field
(the $\vc{e}_1$ component of the vector $\gd\gf^A$ to be precise). Hence it
makes sense to combine the metric and matter scalar quantities into a single
quantity, and it turns out that with the following combination the equations
of motion simplify significantly:
\be
q^A \equiv a \lh \gd\gf^A + \frac{\Psi\gf'}{\cH} e_1^A \rh,
\label{def_q}
\ee
where $\cH \equiv a'/a = aH$. This combination is gauge-invariant (even
if $\gd\gf^A$ and $\Psi$ would not have been made gauge-invariant individually).
It is the multiple-field generalisation of the single-field Sasaki-Mukhanov
variable. It satisfies the following equation of motion (after switching to
spatial Fourier modes $q^A_\vc{k}(\tau)$):
\be
{\cD_\tau^2} q^A_\vc{k} + \lh k^2 + \cH^2 \Omega^A{}_B \rh q^B_\vc{k} = 0 
\label{q_eq}
\ee 
with the ``mass matrix''
\be 
\Omega^A{}_B \equiv \frac{W^A{}_B}{H^2}
-\frac{2\ge}{\gk^2} R^A{}_{CDB} \, e_1^C e_1^D - (2-\ge)\delta^A{}_B
-2\ge\left[(3+\ge)e_1^A e_{1\, B}
+ e_1^A \eta_B + \eta^A e_{1\, B}\right],
\ee 
where as before $W^A{}_B$ should be read as a covariant derivative:
$W^{,A}{}_{;B} = G^{AC}(W_{,C} - \Gamma^D_{BC} W_{,D})$, and $R^A{}_{CDB}$ is the
curvature tensor of the field manifold.

In the sub-horizon limit, where $k^2 \gg \cH^2$, the second term between the
parentheses of (\ref{q_eq}) can be neglected, and the equation looks like a
simple harmonic
oscillator. Moreover, at the level of the Lagrangean we also find the proper
normalization factor for a harmonic oscillator. The covariant
derivative can be dealt with by rewriting $q^A$ as a vector in the special
basis defined in the previous subsection, $q_m = e_{m\,A} q^A$.
The kinetic term $\cD_\tau q^A_\vc{k} G_{AB}
\cD_\tau q^B_\vc{k}$ in the Lagrangean then becomes
$(q'_{m\,\vc{k}} + \cH Z_{mn} q_{n\,\vc{k}})(q'_{m\,\vc{k}} + \cH Z_{mp} q_{p\,\vc{k}})$.
Here the anti-symmetric matrix $Z_{mn}$ is defined as
$Z_{mn} \equiv e_{m\, A} \cD_t e_n^A = e_{m\, A} \cD_\tau e_n^A / \cH$, which means
in particular that $Z_{21} = -Z_{12} = \eta^\perp$. 
Hence we know how to quantize $q^A$, and the initial conditions that would have
been undetermined in a classical setting are now almost completely determined
by 1) imposing the standard commutation relation between $q^A$ and its
canonical momentum, and 2) choosing the vacuum as the minimum of the
inflationary Hamiltonian. The resulting sub-horizon solution is:
\be
q_{m\,\vc{k}}(\tau) = \frac{1}{\sqrt{2k}} \, U_{mn}(\tau) a^\dag_{n\,\vc{k}}
+ \mathrm{h.c.}
\ee
where $a^\dag_{n\,\vc{k}}$ and $a_{n\,\vc{k}}$ are the standard quantum creation
and annihilation operators and h.c.\ denotes the Hermitian conjugate. The
matrix $U_{mn}$ is a unitary matrix that contains the time dependence
$\exp(-\mathrm{i}k\tau)$, a time-dependent rotation matrix due to the $Z_{mn}$
terms in the Lagrangean, and an undetermined constant unitary matrix that is
the only part that would still need to be determined from the initial
conditions. However, its explicit form is unimportant, as it
will drop out from all observables computed from $q_m$.

In the super-horizon region ($k^2 \ll \cH^2$) it is the $k^2$ term in
(\ref{q_eq}) that can be neglected. It turns out that instead of oscillating
solutions we have a growing and a (very rapidly) decaying solution here. In
the following we will always neglect the decaying mode on super-horizon scales
(although it must be taken into account for a proper matching to the
sub-horizon solution).
In between the sub-horizon and the super-horizon regions there is a transition
region where $k \sim \cH$. If we take this region small enough, we can assume
the slow-roll parameters to be constant and then we can solve the differential
equation with constant coefficients exactly in terms of a Hankel function
(of a matrix-valued order). Matching this solution to the sub-horizon solution
is simple using the asymptotic expansion of the Hankel function. In \citep{GNvT2}
this was worked out in detail up to next-to-leading order in slow roll, and it
was matched to the solution in the super-horizon region to finally obtain
the complete solution at the end of inflation. However, in the next chapter
in this thesis another way of obtaining the super-horizon solution that also
works beyond linear order will be detailed, and the only input that it
requires is the linear leading-order slow-roll solution at (or rather slightly
after) horizon crossing. Hence in this introductory section we will only give
that expression (for the growing mode):
\be
q_{m\,\vc{k}} = \frac{a H_*}{\sqrt{2k^3}} \, a^\dag_{m\,\vc{k}}
+ \mathrm{h.c.}
\label{q_lin_sol}
\ee
where all unitary factors that have no impact on the observables have been
omitted.

In the super-horizon region it is more convenient to work with a different
variable than $q^A$, namely $\gz^A$ defined as
\be
\gz^A = - \frac{\gk}{a \sqrt{2\ge}} \, q^A.
\label{def_zeta}
\ee
It is the multiple-field generalisation of the single-field curvature
perturbation\footnote{In the single-field case, and in a gauge where the
  field perturbation is zero (e.g.\ uniform energy density gauge), it is easy
  to see from (\ref{def_q}) and (\ref{def_zeta}) that $\zeta=-\Psi$ (using
  also that $\sqrt{2\ge}=\gk\gf'/\cH$).
  And $\Psi$ is related to the intrinsic spatial curvature on hypersurfaces
  of constant conformal time as $^{(3)}R = 4(\nabla^2\Psi)/a^2$, hence the
  name curvature perturbation for $\gz$.},
and it is just a simple rescaling of $q^A$. When taking components
of $\gz^A$ in the basis defined in the previous subsection, the $\gz_1$
component is called the adiabatic (or curvature) mode, which is the only
one present in single-field inflation, while the $\gz_m$ with $m\geq 2$ are
called isocurvature (or entropy) modes. The latter can only exist in the case
of inflation with multiple fields. The main convenience of $\gz^A$ is that
its adiabatic component satisfies a very simple equation on super-horizon
scales after the decaying solution has disappeared:
\be
\gz'_1 = 2 \cH \eta^\perp \gz_2.
\label{zeta1prime_eq}
\ee
From this equation we can read off the well-known result that in single-field
inflation the adiabatic mode $\gz_1$ is constant on super-horizon scales
(after the decaying solution has disappeared). In multiple-field inflation,
on the other hand, $\gz_1$ can evolve on super-horizon scales, being
sourced by $\gz_2$ (and only $\gz_2$, independent of how many isocurvature
modes there are). The equations for the isocurvature modes $\gz_m$ with
$m\geq 2$ are more complicated, but have the property that they are independent
of the adiabatic mode. To conclude we rewrite the linear leading-order
slow-roll solution at horizon crossing (\ref{q_lin_sol}) in terms of
$\gz^A$, which will be an input for the equations in the next chapter:
\be
\gz_{m\,\vc{k}} = - \frac{\gk H_*}{2 \sqrt{k^3 \ge_*}} \, a^\dag_{m\,\vc{k}}
+ \mathrm{h.c.}
\label{zeta_lin_sol}
\ee
where by convention we have kept the minus sign, even though other unitary
factors were omitted as they have no impact on any observables.

We conclude this section with a few more remarks regarding the isocurvature
modes.
In the universe after inflation, the total isocurvature mode is generally
defined in terms of the total pressure and the total energy density and their
perturbations as (see e.g.~\citep{thesis,vT})
\be
S = \frac{1}{4} \frac{\delta p - c_s^2 \delta\rho}{p - c_s^2 \rho},
\ee
where $c_s^2$ is the sound speed squared, defined as $c_s^2 = p'/\rho'$. The
total isocurvature mode is the specific combination of all the isocurvature
modes that is the source term for the adiabatic mode. In the case of
only photons (with $p_\gamma=\rho_\gamma/3$) and cold dark matter (with
$p_c=0$), it reduces to
\be
S = \frac{\delta\rho_c}{\rho_c} - \frac{3}{4} \frac{\delta\rho_\gamma}{\rho_\gamma}.
\ee
Alternatively this can be written in terms of number densities and their
perturbations as $S=\delta n_c/n_c - \delta n_\gamma/n_\gamma$, showing that
isocurvature perturbations involve relative perturbations in particle
number densities between species (while the adiabatic mode can be viewed as
perturbations in the total energy density or total particle number density,
although this statement, unlike the one for the isocurvature perturbations, is
gauge-dependent). It is clear that this $S$ must be related to the quantity
$\zeta_2$ defined above. It turns out that the relation is
\be
S = - \frac{1}{2} \frac{\ge}{\ge+\getpa} \getpe \zeta_2,
\label{Szeta2rel}
\ee
where we have expressed everything in terms of slow-roll parameters to make the
identification during inflation simpler (alternatively, $\ge/(\ge+\getpa)$
can be written as $(p+\rho)/(p-c_s^2\rho)$).

\section{The cosmic microwave background radiation}

\subsection{Introduction}

Initially the universe was hot enough for protons and electrons, after
they had formed, to remain free (ionized). As free electrons scatter
photons very efficiently (Thomson scattering), photons had a very short mean
free path in the
early universe. However, when the temperature in the universe decreased to
about 3000~K, which happened about 380\,000 years after the Big Bang, protons
and electrons combined into neutral hydrogen atoms, which is called
recombination.\footnote{This is a
  much lower temperature than one would naively expect from the ionisation
  energy of hydrogen of 13.6~eV. The reason is that there are about $10^9$
  times more photons in the universe than protons and electrons. Hence even
  just the high-energy tail of a lower temperature photon distribution can
  keep all hydrogen ionized.}
Suddenly the universe became effectively transparent to
photons. Looking out into the universe, and hence back in time because
of the finite speed of light, we will hit this surface in time before
which the universe was opaque. Hence it seems as if we are surrounded
by a spherical surface in space beyond which we cannot look, and which
radiates as a black body. That surface is called the last-scattering
surface, and the radiation is the cosmic microwave background
radiation (CMB), where ``microwave'' reflects the fact that through
the expansion of the universe those photons have nowadays a
temperature of only 2.725~K, which corresponds to microwaves.

\begin{figure}
  \begin{center}
  \includegraphics[width=0.7\textwidth]{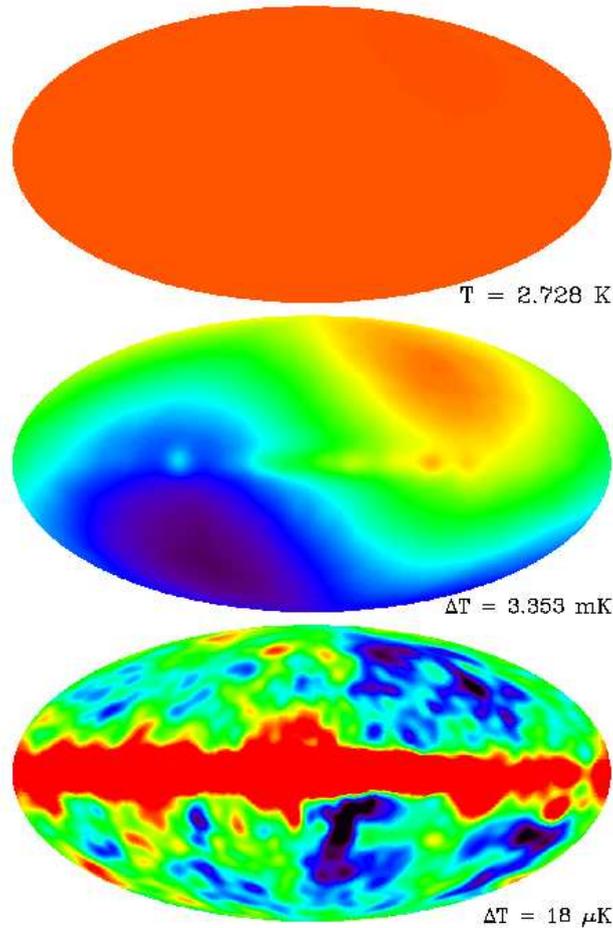}
  \end{center}
  \caption{The CMB temperature monopole, dipole, and other multipoles, as
    measured by the COBE satellite. Figure from
    \url{https://lambda.gsfc.nasa.gov/product/cobe/}.}
  \label{Fig:COBE}
\end{figure}

\begin{figure}
  \begin{center}
    \includegraphics[angle=90,width=0.9\textwidth]{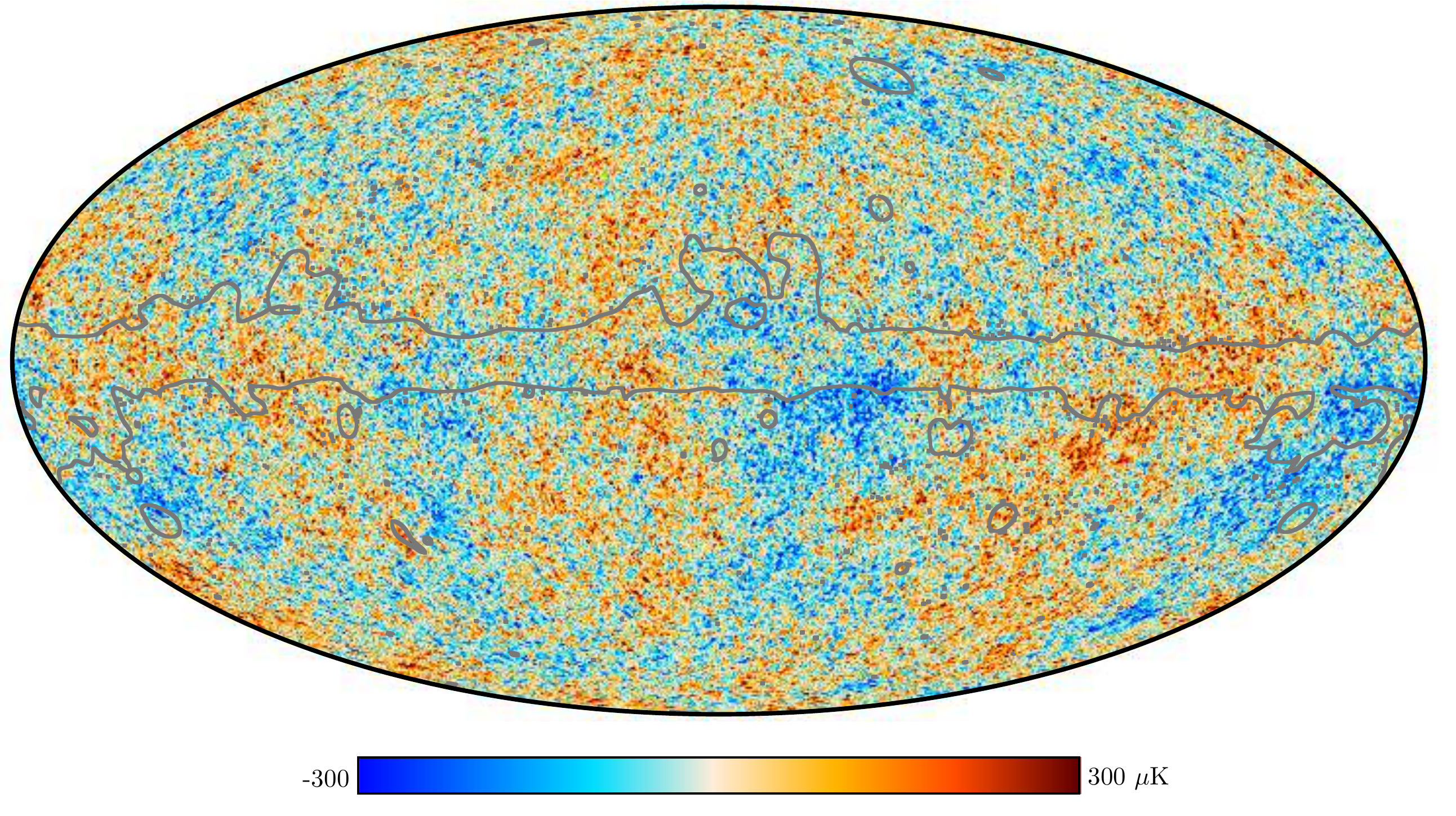}
  \end{center}
  \caption{The CMB temperature fluctuations with $\ell \geq 2$ as measured by
    the Planck satellite (2018 release). The gray outline shows the extent
    of the confidence mask due to (mainly) galactic foregrounds.
    Figure from \citepalias{planck2018-01}.}
  \label{Fig:Planck}
\end{figure}

As illustrated quite nicely by figure~\ref{Fig:COBE} from the COBE satellite,
at a first glance the
CMB appears to be completely isotropic. This is actually one of the motivations
for inflation, as in a pure radiation/matter-dominated universe the CMB
would come from regions that have never been in causal contact, and so it
would be very unlikely for the CMB to be this isotropic (this is called the
horizon problem). Looking more
closely, we see that at the level of $10^{-3}$~K there is a dipole component,
caused by the movement of the Earth with respect to the Hubble flow.
Finally, at the level of $10^{-5}$~K we see
fluctuations in all multipoles $\ell$ (in terms of a spherical harmonic
decomposition).
Of course the latter are now measured at much higher resolution by the Planck
satellite, see figure~\ref{Fig:Planck}.
These are the fluctuations that were presumably generated by
inflation, as explained in the previous section, and that at later times
formed the seeds of structure formation by gravitational collapse. Because
these fluctuations are so small, perturbation theory works very well, which
is one reason why the analysis of CMB data is simpler than the analysis of
large-scale structure data. Linear perturbation theory is a very good
approximation, but because of the high precision of current CMB data we even
have access to second-order corrections, the so-called non-Gaussianities that
will be at the centre of this thesis.

As will be explained in the next section, the primordial fluctuation power
spectrum in terms of $\zeta$ has a very simple, almost flat, shape. For
fluctuations that were still super-horizon at recombination, this translates
into a flat CMB temperature power spectrum as well, as there is no evolution
on super-horizon scales in the standard picture. For fluctuations that had
already re-entered the horizon before recombination, and started to evolve
again, the situation is more complicated. Before recombination
the universe was filled with a plasma of tightly coupled photons and baryons,
which evolves in a landscape of gravitational wells and hills as described by
the fluctuations of $\zeta$. The competition between gravity pulling the
plasma into the gravitational wells and the radiation pressure of the photons
pushing the plasma out of the wells then creates acoustic oscillations in
the plasma. In the CMB we see a snapshot of those oscillations at the time
of recombination. Certain specific wavelengths will have been right at a
maximum or a minimum of the oscillation. Those correspond to peaks in the
power spectrum (as the power spectrum is roughly speaking the square of the
fluctuations, both a maximum (compression) and a minimum (rarefaction)
correspond to a peak). Other wavelengths will have been close to the mean
value of the oscillation, which correspond to the troughs of the power spectrum.
The position and relative height of the peaks is quite sensitive to the
various cosmological parameters (like the amount of dark matter and baryons
in the universe, and the fact if the universe is open, closed or flat). This
is the reason why the CMB is such a gold mine for precision cosmology.

So in the CMB temperature power spectrum (see figure~\ref{Fig:CMBpowspec} in
the next section) we first see an approximately
flat region (called the Sachs-Wolfe plateau) at the smallest multipoles,
corresponding to the largest scales, which were still outside the horizon
at recombination. Because of late-time effects of the dark energy on the
gravitational potential through which the CMB photons travel towards us,
this plateau has
been slightly tilted (integrated Sachs-Wolfe effect, ISW). Then, at
intermediate multipoles we see a series of peaks, caused by the acoustic
oscillations of those scales that re-entered the horizon during the plasma
era before recombination. Finally, at the largest multipoles, we see
an exponential decay, called Silk damping. This is a combination of two
effects that smear out the fluctuations on the smallest scales: the fact that
recombination was not instantaneous, so that the last scattering surface has
a finite thickness, and the fact that before recombination the mean free
path of the photons was not exactly zero, so that they could diffuse out of
gravitational wells (sometimes in the literature only the latter
effect is called Silk damping).

\begin{figure}
  \begin{center}
  \includegraphics[angle=90,width=0.9\textwidth]{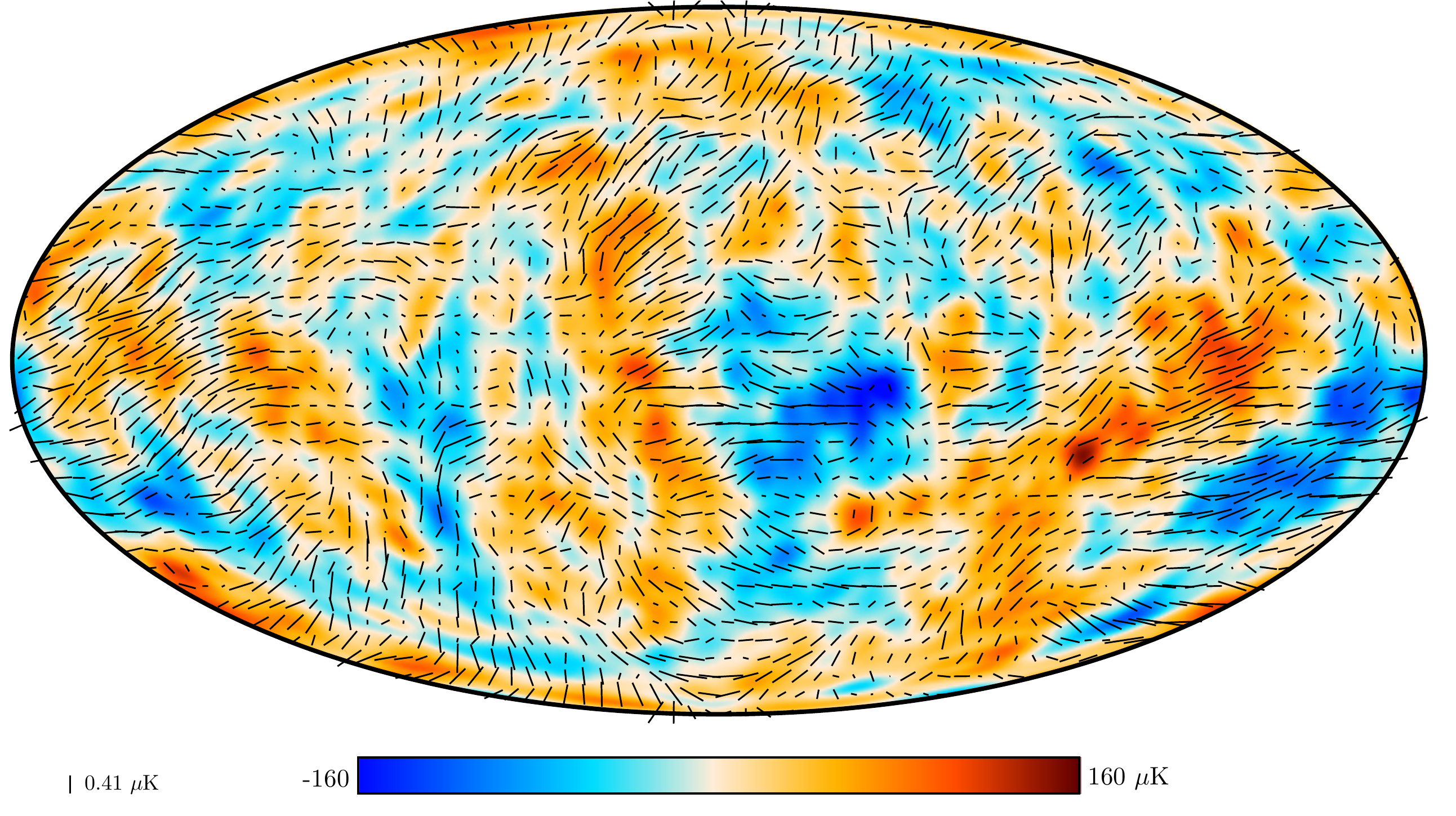}
  \end{center}
  \caption{The 2018 Planck map of the polarized CMB anisotropies, shown as rods
    whose direction and length represent the direction and amplitude of the
    polarized CMB. The coloured background is the map of temperature
    anisotropies, smoothed to 5 degrees. Figure from \citepalias{planck2018-01}.}
  \label{Fig:PlanckPol}
\end{figure}

\begin{figure}
  \begin{center}
  \includegraphics[width=0.5\textwidth]{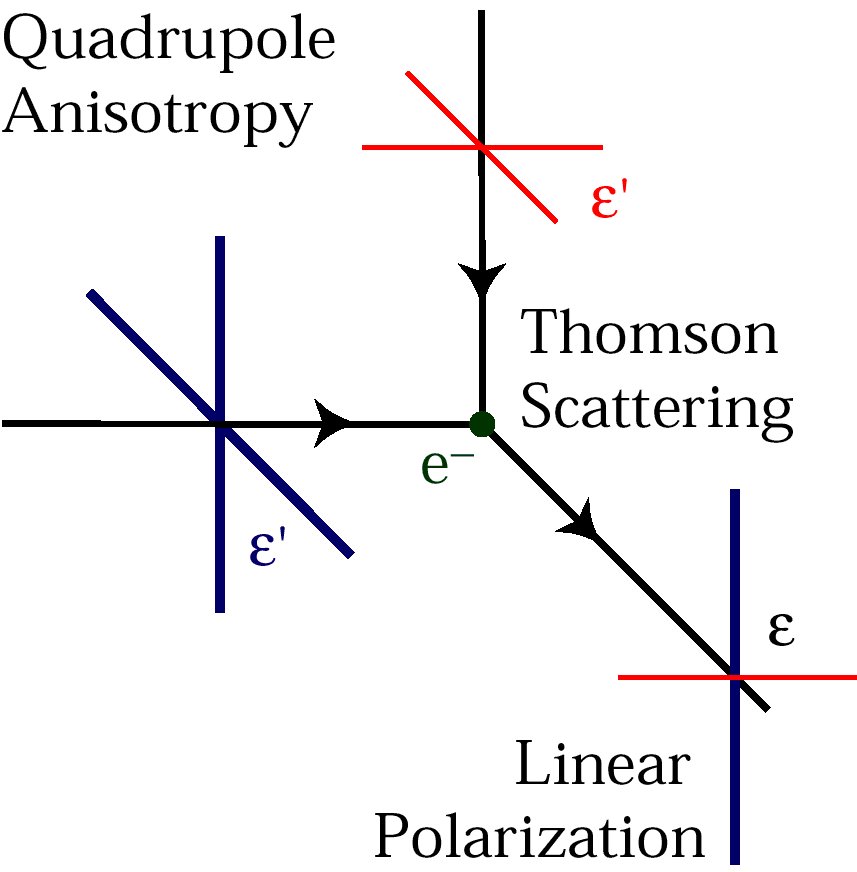}
  \end{center}
  \caption{A figure showing how a quadrupole anisotropy in the incoming
    radiation at last scattering leads to a linear polarization of the CMB.
    Figure from \citep{Hu:1997hv}.}
  \label{Fig:CMBpolarization}
\end{figure}

The CMB is also polarized, see figure~\ref{Fig:PlanckPol}, and its
polarization provides us with additional
information. Thomson scattering of an electromagnetic wave off an electron
creates linear polarization perpendicular to both the incoming and outgoing
wave vectors. If the incoming radiation is isotropic, then a second,
perpendicular, incoming wave will provide the other linear polarization
direction, so that the outgoing wave is again unpolarized. However, if the
incoming radiation is anisotropic, in particular if it has a quadrupole moment,
then the two linear polarization directions will not have the same amplitude,
so that a net linear polarization remains for the outgoing wave, see
figure~\ref{Fig:CMBpolarization}. This is how the temperature anisotropies
at the last-scattering surface lead to linear polarization of the CMB. Note
that it is only the last scattering that is relevant, the effect from earlier
scattering will be averaged out by consecutive scatterings. Normally
linear polarization is described by the $Q$ and $U$ Stokes parameters. As
these are not invariant under rotations, in CMB physics we often use an
alternative description in terms of $E$ (gradient) and $B$ (curl) polarization,
which are invariant under rotations at the price of being non-local (see
next section for precise definitions). Another advantage of the description
in terms of $E$ and $B$ is that scalar perturbations can only create
$E$-polarization, not $B$. Hence the primordial $B$-polarization signal will
be a clear probe of the inflationary tensor perturbations.
In the end the CMB provides us with in principle four power spectra:
the temperature power spectrum $TT$, the $E$-polarization power
spectrum $EE$, the $B$-polarization power spectrum $BB$, and the $TE$
cross spectrum (because of parity conditions the $TB$ and $EB$ cross
spectra are zero). The Planck satellite has given us very precise measurements
of the $TT$, $TE$, and $EE$ spectra. A measurement of the $BB$ spectrum is
still waiting for a future mission.

Finally, let us mention that in addition to the CMB anisotropies that we
consider here, there exist also CMB spectral distortions. CMB anisotropies
assume that the CMB spectrum is a perfect black-body spectrum, just with
a slightly different temperature in each direction. Spectral distortions
on the other hand are deviations from a black-body spectrum. The COBE satellite
confirmed that the CMB spectrum is a black-body spectrum to high precision.
However, theory does predict deviations at some very small level. As
no CMB mission since COBE has had a spectrometer on board, and no
proposed CMB spectrometer mission has been accepted by the various space
agencies to this date, there are currently no high-precision CMB spectral
distortion data, and we will not consider spectral distortions in this thesis.

\subsection{Power spectrum}
\label{CMBpowspecsec}

As the inflationary fluctuations are quantum fluctuations, they are
random. Hence we cannot predict the precise distribution of temperature
fluctuations (or galaxies) on the sky. What inflation does predict are
statistical properties of that distribution, in particular its various
correlation functions. The most important of those is the two-point
correlation function, or its Fourier transform, the power spectrum. If we
look at the CMB on the spherical last scattering surface, the power spectrum
is instead the spherical harmonic transform of the angular two-point
correlation function. If the distribution were Gaussian, then all information
would in fact be contained in the power spectrum, with all odd-point correlation
functions being zero while all even-point correlations functions could be
expressed in terms of the power spectrum. The distribution would be Gaussian
if the generation and evolution of the fluctuations were linear, and as we have
said before, even though that is not true, it is a good approximation because
the fluctuations are so small.

The power spectrum $P_\gz(k)$ of the linear adiabatic curvature perturbation
$\zeta_1$ is defined such that
\be 
\langle 0 | \gz_1(\vc{x}) \gz_1(\vc{y}) | 0 \rangle
  = \int \frac{\d k}{k} \frac{\sin(k|\vc{x}-\vc{y}|)}
  {k|\vc{x}-\vc{y}|}  P_\gz(k),
\ee
or alternatively such that\footnote{Various definitions of the power spectrum
  circulate in the literature. Instead of this so-called dimensionless power
  spectrum, people also consider the so-called dimensionful power spectrum,
  defined as in (\ref{defPzeta}) but without the factor $2\pi^2/k^3$.
  In that case for example
  equation (\ref{Cl_gl}) would have a factor $2k^2/\pi$ instead of $4\pi/k$.
  In addition, the chosen normalization of Fourier transforms also plays
  a role. We have chosen to define both the Fourier transform and its
  inverse with a factor $(2\pi)^{-3/2}$. If instead one takes the Fourier
  transform to have nothing and the inverse transform to have $(2\pi)^{-3}$,
  then an additional factor $(2\pi)^3$ must be introduced on the right-hand
  side of (\ref{defPzeta}), without any impact on equation (\ref{Cl_gl})
  however.}
\be
\langle 0 | \gz_1(\vc{k}_1) \gz_1(\vc{k}_2) | 0 \rangle
  = \frac{2\pi^2}{k_1^3} \delta^3(\vc{k}_1 - \vc{k}_2) P_\gz(k_1).
\label{defPzeta}
\ee
Inserting the solution (\ref{zeta_lin_sol}) gives us the following power
spectrum on super-horizon scales:
\be
P_\gz(k) = \frac{\gk^2 H_*^2}{8 \pi^2 \ge_*}.
\label{Pzetasf}
\ee
Note however that (\ref{zeta_lin_sol}) was only the solution for $\zeta$ at
horizon-crossing. Hence this expression for the power spectrum is only
correct if $\zeta_1$ did not further evolve on super-horizon scales. This would
be the case in single-field inflation, but not in multiple-field inflation,
where a correction term has to be added. This will be worked out in
section~\ref{solpowspecbispecsec}.

The dependence on $k$ of $P_\gz(k)$ only comes via the dependence of $H$
and $\ge$ on $t_*$, the time of horizon crossing of the mode $k$ defined
by $k=aH$. As both $H$ and $\ge$ evolve slowly in slow-roll inflation, this
is a weak dependence. Because of this weak dependence we can expand $P_\gz(k)$
around a pivot scale $k_0$ as
\be
P_\gz(k) = A_s(k/k_0)^{n_s-1}
\ee
with the amplitude
\be
A_s \equiv P_\gz(k_0) = \frac{\gk^2 H_0^2}{8\pi^2 \ge_0}
\ee
and spectral index
\be
n_s-1 \equiv \left.\frac{\partial \ln P_\gz}{\partial \ln k}\right|_{k=k_0}
= -4\ge_0 - 2\eta^\parallel_0
\label{nssf}
\ee
(the subscript $s$ refers to scalar, as this is the power spectrum of the
scalar perturbations). The final equality for $n_s$ can be found using the
definition of $\ge$ and its derivative (\ref{srderivatives}). Again, it must
be stressed that these expressions are only valid for single-field
inflation, the corrections in the multiple-field case will be treated
in section~\ref{solpowspecbispecsec}. So in this way we have reduced the
power spectrum from a full function of $k$ to just two numbers (which can be
extended with higher-order derivatives at the pivot scale, the first of
which is called the running of the spectral index).

In addition to scalar fluctuations, inflation also predicts the presence
of tensor fluctuations. The derivation of the tensor power spectrum is
very similar to, but simpler than, the one of the scalar power spectrum, and
we will only give the final result here:
\be
P_t(k) = \frac{2 \gk^2 H_*^2}{\pi^2}.
\ee
This result is independent of the number of scalar fields, and hence the
same for single-field and multiple-field inflation. As for the scalar
power spectrum we can expand it around a pivot scale and express it in terms
of just two numbers, the tensor amplitude and the tensor spectral index:
\be
A_t = \frac{2 \gk^2 H_0^2}{\pi^2}, \qquad
n_t = -2\ge_0
\ee
(for historical reasons the tensor spectral index is defined without the $-1$
that the scalar spectral index has). The main difference between the scalar and
the tensor amplitudes is that $A_t$ only depends on $H_0$ and not on $\ge_0$,
which means that a measurement of $A_t$ would directly give the inflationary
energy scale. Instead of $A_t$ the tensor-to-scalar ratio $r$ is more
commonly used:
\be
r \equiv \frac{A_t}{A_s} = 16 \ge_0 = -8 n_t.
\label{tensscalratio}
\ee
The last two equalities are only valid in the single-field case (as $A_s$
changes in the multiple-field case). The last
equality $r=-8n_t$ is called the single-field consistency relation. As $A_s$
is larger in multiple-field inflation, it becomes an inequality $r\leq-8n_t$
in the more general case.

To convert these scalar and tensor primordial power spectra into the CMB power
spectrum, we
first need to properly introduce the temperature fluctuations:
\be
\frac{\Delta T(\gth,\gvf)}{T_0} \equiv
\frac{T(\gth,\gvf)-T_0}{T_0}
= \sum_{\ell=2}^{\infty} \sum_{m=-\ell}^{+\ell} a_{\ell m}^T Y_{\ell m}(\gth,\gvf).
\ee
Here $T_0=2.725$~K is the average temperature of the CMB, the $Y_{\ell m}$ are
the spherical harmonics, and $a_{\ell m}^T$ are the temperature mode coefficients
that encode the properties of the temperature fluctuations.
The division by $T_0$ is not always performed, in which case the $a_{\ell m}$
would have the dimension of temperature instead of being dimensionless.
By definition temperature
fluctuations do not have a monopole ($\ell=0$) component. The reason we
also do not consider the dipole ($\ell=1$) component, is because it is too
contaminated by the much larger dipole due to the Earth's movement, as
discussed in the previous section.

To properly define the polarization fluctuations, let us consider for a moment
a single monochromatic electromagnetic plane wave travelling in a given
direction, which without loss of generality we will take to be the $z$
direction. It has the following electric field:
\be
\vec{E}(t,\vec{x}) = \left(\begin{smallmatrix} a_1 e^{i\theta_1}\\ 
  a_2 e^{i\theta_2}\\ 0 \end{smallmatrix}\right) e^{i(\omega t - k z)}.
\ee
Instead of $a_1, a_2, \theta_1, \theta_2$ we can also use the 4 Stokes 
parameters:
\be
I\equiv a_1^2+a_2^2, \quad Q\equiv a_1^2-a_2^2, \quad 
U\equiv 2 a_1 a_2 \cos(\theta_2-\theta_1), \quad
V\equiv 2 a_1 a_2 \sin(\theta_2-\theta_1),
\label{defStokes}
\ee
which are not independent but satisfy the identity $I^2 = Q^2+U^2+V^2$
(reflecting the fact that the overall phase of the wave is unimportant, so that
there were only 3 independent quantities: $a_1$, $a_2$ and $\theta_2-\theta_1$).
$I$ gives the total intensity, $Q$ the horizontal/vertical linear polarization,
$U$ the $\pm 45^\circ$ linear polarization, and $V$ the right/left-handed
circular polarization. Note that the choice of sign of $V$ is related to the
usual
ambiguity in the circular polarization convention, but as circular polarization
is not produced by Thomson scattering and is absent in the CMB, we do not need
to worry about that here and can forget about $V$.
Because of the identity $I^2 = Q^2+U^2+V^2$, it is clear that a single wave
always has some polarization state, it cannot be unpolarized ($Q=U=V=0$).
When we mention unpolarized radiation, we are talking about a superposition
of multiple waves, and the averaged values of the Stokes parameters.

As explained in the previous section, instead of using $Q$ and $U$ to
describe the CMB's polarization, CMB physicists often prefer using $E$ and $B$
defined by
\be
\frac{Q(\gth,\gvf) \pm i U(\gth,\gvf)}{T_0} =
- \sum_{\ell,m}(a_{\ell m}^E \pm i a_{\ell m}^B)\, _{\pm 2}\!Y_{\ell m}(\gth,\gvf)
\ee
with $_{\pm 2}\!Y_{\ell m}$ spin-weighted spherical harmonics of spin $\pm 2$. As
for temperature fluctuations, the division by $T_0$ is performed to create
dimensionless $a_{\ell m}$ coefficients, which is not always done.\footnote{The
  $I$, $Q$, and $U$ defined in (\ref{defStokes}) have the dimension of electric
  field squared. In fact, by multiplying them with the constant
  $\frac{1}{2}c\varepsilon_0$ they get the dimension of flux ($I$ is then
  exactly the intensity of a plane wave, i.e.\ the time-averaged flux), and
  that is their usual dimension (and how they are measured in practice). In
  addition they are sometimes measured as flux per unit solid angle (radiance)
  or as flux per unit solid angle and per unit frequency/wavelength (spectral
  radiance), and the latter is the case for CMB experiments. Finally, to
  convert this measured spectral radiance into a temperature, which is the
  dimension of the $I$, $Q$, and $U$ maps produced by CMB experiments,
  Planck's law giving the spectral radiance of a black body is used,
  $B_\nu = (2h\nu^3/c^2)[\exp(h\nu/(k_BT))-1]^{-1}$. However, this relation
  is linearized for small temperature fluctuations around the average
  temperature $T_0$, leading to the linear relation (with constant of
  proportionality depending on $T_0$) used to convert spectral
  radiance into temperature.}
The $E$ and $B$ fields are then defined from their $a_{\ell m}$ coefficients
using normal spherical harmonics:
\be
\frac{E(\gth,\gvf)}{T_0}
= \sum_{\ell,m} a_{\ell m}^E Y_{\ell m}(\gth,\gvf)
\ee
and similarly for $B$.
The publicly available HEALPix code\footnote{\url{http://healpix.sourceforge.net/}} \citep{Gorski:2004by} can be used to compute
all these types of spherical harmonic transforms, and perform other useful
operations on a pixelized sky as well as display the resulting maps.

The power spectra of the CMB are defined by:
\be
\langle a_{\ell m}^{p_1} a_{\ell' m'}^{p_2\,*} \rangle
= C_\ell^{p_1 p_2} \delta_{\ell \ell'} \delta_{m m'},
\label{Cldef}
\ee
where $p_1,p_2 = T,E,B$. The relation between the power spectrum of $\gz$ and of
the CMB is:
\be
C_\ell^{p_1 p_2} = 4\pi \int \frac{\d k}{k} P_\gz(k) g_\ell^{p_1}(k) g_\ell^{p_2}(k).
\label{Cl_gl}
\ee
In this case $p_1,p_2=T,E$ only, as the scalar perturbations do not contribute
to
$B$ polarization. Here both the evolution effects (like the acoustic peaks) and
the projection effects (going from 3D Euclidean space to a 2D spherical surface)
are encoded in the radiation transfer functions $g_\ell^p(k)$. These functions
are the solution of complicated systems of Boltzmann equations describing
the interactions between all types of particles, and must be determined
numerically. Fortunately public codes to compute them exist. We have always
used CAMB\footnote{\url{http://camb.info/}} \citep{Lewis:1999bs} in our
research.
The contribution of the tensor perturbations to the different CMB power spectra
is given by a similar expression as (\ref{Cl_gl}), just with the tensor
power spectrum $P_t$ instead of $P_\gz$, and with different radiation transfer
functions.

To determine the power spectrum observationally via its definition (\ref{Cldef})
would require performing an ensemble average over a large number of skies.
Obviously we only have one sky that we can measure. The only averaging that
we can do to determine an estimate $\hat{C}_\ell$ of the real $C_\ell$ (assuming
ergodicity) is over the $m$:
\be
\hat{C}_\ell = \frac{1}{2\ell+1}\sum_{m=-\ell}^{+\ell} |a_{\ell m}|^2.
\ee
As the number of $m$ per $\ell$ is limited, especially for low $\ell$, this
leads to a fundamental statistical error on our determination of $C_\ell$,
called cosmic variance, that would be present even in the case of an ideal
experiment with infinite resolution and no noise. It is given by
\be
\Var(C_\ell) = \frac{2}{2\ell+1} \, C_\ell^2.
\ee

\begin{figure}
  \begin{center}
  \includegraphics[width=1.\textwidth]{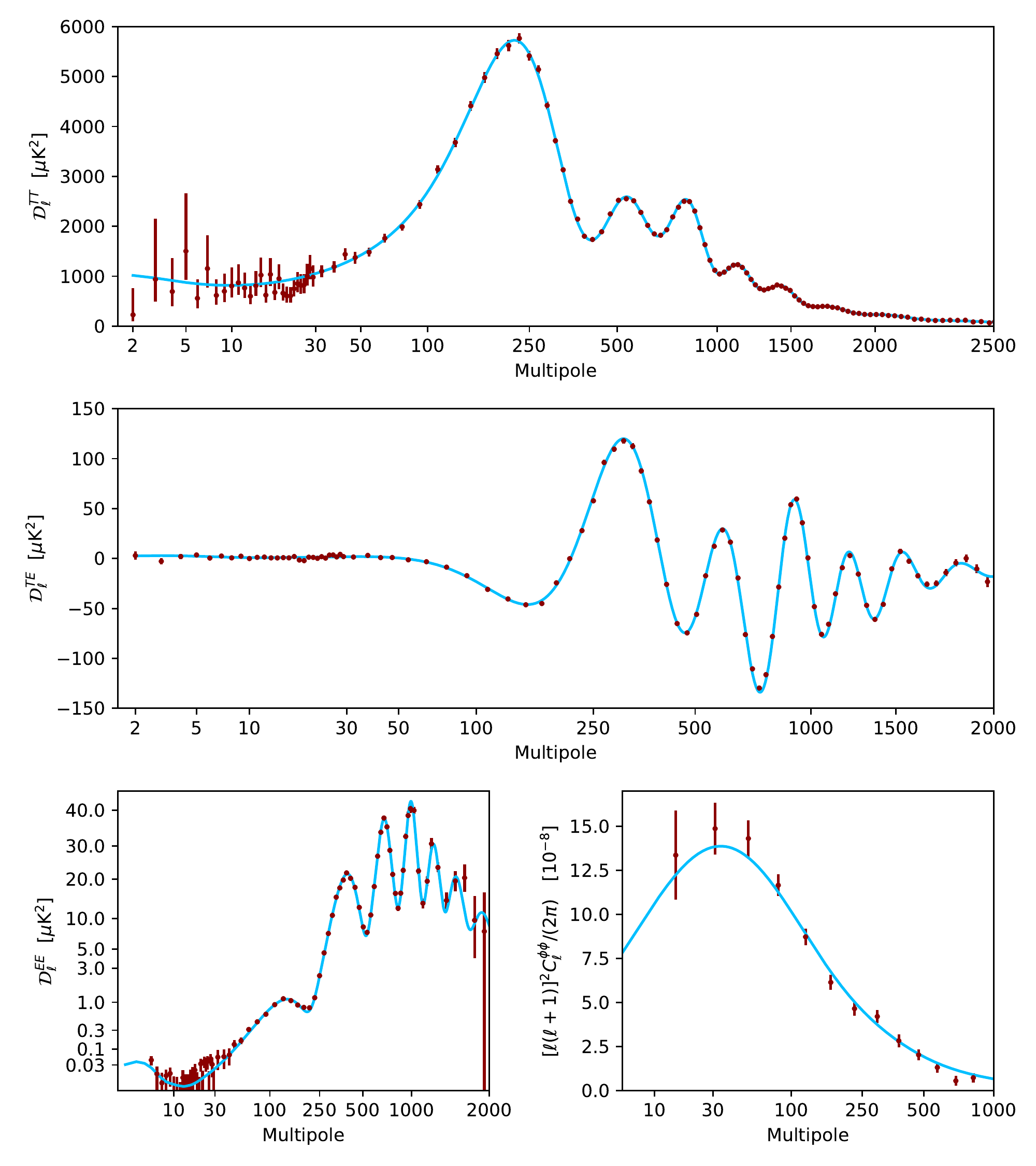}
  \end{center}
  \caption{Planck 2018 CMB power spectra for temperature
(top), the temperature-polarization cross-spectrum (middle), the $E$ mode of
    polarization (bottom left) and the lensing potential (bottom right, which we
    do not discuss in this thesis). 
The quantity $D_\ell$ is defined as $\ell(\ell+1)T_0^2\, C_\ell/(2\pi)$.
The red dots with error bars (that include cosmic variance) indicate the
measurements, while the blue curve is the best-fitting $\Lambda$CDM model.
The anisotropy power spectra use
a standard binning scheme (which changes abruptly at $\ell=30$) and are
plotted here with a multipole axis that goes smoothly from logarithmic at
low $\ell$ to linear at high $\ell$.
  Figure from \citepalias{planck2018-01}.}
  \label{Fig:CMBpowspec}
\end{figure}

The $TT$, $TE$, and $EE$ CMB power spectra (as well as the lensing potential
power spectrum) as measured by the Planck satellite
are shown in figure~\ref{Fig:CMBpowspec}. The red dots with error bars are
the Planck measurements, with the error bars including the contribution from
cosmic variance explained above.
In fact, the uncertainties of the $TT$ spectrum are dominated by cosmic
variance, rather than by noise or foreground residuals, at all scales
below about $\ell=1800$ -- a scale at which the CMB information is essentially
exhausted within the framework of the $\Lambda$CDM model.\footnote{The
  $\Lambda$CDM model is the currently widely accepted model
  of the standard Big Bang cosmology, because it is the simplest model
  that is in agreement with all data. It is named after its two most important
  (but also most mysterious) componenents, the dark energy or cosmological
  constant indicated by $\Lambda$ and the cold dark matter (CDM), but also
  includes ordinary matter and radiation.}
The $TE$ spectrum is about as constraining as the $TT$ one, while the $EE$
spectrum still has a sizeable contribution from noise, and the potential to
be improved by a future CMB mission. The blue line is the best-fit $\Lambda$CDM
model, which is characterized by only 6 parameters: two are the scalar
amplitude and spectral index $A_s$ and $n_s$ defined above, while the other
four are the the baryon and cold dark matter densities, the size of the
sound horizon at last scattering, and the amount of reionization
of the hydrogen gas in the universe due to the emergence of the first stars
(long after recombination and the creation of the CMB).
Other parameters, like the tensor-to-scalar ratio $r$ defined above or the
curvature of the universe, are compatible with zero and hence not required to
describe the basic $\Lambda$CDM model. Yet other parameters, like the age
of the universe and the dark energy density, are not independent and can be
derived from the other six. The best-fit values of all these parameters can
be found in \citepalias{planck2018-06}; here we give
$A_s = (2.100 \pm 0.030) \times 10^{-9}$ and $n_s = 0.9649 \pm 0.0042$.

Everything we discussed in this section is based on the CMB power spectrum,
which would contain all information if the fluctuations were Gaussian.
However, even though the fluctuations are small and linearizing all equations
is a good first approximation, it is not exact. Gravity is inherently
non-linear, and most inflation models also introduce non-linearities.
Hence non-Gaussianity will always be present at some level, the question is
just if it is observable or not. Once we admit that the perturbations are
non-Gaussian, there is in principle information in all correlation functions.
However, given that the $a_{\ell m}$ are small, it makes sense that generically
the next correlation function, the three-point correlation function, will
be where non-Gaussianity can most easily be measured and hence will give the
tightest constraints if it is not detected, as each successive
correlation function will be smaller and smaller. Hence we focus in this
thesis exclusively on the three-point correlator, the Fourier or spherical
harmonic transform of which is called the bispectrum.\footnote{Looking at the
  linear solution (\ref{zeta_lin_sol}) for $\gz$ it is easy to see why
  the three-point correlator would be zero in the linear case: one would
  be computing the vacuum expectation value of an odd number of creation
  and annihilation operators. Once you include second-order terms, however,
  you will be able to take one second-order $\gz$ with two creation/annihilation
  operators and two linear $\gz$ with one each, for a total of four, which
  can have a non-zero vacuum expectation value.}
Just like we introduced the amplitude $A_s$ of the power spectrum, we will
also introduce a bispectrum amplitude parameter called $\fnl$.
For precise definitions of both the bispectrum and $\fnl$ we refer to
section~\ref{solpowspecbispecsec}.

\subsection{The Planck mission}

After the discovery of the CMB by Penzias and Wilson in 1964 (paper
\citep{Penzias:1965wn} published in 1965; Nobel Prize in 1978), its temperature
anisotropies were first detected by NASA's COsmic Background Explorer (COBE)
satellite\footnote{\url{https://lambda.gsfc.nasa.gov/product/cobe/}}.
COBE was launched in 1989 and operated until 1993. It consisted
of three instruments:
FIRAS (Far-InfraRed Absolute Spectrophotometer) to measure the spectrum
of the CMB, which to this date remain the most accurate measurements of
the CMB black-body spectrum and its temperature, DIRBE (Diffuse InfraRed
Background Experiment) to map dust emission, and DMR (Differential Microwave
Radiometer) to map the temperature anisotropies of the CMB at three frequencies:
31.5, 53, and 90~GHz. The detection of these anisotropies by DMR was first
announced in 1992 \citep{Smoot:1992td}, with further publications based on
more data following in the years after. The full four-year data was published
in 1996 \citep{Bennett:1996ce}. COBE's FIRAS and DMR principal investigators,
John Mather and George Smoot, were awarded the Nobel Prize in 2006.

COBE was followed by NASA's Wilkinson Microwave Anisotropy
Probe (WMAP) satellite\footnote{Initially called MAP, it was renamed in 2003 in
honour of David Wilkinson, a member of its science team who died in 2002.
\url{https://lambda.gsfc.nasa.gov/product/map/current/}}
that was launched in 2001 and operated until 2010.
Its instrument was similar to COBE DMR, but with 45 times higher sensitivity
and 33 times higher angular resolution, and with five frequency channels,
at 23, 33, 41, 61, and 94~GHz.
First-year results were published in 2003 \citep{Bennett:2003bz} and are often
credited with starting the era of high-precision cosmology. More releases
followed, with the final nine-year results published in 2012
\citep{Bennett:2012zja}. WMAP also provided the first maps of the $E$-mode
polarization anisotropies of the CMB (the first detection of CMB polarization
anisotropies had been made by the DASI (Degree Angular Scale Interferometer)
telescope at the South Pole in 2002 \citep{Kovac:2002fg}). As WMAP, unlike COBE,
was passively cooled, the experiment had no intrinsic restrictions on its
lifetime, and it continued to run until 2010, when the launch of the Planck
satellite made it obsolete.

\begin{figure}
  \begin{center}
  \includegraphics[width=1.\textwidth]{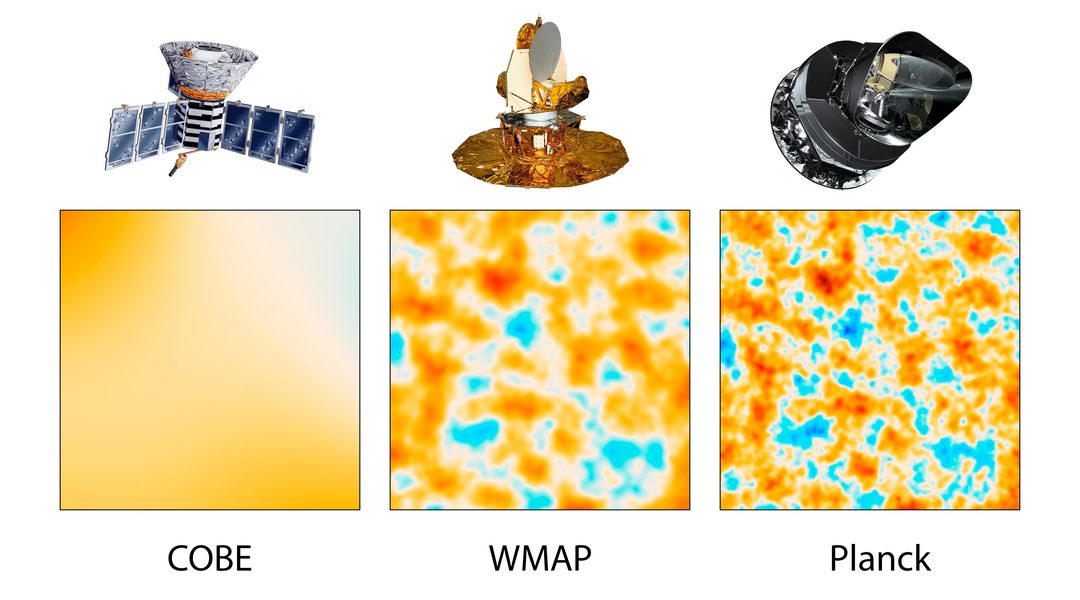}
  \end{center}
  \caption{The three CMB satellites COBE, WMAP, and Planck, with a
    visual comparison of their resolutions on a 10-square-degree patch of the
    CMB sky. {\em Image credit: NASA/JPL-Caltech/ESA.}}
  \label{COBEWMAPPlanck}
\end{figure}
    
The third, and so far last\footnote{Despite many proposals in Europe and in
  the US, the only future CMB satellite proposal that has been accepted by a
  space agency so
  far is the Japanese LiteBIRD satellite (\url{http://litebird.jp/eng/}),
  with a tentative launch date of 2028. LiteBIRD stands for ``Lite (Light)
  satellite for the studies of B-mode polarization and Inflation from
  cosmic background Radiation Detection''.}, generation of CMB satellites is
ESA's Planck
satellite\footnote{\url{https://www.cosmos.esa.int/web/planck}}.
It was launched in 2009 and operated until 2013. It contained two instruments:
LFI (Low Frequency Instrument), based on radiometers like COBE and WMAP and
with three frequency channels (30, 44, and 70~GHz), and HFI (High Frequency
Instrument), with bolometric detectors and six frequency channels (100,
143, 217, 353, 545, and 857~GHz).
Planck also had approximately 10 times higher sensitivity and 2.5 times
higher resolution than WMAP. Beam size decreases with frequency, and the
beam size of Planck's 217~GHz channel is 4.9 arcmin (as compared to 13 arcmin
for WMAP's 94~GHz channel). Planck has the highest sensitivity in its
143 and 217~GHz channels, at 0.55 and 0.78 $\mu\mathrm{K}_\mathrm{CMB}$~deg,
respectively \citepalias{planck2018-01}. A visual comparison regarding resolution
between COBE, WMAP and Planck can be found in figure~\ref{COBEWMAPPlanck}.

Its broad frequency coverage gave Planck
an unprecedented capability of component separation, i.e.\ using the frequency
dependence of various astrophysical foregrounds to identify them and remove
them from the CMB signal. Roughly speaking, 100, 143, and 217~GHz are the
main CMB channels, with the LFI channels serving to determine the
low-frequency foregrounds (most importantly galactic synchrotron radiation)
and the three highest HFI channels serving to determine the high-frequency
foregrounds (mainly galactic thermal dust radiation). The actual four component
separation methods used in the Planck analyses (Commander, Nilc, Sevem, and
Smica, see \citepalias{planck2018-04} for details and references) are more
sophisticated than this simple description implies though. The Planck
collaboration was much larger than the COBE and WMAP ones, which allowed
Planck to use multiple independent pipelines for the various aspects of
its analyses, like the four component separation methods mentioned above, for
extra robustness of its final results. As will be discussed later in this
thesis, there were also four different bispectrum estimator pipelines (based
on three distinct estimators).

HFI required active cooling down to $0.1$~K using the dilution of $^3$He into
$^4$He. After 2.5 years (5 full-sky scans) the $^3$He supply was exhausted
(it lasted almost twice as long as the nominal lifetime), leading to the end of
HFI for scientific purposes. LFI on the other hand only required passive
cooling down to 20~K, and continued functioning for another 1.5 years (3
additional full-sky scans) until it was decided to switch the experiment off
in 2013.
There were three data releases: the first in 2013 \citepalias{planck2013-01}, based
on one year
of data and temperature-only, the second in 2015 \citepalias{planck2015-01}, which
included all data
for both temperature and polarization (with the exception of $\ell\siml 40$
polarization data which were still considered to be insufficiently cleaned
from systematic effects at that time), and a final legacy release in 2018
\citepalias{planck2018-01} containing everything and with an improved treatment of
in particular polarization data.
Each release was accompanied by a suite of papers discussing all primary
science, while so-called intermediate papers on other, mostly astrophysical,
topics were released in between.\footnote{Because of the large size of the
  Planck collaboration, a two-tier system
  regarding authorship of the papers was used. People who had contributed
  enough to the collaboration as a whole (e.g.\ by performing tests of
  general usefulness or producing products or creating tools for general use,
  and not just working on their
  own scientific analysis) obtained the status of Planck Scientist and could
  sign all Planck collaboration papers, while other collaboration members
  could only sign the papers they had contributed to directly. It was also
  decided to keep the order of authors strictly alphabetical for all papers,
  so that without inside information it is impossible to identify the main
  authors of a given paper.} 
Unlike the first two releases, which were
contractual obligations with respect to ESA, the third release was an
initiative of the
principal investigators and hence not as tightly constrained. Because of this,
as well as the reduced manpower, since by this time many people had left the
collaboration to move on to other projects, the primary science papers of
the final release were not all published on the same date, but spread out
over a year, from mid-2018 till mid-2019.

The work in Planck was organised in so-called working groups, each of which
was in charge of a specific part of the analysis, often associated with one
of the main publications. For the non-Gaussianity working
group, most relevant for the work in this thesis, we required several products
produced by other working groups (most of which had their own prerequisites).
In the first place we needed of course the cleaned CMB sky maps, produced
by the component separation working group from the raw sky maps of the
different frequency channels (at an effective 5 arcmin resolution).
In order to compute error bars, the linear correction term (see
chapter~\ref{NGCMBsec}), and for data validation purposes, we also required
simulations. The so-called FFP (Full Focal Plane) simulations were
produced several times (as indicated by different version numbers) over the
course of the Planck data analysis years, including more and more effects and
hence becoming more and more realistic over the years. The final version,
called FFP10 \citepalias{planck2018-02,planck2018-03,planck2018-04} and used in the
2018 data analysis, consisted of a set of CMB-only
map realisations (including the effects of gravitational lensing, satellite
scanning, and beam asymmetries), and a set of noise-plus-systematics
realisations
(of which the input also included a fixed CMB and foreground realisation, which
was subtracted at the end, so that any sky-signal distortion effects are
included as well). These were then passed through the component separation
pipelines in the same way as the real sky map. The required beam transfer
functions and confidence sky masks (for temperature and polarization)
were also provided by the component separation working group.
From the simulations we could determine the CMB power spectrum and noise power
spectrum to be used in the estimator weights.
Finally we required the values of various cosmological parameters to
determine the theoretical
bispectrum templates, which were provided by the power spectrum likelihood and
parameters working group. Conversely, the results produced by the
non-Gaussianity working group
were used to test for example the quality of the component separation products.
This illustrates the important interactivity between the different working
groups in the Planck collaboration.

Apart from space missions, there have also been many ground-based and
balloon-borne CMB experiments. The DASI telescope that first measured the CMB
$E$-polarization was mentioned above. In 2000 the BOOMERanG (Balloon
Observations Of Millimetric Extragalactic Radiation ANd Geophysics)
and MAXIMA (Millimeter Anisotropy eXperiment IMaging Array) balloon-borne
experiments were the first to detect the first peak in the temperature power
spectrum \citep{deBernardis:2000sbo,Hanany:2000qf}.
Most ground-based CMB experiments are either located in the Atacama
Desert in Chile or at the South Pole, due to the good atmospheric conditions
(very low humidity) in those places. With no new CMB satellite data expected
before 2030, it is quite possible that the next great CMB discovery (for
example a detection of the primordial $B$-polarization) will be made from
the ground. However, by their nature ground or balloon experiments cover only
a small fraction of the sky, which makes them less suitable for the detection
of (local) non-Gaussianity, where the correlation between very small and very
large scales is essential, see chapter~\ref{NGCMBsec}.

\chapter{Non-Gaussianity in multiple-field inflation}
\label{NGinflsec}

This chapter treats my work regarding non-Gaussianity in
multiple-field inflation. These are my papers in that
subject area:
\begin{itemize}
\item \citep{RSvT2} --- Introduction of the long-wavelength formalism to compute
  non-linear fluctuations in multiple-field inflation on super-horizon scales.
\item \citep{RSvT1} --- Application to single-field inflation. Introduction
  of a parame\-trization to visually represent the momentum dependence of the
  non-Gaussianity parameter $\fnl$.
\item \citep{RSvT3} --- Derivation of the explicit long-wavelength equations
  for the second-order perturbations for multiple-field inflation and their
  formal solution. Explicit analytic solutions for $\fnl$ in the case of
  constant slow-roll parameters (stronger assumption than usual slow roll).
\item \citep{RSvT4} --- Further refinement of the system of equations
  and their formal solution. Explicit numerical result for $\fnl$ for the
  double quadratic potential.
\item \citep{TvT1} --- The general analytic solution for $\fnl$ for two-field
  inflation is further worked out until just one integral remains. Explicit
  analytic slow-roll solutions in the case of certain classes of potentials are
  derived and compared to the exact numerical results. An example of a potential
  that gives large non-Gaussianity is presented.
\item \citep{TvT2} --- Several remaining technical issues of the long-wavelength
  formalism are sorted out, in particular regarding gauge invariance of the
  adiabatic and isocurvature perturbations at second order. The exact cubic
  action in terms of the gauge-invariant adiabatic, isocurvature en tensor
  perturbations is derived, valid at all scales.
\item \citep{TvT3} --- An investigation of the momentum dependence of the
  bispectrum and $\fnl$ in two-field inflation. Introduction of two
  non-Gaussianity spectral indices, related respectively to rescaling and
  squeezing of the momentum triangle.
\item \citep{TMvT} --- Two-field inflation with non-standard kinetic terms is
  considered. Several technical and definition issues regarding background
  and perturbation quantities in such models are sorted out. The derivation
  of the exact cubic action of \citep{TvT2} is extended to this
  case.\footnote{The work from this paper is not included in this thesis, for
  two reasons: in order not to make this thesis even longer,
  and because this paper is basically only the first step in a generalization of
  our formalism to inflation models with general kinetic terms, which has
  so far not been followed up on.} 
\item \citep{JvT} --- An alternative formulation of the integral term from
  \citep{TvT1} is found, which allows further analytic progress for the
  expression of $\fnl$ in the general two-field case. An extensive analytic
  study of the non-Gaussianity in two-field sum potentials with standard
  kinetic terms is given, both within and beyond the slow-roll approximation,
  showing amongst other things how much the parameter space where
  non-Gaussianity can be large is limited by the Planck constraint on the
  spectral index.
\end{itemize}
The first four papers are collaborations with Paul Shellard and
Gerasimos Rigopoulos.
The next four are collaborations with
Eleftheria Tzavara (as well as Shuntaro Mizuno
for the last of those four). The final one is a collaboration with
Gabriel Jung.

In section~\ref{longwavform} the long-wavelength formalism is defined and
explained, based on \citep{RSvT2}. In section~\ref{longwaveq} the resulting
equations for the perturbations at first and second order and for $\fnl$,
as well as their general solutions, are summarized and discussed, based
on \citep{RSvT3, RSvT4, TvT1, JvT}. The (long) calculations to derive
some of the results are given in appendix~\ref{calcdetailssec}.
Finally, the papers \citep{JvT, TvT2, TvT3},
which are more self-contained extensions or applications of the formalism,
are briefly summarized in section~\ref{longwavother} and included in full
in the appendices~\ref{JvTapp}--\ref{TvT3app}.

\section{The long-wavelength formalism}
\label{longwavform}

\subsection{Introduction}

As discussed in the previous chapter, it is a well-established fact that
the universe on large scales
exhibits a high degree of uniformity. During most cosmological
eras and for a large span of length scales, it can be well
approximated by a Friedmann-Robertson-Walker (FRW) spacetime with
inhomogeneities described as small linear perturbations around the
highly symmetric background. This picture has proved particularly
relevant for the early universe, as the smallness of the cosmic
microwave background (CMB) temperature anisotropies indicates. An
extrapolation of this observational fact suggests that the use of
linear theory would also be justified during inflation when the
perturbations leading to the CMB anisotropies are thought to have
been created. Until about the middle of the first decade of this century,
almost all studies of the generation and
evolution of perturbations in inflation invoked the use of linear
perturbation theory as described in the first chapter. In principle it
offers a tremendous simplification of the task of studying the true
inhomogeneous spacetime.

However, even when attention is focused on the inflationary era,
linear theory cannot be the whole picture. Since gravity is
inherently non-linear and the potential of the inflationary model
is likely to be interacting, some small non-linearity will be
endemic to the perturbations. Given the accuracy of the WMAP and in
particular the Planck CMB observations, it became worthwhile to
investigate whether this non-linearity could be observationally relevant.
The characteristic signatures of non-linear effects are deviations of the
primordial fluctuations from Gaussian statistics. In order for
this non-Gaussianity to be calculated one needs to go to second
order in perturbation theory or develop a fully non-linear
approach.

The issue of calculating non-linearity and the consequent
non-Gaussianity in the primordial universe started attracting
increasing attention from the middle of the first decade of this century,
although some
earlier attempts to calculate it can also be found in the literature
\citep{Salopek:1990re,Gangui:1993tt,1993ApJS...86..333Y}.
A tree-level calculation with a cubic action
for the perturbations was performed in \citep{Maldacena:2002vr} for the
case of slow-roll single-field inflation (with a similar slow-roll
calculation for more general single-field Lagrangeans given in
\citep{Seery:2005wm}). At the level of the equations of motion, various authors
pursued perturbation theory to second order
\citep{Acquaviva:2002ud,Rigopoulos:2002mc,Noh:2003yg,Malik:2003mv,Enqvist:2004bk,
Enqvist:2004ey,Vernizzi:2004nc,Lyth:2005du}, with
\citep{Bartolo:2004if} providing a review of these techniques. Although
interesting results can be obtained at second order, full
exploration of the system of equations suffers from great
computational complexity. On the one hand, the perturbation
equations tend to be rather cumbersome to derive. On the other
hand, gauge-invariant variables, which have proved very useful for
computations and the interpretation of results in linear theory,
are not as simple as their first-order counterparts when
second-order perturbations are considered. 

That was the state of the art when we published our paper \citep{RSvT2}
in early 2005.
In that paper we took a different viewpoint on the study of
non-linear perturbations during inflation.
We used combinations of spatial gradients to construct variables
describing the deviation from a spatially uniform spacetime and
derived equations for these variables on super-horizon scales. These
variables are defined
non-perturbatively and are invariant under changes of the time
coordinate on such scales. They were
first used in \citep{Rigopoulos:2003ak}, where a non-linear
generalisation of the familiar adiabatic conservation law of
linear theory was derived. Later the authors of
\citep{Langlois:2005ii} used similar combinations of gradients in
the context of the covariant formalism \citep{Ellis:1989jt} to also derive
this conservation law. They showed the relation of these simple,
yet fully non-linear, variables to others defined in second-order
perturbation theory. An equivalent conservation law was derived in
\citep{Lyth:2004gb} without using such gradient variables.
At roughly the same time, also in early 2005, another approach to the study
of non-linear perturbations was published in \citep{Lyth:2005fi,Allen:2005ye},
based on the so-called $\delta N$ formalism of \citep{Sasaki:1995aw}.

In order to include the continuous influx of sub-horizon
perturbations to the long-wavelength system, source
terms are added to the long-wavelength equations. Thus, we arrive at a set of
fully non-linear equations which include both matter
and metric perturbations. For actual
multiple-field calculations it is more convenient to use the
explicit field basis \citep{GNvT1,GNvT2} described in the previous chapter.
Since the evolution equations are fully
non-linear on long wavelengths, numerical simulations can be
performed without the need for analytic approximations.
All of this is described in detail in section~\ref{longwavform},
based on \citep{RSvT2}. As discussed in the next section~\ref{longwaveq},
based on \citep{RSvT3, RSvT4, TvT1, JvT}, a perturbative analytic approach can
be applied giving results to second order. At first order, this
perturbative expansion is equivalent to the well-known linear
gauge-invariant perturbation theory. At second order, however, it
is much simpler than the corresponding second-order approaches
pursued before 2005. The full calculation up to and including the computation
of the bispectrum and its amplitude parameter $f_\mathrm{NL}$ is given
in that section~\ref{longwaveq}.

Starting from these pioneering works in 2005, and strengthened from the
end of 2007 when \citep{Yadav:2007yy} claimed a detection of local
non-Gaussianity in the WMAP data, primordial non-Gaussianity became a hot topic
in the literature. This popularity ended in 2013, when the Planck satellite
did not detect any primordial non-Gaussianity \citepalias{planck2013-24} and
hence refuted the earlier claim. But of course even a non-detection puts
important constraints on inflationary models, and so non-Gaussianity,
despite the loss of its ``hot topic'' status, has remained an important research
subject. References to many inflationary non-Gaussianity papers can be found
in sections~\ref{introJvT}, \ref{introTvT2}, and~\ref{intro}, while we give
a few more early ones here:
\citep{Bernardeau:2002jy,Bernardeau:2002jf,Enqvist:2005qu,Vaihkonen:2005hk,
Hattori:2005ac,Kolb:2005ux,Calcagni:2004bb,Bartolo:2005fp,Barnaby:2006cq,
Vernizzi:2006ve,Langlois:2006vv,Weinberg:2005vy,Weinberg:2006ac,Seery:2005gb,
Chen:2006xjb}. 
Finally, some selected references to papers from 2019 on non-Gaussianity
in inflation models to show that the field is still active:
\citep{Fumagalli:2019noh,Ozsoy:2019slf,Bolis:2019fmq,McAneny:2019epy,
  Garcia-Saenz:2019njm,Achucarro:2019lgo,Bjorkmo:2019qno,Fujita:2019tov}.

\subsection{Non-linear equations}

The starting point of what became known as the long-wavelength formalism
(also sometimes referred to as the RSvT formalism, after its
authors) is the
long-wavelength approximation \citep{Salopek:1990jq,Salopek:1990re,Comer:1994np,
Deruelle:1994iz,Parry:1993mw,Khalatnikov:2002kn}:
the fact that on super-horizon scales, spatial
gradients can be neglected with respect to time derivatives.
For any quantity $F(t,\vc{x})$ constructed out of metric and matter variables
typically we will have $\der_i F = \cO(F/L)$ and $\der_t F/N =\cO(HF)$.
From this we see that for length scales $L \mg 1/(aH)$, i.e.\ super-horizon
scales, we can expect
$\left|\frac{1}{a}\der_i F \right| \ml \left| \frac{1}{N}{\der_t F}\right|$.
In practice this means that the second-order spatial gradients in
the various equations of motion can be neglected.\footnote{
More formally this can be viewed as taking only the leading-order terms in the
so-called spatial gradient expansion (see \citep{Salopek:1990jq,Comer:1994np,
Giovannini:2005ev,Tanaka:2006zp} and references therein).
While very plausible, the validity
of this approximation beyond linear order was only assumed at the time of our
first papers. In a later paper \citep{TvT2}, included in appendix~\ref{TvT2app},
we showed by an exact calculation at second order that at least up to and
including second order the approximation is valid on super-horizon scales once
the decaying perturbation mode can be neglected. If slow roll holds at horizon
crossing, this happens within a few e-folds.}

Using the long-wavelength approximation, the most general form of the metric
can be written as
\be
\d s^2 = -N^2(t,\vc{x})\,\d t^2+a^2(t,\vc{x})\,\d \vc{x}^2.
\label{NLmetric}
\ee
The Hubble parameter is defined as $H \equiv \der_t a/(Na)$.
Here we have kept the lapse function $N$ that encodes the choice of time
coordinate (or in other words the choice of time slicing for the spacetime)
but fixed part of the gauge by setting the shift, i.e.\ the $g_{0i}$
component of the metric, to zero. For a proof that this can always be done
in the long-wavelength approximation, see e.g.~\citep{TvT2} included in
appendix~\ref{TvT2app}. In principle
the second term should also contain a factor $h_{ij}(t,\vc{x})$ describing
the tensor perturbations, which beyond linear order no longer decouple from the
scalar perturbations. However, as was shown already in \citep{Salopek:1990re},
in the long-wavelength approximation (after decaying modes have disappeared)
the (non-linear) tensor perturbations are non-dynamical on super-horizon
scales (i.e.\ $h_{ij}(\vc{x})$ does not depend on $t$) and hence uninteresting
for our purposes, so that we set $h_{ij}(\vc{x}) = \delta_{ij}$ for simplicity.
The fully non-linear metric (\ref{NLmetric}) looks identical to the background
metric (\ref{Bmetric}), but the difference is that $N$ and $a$ are now
inhomogeneous functions of time and space, and not just functions of time alone.

On the matter side we assume the same very general multiple-field inflation
model as in section~\ref{sec:background}, so that the Lagrangean matter
density is given by (\ref{matterL}).
We also define the field velocity with respect to proper time, which is also
the conjugate momentum of the field,
\be
\Pi_A \equiv \frac{G_{AB} \der_t \phi^B}{N}.
\ee
However, while our general formalism was indeed set up for this general case,
in most of our explicit calculations we always took a trivial field metric
$G_{AB} = \delta_{AB}$. The only exception is our paper \citep{TMvT}, where we
looked at a Lagrangean even more general than (\ref{matterL}) and were
interested in the new effects due to the non-standard kinetic terms.
For simplicity, we will from now on take $G_{AB}=\delta_{AB}$, and refer the
interested reader to the original papers for the more general expressions
(which are quite similar, just with covariant derivatives instead of normal
derivatives, and an additional curvature tensor of the field manifold next to
the second derivative of the potential).
Similarly, while the formalism can in principle deal with an arbitrary number
of fields, in explicit calculations we always limited ourselves to the case
of two fields. Hence in this thesis we will always assume two fields.
As explained in chapter~\ref{introduction} and below, adding a second
field to a single-field inflation model allows for all kinds of new effects due
to the perturbations no longer necessarily being constant on super-horizon
scales. On the other hand, no such qualitative changes are expected when adding
a third field to a two-field model.

The Einstein and field equations for these metric and matter ingredients are
in the long-wavelength approximation \citep{Salopek:1990jq}:
\bea{\label{H_dynamics}
\der_t H&=-\frac{\gk^2}{2}N\Pi_B\Pi^B\,, \\
\label{momentum_dynamics}
\der_t\Pi_A&=-3NH\Pi_A-NW_A\,,\\
\label{friedmann}
H^2&=\frac{\gk^2}{3}\left(\frac{1}{2}\Pi_B\Pi^B+W\right)\,,\\
\label{0i} 
\partial_iH&=-\frac{\gk^2}{2}\Pi_B\partial_i\phi^B\,,
}
where $W_B\equiv\partial_BW \equiv \der W / \der \gf^B$ and
$\gk^2 \equiv 8\pi G = 8\pi/m^2_\mathrm{pl}$. We will also use the notation
$\Pi\equiv\sqrt{\Pi_B \Pi^B}$.
Just like the metric, the first three equations look identical to the background
equations of motion, but it should be kept in mind that $H$, $\phi^A$ and
$\Pi^A$ are here fully non-linear functions that depend on both time and space,
and not just homogeneous background functions depending only on time.
This explains the notion of ``separate-universe approach'': it seems as if
on super-horizon scales at each spatial point $\vc{x}$ the universe evolves as
an independent FRW universe from some initial condition. The only coupling
between those separate universes comes from the constraint equation (\ref{0i}),
and from the fact that the initial conditions are not arbitrary, but set up
by the sub-horizon evolution where everything is coupled.

\subsection{Gradient variables}

To look at perturbations in such a fully non-linear setup, it makes sense
to define gradient variables, as first advocated by \citep{Ellis:1989jt}, since
by taking a gradient we remove the homogeneous background part, but keep the
full inhomogeneous perturbation. In particular the following combination of
metric and matter gradients is a key variable in our work:
\be\label{zeta_var}
\gz_i^A(t,\vc{x}) \equiv \frac{\Pi^A}{\Pi} \, \der_i \ln a
- \frac{H}{\Pi} \, \der_i \gf^A.
\ee
When linearized, this is exactly the spatial gradient of the $\gz^A$ defined
in (\ref{def_zeta}).
When taking components in the basis defined in (\ref{basis}), as we will do
in the next section, the first (parallel) component is called the adiabatic
perturbation, and the second (perpendicular) one the isocurvature perturbation.
The fully non-linear variable $\gz_i^A$ is gauge-invariant, which in this
context means it is invariant under changes of time slicing (choices of
$N(t,\vc{x})$) within the long-wavelength approximation.
It is also gauge-invariant when linearized. Initially we
thought that this meant the variable was gauge-invariant to any order in
perturbation theory,
but that turned out to be wrong: at second order it is not. In \citep{RSvT4}
we computed the gauge correction required to make contact with the usual
non-Gaussian observables for the adiabatic perturbation, and in \citep{TvT2}
(included in appendix~\ref{TvT2app})
we definitively settled all issues regarding gauge invariance at second order.

From the equations of motion (\ref{H_dynamics}--\ref{0i}), the following
equation can be derived for $\gz_i^A$ (the derivation is mostly straightforward
but can be found in \citep{Rigopoulos:2004gr,RSvT2}; see also the derivation of (\ref{q_eq})
in \citep{GNvT2,thesis}):
\be\label{basic_zeta}
\der_t^2\zeta^A_i - \left(\frac{\der_t{N}}{N}-2NH \left(\frac{3}{2}
+\ge+\eta^\| \right)\right)\der_t\zeta^A_i +
(NH)^2\,\Xi^A{}_B\zeta^B_i=0\,,
\ee
with
\bea{
\Xi^A{}_B  \equiv \:& \frac{W^A{}_B}{H^2} 
+ \left(3\ge+3\eta^\|+2\ge^2+4\ge\eta^\| 
+ (\eta^\perp)^2 + \xi^\| \right)\delta^A{}_B \nonumber \\
&-2\ge\left((3+\ge)\frac{\Pi^A}{\Pi}\frac{\Pi_B}{\Pi}
+\frac{\Pi^A}{\Pi}\,\eta_B+\eta^A\frac{\Pi_B}{\Pi}\right)
}
and the various slow-roll parameters defined in
section~\ref{sec:slowroll}.  It should be stressed that the slow-roll
parameters should just be viewed as short-hand notation here. No
assumption about their size has been made and the equations are
exact (no slow-roll approximation). This equation looks linear, and seems
indeed identical to the first-order perturbation
equation written in terms of $\gz^A$. However, as before this is not correct, as
the coefficients are functions of the fully inhomogeneous $H$,
$\phi^A$ and $\Pi^A$ (and $N$, but that one is fixed by making an
explicit choice of time slicing).  For this reason the system must be
closed with a set of constraint equations expressing the spatial
gradient of those quantities in terms of $\gz_i^A$ and its time
derivative. The explicit form of these constraint equations will be
given later, in (\ref{constr}).
To solve (\ref{basic_zeta}) it is convenient to rewrite this second-order
differential equation as two first-order differential equations:
\be\label{basic_zeta_split} 
\left\{ \begin{array}{l}
\displaystyle\der_t\gz^A_i-\theta^A_i=0\\
\displaystyle \der_t\theta^A_i - \left(\frac{\der_t{N}}{N}-2NH \left(\frac{3}{2}
+\ge+\eta^\| \right)\right)\theta^A_i +
(NH)^2\,\Xi^A{}_B\zeta^B_i=0
\end{array}\right.
\ee
where the velocity $\theta^A_i$ defined by the first equation should now be
viewed as an independent variable.

\subsection{Source terms}

The final ingredient required to complete the system are the initial
conditions, which come from the sub-horizon regime.
To add these explicitly to the equations we proceed as
follows (motivated by the stochastic picture for the generation of
inflationary perturbations \citep{Starobinsky:1986fx,Nakao:1988yi,Kandrup:1988sc,
Salopek:1990re,Stewart:1991dy,Casini:1998wr,Winitzki:1999ve,Afshordi:2000nr,
Matarrese:2003ye,Geshnizjani:2004tf}).
First we note that since $\gz^A_i$ and $\gth^A_i$, solutions
of (\ref{basic_zeta_split}), are valid only on long
wavelengths, we can view them as smoothed long-wavelength versions of
the exact quantities, smoothed using some window function. In Fourier
space this can be written as $\gz^A(\vc{k}) = \cW(k)
\gz^A_\mathrm{ex}(\vc{k})$, and a similar expression for
$\gth^A(\vc{k})$. The window function $\cW$, which we will discuss
further below, filters out short wavelengths (large $k$) below a
certain appropriately chosen smoothing length (which depends on
time). Secondly, while we do not have the fully non-linear equation of
motion on all scales, we do know the equation of motion on all scales
in linear perturbation theory. And as noted before, it looks identical
to the long-wavelength non-linear equation, just with all coefficients
replaced by their homogeneous background version (and after we add the
usual second-order spatial gradient term). So viewing for a moment
(\ref{basic_zeta_split}) as a linear perturbation equation valid on
all scales (after adding the gradient term), we can explicitly apply
the smoothing as defined above to it (most easily done in Fourier space).
Mostly this will just give back the same equation, but with
the exact linear curvature perturbation replaced by its
long-wavelength smoothed version.  However, there will be one extra
term in each equation involving the unsmoothed variable and $\der_t
\cW$, that comes from pulling the window function through the time
derivative. The final step then is to put those extra terms on the
right-hand side of the equations, and return to the long-wavelength
non-linear case by viewing all coefficients again as fully
inhomogeneous functions of space and time (and removing the
second-order spatial gradient term). On the left-hand side of the
equation this is exact, for the reasons explained above.
The assumption here is that this procedure is also correct
on the right-hand side. The final result is:
\be\label{basic_zeta_split_sources} 
\left\{ \begin{array}{l}
\displaystyle\der_t\gz^A_i-\theta^A_i= \cS^A_i\\
\displaystyle \der_t\theta^A_i - \left(\frac{\der_t{N}}{N}-2NH \left(\frac{3}{2}
+\ge+\eta^\| \right)\right)\theta^A_i +
(NH)^2\,\Xi^A{}_B\zeta^B_i= \cJ^A_i
\end{array}\right.
\ee
where the source terms $\cS^A_i$ and $\cJ^A_i$ are given by
\bea{\label{sources}
\cS^A_i(t,\vc{x}) & \equiv \int\frac{\d^3 k}{(2\pi)^{3/2}} \, \der_t{\cW}(t,k)\,
\zeta_\mathrm{lin}^A(t,\vc{k},\vc{x}) \, \mathrm{i}k_i
\,\mathrm{e}^{\mathrm{i} \vc{k}\cdot\vc{x}} +\mathrm{c.c.}\,, \\
\cJ^A_i(t,\vc{x}) & \equiv \int\frac{\d^3 k}{(2\pi)^{3/2}} \, \der_t{\cW}(t,k)\,
\theta_\mathrm{lin}^A(t,\vc{k},\vc{x}) \, \mathrm{i}k_i \,
\mathrm{e}^{\mathrm{i} \vc{k}\cdot\vc{x}} +\mathrm{c.c.}\,,
}
where c.c.\ denotes the complex conjugate and $\gz^A_\mathrm{lin}$ and
$\gth^A_\mathrm{lin}$ are the full, non-smoothed solutions from
linear perturbation theory, that is, incorporating
short-wavelength information. The fact that they depend on
$\vc{x}$ as well as on $\vc{k}$ represents the fact that all
background quantities in these solutions have been made inhomogeneous.

The source terms describe the continuous influx of short-wavelength modes
as they cross the horizon and enter the long-wavelength regime. In this
way the physical initial conditions are set up dynamically and added
explicitly to the equations. The mathematical initial conditions of the
equations are then simply that the smoothed long-wavelength variables
$\gz^A_i$ and $\gth^A_i$ should be zero at early times when all the modes
are sub-horizon, represented schematically by
\be\label{initial}
\lim_{t\rightarrow-\infty} \gz^A_i = 0, \qquad\qquad
\lim_{t\rightarrow-\infty} \gth^A_i = 0.
\ee
While the equations with the source terms are exact at the linear order,
it is only an assumption that by making all background quantities
inhomogeneous in the source terms, they would also be exact at the fully
non-linear level. The fact that this procedure works for the left-hand
side of the equation is no proof. As explained in the next section,
when these source terms are expanded to second order, one does not find the
correct result and a correction must be added. While discouraging, this does not
necessarily mean that the source term cannot be correct fully non-linearly,
as something similar occurs with the gauge invariance of $\gz^A_i$ as
explained above. Anyway, this question remains unanswered as
for the rest of our research we never needed the fully non-linear expression,
only its perturbed version up to first and second order.

Coming back to the window function, we define the smoothing length as
$R = c / (aH)$, where $c$ is a small number. The length $R$ separates
short wavelengths from long wavelengths. It is chosen a bit larger than
the comoving Hubble length so that for wavelengths larger than $R$ we
are already in the long-wavelength regime. For typical situations, where
slow roll holds at horizon crossing (but can be broken later on), a value
of $c=3$ is good enough. $R$ clearly depends on time, as $a$ and $H$ do.
For the window function the requirements are that for short wavelengths
($kR \mg 1$) it goes to zero, while for long wavelengths ($kR \ml 1$)
it goes to one. In our original papers we took (half of) a Gaussian
window function: $\cW(t,k) = \exp(-k^2 R^2 /2)$. Starting from \citep{RSvT4}
we simplified this to a step function (see below for the expression). Of
course the final results should not depend on the explicit form of the
window function, something that we verified explicitly.

The choice of the time slicing, and hence the lapse function $N$, to fix
the final gauge freedom, is somewhat related to the window function.
In our first papers we chose a time variable $t=\ln(aH)$, which corresponds
to $NH = 1/(1-\ge)$. This choice was motivated by the fact that with
this choice, the smoothing length $R$ and hence the window function $\cW$
depend only on time and not on space. It also means that horizon crossing
of a mode $k$ is uniform throughout space. For the Gaussian window function
this gave $\der_t\cW = (kR)^2\cW(k)$. However, starting from our paper
\citep{RSvT4} we switched to the number of e-folds as time variable:
\be
t = \ln a \qquad \Leftrightarrow \qquad NH = 1
\ee
as this simplifies the calculations. Given that (\ref{0i}) can be rewritten
as $\der_i \ln H = \ge \gz_i^A \Pi_A / \Pi$ in this gauge, we see that if
slow roll holds at horizon crossing, the spatial dependence of $H$ (and
hence of $R$) is of
higher order in slow roll, so that it can be neglected during horizon crossing.
Slow roll at horizon crossing is both motivated by observations of the CMB
and a required assumption for the long-wavelength formalism in order for the
decaying mode to disappear rapidly after horizon crossing.
From now on we will always assume
this choice of time slicing, and indicate time derivatives with respect
to this $t$ by a dot. The time derivative of the step function window
function that will be used in all calculations is given by
\be
\dot{\cW}(t,k) = \delta(kR/\sqrt{2}-1).
\ee
This will naturally lead to a time defined by the relation $aH=kc/\sqrt{2}$,
which we call $t_*$, to appear in our relations, where $c\approx 3$ is the
numerical value introduced above. However, because of the assumption of slow
roll at horizon crossing, all quantities vary very slowly at that time, so
that in the end the results do not depend on the value of $c$ and in the final
results one might just as well take $t_*$ to be defined by the usual
horizon crossing condition $aH=k$.

To obtain the linear solutions required for the source terms, one can
either solve the linear perturbation equation exactly numerically, or
use the analytical slow-roll solution determined in \citep{GNvT2}, which to
leading order in slow-roll is given in (\ref{zeta_lin_sol}).
The latter contains quantum creation
and annihilation operators $\hat{a}^{A\,\dagger}(\vc{k})$ and
$\hat{a}^A(\vc{k})$, inherited from the quantum nature of the
fluctuations. As is well known, on super-horizon scales when the
decaying mode can be neglected, the fluctuations become effectively
classical. One can just as well describe them as classical stochastic
fluctuations by replacing the quantum creation and annihilation
operators with Gaussian stochastic quantities $\alpha^A(\vc{k})$ that
satisfy $\langle\alpha^A(\vc{k})\alpha_B^{*}(\vc{k}')\rangle
=\delta^{A}{}_B\,\delta^3(\vc{k}-\vc{k}')$ and
$\langle\alpha^A(\vc{k})\alpha_{B}(\vc{k}')\rangle=0$, 
where $\langle\ldots\rangle$ denotes an ensemble average.
The main purpose of this replacement is numerics: the stochastic quantities
allow for the numerical creation of random realizations.\footnote{The fully
non-linear equations of motion
for $\gz_i^A$ and $\gth_i^A$, with the constraint
equations for the coefficients to close the system and the explicit stochastic
expression for the source terms, is amenable to direct numerical solution on
a multi-dimensional spatial grid to give fully non-linear realizations of
$\gz_i^A(t,\vc{x})$. The methodology for this is briefly described in
\citep{RSvT2}.} 
In the analytical work described in the next sections there is no need
for this replacement, and starting from \citep{RSvT4} we no longer used
the stochastic quantities, but returned to using the quantum creation and
annihilation operators.

\section{Non-Gaussianity}
\label{longwaveq}

\subsection{Perturbative expansion}

To make analytic progress with the system of non-linear equations to compute
the bispectral non-Gaussianity, it makes sense
to expand the system of equations to second order. The resulting equations
and their general solutions are described in this section, based on
\citep{RSvT3, RSvT4, TvT1, JvT}. It should be noted that when we started
our project, the standard way to compute (single-field) non-Gaussianity in
the literature (with the important exception of Maldacena \citep{Maldacena:2002vr})
was to simply generalize the usual linear perturbation calculations by
expanding the metric and fields to second order (with second-order
gravitational potentials etc.)\ and then derive equations of motion for the
second-order perturbations by pushing these through the Einstein and field
equations as in e.g.~\citep{Acquaviva:2002ud}. This led to huge calculations
and expressions, prone to mistakes, where it is hard to identify the proper
gauge-invariant variables and interpret the various contributions. On the
contrary, we started from the fully non-linear equations of motion, which
look almost identical to the linear ones, and derived the second-order
equations of motion for the proper gauge-invariant variables in a very
simple way from those.

Before making the perturbative expansion, we do a few more
manipulations on the long-wavelength system
(\ref{basic_zeta_split_sources}). First we switch to the
multiple-field basis defined in (\ref{basis}), using the indices
$m,n,\ldots$ to specify components in that basis. As a reminder, $m=1$
specifies the component along the field trajectory, and $m=2$ the
perpendicular one. For perturbations that means that $m=1$ is the
adiabatic perturbation and $m=2$ the isocurvature one.\footnote{It should
be noted that $\gth_i^m$ is defined as $\gth_i^m \equiv \der_t(\gz_i^m)$ and
hence is not just the basis vector $e_{m\,A}$ applied to $\gth_i^A$ but
includes a correction term related to the time derivative of the basis vector.
Explicitly one finds
$\gth_i^m = e_{m\,A} \gth_i^A + \eta^\perp \varepsilon_{mn} \gz_i^n$ with
$\varepsilon_{mn}$ the anti-symmetric Levi-Civita symbol.}
As mentioned before, we restrict ourselves to two fields and from now on use the
explicit time slicing $NH=1$ where the time coordinate is the number
of e-folds. We also combine the components of $\gz_i^m$ and $\gth_i^m$
into a vector with components $v_{i\,a}$, and the sources $\cS_i^m$ and
$\cJ_i^m$ into a vector $b_{i\,a}$:
\be
v_{i\,a} \equiv (\gz_i^1, \gz_i^2, \gth_i^2)_a,
\qquad\qquad
b_{i\,a} \equiv (\cS_i^1, \cS_i^2, \cJ_i^2)_a,
\ee
with the indices $a,b,\ldots$ taking the values 1,2,3 (as a reminder,
the indices $i,j$ are spatial indices as our variables are spatial
gradients). The reason that we have not included the adiabatic velocity
$\gth_i^1$ (and its corresponding source) in this vector is that this
component can be eliminated from the equations using the exact
(i.e.\ fully non-linear) long-wavelength identity
\be\label{iso2adi}
\gth_i^1 = 2 \eta^\perp \gz_i^2.
\ee
This important relation was already given and discussed at the linear level in 
(\ref{zeta1prime_eq}),\footnote{Remember that (\ref{zeta1prime_eq}) was given
  in terms of conformal time, which explains the presence of the $\cH$ in that
  equation.} but is valid even fully non-linearly.\footnote{If one derives
  an equation for $\dot{\gth}_i^1$ from (\ref{basic_zeta_split_sources}) in
  the same way as for the other components in (\ref{basic_sym}) (taken without
  the source term that provides the initial conditions coming from the
  sub-horizon region and that only acts around horizon-crossing), see
  e.g.~\citep{RSvT4}, then it is easy to see that
  (\ref{iso2adi}) is a solution of that equation, using (\ref{srderivatives}).
  In \citep{RSvT4} it was also shown (in a rather involved way) that it is the
  only consistent solution in the long-wavelength formalism where we have
  neglected decaying terms.}
The system then becomes
\be\label{basic_sym}
\dot{v}_{i\,a}(t,\vc{x}) + A_{ab}(t,\vc{x}) v_{i\,b}(t,\vc{x})
= b_{i\,a}(t,\vc{x})
\ee
with the matrix $A$ given by
\be\label{A2f}
A = \left( \begin{array}{ccc}
  0 & -2\eta^\perp & 0\\
  0&0&-1\\
  0&3\chi+2\ge^2+4\ge\eta^\parallel+4(\eta^\perp)^2+\xi^\parallel
  &3+\ge+2\eta^\parallel
\end{array} \right).
\ee
For a curved manifold with a nontrivial field metric
the term $-(2\ge/\gk^2) e_2^A R_{ABCD} e_1^B e_1^C e_2^D$ should
be added to the $A_{32}$ component, with $R^A_{\;BCD}$ the
curvature tensor of the field manifold.
Again, we stress that no slow-roll approximation has been made.

To solve the master equation (\ref{basic_sym}) analytically, we
expand the system as an infinite hierarchy of linear perturbation
equations with known source terms at each order. To second order we
obtain
\bea{\label{vimeq1}
\dot{v}^{(1)}_{i\,a} + A^{(0)}_{ab}(t)
v^{(1)}_{i\,b} & = b^{(1)}_{i\,a}(t,\vc{x}), \\
\dot{v}^{(2)}_{i\,a} + A^{(0)}_{ab}(t)
v^{(2)}_{i\,b} & =
- A^{(1)}_{ab}(t,\vc{x})v^{(1)}_{i\,b} + b^{(2)}_{i\,a}(t,\vc{x})\,,
\label{vimeq2}
}
where $v_{i\,a}^{\,}(t,\vc{x})=v^{(1)}_{i\,a}(t,\vc{x})+v^{(2)}_{i\,a}(t,\vc{x})$, and
\be\label{varA}
A_{ab}(t,\vc{x}) = A_{ab}^{(0)}(t) + A_{ab}^{(1)}(t,\vc{x})
\: = \: A_{ab}^{(0)} + \der^{-2} \der^i (\der_i A_{ab})^{(1)}
\equiv A_{ab}^{(0)}(t)
+ \bar{A}^{(0)}_{abc}(t)v^{(1)}_{c}(t,\vc{x}).
\ee
Here we have denoted $v^{(1)}_c \equiv \der^{-2} \der^i
v^{(1)}_{i\,c}$ and $\der_i A_{ab}$ is computed using the
constraint equations (derived from the basic equations
(\ref{H_dynamics})--(\ref{0i}) and the definition of $\gz_i^A$):
\bea{\label{constr}
\der_i \ln H & = \ge \, \gz_i^1,
\qquad
e_{m\,A} \der_i \gf^A \: = \: - \frac{\sqrt{2\ge}}{\gk} \, \gz_i^m,
\nonumber\\
e^A_m \der_i \Pi_A & = - \frac{H\sqrt{2\ge}}{\gk}
\Bigl( \gth_i^m + \eta^\parallel \gz_i^m - \eta^\perp \gz_i^2 \delta_{m1}
+ (\eta^\perp \gz_i^1 + \ge \gz_i^2)\delta_{m2} \Bigr).
}
The explicit expression of the object $\bar{A}^{(0)}_{abc}$ can be found in
(\ref{Abar}). $\tW$ was defined in (\ref{defW}).
The explicit form of the first-order source term can be computed as
\be
b_{i\,a}^{(1)}(t,\vc{x}) = \int\frac{\d^3 k}{(2\pi)^{3/2}} \, \dot{\cW}(k,t)
X_{am}^{(1)}(k,t) \hat{a}^\dagger_m(\vc{k}) \,
\mathrm{i}k_i \,\mathrm{e}^{\mathrm{i} \vc{k}\cdot\vc{x}}
+ \mathrm{c.c.},
\ee
using the slow-roll solution for the linear perturbation at horizon crossing
(\ref{zeta_lin_sol}), and where
\be\label{Xapprox}
X_{am}^{(1)} = - \frac{\gk H}{2 k^{3/2} \sqrt{\ge}}
\left(\begin{array}{cc}1&0\\0&1\\0&-\chi\end{array}\right)
\ee
(which because of the window function will only be evaluated close to $t_*$
where the approximation is valid). To find the last line of this matrix,
for the time derivative of $\gz_2$, expression (\ref{zeta_lin_sol}) is
not enough, as one needs the time dependence of the linear solution
around horizon crossing up to first order in slow roll. We have not
given that more complicated expression in this thesis, but it was first
derived in \citep{GNvT2} and can also be found in \citep{RSvT3}.

Regarding the second-order source term $b^{(2)}_{i\,a}$, in our
initial papers \citep{RSvT1,RSvT3} we computed it as the perturbation
of the source term $b^{(1)}_{i\,a}$ in a similar way as in (\ref{varA}), so
by taking the spatial gradient and using the constraint equations.
As explained in the previous section, unlike the rest of the non-linear
equation, the validity of the source term beyond linear
order was only a conjecture. We found the resulting source term at second
order to be unsatisfactory. In the first place it depended
on the details of the window function. Secondly, in \citep{RSvT1}, where we
studied the single-field limit, we found that while of the correct
order of magnitude, the exact momentum dependence did not agree with
the result by Maldacena \citep{Maldacena:2002vr}. In \citep{RSvT3}, where we
looked at the multiple-field case with constant slow-roll parameters, we
found that the contribution coming from $b^{(2)}_{i\,a}$ was negligible
compared to the contribution from $-A_{ab}^{(1)} v_{i\,b}^{(1)}$ (which is
the true super-horizon contribution that does not depend on the details
of the window function, and which is absent in single-field inflation).
Because of these reasons, we decided to simply neglect the $b^{(2)}_{i\,a}$
term in \citep{RSvT4}. The issue was finally settled
definitively in \citep{TvT1} and \citep{TvT2}, where $b^{(2)}_{i\,a}$ was
computed from the exact cubic action for $\gz_i^m$ (and it was in fact different
from the perturbation of $b^{(1)}_{i\,a}$). The term does indeed
always give negligible non-Gaussianity and is unimportant for our studies,
but its inclusion allows for perfect agreement with e.g.\ the single-field
result of Maldacena. Its expression is
\ba\label{secondsource}
 b_{ia}^{(2)} &= & \int\frac{\d^3\vc{k}}{(2\pi)^{3/2}}\int\frac{\d^3\vc{k}'}{(2\pi)^{3/2}}
\dot{\mathcal{W}}(\mathrm{max}(k',k))\nn\\
&&\times \Bigg\{L_{abc}(t)X_{bm}^{(1)}(k',t)X_{cn}^{(1)}(k,t)
\hat { a }_m^{\dagger}(\vc{k}')
\hat { a }_n^{\dagger}(\vc{k}) i (k'_i+k_i)
e^{i(\vc{k}'+\vc{k})\cdot\vc{x}}\nn\\
&&\ \ \ \ \  +N_{abc}(t)X_{bm}^{(1)}(k',t)X_{cn}^{(1)}(k,t)
\hat { a }_m^{\dagger}(\vc{k}')
\hat { a }_n^{\dagger}(\vc{k}) i k_i e^{i(\vc{k}'+\vc{k})\cdot\vc{x}}
+\mathrm{c.c.}\Bigg\}.
\ea
(Here c.c.\ is an abuse of notation, as there are three additional terms for
each line, given the multiplication of two complex operators.)
The key elements for its derivation can be found in \citep{TvT2}, included
in appendix~\ref{TvT2app}. The term with $L_{abc}$ is a local term that comes
from the redefinition of the perturbations required to remove terms proportional
to the equation of motion from the cubic action, as explained in that paper.
The term with $N_{abc}$ is a non-local term that comes from the gauge
transformation between the uniform energy density gauge and the flat gauge
(see remarks in the next section). The explicit expressions for $L_{abc}$
and $N_{abc}$ can be found in (\ref{LandN}).

\subsection{Solution for power spectrum and bispectrum}
\label{solpowspecbispecsec}

Looking at equations (\ref{vimeq1}) and (\ref{vimeq2}) we see that they
both have the same structure: a linear first-order differential equation
with an inhomogeneous source term. In fact it is easy to show that this
structure is valid to any order in the perturbative expansion, with
the source term at order $(\alpha)$ computable from quantities up to order
$(\alpha-1)$, so that an iterative scheme is possible. The system
is ideal for solving using a Green's function:
\be\label{solviG}
v_{i\,a}^{(\alpha)}(t,\vc{x})=\int_{-\infty}^t \d t' \, G_{ab}(t,t')
\,\tilde{b}_{i\,b}^{(\alpha)}(t',\vc{x}),
\ee
where we use $\tilde{b}_{i\,b}^{(\alpha)}$ to indicate all the terms on the
right-hand side of the equation at order $(\alpha)$ together, and
with the Green's function $G_{ab}(t,t')$ satisfying\footnote{To be
precise, the Green's function is actually defined as the solution
of (\ref{Green}) with $\delta(t-t')$ on the right-hand side instead
of zero. The solution is then a step function times what we call
the Green's function. This step function has been taken into
account by changing the upper limit of the integral in
(\ref{solviG}) from $\infty$ to $t$.}
\be\label{Green}
\frac{\d}{\d t}\, G_{ab}(t,t') + A^{(0)}_{ac}(t) G_{cb}(t,t') = 0,
\qquad\qquad
\lim_{t\rightarrow t'} G_{ab}(t,t') = \delta_{ab}.
\ee
It is important to note that this Green's function is homogeneous,
a solution of a background equation involving only time, not space.
It has to be computed only once, and can then be used to calculate the
solution for $v_{i\,a}^{(\alpha)}$ at each order
using the different source terms as in (\ref{solviG}).
More details about these Green's functions can be found in
appendix~\ref{greensec}.

The first-order solution can then be written as
\be\label{via1sol}
v_{ia}^{(1)}(t,\vc{x})=\int\frac{\d^3 k}{(2\pi)^{3/2}}v_{am}(k,
t)\hat { a }_m^{\dagger}(\vc{k})
\mathrm{i} k_i \mathrm{e}^{\mathrm{i}\vc{k}\cdot\vc{x}}+\mathrm{c.c.},
\ee
introducing the quantities $v_{am}$ given by
\be
v_{am}(k,t) = G_{ab}(t,t_*) X_{bm}^{(1)}(t_*),
\label{vamGreenrel}
\ee
defined for $t \geq t_*$, with the dependence on $k$ entering through $t_*$.
We have not included the explicit first-order indication ${}^{(1)}$ on
$v_{am}$ to lighten the notation, and since that quantity is always first order.
It will also be useful to define a ``reduced'' version of $v_{am}$, indicated
by a bar, without the prefactor of the $X_{bm}^{(1)}(t_*)$ (see (\ref{Xapprox})):
\be
\bar{v}_{am} \equiv v_{am}/\gamma_*,
\qquad\qquad
\gamma_* \equiv -\frac{\gk H_*}{2 k^{3/2} \sqrt{\ge_*}}.
\label{vbar}
\ee

Remembering that $v_{ia} = (\gz_i^1, \gz_i^2, \gth_i^2)$ for $a=1,2,3$ it is
now easy to compute the linear power spectrum. Since at first order $\gz_i$
is gauge-invariant and a total gradient, it is trivial to get rid of the
spatial gradient (the $i$ index). For the adiabatic component we
find (see (\ref{defPzeta}) for the definition of the power spectrum)
\be
P_\gz(k,t) = \frac{k^3}{2\pi^2} (v_{1m})^2
= \frac{\gk^2 H_*^2}{8\pi^2 \ge_*} [ 1 + (\bar{v}_{12})^2].
\label{Pzetamf}
\ee
For the spectral index, defined as $n_s = 1 + \d\ln P_\gz/\d\ln k$, we find
\bea{\label{ns}
n_s-1 = \frac{1}{1-\ge_*}&\left[ -4\ge_* - 2\eta^\|_* + 2 \frac{\bar{v}_{12}}{1+(\bar{v}_{12})^2} \left(-2\eta^\perp_* + \chi_*\bar{v}_{12}\right.\right. \\
  &\left.\left. + G_{13}(t,t_*)\left( -\tW_{221*}
  + 2\ge^2_* + (\eta^\|_*)^2 + (\eta^\perp_*)^2 + 3\ge_{*}(\eta^\|_* - \chi_{*})-2\eta^\|_* \chi_* + \chi_*^2\right)\right)\right]\nonumber
}
with $\tW_{221*} \equiv \sqrt{2\ge_*} W_{221*} / (3\gk H_*^2)$ and
$W_{221*}$ the third derivative of the potential, projected along the
indicated directions of the basis and evaluated at $t_*$ (see the definition
(\ref{defW})). Finally, the tensor-to-scalar ratio $r$ is given by
\be
r = \frac{16\ge_*}{1+(\bar{v}_{12})^2}.
\ee
These expressions should be compared to (\ref{Pzetasf}), (\ref{nssf}),
and (\ref{tensscalratio}), which were computed for the single-field case,
where there is no evolution of $\gz$ on super-horizon scales.\footnote{The
  factor $1/(1-\ge_*)$ is not present in (\ref{nssf}) because that expression
  was given only to leading order in slow roll.}

\begin{figure}
\centering
\includegraphics[width=0.5\textwidth]{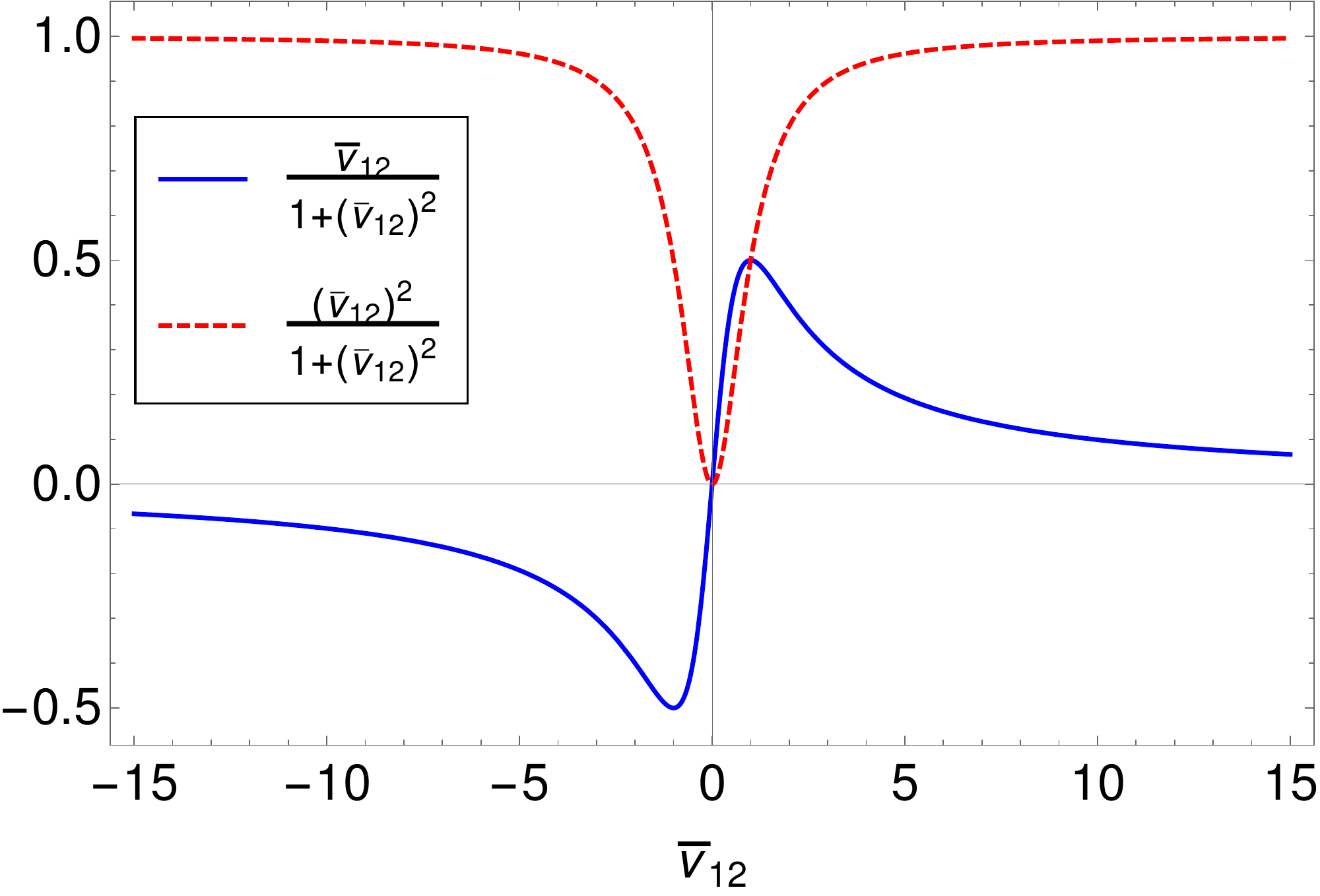}
\caption{$\bv_{12}/(1+\bv_{12}^2)$ and 
$\bv_{12}^2/(1+\bv_{12}^2)$ as a function of $\bv_{12}$.}
\label{fig:factors2}
\end{figure}

An important conclusion can be drawn from the expression of the spectral index.
Given that $G_{13}(t,t_*) \approx \bv_{12}/3$ as we will later show in 
\eqref{proportionality}, 
the relevant factors to study are $\bv_{12}/(1+\bv_{12}^2)$ and 
$\bv_{12}^2/(1+\bv_{12}^2)$, which are shown in figure~\ref{fig:factors2}. 
We see that they are never larger than unity in absolute value and are in fact
of order unity unless $\bv_{12}\approx 0$, which is when multiple-field effects
are negligible and which is not interesting from the point of view of this 
paper.\footnote{The factor $\bv_{12}/(1+\bv_{12}^2)$ also goes to zero for
$|\bv_{12}|\rightarrow\infty$. However, while this term in (\ref{ns}) would then
be compatible with a large $\getpe_*$, that is forbidden by the other terms.}
So barring any fine-tuned cancellations between terms, the observed value of
$n_s$ allows us to conclude that slow roll is a good approximation at horizon
crossing with all first-order slow-roll parameters at $t_*$ at most of order 
$10^{-2}$. However, it is certainly possible for slow roll to be broken
afterwards.

At second order we have to do an additional step. As $\gz_i^m$ is not
gauge-invariant at second order, we have to think if the flat gauge that
we picked (where $a$ does not depend on space) will give us the correct
observable, which should be a gauge-invariant quantity by definition.
In addition, in the flat gauge at second order, $\gz_i^m$ is
not a total gradient, which complicates things. A careful investigation of
gauge invariance at second order, performed in \citep{TvT2} included in
appendix~\ref{TvT2app}, shows that it
is only in the uniform energy density gauge that the expression for
$\gz_i^{m\,(2)}$ coincides with the gauge-invariant expression and hence
gives the correct observable. Moreover, in that gauge the adiabatic component
is a total gradient, so that getting rid of the $i$ index is again trivial.
Hence we have to include in our calculations a gauge-correction term
to convert the result for $\gz_i^{m\,(2)}$ from the flat gauge to the
uniform energy density gauge. For the adiabatic component this correction
to be added is $2\eta^\perp \gz^{1\,(1)} \gz_i^{2\,(1)}$.

By combining the different permutations of
$\langle 0 | \zeta^{1\,(2)}(\vc{k}_1) \zeta^{1\,(1)}(\vc{k}_2) \zeta^{1\,(1)}(\vc{k}_3)| 0 \rangle$
of the Fourier components of the linear and second-order 
adiabatic solutions (first subtracting the average of $\gz^{1\,(2)}(t,\vc{x})$ 
to get rid of the divergent part and not forgetting the gauge correction
explained above), we find the 
bispectrum $B_\gz$ defined as:\footnote{In the literature one 
often sees a factor $(2\pi)^3$ in front of the bispectrum (as well as in 
front of the power spectrum). This is due to a different definition of the 
Fourier transform. We use the convention where both the Fourier transform
and its inverse have a factor $(2\pi)^{-3/2}$.}
\be
\langle 0|\zeta^{1}(\vc{k}_1) \zeta^{1}(\vc{k}_2) \zeta^{1}(\vc{k}_3)|0\rangle^{(2)}
\equiv (2\pi)^{-3/2}\delta^3 (\vc{k}_1 + \vc{k}_2 + \vc{k}_3)B_{\zeta}(k_1,k_2,k_3).
\label{defBzeta}
\ee
From this we can then finally define the (local) $\fnl$ parameter that is
generally used as a measure of (local bispectral) non-Gaussianity:\footnote{
  The left-hand side is called $-\frac{6}{5}\fnl$ and not simply $\fnl$ because
  it was originally defined in terms of the gravitational potential $\Phi$ and
  not $\gz$ as $\Phi = \Phi_\mathrm{L} + \fnl ( \Phi_\mathrm{L}^2 
- \langle \Phi_\mathrm{L}^2 \rangle )$ \citep{Komatsu:2001rj}. 
During recombination (matter domination) the two are 
related by $\gz = - \frac{5}{3} \Phi$. Moreover, when computing the bispectrum
divided by the three permutations of the power spectrum squared using this
expression of $\Phi$, one obtains $2\fnl$ due to the two ways the 
two $\Phi_\mathrm{L}$ inside the second-order solution can be combined with the 
two linear solutions to create the power spectrum. Together these two effects
explain the factor $-6/5$ (see also footnote~\ref{ftnt2} in chapter~\ref{NGCMBsec}).
In the papers before \citep{TvT1} we used a slightly
different definition of $\fnl$ which was larger by a factor $-18/5$, i.e.\
the definition would have $+\fnl/3$ on the left-hand side. \label{ftnt1}}
\be\label{fNL_start}
-\frac{6}{5}\fnl \equiv \frac{B_{\zeta
} (k_1,k_2,k_3)} {\frac{2\pi^2}{k_1^3}P_{\zeta}(k_1)\frac{2\pi^2}{k_2^3}P_{\zeta}(k_2)
  +(k_2\leftrightarrow k_3) + (k_1\leftrightarrow k_3)}.
\ee
Putting all the elements together and doing some calculations that
are specified in appendix~\ref{fNLderivsec}, one finally obtains:\footnote{It
  should be noted that this is only
  the part of $\fnl$ that comes from the three-point correlator of two
  first-order perturbations and one second-order perturbation
  (expressed as products of two first-order ones), sometimes called
  $\fnl^{(4)}$ in the literature (see e.g.~\citep{Vernizzi:2006ve}),
  which is the only contribution on super-horizon scales. It does not
  include the so-called intrinsic non-Gaussianity $\fnl^{(3)}$ due to
  interaction terms in the cubic action, which only play a role before
  and at horizon crossing and are necessarily slow-roll suppressed in
  models with standard kinetic terms.}
\be\label{fNLresult}
-\frac{6}{5}\fnl=\frac{-2\bv_{12}^2}{[1+(\bv_{12})^2]^2}
\Bigg( g_\mathrm{iso}+g_\mathrm{sr}+g_\mathrm{int}\Bigg), 
\ee
where
\bea{
 g_\mathrm{iso}&=(\ge+\eta^\parallel) (\bv_{22})^2 + \bv_{22} \bv_{32},
\qquad
g_\mathrm{sr}=-\frac{\ge_*+\eta^\|_*}{2\bv_{12}^2}+\frac{\eta^\perp_* \bv_{12}}{2}
-\frac{3}{2}\left(\ge_*+\eta^\parallel_*-\chi_*+\frac{\eta^\perp_*}{\bv_{12}}\right),
\nonumber\\
 g_\mathrm{int}&=- \int_{t_*}^t \d t' \Biggl[ 2 (\eta^\perp)^2 (\bv_{22})^2 
	+ (\ge+\eta^\parallel) \bv_{22} \bv_{32} + (\bv_{32})^2
- G_{13}(t,t') \bv_{22} \left( \Xi \bv_{22} + 9 \eta^\perp \bv_{32} \right)\Biggr].
\label{gisosrint}
}
Here we have defined
\be
\Xi \equiv 12 \eta^\perp \chi - 6 \eta^\parallel \eta^\perp
	+ 6 (\eta^\parallel)^2 \eta^\perp + 6 (\eta^\perp)^3
- 2 \eta^\perp \xi^\parallel- 2 \eta^\parallel \xi^\perp
	- \frac{3}{2}(\tW_{211} + \tW_{222}).
\label{Xi}
\ee
The explicit time dependence of all functions has been omitted, except for 
$G_{13}$ since it depends on two times.
Although no slow-roll approximation has been used on
super-horizon scales, we did assume slow roll to hold at horizon crossing,
in order to use the analytic linear solution~(\ref{Xapprox}).
The parameter $\fnl$ is weakly momentum dependent, as most of the momentum
dependence of $B_\gz$ is equal to the momentum dependence of the power
spectrum squared and is divided away in the definition (\ref{fNL_start}).
The (weak) momentum dependence of $\fnl$ is often neglected in the
literature, but strictly speaking the above expression is only valid
for equal momenta. We pointed this out in \citep{TvT1}, where we also gave
the expression for $\fnl$ with general momenta, and studied it further
in \citep{TvT3}, included in appendix~\ref{TvT3app}.

As a reminder, $\bv_{12}$ is the
contribution of the isocurvature mode to the adiabatic mode, which
according to (\ref{iso2adi}) only builds up if 1) the field
trajectory makes a turn ($\eta^\perp \neq 0$) and 2) an isocurvature mode is
present. It remains constant after the isocurvature mode has disappeared
and/or the field trajectory has become straight. Obviously it will be
zero in the single-field case. On the other hand, $\bv_{22}$ and $\bv_{32}$
are directly proportional to the isocurvature mode and will both go to zero
when the isocurvature mode disappears. We will in all our work assume that
this is the case by the end of inflation, so that the $g_\mathrm{iso}$
contribution to $\fnl$ vanishes there. If we relaxed our assumption of
the isocurvature mode going to zero by the end of inflation, it would be easy
to get huge non-Gaussianity at the end of inflation from the $g_\mathrm{iso}$
term (and indeed such papers exist, see e.g.~\citep{Byrnes:2008wi}),
but it would be meaningless from an observational point of view, since one
would have to follow its evolution
explicitly through the rest of the evolution of the universe to get a 
prediction for the observable. Only when the isocurvature mode has gone to
zero and we have returned to an effectively single-field situation, will
the adiabatic perturbation remain necessarily constant on super-horizon
scales, so that we can directly extrapolate the result at the end of
inflation to observables in the CMB.

In the single-field limit, a small slow-roll suppressed part of $g_\mathrm{sr}$
is all that survives and it gives back the $\fnl^{(4)}$ part of the usual 
single-field result of Maldacena \citep{Maldacena:2002vr}. All the other terms
of $g_\mathrm{sr}$ are also slow-roll suppressed, since
they are proportional to slow-roll parameters at horizon crossing. (It is
easy to check that the various functions of $\bv_{12}$ can never become
large, independent of the value of $\bv_{12}$.)
Hence persistent large non-Gaussianity can only come from the integrated
contribution $g_\mathrm{int}$, making this the most important term to study.
Unfortunately it is also the most complicated one, because of the integral
with the integrand depending explicitly on both $t'$ and $t$.

In \citep{TvT1} we showed that this integrated contribution can be computed
explicitly if we make two assumptions: that we can use the slow-roll
approximation and that we have a potential from a specific class, like a
sum or product potential. In particular we showed that the integrated
contribution vanishes, in the slow-roll approximation, for any product
potential (this proof was later repeated in \citep{JvT} which is included in
appendix~\ref{JvTapp}) as well as for any simple equal-power sum potential,
a proof that we reproduce here in the next subsection.
We also found an explicit model giving an
$\fnl$ of order unity (which we call large non-Gaussianity, as it is
two orders of magnitude larger than for standard single-field inflation;
this model was revisited in \citep{JvT} and so can be found in appendix~\ref{JvTapp}).

In \citep{JvT}, we went further and investigated sum potentials in much greater
detail, finding the regions of parameter space where large non-Gaussianity is
possible. In addition we found a way to rewrite the expression for
$g_\mathrm{int}$ that allowed us to draw conclusions even beyond the slow-roll
approximation. As \citep{JvT} is included in appendix~\ref{JvTapp} and
summarized below in section~\ref{summJvT}, we conclude here by giving that
different expression for $g_\mathrm{int}$:
\be
g_\mathrm{int}(t) = A \, G_{12}(t,t_*) + B \, G_{13}(t,t_*)
+ \int_{t_*}^t \d t' P(t').
\label{solgint}
\ee
The derivation is given in appendix~\ref{newgintsec} and is based on
deriving a differential equation for $\dot{g}_\mathrm{int}$ in closed
form, solving it and then integrating one more time.
Note that instead of $G_{12}$ we can also use $\bv_{12}$. Here $A$ and $B$
are constants to be determined from the initial conditions at horizon
crossing (where slow roll is a good approximation and $P(t)$ and its integral
can be computed explicitly in certain classes of potentials) and $P(t)$ is the
particular solution of the complicated differential equation (\ref{equadiff}).
The reason this expression containing an unknown function
is useful, is that it can be shown that under certain conditions that
are valid even beyond the slow-roll approximation, the term involving $P$
is negligible, see appendix~\ref{JvTapp}.

\subsection{Example: equal-power sum potentials}
\label{eqpowsumsec}

While a much more detailed treatment of the $g_\mathrm{int}$ contribution
to $\fnl$ based on (\ref{solgint}) for general sum potentials is given
in appendix~\ref{JvTapp}, we conclude this section with a simple application
of (\ref{gisosrint}) for $g_\mathrm{int}$ to equal-power sum potentials
in the slow-roll approximation, originally given in \citep{TvT1}.

Considering the slow-roll version of equation (\ref{G22eq}) we find
that $G_{22}(t,t')$ satisfies
\be\label{gSReq}
\frac{\d}{\d t} G_{22}(t,t') + \gc(t) \, G_{22}(t,t') = 0.
\ee
Hence $G_{32}(t,t')$, which according to (\ref{greent}) is the derivative
of $G_{22}$, is then given by
\be
G_{32}(t,t') = - \gc(t) G_{22}(t,t').
\ee
We also know from appendix~\ref{greensec} that $G_{x3}=G_{x2}/3$ in the
slow-roll approximation.
Using these results and dropping higher-order terms 
in slow roll, (\ref{gisosrint}) reduces to
\ba
\label{srtot}
g_\mathrm{iso}&\!\!\!\!=&\!\!\!\!(\ge +\getpa-\chi)(\bv_{22})^2,\quad\;\;
g_\mathrm{sr}=-\frac{\ge_*+\getpa_*}{2\bv_{12}^2}+\frac{\get^\perp_* \bv_{12}}{2}
-\frac{3}{2}\lh\ge_*+\get^\parallel_*-\gc_*+\frac{\getpe_*}{\bv_{12}}\rh\!,\\
g_\mathrm{int}&\!\!\!\!=&\!\!\!\!\int_{t_*}^t\!\!\!\!\d t' (\bv_{22})^2 \Bigg[ 2 \get^\perp
\!\lh\! -\! \get^\perp\! +\! \frac{(\ge\!+\!\get^\parallel\!-\!\gc)\gc}
{2\get^\perp}\rh
\!+G_{12} \lh \get^\perp \gc\! -\! 2 \get^\parallel 
\get^\perp\!-\!\frac{1}{2} (\tW_{211} \!+\! \tW_{222}) \rh \Bigg].\nn
\ea

It also proves useful to rewrite $g_\mathrm{int}$ in a different way
using integration by parts.
We use the slow-roll version of relation (\ref{greentprime}),
$2\get^\perp = - \frac{\d}{\d t'} G_{12}(t,t') + \gc G_{12}(t,t')$, to do an
integration by parts, leading to
	\ba\label{Keq}
	g_\mathrm{int} & = & \bv_{12} \lh -\get^\perp_* + 
	\frac{(\ge_* + \get^\parallel_* - \gc_*)\gc_*}{2 \get^\perp_*} \rh
+ \int_{t_*}^t \d t' G_{12} (\bv_{22})^2 
        \Biggl[ 2 \get^\perp \gc
	- \frac{(\ge+\get^\parallel-\gc)\gc^2}{2\get^\perp} \nn\\
	&&\qquad\qquad\; - 2 \get^\parallel \get^\perp
	-\frac{1}{2} (\tW_{211} + \tW_{222}) 
	+ \frac{\d}{\d t'} \lh - \get^\perp 
	+ \frac{(\ge+\get^\parallel-\gc)\gc}{2\get^\perp} \rh
	\Biggr].
	\ea
Using the slow-roll version of the relations (\ref{srderivatives}),
	\be\label{xiparxiperpSR}
	\gx^\parallel = 3 \ge \get^\parallel + (\get^\parallel)^2
	+ (\get^\perp)^2 - \tW_{111}
	\ \ \ \ \ \ \ \mbox{and}\ \ \ \ \ \ \  
	\gx^\perp = 3 \ge \get^\perp + 2 \get^\parallel \get^\perp
	- \get^\perp \gc - \tW_{211},
	\ee
as well as the other time derivatives of the slow-roll parameters in
(\ref{srderivatives}), we can derive that
	\ba
	\frac{\d}{\d t}\!\lh\!-\get^\perp +\frac{(\ge+\get^\parallel-\gc)\gc}
{2\get^\perp} \rh
	&\!\!\!\!\!=&\!\!\!\! \frac{1}{2\get^\perp} \Bigg[ -\gc^3 + (\ge+\get^\parallel)\gc^2
- 4 \lh \ge\get^\parallel+(\get^\perp)^2 \rh \gc\\
	&&+ 4 \lh \ge^2 \get^\parallel + \ge(\get^\parallel)^2
	- \ge(\get^\perp)^2 + \get^\parallel(\get^\perp)^2 \rh \!-\!(\ge+
\get^\parallel-\gc)\tW_{111}\nn\\
	&&+ \lh 2\get^\perp + \frac{(\ge+\get^\parallel-\gc)\gc}
{\get^\perp}\rh \tW_{211} 
	+ (\ge+\get^\parallel-2\gc)\tW_{221}\Bigg].\nn
\ea
Inserting this into expression (\ref{Keq}) for $g_\mathrm{int}$ and including the
remaining terms in the expression for $\fnl$ we finally obtain
\ba\label{fNLresultSR}
-\frac{6}{5}\fnl(t) &\!\!\!\!=&\!\!\!\! \frac{-2(\bv_{12})^2}{[1+(\bv_{12})^2]^2}
\Biggl\{ (\ge+\get^\parallel-\gc) (\bv_{22})^2
-\frac{\ge_*+\getpa_*}{2\bv_{12}^2}-\frac{\get^\perp_* \bv_{12}}{2}+\! \frac{(\ge_*\! +\! \get^\parallel_* \!-\! \gc_*)
\gc_*}{2 \get^\perp_*}\, \bv_{12} \nn\\ 
&& \qquad\qquad\quad
-\frac{3}{2}\lh\ge_*+\get^\parallel_*-\gc_*+\frac{\getpe_*}{\bv_{12}}\rh 
\nn\\
&&+ \int_{t_*}^t \d t' G_{12} (\bv_{22})^2 
\Bigg[2\frac{\ge\get^\parallel}{\get^\perp} 
	\lh\! -\gc\! +\! \ge\! +\! \get^\parallel\! -\!
	\frac{(\get^\perp)^2}{\get^\parallel} \rh\!+\!\frac{1}{2}(\tW_{211}\! 
-\! \tW_{222}\! -\! \frac{\gc}{\get^\perp} \tW_{221})
\nn\\
&& \qquad\qquad\qquad\qquad
- \frac{\ge+\get^\parallel-\gc}{2\get^\perp}
\lh \tW_{111} \!-\! \tW_{221}\! -\! \frac{\gc}{\get^\perp} \tW_{211} \rh
\Bigg] \Biggr\}.
\ea
This is an alternative result for $\fnl$ in the slow-roll approximation.

Equation (\ref{fNLresultSR}), as well as (\ref{srtot}), is characterized by the
same features as the result of the exact formalism (\ref{gisosrint}).
We can easily
distinguish the pure isocurvature $\bv_{22}$ term, which we
assume to vanish before the end of inflation in order for the
adiabatic mode to be constant after inflation, as well as the terms evaluated
at the time of horizon crossing, which are expected to be small. Any
remaining non-Gaussianity at recombination has to originate from the integral.
We will now further work out this expression for the case of the
quadratic sum potential, as well as for a more general equal-power sum
potential.

\subsubsection{Quadratic potential}

The quadratic potential has been widely examined in the past and it is
known that it cannot produce large non-Gaussianity (see for example
\citep{Vernizzi:2006ve}).  Here we use our results to analytically
explain why. While the quadratic potential is a special case of the
more general equal-power sum potential treated below, it is still
interesting to discuss it separately in a different way. We start by
deriving the result that for a quadratic two-field potential
$W(\phi,\sigma) = \frac{1}{2} m_\phi^2 \phi^2 + \frac{1}{2} m_\sigma^2 \sigma^2$
within slow roll,
	\be\label{chirel}
	\gc = \frac{\d}{\d t} \ln \frac{\ge \get^\perp}{\get^\parallel}.
	\ee
Working out the right-hand side, using (\ref{srderivatives}), we find
	\be
	\gc
	= 2\ge + \get^\parallel - \frac{(\get^\perp)^2}{\get^\parallel}
	- \frac{\gx^\parallel}{\get^\parallel} 
	+ \frac{\gx^\perp}{\get^\perp}.
	\ee
Inserting the relations (\ref{xiparxiperpSR}) (with the third derivatives of the
potential equal to zero, since we have a quadratic potential) this reduces to
	\be
	\gc = \ge + \get^\parallel - \frac{(\get^\perp)^2}{\get^\parallel}.
	\ee
It can be checked that this result does indeed satisfy the general equation 
for the time derivative of $\gc$ (\ref{srderivatives}) within the approximations made,
and the remaining integration constant is fixed by realizing that this result
has the proper limit in the single-field case. This concludes the proof of
(\ref{chirel}).

Since the third-order potential derivatives as well as the first term of the
integral in (\ref{fNLresultSR}) are identically zero, we find that for
a quadratic potential the integral completely vanishes in the slow-roll
approximation and no persistent large non-Gaussianity is produced. 
Numerically we find that even for large mass ratios, where during the turn
of the field trajectory slow roll is broken, the integral is still
approximately zero; see appendix~\ref{JvTapp} for a detailed discussion.

\subsubsection{Potentials of the form $W=\alpha \phi^p+\beta \gs^p$}

For a potential of the form
\be
W(\phi,\gs)=\alpha \phi^p+\beta \gs^q
\label{polynomtype}
\ee
we can work out explicitly the form of the integrand in equation
(\ref{fNLresultSR}). We have to use the slow-roll version of the background
field and Friedmann equations as well as of the slow-roll parameters
to easily find after substitution that
\be
 \tg_\mathrm{int}\!=\!-\!\!\int_{t_*}^t\!\!\frac{\alpha\beta p^4 
(y\!-\!1)\phi^{p-3}\gs^{py-3}\left(y(p y\!-\!1)\phi^2
+(p\!-\!1)\gs^2\right)
\left(\alpha^2\phi^{2 p}\gs^2\!+\!\beta^2 y^2\phi^2\gs^{2 p y}\right)^2\!\!}
{2\kappa^4\left(\alpha\phi^p+\beta\gs^{p y}\right)^4
\left(\alpha(p-1)\phi^p\gs^2-\beta y (py-1)\phi^2\gs^{p y}\right)^2}
\d t',
\ee
where $y\equiv q/p$. We use the tilde here to indicate that this is not
exactly the same $g_\mathrm{int}$ as before, since we have split off a part
using integration by parts, see (\ref{Keq}).

From this expression we can derive an important
result: for $y=1$, i.e.\ $p=q$, we immediately
see that the integral is zero. This means that no persistent
non-Gaussianity can be produced after horizon exit for potentials
of the form $W(\phi,\gs)=\alpha \phi^p+\beta\gs^p$, at least within
the slow-roll approximation (with our usual assumption of vanishing
isocurvature modes at the end of inflation). This generalizes the result for
the two-field quadratic potential to any two-field monomial sum potential
with equal powers.

\section{Summary of additional results}
\label{longwavother}

Having discussed the long-wavelength formalism in detail, as synthesized from
several papers, we can now start looking at applications and further
extensions. As these were generally published in self-contained papers, these
papers have simply been added verbatim to the appendices of this chapter,
removing only their conclusions and those sections that contain material
already covered elsewhere in this thesis. In this section we provide a
summary of those papers, based on their conclusions.
The papers in question are \citep{JvT} (section~\ref{summJvT} and
appendix~\ref{JvTapp}) containing specific applications of the formalism,
\citep{TvT2} (section~\ref{summTvT2} and appendix~\ref{TvT2app}) containing
a proper treatment of gauge invariance at second order and a derivation of the
exact cubic action of the second-order perturbations, and \citep{TvT3}
(section~\ref{summTvT3} and appendix~\ref{TvT3app}) discussing momentum
dependence of the bispectrum and $\fnl$.

\subsection{Explicit solutions}
\label{summJvT}

In the article \citep{JvT}, reproduced in appendix~\ref{JvTapp},
we discussed the levels of non-Gaussianity produced
in two-field inflation with a sum potential\footnote{For comparison we also
looked at the case of a product potential.
As was shown before, in that case one cannot get large non-Gaussianity at
all in the slow-roll approximation and with a vanishing isocurvature mode
at the end of inflation.}
$W(\phi,\sigma)=U(\phi)+V(\sigma)$ and standard kinetic terms. We
looked both at the case where the (strong) slow-roll approximation is valid
throughout inflation (meaning that all slow-roll parameters, even the
perpendicular ones, are small), and at the case where slow roll is broken 
during the turn of the field trajectory. An important assumption in our
models is that we impose that the isocurvature mode that is present
during inflation (and whose interaction with the adiabatic mode on
super-Hubble scales generates the non-Gaussianity) has disappeared by
the end of inflation. In that case the super-Hubble adiabatic mode is
constant after inflation and we can extrapolate the results at the end
of inflation directly to the time of recombination and observations of
the CMB without knowing any details about the evolution of the universe
in between. Without this assumption it would be much easier to
create large non-Gaussianity, simply by ending inflation in the middle
of the turn, but the result at the end of inflation would be
meaningless from the point of view of CMB observations without a
proper treatment of the transition at the end of inflation and the
consecutive period of (p)reheating.

We highlighted the tension between a large $\fnl$ (of order unity
or more) and the current observational bounds on the spectral index
$n_s$, both being linked to the second derivative of the potential
$V_{\gs\gs}$, where $\sigma$ is the sub-dominant field at horizon
crossing and until the turn of the field trajectory.  We evaluated
these tensions (within the slow-roll approximation) for monomial
potentials, where it would otherwise be easy, with some fine-tuning,
to reach the requirements for a large $\fnl$.  We showed that a
large part of the parameter space for $\fnl$ of order unity
is simply forbidden because of the constraints on $n_s$. However, we
found that these constraints are very sensitive to the value of $n_s$:
if the lower bound were only smaller by 0.02 ($n_s$ of order 0.94),
the situation would be completely different and most of the parameter
space would be allowed.
This analysis of the monomial potential also
revealed that the duration of inflation after horizon-crossing is
important: a value around fifty e-folds is much more constraining than
the usual sixty e-folds. This also indicates that in the rare working
models, the turn of the field trajectory occurs near the end of
inflation. This raises several issues, the main one being that at that
time, slow-roll parameters generally stop to be small compared to one
and the slow-roll approximation does not work anymore. Moreover, if
the turn occurs too close to the end of inflation, the isocurvature mode
may not have time to vanish. By studying turns where the slow-roll
parameter $\ge$ is still small compared to one we avoid this problem:
the time $\ge$ needs to increase to one and end inflation can give
enough time for the isocuvature mode to vanish.

The natural continuation of this study was to consider what would
happen if we abandoned the slow-roll approximation during the turn and
allowed the slow-roll parameters $\getpa$ and $\getpe$ to become large
there.  On the other hand, we still assume that $\ge$ remains small
during the turn, for several reasons: because of the issue regarding
the vanishing of the isocurvature mode mentioned
above, because we saw numerically in the models we looked at that this
was a good approximation, and because this approximation allowed us to
derive some very interesting analytical results (a potential period of
large $\ge$ right before the turn was taken into account though).  We
identified two different types of models where such a turn can happen, shown
in figure~\ref{fig:slowrollbroken}. Substituting the slow-roll
expression for $\dot{g}_\mathrm{int}$ into \eqref{equadiff} (of which
(\ref{solgint}) is the full solution), we were able to
show (using simple comparisons of the different terms of the
differential equation) that it is also a very good approximation even
if the slow-roll parameters $\getpa$ and $\getpe$ become large during
the turn. The main idea is the following: as long as the
slow-roll approximation is valid, we can compute the particular
solution (see (\ref{solgint})) explicitly, while when it is broken, we can
show that the
particular solution becomes negligible, even though we cannot compute
an analytic expression for it in that case (the fact that $\ge$
remains small is a crucial ingredient in this proof). For the
homogeneous solution we have an analytic expression that is valid
everywhere. We were also able to show that adding the slow-roll
particular solution to the homogeneous solution in the regions where
the exact particular solution is negligible does not introduce a
significant error, which means that we do not have to perform an explicit
matching of the solutions at each transition between a slow-roll and a
non-slow-roll region.

This led us to the conclusion that, within the context of the models studied
and the assumptions mentioned above, the slow-roll expression for $\fnl$
is a very good approximation for the exact value, even in models where
$\getpa$ and $\getpe$ become large during the turn of the field trajectory
and break slow roll. Hence the implications of this expression for having
large non-Gaussianity, discussed in the context of the slow-roll approximation,
mostly apply to this case as well. In particular, the constraints due
to the spectral index $n_s$ remain very important.
A two-field sum potential with large $\fnl$ requires a lot of fine-tuning
(and we showed explicitly in the section with numerical examples how to
construct such a model). Reducing the error bars on the measurements of 
the spectral index could even shrink the parameter region of these models 
where $\fnl$ is of order unity more than reducing the error bars on $\fnl$.

\subsection{Gauge invariance and cubic action}
\label{summTvT2}

In the paper \citep{TvT2}, reproduced in appendix~\ref{TvT2app}, we settled some previously unresolved issues concerning gauge invariance at second order in inflation with 
more than one field. Although the gauge-invariant 
curvature perturbation defined through the energy density had been known for many years, 
the energy density is not the quantity that is used in 
calculations of inflationary non-Gaussianity. These use the scalar fields present during inflation instead of their energy. We found this 
gauge-invariant quantity in terms of the fields and discovered that it contains a non-local term unless slow-roll is assumed.

We also managed to make contact between gauge transformations and the redefinitions of the curvature and 
isocurvature perturbations occurring in the third-order action. 
Since \citep{Maldacena:2002vr} it was known that the redefinition of the curvature 
perturbation in the action, introduced to remove terms proportional to the first-order equations of motion,  
corresponds to its gauge transformation. 
However, these terms  
appeared at first sight to be absent in the flat gauge which would have had as a consequence the absence of 
quadratic contributions of first-order 
curvature perturbations at horizon crossing in this gauge, and hence a gauge dependence of the related horizon-crossing 
non-Gaussianity.  
We extended the calculation for both gauges to second order and proved that in both of them the 
contributions are 
the same. The difference is that, in our perturbative approach, in the uniform energy-density gauge a part of these contributions is due to the first-order corrections 
and the other part to the second-order fields, while in the flat gauge they are all due to the second-order fields. 

In addition to the adiabatic one, we also found the gauge-invariant isocurvature perturbation defined in terms of 
the scalar fields, by studying the relevant fully non-linear 
spatial gradient (\ref{zeta_var}), first defined in \citep{RSvT2}. Usually isocurvature perturbations are studied in terms of 
the pressure of the fields. However, we found 
a definition using the fields themselves that demonstrates the orthogonality of this quantity to the curvature perturbation. While rewriting the action, these 
isocurvature perturbations appear naturally in the form we have defined them, thus showing that this quantity is the relevant 
one to use during inflation.

In order to achieve the above we computed the exact cubic action for the adiabatic and isocurvature perturbations, 
going beyond the 
slow-roll or super-horizon approximations (in the appendix we also gave the tensor part of the action). 
Previously there was no alternative to imposing the slow-roll condition at horizon crossing in 
order to calculate the non-Gaussianity. This was because the only two-field action available was that of the fields given in 
\citep{Seery:2005gb}, thus demanding slow-roll at horizon crossing in order to be able to use the long-wavelength 
formalism or the $\delta N$ formalism to find the curvature perturbation bispectrum. 
The action we provided here can be used directly with the in-in formalism \citep{Weinberg:2005vy} in order to 
calculate the exact non-Gaussianity beyond any restrictions, slow-roll or super-horizon.

\subsection{Momentum dependence}
\label{summTvT3}

In the paper \citep{TvT3}, included in appendix~\ref{TvT3app}, we studied the scale dependence of the local non-Gaussianity
parameter $\fnl$ for 
two-field inflationary models. Multiple-field models with standard kinetic terms
do not exhibit 
the strong scale dependence inherent in models that produce equilateral non-Gaussianity 
at horizon-crossing through quantum mechanical effects. Nevertheless, they are not 
scale independent in general and the interesting question is whether we can  
profit from their scale dependence in order to observationally acquire more 
information about inflation. 

We calculated $\fnl$ using the long-wavelength formalism. This 
constrained us to assume slow roll at horizon-crossing and hence the relevant 
quantities at that time should not vary much, including the 
scale dependence of $\fnl$ for any shape of the momentum triangle. Indeed we confirmed that 
by introducing the conformal spectral index $n_K$ that measures the tilt of $\fnl$
for triangles of the same shape but different size ($K$ is a variable 
proportional to the perimeter of the momentum triangle). For the 
quadratic model with mass ratio $m_{\phi}/m_{\gs}=9$ we found $n_K\simeq0.018$, 
pointing to an almost scale-invariant $\fnl$. 

We also studied the scale dependence of $\fnl$ while varying the shape of the 
triangle and keeping its perimeter constant. $\fnl$ exhibits the opposite behaviour of the full bispectrum, 
i.e.\ it decreases the more squeezed the triangle is (the momentum dependence
of the bispectrum is dominated by that of the products of power spectra, not by
that of $\fnl$). This variation 
is not related to horizon-crossing quantities, but rather to the fact that 
the more squeezed the isosceles triangle under study is, 
the smaller the correlation of its two scales. 
We quantified this effect by introducing the shape spectral index $n_\omega$, which 
for the quadratic model with $m_{\phi}/m_{\gs}=9$ is $n_{\omega}\simeq-0.03$ and has a running of about 
$20\%$ ($\omega$ is defined as the ratio of the
two different sides of an isosceles momentum triangle).

All our calculations were done numerically in the exact background, 
assuming slow roll only at horizon crossing, not afterwards. 
Nevertheless, semi-analytical expressions can be easily produced by directly 
differentiating $\fnl$. If we do assume slow roll, we showed that we can even simplify these expressions further 
and find analytical 
formulas for the final value of $\fnl$ and its spectral indices $n_K$ and 
$n_\omega$, if  
the integral in $\fnl$ and the isocurvature modes vanish by the end of inflation,
which is the case for example for any equal-power sum model. 

We used the two-field quadratic potential in our numerical calculations.
This potential is easy to examine and allows for simplifications in the relevant 
expressions. Although its final non-Gaussianity is small, $\mathcal{O}(\ge_k)$, its 
general behaviour should not be different from other multiple-field 
inflationary models with standard kinetic terms, in the sense 
that the scale dependence of $\fnl$ 
should always depend on horizon-exit quantities and the evolution of the transfer functions during the turning of the fields.  
Indeed we checked that for the potential (\ref{pot}) studied in 
\citep{TvT1}, able to produce $\fnl\sim\mathcal{O}(1)$, the results 
remain qualitatively the same, although the values of the spectral indices are smaller due to the very slow evolution 
of the background at horizon crossing in that model.

Although 
the effect of the magnitude of the triangle on $\fnl$ had been considered before, analytical 
and numerical estimates were not available before this paper. In addition, it was the first time that 
the dependence of $\fnl$ itself (instead of the power spectra in the bispectrum) on the shape of the momentum triangle was studied. Using the long-wavelength 
formalism we managed to study the two different sources of momentum dependence, i.e.\ the slow-roll parameters at horizon 
crossing and the evolution of the transfer functions, and to understand the role of each for the two different triangle deformations 
that we studied. In summary, the later a momentum mode exits the horizon, the larger the slow-roll parameters are at that time and 
the larger $\fnl$ tends to be. In contrast, the final value of $\bv_{12k'}$ and the initial value of $G_{22k'k}$, the 
two transfer functions that are the most important for $\fnl$, are smaller the later the scale exits, which results 
in decreasing values of $\fnl$. These two opposite effects manifest themselves in the two different deformations we studied. When keeping the shape of 
the triangle constant and varying its size, it is the slow-roll parameters at horizon crossing that play the major role in $\fnl$ and result in an increasing $\fnl$ for  
larger $K$. When changing the shape of $\fnl$, it is the correlation between the isocurvature mode at different scales, $G_{22k'k}$, that has the most 
important role, resulting in decreasing values of $\fnl$ when squeezing the triangle (i.e.\ increasing $\omega$). 

We verified that the spectral indices of $\fnl$ ($n_K$ and
$n_\omega$), which we introduced to describe the effect of the two
types of deformations of the momentum triangle, provide a good
approximation over a wide range of values of the relevant scales.  In
the models we studied their values are too small to be detected by
Planck, given that $\fnl$ itself cannot be big (or it would have been
detected by Planck).  Models that break slow roll at horizon crossing
could in principle have larger spectral indices, but in order to study
such models one would need to go beyond the long-wavelength
formalism. Such models could be studied using the
exact cubic action derived in \citep{TvT2}.

\section{Appendix: details of the calculations}
\label{calcdetailssec}

This section contains three appendices. The first discusses in more detail the
Green's functions that are central to solving the equations of motion for
the fluctuations, determined using the long-wavelength formalism, at each order
in a perturbative expansion. The second and third appendices give some of the
calculations to derive the main results in section~\ref{longwaveq}. As these
calculations tend to be rather involved, those appendices can be skipped by
the more casual reader.

\subsection{Green's functions}
\label{greensec}

The functions $G_{xy}(t,t')$ (with $x,y=1,2,3$ and $t \geq t'$) are Green's 
functions introduced to solve the first-order perturbation equations (and
then the same functions also serve to solve the second-order equations). 
They satisfy the following differential equations (simply (\ref{Green})
written in components):
\be
\begin{split}
	&\frac{\d}{\d t} G_{1y}(t,t') = 2\getpe(t) G_{2y}(t,t'),\\
	&\frac{\d}{\d t} G_{2y}(t,t') = G_{3y}(t,t'),\\
	&\frac{\d}{\d t} G_{3y}(t,t') = - A_{32}(t) G_{2y}(t,t') - A_{33}(t) G_{3y}(t,t'),
\end{split}
\label{greent}
\ee 
with the matrix $A$ given in (\ref{A2f}),
as well as the following differential equations in terms of the time $t'$:
\be
\begin{split}
	&\frac{\d}{\d t'} G_{x2}(t,t') = -2\getpe(t')\delta_{x1} + A_{32}(t') G_{x3}(t,t'),\\
	&\frac{\d}{\d t'} G_{x3}(t,t') = -G_{x2}(t,t') + A_{33}(t') G_{x3}(t,t').
\end{split}
\label{greentprime}
\ee
The initial conditions are $G_{xy}(t,t)=\delta_{xy}$. We can also combine
the equations~(\ref{greent}) into a second-order differential equation for
$G_{2y}$ in closed form:
\be
\frac{\d^2}{\d t^2} G_{2y}(t,t') + A_{33}(t) \frac{\d}{\d t}
G_{2y}(t,t') + A_{32}(t) G_{2y}(t,t') = 0.
\label{G22eq}
\ee
For $y=1$, the solutions are: $G_{11} = 1$, $G_{21} = G_{31} = 0$. For $y=2,3$ 
we need to make some approximations to solve the equations analytically.
From (\ref{vbar}) we have the short-hand notation
\be
\bv_{x2}(t) \equiv G_{x2}(t,t_*)-\chi_* G_{x3}(t,t_*).
\label{defvbar}
\ee
This means that $\bv_{12*}=0$, $\bv_{22*}=1$ and $\bv_{32*}=-\chi_*$. 
The functions $\bv_{x2}$ satisfy the same differential equation (\ref{greent}) in terms 
of $t$ as the $G_{x2}$.

In the general case, these equations cannot be solved analytically. Hence, to go further, we will focus on the case $t'=t_*$ and we assume that at horizon-crossing the slow-roll approximation is valid for at least a few e-folds. This means that during these few e-folds, the different slow-roll parameters, which evolve slowly, can be considered as constants at the lowest order. Under these conditions, the differential equation \eqref{G22eq} takes the form:
\be
\ddot{g}(t) + A_{33} \dot{g}(t) + A_{32}g(t)=0,
\ee
where $g$ can be either $G_{22}$, $G_{23}$ or $\bv_{22}$, differing only in initial condition. Here, $A_{32}$ and $A_{33}$ are now constants. The solution of this equation is:
\be
g(t)=\frac{1}{\lambda_- - \lambda_+} \left[ (\lambda_- g_0 - \dot{g}_0) e^{\lambda_+ t} + (-\lambda_+ g_0 + \dot{g}_0)e^{\lambda_- t}\right],
\label{solconstant}
\ee
where $\lambda_+= \frac{1}{2}\lh-A_{33}+\sqrt{(A_{33})^2-4A_{32}}\rh$, $\lambda_-= \frac{1}{2}\lh-A_{33}-\sqrt{(A_{33})^2-4A_{32}}\rh$ and $g_0$, $\dot{g}_0$ are the initial values of $g$ and $\dot{g}$. In the slow-roll regime, $|A_{32}|\ll 1$ while $A_{33}\approx 3$. The direct consequence is that $|\lambda_+|\ll 1$, which implies that the $e^{\lambda_+ t}$ mode does not change much in a few e-folds, while $\lambda_-\approx -3$, which means that the other mode decays exponentially and can be neglected after a few e-folds (three is sufficient).

For two different sets of initial conditions, the ratio between the solutions becomes:
\be
\frac{g_1}{g_2}=\frac{\dot{g}_1}{\dot{g}_2}=\frac{\lambda_- g_{1_0} - \dot{g}_{1_0}}{\lambda_- g_{2_0} - \dot{g}_{2_0}},
\label{propor}
\ee
which is a constant. Hence, $G_{22*}$ (defined as $G_{22}(t,t_*)$), $G_{23*}$ and $\bv_{22}$ become proportional after a few e-folds of slow-roll. Then, after a few more e-folds of inflation, the approximation of constant slow-roll parameters stops to be valid and we can no longer consider $A_{32}$ and $A_{33}$ to be constants. However, by this time the proportionality between $G_{22*}$, $G_{23*}$, $\bv_{22}$ and their derivatives $G_{32*}$, $G_{33*}$, $\bv_{32}$ has been established, and because of the linearity of the differential equation \eqref{G22eq}, they will stay proportional until the end of inflation.

The case of $G_{12}$, $G_{13}$ and $\bv_{12}$ is a little trickier. With $\getpe$ being a constant, these functions are the primitives of $G_{22}$, $G_{23}$, $\bv_{22}$ according to \eqref{greent}. However, one does not obtain the same factor of proportionality \eqref{propor} with a simple integration of \eqref{solconstant} because of the constant of integration. On the other hand, from \eqref{greent} we know these functions stay small compared to one before the turn of the field trajectory, because $\getpe$ is negligible compared to other slow-roll parameters. During the turn, while $\getpe$ is of the same order as other slow-roll parameters or even larger, they can become large. We will see later (in chapter~\ref{JvTapp}) that typical and interesting values of $\bv_{12}$ are larger than order unity. Hence, the only relevant part of the integral is after the beginning of the turn. To compute it, one can just integrate the first equation of \eqref{greent} starting at the beginning of the turn instead of at horizon-crossing. Moreover, once the turn has started, we know that the relations of proportionality between $G_{22*}$, $G_{23*}$ and $\bv_{22}$ are already established, which means that from \eqref{greent} the same relations exist between $\dot{G}_{12*}$, $\dot{G}_{13*}$ and $\dot{\bv}_{12}$ on the only relevant part of the integration interval. Then the common factor is conserved by the integration. During the turn, \eqref{propor} becomes valid for the Green's functions $G_{12*}$, $G_{13*}$ and $\bv_{12}$. In particular this is true for the final values of these functions.
If these functions stay negligible during the turn, or vanish at the end, the result does not hold. However, as already mentioned, this case is not interesting as multiple-field effects will play no role. To summarize, the explicit proportionality relations are:
\be 
\frac{G_{12*}}{G_{13*}}=\frac{G_{22*}}{G_{23*}}=\frac{G_{32*}}{G_{33*}}=-\lambda_-\approx 3 \qquad \text{and} \qquad \frac{\bv_{12}}{G_{12*}}=\frac{\bv_{22}}{G_{22*}}=\frac{\bv_{32}}{G_{32*}}=\frac{\lambda_-+\chi_*}{\lambda_-}\approx 1.
\label{proportionality}
\ee

\subsection{Derivation of expression (\ref{fNLresult}) for $\fnl$}
\label{fNLderivsec}

In the case of equal momenta, equation (\ref{fNL_start})
reduces to
\bea{\label{fNL}
-\frac{6}{5}\fnl =  \frac{-v_{1m}(t) v_{1n}(t)}{\lh v_{1m}(t) v_{1m}(t) \rh^2}
\Bigg\{&\!\!\int_{-\infty}^t\!\!\!\!\!\d t' G_{1a}(t,t') 
\bA_{abc}(t') v_{bm}(t') v_{cn}(t')
- 2 \get^\perp(t) v_{2m}(t) v_{1n}(t)\nn\\
&-G_{1a}(t,t_*)(L_{abc*}+N_{abc*}) v_{bm}(t_*)v_{cn}(t_*)\Bigg\}.
}
We remind the reader that indices $l, m, n$ take the values 1 and 2 (components
in the two-field basis), while indices $a,b,c,\ldots$ take the values 1, 2, 
and 3 (labeling the $\gz^1$, $\gz^2$, and $\gth^2$ components).
The $v_{1m}v_{1n}$ in the numerator comes from the product of two first-order
solutions, while the term between the braces comes from the second-order
solution. The $(v_{1m}v_{1m})^2$ in the denominator comes from the division
by the power spectrum squared. The first and last term between the braces
correspond to the first and last term on the right-hand side of (\ref{vimeq2}),
respectively. The second term is the gauge correction explained in the main
text above (\ref{defBzeta}). The explicit form of the object $\bA$ is:
\ba
\bA_{121} & = & 2\ge\get^\perp
-4\get^\parallel\get^\perp + 2\gxpe, \nn\\
\bA_{122} & = & -6\chi - 2\ge\get^\parallel - 2(\get^\parallel)^2 
- 2(\get^\perp)^2, \nn\\
\bA_{123} & = &	-6 - 2\get^\parallel, \nn\\
\bA_{321}&=&-12\ge\get^\parallel - 12(\get^\perp)^2 
- 6\ge\chi - 8\ge^3 - 20\ge^2 \get^\parallel 
- 4\ge(\get^\parallel)^2 - 12\ge(\get^\perp)^2\nn\\
&&+ 16\get^\parallel(\get^\perp)^2 - 6\ge\gxpa
- 12\get^\perp\gxpe + 3 (\tilde{W}_{111}-\tilde{W}_{221}),\nn\\
\bA_{322}&=&-24\ge\get^\perp - 12\get^\parallel\get^\perp 
+ 24\get^\perp\chi - 12\ge^2\get^\perp + 8(\get^\parallel)^2\get^\perp 
+ 8(\get^\perp)^3\nn\\
&& - 8\ge\gxpe - 4\get^\parallel\gxpe
+ 3 (\tilde{W}_{211}-\tilde{W}_{222}),\nn\\
\bA_{323}&=& 12\get^\perp - 4\ge\get^\perp + 8\get^\parallel\get^\perp
- 4\gxpe, \nn\\
\bA_{331} & = & -2\ge^2 - 4\ge\get^\parallel +
2(\get^\parallel)^2 - 2(\get^\perp)^2 - 2\gxpa, \nn\\
\bA_{332} & = &	-4\ge\get^\perp - 2\gxpe, \nn\\ 
\bA_{333} & = & -2\get^\perp,
\label{Abar}
\ea
while the rest of the elements are zero. The explicit form of the
objects $L$ and $N$ at horizon crossing (where slow roll holds) is:
\ba
&&L_{111*} =\ge_*+\getpa_*,\qquad\qquad\qquad\qquad L_{122*}=-\lh\ge_*+\getpa_*-\chi_*\rh,\nn\\
&&L_{211*} =\getpe_*,\qquad\qquad\qquad\qquad\ \ \ \ \ \ L_{222*}=\getpe_*,\nn\\
&&L_{112*}+L_{121}=2\getpe_*,
\ \ \ \ \qquad\qquad\ \ \  N_{112*}+N_{121*}=-2\getpe_*,\nn\\
&&L_{212*}+L_{221*}=2\lh\ge_*+\getpa_*-\chi_*\rh,
\  N_{212*}+N_{221*}=\chi_*,
\label{LandN}
\ea
with the other elements of $N_{abc}$ being zero. A slow-roll approximation
which expresses $\gth^2$ in terms of $\gz^2$ has 
been used: $\gth^2 = -\chi\gz^2$. This means in particular that the subscripts
$a,b,c$ only take the values 1 and 2, but not 3. However, for consistency in
the notation, we will define here all entries of $L_{abc}$ and $N_{abc}$ to be
equal to zero if one or more of the indices are equal to 3.

To make the expressions a bit shorter, we will drop the
time arguments inside the integrals, but remember that for the Green's
functions the integration variable is the second argument.
One can show that $\bA_{ab1} = - \partial_t A_{ab}$ and hence do an
integration by parts, with the result
	\be\label{fNLint}
	\int_{-\infty}^t\!\!\!\!\!\d t' G_{1a} \bA_{abc}
	v_{bm} v_{cn}
	= 2 \get^\perp v_{2m} v_{1n}
	+ \int_{-\infty}^t\!\!\!\!\! \d t' A_{ab} \frac{\d}{\d t'} 
	[ G_{1a}v_{bm} v_{1n} ]+ 
	\int_{-\infty}^t \!\!\!\!\!\d t' G_{1a} \bA_{ab\bc}
	v_{bm} v_{\bc n},
	\ee
where the index $\bc$ does not take the value 1.
Here we used that the linear solutions $v_{am}$ are zero at $t=-\infty$ 
(by definition), that the Green's function $G_{1a}(t,t) = \gd_{1a}$, and that 
$A_{1b} = -2\get^\perp \gd_{b2}$ (exact).
We see that the first term on the right-hand side exactly cancels 
with the gauge correction (the second term in (\ref{fNL})) that is necessary to 
create a properly gauge-invariant second-order result. 

We start by working out the second term on the right-hand side of
(\ref{fNLint}). We find
\ba
 I & \equiv &\int_{-\infty}^t\d t' A_{ab} \frac{\d}{\d t'} 
	\left [ G_{1a} v_{bm} v_{1n} \right ]
         =\gamma_*^2 \int_{-\infty}^t \d t' A_{ab} \frac{\d}{\d t'} 
	\left [ G_{1a} \bv_{bm} \bv_{1n} \right ]\nn\\
	& = & A_{ab*} G_{1a}(t,t_*) v_{bm*} v_{1n*}
	+ \gamma_*^2 \int_{t_*}^t \d t' A_{ab} \left [ G_{1d} 
	A_{da} \bv_{bm} \bv_{1n}- G_{1a} A_{bd} \bv_{dm} \bv_{1n}\right.\nn\\ 
	&&\qquad\qquad\qquad\qquad\qquad\qquad\qquad\qquad\qquad\qquad\qquad\quad\left. - G_{1a} \bv_{bm} A_{1d} \bv_{dn} \right ]\nn\\
	&=&A_{ab*} G_{1a}(t,t_*) v_{bm*} v_{1n*}
	- \gamma_*^2 \int_{t_*}^t \d t' A_{ab} A_{1d} G_{1a} 
	\bv_{bm} \bv_{dn},
\label{Ainteg}	
\ea
where, as before, a subscript $*$ means that a quantity is evaluated at $t_*$.
Using the explicit form of the matrix $A$ (\ref{A2f}) and the solutions 
$v_{am}$ (\ref{vamGreenrel}) this becomes
	\ba
	I & = & \gamma_*^2 \gd_{m2} \gd_{n1} 
	\lh -2 \get^\perp_* + \chi_* G_{12}(t,t_*) + A_{32*} G_{13}(t,t_*)
- \chi_* A_{33*} G_{13}(t,t_*) \rh\\
	&& + \gamma_*^2 \gd_{m2} \gd_{n2} 
	\int_{t_*}^t \d t' 2\get^\perp \bv_{22} 
	\left [ -2\get^\perp \bv_{22}-G_{12} \bv_{32} 
	+ A_{32} G_{13} \bv_{22} + A_{33} G_{13} \bv_{32} \right ].\nn
	\ea
Realizing that $A_{32}\bv_{22} + A_{33}\bv_{32} = - \frac{\d}{\d t'} \bv_{32}$ 
we can do an integration by parts:
\bea{\label{Ires}
	I = \gamma_*^2 \gd_{m2} \Bigg[ & \gd_{n1} 
	\lh -2 \get^\perp_* + \chi_* G_{12}(t,t_*) + A_{32*}G_{13}(t,t_*)
	- \chi_* A_{33*} G_{13}(t,t_*) \rh \nn\\ 
        & -\delta_{n2}2 \get^\perp_* \chi_*G_{13}(t,t_*)\Bigg]\\
        & \hspace{-2cm}+\gamma_*^2 \gd_{m2} \gd_{n2} \int_{t_*}^t \d t' 2\get^\perp 
\left [ -2\get^\perp (\bv_{22})^2+ 
	\lh - 2 G_{12} + A_{33} G_{13} 
	+ \frac{\dot{\get}^\perp}{\get^\perp} G_{13} \rh \bv_{22} \bv_{32} 
	+ G_{13} (\bv_{32})^2
	\right ].\nn
}
To this result we have to add the final term on the right-hand side of 
(\ref{fNLint}).  We call the sum of these two terms $J$: 
\be
\frac{J}{\gamma_*^2} \equiv \frac{I}{\gamma_*^2} 
+ \int_{t_*}^t \d t' G_{1a} \bA_{ab\bc} \bv_{bm} \bv_{\bc n},
\ee
which is
        \ba
	\frac{J}{\gamma_*^2} & = & \gd_{m2} \gd_{n1} 
	\Big( -2 \get^\perp_* + \chi_* G_{12}(t,t_*)+ A_{32*} G_{13}(t,t_*)
	- \chi_* A_{33*} G_{13}(t,t_*) \Big) \nn\\
	&& - \gd_{m2} \gd_{n2} \, 2 \get^\perp_* \chi_* G_{13}(t,t_*) \\
 && + \gd_{m2} \gd_{n2} \int_{t_*}^t \d t' \Big [ 
\lh \bA_{122} - 4(\get^\perp)^2 + \bA_{322} G_{13} \rh (\bv_{22})^2
+ \lh \bA_{333} + 2\get^\perp \rh G_{13} (\bv_{32})^2 \nn\\
	&&  \qquad\qquad\qquad\; + \lh \bA_{123} - 4\get^\perp G_{12} 
	+ (\bA_{323} + \bA_{332}
	+ 2\get^\perp A_{33} + 2 \dot{\get}^\perp) G_{13} \rh 
	\bv_{22} \bv_{32} 
	\Big] . \nn
	\ea
From (\ref{Abar}) we obtain
	\ba
	\bA_{333} + 2 \get^\perp & = & 0,
	\nn\\
	\bA_{323} + \bA_{332} + 2 \get^\perp A_{33} + 2 \dot{\get}^\perp
	& = & 18 \get^\perp - 4 \dot{\get}^\perp,
	\nn\\
	\bA_{123} & = & -2 A_{33} + 2 \ge + 2 \get^\parallel,
	\nn\\
	\bA_{122} - 4 (\get^\perp)^2 
	& = & -2 A_{32} + 2 \dot{\ge} + 2 \dot{\get}^\parallel,
	\ea
so that we can write
	\ba
	\frac{J}{\gamma_*^2} & = & \gd_{m2} \gd_{n1} 
	\Big( -2 \get^\perp_* + \chi_* G_{12}(t,t_*)+ A_{32*} G_{13}(t,t_*)
	- \chi_* A_{33*} G_{13}(t,t_*) \Big) \nn\\
	&& - \gd_{m2} \gd_{n2} \, 2 \get^\perp_* \chi_* G_{13}(t,t_*) \nn\\
        && + \gd_{m2} \gd_{n2} \int_{t_*}^t \d t' \Big [ 
	2 (\bv_{22})^2  \frac{\d}{\d t'}(\ge+\get^\parallel)
	+ 2 \bv_{22} \frac{\d}{\d t'} \bv_{32}
	- 4 \lh \get^\perp G_{12} + \dot{\get}^\perp G_{13} \rh
	\frac{1}{2} \frac{\d}{\d t'} (\bv_{22})^2 \nn\\
        && \qquad\qquad\qquad\quad + 2 (\ge+\get^\parallel) \bv_{22} \bv_{32}
	+ \bA_{322} G_{13} (\bv_{22})^2
	+ 18\get^\perp G_{13} \bv_{22} \bv_{32} 
	\Big ].
	\ea
Doing integrations by parts on the three terms in the third line we obtain
	\ba
	\frac{J}{\gamma_*^2} & = & \gd_{m2} \gd_{n1} 
	\Big( -2 \get^\perp_* + \chi_* G_{12}(t,t_*)+ A_{32*} G_{13}(t,t_*)
	- \chi_* A_{33*} G_{13}(t,t_*) \Big) \nn\\
	&& + 2 \gd_{m2} \gd_{n2} \Biggl (
	- \get^\perp_* \chi_* G_{13}(t,t_*)
	- (\ge_* + \get^\parallel_*) + \chi_*
	+ \get^\perp_* G_{12}(t,t_*)\nn\\
 	&& \qquad\qquad\quad\,
	+ \dot{\get}^\perp_* G_{13}(t,t_*)
	+ (\ge + \get^\parallel) (\bv_{22})^2
	+ \bv_{22} \bv_{32} \Biggr ) \nn\\
	&& + 2 \gd_{m2} \gd_{n2} \int_{t_*}^t \d t' \Big [ 
	- 2 (\get^\perp)^2 (\bv_{22})^2
	- (\ge+\get^\parallel) \bv_{22} \bv_{32} 
	- (\bv_{32})^2
	+ 9 \get^\perp G_{13} \bv_{22} \bv_{32}\nn\\
	&& \qquad\quad\qquad\qquad\;\;
	+ \frac{1}{2} \lh \bA_{322} + 2 \ddot{\get}^\perp 
	+ 2 \dot{\get}^\perp A_{33} + 2 \get^\perp A_{32} 
	\rh G_{13} (\bv_{22})^2 \Big ] .
	\ea
By computing the derivatives of the slow-roll parameters we find
\ba
	&&\bA_{322} + 2 \ddot{\get}^\perp + 2 \dot{\get}^\perp A_{33}
	+ 2 \get^\perp A_{32}\\
&&\qquad = 24 \get^\perp \chi - 12 \get^\parallel \get^\perp
	+ 12 (\get^\parallel)^2 \get^\perp
        + 12 (\get^\perp)^3
	- 4 \get^\perp \gxpa - 4 \get^\parallel \gxpe
	- 3(\tilde{W}_{211} + \tilde{W}_{222}). \nn
	\ea
We now drop boundary terms that are second order in the slow-roll parameters 
{\em at horizon crossing}, since it would be inconsistent to include them given
that the linear solutions used at horizon crossing are only given up to first 
order. Then the result is
	\bea{
	\frac{J}{\gamma_*^2} = \: & \gd_{m2} \gd_{n1} 
	 ( -2 \get^\perp_* + \chi_* \bv_{12} ) 
	+ 2 \gd_{m2} \gd_{n2} \lh
	- \ge_*\! - \get^\parallel_*\! + \chi_*
	+ \get^\perp_* \bv_{12}
	+ (\ge + \get^\parallel) (\bv_{22})^2
	+ \bv_{22} \bv_{32} \rh \nn\\
 & + 2 \gd_{m2} \gd_{n2} \int_{t_*}^t \d t' \Bigg [ 
	- 2 (\get^\perp)^2 (\bv_{22})^2
	- (\ge+\get^\parallel) \bv_{22} \bv_{32}
	- (\bv_{32})^2 + 9 \get^\perp G_{13} \bv_{22} \bv_{32} \nn\\
 & \qquad\qquad\qquad\qquad 
	+ \Bigg( 12 \get^\perp \chi - 6 \get^\parallel \get^\perp
	+ 6 (\get^\parallel)^2 \get^\perp  + 6 (\get^\perp)^3
	- 2 \get^\perp \gxpa - 2 \get^\parallel \gxpe \nn\\
 & \qquad\qquad\qquad\qquad\qquad
	- \frac{3}{2}( \tilde{W}_{211} + \tilde{W}_{222})
	\Bigg) G_{13} (\bv_{22})^2 \Bigg ] .
	}
Inserting this into (\ref{fNL}) gives the final result for $\fnl$ in
(\ref{fNLresult}).

\subsection{Derivation of expression (\ref{solgint}) for $g_\mathrm{int}$}
\label{newgintsec}

A direct computation of the first, second, and third derivatives of
the definition of $g_\mathrm{int}$ in (\ref{gisosrint}) with respect to $t$ using \eqref{srderivatives} and \eqref{greent}
gives:
\be
\dot{g}_{\mathrm{int}}=  - 2 (\getpe)^2 (\bv_{22})^2 - (\ge + \getpa) \bv_{22}\bv_{32} - (\bv_{32})^2 + 2\getpe \int_{t_{*}}^t \d t' \,  \bv_{22} G_{23} \lh \Xi \bv_{22}+9 \getpe \bv_{32}\rh,
\label{derivgint}
\ee
\begin{align}
\ddot{g}_{\mathrm{int}} = \: & 2\lh\gxpe+\getpe (\ge -2 \getpa)\rh \int_{t_{*}}^t \d t' \,  \bv_{22} G_{23} \lh \Xi \bv_{22}+9 \getpe \bv_{32}\rh \nonumber\\ 
&\!\! + 2\getpe\int_{t_{*}}^t \d t' \,  \bv_{22} G_{33} \lh \Xi \bv_{22}+9 \getpe \bv_{32}\rh \nonumber\\
&\!\! + (\bv_{22})^2 \lh 3 (\ge +\getpa) \chi + 2 \ge^3 + 6\ge^2 \getpa + 4 \ge (\getpa)^2 +12 \getpa(\getpe)^2 + (\ge +\getpa) \gxpa - 4\getpe\gxpe \rh\nonumber\\
&\!\! + \bv_{22} \bv_{32} \lh3\ge + 3\getpa + 6\chi + 3\ge^2 +8\ge\getpa + 3(\getpa)^2 + 3 (\getpe)^2 + \gxpa\rh +(\bv_{32})^2 (6 +\ge+ 3 \getpa),
\end{align}
\begin{align}
\dddot{g}_{\mathrm{int}} = & -(3\getpe-\ge\getpe+6\getpa\getpe-2\gxpe) \int_{t_*}^t \d t' \,  \bv_{22} G_{33} \lh \Xi \bv_{22}+9 \getpe \bv_{32}\rh\nonumber\\
& + \lh 9\ge\getpe + 6\getpa\getpe - 6\getpe\chi -3 \gxpe - 3\tW_{211} + \ge^{2}\getpe - 8 \ge \getpa \getpe + 6 (\getpa)^2\getpe - 6 (\getpe)^3 \right.\nonumber\\
&\left. ~~~ - 4 \getpe \gxpa +(3\ge-2\getpa)\gxpe \rh \int_{t_{*}}^t \d t' \,  \bv_{22} G_{23} \lh \Xi \bv_{22}+9 \getpe \bv_{32}\rh\nonumber\\
& + (\bv_{22})^2 \lh32 \getpa \getpe \gxpe-60 (\getpa)^2 (\getpe)^2-36 \getpa (\getpe)^2-4 (\getpa)^2 \gxpa-3 \getpa \gxpa-12 (\getpa)^2 \chi \right.\nonumber \\
& \qquad\qquad -9 \getpa \chi + 6 (\getpe)^2 \gxpa  +12 \getpe \gxpe+6 (\getpe)^2 \chi +12 (\getpe)^4-6 \gxpa \chi -4 (\gxpe)^{2} \nonumber   \\
& \qquad\qquad -18 \chi ^2-3 \getpa \tW_{111}-3 \ge \tW_{111} +9 \getpe \tW_{211}+3 \getpa \tW_{221} +3 \ge \tW_{221}  - 3 \getpe \tW_{222}\nonumber  \\
& \qquad\qquad +\getpa \gxpa \ge -33 \getpa \chi  \ge +14 \getpa \ge ^3-4 (\getpa)^2 \ge ^2-6 \getpa \ge ^2 -12 (\getpa)^3 \ge -8 \getpe \gxpe \ge  \nonumber\\
& \left. \qquad\qquad -12 (\getpe)^2 \ge ^2-36 (\getpe)^2 \ge +5 \gxpa \ge ^2-3 \gxpa \ge -9 \chi  \ge ^2-9 \chi  \ge +6 \ge ^4-6 \ge ^3\rh \nonumber\\
& + \bv_{22}\bv_{32}\lh-12 \getpa (\getpe)^2 - 24 \getpa \chi -12 (\getpa)^3 -21 (\getpa)^2-9 \getpa+4 \getpe \gxpe-15 (\getpe)^2\right. \nonumber\\
& \qquad\qquad\; -9 \gxpa - 54 \chi -3 \tW_{111} +6 \tW_{221}+14 \getpa \ge ^2-21 (\getpa)^2 \ge -57 \getpa \ge +3 (\getpe)^2 \ge \nonumber\\
& \left. \qquad\qquad\; +9 \gxpa \ge +6 \chi  \ge + 9 \ge ^3 - 30 \ge ^2 - 9 \ge \rh \nonumber\\
& + (\bv_{32})^2 \lh-12 (\getpa)^2-39 \getpa+6 (\getpe)^2+4 \gxpa+6 \chi +3 \getpa \ge +3 \ge ^2-15 \ge -36\rh.
\end{align}
Taking the specific combination of the three expressions above that eliminates
all the terms with integrals then gives the differential equation
\be
\begin{split}
&(\getpe)^2\, \dddot{g}_\mathrm{int} + \getpe\left[3\getpe -\ge \getpe + 6 \getpa\getpe - 2 \gxpe\right]\,\ddot{g}_\mathrm{int} \\
& \quad + \left[(\getpe)^2 \lh - 12\ge + 6\chi + 6(\getpa)^2 + 6 (\getpe)^2 + 4 \gxpa \rh + \getpe \lh 3 \tW_{211} - 8 \getpa\gxpe \rh  + 2(\gxpe)^2 \right]\,\dot{g}_\mathrm{int}\\
&= K_{22} (\bv_{22})^2 + K_{23} \bv_{22}\bv_{32} + K_{33} (\bv_{32})^2,
\end{split}
\label{equadiff}
\ee
with $K_{22}, K_{23}, K_{33}$ given by
\be
\begin{split}
K_{22} = & -18 (\getpe)^2 \chi ^2 + 2 (\getpa)^2 (\getpe)^2 \gxpa - 6 \getpa \getpe \gxpe \chi + 6 (\getpa)^2 (\getpe)^2 \chi- 6 (\getpe)^2 \chi \gxpa -2 (\getpe)^4 \gxpa  \\
& - 6 (\getpe)^4 \chi - 18\ge\getpa(\getpe)^2 \chi  + 12\ge^2\getpa (\getpe)^2+ 12 \ge(\getpa)^2 (\getpe)^2  -6 \ge\getpe \gxpe \chi - 12\ge^2 (\getpe)^2 \chi \\
& - 12 \ge(\getpe)^4 - 3 \getpa (\getpe)^2 \tW_{111}-3 \ge(\getpe)^2 \tW_{111}  +3 (\getpe)^3 \tW_{211}+3 \getpa (\getpe)^2 \tW_{221} \\
& +3\ge (\getpe)^2 \tW_{221} -3 (\getpe)^3 \tW_{222} -2 \getpa \getpe \gxpa \gxpe + 6\ge\getpa (\getpe)^2 \gxpa  - 12\ge ^2\getpa \getpe \gxpe \\
& + 20\ge^3 \getpa (\getpe)^2 + 28 \ge^2(\getpa)^2 (\getpe)^2   - 12\ge \getpa (\getpe)^4  +12\ge(\getpa)^3 (\getpe)^2   -2 \ge\getpe \gxpa \gxpe  \\
& - 8\ge(\getpa)^2 \getpe \gxpe + 4\ge^2 (\getpe)^2 \gxpa  - 4 \ge^3\getpe \gxpe -4 \ge(\getpe)^3 \gxpe    + 4\ge^4(\getpe)^2  - 12 \ge^2(\getpe)^4,
\end{split}
\nonumber
\ee
\be
\begin{split}
K_{23} = &-36(\getpe)^2\chi - 6\ge\getpa(\getpe)^2 - 12\ge^2(\getpe)^2 - 6(\getpe)^4 -6 \ge (\getpe)^2 \chi + 6(\getpa)^2(\getpe)^2  \\ 
& - 6 \getpa \getpe \gxpe + 6 \getpa (\getpe)^2 \chi - 6 (\getpe)^2 \gxpa  -6\ge \getpe \gxpe - 12 \getpe \chi \gxpe - 3(\getpe)^2 \tW_{111} \\
& -3 \getpa \getpe \tW_{211} -3 \getpe \ge \tW_{211}  +6 (\getpe)^2 \tW_{221} -2\getpe\gxpa \gxpe  -2 \getpa (\gxpe)^2 + 2 (\getpa)^2 \getpe \gxpe \\
& + 2 \getpa (\getpe)^2 \gxpa -2 (\getpe)^3 \gxpe -8 \ge\getpa\getpe\gxpe  + 18\ge(\getpa)^2(\getpe)^2 + 24\ge^2\getpa (\getpe)^2   + 4\ge(\getpe)^2\gxpa \\
&  - 6 \ge^2\getpe\gxpe - 6 (\getpe)^4 \ge+ 6\ge^3(\getpe)^2 -2 \ge(\gxpe)^{2}  ,
\end{split}
\nonumber
\ee
\be
\begin{split}
K_{33} = &-18 (\getpe)^2 -6 \ge(\getpe)^2 + 6 \getpa (\getpe)^2 -12 \getpe \gxpe - 3 \getpe \tW_{211}  +6 \ge \getpa (\getpe)^2  +2 \ge ^2 (\getpe)^2  \\
& + 2 \getpa \getpe \gxpe -2 \ge\getpe \gxpe  -2 (\gxpe)^{2} .
\end{split}
\label{defKxy}
\ee

Despite its complicated looks, (\ref{equadiff}) actually admits a completely
exact analytical homogeneous solution:
\be
\dot{g}_\mathrm{int}(t) = 2 A \, \getpe(t) G_{22}(t,t_*) 
+ 2 B \, \getpe(t) G_{23}(t,t_*) + P(t),
\label{solgintdot}
\ee
where $A$ and $B$ are integration constants to be determined from the 
initial conditions and $P(t)$ is a particular solution of the equation.
This expression can then be integrated to give equation (\ref{solgint}),
using the fact that $g_\mathrm{int}(t_*)=0$ to eliminate the additional
integration constant. Note that instead of $2\getpe G_{22}$ we can also use
$2\getpe\bv_{22}$ as independent homogeneous solution, which integrates to
$\bv_{12}$.

\begin{appendices}
\renewcommand{\thechapter}{\ref*{NGinflsec}\Alph{chapter}}  

\chapter{Non-Gaussianity in two-field inflation beyond the slow-roll approximation}
\label{JvTapp}
This appendix contains sections 1, 3, 4, 5 and appendix B (incorporated into
the main text as section~\ref{secprodpot}) of \citep{JvT}, of
which a summary was provided in section~\ref{summJvT}. This paper was written
in collaboration with Gabriel Jung.
Like for the other papers in the next appendices, the conclusions of
the paper are not reproduced here, as their relevant parts
were used in section~\ref{longwavother}. In addition, for this specific
paper, section 2 and appendix A are also not reproduced here, as they were
incorporated into the general description of the long-wavelength formalism in
section~\ref{longwaveq}.

We use the long-wavelength formalism to investigate the level of bispectral
non-Gaus\-sianity produced in two-field inflation models with standard
kinetic terms. Even though the Planck satellite has not detected any
primordial non-Gaussianity, it has tightened the constraints significantly,
and it is important to better understand what regions of inflation model space
have been ruled out, as well as prepare for the next generation of experiments
that might reach the important milestone of $\Delta\fnl^\mathrm{local}=1$.
We apply the long-wavelength formalism to the case of a sum potential and show
that it is very difficult to
satisfy simultaneously the conditions for a large $\fnl$ and the observational
constraints on the spectral index $n_s$. In the case of the sum of two monomial
potentials and a constant we explicitly show in which small region of parameter 
space this is possible, and we show how to construct such a model. Finally, the
general expression for $\fnl$ also allows us to prove that for the sum
potential the explicit expressions derived within the slow-roll approximation
remain valid even when the slow-roll approximation is broken during the turn
of the field trajectory (as long as only the $\ge$ slow-roll parameter remains 
small).

\section{Introduction}
\label{introJvT}

The theory of inflation \citep{Starobinsky:1980te, Guth:1980zm,
  Linde:1983gd} describes a period of rapid and accelerated expansion
which takes place in the very early universe. It solves several issues
of the pre-inflationary standard cosmology like the horizon and the
flatness problems. More remarkably, inflation also gives an
explanation for the origin of the primordial cosmological
perturbations which are the seeds of the large-scale structure in the
universe observed today.

The Cosmic Microwave Background radiation (CMB) is an almost direct window on
these primordial fluctuations and its temperature and polarization
anisotropies have been observed by several missions. The most recent
results come from the Planck satellite \citepalias{planck2015-13, planck2015-20,
planck2015-17}\footnote{Obviously, with the 2018 Planck release, these are no
longer the most recent results. However, I have decided to keep the text
and references from the original paper.}, which, like its predecessors, found
no disagreement with the basic inflationary predictions: the
distribution of primordial density perturbations is almost but not
exactly scale-invariant and it is consistent with Gaussianity. The
main information is encoded in the power spectrum which is the Fourier
transform of the two-point correlation function of CMB
temperature/polarization fluctuations. The most interesting observable
from the point of view of inflation is the spectral index $n_s$ that
describes its slope, or in other words the deviation from exact
scale invariance.  

The Planck satellite also significantly improved the constraints on
any potential deviations from a Gaussian distribution (i.e.\ on
non-Gaussianity) \citepalias{planck2015-17}. Primordial non-Gaussianity is
generally parametrized by the amplitude parameters $\fnl$ of a number
of specific bispectrum shapes that are produced in generic classes of
inflation models. The bispectrum is the Fourier transform of the
three-point correlator and in the case of standard single-field
slow-roll inflation it is known to be unobservably small
\citep{Maldacena:2002vr}. However, this result does not hold in more
general situations and many extensions of that simple case have been
proposed with different predictions for non-Gaussianity, meaning that
observations can in principle be used to constrain them.\footnote{It has 
been pointed out \citep{Tanaka:2011aj, Pajer:2013ana} that the finite size of 
the observable universe leads to gauge corrections, which have to be taken 
into account to convert the inflationary bispectrum to actual observations. 
Indeed in single-field inflation the squeezed limit of the bispectrum vanishes 
identically for a local observer today. In multiple-field inflation, on the 
other hand, these corrections are also of order $1-n_s$ \citep{Tada:2016pmk} 
and hence are expected to be negligible in the case of large $\fnl$.}
For example, models with
higher derivative operators based on the Dirac-Born-Infeld action
\citep{Alishahiha:2004eh, Silverstein:2003hf, Mizuno:2009cv,
Mizuno:2010ag,TMvT} can produce large non-Gausianity of the so-called
equilateral type. Another possibility is to consider multiple fields
during inflation, which adds isocurvature perturbations to the usual
adiabatic perturbation. The isocurvature perturbations can interact with
the adiabatic one on super-Hubble scales (while in single-field inflation the
adiabatic perturbation is constant on super-Hubble scales) which can lead
to so-called local non-Gaussianity. In this case non-Gaussianity can be 
generated long after inflation as in the curvaton scenario
\citep{Lyth:2001nq,Bartolo:2003jx,Enqvist:2005pg,Ichikawa:2008iq,Malik:2006pm,
Sasaki:2006kq,Kobayashi:2013bna,Enqvist:2013paa,
Byrnes:2014xua,Vennin:2015egh,Hardwick:2016whe}, or directly after inflation during (p)reheating
\citep{Zaldarriaga:2003my,Lyth:2005qk,Bernardeau:2004zz,Barnaby:2006km,
Enqvist:2004ey, Jokinen:2005by, Elliston:2014zea}. 
However, in this paper we will be interested in the case where this
local non-Gaussianity is produced on super-Hubble scales during
inflation. Since we will only talk about local non-Gaussianity in the
rest of this paper, $\fnl$ should always be understood as
$\fnl^\mathrm{local}$.  

A large amount of work has been done to study if observably large
non-Gaussianity can be produced during multiple-field inflation. This
involves studying the
large-scale evolution of the perturbations which can be done using different 
formalisms, the
$\delta N$ formalism \citep{Starobinsky:1986fxa, Sasaki:1995aw,
Lyth:2005fi} being the most popular but the long-wavelength
formalism \citep{RSvT2, RSvT3,
RSvT4, TvT1, TvT2, Tzavara:2013yca}
offering an interesting alternative. Many results have been obtained
for two fields, a number sufficient to highlight multiple-field
effects (some of them have then been generalized to more fields). In
the slow-roll approximation, the sum-separable \citep{Vernizzi:2006ve}
as well as the product-separable potential \citep{Choi:2007su} have
been solved analytically, while more general separable potentials have
been studied in \citep{Meyers:2010rg,TvT1}.  The solution
beyond slow-roll for Hubble-separable models was given in
\citep{Byrnes:2009qy, Battefeld:2009ym}. Different conditions for large
non-Gaussianity have been found \citep{Elliston:2011dr,Elliston:2011et}
depending on whether the isocurvature modes have vanished before the
end of inflation or not, the latter case requiring a proper treatment
of the reheating phase to be sure that the results actually persist
until the time of recombination and the CMB, which is generally not
done. The scale dependence of the bispectrum is also an important
topic of study of the last few years. Different aspects have been
studied, like the computation of the bispectrum in the squeezed limit,
the scale-dependence of $\fnl$ or the possible observational effects
\citep{Byrnes:2010ft,Byrnes:2012sc,TvT3,Kenton:2015lxa,
  Byrnes:2015dub,Kenton:2016abp}.
Another related subject that has received much attention in recent years
is the study of features in the effective inflaton potential or kinetic terms
(like changes in the sound speed for the inflaton interactions), possibly 
due to the presence of massive fields, which lead to correlated oscillations
in the power spectrum and the bispectrum \citep{Chluba:2015bqa,Achucarro:2010da,
Flauger:2016idt,Achucarro:2012fd,Hotchkiss:2009pj,Achucarro:2014msa}.
Two codes \citep{Dias:2016rjq, Mulryne:2016mzv} for numerical evaluation of 
the bispectrum have been recently released.

The first aim\footnote{In fact the paper \citep{JvT} had an additional aim,
  the continuation of the work on the long-wavelength formalism, but as that
  part of the paper has been incorporated into section~\ref{longwaveq} and
  is not reproduced here, I have removed the corresponding paragraph from
  this introduction.} of the paper is to understand if it is possible to have
large non-Gaussianity while staying within the slow-roll approximation.
Since Planck has excluded the possibility of large local non-Gaussianity
(of order 10), the reader might wonder what the interest is of looking for
models with large non-Gaussianity. However, it is very important in order
to understand if Planck actually ruled out any significant parts of the
multiple-field model space, or if these models generically
predict small non-Gaussianity. Moreover, with large non-Gaussianity in this 
paper we often mean an $\fnl$ of order 1, which has not yet been ruled out by 
Planck but which might be observable by the next generation of experiments.

For explicitness we assume a two-field sum potential (with standard
kinetic terms), where explicit analytical results within the slow-roll
approximation are possible (and have been derived before). In
particular this question was studied within the $\delta N$ formalism
by the authors of \citep{Elliston:2011dr, Elliston:2011et}, who
concluded that with enough fine-tuning an arbitrarily large $\fnl$ is
possible. However, apart from rederiving those results in another
formalism, the new ingredient here is that we take into account the
constraints from Planck on the other inflationary observables, in
particular $n_s$. And it turns out that satisfying the observational
constraints on $n_s$ while having a large $\fnl$ and staying within
the slow-roll approximation is very hard. In the case of a sum of two monomial
potentials and a constant we explicitly work out the region of the
parameter space (in terms of the powers of the two potentials) where
this is possible. Note that we assume everywhere that the isocurvature
mode has disappeared by the end of inflation. Otherwise it would be
easy to get large non-Gaussianity by ending inflation in the middle of
a turn of the field trajectory, but we feel that in that case the
results at the end of inflation would be meaningless, since they could
not be extrapolated to the time of recombination and the CMB without
properly treating the end of inflation and the consecutive period of
(p)reheating.

The second aim of the paper is to understand the, at first sight
very surprising, numerical observation that even in the case where the
slow-roll approximation is broken during the turn of the field
trajectory, the analytical slow-roll expression for $\fnl$ is often still
a very good approximation of the final exact result. It turns out that we
can understand this using the new formulation of the long-wavelength formalism.
In that formulation $\fnl$ is given by a differential equation and the solution
can be written as the sum of a homogeneous and a particular solution. As
we will show, the homogeneous solution can be given analytically in an exact 
form (without any need of the slow-roll approximation), while the particular
solution is negligible exactly in the regions where slow roll is broken and
we cannot compute it analytically.

This paper is organized as follows. 
Section~\ref{Slow-roll} treats the slow-roll results mentioned in aim one 
above. It uses increasing levels of approximation. First, the slow-roll 
approximation is discussed. Then we add the hypothesis that the potential is
sum-separable to solve the Green's function equations and to obtain
simple expressions for the observables. Finally they are applied to the specific
class of monomial potentials, where the effects of the spectral index 
constraint on the region of the parameter space where $\fnl$ is large are
computed. Some results about product potentials are given at the end of
the section.
In section~\ref{Beyond slow-roll}, we keep the sum-separable potential 
hypothesis to compute $\fnl$ beyond the slow-roll approximation. Two different 
types of generic field trajectories with a turn are discussed. We show that
in the end the slow-roll expression from the previous section also gives a
very good approximation of the exact result for $\fnl$ in this case.
Section~\ref{numerical section} contains several specific
examples to illustrate the different results of the paper. The method
to build a monomial potential that produces a large $\fnl$ while satisfying 
all constraints is detailed, while some examples from existing literature are
also discussed. Each time we compare the exact numerical results in the 
long-wavelength formalism to the approximated analytic expressions derived
in this paper.

\section{Slow roll}
\label{Slow-roll}

In this section, we use several consecutive levels of approximations to 
simplify the main expressions of the long-wavelength formalism from
section~\ref{longwaveq}. We start by applying 
only the (strong) slow-roll approximation to general two-field potentials. 
This means that all slow-roll 
parameters, including $\getpe$ and $\gc$, are assumed to be small, which is a 
stronger approximation than the standard slow-roll approximation where only
parallel slow-roll parameters are assumed to be small. Then, in the next 
subsection, we focus on 
sum-separable potentials where the Green's functions can be computed as well 
as the different observables. Afterwards, in the next two subsections, we 
specialize to the case of monomial sum potentials. The final subsection
contains some results for general product potentials in slow roll.

\subsection{General case}

We apply the slow-roll approximation to all relevant equations, starting by the slow-roll parameters themselves. Using the field equation, we obtain explicit expressions for the basis components. We then perform a first-order slow-roll expansion on the second line of \eqref{srpareq} to obtain $\getpa$ and $\getpe$. For $\gxpa$ and $\gxpe$ we proceed in a similar way on \eqref{srderivatives}. The results are:
\be
\begin{split}
\label{sreta}
&e_{1A} = -\tW_{A},\qquad \getpa = \ge - \tW_{11},\qquad \getpe = -\tW_{21},\\
&\gxpa = 3\ge\getpa + (\getpa)^2 + (\getpe)^2 - \tW_{111}, \qquad \gxpe = 3\ge \getpe + 2\getpa\getpe - \getpe\gc - \tW_{211}.
\end{split}
\ee
The same slow-roll expansion applied to the differential equations for the Green's functions \eqref{greent} and \eqref{greentprime} gives:
\be
\label{greensr}
\frac{\d}{\d t}G_{22}(t,t') + \gc(t) G_{22}(t,t') = 0,
\ee
\be
G_{32}(t,t')=-\gc(t)G_{22}(t,t'),\qquad G_{x3}(t,t')=\frac{1}{3}G_{x2}(t,t').
\label{greensr2}
\ee
For the observables, from \eqref{ns} we get:
\be
n_s - 1  = -4 \ge_* - 2\getpa_* + 2 \frac{\bv_{12}}{1+\bv_{12}^2} \lh -2 \getpe_* + \bv_{12} \chi_* \rh
\ee 
and for the different terms of $\fnl$ in \eqref{fNLresult}:
\be
g_{\mathrm{iso}} = (\ge +\getpa -\gc)\bv_{22}^2, \qquad \qquad g_{\mathrm{sr}}=-\frac{\ge_*+\getpa_*}{2\bv_{12}^2}+\frac{\getpe_*\bv_{12}}{2}-\frac{3}{2} \lh \ge_* + \getpa_* - \gc_* + \frac{\getpe_*}{\bv_{12}}\rh.
\ee
For $g_\mathrm{int}$, the slow-roll approximation is not sufficient to compute the integral. However, we can simplify the differential equation \eqref{equadiff} to:
\be
\label{equadiffsr}
\getpe\,\ddot{g}_{\mathrm{int}} - \left[\getpe(\ge-2\getpa-\gc)+\gxpe\right]\,\dot{g}_{\mathrm{int}}= K_{\mathrm{sr}}(\bv_{22})^2,
\ee
with
\be
\begin{split}
K_{\mathrm{sr}} =&~\getpa \getpe \gxpa+3 (\getpa)^2 \getpe \gc -3 \getpa \getpe \gc ^2- (\getpa)^3 \getpe + \getpa (\getpe)^3-\getpa \gxpe \gc -\getpe \gxpa \gc -(\getpe)^2 \gxpe \\
&+\gxpe \gc ^2+\getpa \getpe \tW_{221}-2 \getpe \gc  \tW_{221} + \ge\getpe\tW_{221} - (\getpe)^2 \tW_{222} + 4 \ge\getpa \getpe \gc   + \ge^2\getpa \getpe   \\
&-4 \ge\getpe \gc ^2  +3\ge^2 \getpe \gc  -2 \ge(\getpe)^3  -\ge\gxpe \gc +\getpe \gc^3 +\ge\getpe \gxpa.
\end{split}
\ee
To obtain \eqref{equadiffsr}, several steps have to be followed. First, on the right-hand side of \eqref{equadiff}, one can use \eqref{greensr2} to eliminate $\bv_{32}$. Then one sees that the lowest-order terms (the first of each $K$ in \eqref{defKxy}) cancel each other. The remaining terms are one or two orders higher than the ones which cancel, so that in the leading-order slow-roll aproximation we only have to keep those one order higher. On the left-hand side of the equation, we also use the fact that a time derivative adds an order in slow roll, so that $\dddot{g}_{\mathrm{int}}$ is one order higher in slow-rol than $\ddot{g}_{\mathrm{int}}$. Hence, we see that the $\dddot{g}_\mathrm{int}$ term disappears completely from the equation. Finally, it is possible to substitute the second line of \eqref{sreta} into the two sides of \eqref{equadiff} to eliminate $\tW_{111}$ and $\tW_{211}$, and after simplifying the common factor $3\getpe$ the result is given in \eqref{equadiffsr}.

This equation can be solved for certain classes of potentials. We will look at the simple case of a sum potential, which was solved initially in \citep{Vernizzi:2006ve, Battefeld:2006sz} and discussed in detail in \citep{Byrnes:2008wi,Elliston:2011et,Elliston:2011dr}. The case of a product potential is treated briefly at the end of the section.

\subsection{Sum potential}

A sum potential has the form
\be
W(\gf,\gs) = U(\gf)+V(\gs).
\ee
An immediate consequence of this form is that all mixed derivatives of the potential are zero. Using this and by writing out 
$\tW_{11}, \tW_{22}, \tW_{21}$ (defined in (\ref{defW})) explicitly in terms
of $\tW_{\gf\gf}, \tW_{\gs\gs}, \tW_{\gf\gs}$ and using the normalization of the 
basis $e_{1\gf}^2 + e_{1\gs}^2=1$, one can show that
\be
e_{1\gf}e_{1\gs}(\tW_{11}-\tW_{22})=(e_{1\gf}^{2}-e_{1\gs}^{2})\tW_{21},
\label{W11eq}
\ee
which using (\ref{srpareq}) and (\ref{defchi}) is equivalent to 
\be
e_{1\gf}e_{1\gs}(\gxpa+3\gc-6\ge)=(e_{1\gf}^{2}-e_{1\gs}^{2})(\gxpe+3\getpe) .
\label{W11eq2}
\ee 
Similarly for third-order derivatives, we can write:
\be
\begin{split}
\label{sumequation}
&e_{1\gf}e_{1\gs}\tW_{221} =  e_{1\gf}e_{1\gs}\tW_{111} + (e_{1\gs}^2-e_{1\gf}^2)\tW_{211},\\
&e_{1\gf}e_{1\gs}\tW_{222} =  e_{1\gf}e_{1\gs}\tW_{211} + (e_{1\gs}^2-e_{1\gf}^2)\tW_{221}.
\end{split}
\ee
Using \eqref{W11eq2}, they are equivalent to
\be
\begin{split}
&(\gxpe+3\getpe)\tW_{221} = (\gxpe+3\getpe)\tW_{111} -(\gxpa +3\gc - 6\ge)\tW_{211},\\
&(\gxpe+3\getpe)\tW_{222} = (\gxpe+3\getpe)\tW_{211} -(\gxpa +3\gc - 6\ge)\tW_{221}.
\end{split}
\ee
Note that these equations are general and not only slow-roll. After a first-order slow-roll expansion, they become:
\be
\label{sumequationsr}
\begin{split}
&\getpe\tW_{221} = \getpe\tW_{111} - (\gc - 2\ge)\tW_{211},\\
&\getpe\tW_{222} = \getpe\tW_{211} - (\gc - 2\ge)\tW_{221}.
\end{split}
\ee
We use this to rewrite the right-hand term of \eqref{equadiffsr} as
\be
K_{\mathrm{sr}}= 2\ge \lh -3 \ge^2 \getpe + 3(\getpa)^2 \getpe - 3 (\getpe)^3 + \ge\getpe\gc - 3\getpa\getpe\gc + \getpe \gxpa +\ge \gxpe - \getpa\gxpe \rh.
\ee
Then, one can show that a particular solution of this equation is $\dot{g}_{\mathrm{int}}=2\ge(\ge+\getpa-\gc)\bv_{22}^2$, which can be integrated into $g_\mathrm{int} = \ge\bv_{22}^2 -\ge_*$.

We also know that $\dot{g}_{\mathrm{int}*}=-2(\getpe_*)^2 + (\ge_*+\getpa_*-\gc_*)\gc_*$ from \eqref{derivgint} and the initial conditions of the Green's functions. Combining this particular solution with the homogeneous solution, we get the full solution for $\dot{g}_{\mathrm{int}}$ and then $g_\mathrm{int}$ after integration, in agreement with the known result from~\citep{TvT1}:
\be
\label{gint}
\begin{split}
\dot{g}_{\mathrm{int}}&= 2\ge(\ge+\getpa-\gc)(\bv_{22})^2 - \frac{e_{1\gf*}^2\tV_{\gs\gs*} - e_{1\gs*}^2\tU_{\gf\gf*}}{e_{1\gf*}e_{1\gs*}} \getpe\bv_{22}, \\
g_\mathrm{int} &= \ge \bv_{22}^2 - \ge_* - \left[ \getpe_* - \frac{1}{2\getpe_*} (\ge_* + \getpa_* - \gc_*)(\gc_* - 2\ge_*)\right]\bv_{12},\\
        &= \ge \bv_{22}^2 - \ge_* - \frac{e_{1\gf*}^2\tV_{\gs\gs*} - e_{1\gs*}^2\tU_{\gf\gf*}}{2e_{1\gf*}e_{1\gs*}} \bv_{12}.
\end{split}
\ee
Here the first two terms on the last line are the particular solution, and the last term the homogeneous solution. It is possible to show that the particular solution and the homogeneous solution are generally of the same order during inflation (this is discussed later in section~\ref{Beyond slow-roll}). However, we are only interested in the final values of the observables $n_s$ and $\fnl$. As discussed before, the only large contribution in $\fnl$ can come from $g_\mathrm{int}$, if we suppose isocurvature modes vanish before the end of inflation, which means in terms of Green's functions that $\bv_{22}$ and $\bv_{32}$ vanish while $\bv_{12}$ becomes constant. Hence in that case, the integrated particular solution is also slow-roll suppressed and only the homogeneous solution matters at the end of inflation.
From now on, the different expressions for the observables are only given at the end of inflation. For every other parameter (like the Green's functions and the slow-roll parameters), if they are evaluated at the end of inflation, it is indicated by the subscript $e$.

Using the result \eqref{gint} with $\bv_{22e}=0$, we can write:
\be
\begin{split}
-\frac{6}{5}\fnl &= \left[ \getpe_* - \frac{1}{2\getpe_*} (\ge_* + \getpa_* - \gc_*)(\gc_* - 2\ge_*)\right]\frac{2(\bv_{12e})^3}{\lh 1+(\bv_{12e})^2\rh ^2} + \mathcal{O}(10^{-2})\\
 &=  \frac{e_{1\gf*}^2\tV_{\gs\gs*} - e_{1\gs*}^2\tU_{\gf\gf*}}{e_{1\gf*}e_{1\gs*}} \frac{(\bv_{12e})^3}{\lh 1+(\bv_{12e})^2\rh ^2} + \mathcal{O}(10^{-2}).
\label{fnlapp}
\end{split}
\ee
This depends on the final value of the Green's function $\bv_{12}$, which describes the contribution of the isocurvature mode to the adiabatic mode. Without computing it, it is possible to determine a necessary condition for $\fnl$ to be of order unity or larger. Indeed it is easy to show that, for any value of $\bv_{12e}$: 
\be
\left|\frac{(\bv_{12e})^3}{\lh 1+(\bv_{12e})^2\rh ^2}\right|\leq\frac{3^{3/2}}{16} \approx 0.325.
\label{numfactor}
\ee
If the slow-roll approximation is valid at horizon-crossing, which is the main assumption in the computation of $\fnl$, we expect that $\tV_{\gs\gs*}$ and $\tU_{\gf\gf*}$ are of order slow-roll (small compared to one). Then, the only possibility to get $\fnl$ of order unity is that one of the basis components is negligible at horizon-crossing. This means one of the fields is dominating at that time, by definition we choose it to be $\gf$. Hence, at horizon-crossing $e_{1\gf*}^2 \approx 1$ and $e_{1\gs*}^2\ll 1$. Using \eqref{sreta}, this also implies that $|U_{\gf*}|\gg |V_{\gs*}|$ and we can simplify:
\be
\frac{e_{1\gf*}^2\tV_{\gs\gs*} - e_{1\gs*}^2\tU_{\gf\gf*}}{e_{1\gf*}e_{1\gs*}} = \frac{e_{1\gf*}\tV_{\gs\gs*}}{e_{1\gs*}} = \frac{\sqrt{2\ge_*}}{\gk} \frac{V_{\gs\gs*}}{V_{\gs*}}.
\label{fnlsrfactor}
\ee
This has to be large to have $\fnl$ non-negligible, which means that the second-order derivative $V_{\gs\gs*}$ is large compared to the first-order derivative $V_{\gs*}$. Hence around $\gs_*$, the potential is very flat in the $\gs$ direction. In terms of slow-roll parameters, this means that $|\getpe_*|\siml |(\ge_* + \getpa_* - \gc_*)(\gc_* - 2\ge_*)|$. For the usual slow-roll order values of $10^{-2}$, $\getpe_*$ is at most of order $10^{-4}$.

\begin{figure}
\centering
\includegraphics[width=0.5\textwidth]{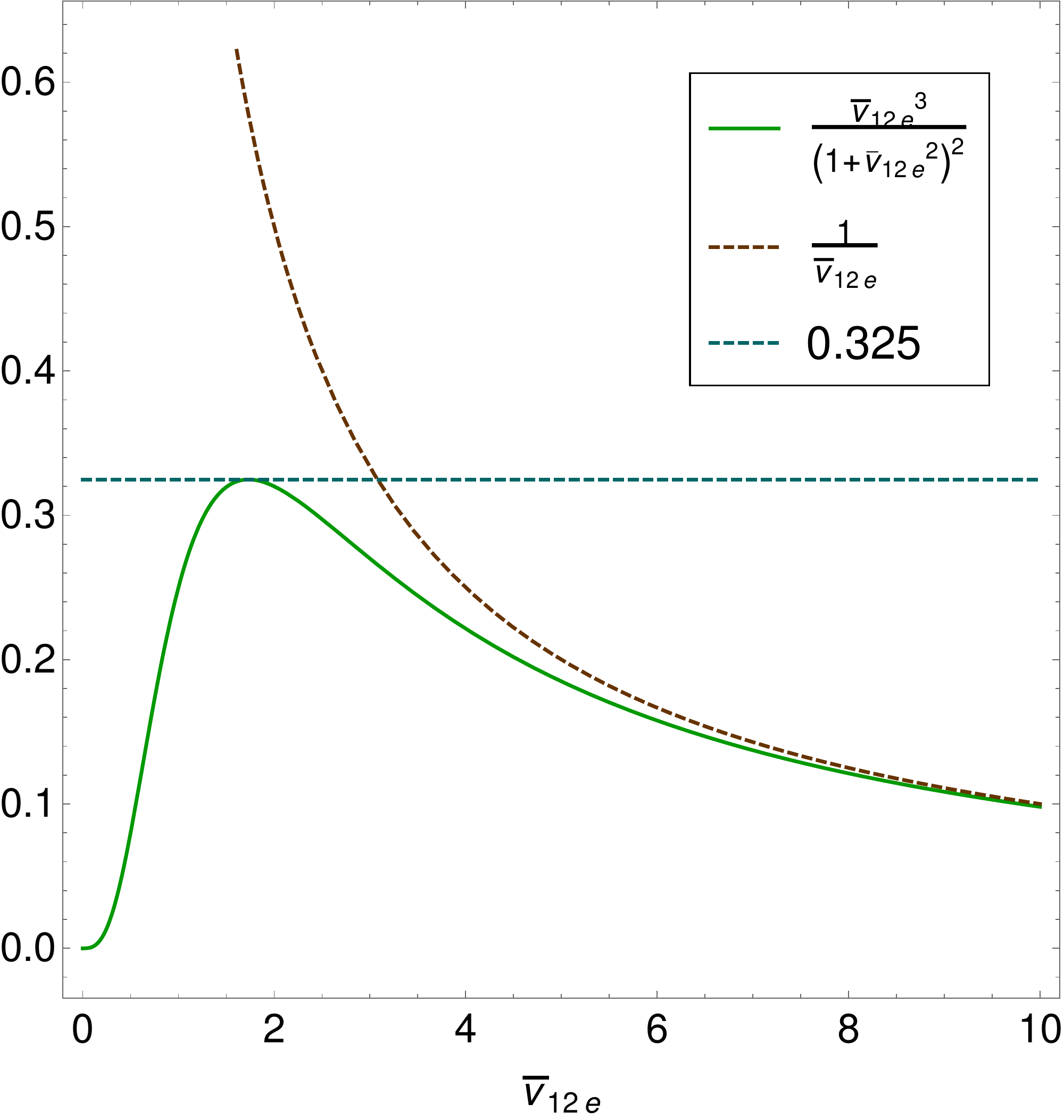}
\caption{$\frac{(\bv_{12e})^3}{\lh(1+(\bv_{12e})^2\rh^2}$ as a function of $\bv_{12e}$ and its two upper limits 0.325 and $1/\bv_{12e}$.}
\label{fig:factors}
\end{figure}

Another useful limit is:
\be 
\label{v12limit}
\left|\frac{\bv_{12e}^3}{(1+\lh\bv_{12e})^2\rh^2}\right|<\left|\frac{1}{\bv_{12e}}\right|,
\ee
which becomes a very good approximation if $|\bv_{12e}|>4$ . These two limits are shown explicitly in figure~\ref{fig:factors}. From \eqref{greent}, if $\bv_{12e}$ is of order unity, this implies that at some time there was a turn of the field trajectory where both the isocurvature mode and $\getpe$ are non-negligible. This turn is then a necessary condition of large non-Gaussianity.

Still using the slow-roll approximation, we can go further by computing the Green's functions. From \eqref{W11eq2}, we get:
\be
\gc = 2\ge +\getpe \frac{e_{1\gf}^2-e_{1\gs}^2}{e_{1\gf}e_{1\gs}}=-\frac{\d}{\d t}\ln{\lh H^2 e_{1\gf}e_{1\gs}\rh}.
\ee
We can then solve \eqref{greensr}:
\be
G_{22}(t,t')=\frac{H(t)^2 e_{1\gf}(t)e_{1\gs}(t)}{H(t')^2 e_{1\gf}(t')e_{1\gs}(t')}.
\ee
Moreover, we have:
\be
\getpe H^2 e_{1\gf}e_{1\gs}= \frac{\gk^2}{6}\frac{\d Z}{\d t},
\label{Z}
\ee
with $Z\equiv V e_{1\gf}^2- U e_{1\gs}^2$ \citep{Vernizzi:2006ve,TvT1}, which gives us:
\be
\bv_{12} = \frac{Z-Z_*}{W_* e_{1\gf*}e_{1\gs*}},\qquad \bv_{22}=\frac{W e_{1\gf} e_{1\gs}}{W_* e_{1\gf*} e_{1\gs*}}.
\label{Green srsolution}
\ee
At the end of inflation, when the fields reach the minimum of the potential, $Z$ tends to zero. Obviously, this can only happen if there is a turn of the field trajectory at some time after horizon-crossing to make both fields evolve. Moreover, if $e_{1\gs*}^2\ll 1$ (necessary condition for $\fnl$ of order unity), $Z_*=V_* e_{1\gf*}^2$. We then obtain, using \eqref{sreta}:
\be
\bv_{12e}=-\frac{V_* e_{1\gf*}}{W_* e_{1\gs*}}= \text{sign}(e_{1\gf*})\sqrt{2\ge_*}\frac{\gk V_*}{V_{\gs*}}.
\label{v12e}
\ee
With a small enough $e_{1\gs*}$, it is easy to obtain $\bv_{12e}$ larger than four or five. In figure~\ref{fig:factors}, this places us on the right where $\frac{(\bv_{12e})^3}{\lh 1+(\bv_{12e})^2\rh^2} \approx \frac{1}{\bv_{12e}}$. The consequence for the potential $V$ is that $\gk V_*\gg V_{\gs*}$.

Substituted into \eqref{fnlapp}, in the case where the slow-roll parameters factor is large, we obtain:
\be
-\frac{6}{5}\fnl \approx \frac{\tV_{\gs\gs*}e_{1\gf*}}{e_{1\gs*}} \frac{1}{\bv_{12e}} =\frac{\tV_{\gs\gs*}e_{1\gf*}}{e_{1\gs*}}\frac{W_*  e_{_1\gs*}}{- V_* e_{1\gf*}} = -\frac{V_{\gs\gs*}}{\gk^2 V_*}.
\label{fnllimit}
\ee
This directly shows that $\fnl$ is of order unity when the second derivative of $V_*$ and $V_*$ itself are of the same order, while its first-order derivative $V_{\gs*}$ is small compared to the two previous quantities because of \eqref{fnlsrfactor} and \eqref{v12e}, a result already highlighted in \citep{Elliston:2011et,Elliston:2011dr}. Larger $\fnl$ is a priori possible, but requires a fine-tuning of the model. Moreover, the sign of $\fnl$ is the sign of $V_{\gs\gs*}$. A negative $\fnl$ corresponds to a potential in the form of a ridge at $t_*$, where $\gs_*$ is very close to the maximum for the potential to be flat enough in the $\gs_*$ direction, while a positive $\fnl$ corresponds to a valley potential.

In the same limit of large $\bv_{12e}$, the spectral index takes the form 
\be
n_s - 1  = -4 \ge_* - 2\getpa_* + 2\chi_*  = -2\ge_* + 2\tV_{\gs\gs*}.
\label{nstwofield}
\ee
The spectral index is close to 1, hence $\tV_{\gs\gs*}=\frac{V_{\gs\gs*}}{\gk^2 W_*}$ is at most of order $10^{-2}$. If it is smaller, this requires a fine-tuning of $\ge_*$. If $\fnl$ is of order unity, then $\frac{V_*}{W_*}$ is also of order $10^{-2}$. 

To summarize, at horizon-crossing, the conditions are $U_* \gg V_*$ and $|U_{\gf*}|\gg |V_{\gs*}|$. The second-order derivative $V_{\gs\gs*}$ is not negligible and can be either smaller, equal or larger than $U_{\gf\gf*}$ but it is not hugely larger or smaller. To be precise, we make a quite general assumption that $|V_{\gs\gs*} U_{\gf*}^2|\gg |U_{\gf\gf*}V_{\gs*}^2|$ and $|V_{\gs\gs*} V_{\gs*}^2|\ll |U_{\gf\gf*}U_{\gf*}^2|$. With these different assumptions for the potential, the expressions for the slow-roll parameters and basis vectors become:
\be
\label{srparametersflat}
\begin{split}
&\ge = \frac{1}{2\gk^2} \lh \frac{U_{\gf}}{U} \rh^2, \qquad
\getpa = - \frac{U_{\gf\gf}}{\gk^2 U} + \ge, \qquad
\getpe = - \frac{V_{\gs}}{U_{\gf}} \frac{U_{\gf\gf} - V_{\gs\gs}}{\gk^2 U},\\
&e_{1\gf}= -\text{sign}(U_{\gf}), \qquad e_{1\gs}=-\frac{V_\gs}{\gk\sqrt{2\ge}\, U}, \qquad \chi = \frac{V_{\gs\gs}}{\gk^2 U} + \ge + \getpa = \frac{U_{\gf}}{V_{\gs}} \getpe + 2\ge.
\end{split}
\ee

At horizon-crossing, the situation is very close to single-field inflation. In the slow-roll regime, by definition everything evolves slowly, hence a legitimate question is to ask when these conditions will stop to be valid. In fact, they will break at the turn of the field trajectory. At that time $V_{\gs}$ stops to be negligible compared to $U_{\gf}$ (or equivalently, $e_{1\gs}$ is not small compared to one). As already discussed, the turn is mandatory to have $\bv_{12e}$ large enough. However, they will also break if $V$ stops to be negligible compared to $U$, this happens when the field $\gf$ is near the minimum of its potential. In this second case, we know the slow-roll approximation will also stop to be valid because $\ge$ is becoming large (similarly to single-field inflation). Hence, if this happens before the turn, as the slow-roll approximation is not valid anymore, we lose the analytical results for the Green's functions and $\fnl$. We have to check if the turn can occur before the first field reaches the minimum of its potential, or in simple terms, is it possible to have $\fnl$ of order unity without breaking the slow-roll approximation? To be able to make progress in answering that question, we will consider a specific class of two-field sum potentials, where both $U$ and $V$ are monomial plus a possible constant.

\subsection{Monomial potentials}
\label{monomialsubsec}

Using the results of the previous section, we want to analytically study inflation between horizon-crossing and the beginning of the turn of the field trajectory. The idea is that the slow-roll approximation is broken when the dominating field $\gf$ gets close to the minimum of its potential, and we want to verify if the turn can occur before that time. This means that the form of the potential does not need to describe the end of inflation. 

We know that $V(\gs)$ has to be very flat around $\gs_*$, hence we can use an expansion in $\gs$ keeping only the largest term to write:
\be
V(\gs) = C + \gb (\gk\gs)^m,
\label{potential V}
\ee
where $C$, $m$ and $\gb$ are constants. Here $m > 1$, while $\gb$ can be either positive or negative. Because of the expansion in $\gs$, this potential is in fact quite general. Depending on the sign of $\gb$, the potential either corresponds to a ridge where $\gs_*$ is near the local maximum $C$ ($\gb<0$) or to a valley with $\gs_*$ near the minimum ($\gb>0$). For the potential $U$, there are many possibilities, we choose to focus on a monomial potential:
\be
U(\gf) = \ga (\gk\gf)^n,
\ee
with $\ga>0$ and $n>1$.

We redefine the fields as being dimensionless: $\tilde{\gf} = \gk \gf$ and $\tilde{\gs} = \gk \gs$ and we will omit the tildes in the redefined fields.
Using the expressions for the slow-roll parameters given at the end of the previous section \eqref{srparametersflat}, we have:
\be
\label{srparmonomial}
\begin{split}
\ge &= \frac{n^2}{2} \frac{1}{\gf^2}, \qquad
\getpa = - \frac{n(n-2)}{2}\frac{1}{\gf^2} = - \frac{n-2}{n} \ge,\\
\getpe &= - \frac{m\gb}{n \ga^2} \frac{\gs^{m-1}}{\gf^{2n-1}}
\lh n(n-1)\ga \gf^{n-2} - m(m-1)\gb \gs^{m-2} \rh\\ 
&= -\frac{m\gb}{n^{2n} \ga^2} 2^{n-\frac{1}{2}} \gs^{m-1} 
\lh n^{n-1}(n-1)2^{1-\frac{n}{2}}\ga \ge^{\frac{n}{2}+\frac{1}{2}}-m(m-1)\gb \gs^{m-2} \ge^{n-\frac{1}{2}}\rh,\\
\chi &= \frac{1}{\ga}\frac{1}{\gf^n} \lh n\ga \gf^{n-2} + m(m-1)\gb \gs^{m-2} \rh 
= \frac{2\ge}{n} + \frac{m(m-1)\gb}{n^{n}\ga}2^{n/2}\ge^{n/2}\gs^{m-2}.
\end{split}
\ee 
It is useful to express the slow-roll parameters as a function of $\ge$ instead of $\gf$ because $\ge$ increases after horizon-crossing, at least until the turn, and with $\ge$ we know exactly when the slow-roll approximation stops to be valid. $\ge$ and $\getpa$ are of the same order except in the case of $n=2$ where $\getpa$ is of order $\ge^2$ as can be checked with a second-order calculation.

The next step is to use the conditions that $\fnl$ should be of order unity and $n_s$ should be within the observational bounds to constrain the free parameters of this potential. With this form of $V$, we have the useful relation:
\be
\label{link12}
(m-1)V_\gs=\gs V_{\gs\gs}.
\ee
We know that $|e_{1\gs*}|\ll 1$ and substituting \eqref{link12} into the expression for $e_{1\gs}$ in \eqref{srparametersflat}, we can write:
\be
e_{1\gs} = -\frac{\tV_{\gs\gs}}{\sqrt{2\ge}}\frac{\gs}{m-1}.
\ee
Combining this with the contraints on the spectral index \eqref{nstwofield} which imply that $\ge_*$ and $\tV_{\gs\gs*}$ are both of order $10^{-2}$ at most, this imposes $\gs_*$ to be small compared to 1. Applying these constraints due to the observables to the potential gives:
\be
\label{explicitconstraints}
\frac{V_{\gs\gs*}}{W_*}=\frac{m(m-1)\gb \gs_*^{m-2}}{\ga \gf^n_* + C + \gb\gs_*^{m}} \sim \mathcal{O}(10^{-2}),\qquad \frac{V_{\gs\gs*}}{V_*}=\frac{m(m-1)\gb\gs_*^{m-2}}{C+\gb\gs_*^{m}}\sim \mathcal{O}(1).
\ee 
Within the limit $\gs_* \ll 1$, we learn from these equations that $\ga \gf_*^n \gg m(m-1)\gb \gs_*^{m-2} \sim C + \gb \gs_*^m$.

We also need to determine the slow-roll parameters at $t_*$, which requires to know $\gf_*$. One way to determine this is to know the amount of inflation due to each field between horizon-crossing and the end of inflation. We can start by solving the field equation:
\be
\dot{\gf} = - \frac{n}{\gf},
\label{eqphi(t)}
\ee
which integrates immediately to:
\be
\label{phi(t)}
\gf(t) = \gf_* \sqrt{1 - \frac{t}{N_\gf}},
\ee
with $N_\gf=\frac{\gf_*^2}{2n}$ the slow-roll approximation of the number of e-folds due to $\gf$ after horizon-crossing. 

The potential is known only before the turn of the field trajectory, especially for $V$ if it is an expansion of some more complicated function. This means that we do not know the value of $N_\gf$, however it is in the range of a few to 60 e-folds. We will test different values. Nevertheless, in the simplest cases $N_\gs$ (number of e-folds due to $\gs$) is small compared to $N_\gf$. As a simple argument here, we consider the case where $\gs$ falls off a ridge, so that $V_*\approx C$. If $V$ keeps the same form almost until the end of inflation, the minimum of the potential ($V=0$) corresponds approximately to $\gs_{e}=(-C/\gb)^{1/m} \sim [m(m-1)]^{1/m}\gs_*^{1-2/m}$, using the second part of \eqref{explicitconstraints}. For $m=2$, this is of order 1, for larger $m$ it becomes smaller (only $m$ close to 1 is problematic). In a pure monomial potential like $U$ without the constant term, having $\gf_*$ of order unity would imply that $N_\gf$ is itself of order unity. $V$ is a bit different because of the constant term, however once $\gs$ starts to fall at a non-negligible pace (the turn), it becomes quite similar and $\gs$ goes from $\gs_*$ negligible to $\gs_e$ of order unity. Hence this also corresponds to $N_\gs$ of order unity which can be neglected in the total number of e-folds compared to $N_\gf$. Note this is not a general proof, just a plausible argument to claim that $N_\gf$ is the dominant contribution. We can also see that $\gs_e$ becomes larger if $V_{\gs\gs*}/V_*$ in \eqref{explicitconstraints} becomes smaller. Hence the fact that $N_\gs$ is small is linked to having $\fnl$ of order unity or more. 

The parameter $\ge_*$ is related to the value of $N_\gf$, hence for these models where $N_\gs \ll N_\gf$, the value of $\ge_*$ is directly fixed by the total number of e-folds after horizon-crossing:
\be
\ge_*=\frac{n}{4N_\gf}.
\ee
When $\ge_*$ is fixed, we can use the spectral index formula \eqref{nstwofield} to constrain $\tV_{\gs\gs*}$:
\be
\tV_{\gs\gs*}=\frac{n_s-1}{2}+\ge_*.
\label{Wss spectral index}
\ee
Using $n_s = 0.968 \pm 0.006$ from the Planck data, table \ref{table:Vss} shows the constraints for integer values of $n$. Note that for $n\geq 5$, the second-order derivative has to be positive. 
\begin{table}
		\centering
		\begin{tabular}{|c|c|c|c|c|c|}
		\hline
		$n$ & 2 & 3 & 4 & 5 & 6\\
		\hline
		$10^3\,\tV_{\gs\gs*}$ & $ -7.7 \pm 3$ & $-3.5 \pm 3$ & $0.7 \pm 3$ & $4.8 \pm 3$ & $9 \pm 3$  \\
		\hline
	\end{tabular}
	\caption{Constraints from the spectral index on $\tV_{\gs\gs*}$ for different $n$ with $N_\gf=60$.}
\label{table:Vss}
\end{table}
 According to \eqref{fnlapp}, we also know that:
\be
\left|-\frac{6}{5}\fnl\right| < 0.65\left| \getpe_* - \frac{1}{2\getpe_*} (\ge_* + \getpa_* - \gc_*)(\gc_* - 2\ge_*)\right|, 
\label{fnllimit2}
\ee
which gave the estimation of $\getpe_*$ of order $10^{-4}$ to get $\fnl$ of order unity. We can neglect the first $\getpe_*$ which is already a few orders of magnitude smaller than the single-field slow-roll typical value of $\fnl$. Then we obtain:
\be
\left|-\frac{6}{5}\fnl \getpe_*\right| < 0.325\left|(\ge_* + \getpa_* - \gc_*)(\gc_* - 2\ge_*)\right|.
\label{etaperconstraint}
\ee
We can rewrite the right-hand side term:
\be
(\ge_* + \getpa_* - \gc_*)(\gc_* - 2\ge_*)= -\tV_{\gs\gs*}\frac{2(1-n)}{n}\ge_* - \tV_{\gs\gs*}^2.
\ee
This is largest for $\tV_{\gs\gs*} = {\textstyle\frac{n-1}{n}\ge_*}$, which corresponds to $n_s = 1-{\textstyle\frac{1}{2N_\gf}}\geq 0.992$ which is outside of the observed value. The maximum of the absolute value in \eqref{etaperconstraint} will then be given by the upper or the lower bound on $n_s$ (because in the interval of the observed value for $n_s$ it can change sign). Table \ref{table:getpe} gives the numerical constraints on $\left|-\frac{6}{5}\fnl \getpe_*\right|$ for integer values of $n$.
\begin{table}
		\centering
		\begin{tabular}{|c|c|c|c|c|c|}
		\hline
		$n$ & 2 & 3 & 4 & 5 & 6\\
		\hline
		$\left|-\frac{6}{5}\fnl \getpe_*\right|$ & $ 6.8 \times 10^{-5}$ & $5.0 \times 10^{-5}$ & $2.6\times 10^{-5}$ & $6.7 \times 10^{-5}$ & $1.2\times 10^{-4}$  \\
		\hline
	\end{tabular}
	\caption{Upper bounds from the spectral index on $\left|-\frac{6}{5}\fnl \getpe_*\right|$ for different $n$ with $N_\gf=60$.}
	\label{table:getpe}
\end{table}
We observe that the maximum value for $\getpe_*$ is two orders of magnitude smaller than $\ge_*$ for $\fnl$ of order unity. Moreover this limit is quite strong since the factor 0.325 \eqref{numfactor} is a limit which asks some fine tuning to be reached. This factor can easily be ten or a hundred times smaller. Hence, in most cases $\getpe_*$ will be a lot smaller than this limit.

To summarize, we know $\ge_*$ once we fix $N_\gf$. We then determine $\tV_{\gs\gs*}$ using $\ge_*$ and the observational constraints on $n_s$. This leads to an upper bound for $|\getpe_*|$ by imposing a value for $\fnl$. However, to see when the turn exactly happens, we need to know the full evolution of $\getpe$, not just its initial value. For this, some work needs to be done on the expression for $\getpe$ given in \eqref{srparmonomial}, where we can eliminate unknown quantities (like the parameters of the potential) by using the expressions for the slow-roll parameters at horizon crossing:
\be
\ge_*=\frac{n^2}{2}\frac{1}{\gf_*^2},\qquad \tV_{\gs\gs*}=\frac{m(m-1)\gb\gs_*^{m-2}}{\ga \gf_*^n}.
\ee
It is then straightforward to compute:
\be
\label{getpeevolution}
\begin{split}
&\tV_{\gs\gs}= \frac{V_{\gs\gs}}{\gk^2 U} =\tV_{\gs\gs*} \lh \frac{\gs}{\gs_*} \rh ^{m-2} \lh \frac{\ge}{\ge_*} \rh ^{n/2},\\
&\getpe = \getpe_* \lh \frac{\gs}{\gs_*} \rh ^{m-1} \lh \frac{\ge}{\ge_*} \rh ^{n/2} \frac{2\frac{n-1}{n}\ge^{1/2}-\tV_{\gs\gs}\ge^{-1/2}}{2\frac{n-1}{n}\ge_*^{1/2}-\tV_{\gs\gs*}\ge_*^{-1/2}}.
\end{split}
\ee
As already discussed, we want to express the time dependence in terms of $\ge$ which is directly related to $\gf$. However, the expression for $\getpe$ also depends on $\gs$, and while a bound for its initial value at horizon-crossing can be given using \eqref{srparmonomial} and the bounds on $\tV_{\gs\gs*}$ and $\getpe_*$, we need to know how it evolves with time. For this we solve the field equation:
\be
\dot{\gs} = - \frac{m\gb}{\ga} \frac{\gs^{m-1}}{\gf^n}.
\ee

Inserting the solution \eqref{phi(t)} for $\gf$ into the equation for $\gs$ we find
the following differential equation:
\be
\frac{\d\gs}{\gs^{m-1}} = - \frac{m\gb}{\ga} \frac{1}{\gf_*^n}
\frac{\d t}{(1-t/N_\gf)^{n/2}}.
\label{eqsig(t)}
\ee
We see that we need to consider the special cases $m=2$ and $n=2$ separately. We start with the most general cas $m\neq 2 $ and $n \neq 2$, where (with $\gs_*$ the initial value of $\gs$):
\be
\begin{split}
\gs &= \gs_* \left[1+\frac{m(2-m)}{n(2-n)}\frac{\gb}{\ga}\frac{\gs^{m-2}}{\gf_*^{n-2}}\lh \lh 1-\frac{t}{N_\gf}\rh^{1-n/2} - 1 \rh \right]^\frac{1}{2-m}\\
    &= \gs_* \left[1+ \frac{1}{2}\frac{m-2}{m-1}\frac{n}{n-2} \frac{\tV_{\gs\gs*}}{\ge_*^{n/2}} \lh \ge^{n/2-1}-\ge_*^{n/2-1}\rh\right]^{\frac{1}{2-m}}.
\end{split}
\ee
In the case $m\neq 2$ and $n=2$, we have:
\be
\begin{split}
\gs &= \gs_* \left[1+\frac{m(2-m)}{4\gs_*^{2-m}}\frac{\gb}{\ga}\ln{\lh 1-\frac{t}{N_\gf}\rh}\right]^{\frac{1}{2-m}}\\
    &= \gs_* \left[1+ \frac{1}{2} \frac{2-m}{m-1} \frac{\tV_{\gs\gs*}}{\ge_*} \ln{\lh\frac{\ge_*}{\ge}\rh}\right]^{\frac{1}{2-m}},
\end{split}
\ee
while for $m=2$ and $n \neq 2$:
\be
\begin{split}
\gs &= \gs_* \exp{\left[\frac{2\gb}{\ga}\frac{\gf_*^{2-n}}{n(2-n)} \lh \lh 1-\frac{t}{N_\gf}\rh^{1-n/2} - 1 \rh\right]}\\
    &= \gs_* \exp{\left[\frac{n}{2(2-n)}\frac{\tV_{\gs\gs*}}{\ge_*^{n/2}}\lh \ge^{n/2-1} - \ge_*^{n/2-1}\rh\right]}.
\end{split}
\ee
Inserting these expressions into \eqref{getpeevolution} gives the ratio $\getpe/\getpe_*$. In the last case $m=2$ and $n=2$, these equations take a nicer form:
\be
\gs =\gs_* \lh 1-\frac{t}{N_\gf}\rh^{\frac{\gb}{2\ga}}=\gs_* \lh 1-\frac{t}{N_\gf}\rh^{\frac{\tV_{\gs\gs*}}{2\ge_*}}= \gs_* \lh \frac{\ge}{\ge_*} \rh^{-\frac{\tV_{\gs\gs*}}{2\ge_*}},\qquad
\frac{\getpe}{\getpe_*}= \lh \frac{\ge}{\ge_*} \rh^{-\frac{\tV_{\gs\gs*}}{2\ge_*}+\frac{3}{2}}.
\ee

\subsection{Discussion}

\begin{figure}
\centering
	\includegraphics[width=0.49\textwidth]{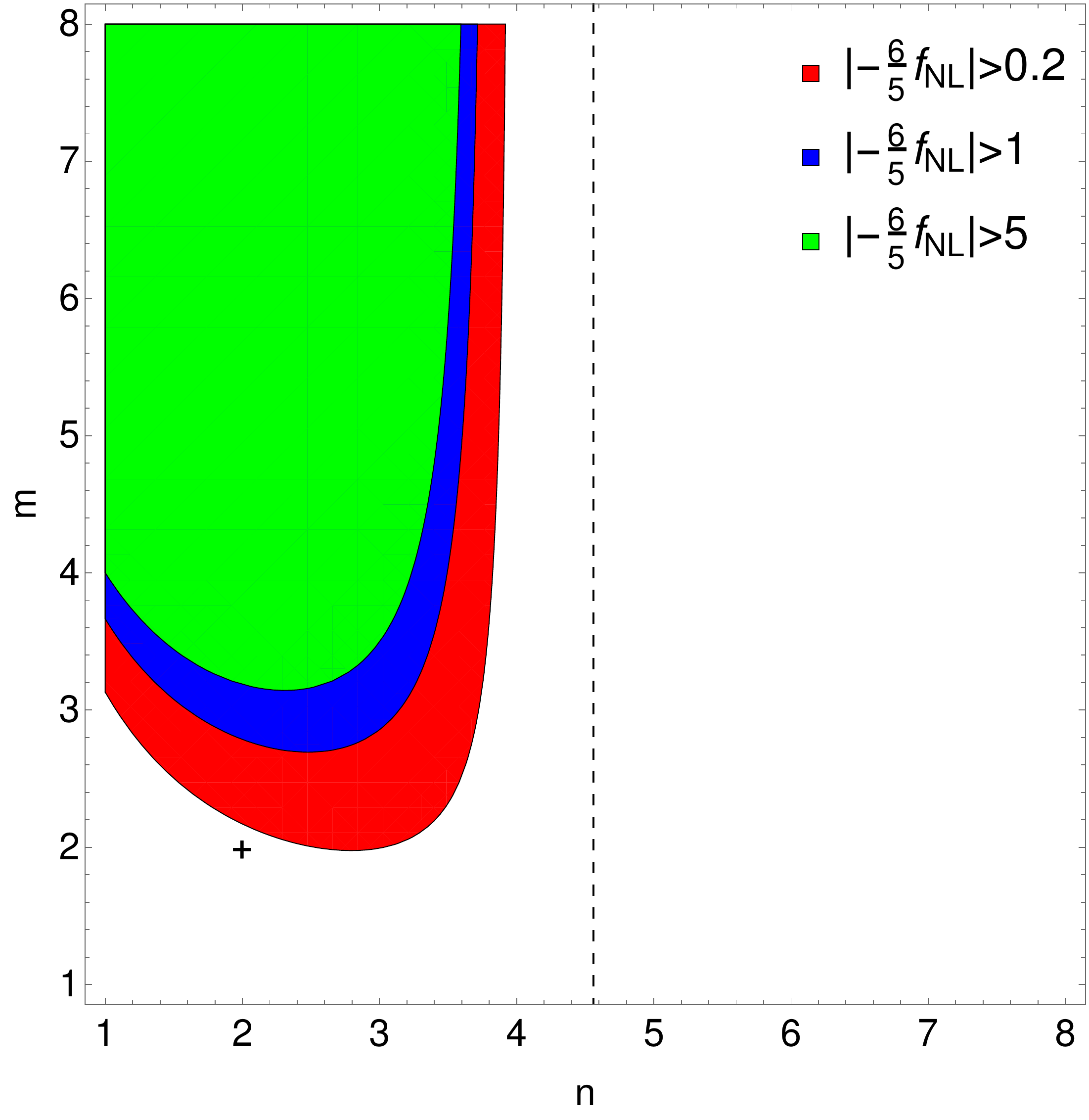}
	\includegraphics[width=0.49\textwidth]{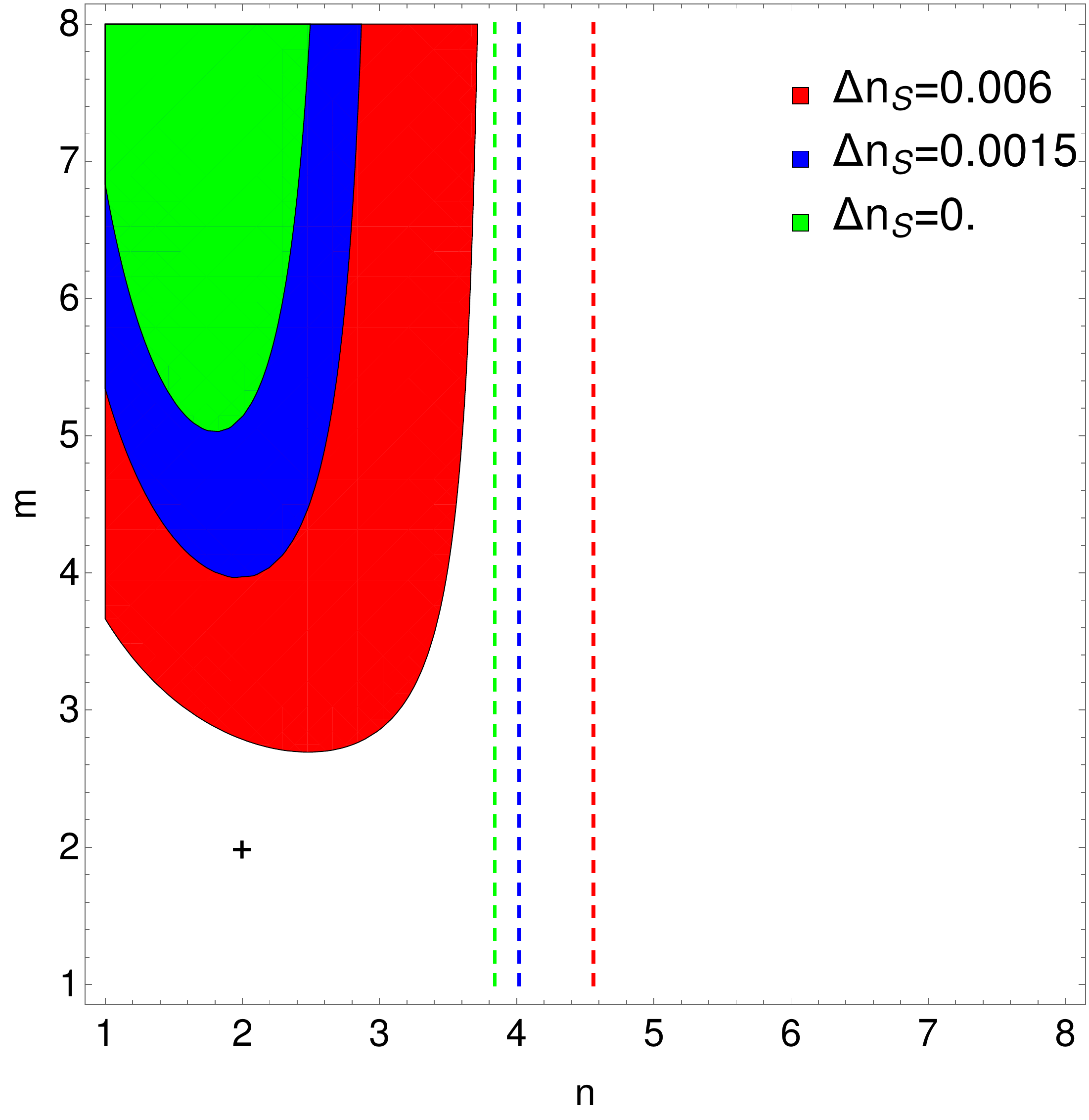}
	\includegraphics[width=0.49\textwidth]{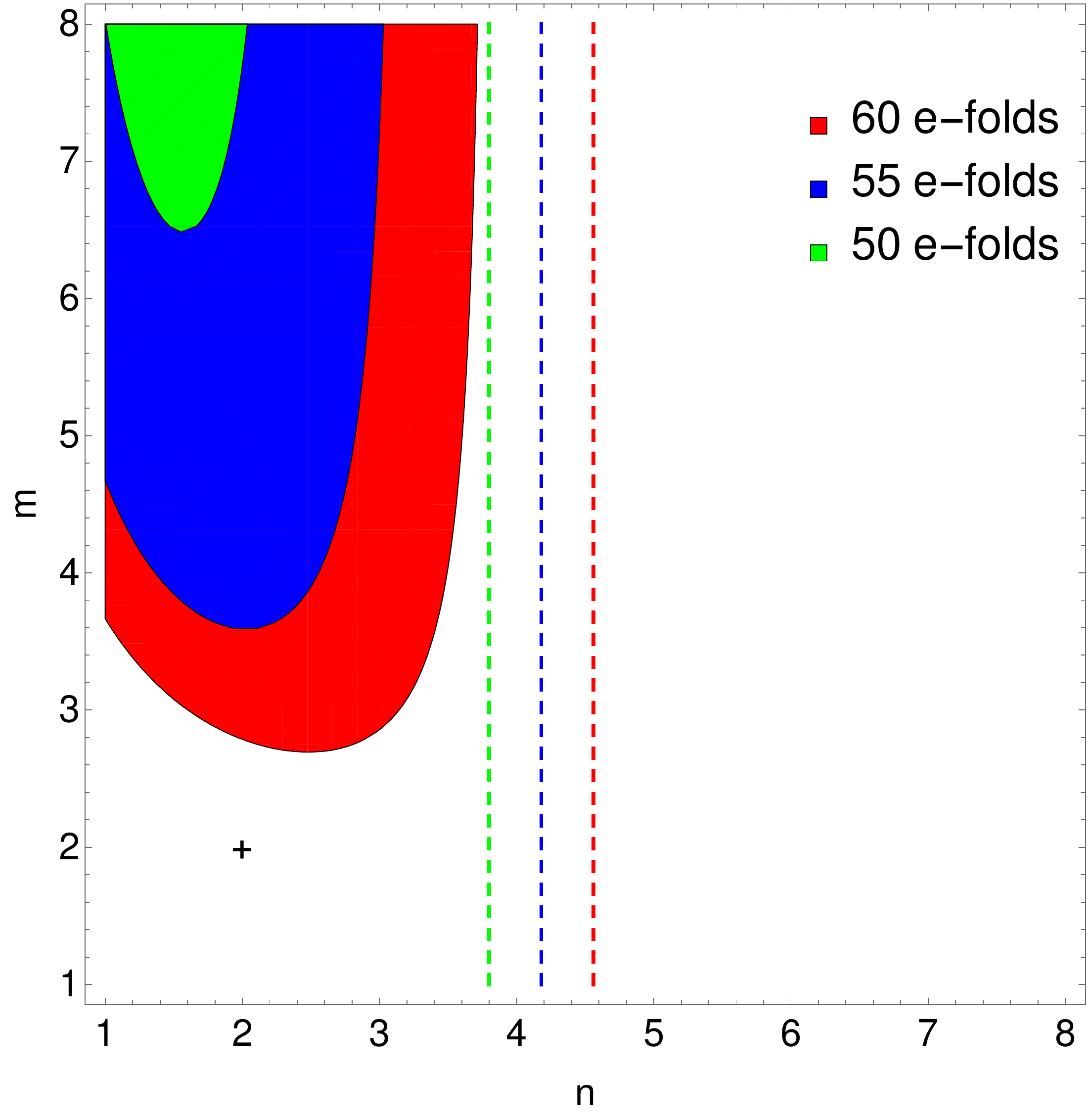}
	\includegraphics[width=0.49\textwidth]{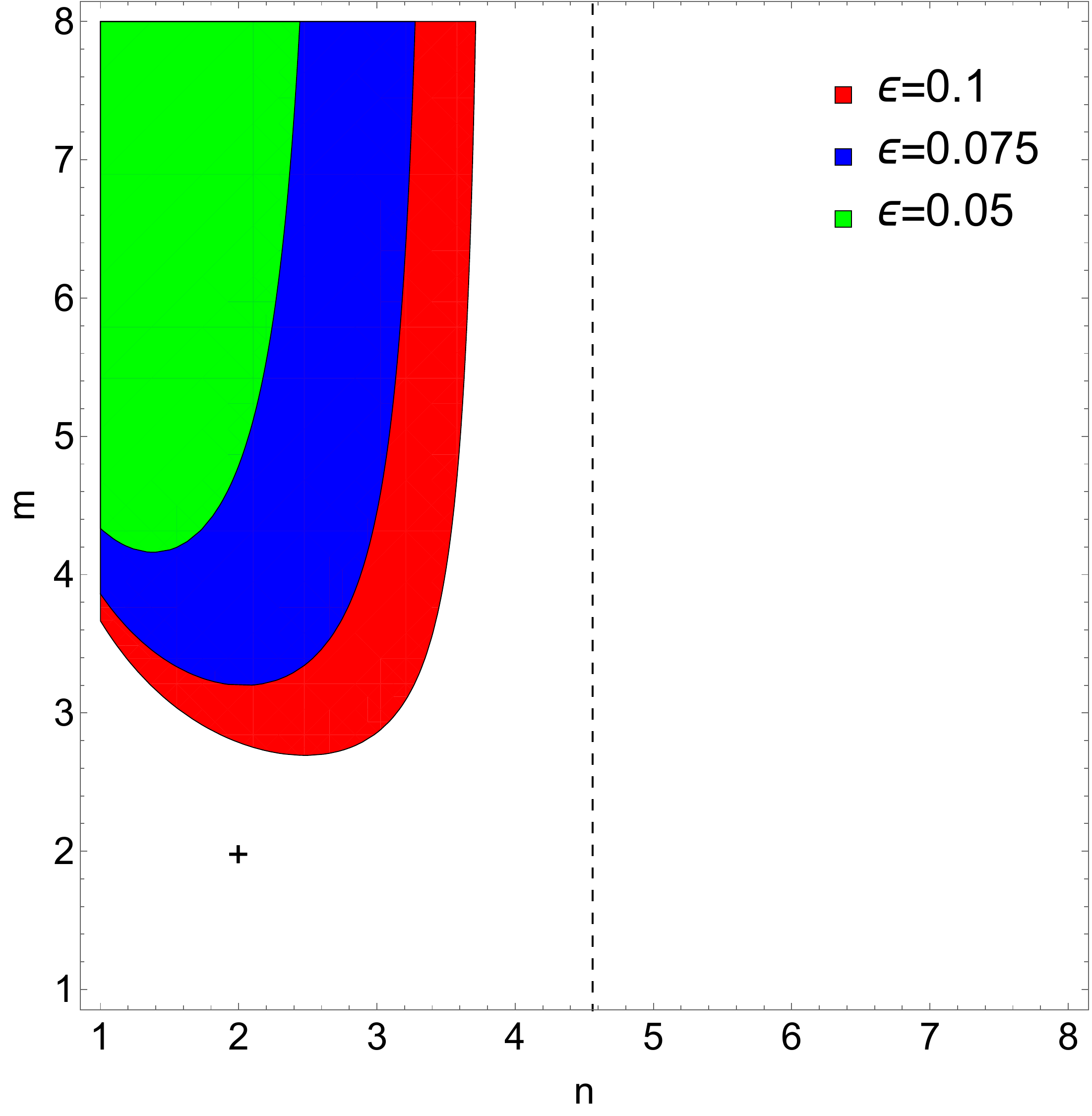}
	\caption{The regions of the parameter space of $m$ and $n$ where the turn of the field trajectory can occur before the end of the slow-roll regime. In the top left figure, these regions are determined for several values of $\fnl$: $|-\frac{6}{5}\fnl|$ larger than 5 (green), 1 (blue) or 0.2 (red). In the top right figure, we explore different error bars of the spectral index: the Planck constraint $n_s = 0.968 \pm 0.006$ (red), $n_s =0.968\pm 0.0015$ in blue (CORE-like experiment) and negligible error bars in green. On the bottom left, different values of $N_\gf$ are represented: $50$ (green), $55$ (blue) and $60$ (red). The last plot changes the constraint on $\ge$ to check when the turn can occur: $\ge=0.05$ (green), $0.075$ (blue) and $0.1$ (red). The dashed vertical lines indicate the change of sign of $\tV_{\gs\gs*}$, which depends on $n$, $N_\gf$ and $n_s$: it is necessarily positive on the right-hand side of this line. The small cross highlights that the double quadratic potential ($m=n=2$) is excluded in all plots.}
	\label{fig:etaper1}
\end{figure}

In figure~\ref{fig:etaper1}, we use the expressions of the previous section to determine the regions of the parameter space of $m$ and $n$ where a turn of the field trajectory might happen before the end of the slow-roll regime. For this we want to verify when multiple-field effects start to play a role or, in terms of slow-roll parameters, we want to find when $\getpe$ becomes of the same order as $\ge$. We choose $\ge$ and not $\getpa$ because $\getpa$ is of the same order as $\ge$ for most cases except if $n\approx 2$ when it is much smaller.

First, we choose the maximum value of $|\getpe_*|$ possible for $|-\frac{6}{5}\fnl|=1$ using the range of values for $\tV_{\gs\gs*}$ determined from the spectral index. Then we compute the maximum value of $|\getpe|$ when $\ge=0.1$. We choose this value of $\ge$ because this is already close to the end of inflation and the slow-roll approximation starts to break down after that point. Moreover, if the turn starts after this time, it is possible that there is not enough time for the isocurvature modes to decay. Finally, we plot the regions of the parameter space of $m$ and $n$ where $\getpe$ is at least as large as $\ge$ at that time, meaning there is a turn of the field trajectory. We also assume that $N_\gf = 60$. These are the default values for the parameters $\fnl$, $N_\gf$ and $\ge$. Next we vary them to test the validity of these choices. We also explore the effects of a future improvement of the spectral index measurements.

The main conclusion of figure~\ref{fig:etaper1} is that for most $m$ and $n$, the turn cannot happen before the end of the slow-roll regime, except in the top left part of the figures (small $n$ and large $m$). For example, the simple quadratic case $m=2$ and $n=2$ (indicated by a small cross) is excluded. 

The first figure shows that obviously the space of allowed parameters decreases if we want $\fnl$ to be larger. In fact, imposing a larger $\fnl$ is the same as imposing a smaller $\getpe_*$. This does not change the evolution of $\getpe$, only its initial condition, so that it will be harder to reach a final value of order $\ge$.

In the second figure, we explore the effects of an improvement of the measurements of the spectral index by comparing the Planck result $n_s=0.968 \pm 0.006$, with the accuracy expected with a CORE-like experiment where the error bar would be of order $\Delta n_s=0.0015$. We also add the case where the error bar becomes negligible. We see that the region where $\fnl$ is at least of order unity is strongly dependent on the spectral index. Decreasing the error bars on $n_s$ decreases the parameter region where $\fnl$ is of order unity. We will see later that in fact it is the lower bound of $n_s$ which matters. If a more accurate measurement would shift the central value of $n_s$, so that its lower bound would be slightly smaller than for Planck, then the size of the top-left region in this plot would increase. This is not indicated in the figure to keep the plot from being too busy, but $n_s = 0.94$ is sufficient to allow most of the parameter region in the figure ($m>2$ and $n<7$).

The third plot shows the effect of the parameter $N_\gf$. We do not know exactly the total duration of inflation; the usual value is between 50 and 60 e-folds. Moreover, we cannot be sure that $N_\gs$ can be neglected, which means that $N_\gf$ is not necessarily the full duration of inflation after horizon-crossing. In this figure, we observe that the surface of the top left region diminishes for smaller $N_\gf$. In fact, for $N_\gf$ smaller than 45 e-folds, it vanishes completely. The smaller $N_\gf$, the harder it will be to build a model where $\fnl$ is large.

The last figure is here to help to determine at what time the turn can occur. In the other figures, the only condition was before the end of the slow-roll regime. However, this regime is valid for most of the time after horizon-crossing. We can see that simply reducing $\ge$ by a factor two reduces a lot the allowed parameter region. This means that having a turn a few e-folds after horizon-crossing is extremely hard to have or even impossible. Most of the time the turn will happen near the end of slow-roll. 

To explain these different behaviours, we first need to discuss $\tV_{\gs\gs*}$. It is determined from the spectral index and $\ge_*$ using equation \eqref{Wss spectral index} which contains two terms: $ \half(n_s-1)$ which is negative and larger in absolute value for the lower bound on the spectral index, and $\ge_*$ which is positive and can be either smaller or larger than the first term. A small $\ge_*$ corresponds to small $n$ and/or large $N_\gf$. This means that in each of the four figures, the left (small $n$) corresponds to a negative $\tV_{\gs\gs*}$, while $\tV_{\gs\gs*}$ is positive on the right (large $n$). The transition happens between $n=4$ and $n=5$ for $N_\gf=60$ for example. If we decrease $N_\gf$, this value decreases and the transition is shifted to the left. The same happens if we increase the lower bound on the spectral index. In every figure this transition is indicated by a dashed vertical line. The sign of $\tV_{\gs\gs*}$ is important because this corresponds to the form of the potential $V$ at horizon-crossing. If it is positive we have a valley, while a negative value describes falling off a ridge.

Now that we have seen the role of the other parameters on $\tV_{\gs\gs}$, we have to explain the different regions by looking at the equations for the evolution of the ratio $\getpe/\getpe_*$ for the different cases. In the valley case ($\tV_{\gs\gs*}>0$), $\gs$ has to decrease to the minimum at $\gs=0$. However, because the potential has to be very flat at horizon-crossing, we start close to the minimum. Even if $\gs$ reaches its minimum before $\gf$, $\getpe$ does not have the time to become large because in $\getpe$, the decrease of $\gs$ is opposed by the increase of $\ge$. Hence, there is no allowed parameter region to the right of the dashed vertical line in the figures.

In the region of negative $\tV_{\gs\gs*}$, the situation is the opposite: $\gs$ increases to fall from the almost flat ridge where it started. Hence in $\getpe$ we have the effect of both $\ge$ and $\gs$ increasing. After inserting $\gs$ for the different cases into \eqref{getpeevolution}, the only dependence on $m$ appears in the ratio $(m-2)/(m-1)$ which tends to $1$ when $m$ increases. This explains the asymptotic behaviour which appears on the right-hand side of the allowed region.
 
Looking at the different expressions for $\gs$, we also see that the largest $\tV_{\gs\gs*}$ in absolute value makes $\gs$ increase the fastest. This implies that the lower bound on the spectral index is the most important to obtain $\tV_{\gs\gs*}$. When $n$ decreases, larger (in absolute value) $\tV_{\gs\gs*}$ are possible, which explains why smaller $m$ are allowed. But in $\gs$ there are also terms which decrease when $n$ becomes smaller and which compensate this effect, which is why for even smaller $n$ the minimum required value of $m$ starts to increase again.

Below equation (\ref{srparametersflat}), the difficulty, or at least the high level of fine-tuning, needed for a model where $\fnl$ is of order unity or more in slow-roll has been highlighted. Here, we showed explicitly that this is even impossible most of the time for simple monomial potentials. However, some examples exist, when $m>4$ and $n<4$ generally. We also showed that $N_\gf$ has to be close to the total number of e-folds after horizon-crossing which should be as large as possible given other constraints (around 60 e-folds), which implies that the turn of the field trajectory is quick. This also means that slow-roll parameters like $\ge_*$ and $\getpa_*$ are exactly the same as in the purely single-field case. However, the observables $n_s$ and $\fnl$ are different. Adding a second field which is responsible for the non-negligible $\fnl$ can help some single-field models which were not working well given the Planck constraints on $n_s$ to go back into the allowed range of parameters. However, this asks a lot of fine-tuning of the potential of the second field. For $\fnl$ to be of order unity or more, this asks even more fine-tuning as only the lowest spectral index values will work. This also means that the improvement of the spectral index measurements expected with a satellite like CORE would seriously constrain the possiblity of having a large $\fnl$, especially if the central value of the spectral index moves closer to the upper bound from Planck.

We have also seen that in the cases that do work, most of the time the turn is near the end of the slow-roll period. This means that $\ge$ and the other parameters are already of order $0.1$ at the start of the turn. Then parameters like $\getpa$ and $\getpe$ can easily become of order 1 or more during the turn when things are getting more violent. The slow-roll approximation is then broken anyway. If the turn happens a bit later, we can expect that isocurvature modes will not have enough time to vanish before the end of inflation (this does not exclude the existence of some cases where they vanish in time, but only a numerical study of such examples is possible). Finally, we can imagine a case where the turn has not started when $\gf$ reaches the minimum of its potential. If this happens, there is a period of large $\ge$ (which would be the end of inflation in the single-field case). Again, during this period the slow-roll approximation is no longer valid. Therefore, these different situations show the need to understand what happens if the very useful slow-roll approximation is not sufficient. This is the topic of the next section, after a brief excursion to product potentials in the final subsection of this section. 

\subsection{Product potential}
\label{secprodpot}

In this subsection, we study the case of product potentials, which take the form $W(\gf,\gs)=U(\gf)V(\gs)$. This case was solved analytically in \citep{Choi:2007su}. Here, we show that the slow-roll version of the $g_\mathrm{int}$ equation \eqref{equadiffsr} takes a simple and nice form which is easy to deal with.

As for the sum-separable case, we start by using the specific form of the potential to find some new relations concerning its derivatives without assuming any approximation. A simple one is $W_{\gf\gs}=\frac{W_{\gf}W_{\gs}}{W}$ which links the second-order mixed derivative of the potential to the first-order ones. Then using the field equation \eqref{fieldeq} and the definitions of $\ge$, $\getpa$ and $\getpe$ given in \eqref{defeps} and \eqref{defeta}, this relation can be rewritten in terms of slow-roll parameters:
\begin{equation}
(3-\ge)\tW_{\gf\gs}=\frac{2}{3}\ge \left[ e_{1\gf}e_{1\gs}\lh(\getpa+3)^2-(\getpe)^2\rh+\getpe(\getpa+3)(e_{1\gf}^{2}-e_{1\gs}^{2})\right].
\end{equation}
We also need the generalized version of \eqref{W11eq}, valid for any two-field potential, which is:
\be
e_{1\gf}e_{1\gs}(\tW_{11}-\tW_{22})=(e_{1\gf}^{2}-e_{1\gs}^{2})\tW_{21} + \tW_{\gf\gs}.
\ee 
Combining the two previous equations and using \eqref{srpareq}, we obtain:
\begin{equation}
\begin{split}
&e_{1\gf}e_{1\gs}\left[-3\gc -\gxpa +\ge\gc-2\ge^2-4\ge\getpa+ \frac{\ge}{3}\lh \gxpa - 2(\getpa)^2 + 2(\getpe)^2\rh\right]\\
&=(e_{1\gf}^{2}-e_{1\gs}^{2})\left[3\getpe+\gxpe+\ge\getpe-\frac{\ge}{3}\lh \gxpe +2\getpa\getpe\rh\right].
\end{split}
\label{product1}
\end{equation}
Similar computations can be done for the third-order derivatives $W_{\gf\gf\gs}=\frac{W_{\gf\gf}W_{\gs}}{W}$ and $W_{\gf\gs\gs}=\frac{W_{\gs\gs}W_{\gf}}{W}$ to show that:
\begin{equation}
\begin{split}
&(3-\ge)\tW_{\gf\gf\gs}=-2\ge\left[(\getpa+3)e_{1\gs}-\getpe e_{1\gf}\right]\tW_{\gf\gf},\\
&(3-\ge)\tW_{\gf\gs\gs}=-2\ge\left[(\getpa+3)e_{1\gf}+\getpe e_{1\gs}\right]\tW_{\gs\gs}.
\end{split}
\label{product2}
\end{equation}
Finally, using the definitions of $\tW_{221}$ and $\tW_{222}$ in terms of third-order derivatives and basis components, substituting them into \eqref{product1} and \eqref{product2} and performing a first-order expansion in terms of slow-roll parameters gives:
\begin{equation}
\begin{split}
\tW_{221} = &-\ge\getpa -\ge\gc +(\getpa)^2 -2\getpa\gc +\gc^2 +(\getpe)^2-\gxpa +\frac{\gc}{\getpe}\gxpe,\\
\tW_{222} = &-\gxpe -\getpe(\ge-2\getpa+2\gc) \\
&-\frac{\gc}{\eta^\perp}\lh-2\ge^2-3\ge\getpa+(\getpa)^2-\gxpa - \ge \gc -2\getpa \gc +\gc^2\rh - \lh\frac{\gc}{\getpe}\rh^2\gxpe.
\end{split}
\end{equation}

These equations can then be used to simplify the right-hand side of \eqref{equadiffsr}, and one easily finds that in fact the right-hand side completely vanishes. Hence, the slow-roll solution consists only of the homogeneous solution and using the initial condition $\dot{g}_{\mathrm{int}*}=-2(\getpe_*)^2 + (\ge_*+\getpa_*-\gc_*)\gc_*$ (from the slow-roll approximation of \eqref{derivgint}) we find:
\begin{equation}
g_\mathrm{int}=-\left[\getpe_{*} - \frac{1}{2\getpe_{*}}(\ge_{*}+\getpa_{*}-\gc_{*})\gc_{*}\right]\bv_{12} 
=- \frac{e_{1\gf*}^2\tV_{\gs\gs*} - e_{1\gs*}^2\tU_{\gf\gf*}}{2e_{1\gf*}e_{1\gs*}} \bv_{12}.
\end{equation}
The most important thing to note here is that the second expression has exactly the same form as the homogeneous part of the sum potential case in \eqref{gint}, without the particular solution. As discussed there, it is that term which can give a large contribution to $\fnl$. The natural question is then if the situation is the same for the product potential. The similarity of the expressions makes it possible to use exactly the same method to answer this question as for the treatment of the sum potential. 

First, we define:
\be
\tilde{g}_{\mathrm{int}}=\frac{-2(\bv_{12})^2}{(1+(\bv_{12})^2)^2} \, g_\mathrm{int} = \frac{e_{1\gf*}^2\tV_{\gs\gs*} - e_{1\gs*}^2\tU_{\gf\gf*}}{e_{1\gf*}e_{1\gs*}} \frac{(\bv_{12})^3}{(1+(\bv_{12})^2)^2},
\label{gint product}
\ee 
which is the entire term depending on $g_\mathrm{int}$ in $\fnl$ \eqref{fNLresult}. As for the sum potential, the only possibility of having this expression larger than order slow-roll is to have one field dominating at horizon crossing: $e_{1\gf*}^2 \approx 1 \gg e_{1\gs*}^2$. But at the same time, it is required that $\bv_{12}$ is at least of order unity (and at least four to obtain the largest $\fnl$). The main difference with the sum potential case comes in fact from the expression for $\bv_{12}$. In the slow-roll approximation, it is possible to solve the Green's function equations. The computation is similar to the sum potential case and is detailed in \citep{TvT1} where it is shown that:
\be
\bv_{12}=\frac{S-S_*}{2e_{1\gs*}e_{1\gf*}},\qquad \bv_{22}=\frac{e_{1\gf}e_{1\gs}}{e_{1\gf*}e_{1\gs*}},
\ee
with $S\equiv e_{1\gf}^2 - e_{1\gs}^2$. These expressions are quite different from \eqref{Green srsolution} for the sum potential.

At horizon crossing, $e_{1\gf*}^2 \approx 1$ meaning $S_* \approx 1$. For the value of $S$ at the end of inflation there are two different situations. As discussed several times in this paper, we want that $\bv_{22}$ goes to zero at the end of inflation to get rid of the isocurvature mode, meaning that the situation is far closer to single-field inflation at the end of inflation than at horizon crossing. Hence, if at the end $\gf$ also dominates (same direction of the field trajectory), $|e_{1\gs}|\ll |e_{1\gs*}|$. This means that $S-S_*\approx e_{1\gs*}^2$, which leads to the fact that $\bv_{12}$ is small compared to 1. In that case $g_\mathrm{int}$ cannot give a large $\fnl$. However, if $\gs$ dominates at the end of inflation (different direction of the field trajectory), we have:
\be
\bv_{12}=\frac{-1}{e_{1\gf*}e_{1\gs*}},
\ee
which is large compared to 1. We can then use that $\frac{(\bv_{12})^3}{(1+(\bv_{12})^2)^2}\approx \frac{1}{\bv_{12}}$ if $|\bv_{12}|\gg 1$ and \eqref{gint product} to write:
\be
\tilde{g}_{\mathrm{int}} \approx  \frac{e_{1\gf*}^2\tV_{\gs\gs*} - e_{1\gs*}^2\tU_{\gf\gf*}}{e_{1\gf*}e_{1\gs*}} \times \frac{e_{1\gf*}e_{1\gs*}}{-1} \approx -\tV_{\gs\gs*},
\ee
which is of order slow roll. Hence also in this case $\fnl$ is small. This is in agreement with the known conclusion that a product potential cannot give a large $\fnl$ in the slow-roll approximation with vanishing isocurvature mode at the end of inflation \citep{Byrnes:2008wi, TvT1}.

\section{Beyond the slow-roll regime} 
\label{Beyond slow-roll}

The previous section showed that it is difficult to have $\fnl$ not be slow-roll suppressed in the slow-roll regime. Is the situation the same if we leave this regime for a short period?
Here we discuss different cases where this can happen and we will show that like in the slow-roll situation, only the homogeneous part of the solution of \eqref{equadiff} is relevant once isocurvature modes have vanished. This means we will use the same quasi-single-field initial conditions at horizon-crossing as above \eqref{srparametersflat}: $V_* \ll U_*$ and $|V_{\gs*}|\ll |U_{\gf*}|$ while $|V_{\gs\gs*} U_{\gf*}^2|\gg |U_{\gf\gf*}V_{\gs*}^2|$ and $|V_{\gs\gs*} V_{\gs*}^2|\ll |U_{\gf\gf*}U_{\gf*}^2|$.

\subsection{Two kinds of turns}

\begin{figure}
	\centering
	\includegraphics[width=0.4\textwidth]{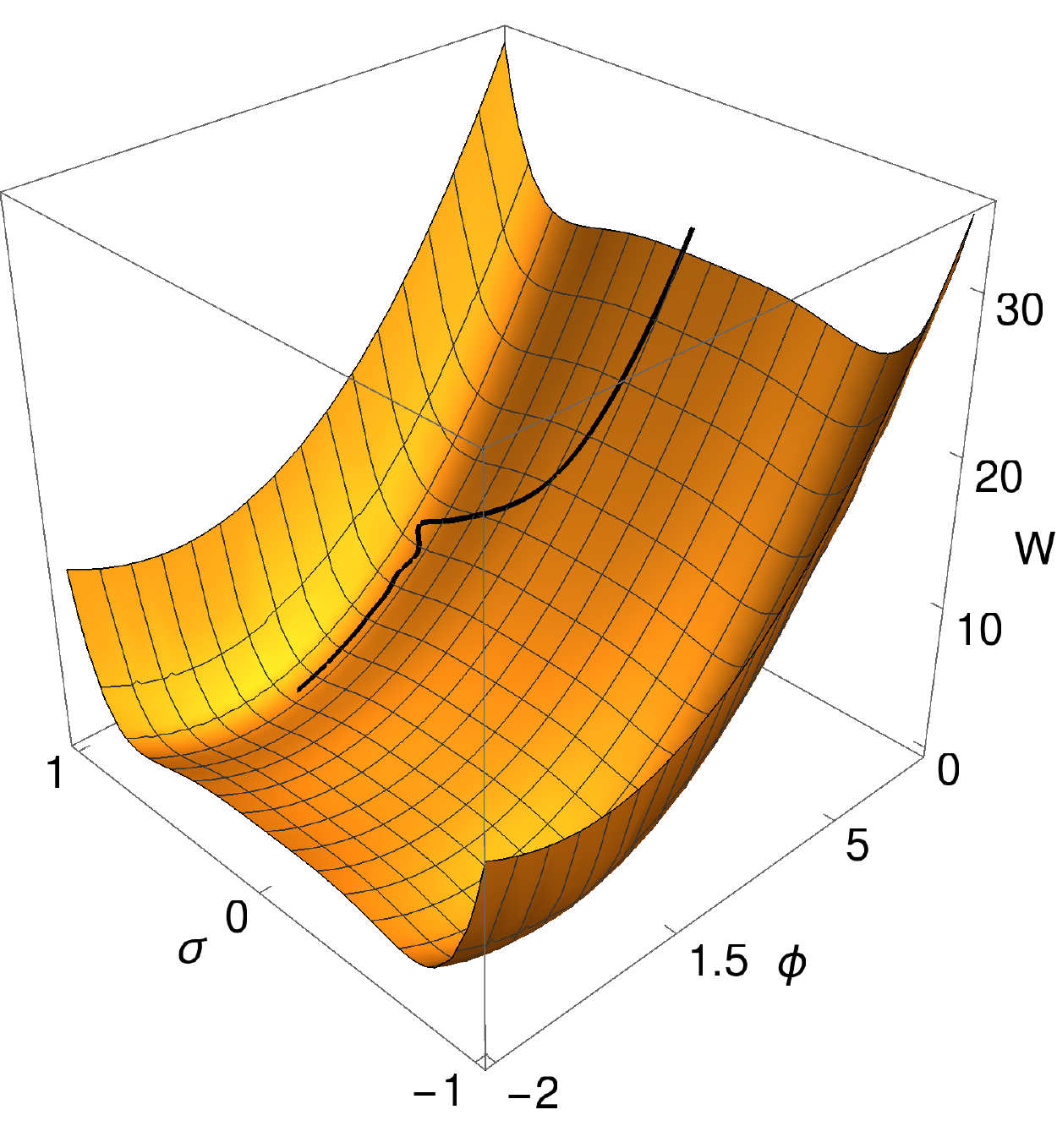}
	\includegraphics[width=0.55\textwidth]{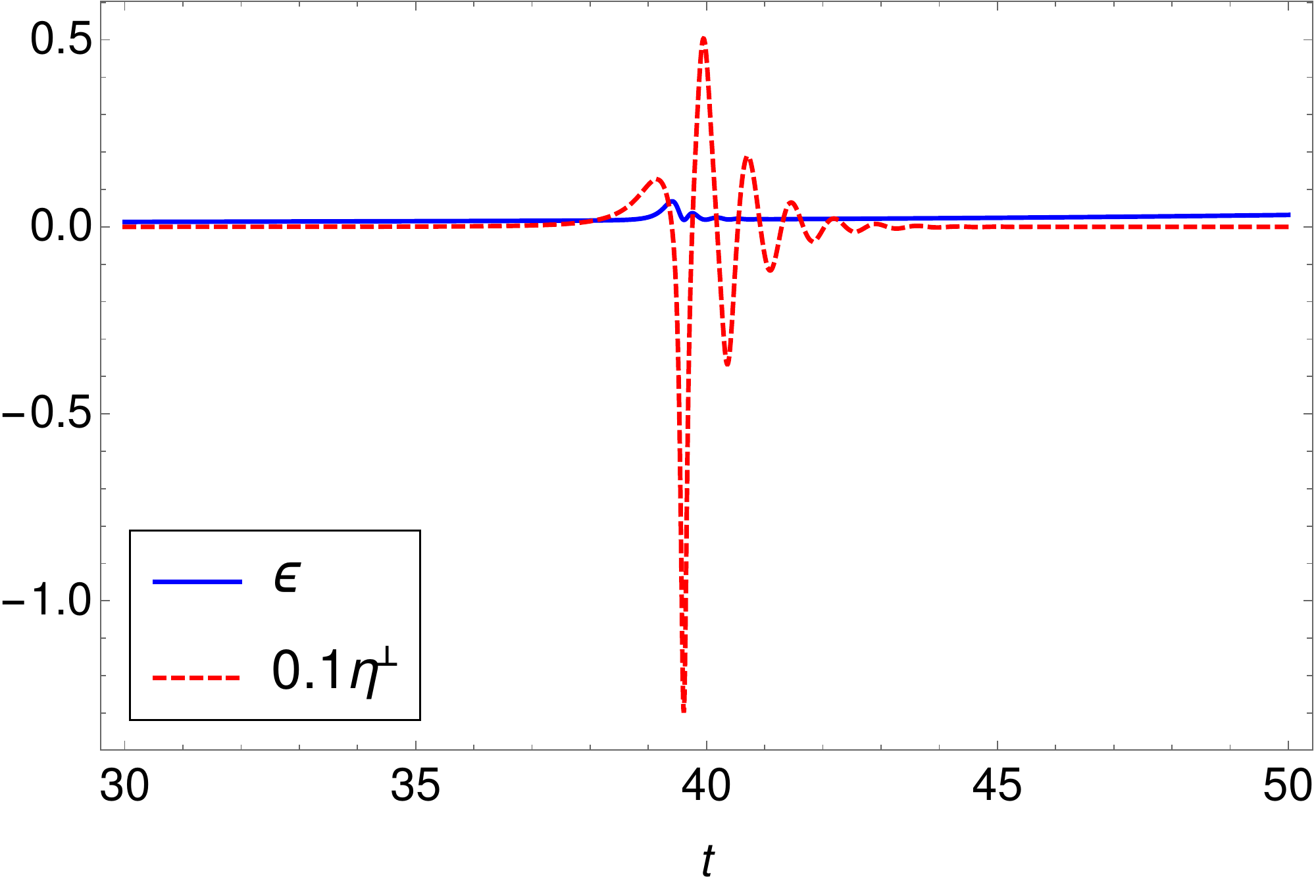}
	\includegraphics[width=0.4\textwidth]{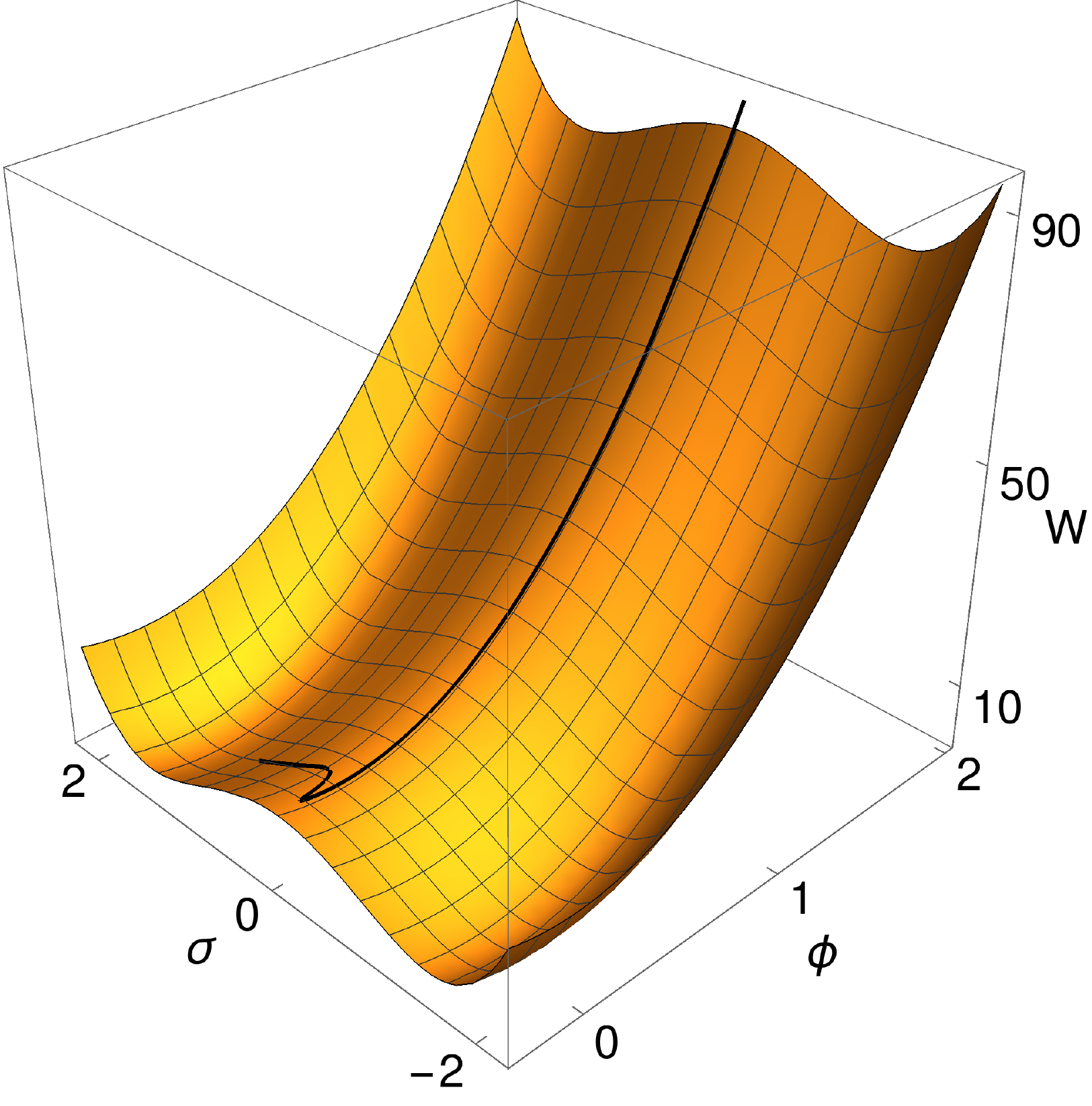}
	\includegraphics[width=0.55\textwidth]{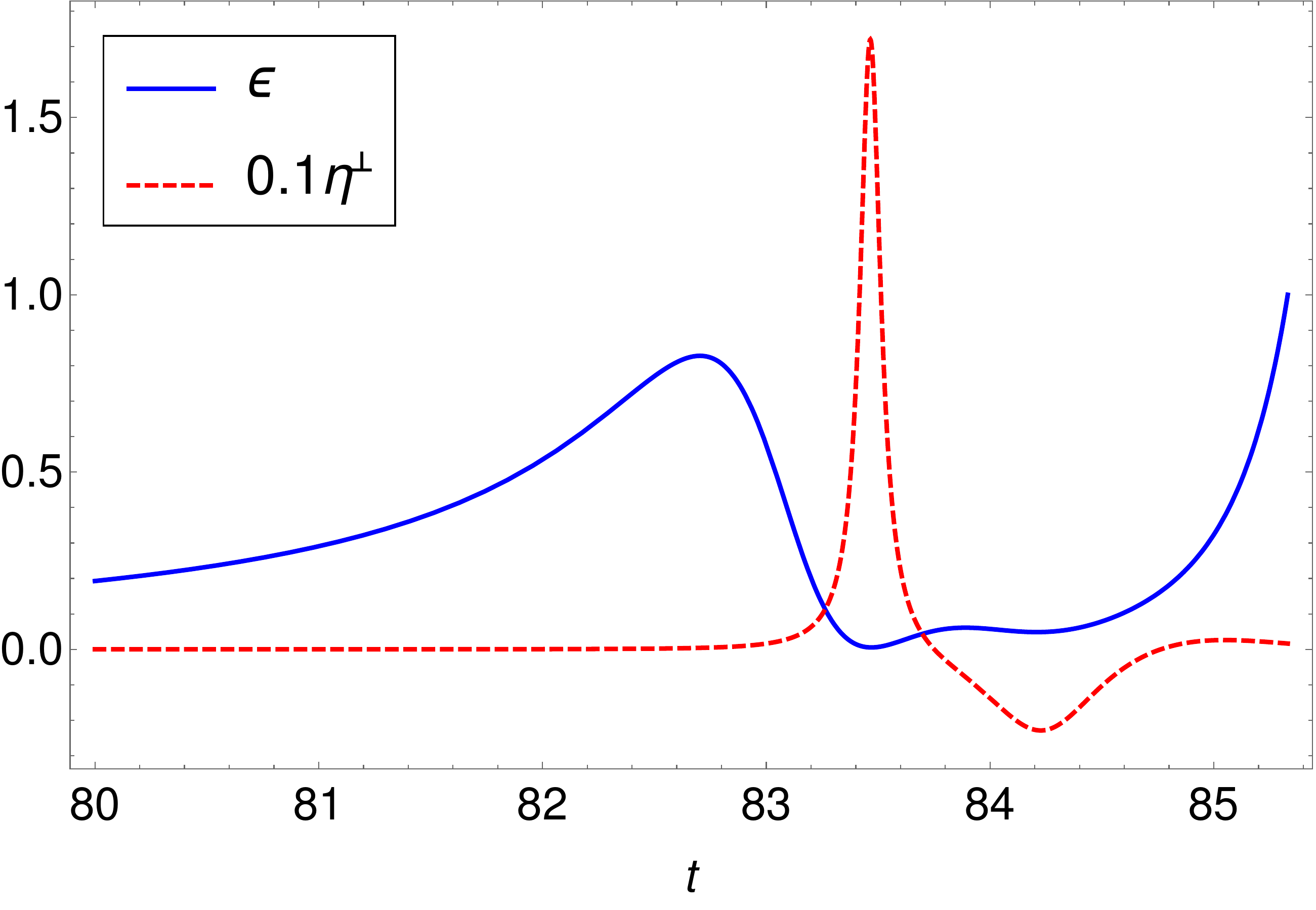}
	\caption{Illustrations of the two different types of turn where the slow-roll approximation is broken. On the left, the field trajectory is displayed in black on the potential while on the right the slow-roll parameters $\ge$ and $\getpe$ are shown for a typical example. The top correspond to what we call the first type, while the plots at the bottom show the second type. Explicit examples of both are given in section~\ref{numerical section}.}
	\label{fig:slowrollbroken}
\end{figure}

We identified two different cases, illustrated in figure \ref{fig:slowrollbroken}, where the slow-roll approximation stops to be valid during the turn. 
In the figure, the main differences of the two situations are highlighted. With potentials of a quite similar form, we have the possibility for two different trajectories depending on the direction before and after the turn. In the previous section, the importance of the parameters $\ge$ and $\getpe$ to study the turn has been highlighted. Graphically they are useful to determine when the turn occurs and when the slow-roll regime is broken.

The first case is the one studied in the previous section. We determined that for a simple monomial potential, if the turn is possible before $\gf$ reaches the minimum of its potential, it is more likely to happen in the last few e-folds when slow-roll parameters are already of order $10^{-1}$, at the limit of the slow-roll approximation. Then, during the turn, $\eta$ parameters may become of order unity or more, which completely invalidates the idea of an expansion in terms of small slow-roll parameters. The turn is still early enough to have $\getpe$ small again at the end of inflation to make the isocurvature mode vanish. In this case the direction of the field trajectory is the same before and after the turn. This is compatible with a monomial potential where we established that $N_\gs$ has to be small compared to $N_\gf$ and the turn is then short. 

In the second case, perpendicular terms are still negligible when $\ge$ becomes of order $10^{-1}$. Then, like in single-field inflation, $\ge$ continues to grow. This is the end of the slow-roll regime. From \eqref{srderivatives} we see that this makes $\getpa$ also become large (in absolute value) and a maximum of $\ge$ is reached when $\getpa= -\ge$. A short time after that point, $\ge$ starts to decrease very fast as the $\getpa$ term dominates in $\dot{\ge}$. A large $\getpa$ also has an effect on the perpendicular parameter $\getpe$ which has been negligible until then. It is possible that $\getpe$ becomes large and that the turn will occur after a few e-folds at most if $\fnl$ is of order unity. Hence, it is possible to have the turn starting with $\ge \ll 1$. This is also motivated by the assumption of isocurvature modes vanishing before the end of inflation. Indeeed, this requires a turn not too close to the end of inflation ($\ge=1$) which is the case if $\ge$ is small compared to one during the turn. In this type of turn, the direction is not the same before and after. Before the turn $\gf$ is dominating but also near the minimum of its potential, while $\gs$ is still at a local maximum. Inflation ends when $\gf$ is still near its minimum but $\gs$ is also evolving towards its own minimum.

In both theses cases, we established that the slow-roll approximation can be broken. We know that solving the equations without any approximation is not possible, even in the simple case of a sum potential. However, we have also seen that $\ge$ is small at the start of the turn simply because of the assumption of vanishing isocurvature modes. Moreover, in $\dot{\ge}$ \eqref{srderivatives}, there is a factor $\ge$ in front. This means that when $\ge$ is small, $\ge$ cannot evolve very fast and will stay small during a short period like the turn, unless the turn is very sharp with $\eta$ parameters becoming very large. Hence during the turn, except in the most extreme cases we do not treat, we still have that $\ge$ is small compared to one which will play an important role in this section.

In the first type of turn, this hypothesis of small $\ge$ has the important consequence that the slow-roll approximation is in fact broken only for the field $\gs$. Indeed, in the field equation \eqref{fieldeq}, each field can only affect the other through $H$ which evolves slowly if $\ge \ll 1$. Hence, even if $\gs$ starts to evolve fast, it is only a small perturbation for $\gf$ which continues to evolve slowly during and after the turn until near the end of inflation when $\ge \approx \gk^2 \dot{\gf}^2 /2$ becomes of order unity. Hence, the derivatives of $\gf$ of order two and more are negligible. This can be used to simplify the slow-roll parameter expressions from \eqref{defeta}, keeping only the terms which are larger than order slow-roll:
\be
\getpa = \frac{\ddot{\gs}\dot{\gs}}{\dot{\gf}^2+\dot{\gs}^2}, \qquad \getpe = -\frac{\ddot{\gs}\dot{\gf}}{\dot{\gf}^2+\dot{\gs}^2}, \qquad \gxpa = \frac{\dddot{\gs}\dot{\gs}}{\dot{\gf}^2+\dot{\gs}^2} \qquad \text{and}\qquad \gxpe = -\frac{\dddot{\gs}\dot{\gf}}{\dot{\gf}^2+\dot{\gs}^2}.
\ee
Using this, a direct computation gives useful relations between the parallel and perpendicular parameters of the same order:
\be 
\label{first case}
e_{1\gf}\getpa = -e_{1\gs}\getpe \qquad\text{and}\qquad e_{1\gf}\gxpa = -e_{1\gs}\gxpe.
\ee

In the second type of turn, the slow-roll approximation is broken for the two fields, so that these relations are then not valid. However, there is also an important approximation we can make in this case. Before the turn, the slow-roll approximation is broken during the period of large $\ge$. Having $\ge$ large for some time also means that $H$ decreases a lot during that period. This means that during the turn, we have:
\be
H^2 \ll H_*^2.
\label{second case}
\ee
A brief remark about the end of inflation is necessary. We use the common definition that the period of inflation finishes when $\ge=1$. However, in the second type of turn, $\ge$ can be larger than 1 for a very small number of e-folds before the turn. A more complete definition of the end of inflation is then that $\ge=1$ with $U\ll U_*$ and $V\ll V_*$, which ensures that the second field as well had time to evolve.

The main tool in this section is the differential equation \eqref{equadiff} which we will  call the $\gint$ equation. We have already solved it during the period of slow-roll which goes from horizon-crossing to the turn or to $\ge$ of order 1. We also know the exact homogeneous solution of the full equation. The only remaining work is to understand what happens to the particular solution beyond the slow-roll approximation. We will each time follow the same method. First we discuss each equation in the more general case, only supposing that $\getpa$ and $\getpe$ are large while $\ge \ll 1$. Then, when needed to go further, we will study separately each case using \eqref{first case} or \eqref{second case} depending on the type of turn considered.

\subsection{Green's functions}
\label{Greensfunctionssubsec}

Beyond the slow-roll regime, we have to solve the second-order differential equation \eqref{G22eq} to compute the Green's functions (recalling that $\bv_{22}(t)$ and $G_{22}(t,t_*)$ obey the same equation). We assume that the solution has the form $\bv_{22} \propto f e_{1\gf}e_{1\gs}$, similar to the slow-roll case (\ref{Green srsolution}). One motivation is that, during the turn, the dominant term will be $(\getpe)^2$ and this is canceled by this form of solution. Substituting this into \eqref{G22eq}, we find a differential equation for the function $f$:
\be
\label{feq}
\begin{split}
e_{1\gf}e_{1\gs}\,\ddot{f}&+\left[2\getpe(e_{1\gs} ^2 - e_{1\gf} ^2) + (3 + \ge + 2\getpa)e_{1\gf}e_{1\gs}\right]\dot{f} \\ &+\left[2\ge\getpe(e_{1\gs} ^2 - e_{1\gf} ^2)+(6\ge+2\ge^{2}+4\ge\getpa)e_{1\gf}e_{1\gs}\right]f=0.
\end{split}
\ee
In the slow-roll regime, a first-order expansion of this equation gives
\be
\label{feqsumsr}
e_{1\gf}e_{1\gs}\,\dot{f} +2\ge\, e_{1\gf}e_{1\gs}f=0,
\ee
and then it is easy to show that $f = H^2$ to find the slow-roll result (\ref{Green srsolution}). During this initial period of slow roll, having a first-order equation as a very good approximation means that the second mode needed to solve the full equation rapidly becomes negligible. Once slow roll is broken, we only need to study how the remaining mode evolves.

In the general case, an analytical solution cannot be found. However, if we take a solution of the form $f=H^\ga$ by inspiration from the slow-roll solution (because that is the form of the solution until the moment when the slow-roll regime is broken), \eqref{feq} becomes:
\be
\begin{split}
&e_{1\gf}e_{1\gs}\left[\ddot{\ga}\ln H + \dot{\ga}^{2} (\ln H)^2-\ga\dot{\ga}\ln H + \dot{\ga}\lh -2\ge+(3+\ge+2\getpa)\ln H\rh \right. \\
&\left. + (\ga-2)(\ga-1)\ge^2 -4(\ga-1)\ge\getpa-3(\ga-2)\ge\right] + 2\getpe(e_{1\gs} ^2 - e_{1\gf} ^2) \left[ \dot{\ga}\ln H +(1-\ga)\ge \right] = 0.
\end{split}
\ee
There are two interesting values for $\ga$ which are 1 and 2. They can be linked to the two regimes already discussed previously where the slow-roll approximation is not valid.

We can see directly that the lowest order term in slow-roll is canceled by $\ga = 2$ as expected. Moreover, the $\ge^2$ term also vanishes with this value. This means that when $\ge$ becomes larger while the other parameters are still small compared to 1, $f=H^2$ is still a good approximation. This is exactly what happens at the end of the slow-roll regime just before the second type of turn, when the first field is near the minimum of its potential. Then, the complete solutions for the Green's functions are:
\be
\bv_{22} = \frac{H^2 e_{1\gf} e_{1\gs}}{H^2_* e_{1\gf*}e_{1\gs *}},\qquad
\bv_{32} =  \frac{H^2}{H^2_* e_{1\gf*}e_{1\gs *}} \lh -2\ge e_{1\gf}e_{1\gs} +  \getpe (e_{1\gs}^2 - e_{1\gf}^2)\rh.
\ee
The same integration as in the slow-roll case works to compute $\bv_{12}$:

\be
\bv_{12} = \frac{Z-Z_*}{W_* e_{1\gf*}e_{1\gs*}},
\ee
with $Z$ previously introduced in \eqref{Z}.

The other interesting value $\ga=1$ cancels every second-order term in the equation. Hence, this is a good solution when $\getpe$ and $\getpa$ are large but $\ge$ is small compared to 1, hence during the turn. The solutions are then, 

\be
\label{v22turn}
\bv_{22} = \frac{H}{\cal{N}}e_{1\gf}e_{1\gs}, \qquad
\bv_{32} =  \frac{H}{\cal{N}} \lh -\ge e_{1\gf}e_{1\gs} +  \getpe (e_{1\gs}^2 - e_{1\gf}^2)\rh,
\ee
where $\cal{N}$ is a constant used to satisfy the continuity of $\bv_{22}$. If we call the time when this solution becomes better than the previous one $t_0$, we have $\cal{N}$ $=\frac{H^2_*}{H(t_0)} e_{1\gf*}e_{1\gs*}$. 

We cannot directly compute $\bv_{12}$ in this regime. However, $\ge$ is supposed to be very small compared to 1 which means that $H$ is almost a constant. We can then write $H(t) = H_0 + \delta H(t)$ where $\delta H(t)$ is only a small correction. Taking the square of this expression and doing a first-order slow-roll expansion gives $\delta H(t)=\half \textstyle\frac{ H^2-H_0^2}{H_0}$. Then it is easy to deduce $H(t) = \half \textstyle\frac{ H^2+H_0^2}{H_0}$. Substituting this into \eqref{v22turn}, we can perform the integration and we get:
\be
\label{v_12}
\bv_{12} = - \frac{H_0^2}{H_*^2} \frac{S-S_0}{4 e_{1\gf*}e_{_1\gs*}} + \frac{Z/2 + Z_0/2 - Z_*}{W_* e_{1\gf*} e_{_1\gs*}},
\ee
with $S\equiv e_{1\gf}^2- e_{1\gs}^2$.

\subsection{The $\gint$ equation during the turn}

A first use of the Green's functions during the turn computed in the previous section is to insert them into \eqref{equadiff} to simplify the right-hand side of the equation: r.h.s.~$\equiv K_{22} (\bv_{22})^2 + K_{23} \bv_{22}\bv_{32} + K_{33}(\bv_{32})^2$. After this step, every term of r.h.s.\  has one factor depending on the basis components: $e_{1\gf}^2 e_{1\gs}^2$, $e_{1\gf}e_{1\gs}(e_{1\gs}^2-e_{1\gf}^2)$ or $(e_{1\gs}^2-e_{1\gf}^2)^2$. We use the relation $(e_{1\gs}^2-e_{1\gf}^2)^2=1-4e_{1\gf}^2 e_{1\gs}^2$ coming from the normalization of the basis to eliminate one of the factors. Having terms with these factors permits us to use equations \eqref{W11eq2} and \eqref{sumequation} to eliminate the slow-roll parameters $\gc$, $\tW_{221}$ and $\tW_{222}$. Finally, we obtain:
\be
\begin{split}
  \mathrm{r.h.s.}= \ge \lh\frac{H}{\cal{N}}\rh ^2 & \Bigl\{ 2(\getpe)^4(\ge+3\getpa) \\
& + e_{1\gf}^2 e_{1\gs}^2 \left[ (\getpe)^4 \lh -18-14\ge-36\getpa\rh - 2(\getpe)^3 \gxpe + 2\getpa (\gxpe)^2 \right. \\
&\left. \qquad\qquad + (\getpe)^2 \lh-3\tW_{111} -18\ge - 6\ge^{2} -24\ge\getpa + 18(\getpa)^2 + 6\gxpa + 2 \ge^2 \getpa \right.\right. \\
&\left.\left. \qquad\qquad\qquad\qquad  + 10 \ge (\getpa)^2 + 12 (\getpa)^3  + 2\ge \gxpa + 12 \getpa \gxpa\rh + 3\getpe \getpa \tW_{211}  \right.\\
&\left. \qquad\qquad+ \getpe\gxpe \lh 6\ge-6\getpa-2\ge\getpa-10(\getpa)^2 -2\gxpa\rh  \right]\\
& + e_{1\gf}e_{1\gs} (e_{1\gs}^2-e_{1\gf}^2)  \left[ -6(\getpe)^5 -4\getpa(\getpe)^2\gxpe \right.\\
& \left. \qquad\qquad\qquad\quad + (\getpe)^3 \lh- 6\ge + 18\getpa + 2\ge^2 + 12 \ge \getpa + 18 (\getpa)^2 + 6 \gxpa \rh \right]\Bigr\}. 
\label{rhs}
\end{split}
\ee 
At first sight, this expression does not look simpler than the original one. However, it has an important new feature which is the $\ge$ factor in front of the whole expression. In fact, in the computation every term without $\ge$ cancels. Recalling that the main assumption we made is that $\ge$ is small during the turn, this indicates that r.h.s.\  might be negligible during the turn, which means that only the homogeneous solution (which is known) is needed. In the rest of this section we will show that this is indeed the case. 

First we have to figure out compared to what r.h.s.\  has to be negligible. One way to answer this question is to use what we already know about the solution: the slow-roll expression given in \eqref{gint} which we write as $\dot{g}_{\mathrm{int}} = P_{\mathrm{sr}} + h_{\mathrm{sr}}$ with:
\be
P_{\mathrm{sr}} = 2\ge(\ge+\getpa)(\bv_{22})^2 + 2\ge\bv_{22}\bv_{32},\qquad
h_{\mathrm{sr}} = - \frac{e_{1\gf*}^2\tV_{\gs\gs*} - e_{1\gs*}^2\tU_{\gf\gf*}}{e_{1\gf*}e_{1\gs*}} \getpe\bv_{22}.
\label{dotgint}
\ee
Here we used that $\bv_{32}=-\gc \bv_{22}$ in the slow-roll regime. $P_{\mathrm{sr}}$ corresponds to the particular solution while $h_{\mathrm{sr}}$ is the homogeneous part. We will study these two parts of the solution in the two next subsections to see how they evolve beyond the slow-roll regime. In the subsection after that, we will discuss why they are sufficient to solve the $\gint$ equation even beyond the slow-roll approximation. We start by focusing on this homogeneous solution.

\subsection{Fate of the slow-roll homogeneous solution}

As already discussed before, the homogeneous slow-roll solution is also a homogeneous solution of the full second-order equation. Hence, we can use it and substitute it into \eqref{equadiff}. Then we look at each term (order1 $\propto \dot{h}_{\mathrm{sr}}$, order2 $\propto \ddot{h}_{\mathrm{sr}}$ and order3 $\propto \dddot{h}_{\mathrm{sr}}$) individually and not at the total sum because that is obviously zero. We want to show that these terms are large compared to r.h.s., so that, during the turn, r.h.s.\  is only a small correction which can be neglected to get a good approximation of $\gint$. To compute the three left-hand side terms, we use the same steps as in deriving (\ref{rhs}) to get:
\be
\begin{split}
\mathrm{order1} = & -\frac{H}{\cal{N}} \frac{e_{1\gf*}^2\tV_{\gs\gs*} - e_{1\gs*}^2\tU_{\gf\gf*}}{e_{1\gf*}e_{1\gs*}} \left\{  (e_{1\gs}^2-e_{1\gf}^2) \left[-6(\getpe)^4 -2(\getpe)^3\gxpe \right] \right. \\
&\left. + e_{1\gf}e_{1\gs}\left[ 6(\getpe)^5 + (\getpe)^3 \lh 6(\getpa)^2 +2\gxpa\rh  -8\getpa\gxpe(\getpe)^2  + 2\getpe (\gxpe)^2 + 3 (\getpe)^2\tW_{211} \right] \right\} \\
\mathrm{order2} = & \frac{H}{\cal{N}} \frac{e_{1\gf*}^2\tV_{\gs\gs*} - e_{1\gs*}^2\tU_{\gf\gf*}}{e_{1\gf*}e_{1\gs*}} \left\{ (e_{1\gs}^2-e_{1\gf}^2) \left[ (\getpe)^4 \lh-3+\ge-6\getpa\rh + 2(\getpe)^3 \gxpe\right] \right. \\
& \left. + e_{1\gf}e_{1\gs} \left[ (\getpe)^3 \lh 6\getpa-2\ge\getpa+12 (\getpa)^2\rh + (\getpe)^2 \gxpe \lh -3 +\ge -10 \getpa\rh + 2\getpe (\gxpe)^2 \right]\right\}\\
\mathrm{order3} = & \frac{H}{\cal{N}} \frac{e_{1\gf*}^2\tV_{\gs\gs*} - e_{1\gs*}^2\tU_{\gf\gf*}}{e_{1\gf*}e_{1\gs*}} \left\{ (e_{1\gs}^2-e_{1\gf}^2) \left[ (\getpe)^4 \lh-3-\ge+6\getpa\rh - 4 (\getpe)^3 \gxpe\right] \right. \\ 
&\left. + e_{1\gf}e_{1\gs} \left[ 6(\getpe)^5 + (\getpe)^3 \lh -6\getpa +2\ge\getpa - 6 (\getpa)^2 +2 \gxpa\rh + (\getpe)^2 \gxpe \lh 3 -\ge + 2 \getpa \rh\right.\right.\\
&\left.\left. \qquad\quad\quad +3 (\getpe)^2 \tW_{211}\right]\right\}.
\label{order}
\end{split}
\ee
We separate our equations into parts easier to compare. We start by comparing the factors in front of the braces of each expression in \eqref{order} and \eqref{rhs} which are:
\be
\label{equadiff_factors}
\frac{H}{\cal{N}} \frac{e_{1\gf*}^2\tV_{\gs\gs*} - e_{1\gs*}^2\tU_{\gf\gf*}}{e_{1\gf*}e_{1\gs*}} \qquad \text{and}\qquad  \ge(\frac{H}{\cal{N}})^2.
\ee
After simplifying the common factor $H/\cal{N}$ and inserting $\cal{N}$ $= H^2_* e_{1\gf *}e_{1\gs *}/H_0$ from \eqref{v22turn}, we use the quasi single-field initial conditions at horizon-crossing to write \eqref{equadiff_factors} as
\be
\tV_{\gs\gs*}\frac{H_*^2}{H^2} \qquad\text{and}\qquad \ge.
\ee
The discussion about the spectral index from around equation \eqref{nstwofield} is still valid, because the only difference from the slow-roll regime is the value of $\bv_{12e}$, but for a large enough value (larger than four) the dependence on $\bv_{12e}$ in \eqref{ns} disappears and \eqref{nstwofield} can be used. Hence, $\tV_{\gs\gs*}$ is typically of order $10^{-2}$, or at least not hugely smaller.

As for the size of $\ge$ and $H_*^2/H^2$, this depends on the type of turn. For the first type, $\ge$ is still of order slow-roll but it can be easily larger than $\tV_{\gs\gs*}$ by an order of magnitude. However $H_*^2 /H^2$ is also larger than one. Moreover, if $\ge$ had enough time to increase since horizon-crossing, the situation is the same for $H_*^2 /H^2$ because $H$ decreases faster if $\ge$ is larger. During a few dozens of e-folds with $\ge$ of order slow-roll, it can also increase by an order of magnitude. This means that both terms will be of the same order during the turn in this case, or at least that neither of them is hugely smaller or larger than the other. For the second type of turn, the situation is different. During the turn, $\ge$ is again of order slow-roll so it is not hugely larger than $\tV_{\gs\gs*}$. However, because of the period of large $\ge$, we know that $H_*^2/H^2 \gg 1$ from \eqref{second case}. Hence the factor in front of order1, order2 and order3 is large compared to the one in r.h.s.\  in this case.

Next we focus on the second part of each expression, which is the part inside the braces and which is a complicated expression depending on basis components and slow-roll parameters. We start with some comments on the factors $e_{1\gf}e_{1\gs}$ and $e_{1\gs}^2 -e_{1\gf}^2$. By definition of the basis, $e_{1\gf}e_{1\gs}$ goes from $-\half$ to $\half$ and $e_{1\gs}^2 - e_{1\gf}^2$ from $-1$ to $1$ and when one is at an extremum, the other one vanishes. When one vanishes, the leftover slow-roll parameter terms are similar in the different expressions. It is also not possible to have both of them small compared to one at the same time, hence the term in r.h.s.\  without a factor depending on the basis is not an issue. Hence, we can forget about these basis component factors which cannot change the conclusion.

The different expressions depend on all the first and second-order slow-roll parameters, except $\gc$, $\tW_{221}$ and $\tW_{222}$ which have been eliminated using the relations specific to sum potentials \eqref{W11eq2} and \eqref{sumequation}. The first step is to study the cancellations of the left-hand side terms. An obvious one is when $\getpe$ vanishes because it multiplies every term in \eqref{order}; the homogeneous solution vanishes in that case. It also multiplies every term in r.h.s.\  except the one term $2\getpa(\gxpe)^2$. However $\gxpe$ is also small when $\getpe$ becomes small. During the turn of the field trajectory, it is usual that the slow-roll parameters oscillate, hence $\getpe$ can vanish several times. At those times our hypothesis that r.h.s.\  is much smaller than the other terms is not valid and we cannot neglect the particular solution. However, we will show in a later subsection that we have a way of dealing with this. Apart from this vanishing of $\getpe$, there is no other possibility to cancel order1, order2 and order3 simultaneously. Indeed the expressions contain similar terms, but with opposite signs or different numerical constants.

Once we know there are no cancellations in the left-hand side terms (apart from the  moments when $\getpe=0$), we can compare their expressions to r.h.s.\  and verify they are of the same order. As the expressions contain terms up to order five in slow-roll parameters, two cases have to be differentiated. First, the slow-roll parameters can be of order unity. Then the powers do not matter and most of the terms have to be taken into account. We remark that the terms are similar on each side of the equation, and that the numerical constants are also of the same order, so that r.h.s.\  cannot be very large compared to the other expressions in this case. However, the slow-roll parameters can also become larger than order unity and this situation requires more discussion. An important remark is that when the slow-roll approximation is broken, the slow-roll cancellations in \eqref{srpareq} disappear which means that $\gxpa$ and $\gxpe$ are of order a few times $\getpa$ and $\getpe$ respectively, and not of order $(\getpa)^2$ and $(\getpe)^2$. Using the expressions for $\dot{\eta}^\parallel$ and $\dot{\eta}^\perp$ in \eqref{srderivatives}, we can see that when $|\getpa|$ is at a maximum, $|\getpe|$ has to be of the same order because the only possibility to cancel the largest term $(\getpa)^2$ in the derivative expression is to have $(\getpe)^2$ of the same order. However, when $|\getpe|$ is at a maximum, we can see in a similar way that $|\getpa|$ must be of the order of a few at most.

Then we can study what happens if the perpendicular parameters are the largest (near the maximum of $|\getpe|$). If $\getpa$ is only a few, the dominant terms in r.h.s.\  and the order1,2,3 are the ones in $(\getpe)^5$ and $(\getpe)^4$ (or the equivalent $(\getpe)^3 \gxpe$). The same terms exist in all the different expressions meaning the part inside the braces has to be of the same order in general. If, on the other hand, the parallel parameters are the largest, there is a term in $(\getpe)^2(\getpa)^3$ in r.h.s.\  which does not exist in the other expressions. However, as discussed a few lines earlier, $\getpe$ is also of the same order as $\getpa$ at that time. Using this, the dominant terms are actually of order $(\getpe)^5$. Again we find similar terms inside the braces for the different expressions which have to be of the same order. Finally, the only term in r.h.s.\  that has no equivalent in the other expressions is $(\getpe)^2 \tW_{111}$. This term, which is only of order three, can never be dominant because $\tW_{111}$ cannot be large enough to make this term a lot larger than the order five ones because this parameter is also in the derivative of $\gxpa$ (see \eqref{srderivatives}).

Hence, we have established that the terms inside the braces are of the same order in the general case for each expression in \eqref{rhs} and \eqref{order}. This is exactly the situation for the second type of turn where the only hypothesis not used \eqref{second case} has no consequence for the terms inside the braces. However, for the first type of turn, the relations \eqref{first case} between the parallel and perpendicular slow-roll parameters of the same order can change the result. To verify this, we substitute them into \eqref{rhs} and \eqref{order}. We also introduce the notation with $\{\}$ in subscript, meaning we consider only the terms inside the braces. The computation gives:
\be
\begin{split}
\mathrm{r.h.s.}_{\{\}}&= -e_{1\gf}^2 e_{1\gs}^2 \left[(\getpe)^2 \lh 3 \tW_{111}+6 \ge^2+18 \ge\rh -6 \getpe \gxpe \ge\right] \\
& - e_{1\gf} e_{1\gs}^3 \left[ 3 (\getpe)^2 \tW_{211}+(\getpe)^3 \lh -2 \ge^2-12 \ge\rh \right]
- e_{1\gf} e_{1\gs} (\getpe)^3 \lh 2 \ge^2-6 \ge\rh +2 e_{1\gf}^2 (\getpe)^4 \ge,\\
\mathrm{order1}_{\{\}}&= \frac{e_{1\gs}}{e_{1\gf}}  \left[ 4 \getpe (\gxpe)^2 + 12(\getpe)^5 + 6(\getpe)^2 \tW_{211}\right]  + \frac{e_{1\gs}^3}{e_{1\gf}} \left[ -4 \getpe (\gxpe)^2-6 (\getpe)^2 \tW_{211}\right] \\
&~~~ + e_{1\gs}^2 \left[ 4 (\getpe)^3 \gxpe-24 (\getpe)^4\right] + 4 (\getpe)^3 \gxpe + 12 (\getpe)^4 ,\\
\mathrm{order2}_{\{\}}& = 12\frac{ e_{1\gs} }{e_{1\gf}}(\getpe)^5 - 48\frac{ e_{1\gs}^3}{e_{1\gf}}(\getpe)^5 - 2 e_{1\gf} e_{1\gs} \left[ 2 \getpe (\gxpe)^2+(\getpe)^2 \gxpe (\ge-3)\right] \\
&~~~-2 e_{1\gs}^2 \left[ 14 (\getpe)^3 \gxpe+(\getpe)^4 (4 \ge-12)\right] +4 (\getpe)^3 \gxpe-6 (\getpe)^4+2 (\getpe)^4 \ge ,\\
\mathrm{order3}_{\{\}}& = \frac{e_{1\gs} }{e_{1\gf}} \left[ (\getpe)^2 (\gxpe (2 \ge-6)-6 \tW_{211})-24 (\getpe)^5\right] \\
&~~~ + \frac{e_{1\gs}^3 }{e_{1\gf}} \left[ 48 (\getpe)^5+(\getpe)^2 (6 \tW_{211}+\gxpe (6-2 \ge))\right] \\
&~~~ + e_{1\gs}^2 \left[ 24 (\getpe)^3 \gxpe+8 (\getpe)^4 \ge\right] -8 (\getpe)^3 \gxpe+(\getpe)^4 (-2 \ge-6).
\end{split}
\label{rhs first case}
\ee
We can directly see that the higher order terms in r.h.s.$_{\{\}}$ have disappeared but are still present in the left-hand side terms. Moreover, most of the remaining terms in r.h.s.$_{\{\}}$ are now proportional to $\ge$, which makes them even smaller. Finally, the divisions by the basis components $e_{1\gf}$ and $e_{1\gs}$, which are smaller in absolute value than one, only appear in order1$_{\{\}}$, order2$_{\{\}}$ and order3$_{\{\}}$. All these observations leads to the conclusion that r.h.s.$_{\{\}}$ is in fact small compared to left-hand side terms for the first type of turn.

\begin{figure}
	\centering
	\includegraphics[width=0.49\textwidth]{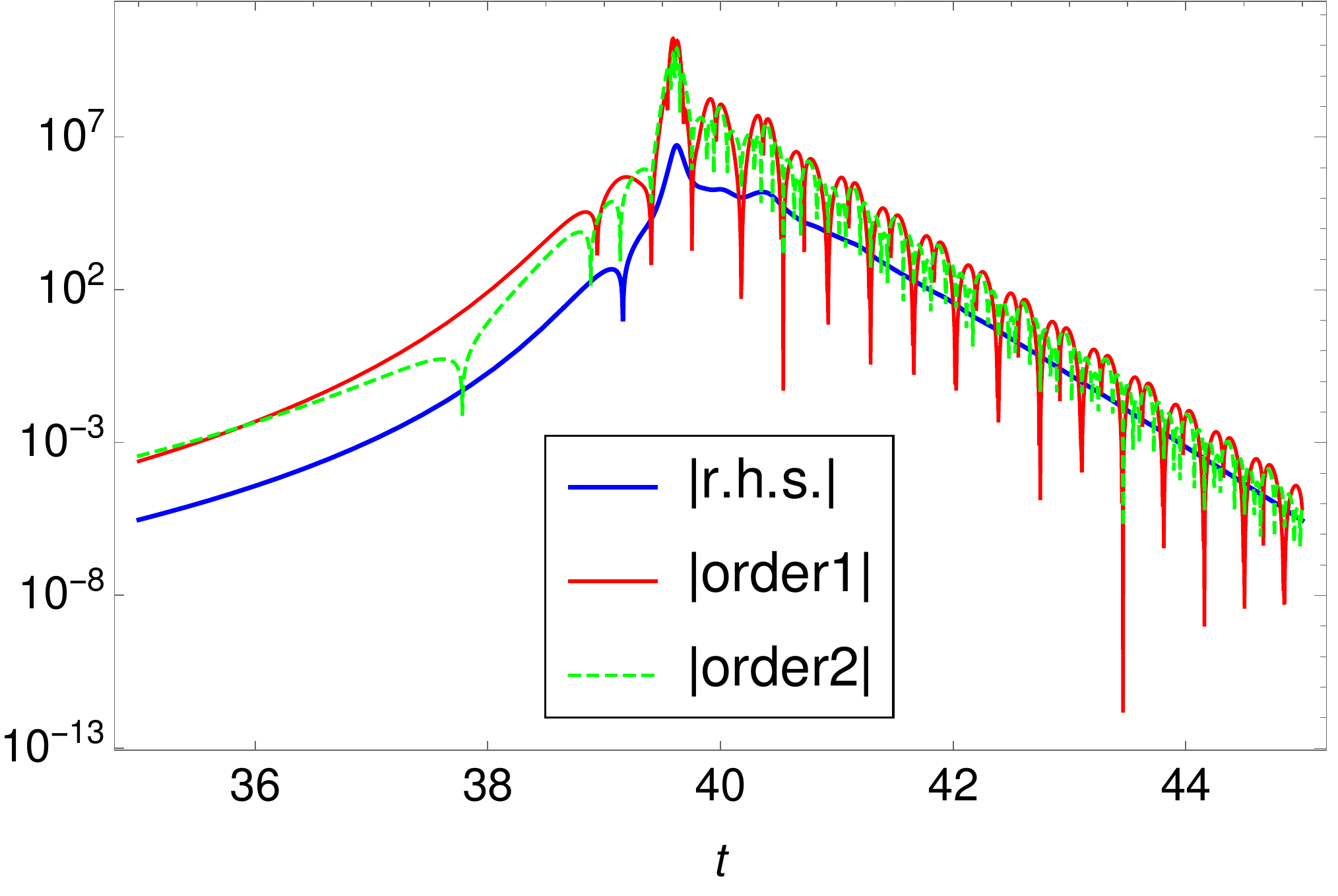}
	\includegraphics[width=0.485\textwidth]{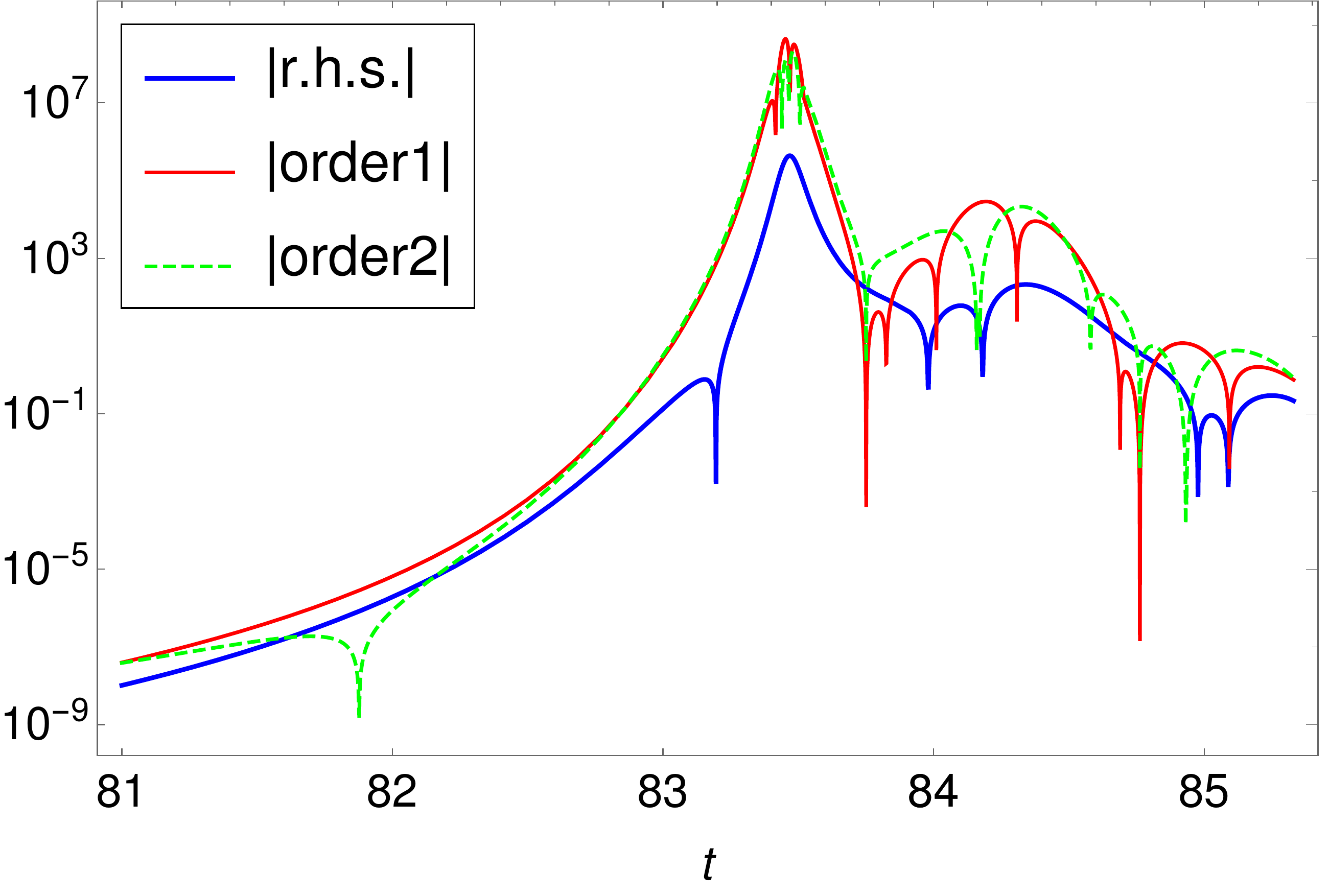}
	\caption{This plot displays $|$r.h.s.$|$ in blue (thick line), $|$order1$|$ in red and $|$order2$|$ in green (dashed) during the turn of the field trajectory for the potentials \eqref{pot n2m4} (on the left) and \eqref{pot n2m2} (on the right), which correspond to the first and second types of turn, respectively. Note the logarithmic scale.}
	\label{fig:rhs}
\end{figure}

To summarize the results of the subsection, we have established that r.h.s.\  is negligible compared to order1, order2 and order3. With the first type of turn, this is due to the cancellations of the dominant terms in r.h.s.\  due to the relations between the parallel and the perpendicular parameters which exist in that case. For the second type of turn, this is simply due to the factor in front of r.h.s.\  which is smaller than the one in order1,2,3 because $H^2 \ll H_*^2$. This means that even if the slow-roll approximation is broken, if the initial condition of that period is the slow-roll homogeneous solution, then the right-hand side of \eqref{equadiff} can be neglected. This is illustrated in figure \ref{fig:rhs} which displays $|$r.h.s.$|$, $|$order1$|$ and $|$order2$|$ (obviously order3 is not needed because it is minus the sum of the two others) for the potentials of each type of turn that are studied in section \ref{numerical section}. This figure (with a logarithmic scale) shows that r.h.s.\  is always several orders of magnitude smaller than the others during the turn (except at the times where $\getpe$ crosses zero, which will be discussed in a later subsection).

From this section we learn that the homogeneous solution, which is known, is sufficient to solve \eqref{equadiff} during the turn when the slow-roll approximation is broken (large $\getpa$ and $\getpe$) as long as $\ge$ remains small, since the particular solution is negligible. 

\subsection{Fate of the slow-roll particular solution}

In the previous subsection, we showed that we only need the homogeneous solution of the $\gint$ equation during the turn when the slow-roll approximation is broken. However, this does not mean that we can forget about the particular solution completely. It is still required  during the slow-roll evolution before and after the turn as we will show explicitly in this section (and potentially during the turn when $\getpe$ crosses zero, see next subsection) and hence plays a role in principle in the determination of the integration constants in the various regions. In fact, to avoid having to perform an explicit matching at every transition it would be very convenient if we could just add the slow-roll particular solution to the homogeneous solution everywhere. We will come back to this point in the next subsection. As a preliminary we will in this subsection investigate the behaviour of the slow-roll particular solution $P_{\mathrm{sr}}$ before and during the turn. We start by comparing $P_{\mathrm{sr}}$ to the homogeneous solution in the different regimes.

First, we focus on the slow-roll regime using the Green's functions determined in \eqref{Green srsolution} when the slow-roll particular solution can be written as:
\be
P_{\mathrm{sr}} = -2\ge\lh e_{1\gf}^2 \tV_{\gs\gs} + e_{1\gs}^2 \tU_{\gf\gf}\rh \frac{H^4 e_{1\gf}^2 e_{1\gs}^2}{H_*^4 e_{1\gf*}^2 e_{1\gs*}^2}
=\frac{2}{3}\ge \lh e_{1\gf}^2 V_{\gs\gs} + e_{1\gs}^2 U_{\gf\gf}\rh e_{1\gf} e_{1\gs} \frac{H^2 e_{1\gf} e_{1\gs}}{H_*^4 e_{1\gf*}^2 e_{1\gs*}^2}.
\ee
Doing the same for the homogeneous part using the quasi single-field initial conditions recalled at the beginning of section~\ref{Beyond slow-roll}, as well as \eqref{srpareq}, we get:
\be
h_{\mathrm{sr}}= -e_{1\gf*}^2\tV_{\gs\gs*} \getpe \frac{H^2 e_{1\gf} e_{1\gs}}{H_*^2 e_{1\gf*}^2 e_{1\gs*}^2}=-e_{1\gf_*}^2 \tV_{\gs\gs*} \lh e_{1\gf}e_{1\gs}(\tV_{\gs\gs}-\tU_{\gf\gf}) -\frac{1}{3}\gxpe \rh H_*^2 \frac{H^2 e_{1\gf} e_{1\gs}}{H_*^4 e_{1\gf*}^2 e_{1\gs*}^2}.
\ee 
In the slow-roll approximation (neglecting the higher-order term $\gxpe$ in $h_{\mathrm{sr}}$), we end up with $\ge \, e_{1\gf} e_{1\gs} \lh e_{1\gf}^2 V_{\gs\gs} + e_{1\gs}^2 U_{\gf\gf}\rh$ and $\tV_{\gs\gs*} e_{1\gf}e_{1\gs}\lh V_{\gs\gs}-U_{\gf\gf}\rh\frac{H_*^2}{H^2}$ to compare, because $e_{1\gf*}^2 \approx 1$ and by definition $\tV_{\gs\gs}= V_{\gs\gs}/(3H^2)$ and $\tU_{\gf\gf}= U_{\gf\gf}/(3H^2)$. As a reminder, we want to see if $P_\mathrm{sr}$ can be negligible compared to $h_\mathrm{sr}$ during the slow-roll regime. First, we look at the terms inside the parentheses which both contain second-order derivatives of the potential. Then, for our models where neither of the derivatives is negligible compared to the other at horizon crossing, we can expect that in general this remains true later, at least up to the turn (it can change during the turn, but at that time the slow-roll approximation is broken and these expressions are not valid as we will discuss later in this subsection). So we conclude that the terms between parentheses in the two expressions are in general of a comparable order (the basis components in $P_{\mathrm{sr}}$ can make it smaller, but not a lot smaller). If there is a difference between the two expressions, it has to come from the remaining factors, which means we have to compare $\ge$ to $\tV_{\gs\gs*}\frac{H_*^2}{H^2}$ like in the previous section. As discussed there, these have to be of the same order because in the slow-roll regime $H$ is still of the same order as $H_*$. There is one exception which corresponds to models where $\ge$ is extremely small compared to $\getpa$ even in the slow-roll regime (Starobinsky-like inflation for example), so that $\ge$ is also small compared to $\tV_{\gs\gs*}$ (in that case, there would be a similarity with the beyond-slow-roll situation studied in this section where $\ge\ll \getpa,~\getpe$ as well). But apart from those specific models, this leads to the conclusion that in general both the particular solution and the homogeneous solution have to be taken into account during the slow-roll regime.

As shown in subsection~\ref{Greensfunctionssubsec}, the slow-roll expressions for the Green's functions are also valid in a region of large $\ge$, which occurs just before a turn of the second type. The same expressions as in the previous paragraph can be used, however $\gxpe$ can no longer be neglected in $h_{\mathrm{sr}}$. On the other hand, there is no reason for $\gxpe$ to become much larger than the other term between the parentheses (which is $\getpe$) either, given that we are still before the turn, so that in the end the conclusion about the terms between parentheses from the previous paragraph still holds. As for the other factors, both the homogeneous and the particular solutions will grow because $\ge$ becomes of order unity, which makes $H_*^2/H^2$ large compared to 1. However, at the end of this period $\ge$ will decrease and becomes of order slow roll again, but the ratio $H_*^2/H^2$ will stay large. This means that the slow-roll particular solution finishes the period of large $\ge$ being small compared to the slow-roll homogeneous solution. We will show below that this is fully consistent with the result for the second type of turn (the type that has a period of large $\ge$ right before the turn) that the slow-roll particular solution is negligible during the turn.

We continue by considering the behaviour of the slow-roll particular solution during the different types of turn. Obviously, it is not an actual solution at that time, but we want to know if it would cause any problems if we were to simply add it to the solution. Again, we follow the same method using the Green's function expressions given in \eqref{v22turn} to write:
\be
P_{\mathrm{sr}} = \ge \frac{H^2 H_0 ^2 e_{1\gf}e_{1\gs}}{H_*^4 e_{1\gf*}^2 e_{1\gs*}^2} \lh \getpa e_{1\gf}e_{1\gs} + \getpe (e_{1\gs}^2-e_{1\gf}^2) \rh , \qquad
h_{\mathrm{sr}} = -e_{1\gf*}^2\tV_{\gs\gs*} \getpe \frac{H H_0 e_{1\gf} e_{1\gs}}{H_*^2 e_{1\gf*}^2 e_{1\gs*}^2}.
\label{srsolutionturn}
\ee
This time we end up with ${\textstyle\frac{H H_0}{H_*^2}}\ge\lh \getpa e_{1\gf}e_{1\gs} + \getpe (e_{1\gs}^2-e_{1\gf}^2) \rh$ and $\tV_{\gs\gs*}\getpe$ to compare. Again the two expressions have a similar form, excluding the factor $H H_0/H_*^2$. As discussed in the previous subsection, $\tV_{\gs\gs*}$ is typically of order $10^{-2}$ and hence cannot be much smaller than $\ge$ which is of order slow-roll. During the turn, the terms depending on the $\eta$ parameters are also of the same order, except in the rare case when $\getpe$ vanishes. Finally, the only large difference can come from the factor in front in the slow-roll particular solution. The two types of turn give different results. In the first type where $\ge$ is of order slow-roll since horizon-crossing, $H$ and $H_0$ are not much smaller than $H_*$. Then the factor is not much smaller than one. Moreover, the cases where it is the smallest are also the cases where $\ge$ has increased the most (and can then be larger than $\tV_{\gs\gs*}$ by an order of magnitude), so that these two effects compensate each other. Hence, the slow-roll particular solution is then typically of the same order as the homogeneous solution during the turn. In the second type of turn, the situation is different, indeed $H$ and $H_0$ are of the same order and we know that $H^2 \ll H_*^2$. This means that this time $P_{\mathrm{sr}}$ is small and negligible during the turn compared to $h_{\mathrm{sr}}$, fully consistent with the result that $P_{\mathrm{sr}}$ has become very small during the period of large $\ge$ just before the turn, as shown above.

More must be said about the slow-roll particular solution during a turn of the first type and we will now show that it becomes in fact proportional to the homogeneous solution of \eqref{equadiff}. To show this, we substitute $P_{\mathrm{sr}}$ in the left-hand side of \eqref{equadiff}, using the Green's function expressions from \eqref{v22turn}, the sum potential relations from \eqref{W11eq2} and \eqref{sumequation} to eliminate $\gc$, $\tW_{221}$ and $\tW_{222}$, and also the relations between parallel and perpendicular parameters \eqref{first case}. We then compare the three terms of the equation corresponding to the three different orders of derivative (called term1, term2 and term3)\footnote{These are the same terms we called order1,2,3 before, however now with the particular solution substituted and not the homogeneous one.} to their sum (called l.h.s.):
\be
\begin{split}
\mathrm{l.h.s.}_{\{\}} & = e_{1\gf}^2 e_{1\gs}^2 \left[ (\getpe)^2 \lh -6\tW_{111} - 12 \ge^2 \rh +12 \ge \getpe \gxpe \right]  \\
& + e_{1\gf} e_{1\gs}^3 \left[ (\getpe)^3 \lh 4 \ge^2+24 \ge\rh -6 (\getpe)^2 \tW_{211}\right]+e_{1\gf} e_{1\gs} (\getpe)^3 \lh 12 \ge-4 \ge^2\rh +4 e_{1\gf}^2 (\getpe)^4 \ge ,\\
\mathrm{term1}_{\{\}} & = -12 e_{1\gf} e_{1\gs} (\getpe)^5 + e_{1\gf}^3 e_{1\gs} \left[ -4 \getpe (\gxpe)^2-6 (\getpe)^2 \tW_{211}\right]  \\
&~~~ +e_{1\gs}^4 \left[ 4 (\getpe)^3 \gxpe-24 (\getpe)^4\right] +36 e_{1\gs}^2 (\getpe)^4 -4 (\getpe)^3 \gxpe-12 (\getpe)^4,\\
\mathrm{term2}_{\{\}} & = -12 e_{1\gf} e_{1\gs} (\getpe)^5 +e_{1\gf}^2 \left[ (\getpe)^4 (6-2 \ge)-4 (\getpe)^3 \gxpe\right] \\
&~~~ + e_{1\gf} e_{1\gs}^3 \left[ 48 (\getpe)^5+36 (\getpe)^3 \ge\right]\\
&~~~ + e_{1\gf}^2 e_{1\gs}^2 \left[ 28 (\getpe)^3 \gxpe + 12 \getpe \gxpe \ge+(\getpe)^4 (20\ge-24)+(\getpe)^2 \lh 6\ge^2-18 \ge\rh \right]\\
&~~~ +e_{1\gf}^3 e_{1\gs} \left[ 4 \getpe (\gxpe)^2+(\getpe)^2 \gxpe (6 \ge-6)+(\getpe)^3 \lh 2 \ge^2-6 \ge \rh \right],\\
\mathrm{term3}_{\{\}} & = e_{1\gf} e_{1\gs}^3 \left[ -48 (\getpe)^5+(\getpe)^2 (\gxpe (6 \ge-6)-12 \tW_{211})+(\getpe)^3 \lh 6 \ge^2 -18 \ge\rh \right]\\
&~~~ + e_{1\gf} e_{1\gs} \left[ 24 (\getpe)^5 + (\getpe)^2 (6 \tW_{211}+\gxpe (6-6 \ge))+(\getpe)^3 \lh 18 \ge-6 \ge^2\rh \right] \\
&~~~+ e_{1\gs}^2 \left[ -32 (\getpe)^3 \gxpe+ (\getpe)^2 \lh -6 \tW_{111}-18 \ge^2 +18 \ge\rh +(\getpe)^4 (-26 \ge-6) \right]\\
&~~~ +e_{1\gs}^4 \left[ 24 (\getpe)^3 \gxpe+(\getpe)^2 \lh 6 \tW_{111}+18 \ge^2 - 18 \ge\rh +20 (\getpe)^4 \ge\right] \\
&~~~ + 8 (\getpe)^3 \gxpe +(\getpe)^4 (6 \ge+6).
\end{split}
\ee
The discussion of these expressions is very similar to the one for \eqref{rhs} and \eqref{order}. We use again the subscript $_{\{\}}$ to indicate that we have left out an overall factor (cf.\ (\ref{rhs}) and (\ref{order})), which is here
the same for all four expressions.
We can see that in l.h.s.$_{\{\}}$ the higher order terms like $(\getpe)^5$ have disappeared. Moreover, most of the terms in l.h.s.\  have an extra factor of $\ge$, which is not the case for the other expressions. This implies that the sum of the three terms is much smaller than the individual terms of \eqref{equadiff} with the slow-roll particular solution. Hence this function is in fact an approximated solution of the homogeneous equation during a turn of the first type when the slow-roll approximation is broken.

If $P_{\mathrm{sr}}$ becomes a homogeneous solution it means that it has to be proportional to a linear combination of the two previously determined exact independent homogeneous solutions $\getpe \bv_{22}$ and $\getpe G_{22*}$. However, using \eqref{proportionality}, these independent solutions have in fact become proportional before the turn. Hence, we simply have that $P_{\mathrm{sr}}$ and $h_{\mathrm{sr}}$ are proportional. Using \eqref{srsolutionturn} and \eqref{first case}, we rewrite the particular solution as:
\be
P_{\mathrm{sr}}= -\ge\frac{H^2 H_0^2 e_{1\gf} e_{1\gs}}{H_*^4 e_{1\gf*}^2 e_{1\gs*}^2}\getpe e_{1\gf}^2.
\ee
We find the same factor $\getpe$ as in the homogeneous solution (\ref{srsolutionturn}), but also another factor $\ge H e_{1\gf}^2$. Hence, the proportionality is true only if $ \ge H e_{1\gf}^2$ is constant during the turn. This happens if $e_{1\gf}^2 \approx 1$, in that case $\gf$ is dominating meaning that $\ge$ and $H$ are purely slow-roll and are almost constant during a short turn. At first, the idea of $\gf$ dominating during the turn might seem odd. However, we recall that this does not have to be during the whole turn, but only when $\getpe$ and $\getpa$ are large enough to break the slow-roll approximation. Looking at the form of trajectory in the top left plot of figure~\ref{fig:slowrollbroken}, the only period when $\gf$ dominates is in fact at the end of the turn when $\gs$ is oscillating around its minimum. This can also be verified with the explicit examples of the next section (see figures \ref{fig:n2m4} and \ref{fig:axion}). Here, we can observe that $\getpe$ becomes large only after the period when $e_{1\gs}$ was not negligible (the turn).

Different behaviours of the slow-roll particular solution depending on the type of turn have been highlighted in this subsection. In the next subsection we will discuss how these results can be used to solve the differential equation \eqref{equadiff} beyond the slow-roll regime.

\subsection{Solution of the $\gint$ equation}

As usual, we will discuss separately the two types of turn, but we start by reminding the reader about the main result of the previous subsections. The solution of \eqref{equadiff} is known until the end of the slow-roll regime and it is composed of a homogeneous solution and a particular solution that both have to be taken into account. When $\getpa$ and $\getpe$ become large, during the turn, only the homogeneous solution (which is exact and does not depend on any slow-roll approximation) is needed to solve the equation. The difficulty is then to ensure the continuity of the solution at the transition between the two regimes. In fact, after the turn, there may also be another period of slow-roll before the end of inflation, and during the turn the slow-roll parameters can oscillate and vanish for a short time, which could lead to a very brief restoration of the slow-roll conditions. So in the end there might be many transitions and it would be very inconvenient if we had to perform an explicit matching of the solutions at each of them. Fortunately, there is another option as we will now show. Finally, we also recall that the slow-roll particular solution evolves differently depending on the type of turn. In the first type, it becomes proportional to the homogeneous solution of \eqref{equadiff}, while in the second type it becomes negligible compared to the homogeneous solution.

It is then easy to see that the case of the first type of turn is most simply treated by keeping the full slow-roll solution at all times. Indeed, at the moment when the slow-roll regime ends and the turn starts, the solution should become only homogeneous, and that is exactly the case because the slow-roll particular solution becomes a homogeneous solution at that time. Continuity at the transition is then automatic, without the need for any explicit matching. Then, if later during the turn or at the end of the turn the slow-roll approximation is re-established, continuity is also ensured since the same solution works on both sides of the transition. Note that if $\getpe$ vanishes, from \eqref{first case}, $\getpa$ has to be of order slow-roll, meaning that the slow-roll approximation is indeed restored during these brief moments.

The second type of turn deserves a longer discussion. Indeed, we do not know the full particular solution during the period of large $\ge$ just before the turn but we know two things: the slow-roll particular solution vanishes (but it is not an exact particular solution at that time) and the right-hand side of \eqref{equadiff} can be neglected once this period has finished, because r.h.s.\  is negligible at the start of the turn. These two ingredients are sufficient to prove that the particular solution during the period of large $\ge$ vanishes, even without having its explicit form. To stay general, we write the particular solution $P$ as $P=P_{\mathrm{sr}} + A h + P_{\perp h}$, where $A$ is a constant, $h$ the homogeneous solution, and $P_{\mathrm{sr}}$ the slow-roll particular solution. $P_{\perp h}$ is the function that, when inserted into \eqref{equadiff}, gives those right-hand side terms that are not given by $P_{\mathrm{sr}}$, and which is zero when these terms vanish (in other words, it does not contain a homogeneous solution). We know that over the course of the period of large $\ge$, $P_{\mathrm{sr}}$ vanishes (see \eqref{srsolutionturn}). The right-hand side of \eqref{equadiff} vanishes during that period too, which means that $P_{\perp h}$ has to vanish by definition. The only remaining term could then be the one proportional to the homogeneous solution, but it has to be zero because of the matching conditions at the start of the period of large $\ge$. Indeed at the end of the slow-roll regime, the particular solution is simply $P_{\mathrm{sr}}$ while $P_{\perp h}$ has to be zero, because the terms of higher order in slow-roll are still negligible and will grow only later during that period of large $\ge$. The function $h$ is not zero at the transition, hence $A$ has to be. Without knowing the exact formula for $P$, we can conclude that it vanishes during that period of large $\ge$. Hence, at the start of the turn, the solution is simply the slow-roll homogeneous solution. 

During the second type of turn, keeping the slow-roll particular solution, even if it is not a particular solution of the exact equation at that time, only induces a negligible error, but it solves any potential issues with matching to later slow-roll periods. When $\getpe$ vanishes, $\getpa$ can be larger than order slow-roll in this type of turn. This is not an issue because then the parameters evolve very fast, meaning that a very short time before $\getpe$ vanishes, the particular solution is still negligible compared to the homogeneous solution, and the same a very short time after. Moreoever, one can verify that at the exact time when $\getpe=0$, the particular solution is $\half P_{\mathrm{sr}}$ and we know that this function is negligible during the rest of the turn. Then it is possible to add this particular solution to the full solution only for these very short periods (without using matching conditions, because at the time of the matchings it it is negligible). It is also important to remember that in the end we are interested in the integrated $\gint$, and when $\getpe$ vanishes, the right-hand side of \eqref{equadiff} is also very small compared to its value a short time before or after (because every term contains $\getpe$ except one which also becomes small), meaning this particular solution is also small at that time compared to its usual value during the turn. In the integral it is then negligible. In fact, when $\getpe$ vanishes, the only thing that happens is that the whole solution almost vanishes (but the particular solution does not vanish at that exact same time), but because the homogeneous solution is zero, it cannot be large compared to the particular solution for once. 

To summarize, we have shown that for both types of turn, the slow-roll solution of \eqref{equadiff} is sufficient to solve this equation even beyond the slow-roll regime, under the condition that $\ge$ stays of order slow-roll during the turn. Of course, knowing the solution $\dot{g}_{\mathrm{int}}$ which is given in \eqref{dotgint} is not sufficient, we also have to integrate it. But the computation is exactly the same as in the slow-roll case even if the slow-roll approximation is not valid, meaning that $\gint$ has again the same form:
\be
\gint = \ge \bv_{22}^2 - \ge_* - \frac{e_{1\gf*}^2\tV_{\gs\gs*} - e_{1\gs*}^2\tU_{\gf\gf*}}{2e_{1\gf*}e_{1\gs*}} \bv_{12}.
\label{gintbsr}
\ee

\subsection{End of inflation and $\fnl$}

Once the form of $\gint$ is known, it is possible to compute $\fnl$ at the end of inflation: 
\be
-\frac{6}{5}\fnl =  \frac{e_{1\gf*}^2\tV_{\gs\gs*} - e_{1\gs*}^2\tU_{\gf\gf*}}{e_{1\gf*}e_{1\gs*}} \frac{(\bv_{12e})^3}{\lh 1+(\bv_{12e})^2\rh ^2} + \mathcal{O}(10^{-2}).
\ee
This expression has the same form as the slow-roll one \eqref{fnlapp}, the difference is hidden in the Green's functions which have been computed in subsection~\ref{Greensfunctionssubsec}. The same discussion of this expression as in section \ref{Slow-roll} holds and the conclusions are the same, see \eqref{fnlsrfactor}. Like in that section, we use the limit \eqref{v12limit} which is a good approximation when $|\bv_{12e}|>4$. Then the only remaining step is to study the value of $\bv_{12}$ at the end of the turn using \eqref{v_12}, when the slow-roll approximation is valid again, which is equal to $\bv_{12e}$.

As usual, we need to distinguish the two types of turn because they have different initial and final conditions. In the first case, the turn occurs early which means that $U_0\gg V_0$ (as defined before, the subscript 0 indicates that the function is evaluated at $t_0$ when the slow-roll approximation stops to be valid). However because there is a turn, we cannot neglect $e_{1\gs0}$ anymore. We can then write $S_0=e_{1 \gf0}^2 -e_{1\gs0}^2 = 1-2e_{1\gs0}^2$ and $Z_0\approx -U_0 e_{1\gs0}^2$. Moreover, before the turn we are still in slow-roll, meaning that $\ge_0\ll 1$ and we can use the slow-roll expression $H_0^2=\gk^2 U_0/3$. At the end of the turn, the situation is similar to single-field inflation in the direction $\gf$ meaning that $Z\approx 0$ and $S \approx 1$. Inserting this into \eqref{v_12}, we obtain:
\be
\bv_{12e}=\frac{U_0}{W_*}\frac{2e_{1\gs0}^2}{4e_{1\gf*}e_{1\gs*}} +\frac{-U_0 e_{1\gs_0}^2}{2W_*e_{1\gf*}e_{1\gs*}}+\frac{-V_*}{W_*e_{1\gf*}e_{1\gs*}}=\frac{-V_*}{W_*e_{1\gf*}e_{1\gs*}}.
\ee
This is exactly the same limit as in the slow-roll situation. Hence for this first type of turn, we get the same result:
\be 
-\frac{6}{5}\fnl  = -\frac{V_{\gs\gs*}}{\gk^2 V_*}.
\ee
The implications of this result were already discussed in section \ref{Slow-roll}.

In the second type of turn, the situation is slightly different. Firstly, the slow-roll approximation is not valid at the time $t_0$, at the end of the period of large $\ge$. Moreover, at that time we are still in a single-field case ($\gf$ dominates), hence $S_0\approx 1$ and $Z_0 \approx -V_0\approx V_*$ (because even if $U_0$ is not zero, it cannot be large compared to $V_0$ because we are near the moment when $\gf$ reaches the minimum of $U$). After the turn, the single-field situation is now in the $\gs$ direction, hence $S \approx -1$. At the end of inflation, the situation is:
\be
\bv_{12e}= -\frac{H_0^2}{2 H_*^2 e_{1\gf*}e_{_1\gs*}} + \frac{- Z_*}{2W_* e_{1\gf*} e_{_1\gs*}}=  \frac{-\frac{3}{\gk^2} H_0^2 - Z_*}{2W_* e_{1\gf*} e_{_1\gs*}}.
\ee
Substituting this into $\fnl$, we obtain:
\be
-\frac{6}{5}\fnl  = -\frac{2V_{\gs\gs*}}{3 H_0^2 + \gk^2 V_*}.
\ee
However, we can add that $H_0^2 >\gk^2 W_0/3$ because $\ge$ is not negligible (equality in the slow-roll case). Moreover, $W_0=U_0 + V_0 \approx U_0 + V_* > V_*$. We can then write:
\be
|\bv_{12e}|> \left|\frac{-V_*}{W_*e_{1\gf*}e_{1\gs*}}\right|,
\ee
which has an immediate consequence for $\fnl$:
\be 
\left|-\frac{6}{5}\fnl\right|  < \left|-\frac{V_{\gs\gs*}}{\gk^2 V_*}\right|.
\ee
In this case, the value of $\fnl$ is smaller than the slow-roll result. However, it is easily of the same order because $U_0$ and $V_0$ are of the same order while even if $\ge_0=1$, it only changes the factor between $H_0^2$ and $\gk^2 W_0$ from $1/3$ to $1/2$.

So in the end we have derived the rather surprising result that in the class of models considered (two-field sum potentials), the slow-roll expression for $\fnl$ gives a very good approximation of the exact result, even in the case where the slow-roll approximation breaks down during the turn. Allowing for the break-down of slow-roll does however increase the region of the parameter space where large non-Gaussianity can occur compared to the results shown in figure~\ref{fig:etaper1}, because we no longer have the constraint that the turn has to happen before the end of the slow-roll regime.

\section{Numerical examples}
\label{numerical section}

Here, we provide several explicit examples to illustrate the different results of the previous sections. We also show how to explicitly construct a model that produces $\fnl$ of order unity while satisfying all observational constraints.

\subsection{Double quadratic potential}

The double quadratic potential has the form:
\be
W({\gf,\gs})=\frac{1}{2}m_\gf^2 \gf^2 + \frac{1}{2}m_\gs^2 \gs^2.
\label{pot quad}
\ee 
It has been studied and discussed in many papers, see e.g.~\citep{Vernizzi:2006ve,RSvT4,TvT1}. However, it is always a good introductory example.

Without taking into account the exact constraints of the monomial potential yet, we keep the main idea that the second field has a negligible effect at the time of horizon-crossing. This can be achieved by taking $m_\gf^2 \gg m_\gs^2$ and we will use the same values as in \citep{TvT1}: $m_\gf = 20 m_\gs$ and $m_\gs = 10^{-5}\gk^{-1}$. As initial conditions, we use $\gf_i=13\gk^{-1}$ and $\gs_i=13\gk^{-1}$, while their derivatives $\dot{\gf}_i$ and $\dot{\gs}_i$ are determined by the slow-roll approximation.
\begin{figure}
	\centering
	\includegraphics[width=0.49\textwidth]{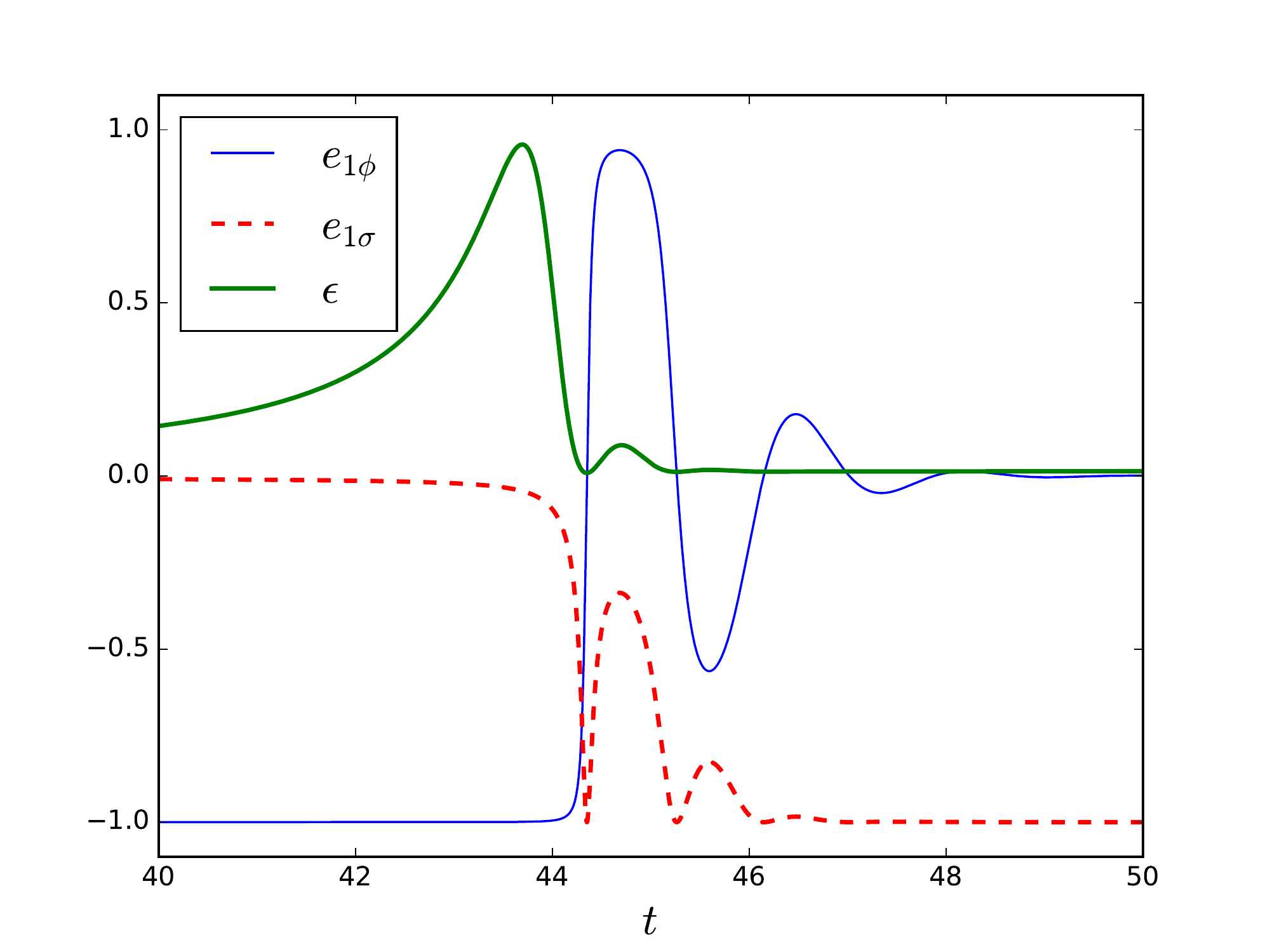}
	\includegraphics[width=0.49\textwidth]{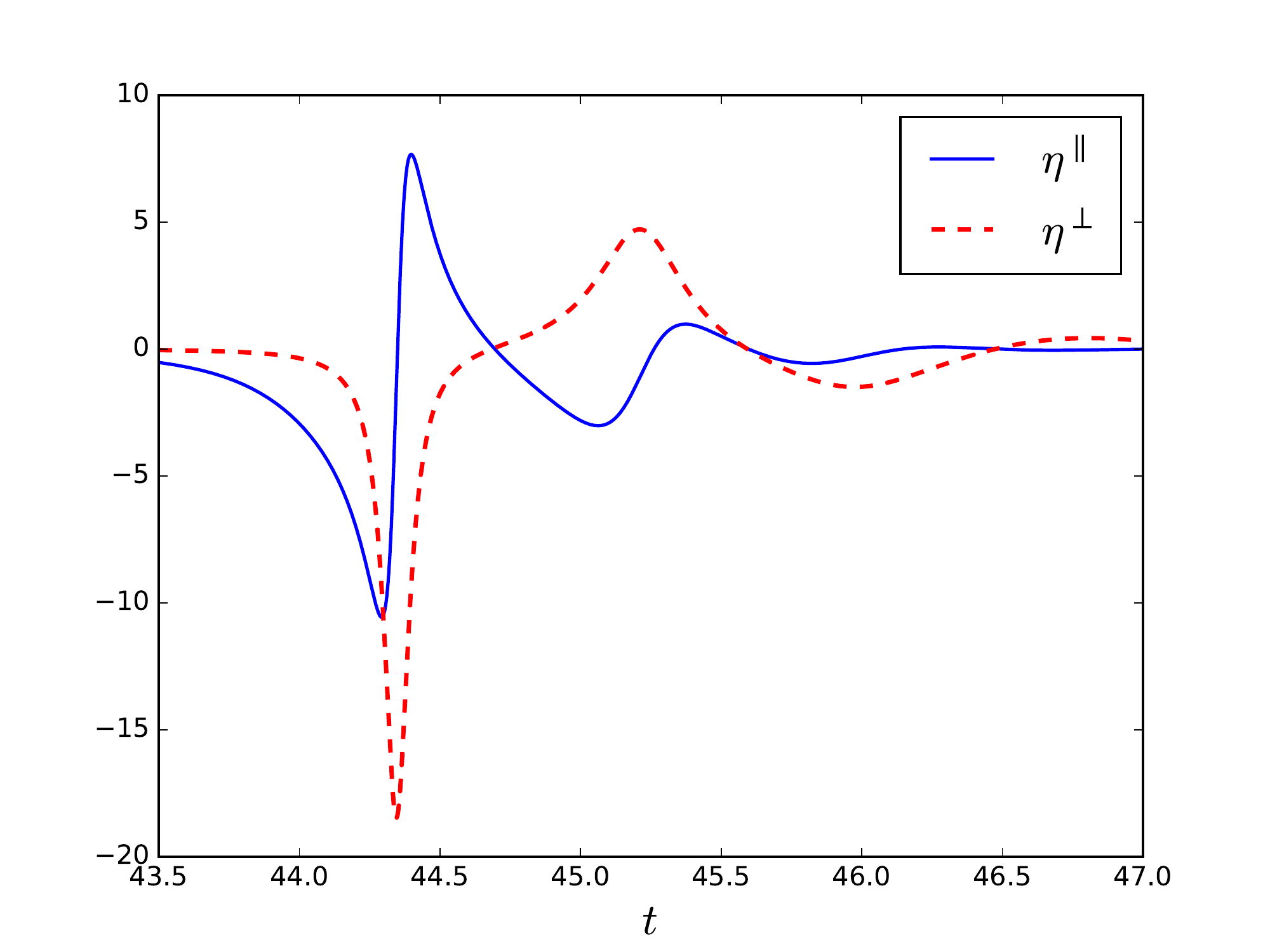}
	\includegraphics[width=0.49\textwidth]{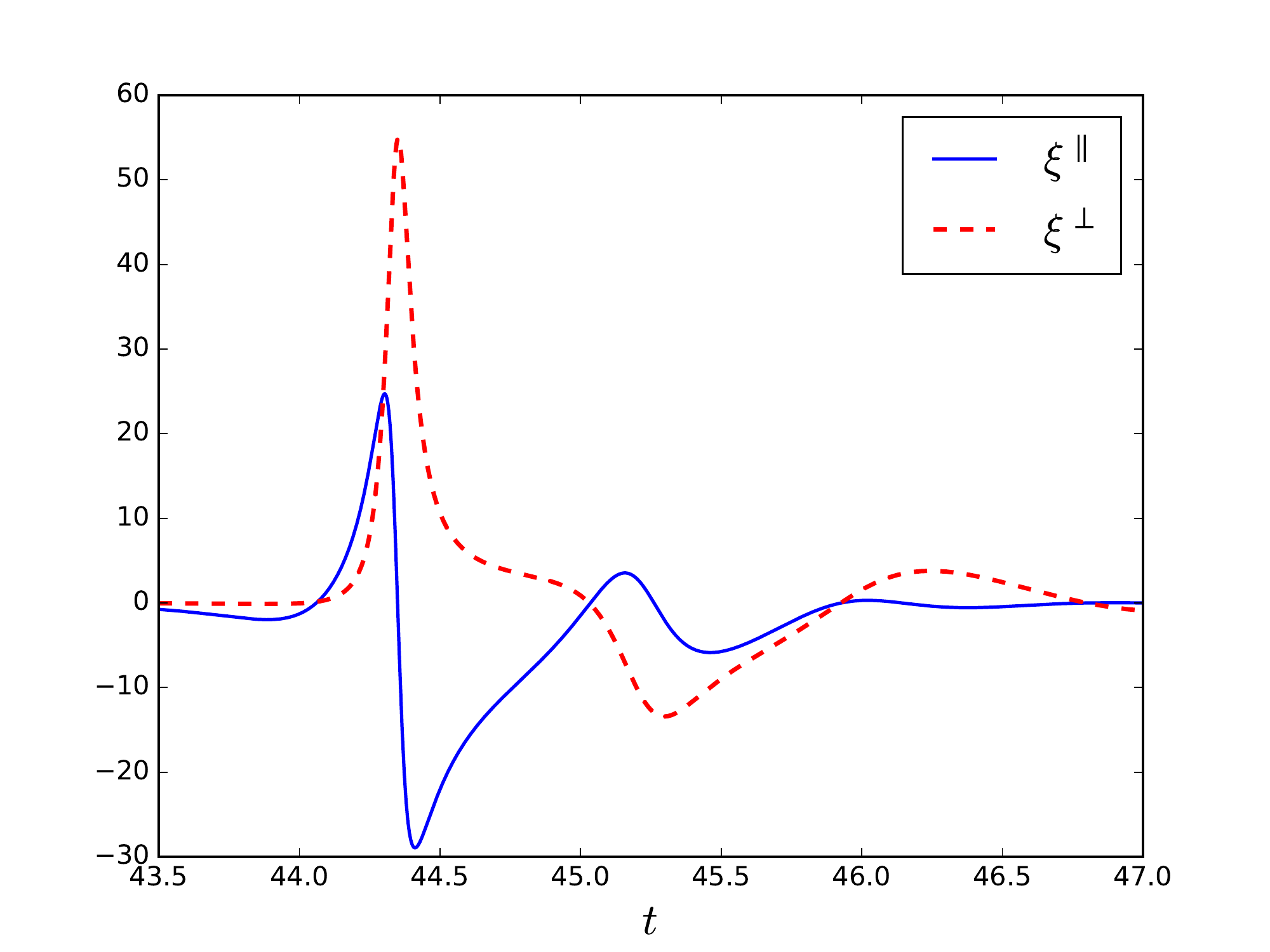}
	\includegraphics[width=0.49\textwidth]{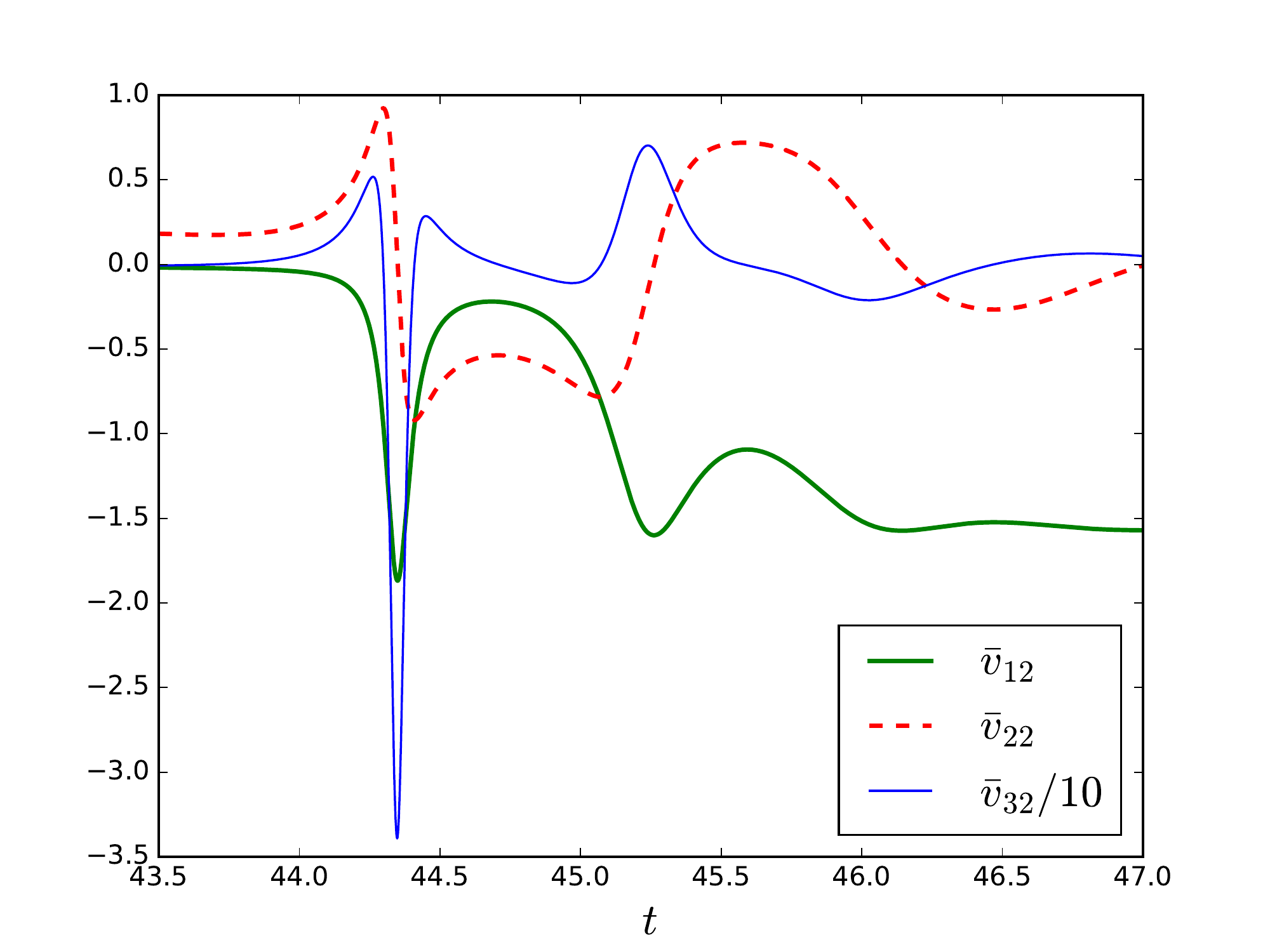}
	\includegraphics[width=0.49\textwidth]{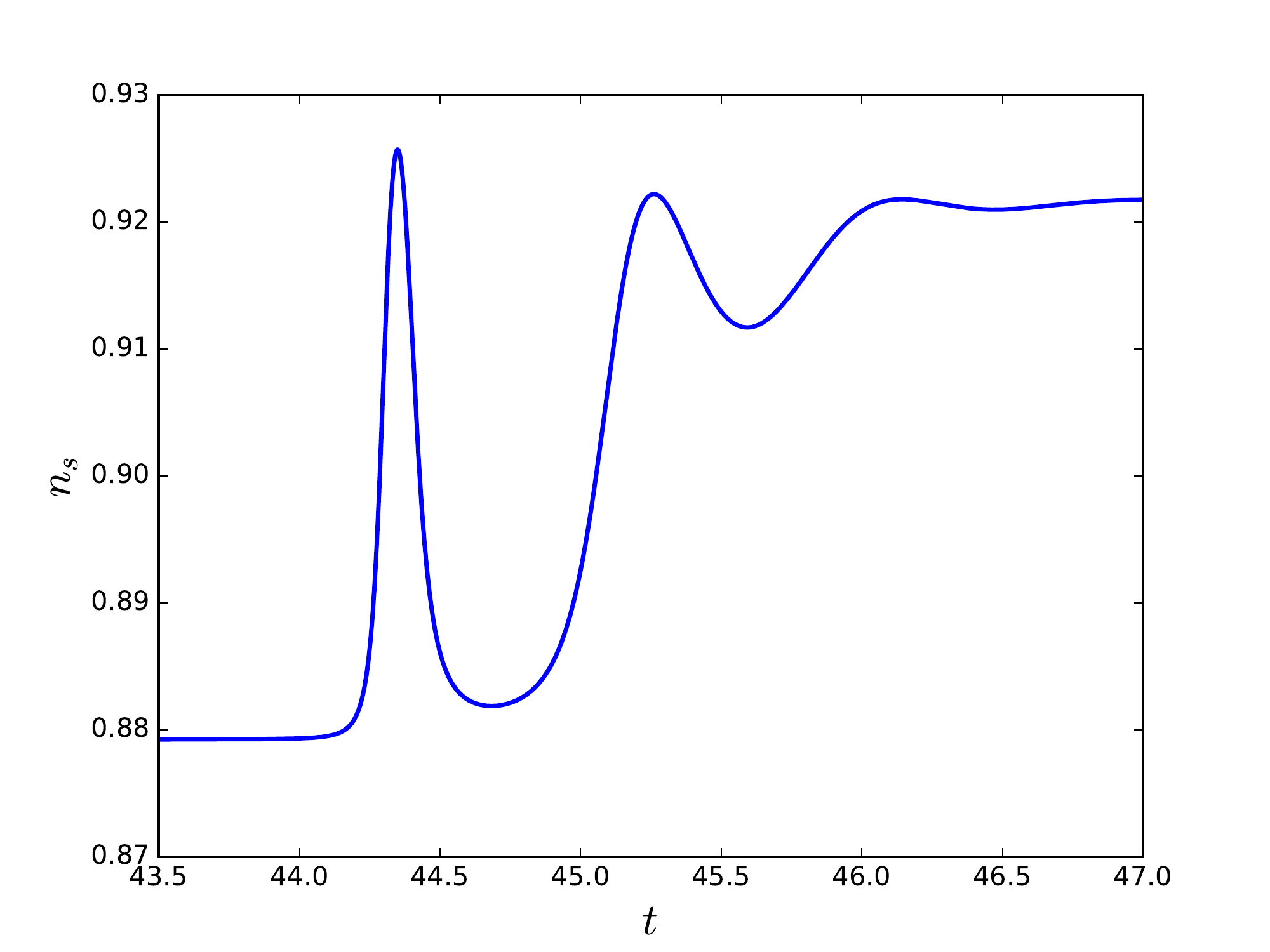}
	\includegraphics[width=0.49\textwidth]{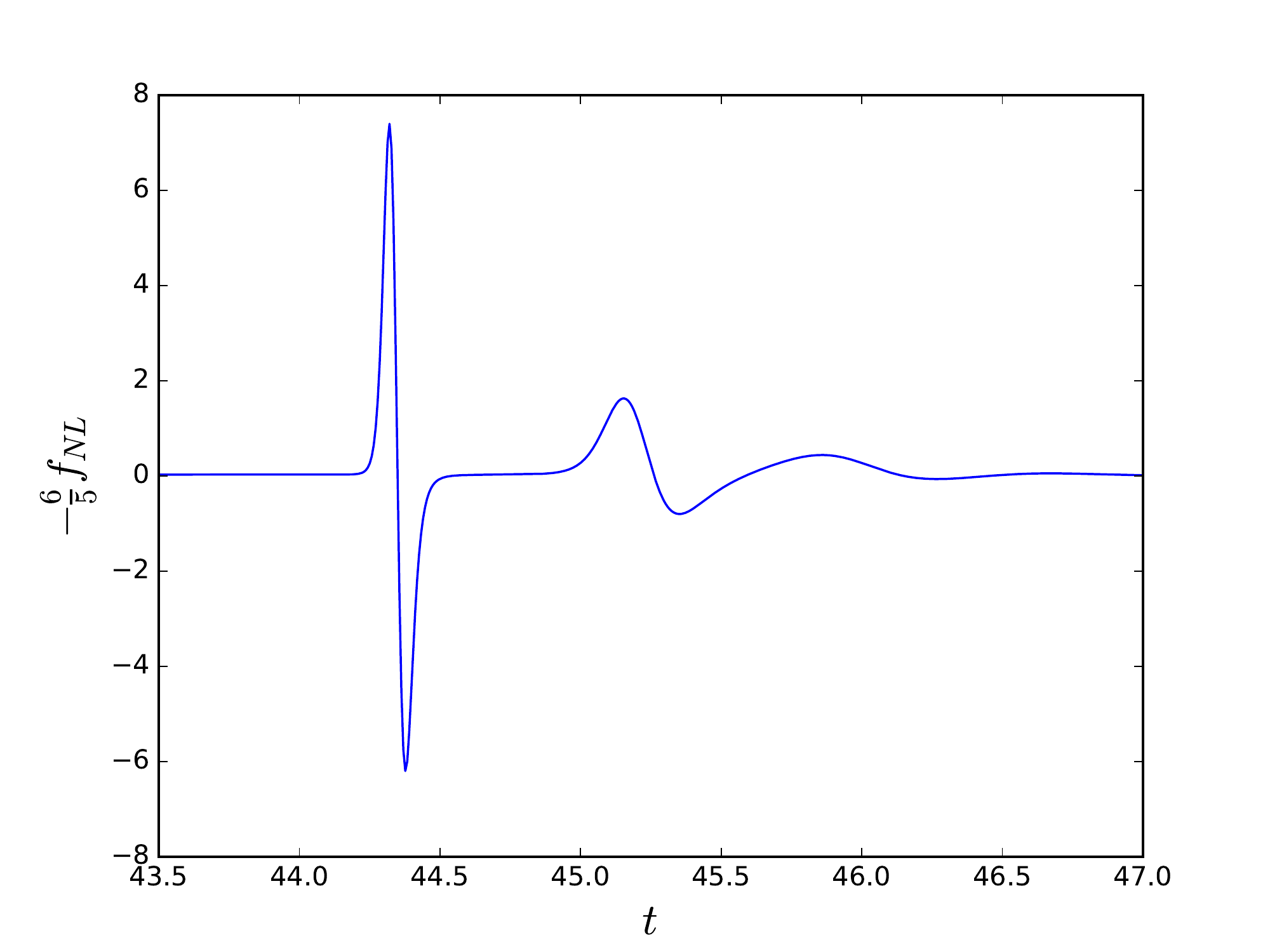}
	\caption{The exact numerical solutions for the different interesting parameters (basis components, slow-roll parameters, Green's functions, the spectral index and $\fnl$) during the turn for the double quadratic potential \eqref{pot quad}.}
	\label{fig:quad}
\end{figure}
In figure \ref{fig:quad} we show how the various relevant quantities evolve during the turn of the field trajectory. First, one can see clearly when the turn occurs: $\getpe$ becomes large and $e_{1\gs}$ becomes of the same order as $e_{1\gf}$. We also see that this example corresponds to the second type of turn where $\gf$ reaches the minimum of its potential and $\ge$ is of order unity before the turn. Another remark is that the second-order parameters $\gxpa$ and $\gxpe$ do not give new information compared to the first-order parameters  $\getpa$ and $\getpe$, at least not by eye.

However, in this model the two most important constraints and goals, concerning the two observables $n_s$ and $\fnl$, are not achieved. The spectral index, which is 0.92, is clearly outside the bounds from the Planck observations. $\fnl$ is slow-roll suppressed and far from the goal of $\fnl$ of order unity. Moreover, $\bv_{12e}$ is only $-1.5$ which is smaller in absolute value than the value 4 needed to use the approximations \eqref{fnllimit} and \eqref{nstwofield} for $\fnl$ and $n_s$.

The main result of the previous section was the validity of the slow-roll expressions in cases beyond slow-roll, like this one, at least to give an estimation of the Green's functions. Hence, we can use this approximation to compute $\bv_{12e}$ to see how the situation can be improved. Using dimensionless fields, \eqref{srparametersflat}, the slow-roll expression for $\bv_{12e}$ given by \eqref{v12e} becomes: 
\be
\bv_{12e}=-\frac{V_*}{W_*e_{1\gf*}e_{1\gs*}}=-\sqrt{2\ge}\frac{V_*}{V_{\gs*}}=-\frac{\gs_*}{\gf_*}.
\ee
This shows that $\bv_{12e}$ can be increased only by changing the initial conditions. Assuming that now we have $\bv_{12e}$ large enough, $\fnl$ takes the form:
\be
-\frac{6}{5}\fnl=-\frac{V_{\gs\gs*}}{V_*}=\frac{2}{\gs_*^2}.
\ee
The value of $\fnl$ becomes smaller if we increase $\gs_*$. Hence, it is impossible to increase both $\bv_{12e}$ and $\fnl$ at the same time. One can also verify there is no optimal value of $\gs_*$ where $\fnl$ would be larger than order slow-roll, meaning that this potential cannot produce large persistent non-Gaussianity. 

Instead of looking directly at $\fnl$, we could also have used the conclusion that for a monomial potential $N_\gs \propto \gs^2$ has to be of order unity to have $\fnl$ large, which requires here to decrease $\gs_*$ and $\bv_{12e}$. The solution is then to add an extra parameter in the potential.

\subsection{How to build a monomial potential model that produces $\fnl$ of order unity}

The form of the potentials we are interested in is:
\be
W(\gf,\gs)= \ga(\gk\gf)^n + C + \gb(\gk\gs)^m \lh + \lambda (\gk \gs)^{m'}\rh,
\ee
which is the one studied in section~\ref{monomialsubsec}. There is an extra term with $m'>m$ inside the parentheses to complete the model (i.e.\ make sure it has a minimum) and we will choose it to be negligible until after the turn. Hence this does not change the different expressions determined for a monomial potential. 

A first step is to choose the value of $m$ and $n$ using figure \ref{fig:etaper1} to be in the region where $\fnl$ of order unity is possible. $\ga$ can be put as an overall factor of the whole potential, hence it does not count in the number of parameters. $\gf_*$ is given by $N_{\gf*}\approx 60$ and this also determines $\ge_*$ because it only depends on $\gf_*$. Once $\ge_*$ is known, it is possible to determine $\gs_*$, $\gb$ and $C$ using the three constraints we have ($\fnl$, $n_s$ and $\bv_{12e}$) as follows.

We can start by choosing the value of $\fnl$ and \eqref{fnllimit} takes the following form for a monomial potential:
\be
\label{constraint1}
-\frac{6}{5}\fnl =  -\frac{V_{\gs\gs*}}{\gk^2 C}.
\ee
Using \eqref{nstwofield} and the lower bound on the spectral index $n_s=0.962$, as this is the easiest way to get a large $\fnl$, we have:
\be
\label{constraint2}
V_{\gs\gs*}= \gk^2 U_*\lh\frac{n_s-1}{2}+\ge_*\rh.
\ee
Finally, we need $\bv_{12e} > 4$. Using the slow-roll expression for $\bv_{12e}$ in \eqref{v12e}, \eqref{srparametersflat} and \eqref{link12}, we get:
\be
\label{constraint3}
\bv_{12e}=-\frac{V_*}{W_*e_{1\gf*}e_{1\gs*}}=-\sqrt{2\ge_*}\frac{\gk C}{V_{\gs*}}=-\sqrt{2\ge_*}\frac{m-1}{\gs_*}\frac{\gk C}{V_{\gs\gs*}}.
\ee
A last step is to determine $\lambda$, this is done using the fact that the minimum of the potential has to be zero. Then it is possible to verify if the last term is really negligible at horizon-crossing, if not it is possible to increase $m'$ to decrease it because $\gs_*$ is small compared to one. We will now apply this to two different potentials with a turn of the first type.

\subsection{First type of turn}
\subsubsection{First example: $n=2$ and $m=4$}

This first example corresponds to the case where the turn occurs early enough to have a trajectory with the same direction before and after the turn, see the top right plot of figure \ref{fig:slowrollbroken} for an illustration of the field trajectory. The potential is:
\be
W(\gf,\gs)= \ga \gf^2 + C + \gb \gs^4 + \lambda \gs^{6},
\label{pot n2m4}
\ee
with $\ga=\half\gk^{-2}$, $C= {\textstyle\frac{4}{27}\frac{\gb^3}{\lambda^2}}$, $\gb=-12.5$, and $\lambda=-{\textstyle\frac{4}{3}}\gb\gk^2$. The intial conditions are $\gf_i= 16\gk^{-1}$ and $\gs_i=0.09\gk^{-1}$ and, as usual, $\dot{\gf}_i$ and $\dot{\gs}_i$ are determined by the slow-roll approximation.

With this, it is possible to obtain an analytical estimate of the observables. First, we need to compute $\gf_*$ and $\gs_*$, using the solutions of equations \eqref{eqphi(t)} and \eqref{eqsig(t)} determined for monomial potentials. These solutions were computed assuming that $\gf_*$ and $\gs_*$ were the initial conditions, one has just to replace them by $\gf_i$ and $\gs_i$ here. This quick computation gives:
\be
\gf_* = 15.2\gk^{-1} \qquad \text{and} \qquad \gs_* = 0.092\gk^{-1}.
\ee
Using these values and the different expressions \eqref{constraint1}, \eqref{constraint2} and \eqref{constraint3}, we obtain:
\be
\bv_{12e}= -\frac{2}{\gf_*}\frac{3}{\gs_*}\frac{\gk C}{12\gb \gs_*^2}=3.52,\quad n_s =  1-\frac{4}{\gf_*^2}-2\frac{12\gb\gs_*^2}{\gk^2\ga\gf_*^2}=0.961, \quad -\frac{6}{5}\fnl=-\frac{12\gb\gs_*^2}{\gk^2 C}=1.2.
\label{predictions 1}
\ee
In these calculations, there are different approximations. First we use the monomial expressions to compute $\gf_*$ and $\gs_*$ (we refer the reader to section \ref{monomialsubsec} for the details, but they require the slow-roll approximation and a quasi single-field situation, at least until horizon-crossing). Second, we use the limit of large $\bv_{12e}$ to compute the observables, the validity of this limit is explained in detail in section~\ref{Slow-roll} as well.
Hence, an error of order slow-roll (at horizon-crossing) is expected compared to the exact numerical results, which can be larger here since $\bv_{12e}$ is a little smaller than four.

\begin{figure}
	\centering
	\includegraphics[width=0.49\textwidth]{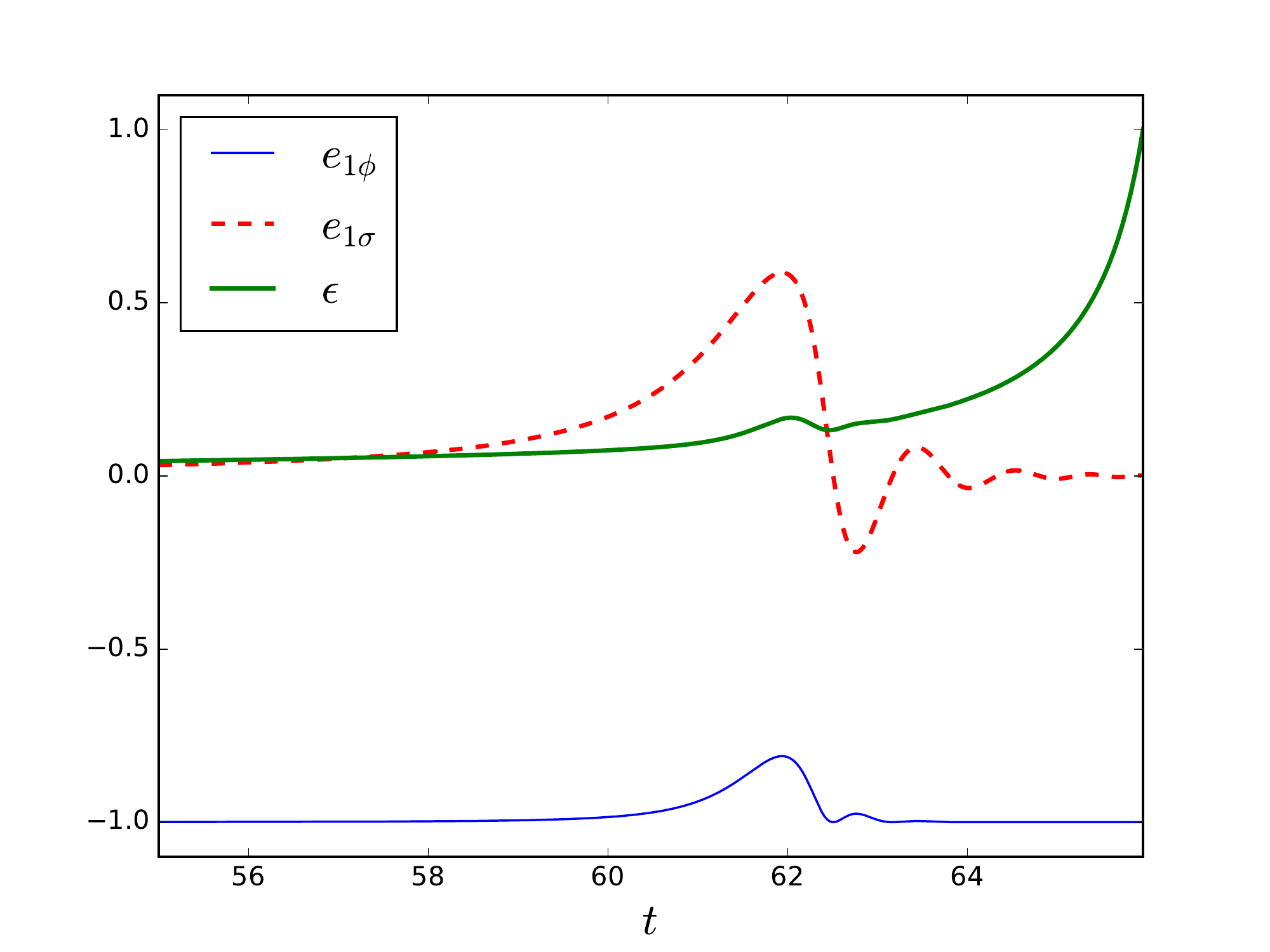}
	\includegraphics[width=0.49\textwidth]{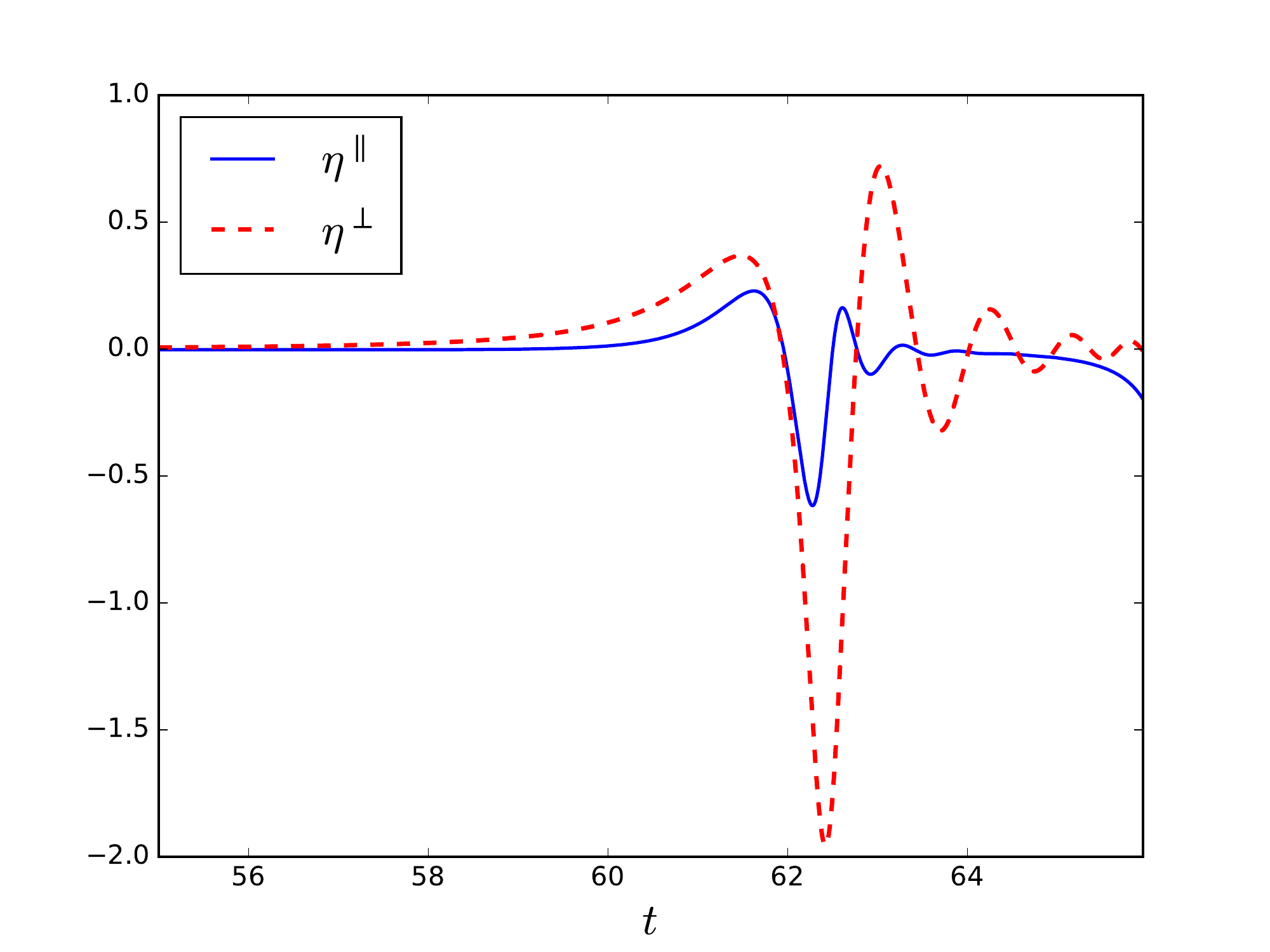}
	\includegraphics[width=0.49\textwidth]{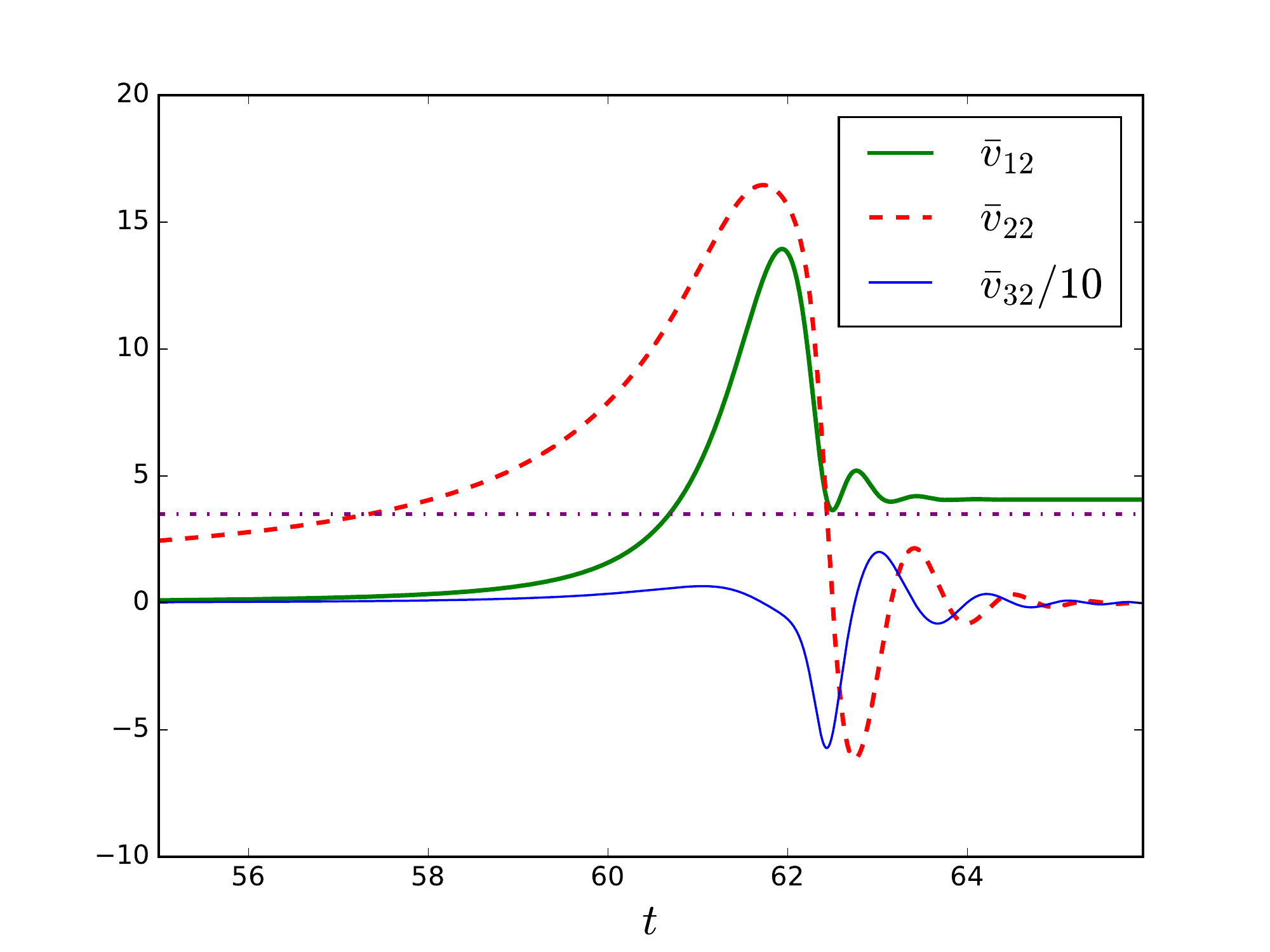}
	\includegraphics[width=0.49\textwidth]{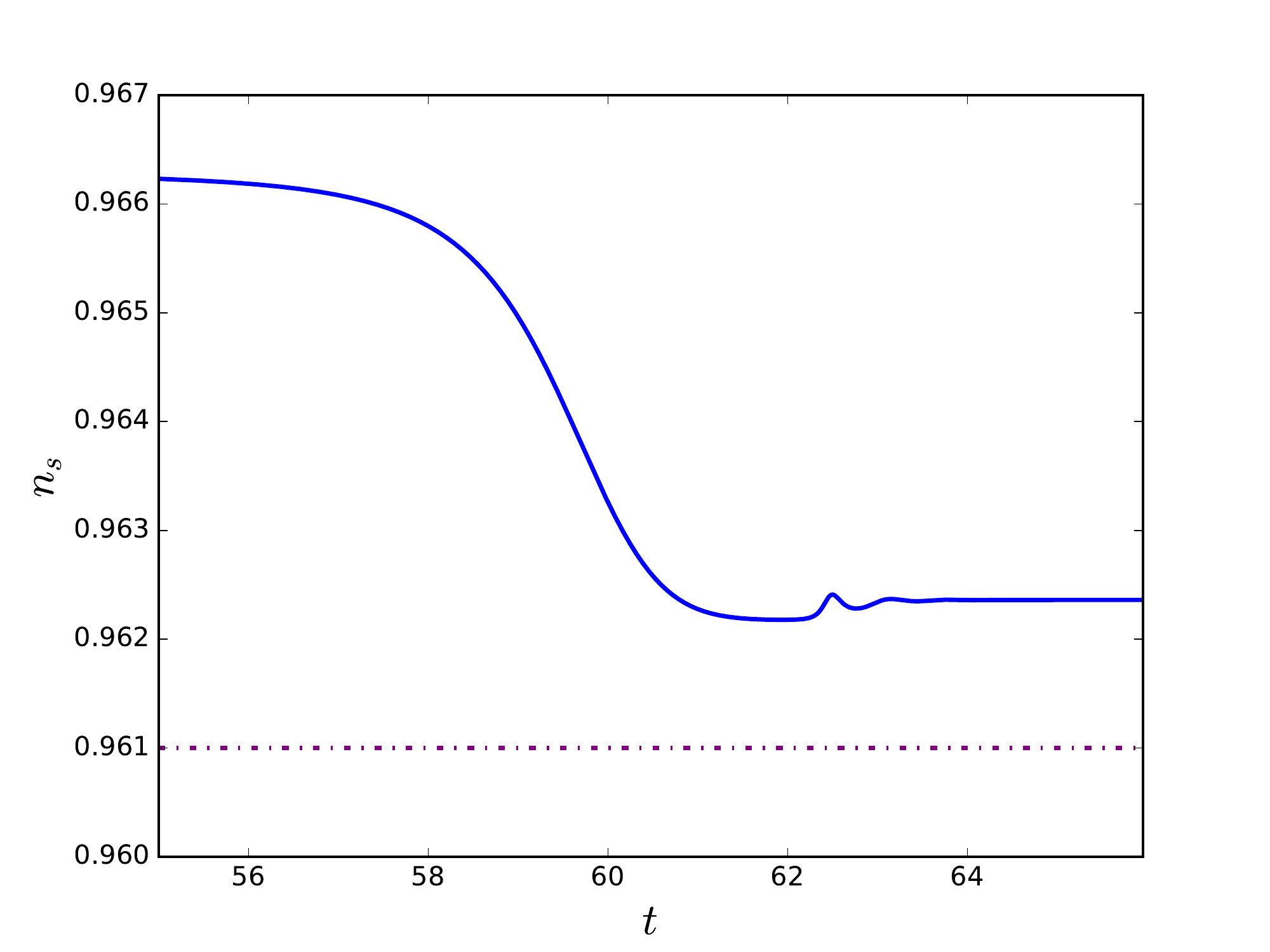}
	\includegraphics[width=0.49\textwidth]{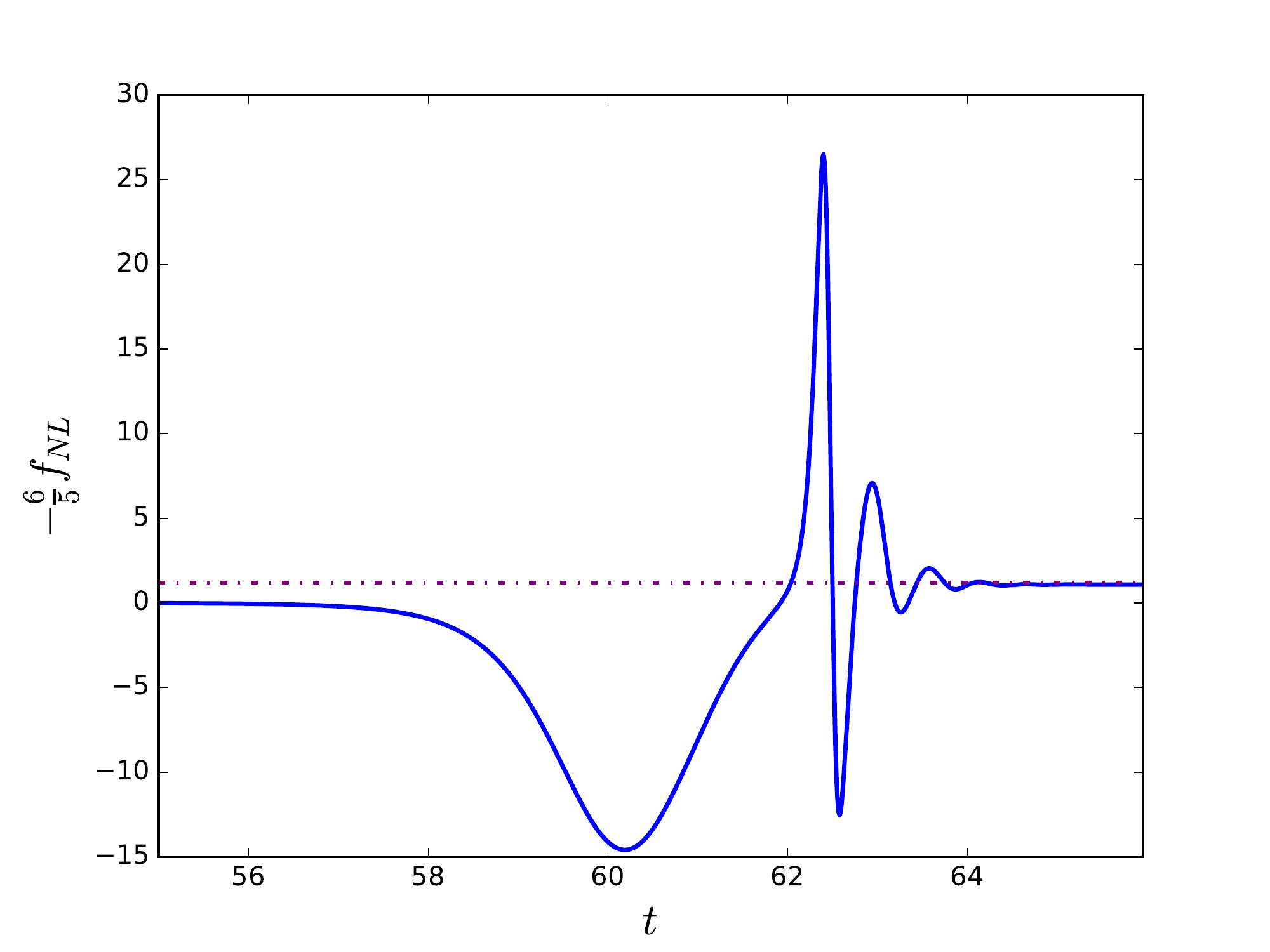}
    \includegraphics[width=0.49\textwidth]{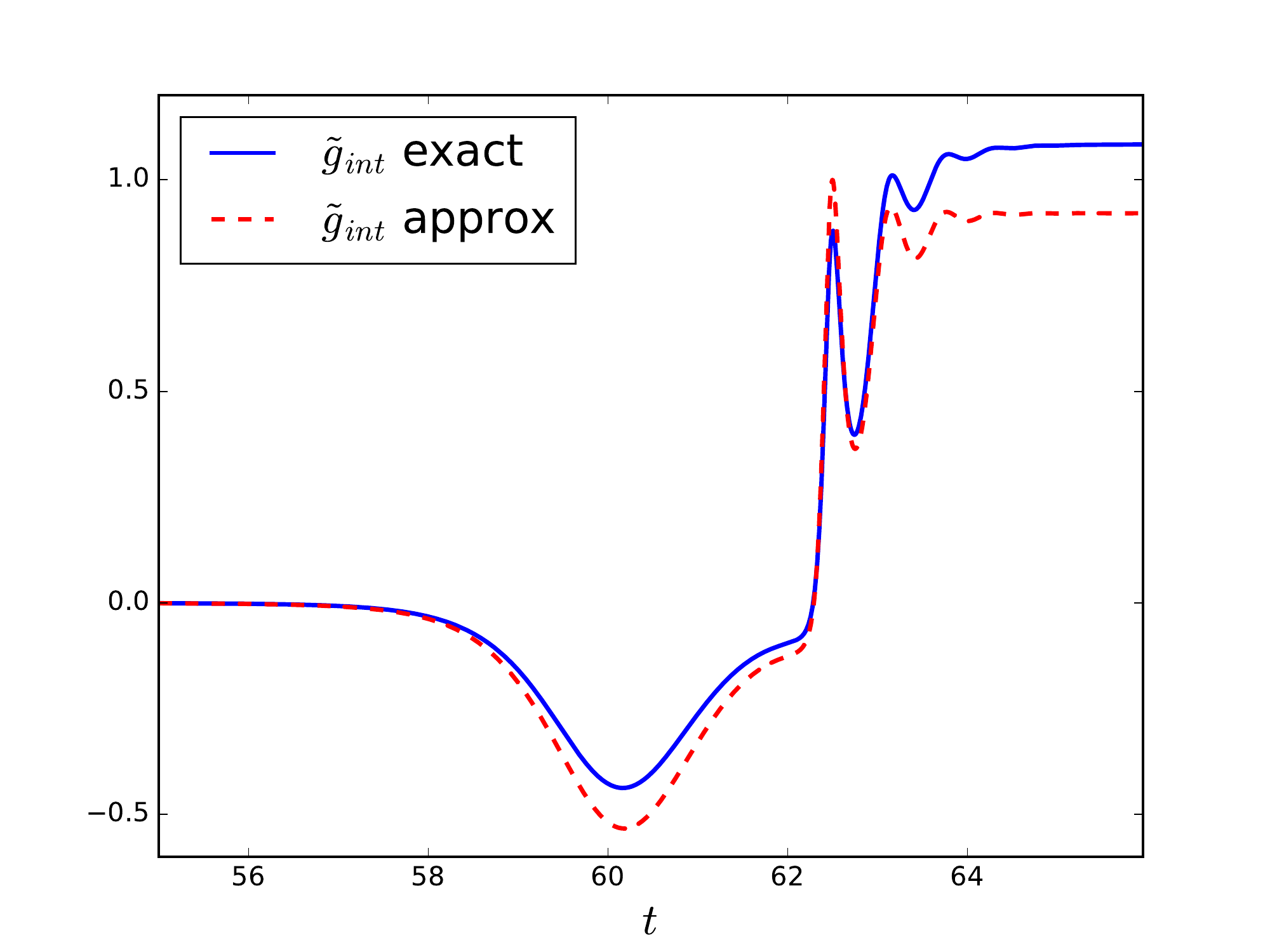}
	\caption{The exact numerical solutions for the different interesting parameters (basis components, slow-roll parameters, Green's functions, the spectral index and $\fnl$) during the turn for the first example of a monomial potential \eqref{pot n2m4} with $n=2$ and $m=4$. The last figure shows both the exact numerical solution for $\tilde{g}_{\mathrm{int}}$ and its analytical approximation. The horizontal purple dash-dot lines are the analytical predictions for $\bv_{12e}$, $n_s$ and $\fnl$.}
	\label{fig:n2m4}
\end{figure}

Figure \ref{fig:n2m4} contains the same plots as shown for the double quadratic potential except that we have removed the plot of $\gxpa$ and $\gxpe$ which does not provide any additional information, and added a plot of $\gint$. The different analytical predictions in \eqref{predictions 1} are reasonable estimations of the different parameters but the difference is larger than expected, especially on the new plot concerning $\gint$. This plot displays both the exact numerical $\tilde{g}_{\mathrm{int}}$ and its analytical prediction from \eqref{gintbsr} (more precisely, the analytical form of the approximated solution, with the different parameters determined numerically), using the definition: 
\be
\tilde{g}_{\mathrm{int}}=-\frac{2(\bv_{12})^2}{\lh 1 + (\bv_{12})^2\rh^2}\gint.
\ee
As one can see, both curves have a similar form, but there is a difference of around 15\%. The reason is that the turn occurs late with $\ge\approx 0.2$ when it starts. This value is already too large to have the slow-roll approximation working perfectly, but not enough for it to totally break down. In fact, this problem is quite general with the monomial potential because the turn has to occur late to get $\fnl$ of order unity, as we have shown. 

\begin{figure}
	\centering
	\includegraphics[width=0.49\textwidth]{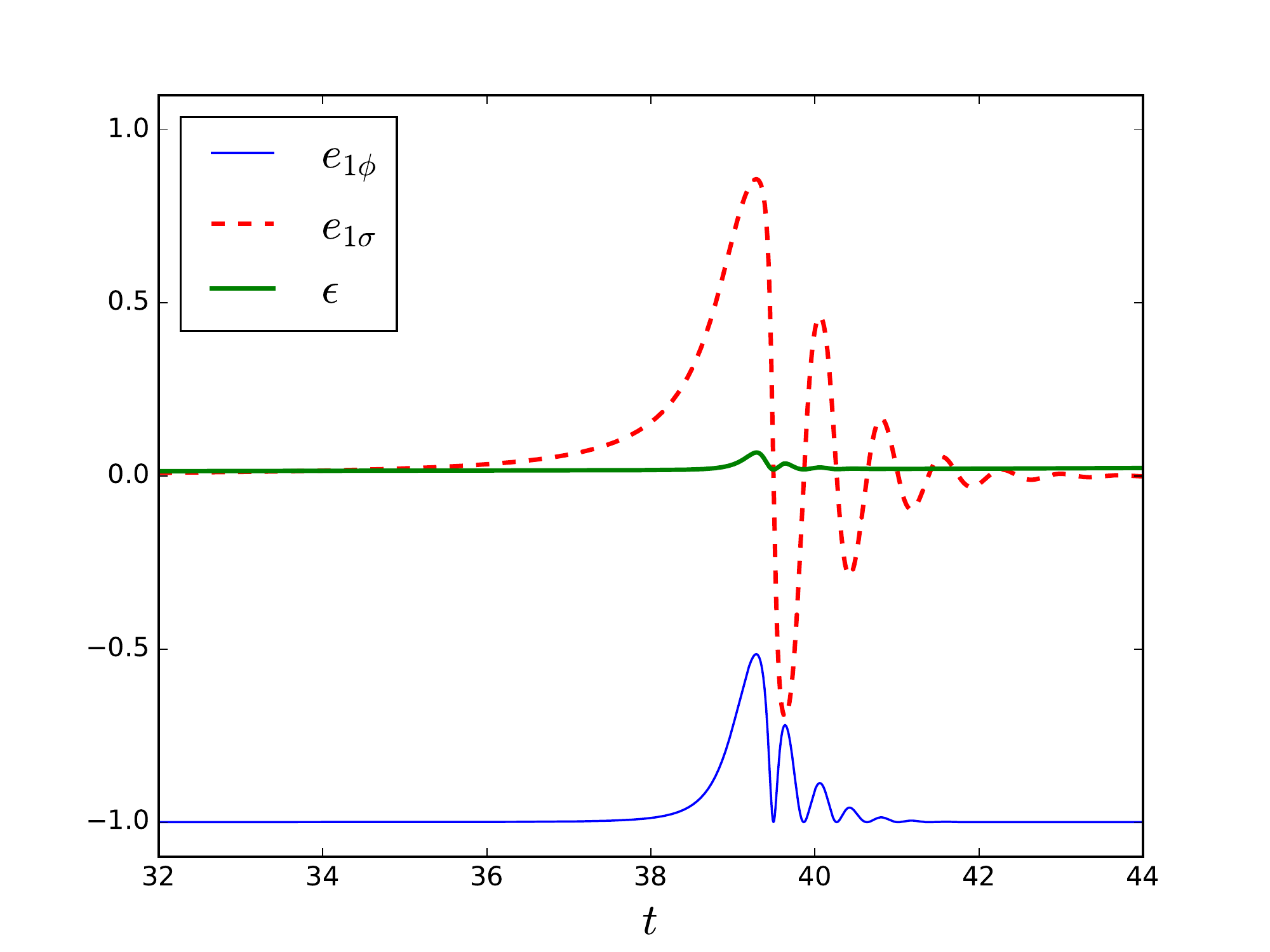}
	\includegraphics[width=0.49\textwidth]{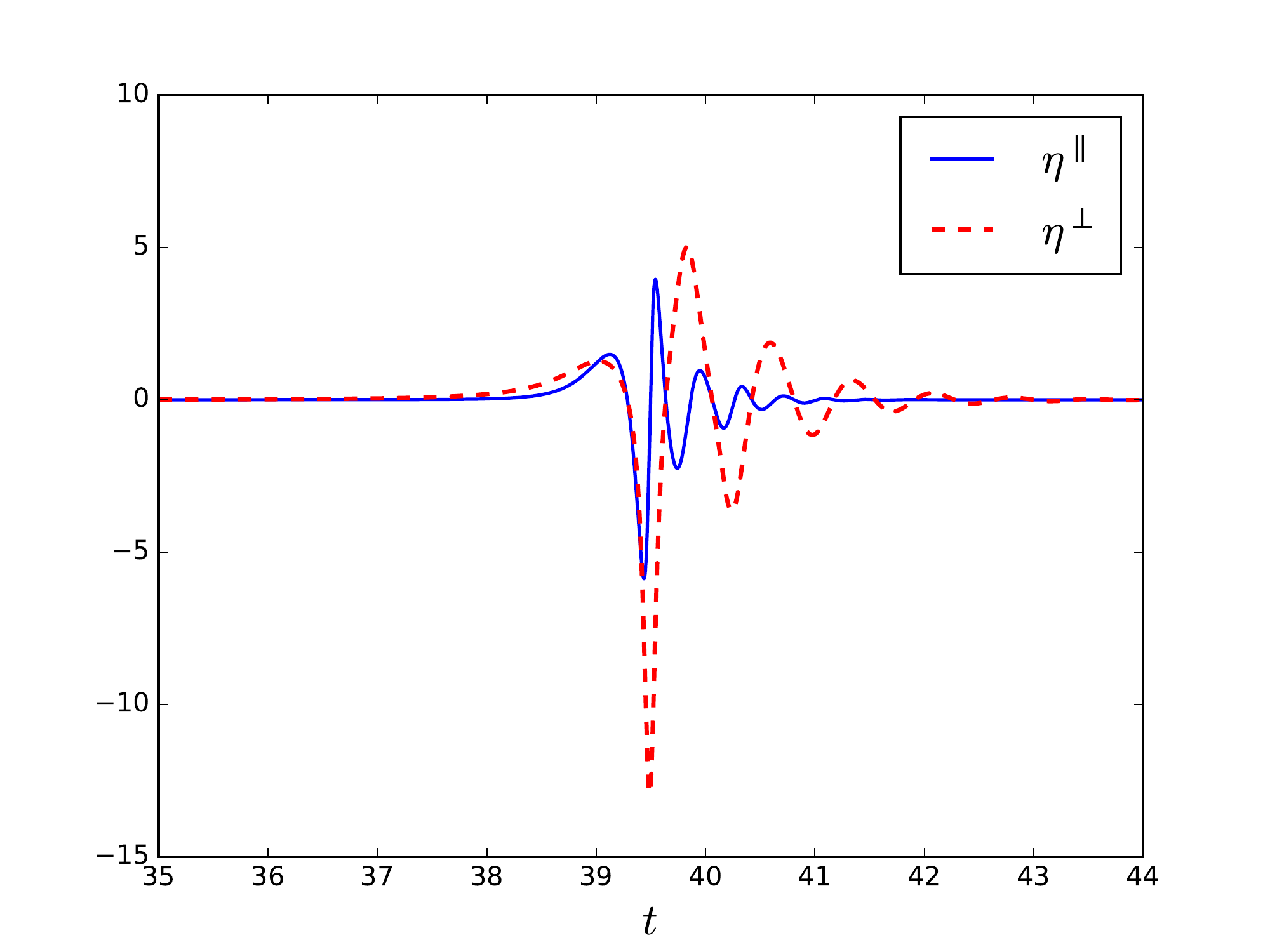}
	\includegraphics[width=0.49\textwidth]{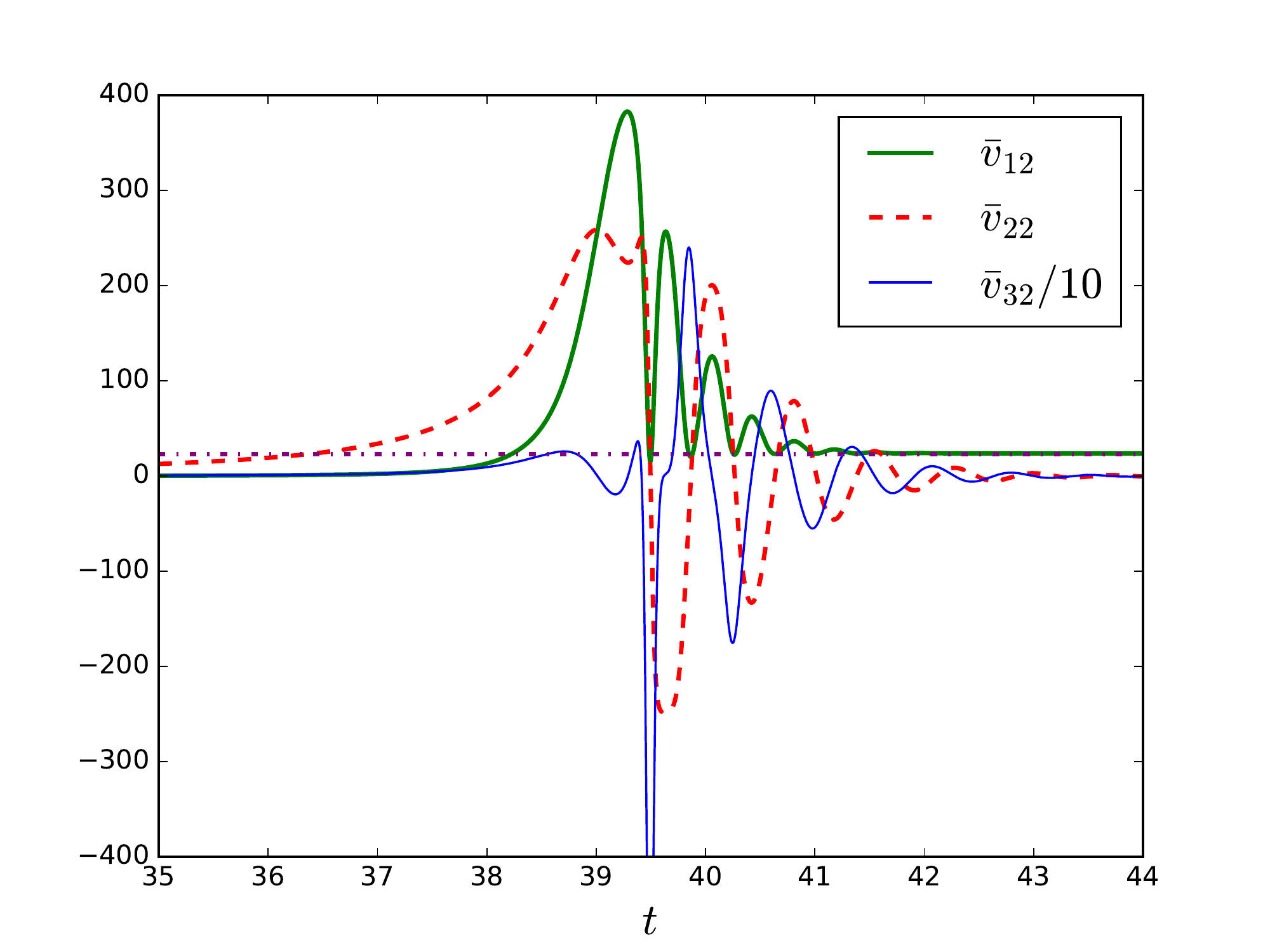}
	\includegraphics[width=0.49\textwidth]{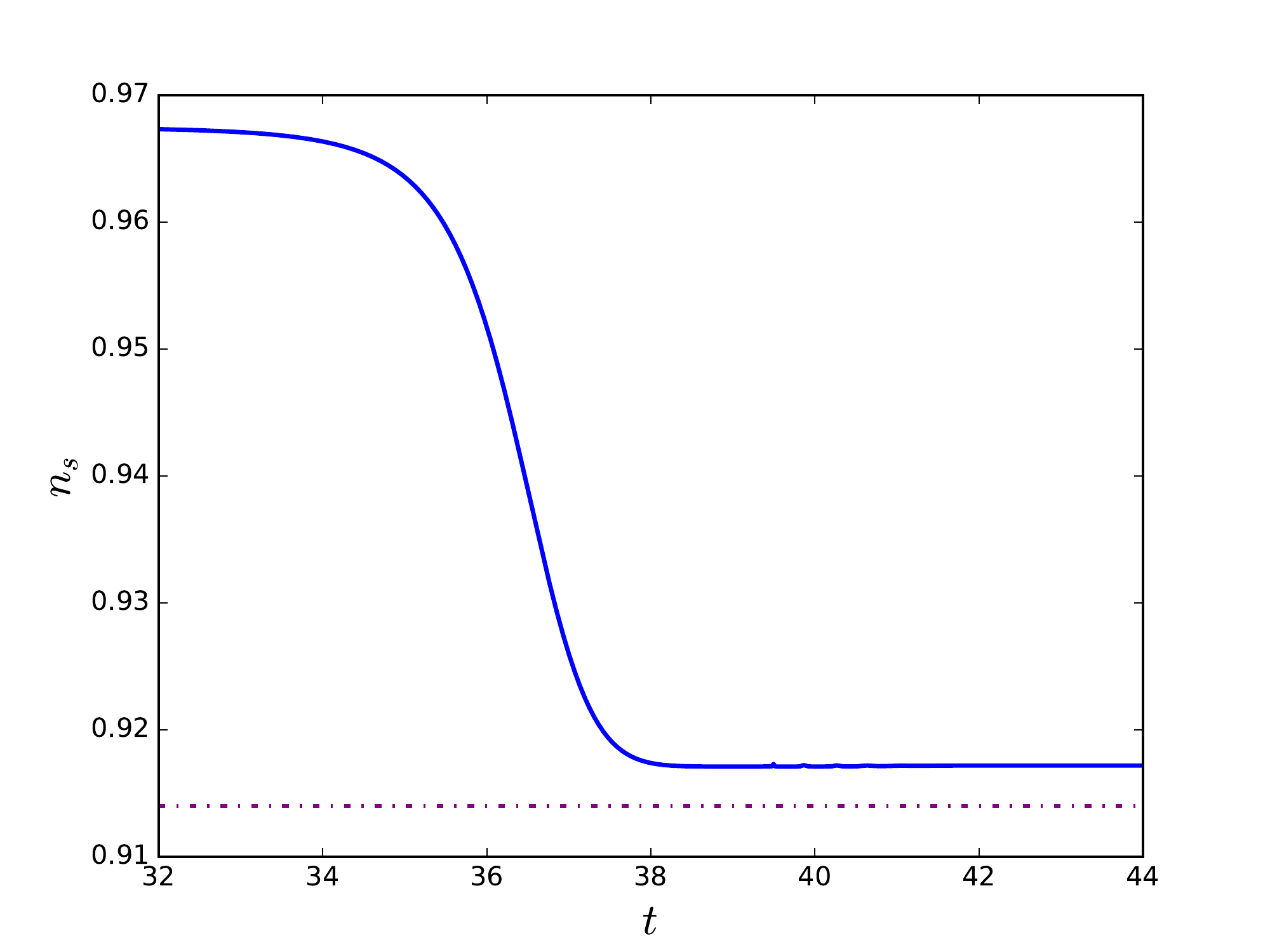}
	\includegraphics[width=0.49\textwidth]{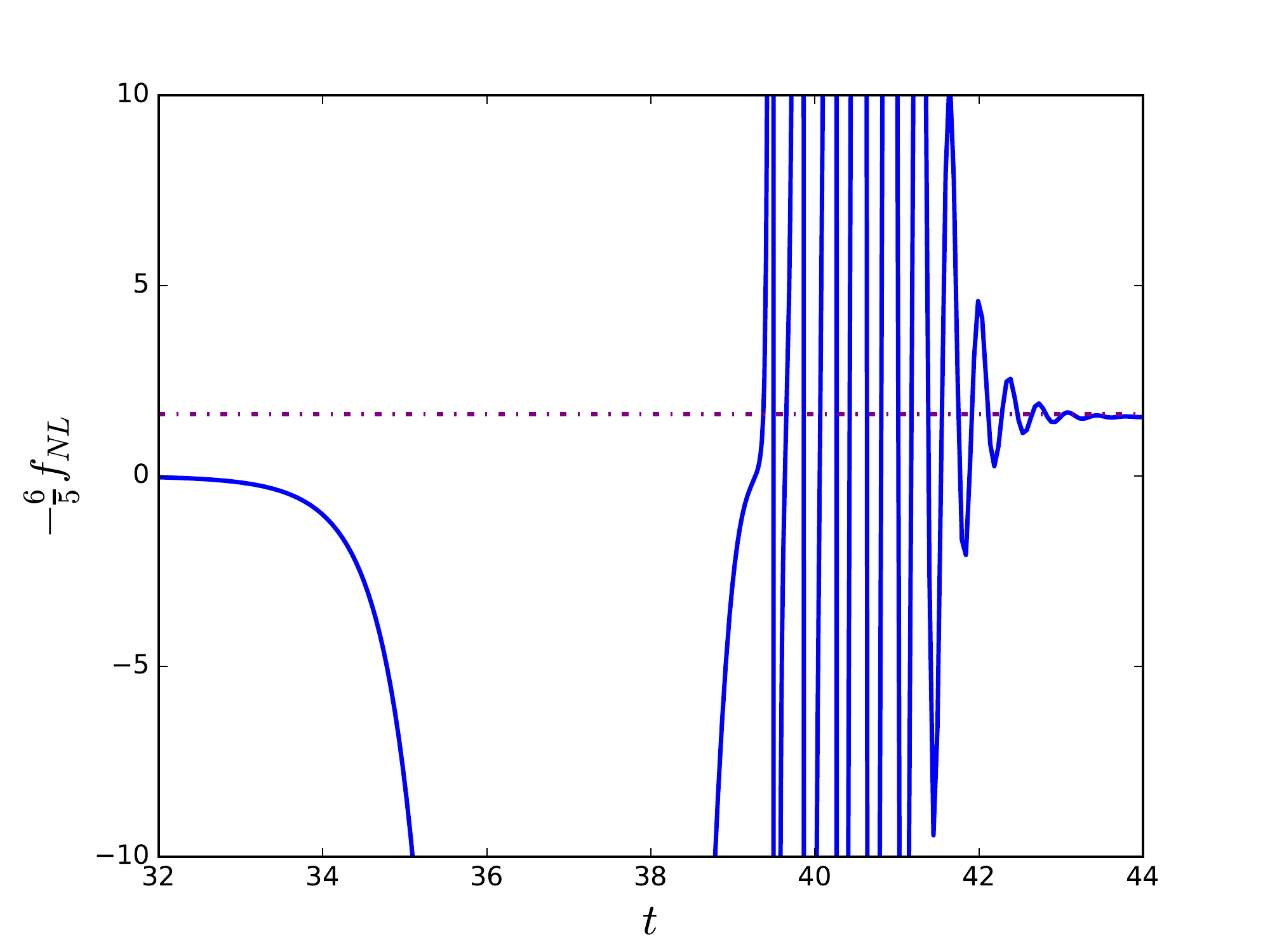}
    \includegraphics[width=0.49\textwidth]{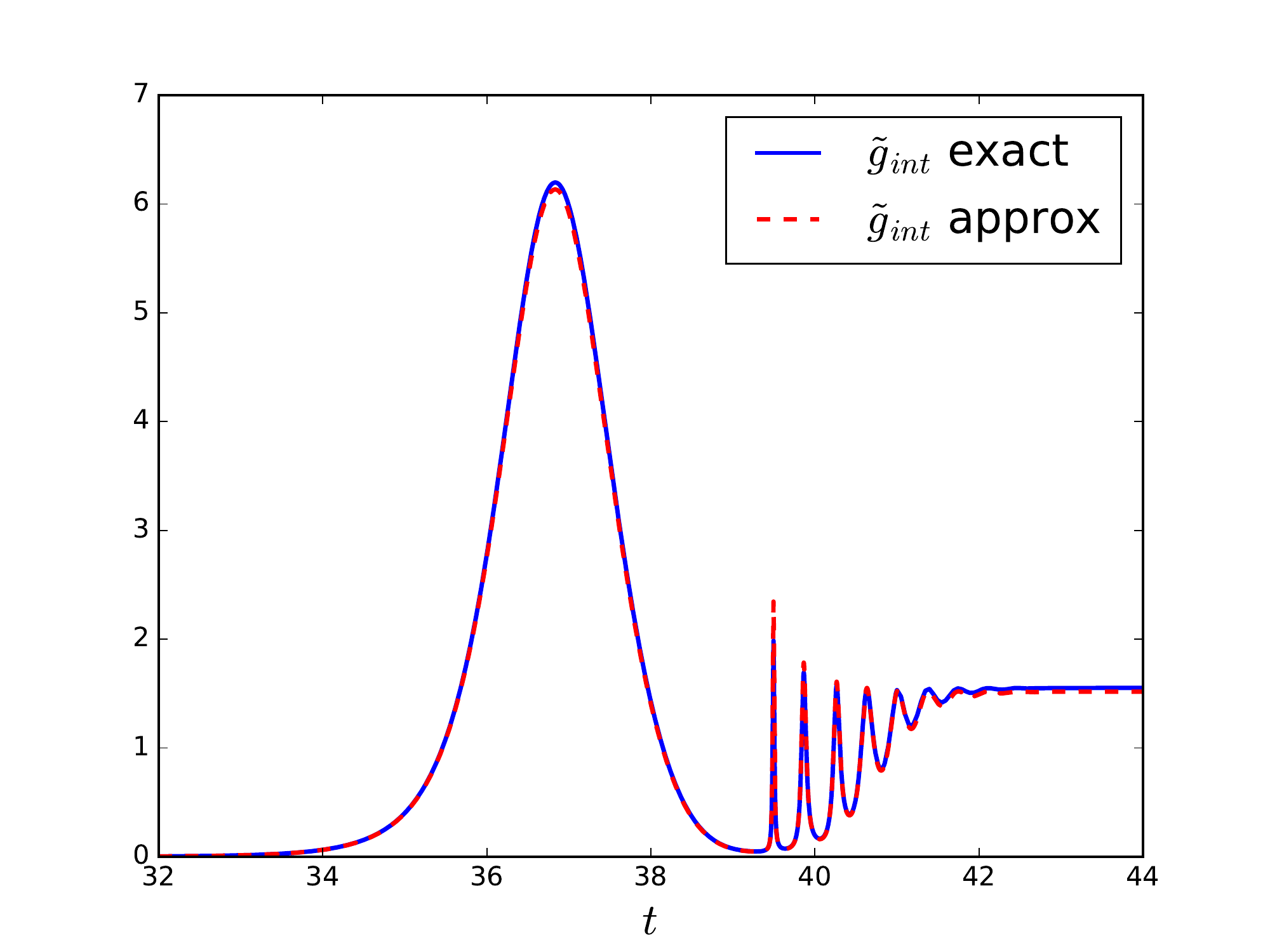}
	\caption{Same as figure \ref{fig:n2m4} but for the second example of monomial potential \eqref{pot n2m4} with $n=2$ and $m=4$ (with the parameter values
given just above (\ref{gfgsstarvalues})).}
	\label{fig:n2m42}
\end{figure}

However, if we forget momentarily about the observational constraint on the spectral index, only for one example to illustrate the validity of the analytical expressions, it is possible to have the turn occuring earlier.
The second set of values is: $\ga=\half\gk^{-2}$, $C= {\textstyle\frac{4}{27}\frac{\gb^3}{\lambda^2}}$, $\gb=-2000$, and $\lambda=-{\textstyle\frac{40}{3}}\gb\gk^2$ with the initial conditions $\gf_i= 16\gk^{-1}$ and $\gs_i=0.01\gk^{-1}$.  This time, the analytical predictions are:
\be
\gf_* = 15.3\gk^{-1} \qquad \text{and} \qquad \gs_* = 0.0106\gk^{-1}.
\label{gfgsstarvalues}
\ee
which leads to:
\be
\bv_{12e}= 23,\quad n_s = 0.914 \quad \text{and} \quad -\frac{6}{5}\fnl=1.62.
\ee
Figure \ref{fig:n2m42}, which contains the same plots as figure \ref{fig:n2m4} but for the new parameters, shows that $\ge$ is of order $10^{-2}$ during the turn, which is in the domain of validity of the main hypothesis $\ge \ll 1$. During the turn, $\getpe$ is of order 10 at most, which shows that the slow-roll regime is broken. As expected, analytical predictions are now a very good estimation. However, the spectral index is 0.917, which is outside the observational bounds. This example is also used in the previous section in figure~\ref{fig:rhs} to illustrate that r.h.s.\ is several orders of magnitude smaller than the left-hand side terms of \eqref{equadiff}.

\subsubsection{Second example: Axion}

The next example is the axion-quartic model originally introduced in \citep{Elliston:2012wm} and discussed more recently in \citep{Dias:2016rjq}. The potential is:
\be
W(\gf,\gs)=\frac{1}{4}g\gf^4 + \Lambda^4 \left[ 1-\cos{\lh \frac{2\pi\gs}{f}\rh}\right],
\label{pot axion}
\ee
with $g=10^{-10}$, $\Lambda^4 = \lh\frac{25}{2\pi}\rh^2 g \gk^{-4}$ and $f=\gk^{-1}$. The initial conditions are $\gf_i = 23.5 \gk^{-1}$ and $\gs_i= \frac{f}{2} - 10^{-3}\gk^{-1}$. Defining $\gs'={\textstyle\frac{f}{2}}-\gs$, we have $\gs'\ll \gk^{-1}$. This will stay true until the turn, hence it is possible to perform an expansion of the potential in terms of this small parameter. At first order, we have $\cos{\lh \frac{2\pi\gs}{f}\rh}= -\cos{\lh \frac{2\pi\gs'}{f}\rh} =-1+ \half\lh {\textstyle\frac{2\pi\gs'}{f}} \rh^2$ which substitued into the potential gives:
\be 
W(\gf,\gs')=\frac{1}{4}g \gf^4 + 2 g \lh \frac{25}{2\pi} \rh^2 \gk^{-4} -\frac{1}{2}g \lh\frac{25}{f}\rh ^2  \gk^{-4} (\gs')^2.
\ee
This is a monomial potential with $n=4$ and $m=2$, hence in the region of parameters where the spectral index constraints cannot be satisfied. This is verified by computing  the analytical predictions like for the previous example. The fields at horizon-crossing are:
\be
\gf_* = 21.8\gk^{-1} \qquad \text{and} \qquad \gs'_* = -1.1 \times 10^{-3}\gk^{-1}.
\ee 
which leads to:
\be 
\bv_{12e}=-\frac{8\gk}{\gf_*\gs'_*}\lh\frac{f}{2\pi}\rh^2=-8.4, \quad n_s = 1 - \frac{16}{\gf_*^2} - 8\frac{25^2}{\gk^2 f^2 \gf_*^4} = 0.944 \quad \text{and} \quad -\frac{6}{5}\fnl = 2\pi^2.
\ee
This model gives $\fnl$ of order ten, however the spectral index is lower than the Planck constraints. 

\begin{figure}
	\centering
	\includegraphics[width=0.49\textwidth]{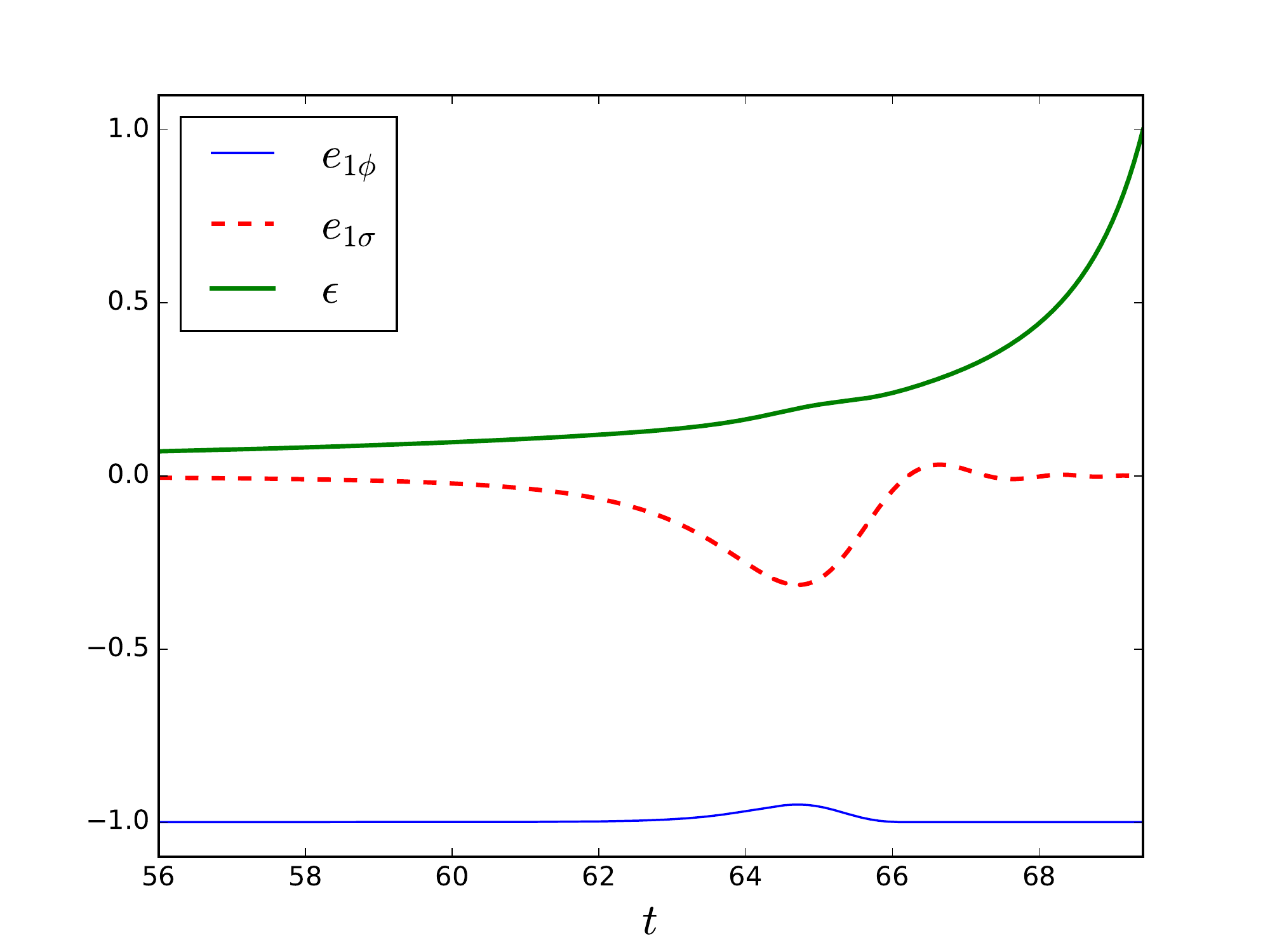}
	\includegraphics[width=0.49\textwidth]{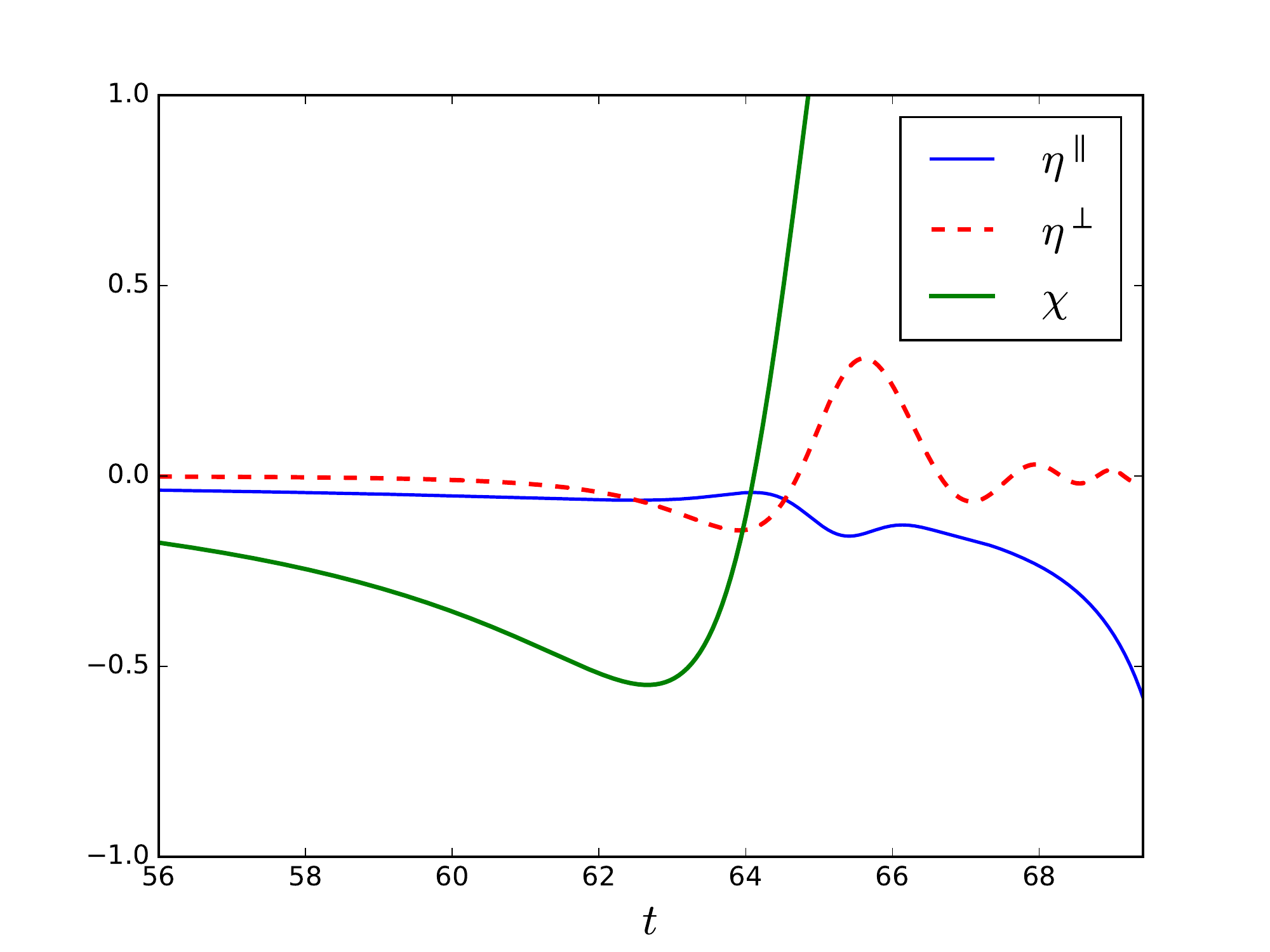}
	\includegraphics[width=0.49\textwidth]{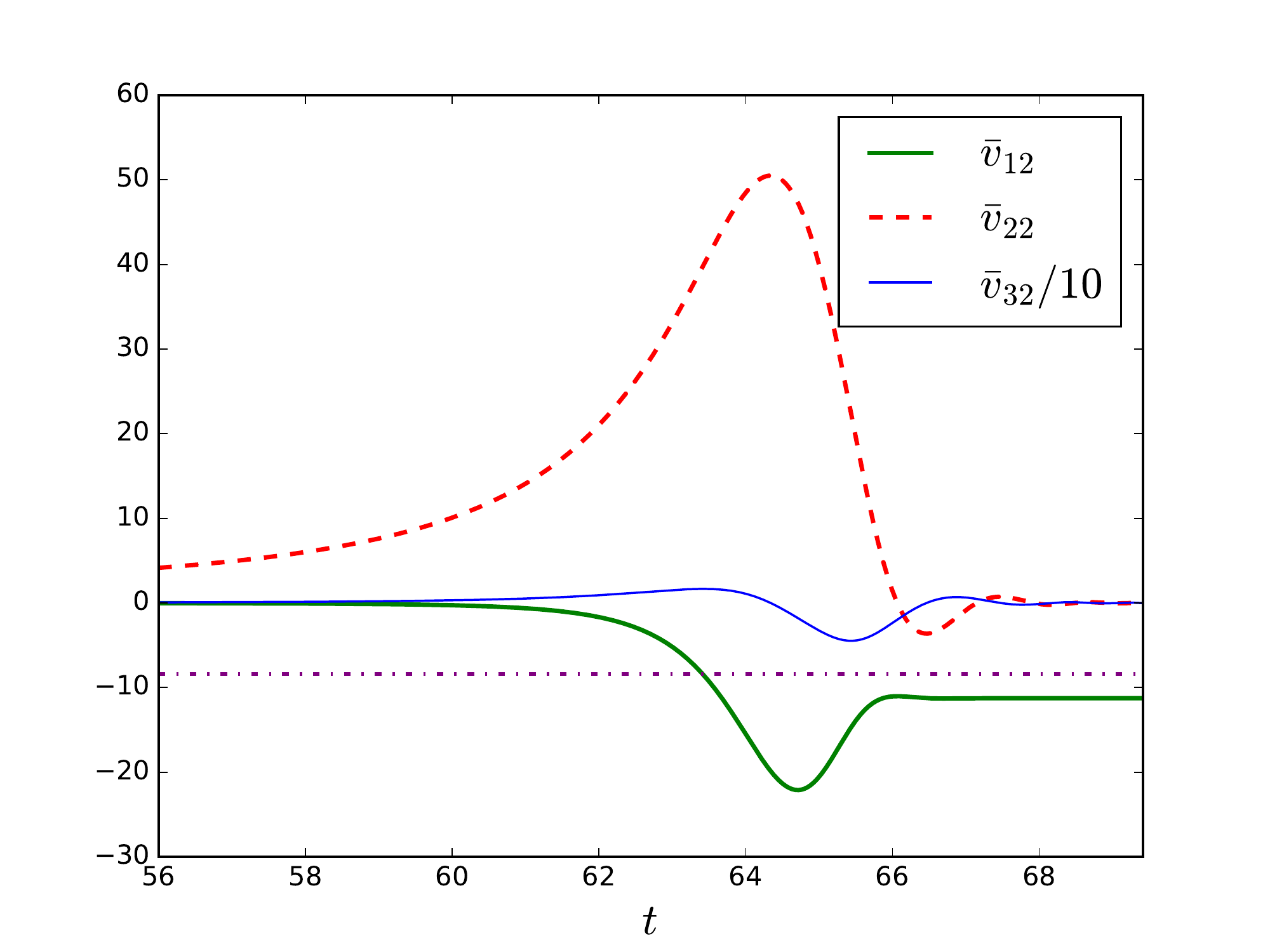}
	\includegraphics[width=0.49\textwidth]{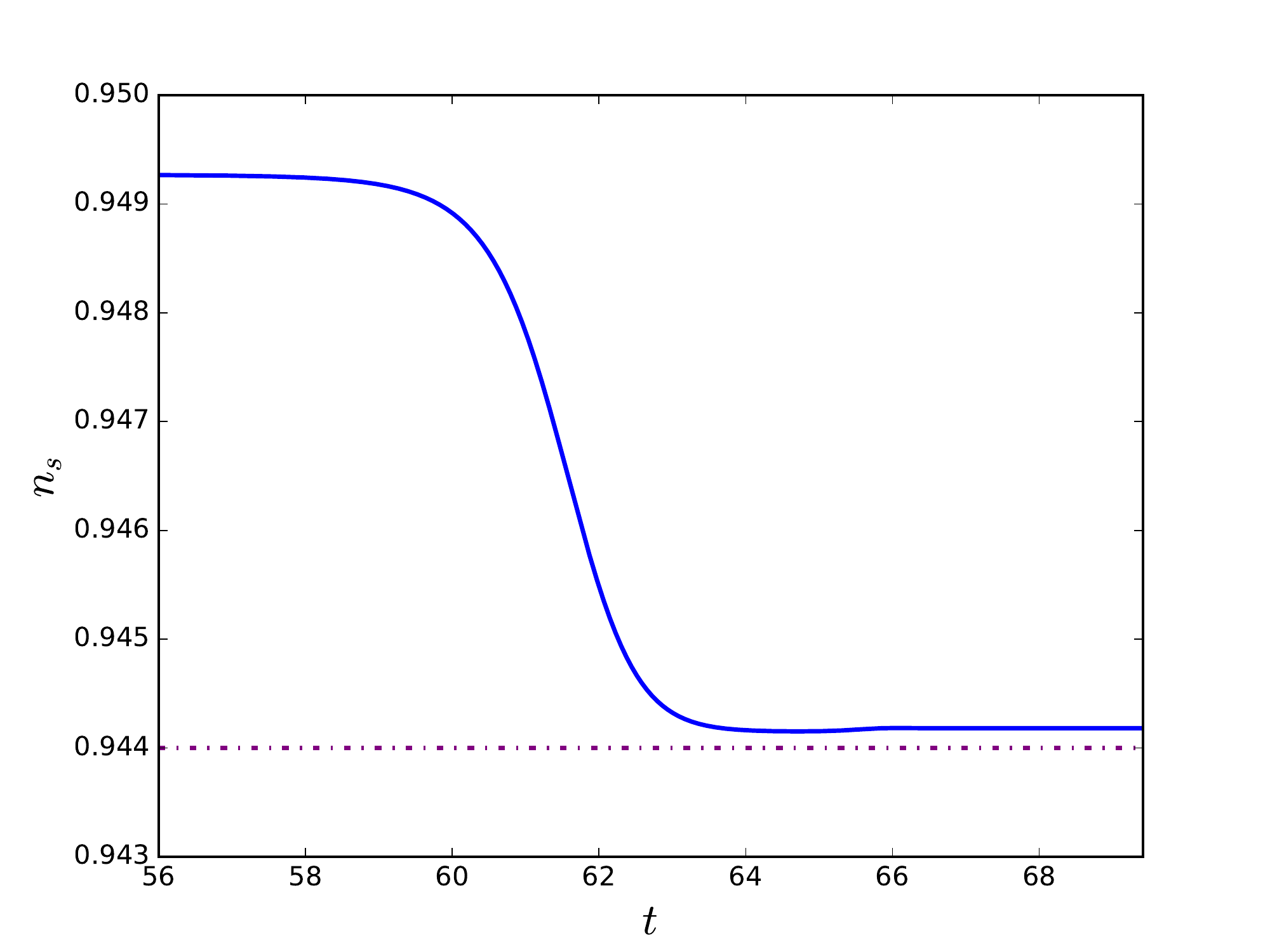}
	\includegraphics[width=0.49\textwidth]{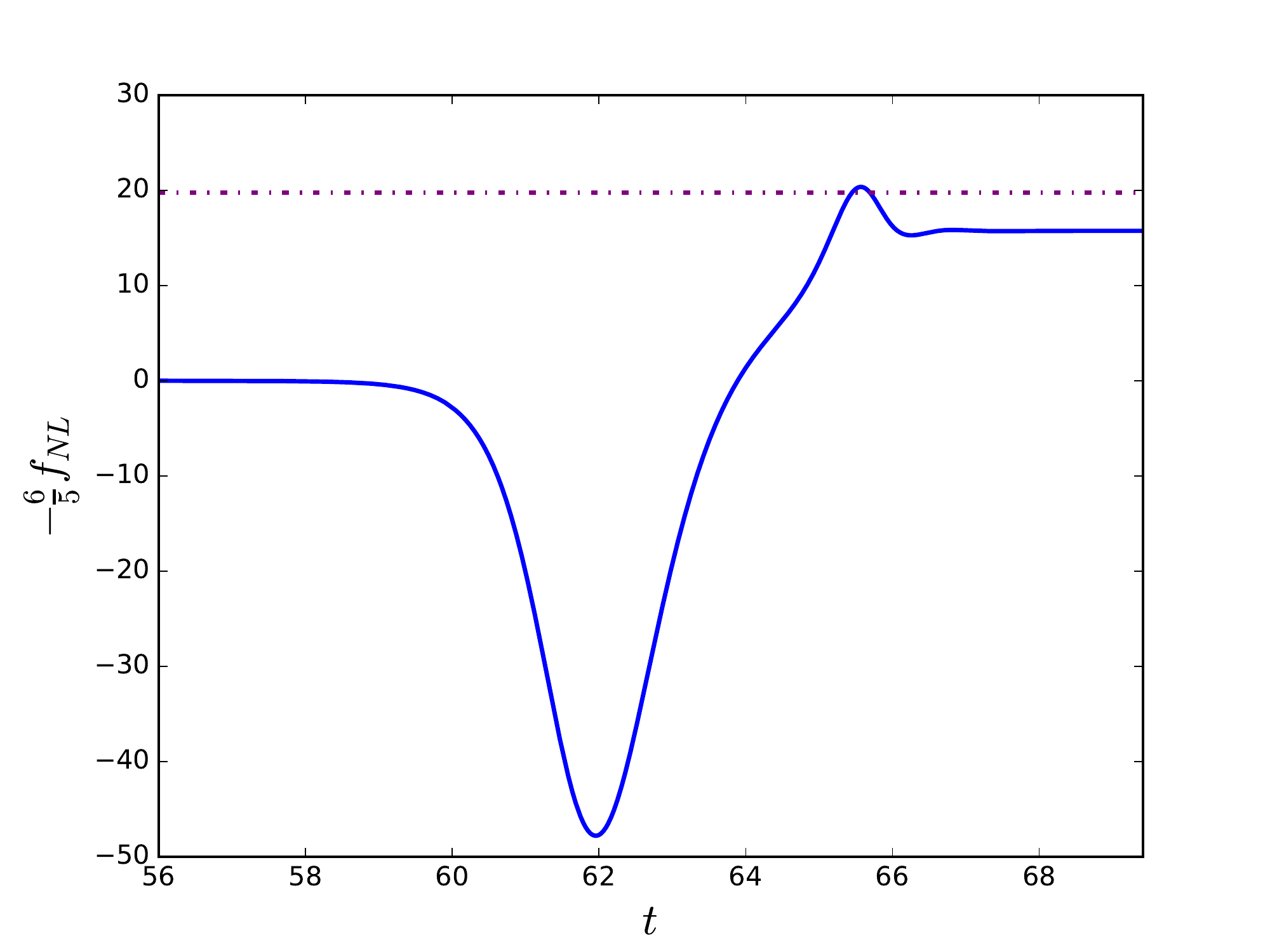}
    \includegraphics[width=0.49\textwidth]{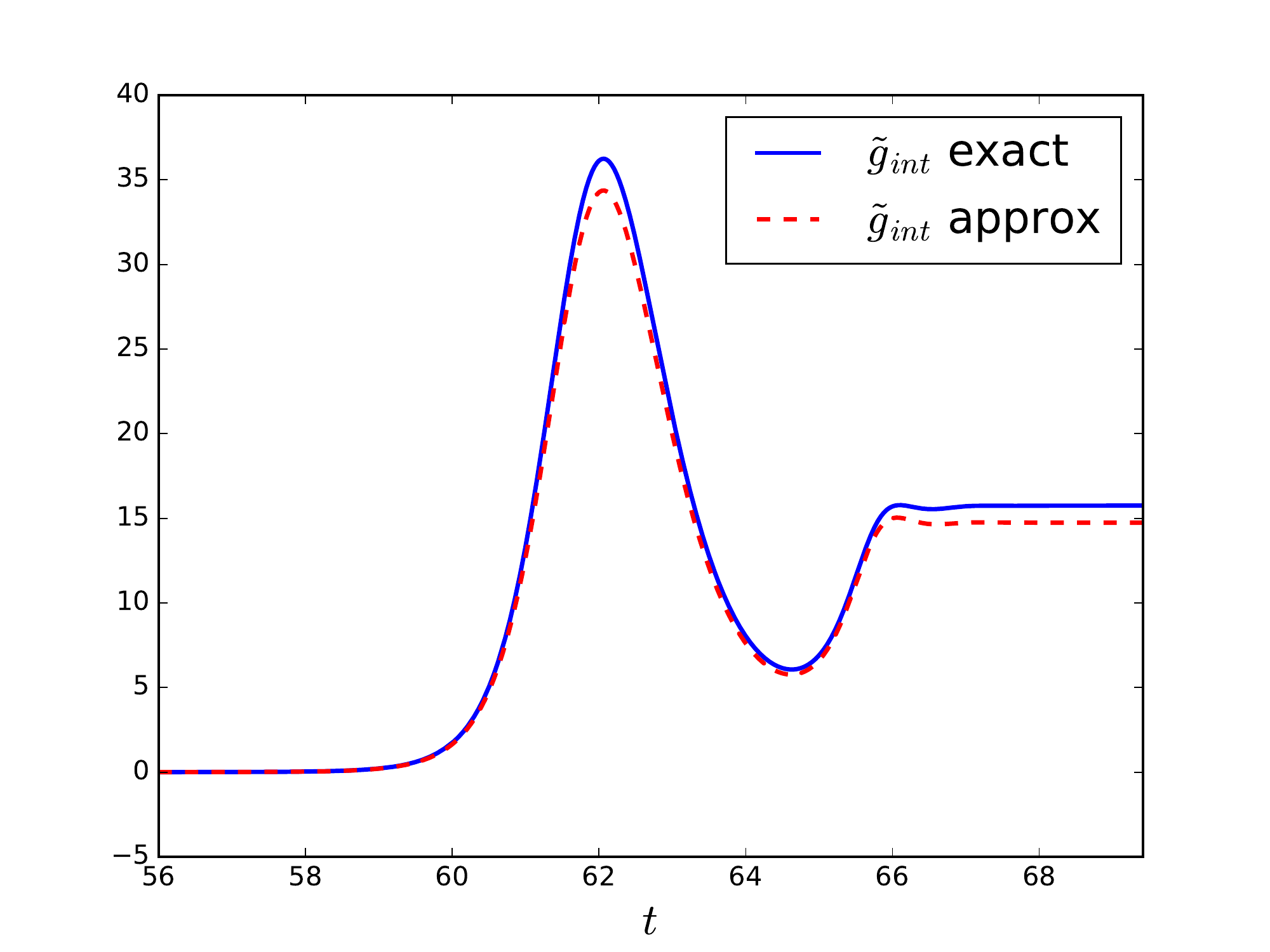}
	\caption{Same as figure \ref{fig:n2m4} but for the quartic-axion potential \eqref{pot axion}.} 
	\label{fig:axion}
\end{figure}

Figure \ref{fig:axion} confirms these results. Again in this model the turn occurs very late and there is a shift between the prediction and the exact result even if $\ge$ is still small enough during the turn. Moreover, $\getpa$ and $\getpe$ stay smaller than one during the turn, but $\gc$, which is displayed on the same plot, becomes large. This is another regime than the ones studied in subsection~\ref{Greensfunctionssubsec}. This has a direct impact on the Green's functions because $\gc$ appears in \eqref{G22eq} which explains the difference between the slow-roll prediction for $\bv_{12e}$ and the exact value. However, one interesting point is that the analytical form of $\gint$ stays valid. This case of large $\gc$ when other slow-roll parameters are small is not common and is due here partially to the fact that $\tV_{\gs\gs*}$ is too large to respect the Planck constraint (because as discussed in section \ref{Slow-roll}, $\gc_*=\ge_* + \getpa_* + \tV_{\gs\gs*}$).

\subsection{Second type of turn}
\subsubsection{$m=2$ and $n=2$}

Figure \ref{fig:etaper1} shows that a turn of the first type respecting observational constraints is not possible for a monomial potential with $n=2$ and $m=2$. However, if we do not keep the constraint that the turn must start before the end of the slow-roll regime, this model can have a turn of the second type. This example was published originally in \citep{TvT1} and is here adapted to be in agreement with the latest Planck constraints. See the second line of plots in figure \ref{fig:slowrollbroken} for an illustration of the field trajectory. The potential has the form:
\be
W(\gf,\gs) = \ga \gf^2 + C + \gb \gs^2 + \lambda \gs^4,
\label{pot n2m2}
\ee
with $\ga=20 \gk^{-2}$, $C= {\textstyle\frac{\gb^2}{4\lambda}}$, $\gb= -9\gk^{-2}$ and $\lambda=2$. The initial conditions are $\gf_i=18 \gk^{-1}$ and $\gs_i = 0.01 \gk^{-1}$ with $\dot{\gf}_i$ and $\dot{\gs}_i$ determined by the slow-roll approximation. At horizon-crossing, we have:
\be
\gf_* = 14.9\gk^{-1} \qquad \text{and} \qquad \gs_* = 0.011 \times 10^{-3}\gk^{-1}.
\ee
Substituted into \eqref{constraint1}, \eqref{constraint2} and \eqref{constraint3} this gives:
\be 
\bv_{12e} = -\frac{2\gk}{\gf_*\gs_*}\frac{C}{2\gb} = 6.9, \quad n_s = 1 - \frac{4}{\gf_*^2} +\frac{4\gb}{\gk^2 \ga\gf_*^2} = 0.974 \quad \text{and} \quad -\frac{6}{5}\fnl = -\frac{2\gb}{\gk^2 C} = 1.8 .
\ee

\begin{figure}
	\centering
	\includegraphics[width=0.49\textwidth]{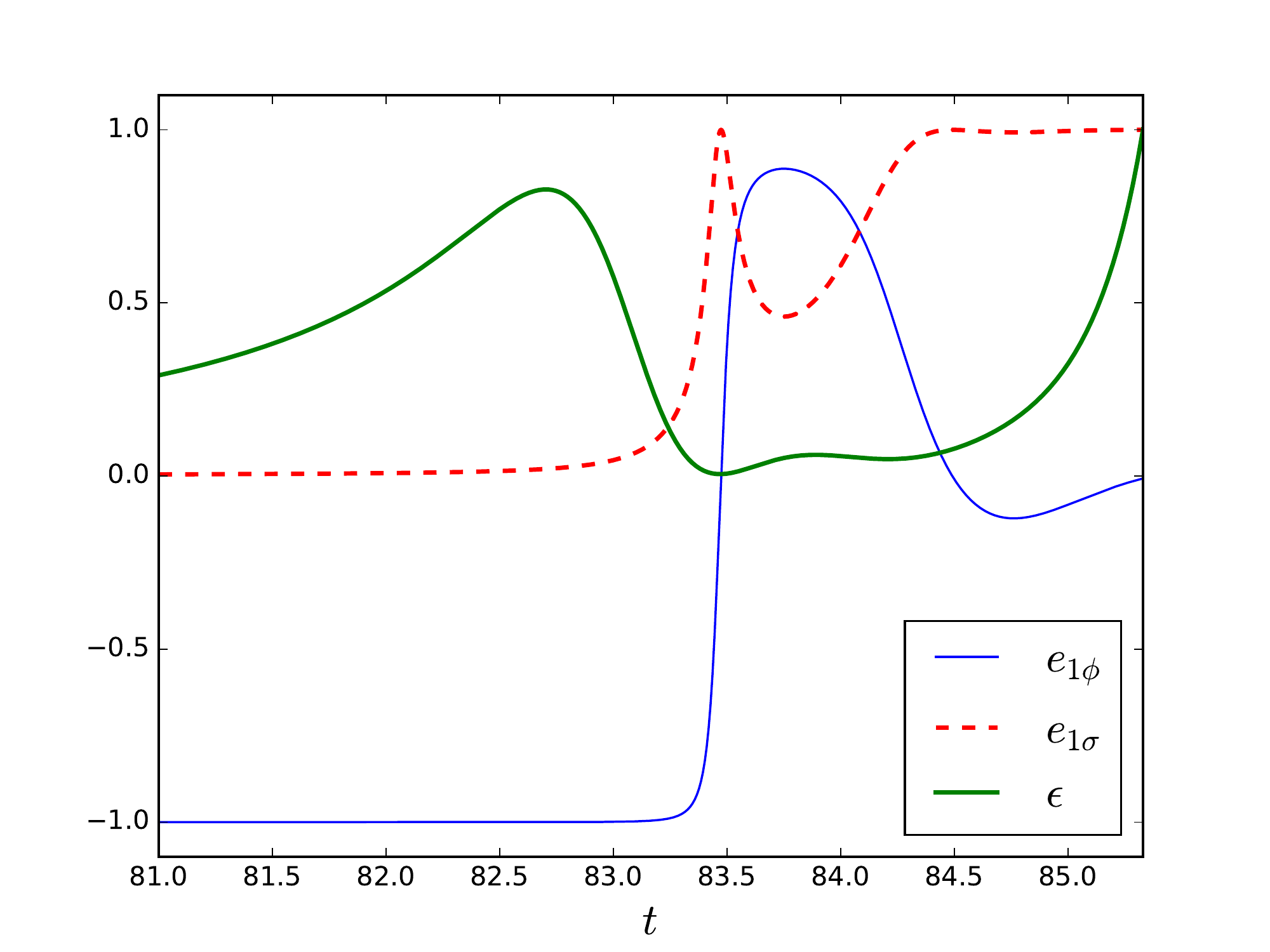}
	\includegraphics[width=0.49\textwidth]{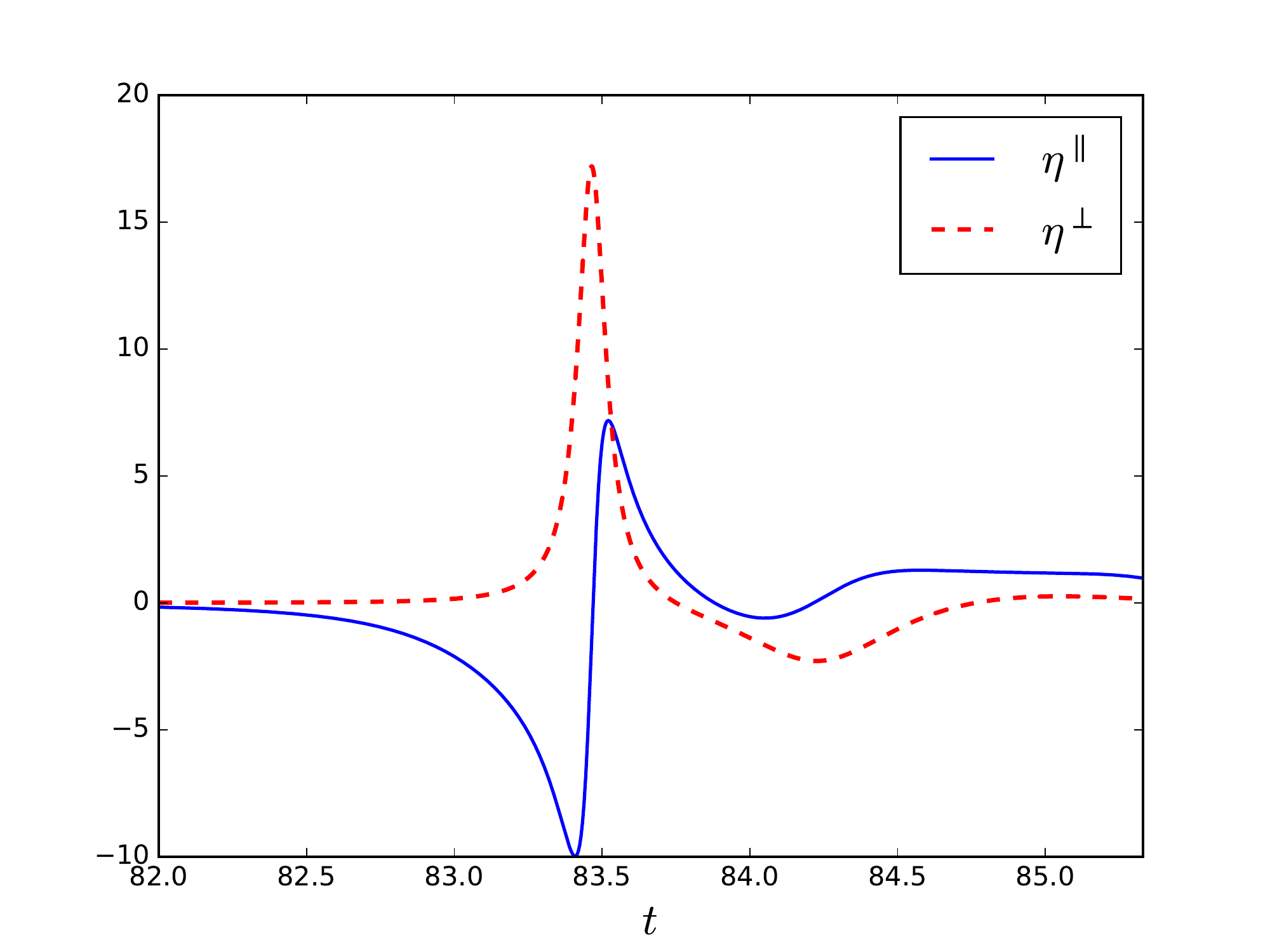}
	\includegraphics[width=0.49\textwidth]{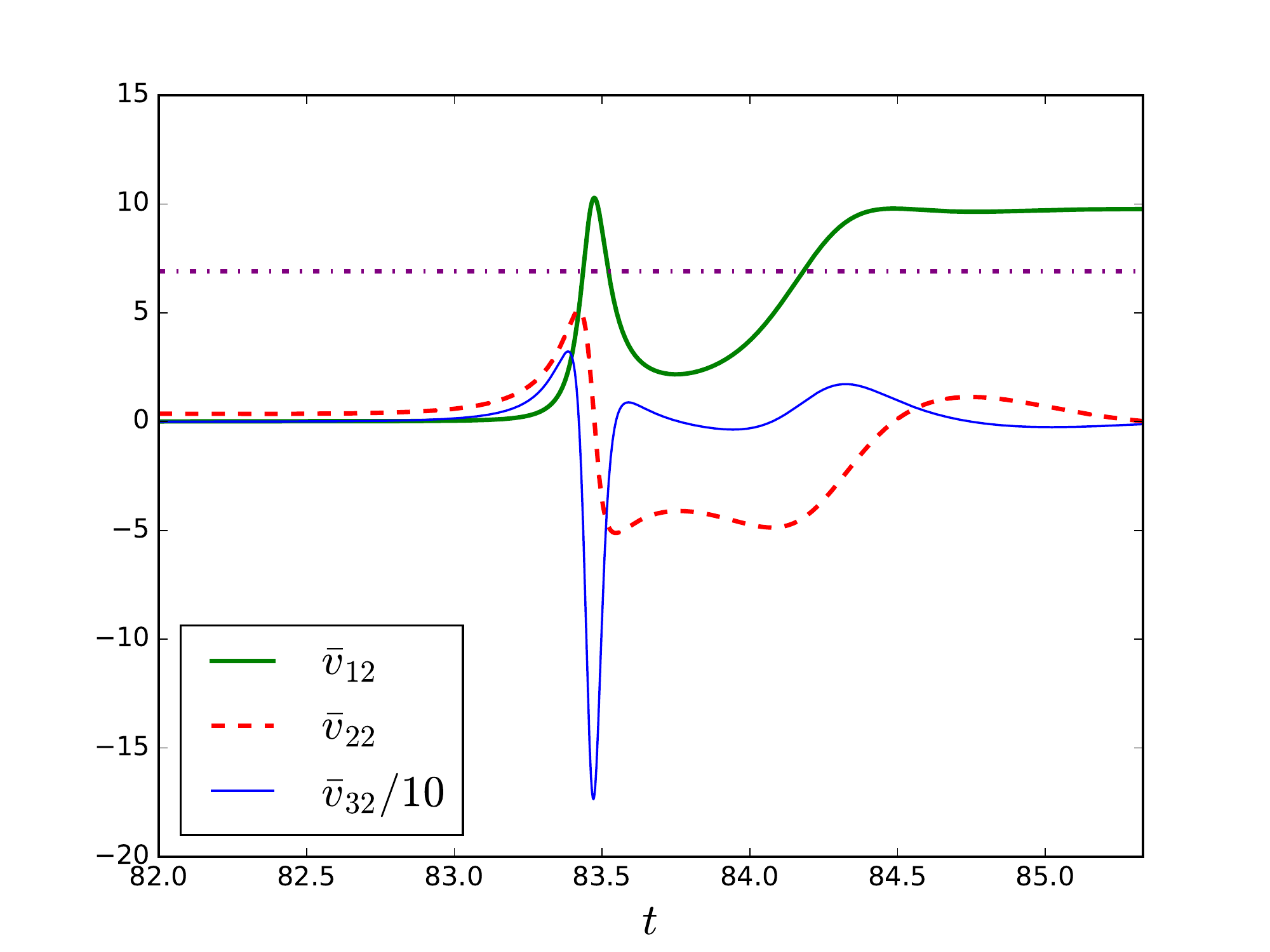}
	\includegraphics[width=0.49\textwidth]{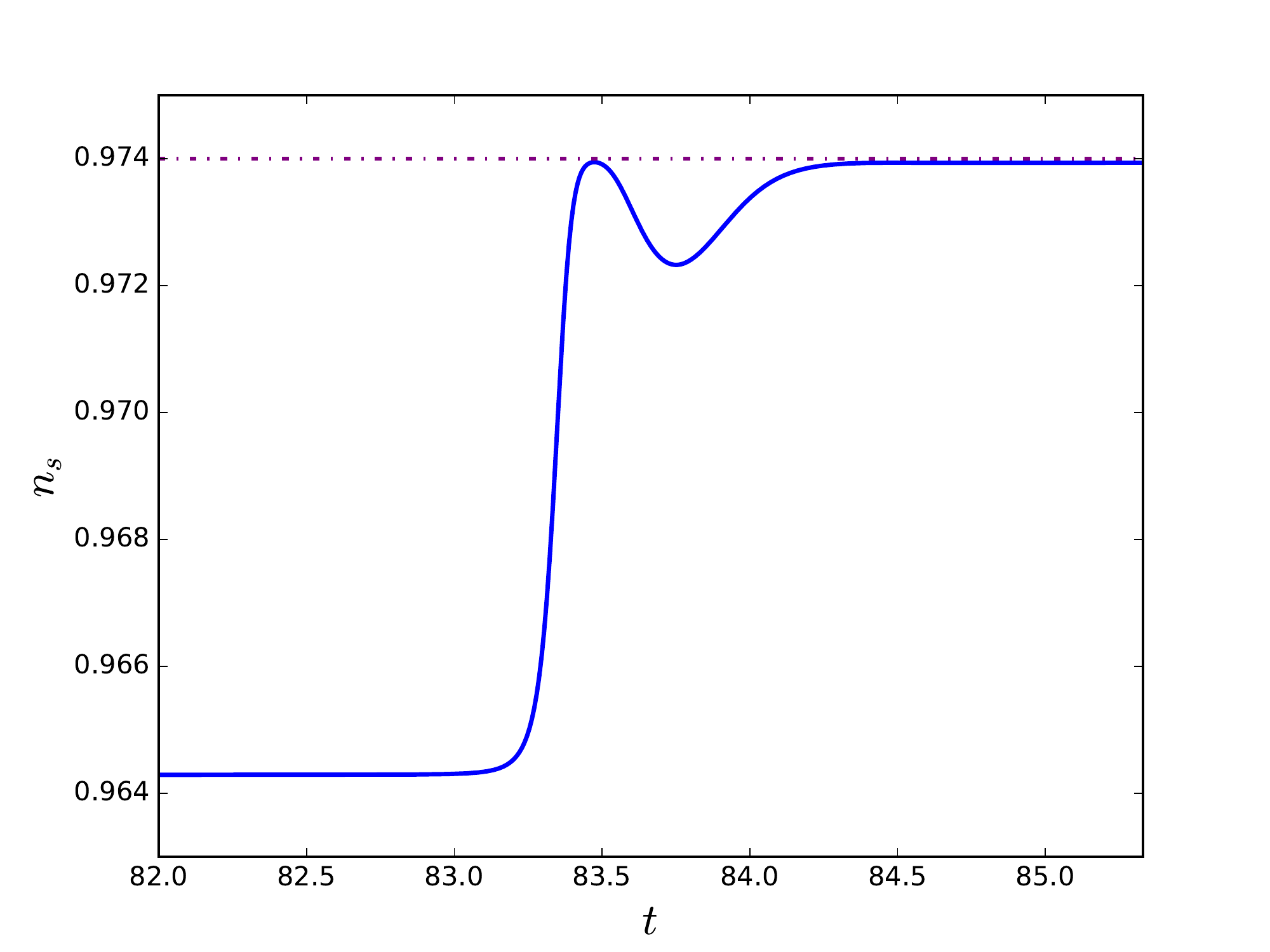}
	\includegraphics[width=0.49\textwidth]{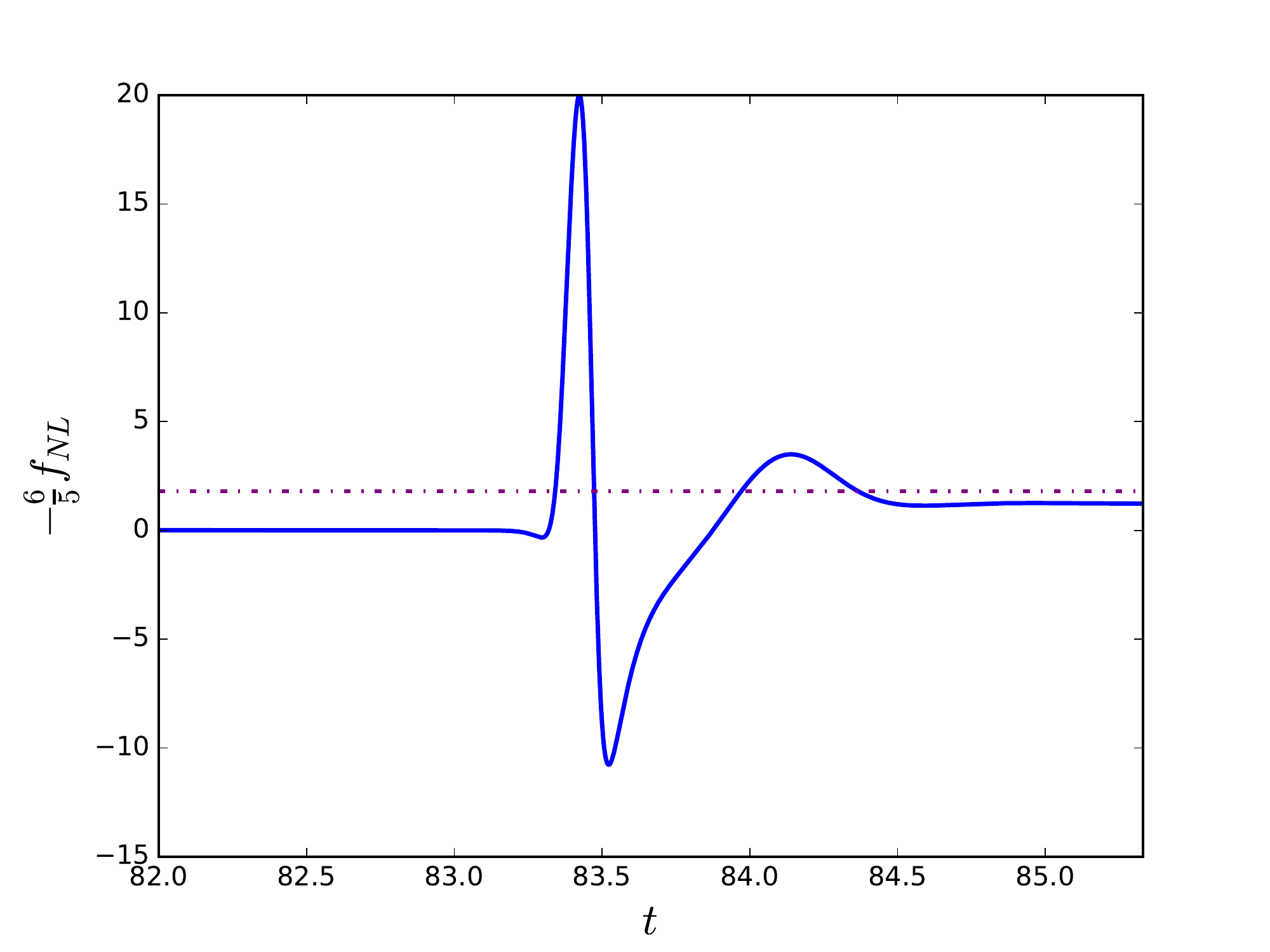}
    \includegraphics[width=0.49\textwidth]{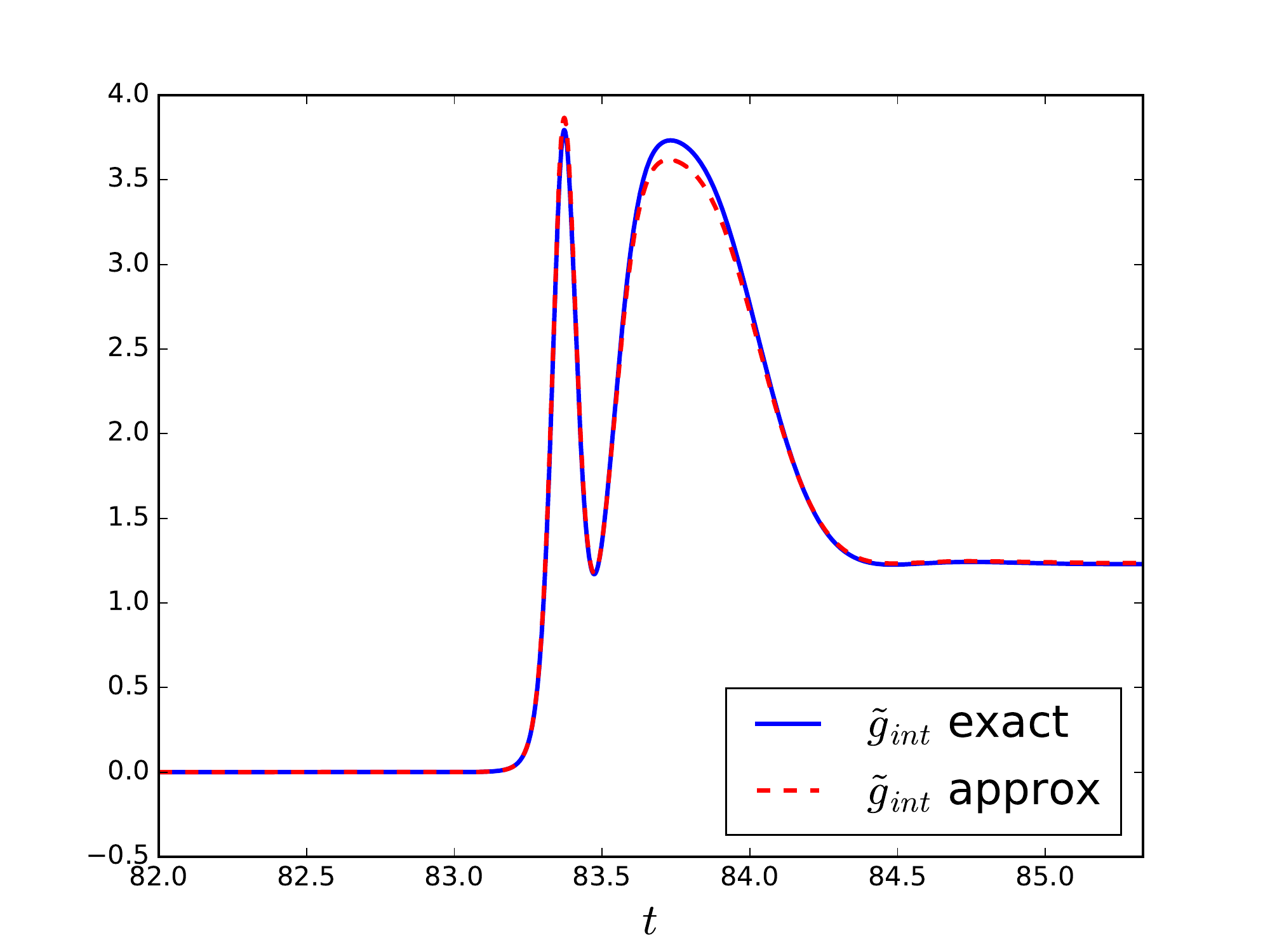}
	\caption{Same as figure \ref{fig:n2m4} but for the monomial potential with $n=2$ and $m=2$ \eqref{pot n2m2}.} 
	\label{fig:n2m2}
\end{figure}

Figure \ref{fig:n2m2} confirms that in this example the turn occurs after the field $\gf$ reaches the minimum of its potential. The Green's function $\bv_{12e}$ is larger than the slow-roll value, hence $\fnl$ is a little smaller than expected. This is in agreement with the discussion of the second type of turn at the very end of section \ref{Beyond slow-roll}. However, this does not have any impact on the spectral index because the dependence on $\bv_{12e}$ disappears when it is larger than 4. Hence, this model is allowed by the Planck constraints.

\subsubsection{A non-monomial example}

This last example is in the vein of the previous one in terms of the form of the field trajectory. However, there are several supplementary terms to show the validity of some analytical results beyond simple monomial potentials. The model has the following potential:
\be
W(\gf,\gs)= \frac{1}{4}\lambda \lh \gf^4 + \gs^4 + m^4 - 2 m^2  \gf^2 - 2 m^2  \gs^2 \rh + \nu  (m-\gf)^3 + W_0,
\label{last}
\ee
with $\lambda = 1200$, $\nu = 100\gk^{-1}$, $m=2\gk^{-1}$ and $W_0={\textstyle\frac{1}{4}}\lambda m^4$. The initial conditions are $\gf_i=25\gk^{-1}$ and $\gs_i=0.05 \gk ^{-1}$. We cannot use the monomial potential equations to determine $\gf_*$ and $\gs_*$, however the slow-roll estimation of $\fnl$ does not require them:
\be
-\frac{6}{5}\fnl=\frac{4}{\gk^2 m^2} = 1.
\ee

\begin{figure}
	\centering
	\includegraphics[width=0.49\textwidth]{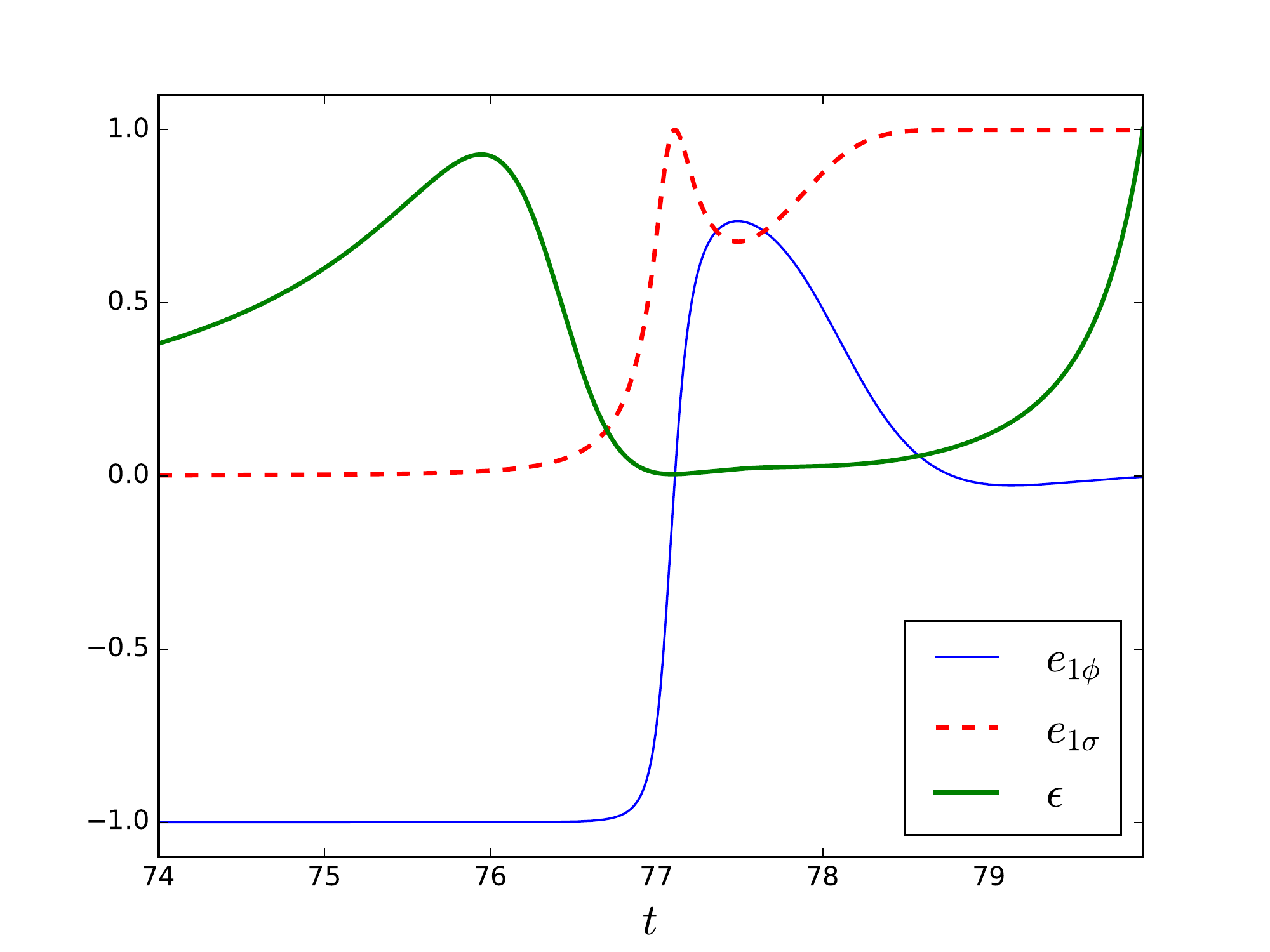}
	\includegraphics[width=0.49\textwidth]{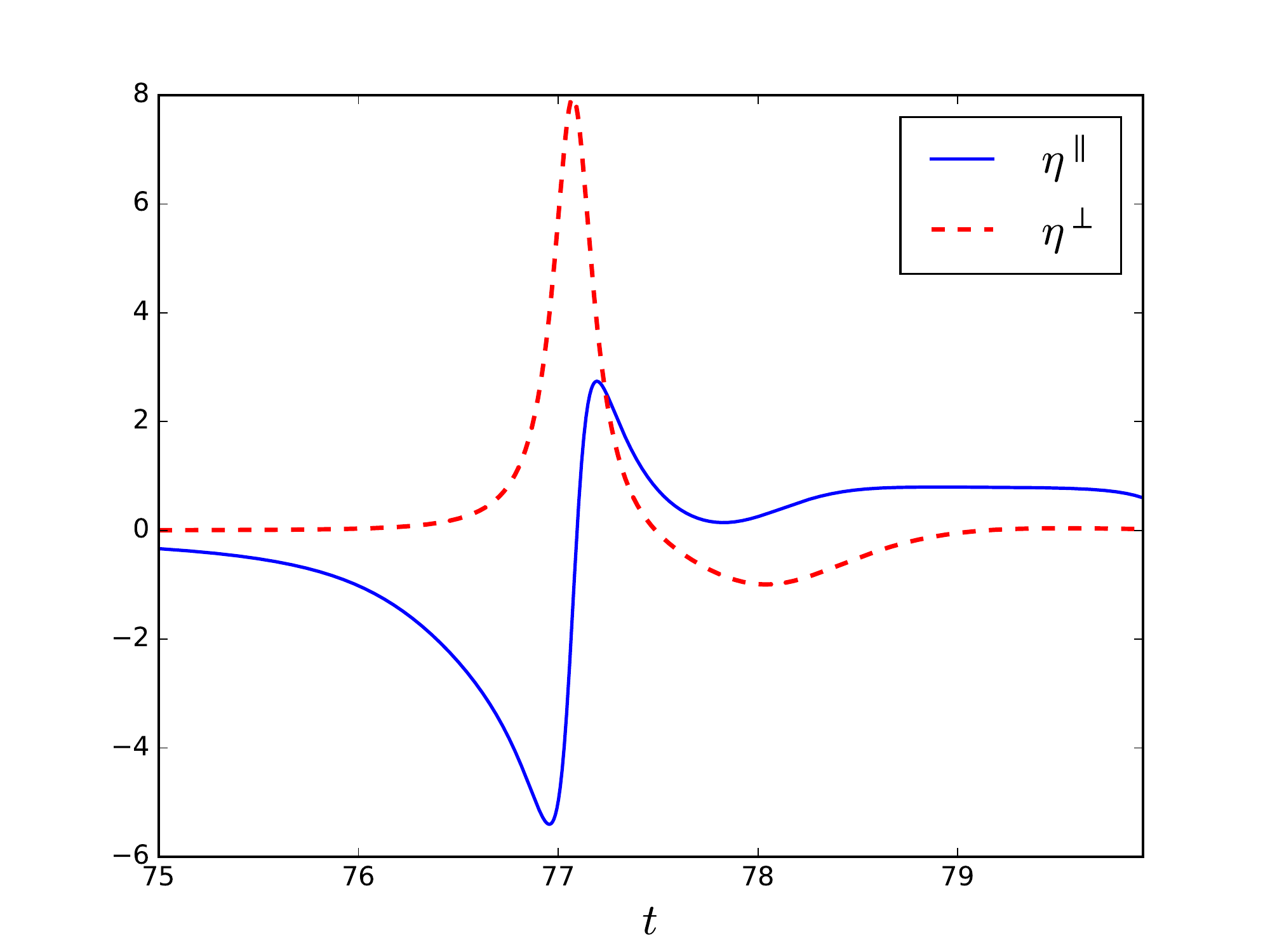}
	\includegraphics[width=0.49\textwidth]{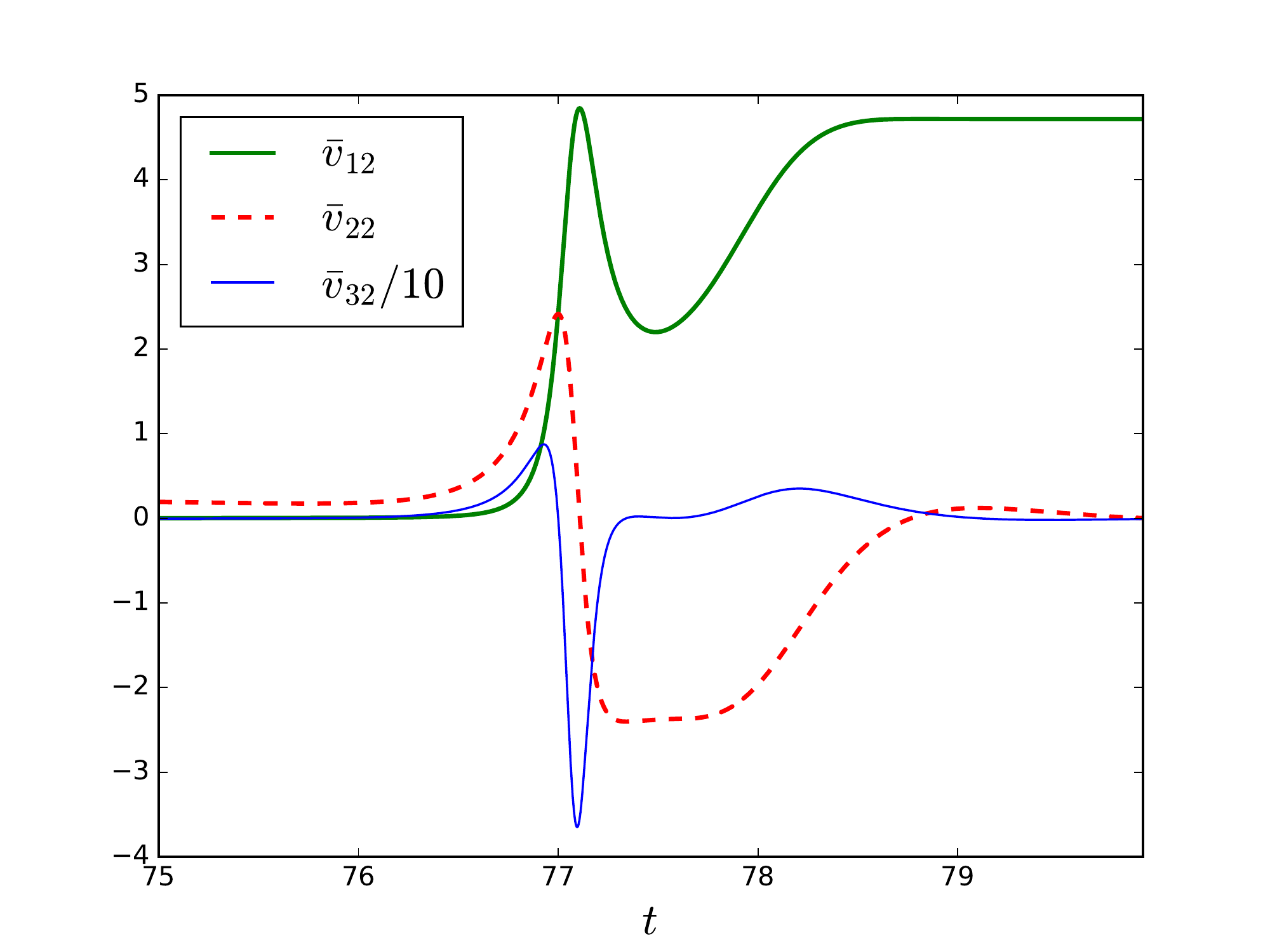}
	\includegraphics[width=0.49\textwidth]{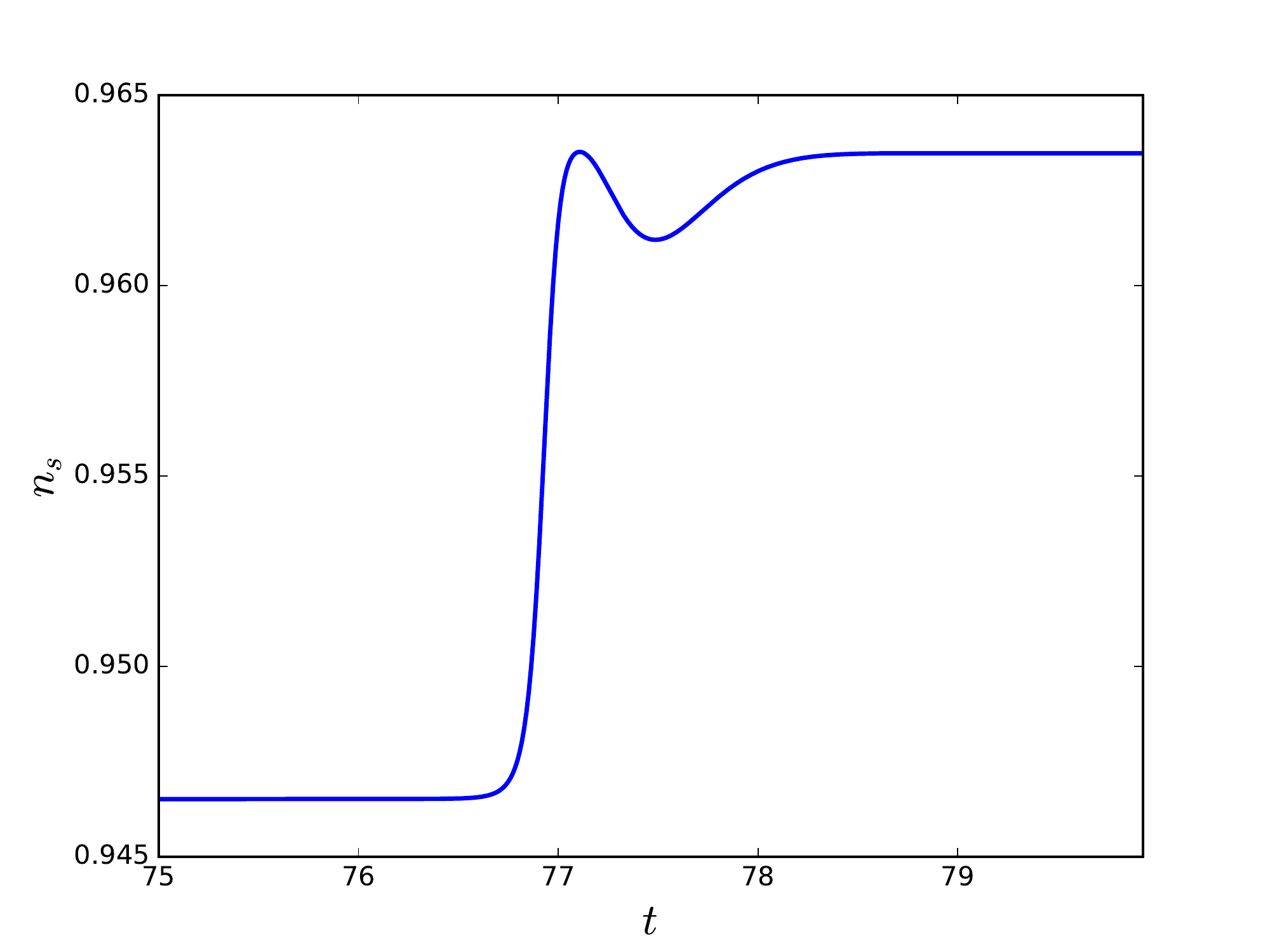}
	\includegraphics[width=0.49\textwidth]{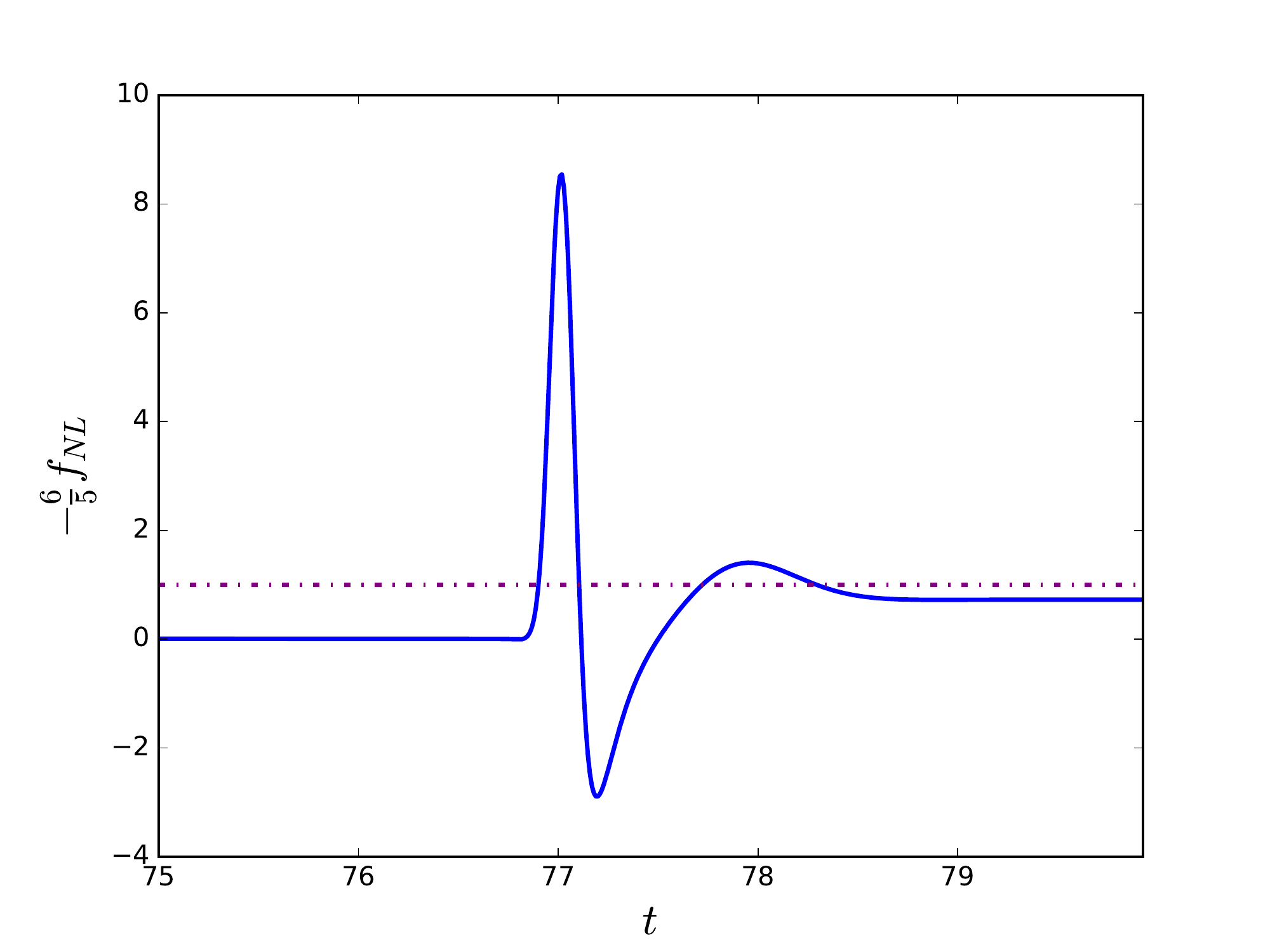}
    \includegraphics[width=0.49\textwidth]{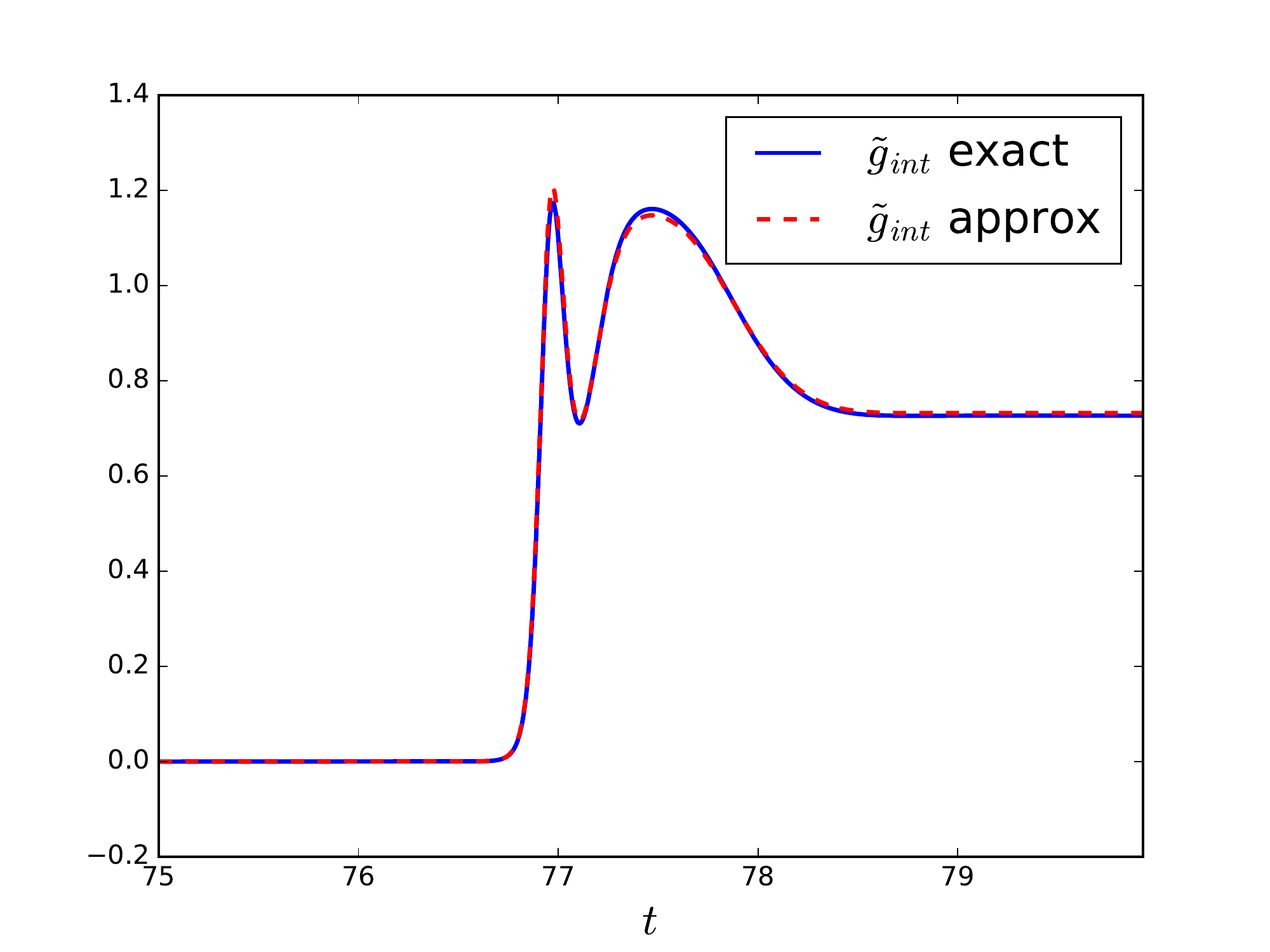}
	\caption{Same as figure \ref{fig:n2m2} but for the non-monomial potential \eqref{last} (without analytical predictions for $n_s$ and $\bv_{12e}$).} 
	\label{fig:last}
\end{figure}

Figure \ref{fig:last} shows a similar behaviour as for the previous example. Again $\fnl$ is smaller than its slow-roll prediction. The reason is still the same, the period of large $\ge$ makes $\bv_{12e}$ larger by a factor of order unity than in the slow-roll approximation and the direct consequence is that $\fnl$ is reduced by the same factor.

\chapter{Gauge-invariant perturbations at second order in two-field inflation}
\label{TvT2app}

This appendix contains the full paper \citep{TvT2}, except for the conclusions
that were used as a summary in section~\ref{summTvT2}. It was written in
collaboration with Eleftheria Tzavara.

We study the second-order gauge-invariant adiabatic and isocurvature perturbations in terms of the scalar fields 
present during inflation, 
along with the related fully non-linear spatial gradient of these quantities. 
We discuss the relation with other perturbation quantities defined in the literature. 
We also construct the 
exact cubic action of the second-order perturbations (beyond any slow-roll or super-horizon approximations and including 
tensor perturbations), both in the uniform energy-density gauge and the flat gauge in order to settle various 
gauge-related issues. 
We thus provide the tool to calculate the exact non-Gaussianity beyond slow-roll and at any scale.

\section{Introduction}
\label{introTvT2}

The concept of inflationary curvature perturbations was first invoked in order to explain the primordial fluctuations that 
source the CMB anisotropy and structure formation 
\citep{paper1,paper2,paper3,paper4}. The inflationary paradigm has been observationally tested for more than 
10 years and its prediction for an almost scale invariant 
spectrum of the first-order curvature perturbations has been verified by the data of many experiments (see for example 
\citep{Komatsu:2010fb}).
Since the definition of perturbations depends in general on the gauge choice, a gauge-invariant definition of the 
cosmological perturbations 
is of vital importance to make contact with physical observables, which are obviously gauge-invariant. 
That was investigated in detail in \citep{Bardeen:1980kt} and later in \citep{Mukhanov:1990me}. 
In the mean time the need for more observational quantities than just those based on linear perturbation theory  
has become clear, in order to break the degeneracy of the immense number of 
inflationary models. One of the most fruitful has proven to be the non-Gaussian 
characteristics of the perturbations. This has led to the development of new methods 
to study the combination of the scalar field and metric 
perturbations, the one sourcing the other, in a gauge-invariant way beyond first order.

It was not until 2003 that Malik and Wands in \citep{Malik:2003mv} defined the gauge-invariant quantity at second order that reduces to the 
curvature perturbation in the uniform energy-density gauge. In \citep{Malik:2005cy} the super-horizon equations of motion of 
these quantities were derived (but see also \citep{Noh:2003yg} for a gauge-ready formulation of the 
perturbations and their equations). Another way to deal with perturbations at second order is the space gradients approach 
first introduced in \citep{Ellis:1989jt} and specifically the gradient of the fully non-linear curvature perturbation used in 
\citep{RSvT2} and defined by the same authors in \citep{Rigopoulos:2004gr}. The advantage of the method is that when 
the space gradients are expanded to first order they are automatically gauge-invariant. Some years later the gradient of the  
curvature perturbation was redefined in a covariant way in \citep{Langlois:2005qp}. In this paper it was shown that 
when expanded to second order, this quantity reduces to the gauge-invariant curvature perturbation defined in \citep{Malik:2003mv} 
plus a gauge transformation term. 

In this paper we generalize the definition of the gauge-invariant curvature perturbation (or the gradient 
of the relevant fully non-linear quantity) in terms of the energy density to a definition in terms of the scalar fields present 
during inflation and study the consequences of this change at second order. 
Since the scalar fields are the principal 
quantities during inflation, it makes more sense to use these as a starting point, especially in the case of multiple-field 
inflation.
Our original need for such a definition was to find the 
horizon crossing contribution to the second-order curvature perturbation in terms of the first-order ones 
in the long-wavelength formalism \citep{TvT1}.
Indeed, such a definition helps to fully understand and potentially generalize the two formalisms used to compute non-Gaussianity 
during inflation, 
i.e.\ the long-wavelength formalism \citep{RSvT2,Rigopoulos:2011eq,TvT1} and the $\delta N$ formalism 
\citep{Starobinsky:1986fxa,Sasaki:1995aw,Sasaki:1998ug,Lyth:2004gb,Lyth:2005fi}, 
where instead of the energy density, the values of the fields themselves are used. 

In the case of multiple-field inflation not only an adiabatic curvature perturbation is produced, but also one or more 
isocurvature perturbations. 
We would like to provide the same type 
of study for the isocurvature perturbation as for the curvature one, using the generalized quantity defined in \citep{RSvT2}, and 
deduce from that 
the second-order gauge-invariant analogue. 
This definition of the isocurvature perturbation makes direct contact with the scalar fields during 
inflation (instead of using their pressure), which we find more useful during the period of scalar field domination of the universe. 
It has a simple physical meaning, 
that is the combination of the fields that remains orthogonal to the field trajectory, as opposed to the adiabatic perturbation that is 
parallel to the field trajectory (and proportional to the energy density).

On a related subject, Maldacena in \citep{Maldacena:2002vr} found the third-order action for the first-order 
adiabatic perturbation in a single-field dominated universe, both in the flat gauge and in the uniform energy-density gauge. 
In order to rewrite the action in a gauge-invariant form starting from the uniform energy-density gauge, he needed to introduce 
a redefinition of the first-order perturbations, hence changing their ground state. This redefinition 
corresponds to part of the second-order gauge-invariant curvature perturbation and contributes to the local non-Gaussianity. 
His work was followed by \citep{Chen:2006nt} introducing general kinetic terms, \citep{Seery:2005gb} for two fields and 
\citep{Langlois:2008qf,Arroja:2008yy,Gao:2008dt} for multiple-field models with a generalized kinetic term. 
In \citep{Seery:2005gb} the treatment of two fields in the flat gauge showed that no field redefinitions occur (see also \citep{Rigopoulos:2011eq}). 
Nevertheless, the absence of redefinitions in this case does not mean the absence of local non-Gaussianity, because the action 
was computed in terms of the scalar fields and not in terms of the adiabatic and isocurvature perturbations. This means that in the method of 
\citep{Seery:2005gb} the $\delta N$ formalism or the long-wavelength formalism is needed to compute the final non-Gaussianity, which requires 
that the slow-roll approximation is imposed at horizon crossing.

Here we generalize the above results and write the two-field action in terms of the gauge-invariant perturbations themselves, 
both in the uniform energy-density gauge and the flat gauge in order to compare the results.
We expand the calculation to include second-order perturbations and tensor modes, and study the 
various contributions that occur. Hence we derive the exact third-order action, going beyond the slow-roll or the 
super-horizon approximation. We thus provide the missing tool that will enable people to calculate non-Gaussianity, using the in-in 
formalism \citep{Weinberg:2005vy}, beyond these standard approximations used by both the long-wavelength formalism and the $\delta N$ formalism.

This paper is organized as follows. In section \ref{back} we provide the gauge-invariant definitions and conventions for the metric 
and the field perturbations, along with the description of the space-time of the universe, using the ADM formalism. 
In the first part of section \ref{gic} we study the gauge-invariant curvature and isocurvature perturbations in terms of the fields, 
while in the second part we make the connection to the fully non-linear spatial gradients of the relevant quantities. 
In the whole of section \ref{gic} we use the long-wavelength approximation to keep the calculations short and tractable, but we present 
the generalization of the results beyond this approximation in the first appendix in section~\ref{app}.  
In section 
\ref{cubic} we construct the exact cubic action, going beyond the long-wavelength approximation, to find the redefinitions of the 
perturbations and compare their contributions to the gauge-invariant 
quantities found in section \ref{gic}. To keep the main text more accessible, many of the details of the 
calculations have been moved to the appendices in section~\ref{app}. 
Finally we summarize the results at the end of section \ref{cubic}.

\section{Preliminaries}\label{back}

In this section we give the basic elements required for the calculations in this paper. We start by 
summarizing in the first subsection the ADM formalism, along with the definitions of the cosmological 
quantities, the slow-roll parameters and the field basis we use. In the second subsection we provide the conventions of cosmological perturbation theory and clarify 
different approaches in the literature.

\subsection{The ADM formalism}

We will consider a universe filled with two scalar fields with a trivial
field metric. The generalization to more fields and a non-trivial
field metric is conceptually straightforward (see \citep{RSvT2}), but involves more complicated 
expressions and calculations.   
 The energy-momentum tensor for the two fields $\varphi^A$ ($A,B = 1,2$) is
\be
T_{\mu\nu}=\gd_{AB}\partial_{\mu}\varphi^A\partial_{\nu}\varphi^B-g_{\mu\nu}
\left(\frac{1}{2}\gd_{AB}g^{\lambda\kappa}\partial_{\kappa}\varphi^A\partial_{\lambda}\varphi^B
+W\right),
\ee
where $W$ is the field potential. We will denote the homogeneous part of the fields by $\phi^A$. 
The Einstein summation convention is assumed
throughout this paper.
We shall work in the ADM formalism and write the metric $g_{\mu\nu}$ as
\be
ds^2=-\bN^2dt^2+h_{ij}(dx^i+N^idt)(dx^j+N^jdt),\label{metricexact}
\ee
where $\bN$ is the lapse function and $N^i$ the shift. The action takes the form \citep{Misner:1974qy}
\be
S=\frac{1}{2}\int\d^4x\sqrt{h}\Big[-2\bN W+\kappa^{-2}\bN^{-1}(E_{ij}E^{ij}-E^2)+\bN \bar{\Pi}^2
+\kappa^{-2}\bN R^{(3)}-\bN h^{ij}\partial_i\varphi^A\partial_j\varphi_A\Big],\label{actionexact}
\ee
where $\kappa^2\equiv8\pi G=8\pi/m_{pl}^2$, $h$ is the determinant of the space metric $h_{ij}$, $R^{(3)}$ is the intrinsic 
3-curvature, the tensor $E_{ij}$ (proportional to the extrinsic curvature $K_{ij}=-\bN^{-1}E_{ij}$) is
\be
E_{ij}=\frac{1}{2}\left(\dot{h}_{ij}-\nabla_iN_j-\nabla_jN_i\right)
\ee
and $\bar{\Pi}$ is the length of the canonical momentum of the fields 
\be
\bar{\Pi}^A=(\dvphi^A-N^i\partial_i\varphi^A)/\bN.
\ee
Variation of the action with respect to $\bar{N}$ and $N^i$ gives the energy and momentum constraints
\ba
&&\kappa^{-2}R^{(3)}-2W-\kappa^{-2}\bN^{-2}(E_{ij}E^{ij}-E^2)-\bar{\Pi}^2-h^{ij}\partial_i\varphi^A\partial_j\varphi_A=0,\label{energy}\\
&&\nabla_j\Big[\frac{1}{\bN}(E_i^j-E\delta_i^j)\Big]=\kappa^2\bar{\Pi}^A\partial_i\varphi_A,\label{momentum}
\ea
where $\nabla_j$ denotes the covariant derivative with respect to the space metric and $E$ is the trace of $E_{ij}$.

Following \citep{Maldacena:2002vr} we decompose the space metric as
\be
h_{ij}=a(t)^2e^{2\alpha(t,x)}e^{\gamma_{ij}(t,x)},\qquad \partial_i\gamma^{ij}=0,\qquad\gamma_i^i=0.\label{metric}
\ee
From now on contravariant tensors should be understood as $T^j=\eta^{ij}T_i$, where $\eta^{ij}=\mathrm{diag}(1,1,1)$, since 
in the calculations we are showing we have already taken into account explicitly the $h^{ij}$ part of the initial 
contravariant tensors. 
The generalized Hubble parameter is defined as 
\be
\bar{H}\equiv\frac{E}{3\bar{N}}.
\ee 
We use the bar for the lapse function, the Hubble parameter and the canonical momentum to distinguish these fully non-linear quantities 
from their background values $N(t)$, $H(t)=\dot{a}/(a N)$ and $\Pi(t)=\dphi/N$, respectively. 
In this paper we will use as time variable the number of e-folds, meaning that $\dot{a}=a$, so that the background 
value of the lapse function $N(t)$ is just $1/H(t)$.

The background field equation and the background Einstein equations are
\ba
\dot{\Pi}^A= -3HN\Pi^A-N W^{A},\qquad \dot{H}=-\frac{\kappa^2}{2}N\Pi^2,\qquad H^2=\frac{\kappa^2}{3}\lh\frac{\Pi^2}{2}+W\rh,
\label{fieldeqTz}
\ea
with $W_{A}\equiv\der W/\der\gf^A$. 
We construct an orthonormal basis ${e_m^A}$ in field space, consisting of $e_1^A\equiv\Pi^A/\Pi$, parallel to the field velocity, 
and $e_2^A$ parallel to the part of the field acceleration perpendicular to the field velocity  
 \citep{GNvT2}. The $m=1$ component of physical quantities describes the single-field (adiabatic) part, while 
the $m=2$ component captures the multiple-field (isocurvature) effects. 
One can show that for the two-field case the basis vectors are related through 
\citep{TvT1}
\be
\epsilon_{AB}e_1^Ae_2^B=-1,
\label{antisym}
\ee 
where $\ge^{AB}$ is the antisymmetric tensor. The background slow-roll parameters then take the form 
\ba
\ge(t)  &\equiv& - \frac{\dot{H}}{NH^2},
\quad
\getpa(t) \equiv \frac{e_{1A}\dot{\Pi}^A}{NH\Pi},
\quad 
\getpe(t) \equiv \frac{e_{2A}\dot{\Pi}^A}{NH\Pi},\quad
\chi(t) \equiv \frac{W_{22}}{3H^2}+\ge+\getpa,\nn\\
\gx^\parallel(t) &\equiv& \frac{e_{1A}\ddot{\Pi}^A}{N^2H^2\Pi}-\frac{\dot{N}}{N^2H}\getpa,
\qquad\!\!
\gx^\perp(t) \equiv \frac{e_{2A}\ddot{\Pi}^A}{N^2H^2\Pi}-\frac{\dot{N}}{N^2H}\getpe,
\label{srvar}
\ea
where $W_{mn}\equiv e_m^Ae_n^BW_{,AB}$. 
Throughout this paper the indices $m, n$ will indicate components in the
basis defined above, taking the values 1 and 2, while $i,j$ are spatial indices and $A,B$ are indices of the original fields. 
In a slow-roll approximation one can think of $\getpa$ being related to $W_{11}$, $\getpe$ to $W_{21}$ and 
$\chi$ to $W_{22}$.
The $\xi$ parameters are second-order slow-roll parameters (in a slow-roll approximation they are related to 
the third derivatives of the potential). However, we emphasize that we have not made any slow-roll approximations; the above 
quantities should be viewed as short-hand notation and can be large.
We also give the time derivatives of the background slow-roll parameters and of the unit vectors,
\ba
&&\dot{e}_1^A= NH\getpe e_2^A,\qquad\qquad \dot{e}_2^A=- NH\getpe e_1^A,
\qquad\qquad \dot{\ge} = 2 NH \ge ( \ge + \get^\parallel ),\nn\\
&& \dot{\get}^\parallel\!= NH\!\lh\gx^\parallel \!+ (\get^\perp)^2 + (\ge - \get^\parallel) \get^\parallel\rh, \qquad\qquad
\dot{\get}^\perp\! =NH\!\lh \gx^\perp \!+ (\ge - 2\get^\parallel) \get^\perp\rh,\nn\\
&&\dot{\gc} =NH\lh \ge \get^\parallel + 2 \ge \gc - (\get^\parallel)^2
+ 3 (\get^\perp)^2 + \gx^\parallel + \frac{2}{3} \get^\perp \gx^\perp
+ \frac{\sqrt{2\ge}}{\gk}\frac{W_{221}}{3H^2}\rh,\nn\\
&&\dot{\xi}^\parallel=NH\lh -\frac{\sqrt{2\ge}}{\gk}\frac{W_{111}}{H^2}+2\getpe\gx^\perp+2\ge\gx^\parallel-3\gx^\parallel+9\ge\getpa
+3(\getpe)^2+3(\getpa)^2\rh,\nn\\
&&\dot{\xi}^\perp=NH\lh -\frac{\sqrt{2\ge}}{\gk}\frac{W_{211}}{H^2}-\getpe\gx^\parallel+2\ge\gx^\perp-3\gx^\perp+9\ge\getpe+6\getpe\getpa-3\getpe\chi\rh,
\label{dere}
\ea
where $W_{lmn}\equiv e_m^Ae_n^Be_l^CW_{,ABC}$.

\subsection{Second-order perturbations and gauge transformations}

In the context of perturbation theory around an homogeneous background any quantity 
$\bar{A}$ can be 
decomposed into an homogeneous part and an infinite series of perturbations as 
\be
\bar{A}(t,\vc{x})=A(t)+A_{(1)}(t,\vc{x})+\frac{1}{2}A_{(2)}(t,\vc{x})+\dots, 
\ee
where the subscripts in the parentheses denote the order of the perturbation. Up to first order the scalar part of the 
space metric element of (\ref{metric}) is equal to
\be
h_{ij}=a^2(t)(1+2\ae)\delta_{ij}.\label{first}
\ee
When one wants to expand perturbation theory up to second order there are two choices found in the literature: either expand (\ref{first}) 
as Malik and Wands do in \citep{Malik:2003mv} to find
\be
h_{ij}=a^2(t)(1+2\ae+\aee)\delta_{ij} 
\ee
or expand directly the space part of (\ref{metric}) as for example Lyth and Rodriguez do in 
\citep{Lyth:2005du} to find
\be
h_{ij}=a^2(t)(1+2\ae+2\ae^2+\aee)\delta_{ij}.\label{second}
\ee
We will take this second approach and use the exponent of the perturbation in our calculations.

Since perturbations depend on the gauge choice we make, we need to construct quantities that are invariant 
under gauge transformations. Under an arbitrary second-order coordinate transformation
\be
\widetilde{x}^{\mu}=\hat{x}^{\mu}+\beta_{(1)}^{\mu}+\frac{1}{2}\lh\beta_{(1),\nu}^{\mu}\beta^{\nu}_{(1)}+\beta_{(2)}^{\mu}\rh,
\ee
the perturbations of a tensor transform as \citep{Bruni:1996im}
\ba
&&\widetilde{A}_{(1)}=\hat{A}_{(1)}+L_{\beta_{(1)}}A,\nn\\
&&\widetilde{A}_{(2)}=\hat{A}_{(2)}+L_{\beta_{(2)}}A+L_{\beta_{(1)}}^2A+2L_{\beta_{(1)}}\hat{A}_{(1)},\label{trans}
\ea
where $L_{\beta}$ is the Lie derivative along the vector $\beta$
\be
\lh L_{\beta}A\rh^{\mu_1\mu_2...}_{\nu_1\nu_2...}=\beta^\kappa\partial_\kappa A^{\mu_1\mu_2...}_{\nu_1\nu_2...}
-\partial_\kappa\beta^{\mu_1}
A^{\kappa\mu_2...}_{\nu_1\nu_2...}-\dots +\partial_{\nu_1}\beta^{\kappa}A^{\mu_1\mu_2...}_{\kappa\nu_2...}+\dots .
\ee
Note here that spatial gradients, having vanishing background 
values, are automatically gauge-invariant at first order, while at second order they transform as
\be
\partial_i\widetilde{A}_{(2)}=\partial_i\hat{A}_{(2)}+2L_{\beta_{(1)}}\partial_i\hat{A}_{(1)}.\label{zerotr}
\ee

\section{Super-horizon gauge transformations}\label{gic}

In this section we derive first and second-order super-horizon gauge-invariant combinations.  
We study these during an inflationary period and thus, though we start 
from the energy-density definitions of the perturbations, we naturally end up with field definitions for the gauge-invariant 
perturbations. 
Our goal is to find the second-order adiabatic and isocurvature perturbations in terms of the first-order ones and 
the slow-roll parameters. 

In this section we restrict ourselves to super-horizon calculations for simplicity, though in the next section we will 
abandon this approximation and study the full action of the cosmological perturbations. 
However, in the first appendix in section~\ref{app} we present the generalization of the results of this section 
beyond the long-wavelength 
approximation. We note that the long-wavelength (or super-horizon) approximation 
is equivalent to the zeroth order space gradient approximation and is valid once the decaying mode has disappeared (which happens 
rapidly if slow-roll holds during horizon exit), even if there is a subsequent non slow-roll phase.

In the super-horizon regime one can choose to work in the time-orthogonal gauge where $N^i=0$ 
(proof for that choice is given in the next section) and employ the long-wavelength approximation to simplify calculations. 
The latter boils down to ignoring second-order spatial derivatives when compared to time derivatives. 
As a consequence the traceless part of the extrinsic curvature quickly decays and 
can be neglected \citep{Salopek:1990re}. Hence the space part of the metric can be described by 
\be
h_{ij}=a(t)^2e^{2\alpha(t,x)}\delta_{ij}.
\ee
The field and Einstein equations in that case are identical to (\ref{fieldeqTz}), but now the quantities involved are fully non-linear. 
Additionally the momentum constraint (\ref{momentum}) can be written as \citep{RSvT2}
\be
\partial_i\bar{H}=-\frac{\kappa^2}{2}\bar{\Pi}_A\partial_i\varphi^A.\label{fieldeqsu}
\ee 

\subsection{Gauge-invariant quantities}

The well-known first-order adiabatic gauge-invariant curvature perturbation has the form 
\be
\zeta_{1(1)}\equiv\ae-\frac{NH}{\dot{\rho}}\rho_{(1)},\label{z110}
\ee
where $\rho$ is the energy density. The subscript without parentheses corresponds to the first component in our basis, 
which is exactly the adiabatic component, while the 
subscript between parentheses denotes the order in the perturbation series. 
Notice that in the literature it is common to work with cosmic time, i.e.\ $N=1$, while the space part of the metric is 
decomposed using a quantity $\psi=-\alpha$ (not to be confused with the $\psi$ introduced in the appendices), 
so that the first-order curvature perturbation becomes in that case 
$-\zeta_{1(1)}=\psi_{(1)}+(H/\dot{\rho})\rho_{(1)}$.
Here we choose to work with the number of e-folds as time variable so that the first-order curvature perturbation is
\be
\zeta_{1(1)}=\ae-\frac{\rho_{(1)}}{\dot{\rho}}.\label{z11}
\ee 

The gauge-invariant combination (\ref{z110}) is calculated via the requirement that it coincides with the curvature 
perturbation $\zeta_{1(1)}=\tilde{\alpha}_{(1)}$ in the uniform energy-density gauge where $\tilde{\rho}_{(1)}=0$. 
 From now on tilded quantities 
will denote the uniform energy-density gauge, 
while hatted quantities will denote the flat gauge. One has to use the gauge 
transformations (\ref{trans}) for a scalar (here the energy density and the logarithm of the space dependent scale factor 
$\alpha$) and require that in the 
uniform energy-density gauge the first-order energy perturbation is zero (for details see the appendix). That way one can 
determine the first-order time  
shift and hence find the gauge-invariant combination corresponding to the curvature perturbation. Notice 
that in the flat gauge, i.e.\ $\hat{\alpha}_{(1)}=0$, $\zeta_{1(1)}=-\hat{\rho}_{(1)}/\dot{\rho}$. 

Keeping in mind the expansion (\ref{second}), one can repeat the above considerations at second order. 
We find that for super-horizon scales (where we neglect second-order space derivatives when compared to second-order time derivatives) 
the second-order gauge-invariant adiabatic perturbation takes the form (see the appendix)
\be
\frac{1}{2}\zeta_{1(2)}\equiv\frac{1}{2}\tilde{\alpha}_{(2)}
=\frac{1}{2}\alpha_{(2)}-\frac{1}{2}\frac{\rho_{(2)}}{\dot{\rho}}
+\frac{\dot{\rho}_{(1)}\rho_{(1)}}{\dot{\rho}^2}
-\frac{\rho_{(1)}}{\dot{\rho}}\dot{\alpha}_{(1)}
-\frac{1}{2}\frac{\rho_{(1)}^2}{\dot{\rho}^2} \frac{\ddot{\rho}}{\dot{\rho}}.\label{defmalik}
\ee
If we chose the second gauge to be flat, i.e.\ $\ha_{(i)}=0$, we find
\be
\frac{1}{2}\zeta_{1(2)}=\frac{1}{2}\tilde{\alpha}_{(2)}=-\frac{1}{2}\frac{\hr_{(2)}}{\dot{\rho}}
+\dot{\zeta}_{1(1)}\zeta_{1(1)}+\frac{1}{2}\frac{\hr_{(1)}^2}{\dot{\rho}^2} \frac{\ddot{\rho}}{\dot{\rho}}.\label{36}
\ee

During inflation we find it more useful to work directly with the fields and not their energy density, since both 
the long-wavelength formalism and the $\delta N$ formalism make use of the field values to compute $f_\mathrm{NL}$. Using the fields, 
the first-order adiabatic perturbation becomes
\be
\zeta_{1(1)}=\tilde{\alpha}_{(1)}=\alpha_{(1)}-\frac{H}{\Pi}e_{1A}\varphi^A_{(1)},\label{38}
\ee
since the energy-density constraint $\tilde{\rho}_{(1)}=0$ is equivalent to $e_{1A}\tvarphi^A_{(1)}=0$. The detailed calculation 
is shown in the second appendix in section~\ref{app}. Notice that in (\ref{38}) we have kept the lapse 
function $N$ arbitrary, as we will also do in all definitions hereafter, but in our calculations 
$H/\Pi$ is just $1/\dphi$ for the choice $N=1/H$.

The second-order calculation turns out to be more complicated. The details are given in the second appendix. Here we give the result for 
the gauge-invariant adiabatic perturbation in the uniform energy-density gauge and in the flat gauge:
\ba
\frac{1}{2}\zeta_{1(2)}=\frac{1}{2}\tilde{\alpha}_{(2)}&=&\frac{1}{2}\hat{Q}_{1(2)}
+\frac{\ge+\getpa}{2}\lh\zeta_{1(1)}^2-\zeta_{2(1)}^2\rh-\getpe\zeta_{1(1)}\zeta_{2(1)}+\dot{\zeta}_{1(1)}\zeta_{1(1)}\nn\\
&&-\partial^{-2}\partial^i\lh\dot{\zeta}_{2(1)}\partial_i\zeta_{2(1)}\rh,\label{z1ft}
\ea
where we introduced the auxiliary quantities
\be
Q_{m(i)}\equiv-\frac{H}{\Pi}e_{mA}\varphi^A_{(i)}\
\ee
and the new combination \citep{RSvT2}
\be
\zeta_{2(1)}\equiv-\frac{H}{\Pi}e_{2A}\varphi^A_{(1)}=Q_{2(1)},\label{z21}
\ee
that represents the isocurvature perturbation to first order. 
Usually the isocurvature perturbation is described in terms of the gradient of the pressure of the matter content of the universe, 
as for example in \citep{Langlois:2005qp}. Here we choose to characterize it in terms of the fields themselves and the vector $e_{2A}$. 
The latter indicates we are dealing with a purely multiple-field effect and hence it is an appropriate quantity to use 
during the inflationary 
period to describe the non adiabatic perturbations. Starting from the long-wavelength definition of the pressure 
$\bar{p}=\bar{\Pi}^2/2-W$, one can show that the gradient of the isocurvature perturbation defined in \citep{Langlois:2005qp} is equal to
\be
\Gamma_i\equiv\partial_i\bar{p}-\frac{\dot{\bar{p}}}{\dot{\bar{\rho}}}\partial_i\bar{\rho}=-2\bar{\eta}^{\perp}\bar{\Pi}^2\zeta_{2i},
\ee
where $\zeta_{2i}$ is the fully non-linear gradient of the isocurvature perturbation 
(for more details see the next subsection) and $\bar{\eta}^{\perp}$ the fully non-linear generalization 
of $\getpe$, i.e.\ as it is defined in (\ref{srvar}) but with barred quantities \citep{RSvT2}. 
So our definition of $\zeta_2$ agrees with the pressure definition of the isocurvature perturbation. 
The next logical step would be to define the second-order isocurvature perturbation $\zeta_{2(2)}$. However, 
the above equation 
shows that there is a non-trivial relation between $\zeta_2$ and the pressure $p$, involving the non-linear quantities 
$\bar{\eta}^{\perp}$ and $\bar{\Pi}$, 
which makes a derivation using the methods of this subsection rather complicated. 
For that reason we prefer to find $\zeta_{2(2)}$ in an easier way in the next subsection 
using gradients.

Notice that unlike in the original definition of $\zeta_{1(2)}$ in terms of $\rho_{(2)}$, a non-local term appears in (\ref{z1ft}) 
when one uses the fields 
instead of the energy density, because of (\ref{alp_flat}). The time derivatives of the 
fully non-linear gradients of the perturbations (see next subsection) were found in \citep{RSvT2}. Expanding  
to first order these yield
\be
\dot{\zeta}_{1(1)}=2\getpe \zeta_{2(1)}
\ee   
for the adiabatic perturbation and
\be 
\dot{\zeta}_{2(1)}=-\chi \zeta_{2(1)}
\ee    
for the isocurvature perturbation, the latter valid only in the slow-roll regime.
Then we find
\be
\frac{1}{2}\zeta_{1(2)}=\frac{1}{2}\tilde{\alpha}_{(2)}=\frac{1}{2}\hat{Q}_{1(2)}
+\frac{\ge+\getpa}{2}\zeta_{1(1)}^2-\frac{\ge+\getpa-\chi}{2}\zeta_{2(1)}^2+\getpe\zeta_{1(1)}\zeta_{2(1)},
\ee
i.e.\ without a non-local term.
However, we will not use the slow-roll approximation in this paper.

\subsection{The gradient of the perturbations}

As an alternative to the $\zeta_m$ defined in the previous section, one can use the gradient quantity $\zeta_{1i}$ 
along with the isocurvature analogue $\zeta_{2i}$, both defined in \citep{RSvT2} and later in 
\citep{Langlois:2005qp} in a covariant way, to construct a gauge-invariant quantity.  
These gradient quantities (not gauge-invariant to all orders) are given by
\be
\zeta_{mi}=\delta_{m1}\partial_i\alpha-\frac{\bar{H}}{\bar{\Pi}}\bar{e}_{mA}\partial_i\varphi^A\label{defgrad},
\ee
where now $\bar{e}_{mA}$ represents the fully non-linear super-horizon version of the orthonormal basis vectors, e.g.\  
$\bar{e}_{1A}=\bar{\Pi}_A/\bar{\Pi}$, with $\bar{\Pi}_A=\dot{\varphi}_A/\bar{N}$ since we are working in the super-horizon regime. 
Notice that the basis vectors still obey (\ref{antisym}) as was shown in 
\citep{TvT1}. 
$\zeta_{mi}$ is by construction gauge-invariant at first order, since it has no background value: it is just the gradient of 
the gauge-invariant $\zeta_{m(1)}$ defined before.

Expanding to second order we find for the adiabatic perturbation
\be
\frac{1}{2}\zeta_{1i(2)}=\frac{1}{2}\partial_i\lh\alpha_{(2)}+Q_{1(2)}\rh
-\frac{1}{\dphi^2}\dvphi^A_{(1)}\partial_i\varphi_{A(1)}
-2\frac{1}{\dphi}e_{1A}\dvphi^A_{(1)} \partial_iQ_{1(1)}.\label{akoma}
\ee
In the uniform energy-density gauge this gives (see the second appendix
in section~\ref{app})
\be
\frac{1}{2}\tilde{\zeta}_{1i(2)}=\frac{1}{2}\partial_i\tilde{\alpha}_{(2)},\label{zai}
\ee
while in the flat gauge where $\partial_i\ha=0$ and $\zeta_{1(1)}=\hat{Q}_{1(1)}$, we find 
\ba
\frac{1}{2}\hat{\zeta}_{1i(2)}
& = & \partial_i\Bigg[\frac{1}{2}\hat{Q}_{1(2)}+\frac{\ge+\getpa}{2}\lh\zeta_{1(1)}^2-\zeta_{2(1)}^2\rh-\getpe\zeta_{1(1)}\zeta_{2(1)}
+\dot{\zeta}_{1(1)}\zeta_{1(1)}\Bigg]-\dot{\zeta}_{2(1)}\partial_i\zeta_{2(1)}\nn\\
&&-\zeta_{1(1)}\partial_i\dot{\zeta}_{1(1)},
\label{z1fg}
\ea
where we used the basis completeness relation and (\ref{difzflat}) to rewrite the terms. 
$\tilde{\zeta}_{1i(2)}$ in the uniform energy-density gauge (\ref{zai}) coincides with the gradient of the 
gauge-invariant second-order adiabatic perturbation. 
However, by comparing (\ref{z1ft}) to (\ref{z1fg}) we see 
that in the flat gauge $\hat{\zeta}_{1i(2)}$ is the gradient 
of the gauge-invariant curvature perturbation $\zeta_{1(2)}$ expressed in the flat gauge plus a new non-local term. 
This is in agreement 
with the findings in \citep{Langlois:2005qp}. This new term is nothing else but the
gauge transformation of $\zeta_{1i(2)}$. A quantity with zero background value as $\zeta_i$ is, transforms as (\ref{zerotr}). 
One can check, using the gauge transformations (\ref{trans}) for $\rho$ and for $\alpha$ 
and requiring that $\tilde{\rho}_{(1)}=0$, 
that starting from a flat gauge and transforming to the uniform energy-density gauge the time shift is 
$T_{(1)}=\zeta_{1(1)}$ (see the first appendix), so that
\be
\frac{1}{2}\tilde{\zeta}_{mi(2)}=\frac{1}{2}\hat{\zeta}_{mi(2)}+\zeta_{1(1)}\dot{\zeta}_{mi(1)}.\label{gat}
\ee

Next we try to find the second-order gauge-invariant part of the isocurvature perturbation $\zeta_{2i}$ by expanding (\ref{defgrad}),
\be
\frac{1}{2}\zeta_{2i(2)}=\frac{1}{2}\partial_iQ_{2(2)}
+\frac{1}{\dphi^2}\ge^{BA}\dvphi_{B(1)}\partial_i\varphi_{A(1)}-2\frac{1}{\dphi}e_{1A}\dvphi^A_{(1)} \partial_i\zeta_{2(1)},
\ee
or equivalently,
\be
\frac{1}{2}\zeta_{2i(2)}=\frac{1}{2}\partial_iQ_{2(2)}
-\frac{1}{\dphi}e_{2A}\dvphi^A_{(1)} \partial_iQ_{1(1)}-\frac{1}{\dphi}e_{1A}\dvphi^A_{(1)} \partial_i\zeta_{2(1)},\label{another}
\ee
where we used (\ref{antisym}) to express $\vc{e}_{2}$ in terms of $\vc{e}_1$. 
The first-order uniform energy constraint $\partial_i\tilde{Q}_{1(1)}=0$ alone implies
\be
\frac{1}{2}\tilde{\zeta}_{2i(2)}=\partial_i\Bigg[\frac{1}{2}\tilde{Q}_{2(2)}-\frac{\getpe}{2}\zeta_{2(1)}^2\Bigg]
+\dot{\zeta}_{1(1)}\partial_i\zeta_{2(1)}\label{z2u},
\ee
where we used (\ref{difzue}) for the two last terms in (\ref{another}). 
For the flat gauge $\partial_i\ha=0$ we find using (\ref{difzflat})
\ba
\frac{1}{2}\hat{\zeta}_{2i(2)}&=&\partial_i\Bigg[\frac{1}{2}\hat{Q}_{2(2)}
+\frac{\getpe}{2}\lh\zeta_{1(1)}^2-\zeta_{2(1)}^2\rh
+(\ge+\getpa)\zeta_{1(1)}\zeta_{2(1)}+\dot{\zeta}_{2(1)}\zeta_{1(1)}\Bigg]+\dot{\zeta}_{1(1)}\partial_i\zeta_{2(1)}\nn\\
&&-\zeta_{1(1)}\partial_i\dot{\zeta}_{2(1)}\label{z2f}.
\ea
We notice that the term in the second line corresponds again to a gauge transformation familiar from the curvature perturbation case 
studied earlier (\ref{gat}). In the first appendix we verify that the rest of the expression is a gauge-invariant quantity 
corresponding to the one in (\ref{z2u}). Indeed this expression is gauge-invariant beyond the long-wavelength approximation 
as shown in that appendix.

We conclude that the gradients of the perturbations are in some sense equivalent to the perturbations themselves, since both 
allow for the definition  of gauge-invariant second-order adiabatic and isocurvature quantities. However, since the 
gradients are defined using fully non-linear quantities, their equations of motion can be treated 
more easily, as was shown in \citep{RSvT2}.

\section{The cubic action}\label{cubic}

An alternative way to calculate the second-order gauge-invariant quantities and reconsider their meaning, is to compute the third-order 
action for the interacting fields. 
Maldacena \citep{Maldacena:2002vr} was the first to perform that calculation for a single field, in the uniform 
energy-density gauge. 
In this way he managed to find the cubic interaction terms due to non-linearities of the Einstein action as well as 
of the field potential, which among 
other consequences change the ground state of the adiabatic perturbation $\zeta_{1(1)}$. 
This change can be quantified through a redefinition of the 
form \citep{Maldacena:2002vr}
\be
\zeta_{1(1)}=\zeta_{1c(1)}+\frac{\ge+\getpa}{2}\zeta_{1(1)}^2,
\ee
where $\zeta_{1c}$ is the redefined perturbation. One sees that the correction term of the redefinition coincides with the 
surviving quadratic term of the single-field limit of the transformation (\ref{z1ft}), taking into account 
that the super-horizon adiabatic perturbation is constant in that case.
In \citep{Maldacena:2002vr} the curvature perturbation was considered a first-order quantity, 
while the second-order curvature perturbation was not taken into account, since its contribution in the uniform 
energy-density gauge 
is trivial: it introduces a redefinition of the form $\zeta_{1(1)}+\zeta_{1(2)}/2=\zeta_{1c(1)}$ 
(for proof, see later in this section). 
Seery and Lidsey \citep{Seery:2005gb} performed the same calculation for the multiple-field case in the flat gauge 
in terms of the scalar fields $\varphi^A$ and not of the adiabatic and isocurvature perturbations $\zeta_m$. They found 
no redefinitions, but as mentioned before their results would have to be supplemented 
by the $\delta N$ formalism (with its associated slow-roll approximation at horizon-crossing) to say anything about the non-Gaussianity 
of the gauge-invariant perturbations $\zeta_m$.

In this section we generalize the above calculations to second order in the expansion of the curvature perturbation in both the uniform 
energy-density gauge and the flat gauge. Doing so we compute the full form of the third-order action. The latter not only 
consists of the cubic interactions of the first-order curvature perturbations, but also of lower order interaction terms of the second 
order quantities. We first perform the calculation relevant to the first-order quantities and then add the second-order 
effects. In this section we only present the scalar part of the action, but in the appendices the 
tensor part can be found as well. We emphasize that in this section we no longer make the long-wavelength approximation, so that 
the results are valid at any scale.

\subsection{The second-order action}

We start by performing our calculation in the gauge $e_{1A}\tilde{\varphi}^A_{(1)}=0$. This constraint reduces to the 
uniform energy-density 
gauge outside the horizon, which is why we will continue to refer to the tilded gauge as the uniform energy-density gauge. 
From now on we drop the explicit 
subscript $(1)$ on first-order quantities. We will keep this part brief since its results are 
already known, but we give the basic elements of the calculation in the third
appendix in section~\ref{app}. 
The second-order action takes the form 
\ba
\tilde{S}_2&=&\!\!\int\!\! \d^4x L_2\nn\\
&=&\!\!\int\!\! \d^4x\ \ge \Bigg\{\!\!-a\frac{1}{H}\Big( (\partial\zeta_1)^2+(\partial\zeta_2)^2\Big)
 +a^3H\Big(\dot{\zeta}_1^2+\dot{\zeta}_2^2-4\getpe \dot{\zeta}_1\zeta_2
+2\chi \dot{\zeta}_2\zeta_2\Big)\label{s2}\\
&&\qquad\!\!+a^3H\Big(\!\sqrt{\frac{2\ge}{\kappa}}\frac{W_{221}}{3H^2}-2\ge^2-(\getpa)^2+3(\getpe)^2+\frac{2}{3}\getpe\gx^\perp
-3\ge(\getpa-\chi)+2\getpa\chi\Big) \zeta_2^2\Bigg\}\nn,
\ea
where $L_2$ is the second-order Lagrangian. 
While we have started from an action describing the evolution of the fields $\tilde{\alpha}$ and $e_{2A}\tvarphi^A$ we have now 
constructed an action 
in terms of the adiabatic and isocurvature perturbations $\zeta_1$ and $\zeta_2$.
The equations of motion that $\zeta_1$ and $\zeta_2$ obey are ($\delta L_2/\delta\zeta_m$ being a short-hand notation for the 
relevant variations of the Lagrangian)
\ba
\frac{\delta L_2}{\delta\zeta_1}&=&-2a^3\ge H\Big[\ddot{\zeta}_1+\lh3+\ge+2\getpa\rh \dot{\zeta}_1-2\getpe \dot{\zeta}_2
+2(-\gx^\perp-2\ge\getpe-3\getpe)\zeta_2\Big]+2a\frac{\ge}{H}\partial^2\zeta_1\nn\\
&=&-\frac{\d}{\d t}(2a^3H\partial^2\lambda)+2a\frac{\ge}{H}\partial^2\zeta_1=0,\label{eq}\\
\frac{\delta L_2}{\delta\zeta_2}&=&-2a^3\ge H\Big[\ddot{\zeta}_2+\!\lh3+\ge+2\getpa\rh \dot{\zeta}_2+2\getpe \dot{\zeta}_1+(\gx^\parallel+2\ge^2+4\ge\getpa+3\chi)\zeta_2
  \Big]\!+\!2a\frac{\ge}{H}\partial^2\zeta_2\nn\\
&=&0,\nn
\ea
where $\partial^2\lambda=\ge\dot{\zeta}_1-2\ge\getpe\zeta_2$ (for the reason of introducing $\lambda$ see the appendix).
Thus we have found the evolution equations for the first-order adiabatic and isocurvature perturbations. Their super-horizon limit 
coincides with the equations derived in \citep{RSvT2} for the gradient of the perturbations, since up to first 
order $\zeta_{mi(1)}=\partial_i\zeta_{m(1)}$. 
One can show that the first-order energy constraint, which outside the horizon reduces to 
\be
\dot{\zeta}_1-2\getpe\zeta_2=0,\label{mc}
\ee
is the first integral of the super-horizon part of the first equation of (\ref{eq}), i.e.\ without the space gradient. 
In fact it was shown in \citep{RSvT4} that this is the case at all orders. 
In the same paper it was found that assuming the slow-roll limit, $\dot{\zeta}_2=-\chi\zeta_2$ is the super-horizon first integral of 
the equation for $\zeta_2$, which can be easily verified. 

While working in the flat gauge we find the same action (\ref{s2}) (see the appendix). So the curvature perturbations 
$\zeta_m$ 
satisfy to first order the same equations in both gauges as expected, due to the gauge invariance of $\zeta_m$ 
(or equivalently the gauge invariance of the action).

\subsection{The third-order action}

In this section we compute the third-order action. 
Again we present only the final results, while in the fourth appendix in section~\ref{app} we give the intermediate steps of the 
calculation. In the same appendix we also give the tensor-scalar part of the action. 
The scalar cubic action in the uniform energy-density gauge due to the first-order perturbations $\zeta_m$ takes the form
\ba
\tilde{S}_{3(1)}\!&=&S_{3(1)}-\int\d^4x\frac{\delta L_2}{\delta\zeta_m}f_m
\ea
with 
\ba
&&f_1=\frac{\ge+\getpa}{2}\zeta_1^2-\getpe\zeta_1\zeta_2
+\dot{\zeta}_1\zeta_1-\frac{1}{4a^2H^2}\Big((\partial\zeta_1)^2
-\partial^{-2}\partial^i\partial^j(\partial_i\zeta_1\partial_j\zeta_1)\Big)
\nn\\
&&\qquad
+\frac{1}{2}\Big(\partial^i\lambda\partial_i\zeta_1-\partial^{-2}
\partial^i\partial^j(\partial_i\lambda\partial_j\zeta_1)\Big),\nn\\
&&f_2=(\ge+\getpa)\zeta_1\zeta_2+\dot{\zeta}_2\zeta_1+\frac{\getpe}{2}\zeta_1^2\label{rede1}.
\ea
The exact form of $S_{3(1)}$ can be found in the appendix or equivalently it is the cubic part of (\ref{fin}). 
The reason for introducing $S_{3(1)}$ without the tilde will become clear below. 

The terms proportional to 
$\delta L_2/\delta\zeta_m$, i.e.\ the first-order equations of motion, 
can be removed  by a redefinition of $\zeta_m$
\citep{Maldacena:2002vr} and lead to a change in the ground state of the perturbations. This works as follows.  
The cubic terms of the action (i.e.\ $\tilde{S}_{3(1)}$) are not affected by the redefinition, because the redefinition always 
involves terms proportional to $\zeta_m^2$, which would give quartic and not cubic corrections. 
It is only the second-order 
terms (i.e.\ $\tilde{S}_2$) that change. 
Indeed one can show that under a redefinition of the form $\zeta_m=\zeta_{mc}+f_m$, the second-order action changes as 
$S_2=S_{2c}+(\delta L_2/\delta\zeta_m)f_m$. These new terms cancel out the relevant terms coming from the cubic action 
(remember that the total action up to cubic order is the sum of the second and third-order action) and we are left 
with
\be
\tilde{S}_{3(1)}=S_{3(1)}(\zeta_{mc}).\label{ss3}
\ee

If we repeat the same calculations for the flat gauge (see the appendix), 
performing several integrations by part, we find that 
\be
\hat{S}_{3(1)}=S_{3(1)}(\zeta_{m}).
\ee 
This is a consequence of the action staying invariant under a gauge transformation. 
Nevertheless if one associates the redefinition appearing in the uniform energy-density gauge to a change in the ground state 
of $\zeta_m$, it would mean that directly after horizon crossing, when super-horizon effects have not yet 
been switched on, the second-order contribution to $\zeta_1$ would be zero for the flat gauge and non-zero for the uniform energy-density 
gauge. In terms of non-Gaussianity, this can be restated as: the non-Gaussianity present after horizon-crossing is different for the 
two gauges. 
Indeed if one was to calculate the three-point functions for the above action, one would need to perform 
two steps. First, change to the interaction picture, where it can be proved that the interaction Hamiltonian up to and 
including cubic order is just $H_{int}=-L_{int}$, where $L_{int}$ are the cubic terms of the Lagrangian, and compute the 
expectation value $\langle\zeta_c\zeta_c\zeta_c\rangle$ as in \citep{Maldacena:2002vr}. Second, take into account that the 
fields have been redefined as $\zeta=\zeta_c+\lambda\zeta_c^2$. Then the three-point correlation function can be 
written as
\be
\langle\zeta\zeta\zeta\rangle=\langle\zeta_c\zeta_c\zeta_c\rangle
+2\lambda[\langle\zeta_c\zeta_c\rangle
\langle\zeta_c\zeta_c\rangle+\mathrm{cyclic}].\label{cyclic}
\ee
These new terms, products of the second-order correlation functions, are only present in the uniform energy-density gauge and not in the flat gauge,
if we restrict ourselves to $S_{3(1)}$. 

In order to cure this bad behaviour we need to add to the above results the effect of the second-order fields. 
We find (see the appendix)
\ba
\tilde{S}_{3(2)}=\int\d^4x\Bigg\{\frac{\delta L_2}{\delta\zeta_1}\lh\frac{\tilde{Q}_{1(2)}}{2}+\frac{\zeta_{1(2)}}{2}\rh+
\frac{\delta L_2}{\delta\zeta_2}\frac{\tilde{Q}_{2(2)}}{2}\Bigg\}.
\ea
Since all terms in $\tilde{S}_{3(2)}$ are proportional to $\delta L_2/\delta\zeta_m$, $\tilde{S}_{3(2)}$ only contains 
redefinitions of $\zeta_m$. 
Notice that the second-order lapse and shift functions do not appear in the final action, since these two 
are multiplied by a factor equal to the energy and momentum constraint equations (\ref{n1ni}). 
On the other hand, the second-order field perturbations are dynamical variables that obey second-order equations of motion that cannot 
be set to zero in the action. The single-field limit of this 
action is just the term proportional to $\zeta_{1(2)}$, since $\tilde{Q}_{m(i)}=0$ identically in that case for 
the uniform energy-density gauge. 
The term proportional to $\zeta_{1(2)}$ in $\tilde{S}_{3(2)}$, along with the terms proportional to $\lambda$ in $\tilde{S}_{3(1)}$, 
originate from the contribution of $N^i$ in the action. The latter vanish outside the horizon since then $\partial^2\lambda$ 
coincides with the super-horizon energy constraint and hence is identically zero. 
So if we were to study only the quadratic contributions of the first-order perturbations outside the horizon, 
we would be allowed not only to ignore the tensor parts of the metric \citep{Salopek:1990re}, 
but also work in the time-orthogonal gauge $N^i=0$.
 
Coming back to the redefinition, its final form, including the tensor parts (see the appendix), is
\ba
\zeta_1&=&\zeta_{1c}-\frac{\zeta_{1(2)}}{2}-\frac{\tilde{Q}_{1(2)}}{2}+\dot{\zeta_1}\zeta_1+\frac{\ge+\getpa}{2}\zeta_1^2
-\getpe\zeta_1\zeta_2\nn\\
&&-\frac{1}{4a^2H^2}\Big((\partial\zeta_1)^2
-\partial^{-2}\partial^i\partial^j(\partial_i\zeta_1\partial_j\zeta_1)\Big)\nn\\
&&+\frac{1}{2}\Big(\partial^i\lambda\partial_i\zeta_1-\partial^{-2}\partial^i\partial^j(\partial_i\lambda\partial_j\zeta_1)\Big)
-\frac{1}{4}\partial^{-2}(\dot{\gamma}_{ij}\partial^i\partial^j\zeta_1),\nn\\
\zeta_2&=&\zeta_{2c}-\frac{\tilde{Q}_{2(2)}}{2}+\dot{\zeta_2}\zeta_1+\frac{\getpe}{2}\zeta_1^2+(\ge+\getpa)\zeta_1\zeta_2.\label{rede2}
\ea

Finally we perform the above calculations for the flat gauge and find the action
\ba
\hat{S}_{3(2)}=\int\d^4x\Bigg\{\frac{\delta L_2}{\delta\zeta_1}\frac{\hat{Q}_{1(2)}}{2}
+\frac{\delta L_2}{\delta\zeta_2}\frac{\hat{Q}_{2(2)}}{2}\Bigg\}.
\ea
The redefinitions in the flat gauge take the simple form
\ba
&&\zeta_1=\zeta_{1c}-\frac{\hat{Q}_{1(2)}}{2}\nn\\
&&\zeta_2=\zeta_{2c}-\frac{\hat{Q}_{2(2)}}{2}.\label{red}
\ea

We want to write these redefinitions as well as the action itself in terms of gauge-invariant quantities and compare them. We would also like 
to compare with the definitions of the second-order gauge-invariant perturbations found in the first appendix and section \ref{gic}. 
After using the second-order uniform energy constraint (\ref{con}) and the uniform energy gauge definition of $\zeta_{2(2)}$ 
(\ref{z2u}) we can rewrite (\ref{rede2}) as
\ba
\zeta_1+\frac{\zeta_{1(2)}}{2}&\!\!\!\!=&\!\!\!\!\zeta_{1c}+\dot{\zeta_1}\zeta_1+\frac{\ge+\getpa}{2}\lh\zeta_1^2-\zeta_2^2\rh-\getpe\zeta_1\zeta_2
-\partial^{-2}\partial^i\lh\dot{\zeta}_2\partial_i\zeta_2\rh
-\frac{1}{4}\partial^{-2}(\dot{\gamma}_{ij}\partial^i\partial^j\zeta_1)
\nn\\
&&\!\!\!\!-\frac{1}{4a^2H^2}\Big((\partial\zeta_1)^2
-\partial^{-2}\partial^i\partial^j(\partial_i\zeta_1\partial_j\zeta_1)\Big)
+\frac{1}{2}\Big(\partial^i\lambda\partial_i\zeta_1-\partial^{-2}\partial^i\partial^j(\partial_i\lambda\partial_j\zeta_1)\Big)\nn\\
\zeta_2+\frac{\zeta_{2(2)}}{2}&\!\!\!\!=&\!\!\!\!\zeta_{2c}+\dot{\zeta_2}\zeta_1+\zeta_2\dot{\zeta}_1+\frac{\getpe}{2}\lh\zeta_1^2-\zeta_2^2\rh+(\ge+\getpa)\zeta_1\zeta_2
-\partial^{-2}\partial^i\lh\zeta_2\partial_i\dot{\zeta}_1\rh.\label{finu}
\ea
When comparing the first equation of (\ref{finu}) with (\ref{adi}), we see that we recover (\ref{red}). The same is true for the isocurvature 
part of the redefinition: comparing the second equation of (\ref{finu}) with (\ref{z2f}), we recover the redefinition 
for $\zeta_2$ (\ref{red}). 
Hence the two redefinitions are the same, as is necessary for the action to be gauge-invariant. 
Notice that the single-field limit of (\ref{finu}) is 
$\zeta_1+\zeta_{1(2)}/2=\zeta_{1c}+(\ge+\getpa)\zeta_1^2/2$ 
in agreement with the total redefinition found in the uniform energy-density gauge. 

Equation (\ref{finu}) is the implicit definition of the redefined, gauge-invariant $\zeta_{mc}$. One can see that up to 
and including second order, it is a function of only the combination $\zeta_{m(1)} + \zeta_{m(2)}/2$. One can also notice 
that the purely second-order perturbation $\zeta_{m(2)}$ does not occur explicitly in the cubic action (see e.g.\ (\ref{fin}) below). 
Hence one could in principle
consider the quantities $\zeta_{m(2)}$ (and similarly $Q_{m(2)}$) as auxiliary quantities and try to avoid introducing them in the 
first place, but consider the quadratic first-order terms directly as a correction to the first-order perturbations, as is done for 
the single-field case in \citep{Rigopoulos:2011eq}. While the calculations would be roughly equivalent, we have chosen not to follow this route 
for two reasons. In the first place it seems conceptually simpler to us to expand the perturbations and the action consistently up to 
the required order, and more logical to view quadratic first-order terms as a correction to a second-order quantity than to a first-order 
one. Secondly, in the multiple-field case (as opposed to the single-field case), one would have to introduce the second-order quantities at 
some intermediate steps anyway in order to find the correct non-linear relation between the $Q_m$ and $\zeta_m$ (which is derived from the 
second-order gauge transformation performed in the first appendix). 

So in the end we have managed to find the source of the non-Gaussianity present at horizon crossing due to first-order perturbations 
and identify it with the quadratic 
terms of (\ref{finu}). With source here we mean the second-order perturbation that, when contracted with two first-order 
perturbations, gives the bispectrum. The super-horizon limit of (\ref{finu}) was derived and used in our previous paper \citep{TvT1}, 
but here we have not only generalized the result, but also have obtained a much better understanding of the gauge issues. 
Equation (\ref{finu}) is gauge-invariant, as it should be. 
Additionally, the redefinition of the perturbations  
that we perform is essential not only to simplify calculations but also to find the gauge-invariant form of the 
action itself. 
We clearly see that the quadratic corrections in the flat gauge seem to be zero if one takes into account only the first-order fields. 
In that gauge all of the second-order contributions are hidden in the second-order fields as opposed to the uniform energy-density 
gauge where part of the quadratic contributions is attributed to the redefinition of the first-order $\zeta_m$ and the rest of them lie 
in the second-order field. 

\subsection{Summary}

Let us summarize the results of this section. Cosmological gauge-invariant perturbations 
should obey a gauge-invariant action. Using first-order perturbations the action up to third order is the same in the uniform 
energy-density gauge and the flat gauge only after a redefinition of $\zeta_m$ in the uniform energy-density gauge 
$\zeta_m=\zeta_{mc_1}+f_{m1}$ (\ref{rede1}) (the subscript $1$ indicating the use  of only first-order perturbations)  and takes the form 
\ba
S&=&\hat{S}(\zeta_{m})=\tilde{S}(\zeta_{mc_1})\nn\\
&=&\!\!\int\!\!\d^4x\frac{a\ge}{H}(\ge\zeta_1-1)\Big((\partial\zeta_1)^2+(\partial\zeta_2)^2\Big)
\nn\\
&&+\!\!\int\!\!\d^4x\Bigg\{\!a^3\!\ge H\Bigg[\lh1+\ge\zeta_1\rh\lh\dot{\zeta}_1^2+\dot{\zeta}_2^2\rh
-2\partial^i\lambda\lh\dot{\zeta}_2\partial_i\zeta_2+\dot{\zeta}_1\partial_i\zeta_1\rh\nn\\
&&\qquad\qquad\qquad-2(\ge+\getpa)\zeta_2\partial^i\lambda\partial_i\zeta_2+4\getpe\zeta_2\partial^i\lambda\partial_i\zeta_1
+\frac{1}{2}\zeta_1\lh\partial^i\partial^j\lambda\partial_i\partial_j\lambda-\lh\partial^2\lambda\rh^2\rh\nn\\
&&\qquad\qquad\qquad+2\dot{\zeta}_2\lh\chi\zeta_2+\ge\zeta_1\lh(\ge+\getpa)\zeta_2+\getpe\zeta_1\rh\rh
+\dot{\zeta}_1\lh-4\getpe\zeta_2+\ge\zeta_1^2(3\getpa+2\ge)\rh\nn\\
&&\qquad\qquad\qquad+\zeta_1^2\lh \lh\sqrt{\frac{2\ge}{\kappa}}\frac{W_{211}}{H^2}-2\ge\lh\ge\getpe+\getpa\getpe+\gx^\perp+3\getpe\rh\rh\zeta_2\right.\nn\\
&&\qquad\ \left.\qquad\qquad\qquad +\lh\sqrt{\frac{2\ge}{\kappa}}\frac{W_{111}}{3H^2}-\ge\lh\gx^\parallel+3\getpa-(\getpe)^2-(\getpa)^2\rh\rh\zeta_1
\rh\nn\\
&&\qquad\qquad\qquad+\zeta_2^2\lh
\sqrt{\frac{2\ge}{\kappa}}\frac{W_{221}}{3H^2}-2\ge^2-(\getpa)^2+3(\getpe)^2+\frac{2}{3}\getpe\gx^\perp
-3\ge(\getpa-\chi)+2\getpa\chi\right.\nn\\
&&\qquad\qquad\qquad\qquad\ +
\Bigg(\sqrt{\frac{2\ge}{\kappa}}\frac{W_{221}}{H^2}+\ge\lh-3(\getpe)^2+(\ge+\getpa)^2\rh+3\ge(\chi-\ge-\getpa)\Bigg)\zeta_1\nn\\
&&\left.\qquad\qquad\qquad \ \qquad+\sqrt{\frac{2\ge}{\kappa}}\frac{W_{222}}{3H^2}\zeta_2\rh\Bigg]\Bigg\}
\label{fin}
\ea
where we have kept the notation $\partial^2\lambda=\ge\dot{\zeta}_1-2\ge\getpe\zeta_2$ in order to mark clearly the terms 
that vanish outside the horizon, namely the terms proportional to $\lambda$ along with the terms involving second-order space derivatives. 
This is one of our main results. 
We managed to compute the cubic action for adiabatic and isocurvature perturbations in the exact theory, beyond any super-horizon or 
slow-roll approximation. 
Its single-field limit coincides with the action computed by Maldacena in \citep{Maldacena:2002vr} or by Rigopoulos in 
\citep{Rigopoulos:2011eq}. 
Let us examine the implications of this action. 
Forgetting about the redefinition of the perturbations in the uniform energy-density gauge, the form of the action is gauge-invariant. 
One can use it to easily calculate the non-Gaussianity related to the interaction terms as is explained in detail 
in \citep{Seery:2005wm,Weinberg:2005vy}. This is known in the literature as $f_{NL}^{(3)}$, the parameter of non-Gaussianity 
related to the three-point correlation function of three first-order perturbations, which is only non-zero in the case of intrinsic 
non-Gaussianity.

However, taking into account the need for a redefinition of the perturbations in the uniform energy-density gauge, one might worry that 
the action is not actually gauge-invariant. 
The action in the uniform energy-density gauge before the redefinition  
has extra terms that are proportional to the second-order equations that the perturbations obey. 
This means that when calculating the non-Gaussianity in the uniform energy-density gauge, one not only has contributions due to the the 
interaction terms in the cubic action, but also ones due to the redefinition of $\zeta_m$, which contribute as explained in (\ref{cyclic}). 
They are part of what is known in the literature as $f_{NL}^{(4)}$, the parameter of non-Gaussianity related to the 
three-point correlation function of a second-order perturbation (in terms of products of first-order ones) and two first-order perturbations, 
which reduces to products of two-point functions of the first-order perturbations. 

This would mean that the non-Gaussianity calculated in the two gauges would not be the same due to the lack of any redefinition in the flat 
gauge. 
However, if one takes into account only the corrections coming from first-order perturbations, the redefinition associated to 
the second-order 
perturbation is not complete as one can check by comparing the super-horizon version of the adiabatic part of (\ref{rede1}) with (\ref{z1ft}).  
As we showed, the solution of this issue is to include second-order fields since they also contribute to the cubic 
action. 
As one would expect these do not change the action itself, so that 
(\ref{fin}) still holds. 
The effect of the new terms is to redefine the perturbations in both gauges. 
It should be noted that, if one had incorporated all quadratic first-order terms (found by a second-order gauge transformation 
as in the first appendix) directly as a correction to the first-order perturbations, one would have found the two contributions 
$S_{3(1)}$ and $S_{3(2)}$ together and hence there would have been no initial discrepancy between the two gauges. 
However, we explained at the end of the previous subsection our reasons for proceeding in this way. 
So in any case we finally obtain
\be
S=\hat{S}(\zeta_{mc})=\tilde{S}(\zeta_{mc}),\label{finale}
\ee
where $\zeta_{mc}$ is given in (\ref{finu}).  
Now the two redefinitions as well as the action itself are the same for the two gauges, 
hence the action is truly gauge-invariant and the $f_{NL}^{(4)}$, related to the products of first-order $\zeta_m$ in the redefinitions, 
is the same in the two gauges. 

This exact action allows one to compute $f_{NL}^{(4)}$ without the need for 
the slow-roll approximation at horizon crossing that is essential for both the long-wavelength formalism and the $\delta N$ formalism: 
the long-wavelength 
formalism needs slow-roll at horizon crossing in order to allow for the decaying mode to vanish rapidly, while the $\delta N$ 
formalism requires it in order to ignore the derivatives with respect to the canonical momentum. 
Additionally, up to now only 
the slow-roll \textit{field} action \citep{Seery:2005gb}  (and not the action of the $\zeta_m$ themselves) was known, 
so in order to compute the non-Gaussianity at horizon crossing one had to use the long-wavelength 
or $\delta N$ formalism to transform to $\zeta_m$ 
and hence one was in any case required to  make the assumption of slow-roll, 
even if the exact action for the fields would have been known. It will be interesting to investigate models that do not satisfy 
the conditions for the long-wavelength or $\delta N$ formalism using the action (\ref{fin}).

In order to connect the redefinitions to some previously derived results in the literature we assume the super-horizon and slow-roll approximations. The 
super-horizon approximation is already assumed in (\ref{finu}) and it can be supplemented by the condition $\dot{\zeta}_1=2\getpe\zeta_2$. 
The slow-roll assumption translates into $\dot{\zeta}_2=-\chi\zeta_2$. Then 
the quadratic part of the redefinitions, relevant to $f_{NL}^{(4)}$, takes the form 
\ba
&&\zeta_1=\zeta_{1c}+\frac{\ge+\getpa}{2}\zeta_1^2+\getpe\zeta_2\zeta_1-\frac{\ge+\getpa-\chi}{2}\zeta_2^2\nn\\
&&\zeta_2=\zeta_{2c}+\frac{\getpe}{2}\zeta_1^2+(\ge+\getpa-\chi)\zeta_1\zeta_2+\frac{\getpe}{2}\zeta_2^2.
\ea
The redefinitions in this form were used in \citep{TvT1} to find the second-order source term of the evolution equations for the 
super-horizon perturbations. Their contribution to the super-horizon $f_{NL}^{(4)}$ was calculated in that paper using the 
long-wavelength formalism. In the equal-momenta case it was shown to be
\be
-\frac{6}{5}f_{NL,h.c.}^{(4)}=\frac{\ge_*+\getpa_*+\getpe_*\bv_{12}}{1+(\bv_{12})^2},\label{difer}
\ee
where the index $*$ indicates the time when the scale exits the horizon and $\bv_{12}$ is essentially a transfer function 
showing how the isocurvature mode $\zeta_2$ sources the adiabatic mode $\zeta_1$ (see \citep{TvT1} for details, where this 
term is part of what is called $g_\mathrm{sr}$). 
Directly after horizon crossing or equivalently in the 
single-field limit, when $\bv_{12}=0$, this reduces to the well-known result by Maldacena $-6/5f_{NL}^{(4)}=\ge_*+\getpa_*$.

\section{Appendices}
\label{app}

This section contains the four appendices of the paper.

\subsection{Gauge transformations}

From the infinite number of possible gauge-invariant combinations, we choose to work with quantities constructed from 
the energy density and the logarithm of the space dependent scale factor $\alpha$. We will consider a gauge transformation 
$\beta_{(i)}=(T_{(i)},\vec{q}_{(i)})$ 
from the hatted gauge 
to the tilded gauge, where for the moment both gauges are taken to be arbitrary (not yet the flat and uniform energy 
density gauge). Notice though, that the space part of the 
transformation is not relevant outside the horizon, since when introduced in the relations below, it is connected to a second-order 
space derivative \citep{Malik:2003mv}. Within the super-horizon approximation, we find using (\ref{trans})
\ba
&&\tilde{\rho}_{(1)}=\hat{\rho}_{(1)}+\dot{\rho}T_{(1)},\qquad 
\tilde{\rho}_{(2)}=\hat{\rho}_{(2)}+\dot{\rho}T_{(2)}+T_{(1)}\lh
2\dot{\hat{\rho}}_{(1)}+\dot{\rho}\dot{T}_{(1)}+\ddot{\rho}T_{(1)}\rh,\nn\\
&&\tilde{\alpha}_{(1)}=\hat{\alpha}_{(1)}+T_{(1)},\qquad
\tilde{\alpha}_{(2)}=\hat{\alpha}_{(2)}+T_{(2)}+T_{(1)}\lh 2\dot{\hat{\alpha}}_{(1)}+\dot{T}_{(1)}\rh.\label{a1}
\ea
We want to construct a gauge-invariant quantity that reduces to $\alpha_{(i)}$ in the uniform energy-density gauge, 
which we now identify with the tilded gauge so that $\tilde{\rho}_{(i)}=0$. This way we find 
\be
T_{(1)}=-\frac{\hat{\rho}_{(1)}}{\dot{\rho}},\qquad\qquad
T_{(2)}=-\frac{\hat{\rho}_{(2)}}{\dot{\rho}}+\frac{\hat{\rho}_{(1)}\dot{\hat{\rho}}_{(1)}}{\dot{\rho}^2}\label{uetr}
\ee
and obtain
\ba
&&\zeta_{1(1)}\equiv\tilde{\alpha}_{(1)}=\hat{\alpha}_{(1)}-\frac{\hat{\rho}_{(1)}}{\dot{\rho}},\nn\\
&&\frac{1}{2}\zeta_{1(2)}\equiv\frac{1}{2}\tilde{\alpha}_{(2)}
=\frac{1}{2}\hat{\alpha}_{(2)}-\frac{1}{2}\frac{\hat{\rho}_{(2)}}{\dot{\rho}}
+\frac{\dot{\hat{\rho}}_{(1)}\hat{\rho}_{(1)}}{\dot{\rho}^2}
-\frac{\hat{\rho}_{(1)}}{\dot{\rho}}\dot{\hat{\alpha}}_{(1)}
-\frac{1}{2}\frac{\hat{\rho}_{(1)}^2}{\dot{\rho}^2} \frac{\ddot{\rho}}{\dot{\rho}}.
\ea
Notice that the initial hatted gauge is still arbitrary, but if one was to associate it with the flat gauge 
$\hat{\alpha}_{(i)}=0$, then the time shift would become $T_{(1)}=\zeta_{1(1)}$.

Next, we derive the exact gauge-invariant adiabatic and isocurvature perturbations, going beyond the super-horizon approximation. 
We use (\ref{trans}) for the scalar fields and the space part of the metric tensor (\ref{metric}), and find
\ba
\tilde{\varphi}^A_{(1)}&\!\!\!\!=&\!\!\!\!\hat{\varphi}^A_{(1)}+\dot{\phi}^AT_{(1)}\label{ftrans}\\
\tilde{\alpha}_{(1)}\delta_{ij}+\frac{1}{2}\tilde{\gamma}_{(1)ij}
&\!\!\!\!=&\!\!\!\!\hat{\alpha}_{(1)}\delta_{ij}+\frac{1}{2}\hat{\gamma}_{(1)ij}+T_{(1)}\delta_{ij}+\partial_i\partial_jq_{(1)}
+\frac{1}{2}\lh\partial_iq^{\perp}_{(1)j}+\partial_jq^{\perp}_{(1)i}\rh.\label{atrans}
\ea
Here we followed \citep{Bruni:1996im} and split the component $i$ of the space shift as $q_i=\partial_iq+q^{\perp}_i$, where 
$\partial^iq^{\perp}_i=0$. We choose the uniform energy-density gauge, defined by $e_{1A}\tilde{\varphi}^A_{(1)}=0$ and use (\ref{ftrans}) to 
find the first-order time shift to be $T_{(1)}=-e_{1A}\hat{\varphi}^A_{(1)}/\dphi$. Then the trace and the traceless part of 
(\ref{atrans}) give
\ba
&&\tilde{\alpha}_{(1)}=\hat{\alpha}_{(1)}+T_{(1)}+\frac{1}{3}\partial^2q_{(1)}\label{atrans1}\\
&&\tilde{\gamma}_{(1)ij}=\hat{\gamma}_{(1)ij}+2\lh\partial_i\partial_j-\frac{1}{3}\delta_{ij}\partial^2\rh q_{(1)}+
\partial_iq^{\perp}_{(1)j}+\partial_jq^{\perp}_{(1)i}.
\ea
In order to make the definition of the super-horizon adiabatic perturbation at first order (\ref{a1}) to agree with (\ref{atrans1}), 
we choose $q_{(1)}=0$ (any choice of $q_{(1)}$ is a gauge-invariant quantity, but only $q_{(1)}=0$ corresponds with the 
adiabatic perturbation $\zeta_{1(1)}$ in the literature). 
Note that while working with the super-horizon approximation, no such choice needs to be made, and 
$q_{(1)}$ remains arbitrary in that case. Similarly, we also assume that 
$q^{\perp}_{(1)i}=0$, so we find
\be
\zeta_{1(1)}=\tilde{\alpha}_{(1)}=\hat{\alpha}_{(1)}-\frac{1}{\dphi}e_{1A}\hat{\varphi}^A_{(1)}\qquad\mathrm{and}\qquad 
\gamma_{(1)ij}\equiv\tilde{\gamma}_{(1)ij}=\hat{\gamma}_{(1)ij}.
\ee
Using (\ref{ftrans}), one easily finds that the isocurvature perturbation at first order is gauge-invariant since $e_{2A}\dphi^A=0$
\be
\zeta_{2(1)}=-\frac{1}{\dphi}e_{2A}\tilde{\varphi}^A_{(1)}=-\frac{1}{\dphi}e_{2A}\hat{\varphi}^A_{(1)}.
\ee 
We now fix the hatted gauge to be the flat one, $\hat{\alpha}_{(i)}=0$, in order to lighten the calculations.  
This implies that $T_{(1)}=\zeta_{1(1)}$. 

At second order we find
\ba
\tilde{\varphi}^A_{(2)}=\hat{\varphi}^A_{(2)}+T_{(2)}\dphi^A+\zeta_{1(1)}\lh\dot{\zeta}_{1(1)}\dphi^A+\zeta_{1(1)}\ddot{\phi}^A
+2\dot{\hat{\varphi}}_{(1)}^A\rh,\label{ftrans1}
\ea
where we have omitted a term proportional to $q_{(1)}$, which as mentioned above is chosen to be zero. At second order we choose the gauge 
$\frac{1}{2}\tilde{Q}_{1(2)}=
\frac{\ge+\getpa}{2}\zeta_{2(1)}^2+\partial^{-2}\partial^i\lh\dot{\zeta}_{2(1)}\partial_i\zeta_{2(1)}\rh$, see (\ref{con}), that reduces to the 
uniform energy-density gauge on super-horizon scales (for the definition of $Q_{1(2)}$ in (\ref{b12}) and details about that 
gauge choice, the reader can refer to the next appendix). Using (\ref{difzflat}) we find from (\ref{ftrans1})
\be
T_{(2)}=\hat{Q}_{1(2)}+(\ge+\getpa)\lh\zeta_{1(1)}^2-\zeta_{2(1)}^2\rh+\dot{\zeta}_{1(1)}\zeta_{1(1)}-2\getpe\zeta_{1(1)}\zeta_{2(1)}
-2\partial^{-2}\partial^i\lh\dot{\zeta}_{2(1)}\partial_i\zeta_{2(1)}\rh.\label{deft2}
\ee 

Before turning to the adiabatic perturbation, let us prove that the first line of (\ref{z2f}) is a gauge-invariant quantity 
corresponding to the one in (\ref{z2u}). This is true in the exact theory, beyond the long-wavelength approximation, 
for $q_{(1)}=0$. Multiplying (\ref{ftrans1}) with $-e_{2A}/(2\dphi)$, noticing that $e_{2A}\dphi^A=0$, and using 
(\ref{difzflat}), one finds 
\be
\frac{1}{2}\tilde{Q}_{2(2)}=\frac{1}{2}\hat{Q}_{2(2)}+\frac{\getpe}{2}\zeta_{1(1)}^2+\dot{\zeta}_{2(1)}\zeta_{1(1)}
+(\ge+\getpa)\zeta_{1(1)}\zeta_{2(1)}\label{x2}
\ee
and indeed by comparing the total gradient of (\ref{z2u}) and (\ref{z2f}) we see that it corresponds to the second-order gauge-invariant 
isocurvature perturbation. 

For the second-order adiabatic perturbation we need to perform the gauge transformation (\ref{trans}) of the space part of the metric tensor 
between the uniform energy-density gauge and the flat gauge,
\ba
\!\!\!\!\!\!\!\!\!\!\!\!\tilde{\alpha}_{(2)}\delta_{ij}+\frac{1}{2}\tilde{\gamma}_{(2)ij}&\!\!\!\!=&\!\!\!\!\frac{1}{2}\hat{\gamma}_{(2)ij}+T_{(2)}\delta_{ij}+\partial_i\partial_j
q_{(2)}+\frac{1}{2}\lh\partial_iq^{\perp}_{(2)j}+\partial_jq^{\perp}_{(2)i}\rh+\zeta_{1(1)}\dot{\gamma}_{(1)ij}\nn\\
&&\!\!\!\!+\partial_i\zeta_{1(1)}\partial_j\lambda+\partial_j\zeta_{1(1)}\partial_i\lambda+\zeta_{1(1)}\dot{\zeta}_{1(1)}\delta_{ij}
-\frac{1}{a^2H^2}\partial_i\zeta_{1(1)}\partial_j\zeta_{1(1)},\label{atrans2}
\ea
where we substituted $T_{(1)}=\zeta_{1(1)}$ and $T_{(2)}$ is given in (\ref{deft2}). In order to find 
$\zeta_{1(2)}$ we take the trace and subtract the $\partial^{-2}\partial^i\partial^j$ of (\ref{atrans2}) to eliminate the terms proportional to 
$q_{(2)}$ and obtain the result
\ba
\frac{1}{2}\zeta_{1(2)}&\!\!\!\!=&\!\!\!\!\frac{1}{2}\tilde{\alpha}_{(2)}\nn\\
&\!\!\!\!=&\!\!\!\!\frac{1}{2}\hat{Q}_{1(2)}+\dot{\zeta}_{1(1)}\zeta_{1(1)}+\frac{\ge+\getpa}{2}\lh\zeta_{1(1)}^2-\zeta_{2(1)}^2\rh
-\getpe\zeta_{1(1)}\zeta_{2(1)}
-\partial^{-2}\partial^i\lh\dot{\zeta}_{2(1)}\partial_i\zeta_{2(1)}\rh
\nn\\
&&\!\!\!\!-\frac{1}{4}\partial^{-2}(\dot{\gamma}_{(1)ij}\partial^i\partial^j\zeta_{1(1)})
-\frac{1}{4a^2H^2}\Big((\partial\zeta_{1(1)})^2-\partial^{-2}\partial^i\partial^j(\partial_i\zeta_{1(1)}\partial_j\zeta_{1(1)})\Big)\nn\\
&&\!\!\!\!+\frac{1}{2}\Big(\partial^i\lambda\partial_i\zeta_{1(1)}-\partial^{-2}\partial^i\partial^j(\partial_i\lambda\partial_j\zeta_{1(1)})\Big).
\label{adi}
\ea
This is the second-order adiabatic gauge-invariant perturbation in the exact theory, the generalization of (\ref{z1ft}).

\subsection{Super-horizon calculations}

In this appendix we give the detailed calculations of section \ref{gic}.

An important property of the long-wavelength assumption is that outside the horizon the uniform energy density can be 
recast in terms of the fields at least at first order: one can show 
that the exact 0i-Einstein equation (\ref{fieldeqsu}) outside the horizon can be rewritten as 
\be
\partial_i\bar{\rho}=-3\bH\bar{\Pi}_B\partial_i\varphi^B.\label{0iTz}
\ee
Again $\bar{\rho}$ denotes the fully non-linear energy density $\bar{\rho}=\bar{\Pi}^2/2+W$. 
Expanding (\ref{0iTz}) to first 
order and using the background equations to prove that
\be
\dot{\rho}=-3\Pi^2NH,
\ee
one can show that outside the horizon
\be
\frac{\rho_{(1)}}{\dot{\rho}}=\frac{1}{\Pi N}e_{1A}\varphi^A_{(1)}
\label{ratio1}
\ee
so that 
\be
\zeta_{1(1)}=\tilde{\alpha}_{(1)}=\alpha_{(1)}-\frac{H}{\Pi}e_{1A}\varphi^A_{(1)}.
\ee
and thus the energy-density constraint $\tilde{\rho}_{(1)}=0$ is equivalent to $e_{1A}\tvarphi^A_{(1)}=0$.
 
Unfortunately the nice property described by (\ref{ratio1}) does not hold anymore at second order. After expanding up to 
second order, combining the second-order Einstein equations and using the completeness relation of the field basis, 
one can show that (note that the zeroth order lapse function is taken from now on to be $N(t)=1/H(t)$) 
\be
\frac{\rho_{(2)}}{\dot{\rho}}-\frac{1}{\dphi}e_{1A}\varphi^A_{(2)}
=\lh\frac{1}{\dphi}e_{1A}\varphi_{(1)}^A\rh^2(\getpa-\ge)+\frac{1}{\dphi^2}A,
\label{ratio2}
\ee
with 
\ba
\frac{1}{2}\partial_iA&=&\getpe e_{2B}\varphi^B_{(1)}e_{1A}\partial_i\varphi^A_{(1)}+e_{2B}\partial_i\varphi^B_{(1)}e_{2A}\dvphi^A_{(1)},
\label{alp}
\ea
where we used (\ref{dere}) to simplify the expressions in terms of the slow-roll parameters.
We see that the purely second-order contribution of $\rho_{(2)}$ is recast as a second-order contribution of $\varphi^A_{(2)}$, 
some quadratic first-order terms and a non-local term arising essentially from the 0i-Einstein equation.
In the flat gauge $\partial_i\ha=0$ this non-local term can be written as
\be
\frac{1}{2\phi^2}\partial_i\hat{A}=\partial_i\Bigg[\getpe\zeta_{1(1)}\zeta_{2(1)}+\frac{\ge+\getpa}{2}\zeta_{2(1)}^2\Bigg]
+\dot{\zeta}_{2(1)}\partial_i\zeta_{2(1)},\label{alp_flat}
\ee
where when needed we employed the following useful relations (valid for the flat gauge beyond the long-wavelength approximation):
\ba
-\frac{1}{\dphi}e_{1A}\dot{\hat{\varphi}}^A_{(1)}&=&\dot{\zeta}_{1(1)}+(\ge+\getpa)\zeta_{1(1)}-\getpe\zeta_{2(2)},\nn\\
-\frac{1}{\dphi}e_{2A}\dot{\hat{\varphi}}^A_{(1)}&=&\dot{\zeta}_{2(1)}+(\ge+\getpa)\zeta_{2(1)}+\getpe\zeta_{1(2)},\label{difzflat}
\ea
derived by differentiating the adiabatic perturbation and the  
new combination
\be
\zeta_{2(1)}\equiv-\frac{H}{\Pi}e_{2A}\varphi^A_{(1)},
\ee
that represents the isocurvature perturbation to first order.
Putting everything together in (\ref{36}) we find the second-order gauge-invariant curvature 
perturbation in the flat gauge to be
\ba
\frac{1}{2}\zeta_{1(2)}=\frac{1}{2}\tilde{\alpha}_{(2)}&=&\frac{1}{2}\hat{Q}_{1(2)}
+\frac{\ge+\getpa}{2}\zeta_{1(1)}^2-\frac{\ge+\getpa}{2}\zeta_{2(1)}^2-\getpe\zeta_{1(1)}\zeta_{2(1)}+\dot{\zeta}_{1(1)}\zeta_{1(1)}\nn\\
&&-\partial^{-2}\partial^i\lh\dot{\zeta}_{2(1)}\partial_i\zeta_{2(1)}\rh,
\ea
where we defined the auxiliary quantities
\be
Q_{m(i)}\equiv-\frac{H}{\Pi}e_{mA}\varphi^A_{(i)}.\label{b12}
\ee

We turn now to the calculations relevant to the gradient of the perturbations. 
In the uniform energy-density gauge we can use (\ref{0iTz}) to find the constraints
\ba
&&\partial_i\tilde{\rho}_{(1)}=-3\dphi_A\partial_i\tvarphi^A_{(1)}=0 \qquad \mathrm{or} \qquad \partial_i\tilde{Q}_{1(1)}=0,\label{con1}\\
&&\partial_i\tilde{\rho}_{(2)}=-\frac{3}{2}\dphi_A\partial_i\tvarphi^A_{(2)}-3\tdvphi_{A(1)}\partial_i\tvarphi^A_{(1)}=0\qquad \mathrm{or}\nn\\
&&\frac{1}{2}\partial_i\tilde{Q}_{1(2)}=\frac{1}{\dphi^2}\tdvphi_{A(1)}\partial_i\tvarphi^A_{(1)}=
\frac{\ge+\getpa}{2}\partial_i\zeta_{2(1)}^2+\dot{\zeta}_{2(1)}\partial_i\zeta_{2(1)},
\label{con}
\ea
so that when inspecting (\ref{akoma}) we see that
\begin{displaymath}
\frac{1}{2}\tilde{\zeta}_{1i(2)}=\frac{1}{2}\partial_i\tilde{\alpha}_{(2)}.
\end{displaymath}
To derive the last equality in (\ref{con}) we used the completeness relation of the field basis along with the following relations 
valid beyond the long-wavelength approximation:
\ba
&&-\frac{1}{\dphi}e_{1A}\dot{\tilde{\varphi}}^A_{(1)}=-\getpe\zeta_{2(1)},\nn\\
&&-\frac{1}{\dphi}e_{2A}\dot{\tilde{\varphi}}^A_{(1)}=\dot{\zeta}_{2(1)}+(\ge+\getpa)\zeta_{2(1)},\label{difzue}
\ea
derived by differentiating the definition of the isocurvature perturbation and the first-order uniform energy-density gauge constraint 
$e_{1A}\tilde{\varphi}^A_{(1)}=0$.

\subsection{Second-order action calculation}

In order to rewrite the action we first need to calculate the 
extrinsic curvature. To do that we decompose $\bar{N}=1/H+N_1$, $N^i=\partial^i\psi+N^i_{\perp}$, 
where $\partial_iN^i_{\perp}=0$. 
 From now on we drop the explicit 
subscript $(1)$ on first-order quantities and set $\kappa^2=1$ to lighten the notation 
(notice though that the final results remain unchanged when we restore $\kappa^2=8\pi G$, 
since all $\kappa^2$ are absorbed in $\ge$ when rewriting the fields in terms of $\zeta_m$). We start by performing the calculation in the gauge 
\be
e_{1A}\tilde{\varphi}^A=0,\label{ugauge}
\ee
which we call uniform energy-density gauge, 
since the above constraint reduces to zero energy perturbation outside the horizon.
We first use the energy and momentum constraint (\ref{energy}), (\ref{momentum}) to find that to first order
\ba
&&\tilde{N}_1=\frac{\dot{\tilde{\alpha}}}{H}=\frac{\dot{\zeta}_1}{H}, \quad\ \ \qquad \tilde{N}^i_{\perp}=0,\nn\\
&&\tilde{\psi}=-\frac{1}{a^2}\frac{\zeta_1}{H^2}+\lambda, \qquad \partial^2\lambda=\ge\dot{\zeta}_1-2\ge\getpe\zeta_2.
\label{n1ni}
\ea
It turns out that we do not need to calculate the shift or the lapse function to higher order, since in the action 
those terms are multiplied by constraint relations and hence vanish.

We start by working out the scalar part of the action. 
Keeping in mind the gauge constraint (\ref{ugauge}) we perturb (\ref{actionexact}) to second order
\ba
\tilde{S}_2&=&\frac{1}{2}\int\d^4x\Bigg\{a^3 e^{3\tilde{\alpha}}
\Big[(\frac{1}{H}+\tilde{N}_1+\frac{\tilde{N}_2}{2})\Big(-2W-2W_A\tvarphi^A-W_{AB}\tvarphi^A\tvarphi^B\Big)\nn\\
&&\qquad\qquad\qquad\quad+H(1-H\tilde{N}_1+H^2\tilde{N}_1^2-H\frac{\tilde{N}_2}{2})
\Big(-6(1+\dot{\tilde{\alpha}})^2+\dot{\phi}^2+2\dot{\phi}^A\dot{\tvarphi}_A+\dot{\tvarphi}^2\Big)\Big]\nn\\
&&\qquad\qquad-a e^{\tilde\alpha}\Big[(\frac{1}{H}+\tilde{N}_1)2
\lh(\partial\tilde{\alpha})^2+2\partial^2\tilde{\alpha}\rh
+\frac{1}{H}\partial_i\tilde{\varphi}_A\partial^i\tilde{\varphi}^A\Big]\Bigg\},
\ea
where we have omitted a total derivative with respect to $\tilde{\psi}$. 
We then use the background Einstein and field equations to eliminate some terms and find that the term 
proportional to $\tilde{N}_2$ vanishes.
Now the second-order action can be written as
\ba
\tilde{S}_2&\!\!\!\!=&\!\!\!\!\frac{1}{2}\int\!\!\d^4x\Bigg\{a^3H\Big[
\dot{\zeta}_1\Big(\!-4\dot{\phi}^A\dot{\tvarphi}_A
+\dot{\phi}^2\dot{\zeta}_1\Big)+9\zeta^2_1(-6+\dot{\phi}^2)-\frac{1}{H^2}W_{AB}\tvarphi^A\tvarphi^B-36\zeta_1\dot{\zeta}_1
+\dot{\tvarphi}^2\Big]\nn\\
&&\qquad\quad-a\frac{1}{H}\Big[2\ge(\partial\zeta_1)^2+\partial^i\tilde{\varphi}_A\partial_i\tilde{\varphi}^A\Big]\Bigg\}.
\label{sin}
\ea
The terms of (\ref{sin}) proportional to $\tilde{\varphi}^A$ can be recast in terms of the curvature perturbations by applying the 
completeness property of the field basis and (\ref{difzue}),
so that after integrating by parts and using $\dot{H}=-\ge H$ the action can be written as
\ba
\tilde{S}_2&=&\int\d^4x\ \ge\Bigg\{a^3H\Big[\dot{\zeta}_1^2+\dot{\zeta}_2^2-4\getpe\dot{\zeta}_1\zeta_2
-3(\chi-\ge-\getpa)\zeta_2^2
+2(\ge+\getpa)\zeta_2\dot{\zeta}_2\nn\\
&&\qquad\qquad\qquad\;+(\getpe)^2\zeta_2^2+(\ge+\getpa)^2\zeta_2^2\Big]-
a\frac{1}{H}\Big[ (\partial\zeta_1)^2+(\partial\zeta_2)^2\Big]
\Bigg\},\label{simple}
\ea
or after further integration by parts as
\ba
\tilde{S}_2&\!\!\!=&\!\!\!\int\!\!\d^4x\ \ge \Bigg\{\!\!-a\frac{1}{H}\Big( (\partial\zeta_1)^2+(\partial\zeta_2)^2\Big)
 +a^3H\Big(\dot{\zeta}_1^2+\dot{\zeta}_2^2-4\getpe \dot{\zeta}_1\zeta_2
+2\chi \dot{\zeta}_2\zeta_2\Big) \label{s22}\\
&&\qquad\quad+a^3H\Bigg(\!\sqrt{\frac{2\ge}{\kappa}}\frac{W_{221}}{3H^2}\!-\!2\ge^2\!-\!(\getpa)^2\!+\!3(\getpe)^2\!+\!\frac{2}{3}\getpe\gx^\perp
\!-\!3\ge(\getpa\!-\!\chi)\!+\!2\getpa\chi\Bigg) \zeta_2^2\Bigg\}.\nn
\ea

We can reach the same result while working in the flat gauge $\partial_i{\hat{\alpha}}=0$.  
One can prove that $\hat{N}_1=-\ge\zeta_1/H,\ \hat{N}^i_{\perp}=0$ and 
$\partial^2\hat{\psi}=\partial^2\lambda=\ge\dot{\zeta}_1-2\ge\getpe\zeta_2$. The $\hat{\psi}$ terms cancel out and the second-order 
action takes the form
\ba
 \hat{S}_2&\!\!\!\!=&\!\!\!\!\frac{1}{2}\int\d^4x\Big\{ a^3\Big[H\dot{\hat{\varphi}}^2-\frac{1}{H}W_{AB}\hat{\varphi}^A\hat{\varphi}^B
+\hat{N}_1(-2W_A\hat{\varphi}^A-2H^2\dphi^A\dot{\hat{\varphi}}_A)+\hat{N}_1^2H^3(-6+\dphi^2)\Big]\nn\\
&&\qquad\qquad\!\!-a\frac{1}{H}\partial^i\hat{\varphi}_A\partial_i\hat{\varphi}^A\Big\}.
\ea
Using the definition of $\zeta_m$, along with the background equations (\ref{fieldeqTz}), (\ref{srvar}) 
and (\ref{difzflat}) this can be rewritten as (\ref{s22}). 

The second-order tensor part of the action in both gauges takes the form 
\be
S_{2\gamma}=\int\d^4xL_{2\gamma}=\frac{1}{2}\int\d^4x\Big\{\frac{a^3}{4}H(\dot{\gamma}_{ij})^2
-\frac{a}{4H}(\partial_k\gamma_{ij})^2\Big\},
\ee 
where $L_{2\gamma}$ is the second-order Lagrangian for the tensor modes. 
We also give the equation of motion of the gravitational waves
\be
\frac{\delta  L_{2\gamma}}{\delta\gamma_{ij}}=-\frac{1}{4}\frac{\d}{\d t}(a^3H\dot{\gamma}_{ij})
+\frac{1}{4}\frac{a}{H}\partial^2\gamma_{ij}=0,
\ee
which we are going to use in the next section. In this paper we will not discuss the evolution and physics of gravitational 
waves, but at linear order this is 
a standard subject in 
the literature, for a discussion see for example \citep{Misner:1974qy}.

\subsection{Third-order action calculation}

In order to compute $S_3$ we follow the same procedure starting from the uniform energy-density gauge. 
Notice that $\tilde{N}_3$ will multiply $(-2W+6H^2-\Pi^2)$ in exact 
analogy with $\tilde{N}_2$ in $S_2$, so it vanishes. Moreover, the overall factor multiplying $\tilde{N}_2$ is 
the first-order energy constraint (\ref{n1ni}), so it can be consistently set to zero as well.

We start by computing the cubic action of the first-order curvature perturbations up to $\tilde{N}_1$ involving only 
scalar quantities
\ba
\tilde{S}_{3(1)}&\!\!\!=&\!\!\!\frac{1}{2}\int\!\!\d^4x\Bigg\{\!a^3e^{3\tilde{\alpha}}\Bigg[ \Big(\frac{1}{H}+\tilde{N}_1\Big)
\Big(-2W-2W_A\tvarphi^A-W_{AB}\tvarphi^A\tvarphi^B-\frac{1}{3}W_{ABC}\tvarphi^A\tvarphi^B\tvarphi^C\Big)\nn\\
&&\qquad\qquad\qquad\;+H\Big[(1-H\tilde{N}_1+H^2\tilde{N}_1^2
-H^3\tilde{N}_1^3
)\Big(-6(1+\dot{\tilde{\alpha}})^2
+\dot{\phi}^2+2\dot{\phi}^A\dot{\tvarphi}_A+\dot{\tvarphi}^2\Big)\nn\\
&&\qquad\qquad\qquad\;+\Big(\partial^i\partial^j\tilde{\psi}\partial_i\partial_j\tilde{\psi}
-(\partial^2\tilde{\psi})^2\Big)(1-H\tilde{N}_1)-4\partial^i\tilde{\psi}\partial_i\zeta_1\partial^2\tilde{\psi}
-2\dot{\tilde{\varphi}}_A\partial^i\tilde{\psi}\partial_i\tilde{\varphi}^A\Big]\Bigg]\nn\\
&&\qquad\quad\;-ae^{\tilde{\alpha}}\Big[(\frac{1}{H}+\tilde{N}_1)\Big(\partial^i\tilde{\varphi}_A\partial_i\tilde{\varphi}^A
+4\partial^2\zeta_1+2(\partial\zeta_1)^2\Big)
  \Big]\Bigg\}.\label{s311}
\ea
After using the background equations and the definitions of the perturbations, (\ref{s311}) takes the form
\ba
\tilde{S}_{3(1)}&\!\!\!\!\!=&\!\!\!\!\!\!\int\!\!\d^4x\Bigg\{\!a^3e^{3\zeta}H\Big[\ge(1-\dot{\zeta}_1)
\Big(\!\dot{\zeta}_1^2+\dot{\zeta}_2^2+2(\ge+\getpa)\dot{\zeta}_2\zeta_2+\!\lh\!(\getpe)^2\!+\!(\ge+\getpa)^2\rh\!\zeta_2^2
-2\getpe \zeta_2\dot{\zeta}_1\!\Big)\nn\\
&&\qquad\qquad\quad\;-2\ge\getpe \zeta_2\dot{\zeta}_1
-3(1+\dot{\zeta}_1)\ge(\chi-\ge-\getpa)\zeta_2^2+\ge\sqrt{\frac{2\ge}{\kappa}}\frac{W_{222}}{3H^2}\zeta_2^3
-2\partial^i\tilde{\psi}\partial_i\zeta_1\partial^2\tilde{\psi}\nn\\
&&\qquad\qquad\quad\;+\frac{1}{2}\Big(\partial^i\partial^j\tilde{\psi}\partial_i\partial_j\tilde{\psi}
-(\partial^2\tilde{\psi})^2\Big)(1-\dot{\zeta}_1)
-2\ge\partial^i\tilde{\psi}\Big((\ge+\getpa)\zeta_2\partial_i\zeta_2+\dot{\zeta}_2\partial_i\zeta_2\Big)\Big]\nn\\
&&\qquad
-a\frac{1}{H}(\zeta_1+\dot{\zeta}_1)\Big[2\partial^2\zeta+(\partial\zeta_1)^2+\ge(\partial\zeta_2)^2\Big]\Bigg\}.\label{s31}
\ea
By performing integrations by parts in (\ref{s31}) we find 
\ba
\!\tilde{S}_{3(1)}&\!\!\!\!\!=&\!\!\!\!\!\!\int\!\!\d^4x\Bigg\{\!a^3\ge H\Bigg[\ge\zeta_1(\dot{\zeta}_1^2+\dot{\zeta}_2^2)
-2\dot{\zeta}_1\partial^i\lambda\partial_i\zeta_1-2\dot{\zeta}_2\partial^i\lambda\partial_i\zeta_2
-2\Big(\ge\getpa\!+\!(\getpa)^2\!+\!(\getpe)^2\!\Big)\zeta_1\zeta_2\dot{\zeta}_2\nn\\
&&
\qquad\qquad +(3\ge\getpe+\gx^\perp)\zeta_1^2\dot{\zeta}_2
+2(\ge\getpe+\gx^\perp)\zeta_1\zeta_2\dot{\zeta}_1
-\Big(\ge^2+2\ge\getpa+(\getpa)^2+(\getpe)^2\Big)\zeta_2^2\dot{\zeta}_1 \nn\\
&&\qquad\qquad +\Big(2\ge^2+3\ge\getpa-\!(\getpa)^2-\!(\getpe)^2+\gx^\parallel\Big)\zeta_1^2\dot{\zeta}_1
+\sqrt{\frac{2\ge}{\kappa}}\frac{W_{222}}{3H^2}\zeta_2^3 \nn\\
&&\qquad\qquad +\Bigg(2(\ge+\getpa)\gx^\perp+\getpe\Big(2\ge(3+\ge)+6\getpa-\gx^\parallel-3\chi\Big)\Bigg)\zeta_1^2\zeta_2
\nn\\
&&\qquad\qquad +\Bigg(\!\!-\ge\Big(4\ge^2+6\ge+12\ge\getpa+\getpa(9+8\getpa)+8(\getpe)^2+2\gx^\parallel-3\chi\Big)
\!+\!\sqrt{\frac{2\ge}{\kappa}}\frac{W_{221}}{H^2}\nn\\
&&\qquad\qquad\quad -3(\getpa)^2-3(\getpe)^2-2\getpa\gx^\parallel-2\getpe\gx^\perp\Bigg)\zeta_2^2\zeta_1
-2(\ge+\getpa)\zeta_2\partial^i\lambda\partial_i\zeta_2\nn\\
&&\qquad\qquad +4\getpe\zeta_2\partial^i\lambda\partial_i\zeta_1
+\frac{1}{2}\zeta_1(\partial^i\partial^j\lambda\partial_i\partial_j\lambda-(\partial^2\lambda)^2)\Bigg]
+a\ge^2\frac{1}{H}\zeta_1\Big[(\partial\zeta_1)^2\!+\!(\partial\zeta_2)^2
\Big]\nn\\
&&\qquad -\frac{\delta L_2}{\delta\zeta_1}\Big(\frac{\ge+\getpa}{2}\zeta_1^2-\getpe\zeta_1\zeta_2
+\dot{\zeta}_1\zeta_1-\frac{1}{4a^2H^2}(\partial\zeta_1)^2
+\frac{1}{4a^2H^2}\partial^{-2}\partial^i\partial^j(\partial_i\zeta_1\partial_j\zeta_1)\nn\\
&&\qquad\qquad\quad +\frac{1}{2}\partial^i\zeta_1\partial_i\lambda
-\frac{1}{2}\partial^{-2}\partial^i\partial^j(\partial_i\lambda\partial_j\zeta_1)\Big)\!-\!\frac{\delta L_2}{\delta\zeta_2}
\lh\!(\ge+\getpa)\zeta_1\zeta_2+\dot{\zeta}_2\zeta_1+\frac{\getpe}{2}\zeta_1^2\rh\!\!\Bigg\}\nn\\
&&
\ea
where $\delta L_2/\delta\zeta_m$ are the first-order equations of motion. 
We can further integrate by parts the rest of the action to simplify it and prove that it takes the form of the 
flat gauge action (\ref{sflatap}), as expected since the action should be gauge-invariant.
The terms involving $\lambda$ along with the terms with 
space gradients vanish outside the horizon in the long-wavelength approximation, since $\lambda$ is equal to the 
first-order super-horizon energy constraint (\ref{mc}).

Finally we include the second-order fields. The extra terms in the action are
\ba
\tilde{S}_{3(2)}&\!\!\!\!\!=&\!\!\!\!\!\frac{1}{2}\!\int\!\!\d^4x\Bigg\{\!a^3e^{3\zeta_1}\!\Big[(\frac{1}{H}\!+\!\tilde{N}_1)
(-W_A\tvarphi^A_{(2)}\!-\!W_{AB}
\tvarphi^A_{(2)}\tvarphi^B)\!+\!H(1\!-\!H\tilde{N}_1)(\dphi_A\tdvphi^A_{(2)}\!+\!\tdvphi_A\tdvphi^A_{(2)})\nn\\
&&\qquad\qquad\qquad -H\dot{\phi}^A\partial^i\tilde{\psi}\partial_i\tilde{\varphi}_{(2)}+2H\dot{\zeta}_{1(2)}\partial^2\tilde{\psi}\Big]\nn\\
&&
\qquad\quad -a\frac{1}{H}\Big[-\partial^i\zeta_1\partial_i\zeta_{1(2)}+\dot{\zeta}_1\partial^2\zeta_{1(2)}
+\partial^i\tilde{\varphi}^A\partial_i\tilde{\varphi}_{A(2)}\Big]\Bigg\},\label{s32}
\ea
where $\varphi^A$ without a subscript always denotes the first-order perturbation.  
After performing integrations by parts we find
\ba
\tilde{S}_{3(2)}=\int\d^4x\Bigg\{\frac{\delta L_2}{\delta\zeta_1}\lh\frac{\zeta_{1(2)}}{2}+\frac{\tilde{Q}_{1(2)}}{2}\rh
+\frac{\delta L_2}{\delta\zeta_2}\frac{\tilde{Q}_{2(2)}}{2}\Bigg\}.\label{s32s}
\ea

Next, we perform the same calculation for the flat gauge, starting from
\ba
\hat{S}_{3(1)}&\!\!\!\!\!=&\!\!\!\!\!\frac{1}{2}\!\int\!\d^4x\Bigg\{\!a^3\Big[ \Big(\frac{1}{H}+\hat{N}_1\Big)
           \Big(-2W-2W_A\hat{\varphi}^A-W_{AB}\hat{\varphi}^A\hat{\varphi}^B-\frac{1}{3}W_{ABC}\hat{\varphi}^A\hat{\varphi}^B
\hat{\varphi}^C\Big)\nn\\
 &&\qquad\qquad\;+H(1-H\hat{N}_1+H^2\hat{N}_1^2
-H^3\hat{N}_1^3
)\Big(-6+\dot{\phi}^2+2\dot{\phi}^A\dot{\hat{\varphi}}_A+\dot{\hat{\varphi}}^2
+4\partial^2\hat{\psi}\nn\\
&&\qquad\qquad\qquad\qquad\qquad\qquad+\partial^i\partial^j\hat{\psi}\partial_i\partial_j\hat{\psi}-(\partial^2\hat{\psi})^2
-2\partial^i\hat{\psi}\dot{\phi}^A\partial_i\hat{\varphi}_A
-2\partial^i\hat{\psi}\dot{\hat{\varphi}}^A\partial_i\hat{\varphi}_A\Big)\Big]\nn\\
&&\qquad\quad-a\hat{N}_1\partial^i\hat{\varphi}^A\partial_i\hat{\varphi}_A\Bigg\},
\ea
again taking into account that $\hat{N}_2$ multiplies the first-order energy constraint and thus we set it to zero. 
We find using the definition of $\zeta_m$, along with (\ref{fieldeqTz}), (\ref{srvar}) and (\ref{difzflat})
\ba
\hat{S}_{3(1)}&=&\!\!\int\!\!\d^4x\Bigg\{\!a^3\!\ge H\Bigg[\ge\zeta_1(\dot{\zeta}_1^2+\dot{\zeta}_2^2)
-2\dot{\zeta}_2\partial^i\lambda\partial_i\zeta_2-2\dot{\zeta}_1\partial^i\lambda\partial_i\zeta_1
\nn\\
&&\qquad\qquad\quad+2\ge(\ge+\getpa)\zeta_1\zeta_2\dot{\zeta}_2
+2\ge\getpe\zeta_1^2\dot{\zeta}_2 
+\ge(3\getpa+2\ge)\zeta_1^2\dot{\zeta}_1 \nn\\
&&\qquad\qquad\quad+\Bigg(\!\!\sqrt{\frac{2\ge}{\kappa}}\frac{W_{211}}{H^2}-2\ge(\ge\getpe+\getpa\getpe+\gx^\perp+3\getpe)\Bigg)\zeta_1^2\zeta_2
+\sqrt{\frac{2\ge}{\kappa}}\frac{W_{222}}{3H^2}\zeta_2^3\nn\\
&&\qquad\qquad\quad+\Bigg(\!\!\sqrt{\frac{2\ge}{\kappa}}\frac{W_{221}}{H^2}+\ge\lh-3(\getpe)^2+(\ge+\getpa)^2\rh+3\ge(\chi-\ge-\getpa)\Bigg)\zeta_2^2\zeta_1\nn\\
&&\qquad\qquad\quad+\lh\!\!\sqrt{\frac{2\ge}{\kappa}}\frac{W_{111}}{3H^2}-\ge\lh\gx^\parallel+3\getpa-(\getpe)^2-(\getpa)^2\rh\rh\zeta_1^3
\nn\\
&&\qquad\qquad\quad-2(\ge+\getpa)\zeta_2\partial^i\lambda\partial_i\zeta_2+4\getpe\zeta_2\partial^i\lambda\partial_i\zeta_1
+\frac{1}{2}\zeta_1(\partial^i\partial^j\lambda\partial_i\partial_j\lambda-(\partial^2\lambda)^2)\Bigg]
\nn\\
&&\qquad\quad +\frac{a\ge^2}{H}\zeta_1\Big((\partial\zeta_1)^2+(\partial\zeta_2)^2\Big)\Bigg\}.\label{sflatap}
\ea

Finally we include the second-order fields. The surviving terms in the action are
\ba
\hat{S}_{3(2)}&\!\!\!\!\!=&\!\!\!\!\!\frac{1}{2}\!\int\!\!\d^4x\Bigg\{\! a^3\Big[(\frac{1}{H}+\hat{N}_1)(-W_A\hat{\varphi}^A_{(2)}-W_{AB}
\hat{\varphi}^A_{(2)}\hat{\varphi}^B)
+H(1-H\hat{N}_1)(\dphi_A\dot{\hat{\varphi}}^A_{(2)}+\dot{\hat{\varphi}}_A\dot{\hat{\varphi}}^A_{(2)})\nn\\
&&\qquad\qquad\; -H\dot{\phi}^A\partial^i\hat{\psi}\partial_i\hat{\varphi}_{A(2)}\Big]
-a\frac{1}{H}\partial^i\hat{\varphi}^A\partial_i\hat{\varphi}_{A(2)}\Bigg\}
\ea
and they can be rewritten as 
\ba
\hat{S}_{3(2)}=\int\d^4x\Bigg\{\frac{\delta L_2}{\delta\zeta_1}\frac{\hat{Q}_{1(2)}}{2}
+\frac{\delta L_2}{\delta\zeta_2}\frac{\hat{Q}_{2(2)}}{2}\Bigg\}.
\ea

In the last part of this appendix we consider the tensor scalar part of the action. There will be no contributions from the 
second-order fields, since these 
cancel due to $\gamma_{ij}$ being transverse. We start from the action for two scalar and one tensor modes in the 
uniform energy-density gauge
\ba
\tilde{S}_{\zeta\zeta\gamma}&=&\int\d^4x\Big\{\frac{a}{H}\Big[-2\gamma_{ij}\partial^i\dot{\zeta}_1\partial^j\zeta_1
-\gamma_{ij}\partial^i\zeta_1\partial^j\zeta_1+\ge\gamma_{ij}\partial^i\zeta_2\partial^j\zeta_2\Big]\nn\\
&&\qquad\quad +\frac{1}{2}a^3H\Big[-(3\zeta_1-\dot{\zeta}_1)\dot{\gamma}_{ij}\partial^i\partial^j\tilde{\psi}
+\partial_k\gamma_{ij}\partial^i\partial^j\tilde{\psi}\partial^k\tilde{\psi}\Big]\Big\},
\ea
which after integrations by parts becomes
\ba
\tilde{S}_{\zeta\zeta\gamma}&\!\!\!=&\!\!\!\int\!\!\d^4x\Big\{\!a^3H\Big[
\frac{\ge}{2}\dot{\gamma}_{ij}\partial^i\zeta_1\partial^j\lambda
+\frac{1}{4}\partial^2\gamma_{ij}\partial^i\lambda\partial^j\lambda\Big]
+\frac{a}{H}\ge\gamma_{ij}\Big[\partial^i\zeta_1\partial^j\zeta_1+
\partial^i\zeta_2\partial^j\zeta_2\Big]\nn\\
&&\qquad\;+\frac{\delta L_{2\gamma}}{\delta\gamma_{ij}}\Big(\frac{1}{a^2H^2}\partial_i\zeta_1\partial_j\zeta_1
-(\partial_i\zeta_1\partial_j\lambda+\partial_j\zeta_1\partial_i\lambda)\Big)
+\frac{\delta L_2}{\delta\zeta_1}\frac{1}{4}\partial^{-2}(\dot{\gamma}_{ij}\partial^i\partial^j\zeta_1)\Big\}.\nn\\
&&\label{szzg}
\ea
In the flat gauge one finds directly after substitution into (\ref{actionexact}) the first line of (\ref{szzg}), so that 
there are no redefinitions.

Finally we calculate the part of the action consisting of one scalar and two tensor modes, starting from the uniform 
energy-density gauge:
\be
\tilde{S}_{\zeta\gamma\gamma}=\frac{1}{2}\int\d^4x\Big\{a^3H\Big[\frac{1}{4}(3\zeta_1-\dot{\zeta}_1)
(\dot{\gamma}_{ij})^2-\frac{1}{2}\dot{\gamma}_{ij}
\partial_k\gamma^{ij}\partial^k\tilde{\psi}\Big]-\frac{a}{4H}(\zeta_1+\dot{\zeta}_1)(\partial_k\gamma_{ij})^2\Big\}
\ee
or equivalently
\be
\tilde{S}_{\zeta\gamma\gamma}=\int\d^4x\Big\{-\zeta_1\dot{\gamma}_{ij}\frac{\delta L_{2\gamma}}{\delta\gamma_{ij}}
+a^3H\Big[\frac{\ge}{8}\zeta_1 (\dot{\gamma}_{ij})^2-\frac{1}{4}\dot{\gamma}_{ij}\partial_k\gamma^{ij}\partial^k\lambda\Big]
+\frac{a}{8H}\ge\zeta_1(\partial_k\gamma_{ij})^2\Big\}.
\ee
In the flat gauge we find directly
\be
\hat{S}_{\zeta\gamma\gamma}=\frac{1}{2}\int\d^4x\Big\{a^3H\Big[\frac{\ge}{4}\zeta_1(\dot{\gamma}_{ij})^2-\frac{1}{2}\dot{\gamma}_{ij}
\partial_k\gamma^{ij}\partial^k\lambda\Big]+\frac{a}{4H}\ge\zeta_1(\partial_k\gamma_{ij})^2\Big\}.
\ee

The three tensor modes action does not contain any redefinitions. For details the reader may look in \citep{Maldacena:2002vr}.

\chapter{Momentum dependence of the bispectrum in two-field inflation}
\label{TvT3app}

This appendix contains the paper \citep{TvT3}, except for the conclusions
that were used as a summary in section~\ref{summTvT3}. In addition, most
of section 2 of the paper has also been removed (and the remainder combined
with section 3), as it only summarized equations
that have been given earlier in this thesis. The paper was written in
collaboration with Eleftheria Tzavara.

We examine the momentum dependence of the bispectrum of two-field 
inflationary models within the long-wavelength formalism. We determine 
the sources of scale dependence in the expression for the parameter of 
non-Gaussianity $\fnl$ and study two types of variation of the momentum 
triangle: changing its size and changing its shape. We introduce two 
spectral indices that quantify the possible types of momentum dependence 
of the local type $\fnl$ and illustrate our results with examples.

\section{Introduction}\label{intro}

The study of inflationary non-Gaussianities and their impact on 
the cosmic microwave background has been an important 
subject of cosmological research in recent years. In nine years of WMAP data 
\citep{Bennett:2012zja} and one year of data from the Planck satellite 
\citepalias{planck2013-24} no primordial non-Gaussianity of the local and equilateral
types (see below) was observed, and constraints have tightened considerably.
Next year's Planck release\footnote{Again I have decided to keep the original
  text and references from the paper, which dates from 2012-13.} is expected
to put even tighter constraints on 
those types of non-Gaussianity, as well as investigate many additional types
(which differ in their momentum dependence).
The importance of the current constraints and a future possible detection
or further improvements of the constraints lies in the fact that they allow
us to discriminate between different classes of models of inflation, since these
predict different types and amounts of non-Gaussianity.

There are basically two distinct types of non-Gaussianity that are most important from the point of view of inflation: the equilateral type 
produced at horizon-crossing, which has a quantum origin and is  
maximal for equilateral triangle configurations \citep{Creminelli:2005hu}, 
and the local type produced outside the inflationary horizon due to the existence of interacting fields. The latter is maximized for squeezed triangles, 
i.e.\  isosceles triangles with one side much smaller than the other two \citep{Komatsu:2001rj,Babich:2004yc}. 
The first type is known to be slow-roll suppressed for single-field models with standard kinetic terms and trivial field metric 
\citep{Maldacena:2002vr}. 
On the other hand, some models with non-standard kinetic terms coming from higher-dimensional cosmological models are known to produce non-Gaussianity 
of the equilateral type so large that it is not compatible with WMAP and Planck observations 
\citep{Alishahiha:2004eh,Silverstein:2003hf,Mizuno:2009cv,Mizuno:2010ag}, thus leading people to consider an extra field 
in order to achieve smaller values of the parameter $\fnl$ of non-Gaussianity,
see e.g.~\citep{RenauxPetel:2009sj}.

Non-Gaussianity of the squeezed type can be found naturally in multiple-field models of inflation \citep{RSvT4,Bernardeau:2002jy}, 
due to the sourcing of the adiabatic mode by the isocurvature components outside the horizon. 
For single-field models this is obviously impossible due 
to the absence of isocurvature modes. There has been much study of two-field models 
\citep{Seery:2005gb,Kim:2006te,Battefeld:2006sz,Battefeld:2007en,Langlois:2008vk,
  Cogollo:2008bi,Vernizzi:2006ve,RenauxPetel:2009sj,Peterson:2010mv,TvT1},
being the easiest to investigate, in the hope of 
finding a field potential that can produce local non-Gaussianity large enough to be measurable in the near-future. It proves to be non-trivial to 
sustain the large non-Gaussianity produced during the turn of the fields until the end of inflation.

Non-Gaussianity produced at horizon-crossing is known to be momentum-dependent. 
The scale dependence of the equilateral $\fnl$ produced for example from DBI inflation \citep{Langlois:2008qf,Arroja:2008yy,Mizuno:2009cv,Cai:2009hw,Senatore:2010wk}, has been examined both theoretically 
\citep{Chen:2005fe,Khoury:2008wj,Byrnes:2009qy,Leblond:2008gg} and in terms of observational forecasts \citep{LoVerde:2007ri,Sefusatti:2009xu}. 
In this paper we are going to study the scale dependence of local-type models that has not been studied as much. 
Squeezed-type non-Gaussianity, produced outside the horizon, is usually associated with a parameter of non-Gaussianity 
$\fnl^\mathrm{local}$ that is local in real space, and therefore free of any explicit momentum dependence, defined through 
$\zeta(x)=\zeta_{L}(x)-(3/5)\fnl^\mathrm{local}(\zeta_{L}(x)^2-\langle\zeta_L(x)\rangle^2)$, where $\zeta_{L}$ is the linear Gaussian part. 
Nevertheless, calculations of $\fnl$ for several types of multiple-field models (see e.g.\  \citep{Vernizzi:2006ve,Byrnes:2008wi,
TvT1}) show that there is always a momentum dependence inherited from the horizon-crossing era, which can in principle result in 
a tilt of $\fnl$.  
When a physical quantity exhibits such a tilt one usually introduces a spectral index, as for example in the case of the power spectrum.  
The observational prospects of the detection of this type of scale dependence of local $\fnl$ were studied in \citep{Sefusatti:2009xu}.  
Only recently spectral indices for $\fnl$ were defined in \citep{Byrnes:2010ft,Byrnes:2009pe,Byrnes:2012sc}, keeping constant the shape of the 
triangle or two of its sides, within the 
$\delta N$ formalism.  
Note, however, that most theoretical predictions have considered equilateral triangles for simplicity, even though the local-type 
configuration is maximal on squeezed triangles. 
If one were to calculate a really squeezed triangle, then $\fnl^\mathrm{local}$ acquires some intrinsic momentum dependence 
due to the different relevant scales, as was shown in \citep{TvT1}. 

It is both these effects we want to study in this paper: 
on the one hand the tilt of $\fnl$ due to the background evolution at horizon-crossing and on the other hand the impact of the shape of the triangle on $\fnl$.  
In order to do that in a concrete way, such that these effects do not mix, we define two independent spectral indices, each one quantifying different deformations of the momentum triangle. 
Moreover, having an exact expression of $\fnl$ for an isosceles triangle, we are able to study and understand for the first time the origin of both types 
of momentum dependence of $\fnl$. We also provide analytical estimates for the quadratic model (which actually hold for any equal-power sum model) that we use in this paper to illustrate our results. 

The paper is organised as follows. 
In section \ref{sour} 
we present the long-wavelength formalism results and discuss the sources of scale 
dependence in the expression for $\fnl$. We also introduce two  
spectral indices, able to quantify the effects of different triangle 
deformations. 
In section \ref{conf} we study the scale dependence for triangles of constant shape but of varying size, which is mainly due
to horizon-crossing quantities, while 
in section \ref{shape} we study the scale dependence related to the shape of 
the triangle.

\section{Sources of scale dependence}\label{sour}

In addition to the power spectrum we can gain more information from the CMB 
by studying the Fourier transform of the three-point correlation function, 
\ba
\langle \zeta_{1\vca{k_1}} \zeta_{1\vca{k_2}} \zeta_{1\vca{k_3}}\rangle
& \equiv & (2\pi)^{-3/2}\delta^3 (\sum_s\vca{k_s})B_{\zeta}(k_1,k_2,k_3),
\ea
where $B_{\zeta}$ is the bispectrum. Because of the overall $\delta$-function we see that the vectorial sum of the 
three $k$-vectors has to be zero. In other words, the three $k$-vectors form a triangle. 
The amplitude of the bispectrum can 
provide additional constraints on the slow-roll parameters of a given type 
of inflationary model. The profile of the bispectrum, i.e.\  the shape of the momentum triangle, 
gives information on the type of the 
inflationary model itself. For example, models with higher-order kinetic 
terms produce a bispectrum of the equilateral type (see e.g.~\citep{Komatsu:2010hc}), mainly due to quantum 
interactions at horizon crossing. By equilateral type we mean a 
bispectrum that becomes maximal for equilateral triangles. On the other hand,  
canonical multiple-field inflation models predict a bispectrum of the local type. 
This arises from non-linearities of the form $\zeta_1=\zeta_{1L}-(3/5)\fnl(\zeta_
{1L}^2-\langle\zeta_{1L}\rangle^2)$ ($\zeta_{1L}$ being the first-order adiabatic perturbation) that 
are created classically outside the horizon, leading to a bispectrum of the form
\be
B_{\zeta}(k_1,k_2,k_3)=-\frac{6}{5}\fnl\lh \frac{2\pi^2}{k_1^3}
\mathcal{P}_{\zeta}(k_1)\frac{2\pi^2}{k_2^3}\mathcal{P}_{\zeta}(k_2)
+\lh k_2\leftrightarrow k_3\rh
+\lh k_1\leftrightarrow k_3\rh\rh,\label{bisp}
\ee
where $\fnl$ is usually assumed to be constant.
This bispectrum becomes maximal for a squeezed triangle, i.e.\  
a triangle with two sides almost equal and much larger than the 
third one. 
As we will discuss in the rest of the paper, $\fnl$ is not actually a constant,
but depends on the size and shape of the momentum triangle.

\begin{figure}
\begin{center}
\includegraphics[trim=0 0 400 0,width=0.9\textwidth]{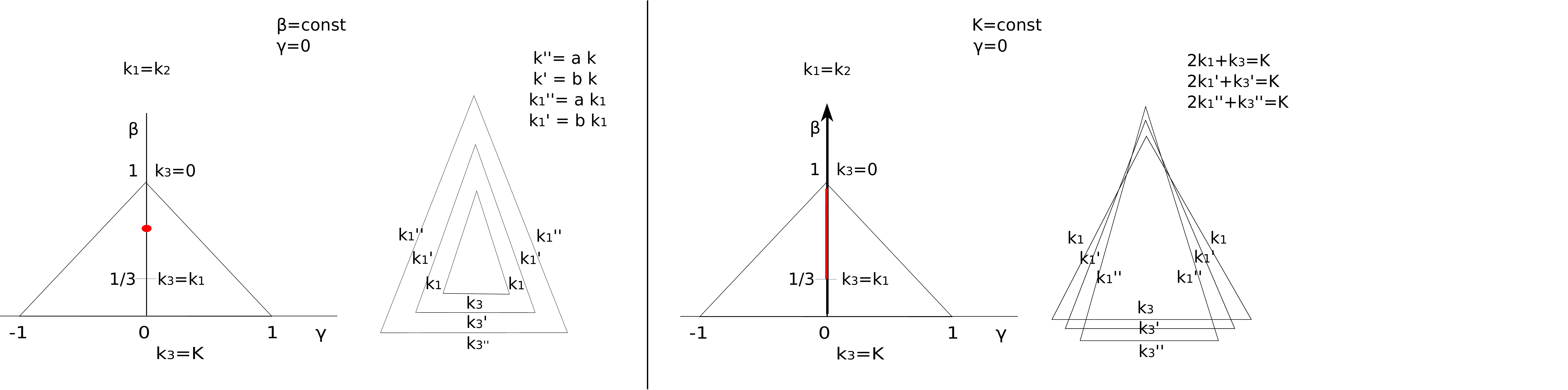}
\end{center}
\caption{The types of deformations of the momentum triangle we are 
considering. Left: Conformal transformation of the 
triangle. In the $\gamma,\beta$ plane this corresponds to a 
constant (red) point. Right: Keeping the perimeter of the triangle 
$K$ constant we change the shape of an isosceles 
$\gamma=0$ triangle, moving along the bold red line for 
$k_1=k_2\geq k_3$.}
\label{fig01}
\end{figure}

In order to study the dependence of the non-Gaussianity on the 
shape of the triangle, instead 
of using $k_1,k_2$, and $k_3$ we will use the variables introduced in 
\citep{RSvT1,Fergusson:2008ra}, 
\ba
K=\frac{k_1+k_2+k_3}{2},\qquad\gamma=\frac{k_1-k_2}{K},
\qquad\beta=-\frac{k_3-k_1-k_2}{2K},\label{newvar}
\ea
which correspond to the perimeter of the triangle and two scale ratios 
describing effectively the angles of the triangle. They have the following 
domains: $0\leq K\leq \infty,\ 0\leq\beta\leq1$ and 
$-(1-\beta)\leq\gamma\leq1-\beta$, see figure~\ref{fig01}. 
As one can check from 
the above equations,  
the local bispectrum becomes maximal for $\beta=1$ and $\gamma=0$,   
or $\beta=0$ and $\gamma=\pm 1$, i.e.\  for a squeezed triangle. 
In this paper we always assume 
$k_1=k_2$, dealing only with equilateral or isosceles triangles
(note that the relation $k_1=k_2$ is satisfied by definition for both 
equilateral and squeezed triangles).  
The two scales of the triangle $k_3 \equiv k \le k' \equiv k_1 = k_2$ can be expressed 
in terms of the new parameters $\beta$ and $K$ as 
\be
k=(1-\beta)K
\qquad\mathrm{and}
\qquad k'=\frac{1+\beta}{2}K,\label{par}
\ee
while $\gamma=0$. The condition $k\le k'$ means that we only have to study acute isosceles triangles $1/3\leq\beta\leq1$.

\subsection{Long-wavelength results}

In this paper we use the long-wavelength formalism to study the 
parameter of 
non-Gaussianity $\fnl$ and its scale dependence. 
The non-Gaussianity parameter for an isosceles triangle of 
the form $k_1=k_2\equiv k'\geq k_3\equiv k$ was found in \citep{TvT1} to be
\ba
	-\frac{6}{5}\fnl&=&\frac{-2\bv_{12k'}/[1+(\bv_{12k'})^2]}
{1+(\bv_{12k'})^2+2\frac{\gamma_k^2}{\gamma_k'^2}[1+(\bv_{12k})^2]}
	\Bigg[\!\bv_{12k'}\!\Bigg
(\!g_\mathrm{sr}
	(k',k')\!
	+\!g_\mathrm{iso}(k',k')\!+\!g_\mathrm{int}(k',k')\!\Bigg) \nn\\	
&&\qquad\qquad\qquad\qquad\;+2\frac{\gamma_k^2}{\gamma_k'^2}\bv_{12k}\Bigg(g_\mathrm{sr}(k',k)\!+\!g_\mathrm{iso}(k',k)
\!+\!g_\mathrm{int}(k',k)\!\Bigg)\Bigg],
\label{fNLgeni}
\ea
where $\fnl=\fnl(t;t_{k'},t_k)$ depends on $t_{k'}$ and $t_k$, denoting 
the horizon-crossing times of the two scales $k'$ and $k$ of the triangle,
respectively. 
This result is exact and valid beyond the slow-roll approximation after horizon-crossing. 
All quantities appearing in this formula will be explained below.

The quantity $\bv_{12}$ is a 
transfer function showing how the isocurvature mode (denoted by the subscript $2$) sources  
the adiabatic component $\zeta_1$. 
In the following two more transfer 
functions will appear, namely $\bv_{22}$ and $\bv_{32}$, showing how the isocurvature mode sources the isocurvature 
component $\zeta_2$ and the velocity of the isocurvature component 
$\theta_2 \equiv \dot{\zeta_2}$, respectively. $\bv_{a2}$ is a function of the 
horizon-exit time $t_k$ of the relevant perturbation of scale $k$ and it 
also evolves with time $t$, at least during inflation. In (\ref{fNLgeni}) as 
well as in the formulas that follow, $\bv_{a2k}\equiv\bv_{a2}(t,t_k)$.  
The indices $a,b$ take the values $1,2,3$, indicating respectively the 
adiabatic perturbation $\zeta_1$, the isocurvature perturbation $\zeta_2$,
and the isocurvature velocity $\theta_2$.\footnote{Due to the exact relation
$\theta_1 = 2 \eta^\perp \zeta_2$, there is no need to consider the velocity
of the adiabatic perturbation $\theta_1$ as an additional variable
\citep{RSvT4}.} 
$\bv_{a2}$ comes from the 
combination of the Green's functions $G_{a2}$ and $G_{a3}$ of the system of equations for the 
super-horizon perturbations 
(for the system of equations that 
the Green's functions obey see (\ref{greent})):
\be
\bv_{a2}(t,t_k) = G_{a2}(t,t_k)-\chi_k G_{a3}(t,t_k). 
\label{green}
\ee
The quantity $\gamma_k$ in (\ref{fNLgeni}) is defined as 
$\gamma_k\equiv-\kappa H_k/(2k^{3/2}\sqrt{\ge_k})$ (and is not related to the $\gamma$ defined in (\ref{newvar})).

Except for the overall factor, $\fnl$ has been split into three 
contributions: $g_\mathrm{sr}$, $g_\mathrm{iso}$ and $g_\mathrm{int}$.\footnote{In 
\citep{TvT1} we had also a fourth contribution $g_\mathrm{k}$, denoting the terms that vanish for an equilateral triangle. 
Here we have incorporated these terms in $g_\mathrm{sr}$ (the last two lines), since they are also slow-roll suppressed.}
$g_\mathrm{sr}$ is a term that is slow-roll suppressed, since it depends only on horizon-exit quantities, where by assumption slow-roll holds,
\ba
g_\mathrm{sr}({k}_1,{k}_2)&\!\!\!\!=&\!\!\!\!\getpe_{k_1}\Bigg(\frac{G_{22k_1k_2}\bv_{12k_1}}{2}   
-\frac{1}{\bv_{12k_2}}-\frac{G_{22k_1k_2}}{2\bv_{12k_1}} 
\Bigg)+\frac{3\chi_{k_2}}{4}G_{33k_1k_2}
-\frac{3}{2}(\ge_{k_1}+\getpa_{k_1})G_{22k_1k_2}\nn\\
&&\!\!\!\!
+\frac{\chi_{k_1}}{4}\Bigg(2\frac{\bv_{12k_1}}{\bv_{12k_2}}+
G_{22k_1k_2}\Bigg)-\frac{\ge_{k_1}+\getpa_{k_1}}{2(\widetilde{v}_{12})^2}
\\
&&\!\!\!\!
+\frac{G_{13}(t,t_{k_1})}{2}\Bigg[\frac{3\lh\chi_{k_1}G_{22k_1k_2}
\!-\!\chi_{k_2}G_{33k_1k_2}\rh}{2\bv_{12k_1}}
+G_{32k_1k_2}\lh\frac{3+\ge_{k_1}+2\getpa_{k_1}}{2\bv_{12k_1}}
+\getpe_{k_1}\!\rh\!\Bigg]\nn\\
&&\!\!\!\!
-\frac{3}{4}G_{32k_1k_2}
-\frac{1}{2}G_{12k_1k_2}\lh\!
\ge_{k_1}+\getpa_{k_1}+2\getpe_{k_1}-\frac{\chi_{k_1}}{2}\lh1+\bv_{12k_1}\rh
+\frac{\ge_{k_1}+\getpa_{k_1}}{\bv_{12k_1}}\!\rh.\nn
\ea
Here we introduce some new notation,
\begin{displaymath}
\begin{array}{l}
(\widetilde{v}_{12})^2\equiv\bv_{12k_1}\bv_{12k_2},\qquad 
(\widetilde{v}_{22})^2\equiv\bv_{22k_1}\bv_{22k_2},\qquad
(\widetilde{v}_{32})^2\equiv\bv_{32k_1}\bv_{32k_2},\\  
\widetilde{v}_{22}\widetilde{v}_{32}\equiv\frac{1}{2}(\bv_{22k_1}\bv_{32k_2}
+\bv_{22k_2}\bv_{32k_1}),
\end{array}
\end{displaymath}
and also $G_{abk_1k_2}\equiv G_{ab}(t_{k_1},t_{k_2})$. Moreover, we assume
$k_1 \geq k_2$.
$g_\mathrm{sr}$ is the only term from which a (small) part survives in the 
single-field limit, i.e.\  in the 
limit where $\bv_{12}=0$ at all times. 
For the equilateral case $k'=k$ the two last lines of $g_\mathrm{sr}$ are zero,%
\footnote{In addition, in the equilateral case $k'=k$ the $\gamma_k$ ratios 
in (\ref{fNLgeni}) reduce to $1$ and the two terms in the brackets of (\ref{fNLgeni}) become 
identical (apart from the factor $2$). See (\ref{fNLresult}--\ref{gisosrint})
for the full expression of $\fnl$ in the equilateral case.} 
since the Green's functions satisfy
\be
G_{ab}(t,t)=\delta_{ab}.
\label{Green_norm}
\ee

The contribution $g_\mathrm{iso}$ is a term that survives as long as the isocurvature modes are alive,
\be
g_\mathrm{iso}({k}_1,{k}_2)=(\ge+\getpa)(\widetilde{v}_{22})^2
+\widetilde{v}_{22}\widetilde{v}_{32}.\label{giso}
\ee
If at the end of inflation these are non-zero, $\fnl$ can still evolve 
afterwards and we cannot be sure that its value survives until today. Finally, 
$g_\mathrm{int}$ is given by
\be
g_\mathrm{int}({k}_1,{k}_2)=\!-\!\int_{t_{k_1}}^t\!\!\!\!\d t'\Big[2(\getpe)^2
(\widetilde{v}_{22})^2
\!+\!(\ge+\getpa)\widetilde{v}_{22}\widetilde{v}_{32}\!+\!(\widetilde{v}_{32})^2
\!-\!G_{13}(t,t')\widetilde{v}_{22}(\Xi\widetilde{v}_{22}+9\getpe\widetilde{v}_{32})
\Big]\label{gintTz2}
\ee
with   
\be
\Xi \equiv 12 \get^\perp \gc - 6 \get^\parallel \get^\perp
	+ 6 (\get^\parallel)^2 \get^\perp + 6 (\get^\perp)^3
- 2 \get^\perp \gx^\parallel- 2 \get^\parallel \gx^\perp
	- \sqrt{\frac{\ge}{2}}\frac{1}{\kappa H^2}(W_{211} + W_{222}),
\label{Ci}
\ee
where $W_{mnl}=W^{,ABC}e_{mA}e_{nB}e_{lC}$. 
It is from this integrated effect that any large, persistent non-Gaussianity 
originates, if we consider only models where the isocurvature modes have
vanished by the end of inflation.
For the analytical approximations that we will provide (in addition to the
exact numerical results), it is useful to note that within the slow-roll
approximation $g_\mathrm{int}$ can be rewritten as 
\be
g_\mathrm{int}(k_1,k_2) = \bv_{12k_1} G_{22 k_1 k_2} \left( -\eta^\perp_{k_1} 
+ \frac{(\epsilon_{k_1}
+\eta^\parallel_{k_1} -\chi_{k_1})\chi_{k_1}}{2\eta^\perp_{k_1}} \right) 
+ \tg_\mathrm{int}(k_1,k_2),
\label{tgint}
\ee
where $\tg_\mathrm{int}$ is another integral that is identically zero for the two-field
quadratic model, or even more generally for any two-field equal-power sum model 
(see section~\ref{eqpowsumsec} for details).

\subsection{Discussion}

Inspecting (\ref{fNLgeni}) ones sees that there are two sources of momentum 
dependence for $\fnl$: the slow-roll parameters 
at horizon-crossing and the Green's functions $G_{ab}$ or their combinations 
$\bv_{a2}$. In order to study their impact 
we shall use the quadratic model 
\be
W=\frac{1}{2}m_{\phi}^2\phi^2+\frac{1}{2}m_{\gs}^2\gs^2,\label{qua}
\ee
with $m_{\phi}/m_{\gs}=9$. The procedure to follow is to solve for the 
background quantities and then for the Green's functions 
in order to apply the formalism. The quadratic model's Green's functions can be found 
numerically, or even analytically within the slow-roll approximation, which is valid for a 
small mass ratio like the one we chose here. However, all our calculations in this paper 
are numerical and exact, without assuming the slow-roll approximation after horizon 
crossing. We only use the slow-roll approximation after horizon crossing for the
analytical approximations that we provide (e.g.\ eq.~(\ref{fnlfin})) and 
sometimes to clarify the physical interpretation of results (e.g.\ the use of 
(\ref{difu}) below to explain the behaviour of $\bv_{12}$).
Inflation ends at $t_f$ defined as the time when $\ge_f=1$. 
From now on a subscript $f$ 
will denote quantities evaluated at the end of inflation. 
We also define the scale that exited the 
horizon $60$ e-folds before the end of inflation as $k_{60}$ and use it as a reference 
scale, around which we perform our computations ($k_{60}$ being the scale that 
corresponds to the text books' minimal necessary amount of inflation). 

\begin{figure}
\begin{tabular}{cc}
\includegraphics[width=0.45\textwidth]{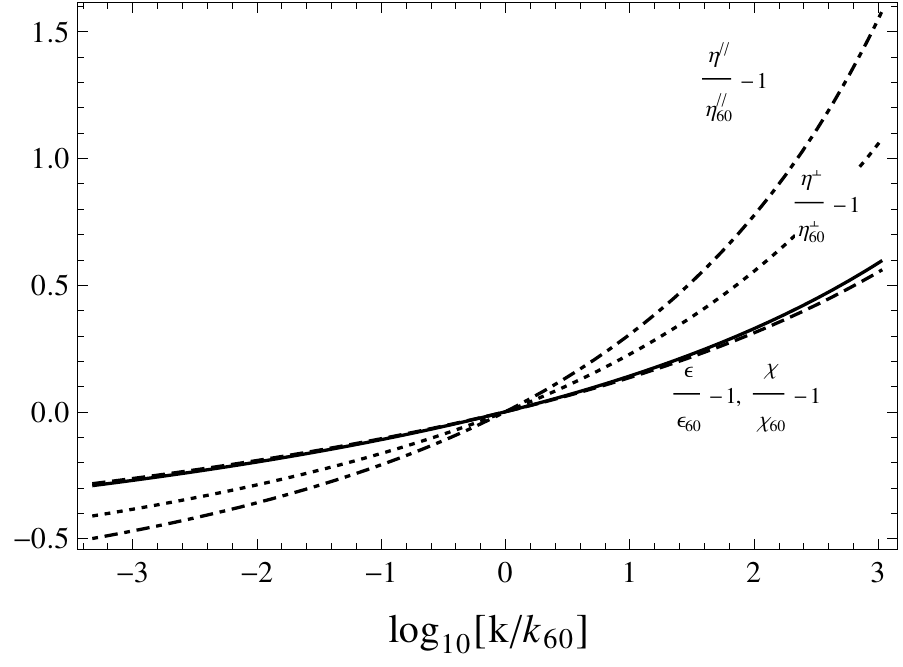}
&\includegraphics[width=0.48\textwidth]{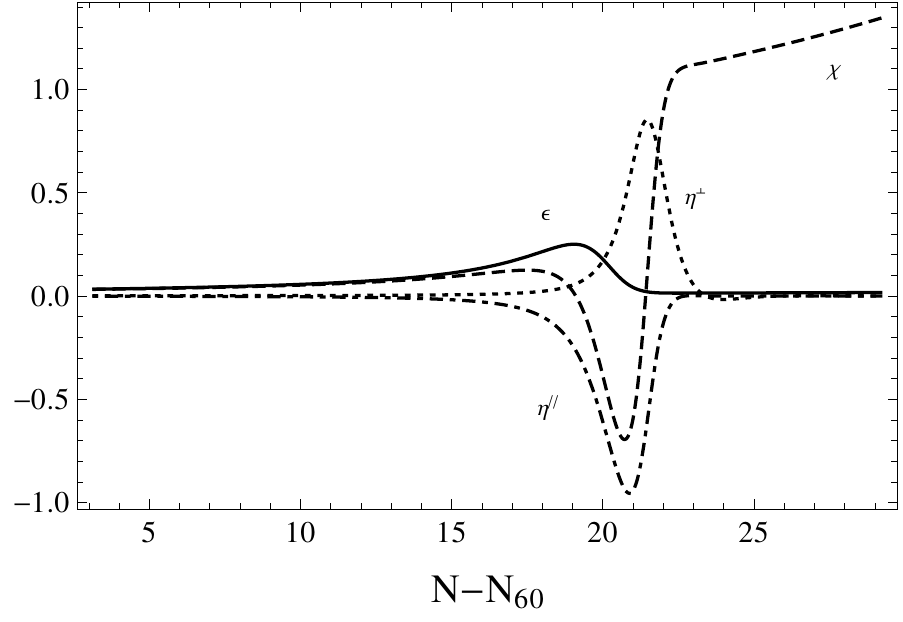}
\end{tabular}
\caption{Left: The relative change of the horizon-crossing first-order slow-roll 
parameters $\ge$ (solid curve), $\getpa$ (dot-dashed curve), $\getpe$ (dotted curve) and $\chi$ (dashed curve) 
at $t_k$ as a function of the 
ratio $k/k_{60}$ of the horizon-exit scale to the scale that left the horizon $60$ e-folds 
before the end of inflation, 
for the model (\ref{qua}) with mass ratio $m_{\phi}/m_{\gs}=9$. 
Right: The evolution of the first-order slow-roll parameters $\ge$ (solid curve), $\getpa$ (dot-dashed curve), $\getpe$ (dotted curve) and $\chi$ (dashed curve) as a function of the number of e-foldings $N-N_{60}$ for the time interval around the turning of the fields, for the same model.
}
\label{fig2}
\end{figure}

In figure \ref{fig2} we plot the first-order slow-roll parameters for a 
range of horizon-crossing times around $k_{60}$. While the heavy field rolls 
down its potential, the slow-roll parameters increase, reflecting the 
evolution of the background. This implies that 
$\fnl$, which is in general proportional to the 
slow-roll parameters evaluated at $t_k$ and $t_{k'}$, 
should increase as a function of $k$ and $k'$. 
This can easily be verified for the initial value of $f_{\mathrm{NL},in}$ at 
$t=t_{k'}$, which according to (\ref{fNLgeni}) with $\bv_{12k'}=0$ takes 
the value 
\be
-\frac{6}{5}f_{\mathrm{NL},in}=\ge_{k'}+\getpa_{k'}+\frac{2\frac{\gamma_k^2}{\gamma^2_{k'}}G_{12k'k}}{1+2\frac{\gamma_k^2}{\gamma^2_{k'}}
[1+(G_{12k'k})^2]}\getpe_{k'}G_{22k'k}.
\label{fnlin}
\ee

Apart from the slow-roll parameters the other source of momentum dependence for $\fnl$ lies in the Green's functions and 
particularly how their time evolution depends on the relevant horizon-crossing scale. The two main quantities that we 
need to study in order to understand their impact on $\fnl$ are the transfer functions $\bv_{12}$ and $\bv_{22}$. This is due to the fact 
that $\bv_{32}$ is slow-roll suppressed and the rest of the Green's functions appearing in (\ref{fNLgeni}) can 
be rewritten in terms of $\bv_{12}$ and $\bv_{22}$ within the slow-roll approximation (for details, see section~\ref{greensec}).  
In particular  $G_{a3}=G_{a2}/3$, $G_{32}(t,t_k)=-\chi(t) G_{22}(t,t_k)$ and hence $G_{a2} \approx \bv_{a2}$. Note that except for the era of 
the turning of the fields, the slow-roll assumption is a good approximation during inflation in this particular model.   
The slow-roll evolution equations for $\bv_{12k}$ and $\bv_{22k}$ are
\be
\frac{\d}{\d t}\bv_{12k}=2\getpe\bv_{22k}\qquad\mathrm{and}\qquad\frac{\d}{\d t}\bv_{22k}=-\chi\bv_{22k}.\label{difu}
\ee

As was discussed above, $\bv_{12}$ describes how the isocurvature mode sources the 
adiabatic one, while $\bv_{22}$ describes how the 
isocurvature mode sources itself.  
By definition $\bv_{12}(t_k,t_k)=0$ and $\bv_{22}(t_k,t_k)=1$ at horizon crossing, 
since no interaction of the different modes has yet occurred (see also 
(\ref{green}) and (\ref{Green_norm})). 
For the transfer functions of the adiabatic mode one finds that $\bv_{11}=1$ 
and $\bv_{21}=0$, since the curvature perturbation is conserved for purely 
adiabatic perturbations and adiabatic perturbations cannot source entropy 
perturbations. 
In order to better understand the 
role of the transfer functions, we can use the Fourier transformation of the 
perturbations (\ref{via1sol}) along with these last identities, to find
\ba
&&\zeta_{1}(t)=\int\frac{\d^3\vca{k}}{(2\pi)^{3/2}}\gamma_k\bv_{1m}\hat{a}^{\dagger}_m(\vca{k})
 e^{i \vca{k}\cdot \vca{x}}=\zeta_{1}(t_k)+\bv_{12}(t,t_k)\zeta_{2}(t_k),\nn\\
&&\zeta_{2}(t)=\int\frac{\d^3\vca{k}}{(2\pi)^{3/2}}\gamma_k\bv_{2m}\hat{a}^{\dagger}_m(\vca{k})
 e^{i \vca{k}\cdot \vca{x}}=\bv_{22}(t,t_k)\zeta_{2}(t_k),\label{ze}
\ea
where $\zeta_{m}$ with $m=1,2$ are the first-order adiabatic and isocurvature 
perturbation. 
 
\begin{figure}
\begin{tabular}{cc}
\includegraphics[width=0.46\textwidth]{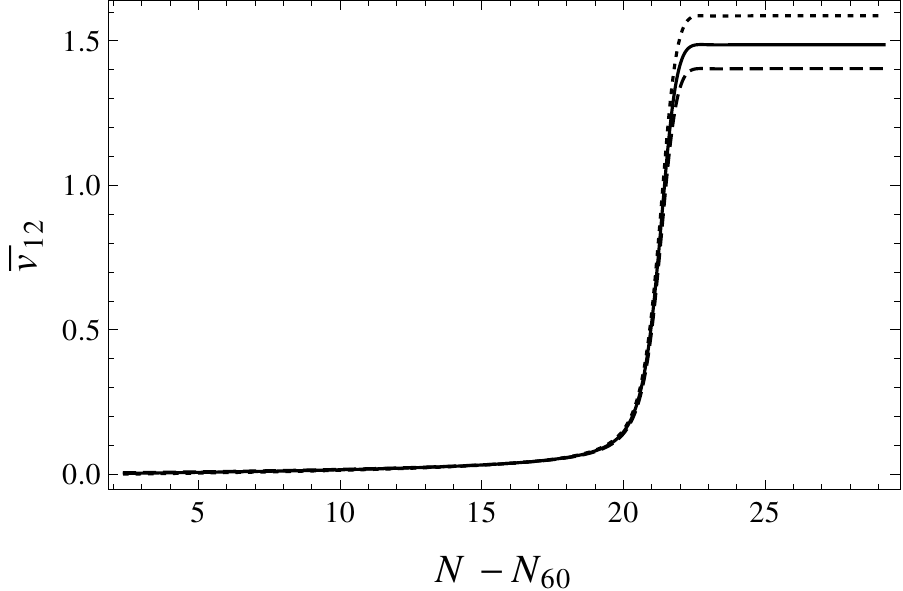}
&\includegraphics[width=0.46\textwidth]{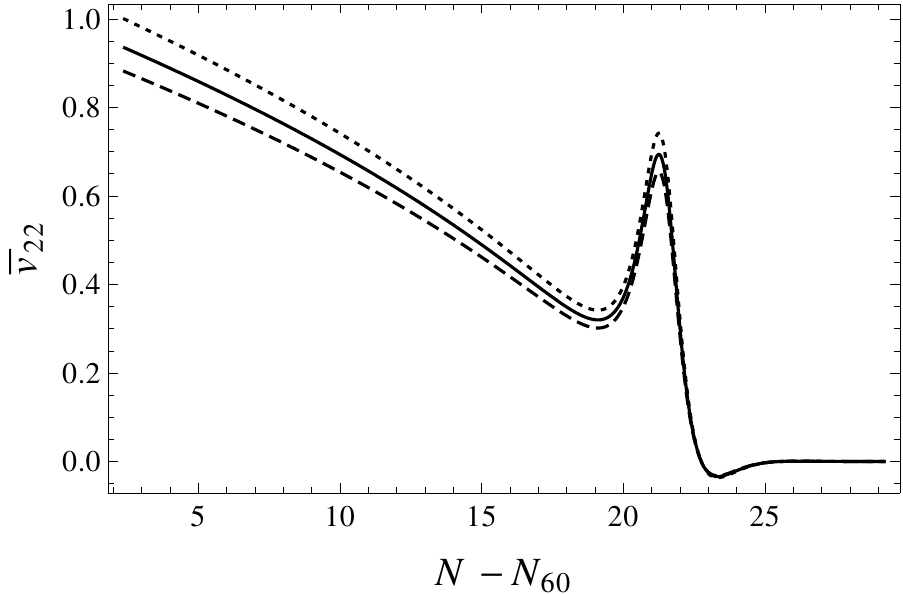}
\end{tabular}
\caption{The evolution of the transfer functions $\bv_{12}$ (left) and 
$\bv_{22}$ (right) as a function of the number 
of e-foldings $N-N_{60}$ for the time interval around the turning of the fields 
and for different horizon exit scales, varying 
from top to bottom as $k_{60}\times10$ (dotted curve), $k_{60}$ (solid curve) and $k_{60}/10$ (dashed curve), 
for the model (\ref{qua}) with mass ratio $m_{\phi}/m_{\gs}=9$.}
\label{fig1}
\end{figure}

Let us start by discussing the time evolution of $\bv_{12}$. Each one of the 
curves on the left-hand side of figure \ref{fig1} corresponds to the 
time evolution of $\bv_{12}$ for a different horizon-exit scale. 
At $t=t_k$, i.e.\  when the relevant mode $k$ exits the horizon, $\bv_{12k}=0$   
since the isocurvature mode has not had time to affect the adiabatic one. Outside 
the horizon and well in the slow-roll regime of the sole dominance of the heavy field, the
isocurvature mode sources the adiabatic one and the latter slowly increases. 
As time goes by, the heavy field rolls down its potential 
and the light field becomes more important. 
During this turning of the field trajectory, the slow-roll parameters suddenly
change rapidly, with important consequences for the evolution of the adiabatic
and isocurvature mode. The transfer function
$\bv_{12k}$ grows substantially during that era because of the increasing 
values of $\getpe$ in (\ref{difu}) as well as the growing contribution of $\bv_{22k}$, 
to become constant afterwards  
when the light field becomes dominant in an effectively single-field universe. 

Note that the earlier the mode exits the horizon, the smaller is the final $\bv_{12k}$. This is opposite to the behaviour of the initial 
value, just after horizon-crossing, 
when the earlier the scale exits the horizon the more has its adiabatic mode been sourced by the isocurvature one at a given time $t$, and hence the 
larger is its $\bv_{12k}$. 
This can be understood by the evolution equations of $\bv_{12k}$ and $\bv_{22k}$ 
in (\ref{difu}), 
showing that $\bv_{12k}$ is sourced by $\bv_{22k}$, which itself is a decreasing function 
of time, 
at least during eras when the universe is dominated by a single field (see 
the right-hand side of figure \ref{fig1}). If the equation (\ref{difu}) for $\bv_{12k}$ did not depend on $\getpe$, the $\bv_{12k}$ curves 
would never cross each other since they would be similar and only boosted by their horizon-crossing time shift. It is the increasing 
value of $\getpe$ 
that results in the larger values of $\bv_{12k}$ for larger $k$.

On the right-hand side of figure \ref{fig1} we show the evolution of $\bv_{22}$. 
According to (\ref{difu}) $\bv_{22}$ and hence the isocurvature mode evolves independently from the adiabatic mode. 
At horizon-crossing $t=t_k$, the transfer function $\bv_{22k}=1$.
Once outside the horizon, the isocurvature mode decays due to the small but positive value of $\chi$ (defined in (\ref{defchi})).
During the turning of the fields the slow-roll parameters evolve rapidly, thus leading to first an enhancement of $\bv_{22k}$ and then a diminution due to the varying value of $\chi$ in (\ref{difu}). 
As can be seen from the right-hand side plot in figure \ref{fig2}, during the turning $\chi$ first becomes negative and then positive. 
After the turning of the fields, the remnant isocurvature modes again decay and (for this model) at the end of inflation none are left.
The parameter $\chi$ plays a crucial role in the evolution of the isocurvature mode. 
It represents effectively the second derivative of the potential in the $22$ direction. 
Before the turning of the fields the trajectory goes down the potential in the 
relatively steep $\phi$ direction, which means that $W_{22}$ then corresponds
to the relatively shallow curvature in the direction of the light field 
$\sigma$ and hence $\chi$ is small. After the turning the trajectory goes along
the bottom of the valley in the $\sigma$ direction and $W_{22}$ corresponds with
the large curvature of the potential in the perpendicular direction, leading 
to large values of $\chi$. The negative values of $\chi$ during the turn come
from the contribution of $\getpa$ (see (\ref{defchi})).

\begin{figure}
\begin{tabular}{cc}
\includegraphics[width=0.45\textwidth]{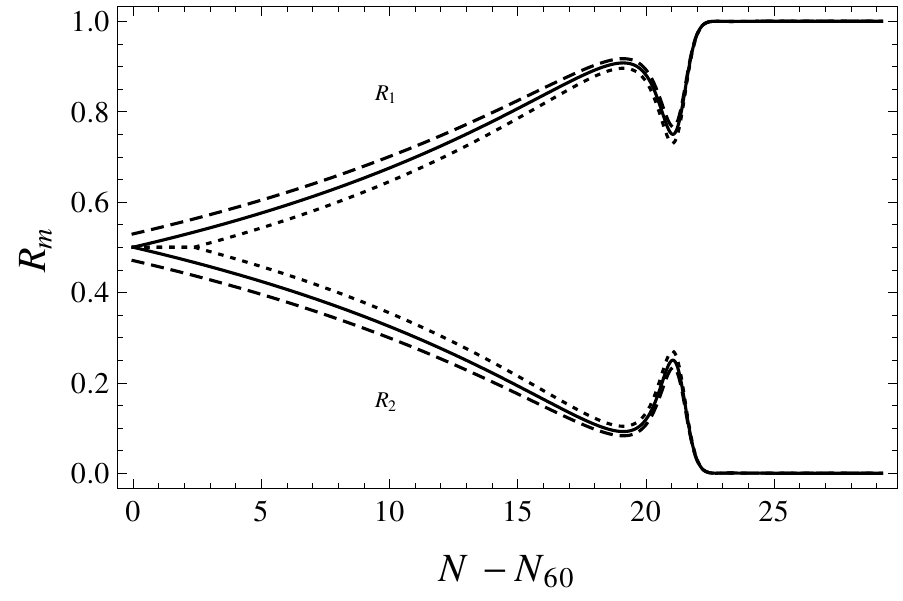}
&\includegraphics[width=0.475\textwidth]{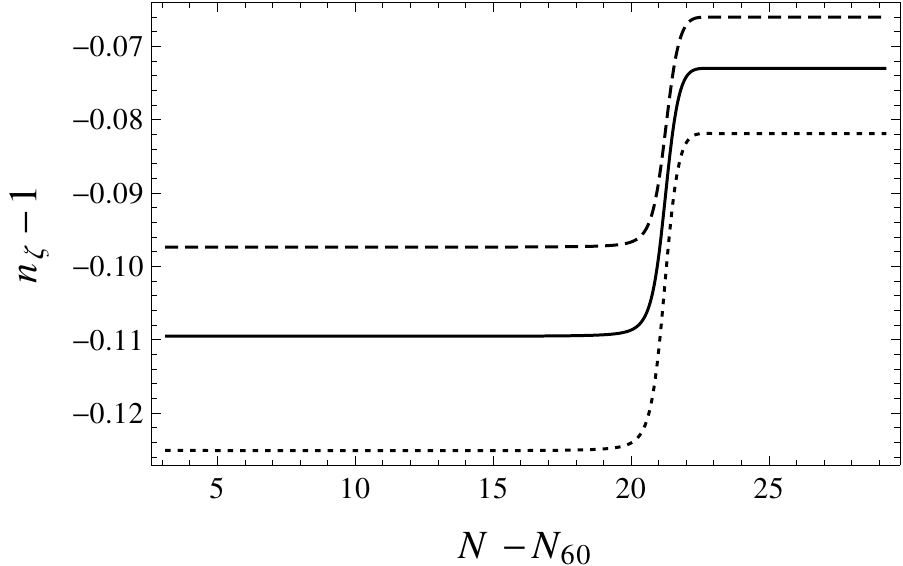}
\end{tabular}
\caption{Left: The time evolution of the adiabatic $R_1$ and the isocurvature 
$R_2$ ratios of power spectra as a function of the number of e-foldings $N-N_{60}$ 
for scales $k_{60}/10$ (dashed line), $k_{60}$ (solid line) and 
$k_{60}\times10$ (dotted line). Right: The time evolution of the spectral index 
as a function of the number of e-foldings $N-N_{60}$ 
for scales $k_{60}/10$ (dashed line), $k_{60}$ (solid line) and 
$k_{60}\times10$ (dotted line). Both plots are made 
for the model (\ref{qua}) with mass ratio $m_{\phi}/m_{\gs}=9$.}
\label{fig11}
\end{figure}

Instead of looking at the tranfer functions, using (\ref{ze}) one can also 
construct more physical quantities from the operators $\zeta_{m}$ 
and hence from the $\bv_{m2}$, namely the ratios 
of the adiabatic and isocurvature power spectrum to the total power spectrum:
\ba
&&R_1\equiv\frac{\langle\zeta_{1}\zeta_{1}\rangle}{\langle\zeta_{1}\zeta_{1}\rangle+
\langle\zeta_{2}\zeta_{2}\rangle}=\frac{1+\lh\bv_{12}\rh^2}{1+\lh\bv_{12}\rh^2+\lh\bv_{22}\rh^2},\nn\\
&&R_2\equiv\frac{\langle\zeta_{2}\zeta_{2}\rangle}{\langle\zeta_{1}\zeta_{1}\rangle+
\langle\zeta_{2}\zeta_{2}\rangle}=\frac{\lh\bv_{22}\rh^2}{1+\lh\bv_{12}\rh^2+\lh\bv_{22}\rh^2}.
\ea
These are plotted on the left-hand side of figure \ref{fig11} as a function 
of the number of e-foldings for different scales. 
One can clearly see that both 
ratios start as equal to $1/2$ when the scale exits the horizon, while 
afterwards the adiabatic ratio $R_1$ increases to reach $1$ at the end 
of inflation and the isocurvature $R_2$ decreases to reach $0$, for this 
particular model. During the turning of the fields we see that the temporary
increase in the isocurvature mode due to the negative value of $\chi$ is 
reflected in $R_2$, while the adiabatic $R_1$ necessarily has the opposite 
behaviour. 

On the right-hand side of figure \ref{fig11} we plot the time evolution of the spectral index of the power spectrum. 
The spectral index measures by construction the tilt of the power spectrum for different horizon-crossing scales and hence it 
depends on the horizon-crossing slow-roll parameters. For multiple-field models the power spectrum evolves during inflation even after horizon-crossing, 
and so does the spectral index.   
During the turning of the fields the spectral index increases, to remain constant afterwards. 
The earlier 
a scale exits the horizon the less negative is its spectral index $n_s-1$. This implies that the power spectrum itself decreases faster 
for larger horizon-crossing scales. This is due to the fact that except for the factor $1+(\bv_{12k})^2$ in the expression for the power 
spectrum there is also an inverse power of $\ge_k$ (see (\ref{Pzetamf})).

\subsection{Spectral indices}

Finally let us discuss the scale dependence of the local $\fnl$ in terms of the relevant spectral indices. 
Equation (\ref{fNLgeni}) for an isosceles triangle implies that 
\ba
-\frac{6}{5}\fnl&=&\frac{1}{(2\pi^2)^2}\frac{f(k',k')+2(\frac{k'}{k})^3f(k',k)}
{\mathcal{P}_\zeta(k')^2+2(\frac{k'}{k})^3\mathcal{P}_\zeta(k')\mathcal{P}_\zeta(k)},
\ea
where 
\be
f(k',k) = -2 (k'k)^3 \gamma_{k'}^2 \gamma_k^2 \, \bv_{12k'} \bv_{12k}
\left( g_\mathrm{sr}(k',k) + g_\mathrm{iso}(k',k) + g_\mathrm{int}(k',k) \right).
\ee 
For an arbitrary triangle configuration this is generalized as
\ba
-\frac{6}{5}\fnl&=&\frac{1}{(2\pi^2)^2}\frac{k_3^3f(k_1,k_2)+\mathrm{perms.}}
{k_3^3\mathcal{P}_\zeta(k_1)\mathcal{P}_\zeta(k_2)+\mathrm{perms.}}.
\ea
The local $\fnl$ depends on a two-variable function $f(k_1,k_2)$, with $k_1\geq k_2$. This is due to its super-horizon origin, which 
yields classical non-Gaussianity proportional to products of two power spectra. 
Hence one expects that the scale dependence of 
$\fnl$ can be expressed in terms of only two spectral indices, characterizing the function $f$. 
Notice that this is particular to the local case. In general the bispectrum 
cannot be split as a sum of two-variable functions and one anticipates that three spectral indices would be needed. 

The next issue to be resolved is which are the relevant spectral indices for $f$. The naive guess would be 
$f(k_1,k_2)=f(k_{1,0},k_{2,0})(k_1/k_{1,0})^{\tn_{k_1}}(k_2/k_{2,0})^{\tn_{k_2}}$. 
We tested this parametrization and we did not find good agreement with the exact value of $f$. 
Instead of that, we found that $f$ is best approximated by keeping either the shape or the magnitude of the triangle constant. 
This statement can be expressed as 
\be
f(k_1,k_2)=f_0\lh\frac{K}{K_0}\rh^{\tn_{K}}\lh\frac{\omega}{\omega_0}\rh^{\tn_{\omega}},
\label{ind}
\ee
where
\be
\tn_{K}\equiv\frac{\d\ln f}{\d\ln K}\qquad\mathrm{and}\qquad \tn_{\omega}\equiv\frac{\d\ln f}{\d\ln\omega}\label{inddef}
\ee
and
\be
\omega\equiv\frac{k_1}{k_2}=\frac{1+\beta}{2(1-\beta)}. 
\ee
The last equality is valid only for the isosceles case $\gamma=0$ (see
(\ref{newvar})). We dropped 
the $-1$ of the power spectrum spectral index definition to follow the 
definitions in \citep{Byrnes:2009pe,Byrnes:2010ft}. 
We added a tilde to indicate that these spectral indices are defined for the
function $f$, not yet for the full $\fnl$.
In the next two sections we are going to examine the scale-dependence of $\fnl$, 
changing the magnitude and the shape of the triangle separately, 
and verify assumption (\ref{ind}).

\section{Changing the magnitude of the triangle}\label{conf}

\begin{figure}
\begin{center}
\includegraphics[scale=0.8]{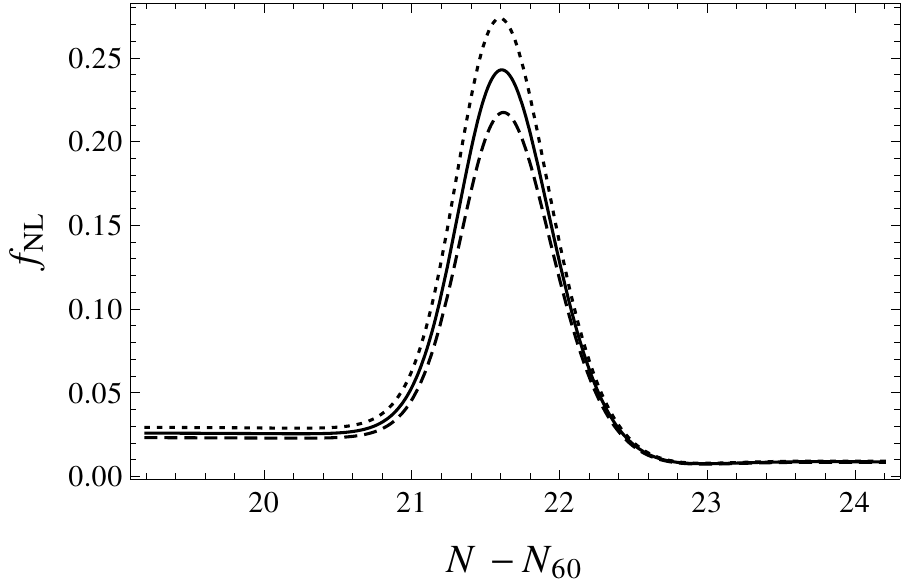}
\end{center}
\caption{The time evolution of $\fnl$ in terms of the number of e-foldings 
$N-N_{60}$ around the time of the turning of the fields,
for equilateral ($\omega=1$) triangles with $K=(3/2)k_{60}$ (solid curve), 
$K=(3/2)k_{60}/10$ (dashed curve) and $K=(3/2)k_{60}\times10$ 
(dotted curve),
for the model (\ref{qua}) with mass ratio $m_{\phi}/m_{\gs}=9$.}
\label{fig56}
\end{figure}

\begin{figure}
\begin{tabular}{cc}
\includegraphics[width=0.46\textwidth]{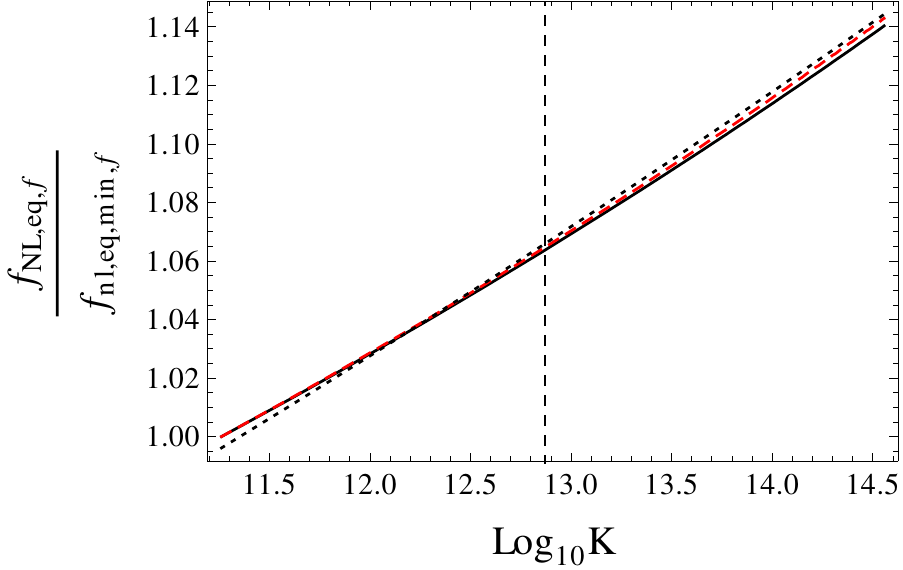}
&\includegraphics[width=0.46\textwidth]{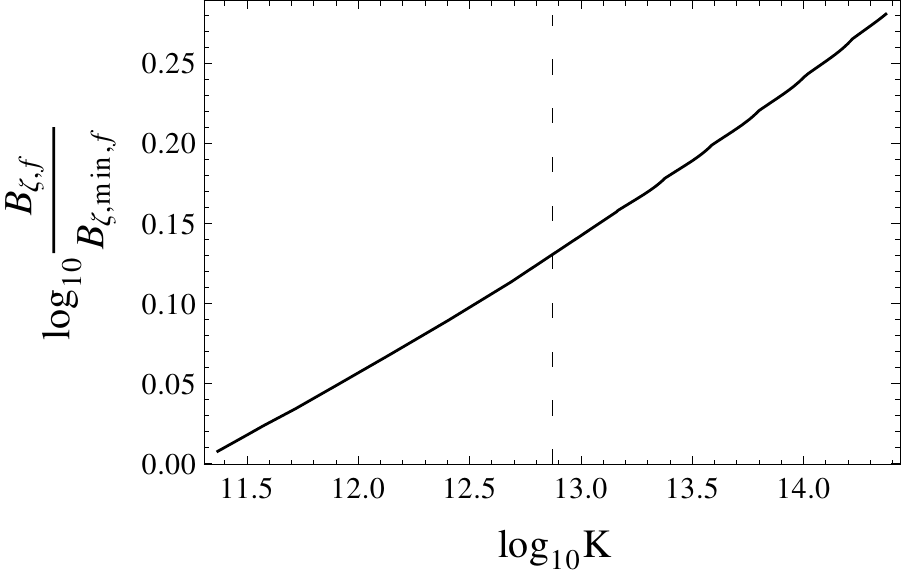}
\end{tabular}
\caption{Left: The relative change of the final value of $\fnl$ arbitrarily normalized to one at the smallest value of $K$ on the figure, 
as a function of $K$ for equilateral triangles ($\omega=1$), calculated exactly (solid curve), using the analytical approximation (\ref{fnlfin}) (dashed red curve) 
and using the shape index (\ref{nkeq}) (dotted black curve).  
Right: The logarithm of the final value of the exact bispectrum, similarly normalized,  
as a function of 
$K$ for equilateral triangles ($\omega=1$).  
Both figures are for the quadratic model (\ref{qua}) with mass ratio 
$m_{\phi}/m_{\gs}=9$. The vertical dashed line corresponds to $K=(3/2)k_{60}$.}
\label{fig31}
\end{figure}

In this section we shall study the behaviour of $\fnl$ for triangles 
of the same shape but different size, see the left-hand side of 
figure~\ref{fig01}. In figure 
\ref{fig56} we plot the time evolution of $\fnl$ for equilateral 
triangles (the result would remain qualitatively the same for any isosceles 
triangle) of perimeter $K=(3/2)k_{60}\times10$ (top curve), 
$K=(3/2)k_{60}$ (middle curve) and $K=(3/2)k_{60}/10$ (bottom curve). The later 
the relevant scale exits the horizon the larger is its initial $\fnl$ as 
explained in the previous section. $\fnl$ grows during the turning of the 
fields due to isocurvature effects as described by (\ref{giso}) and 
(\ref{gintTz2}), but by the end of inflation, when isocurvature modes vanish, it 
relaxes to a small, slow-roll suppressed value (see e.g.\  \citep{Vernizzi:2006ve,TvT1}). 
In figure \ref{fig31} we plot the final value of $\fnl$ (left) and the final 
value of the bispectrum (right) for equilateral triangles, varying $K$ for 
values around $K=(3/2)k_{60}$, within the Planck satellite's resolution 
($k'/k\sim1000$). The later the scale exits the horizon, i.e.\  the larger
$K$, the larger is the final value of $\fnl$ and of the bispectrum. 

The final value of $\fnl$ can be found analytically for the quadratic model 
within the slow-roll approximation. 
By the end of inflation $\bv_{22,f}=0$ so that 
(\ref{giso}) vanishes, while (\ref{gintTz2}) can be further 
simplified to give some extra horizon-crossing terms and a new integral 
$\tg_\mathrm{int}$ that is identically zero for the quadratic potential 
(see (\ref{tgint}) and section~\ref{eqpowsumsec}). 
For simplicity we give here the final value of 
$\fnl$ for equilateral triangles, 
\be
f_{\mathrm{NL},eq,f}(k)=\frac{
3\lh\bv_{12k}\rh^2\lh\ge_k+\getpa_k-\chi_k+\frac{\getpe_k}{\bv_{12k}}\rh
+\lh\bv_{12k}\rh^3\lh\getpe_k-
\frac{\lh\ge_k+\getpa_k-\chi_k\rh\chi_k}{\getpe_k}\rh+\ge_k+\getpa_k}
{\lh1+\lh\bv_{12k}\rh^2\rh^2}.\label{fnlfin}
\ee
This formula is actually valid for any two-field model for which isocurvature modes
vanish at the end of inflation and for which $\tg_\mathrm{int}=0$, 
like for example equal-power sum models. 
Inspecting the various terms, it turns out that although $\bv_{12k}$ tends to decrease 
the value of $f_{\mathrm{NL},f}$ as a function of $k$, it is the contribution of the horizon-crossing  
slow-roll parameters that wins and leads to an increase of the parameter 
of non-Gaussianity for larger horizon-crossing scales.
Note that for equilateral triangles $K$ is simply $3k/2$.

We turn now to the spectral index $n_{K}$. Using (\ref{inddef}) with (\ref{par}) and assuming 
that $\gamma=0$ and $\beta=\mathrm{const.}$, we can express $\tn_{K}$ in terms of the horizon-crossing time derivatives as 
\be 
\tn_{K}(t;t_{k_1},t_{k_2})=\frac{\partial\ln{f}}{\partial t_{k_1}}\frac{1}{1-\ge_{k_1}}
+\frac{\partial\ln{f}}{\partial t_{k_2}}\frac{1}{1-\ge_{k_2}}.
\label{conTz2}
\ee 
Then $\fnl$ takes the form
\be
-\frac{6}{5}\fnl=\frac{1}{(2\pi^2)^2}\frac{f(k'_0,k'_0)\lh\frac{K}{K_0}\rh^{\tn_{K}(t_{k'_0},t_{k'_0})}+
2\omega^3f(k'_0,k_0)\lh\frac{K}{K_0}\rh^{\tn_{K}(t_{k'_0},t_{k_0})}}
{\mathcal{P}_\zeta(k'_0)^2\lh\frac{k'}{k'_0}\rh^{2(n_s(t_{k'_0})-1)}
+2\omega^3\mathcal{P}_\zeta(k'_0)\mathcal{P}_\zeta(k_0)\lh\frac{k'}{k'_0}\rh^{n_s(t_{k'_0})-1}\lh\frac{k}{k_0}\rh^{n_s(t_{k_0})-1}}.
\label{nkf}
\ee
Note that the ratios $k'/k'_0=k/k_0=K/K_0$, since $\beta=\mathrm{const}$. 

The above formula can be simplified in the limit of squeezed-triangle configurations, as well as in the equilateral limit. 
When one takes the squeezed limit $\omega^3\mg1$ (note that this would be true for $\beta\simg 2/3$), one finds:
\ba
-\frac{6}{5}\fnl&=&\frac{1}{(2\pi^2)^2}\frac{f(k'_0,k_0)}{\mathcal{P}_\zeta(k'_0)\mathcal{P}_\zeta(k_0)}
\lh\frac{K}{K_0}\rh^{\tn_{K}(t_{k'_0},t_{k_0})-n_s(t_{k'_0})-n_s(t_{k_0})+2}\nn\\
&\equiv&-\frac{6}{5}f_{\mathrm{NL},0}
\lh\frac{K}{K_0}\rh^{n_{K}(t_{k'_0},t_{k_0})},
\ea
where
\be
n_K(t;t_{k'},t_k)\equiv\frac{\d \ln\fnl}{\d\ln K}=\frac{\partial\ln{\fnl}}{\partial t_{k'}}\frac{1}{1-\ge_{k'}}
+\frac{\partial\ln{\fnl}}{\partial t_{k}}\frac{1}{1-\ge_{k}}.\label{nkt}
\ee
For the equilateral case $\omega=1$, (\ref{nkf}) becomes
\ba
-\frac{6}{5}\fnl&=&\frac{1}{(2\pi^2)^2}\frac{f(k'_0,k'_0)}{\mathcal{P}_\zeta(k'_0)^2}
\lh\frac{K}{K_0}\rh^{\tn_{K}(t_{k'_0},t_{k'_0})-2n_s(t_{k'_0})+2}\nn\\
&\equiv&-\frac{6}{5}f_{\mathrm{NL},0}
\lh\frac{K}{K_0}\rh^{n_{K}(t_{k'_0},t_{k'_0})}.\label{nkeq}
\ea

The conformal spectral index $n_K$ measures the change of $\fnl$ due to the 
overall size of the triangle, namely due to a conformal transformation of the triangle. 
For an isosceles triangle this is conceptually sketched on the left-hand side of 
figure \ref{fig01}, but it can be generalized for any shape. $n_K$ coincides with the 
$n_{\fnl}$ of \citep{Byrnes:2009pe,Byrnes:2010ft} and grossly speaking it describes the 
tilt of $\fnl$ due to the pure evolution of the inflationary background (note that for 
an equilateral triangle this statement would be exact).

\begin{figure}
\begin{tabular}{ll}
\includegraphics[width=0.46\textwidth]{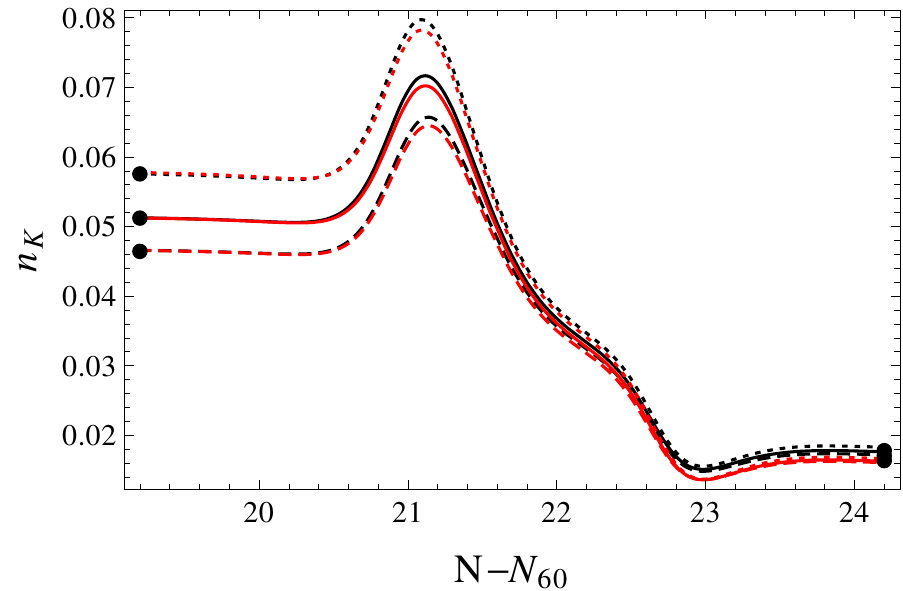}
&\includegraphics[width=0.46\textwidth]{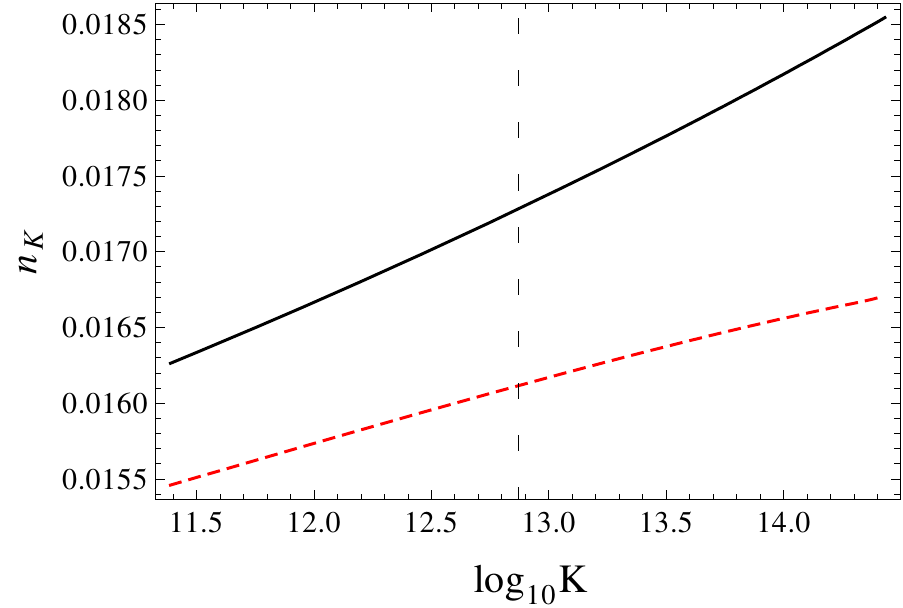}
\end{tabular}
\caption{Left: The time evolution of the conformal index 
$n_K$ (\ref{conTz2}) around the time of the turning of the fields,
for triangles with $\omega=1$ (black curves) and 
$\omega=5/2$ (red curves, below the $\omega=1$ curves), 
with perimeter $K=(3/2)k_{60}$ (solid curve), 
$K=(3/2)k_{60}/10$ (dashed curve) and $K=(3/2)k_{60}\times10$ (dotted curve). 
The three points on the left and on the right correspond to the analytical values 
of the index as calculated from (\ref{ind1}) and (\ref{A1}) respectively. 
Right: The final value of the conformal index $n_K$ 
(\ref{conTz2}) for triangles with $\omega=1$ (black curve) and $\omega=5/2$ (red dashed curve)  
as a function of $K$. 
Both figures are for the model (\ref{qua}) with mass 
ratio $m_{\phi}/m_{\gs}=9$.}\label{fig51}
\end{figure}

On the left-hand side of figure \ref{fig51} we plot the time evolution of the conformal spectral index for an equilateral 
$\omega=1$ and an isosceles $\omega=5/2$ triangle that exited the horizon 
at three different times, namely for $K=(3/2)k_{60}$ (solid curve), 
$K=(3/2)k_{60}/10$ (dashed curve) and $K=(3/2)k_{60}\times10$ (dotted curve). We plot 
the $\omega=5/2$ case only to demonstrate that the results remain qualitatively the same; 
we shall study the effect of different triangle shapes in the next section. The characteristic peaks that $n_K$ exhibits during the 
turning of the fields are inherited from the behaviour of $\fnl$ at that time and it is a new feature that is absent 
in the time evolution of the power spectrum spectral 
index $n_s-1$ (see the right-hand side of figure \ref{fig11}).

In the context of the long-wavelength formalism we are 
restricted to work with the slow-roll approximation at horizon 
exit, so that the slow-roll parameters at that time should be small and vary 
just a little. This should be reflected in the initial value of 
the spectral index, which should be $\mathcal{O}(\ge_{k'})$. 
The earlier 
the scale exits, e.g.\  the dashed curve, the smaller are the slow-roll 
parameters evaluated at horizon crossing and hence the 
smaller is the initial $n_K$.  
Indeed, using the definition (\ref{nkt}) with (\ref{fnlin}) for the initial 
value of $f_{\mathrm{NL},in}$, we find for equilateral triangles
\be
n_{K,in}=\frac{2\ge_k^2+3\ge_k\getpa_k+(\getpe_k)^2-(\getpa_k)^2+\xi^{\parallel}_k}
{\ge_k+\getpa_k},\label{ind1}
\ee
which confirms the above statement. 

We notice that the initial, horizon-crossing, differences between the values of $n_K$ for the different horizon-crossing scales
mostly disappear by the end of inflation, after peaking during the turning of the fields. 
The final value of the spectral index is plotted on the right-hand side of figure 
\ref{fig51} and is smaller than its initial value. 
It exhibits a small running of $\mathcal{O}(10\%)$ within the 
range of scales studied, inherited from the initial dispersion of its values at horizon-crossing. 
To verify that $n_K$ describes well the behaviour of $\fnl$, we have plotted the
approximation (\ref{nkeq}) in figure \ref{fig31} where it can be compared with
the exact result.
We have also verified this for other inflationary models, including the potential 
\be
W=b_0-b_2\gs^2+b_4\gs^4+a_2\phi^2,\label{pot}
\ee
studied in 
\citep{TvT1}, able to produce $\fnl$ of ${\mathcal O}(1)$. 
The final value of the spectral index in that model is two orders of magnitude smaller than the value for the quadratic model. 
This is related essentially to the fact that for the potential (\ref{pot}) the turning of the fields, 
and hence the slow-roll breaking, occurs near the end of inflation. This means that at the horizon-crossing times of the scales 
of the triangle, slow-roll parameters change very slowly and as a result the initial variation of $\fnl$ is much smaller than the one 
for the quadratic potential. 
As a consequence, the final tilt of $\fnl$ will be smaller.

By differentiating (\ref{fnlfin}) and using (\ref{conTz2}) we can find the final value of $n_K$ for equilateral triangles in the slow-roll approximation, 
assuming that isocurvature modes have vanished 
for an equal-power sum potential (for which the $\tg_\mathrm{int}$ contribution is zero, see (\ref{tgint}) and section~\ref{eqpowsumsec}):
\begin{multline}
n_{K,eq,f}=\\
-4\frac{\bv_{12k}(\bv_{12k}\chi_k-2 \getpe_k)}{1+(\bv_{12k})^2}
-\frac{1}{f_{\mathrm{NL},eq,f}(1+(\bv_{12k})^2)^2} \Bigg[-2\ge_k^2-3\ge_k\getpa_k+(\getpa_k)^2+5(\getpe_k)^2-\xi^\parallel_k\\
+3\bv_{12k}\Bigg(\getpe_k(3\ge_k+6\getpa_k-5\chi_k)-\xi^\perp_k\Bigg)+3(\bv_{12k})^2\Big(\tilde{W}_{221k}+4(\getpe_k)^2
-2(\ge_k+\getpa_k-\chi_k)(\ge_k+2\chi_k)\Big)\\
+\frac{(\bv_{12k})^3}{\getpe_k} 
\Bigg(\chi_k\Big(3\ge_k^2-2\tilde{W}_{221k}+4 \ge_k\getpa_k +3 (\getpa_k)^2 - 8(\getpe_k)^2 +\getpa\chi_k -3\chi_k^2\Big)
+\xi^\parallel_k(\ge_{k}+\getpa_{k}-\chi_{k})\\
+\getpa_k(\ge_k^2 - (\getpa_k)^2)+\tilde{W}_{221k}(\ge_k+\getpa_k)+(\getpe_k)^2(2\ge_k+5\getpa_k)-\xi^\perp_k\lh\getpe_{k}+\frac{(\ge_{k}+\getpa_{k}+\chi_{k})\chi_{k}}{\getpe_{k}}\rh
\Bigg)
\Bigg],\label{A1}
\end{multline}
where $\tilde{W}_{221}=(\sqrt{2\ge}/\kappa)W_{221}/(3H^2)$. 
We have checked this approximation and we find good agreement with the exact conformal index for equilateral triangles.

\section{Changing the shape of the triangle}\label{shape}

\begin{figure}
\begin{center}
\includegraphics[scale=0.8]{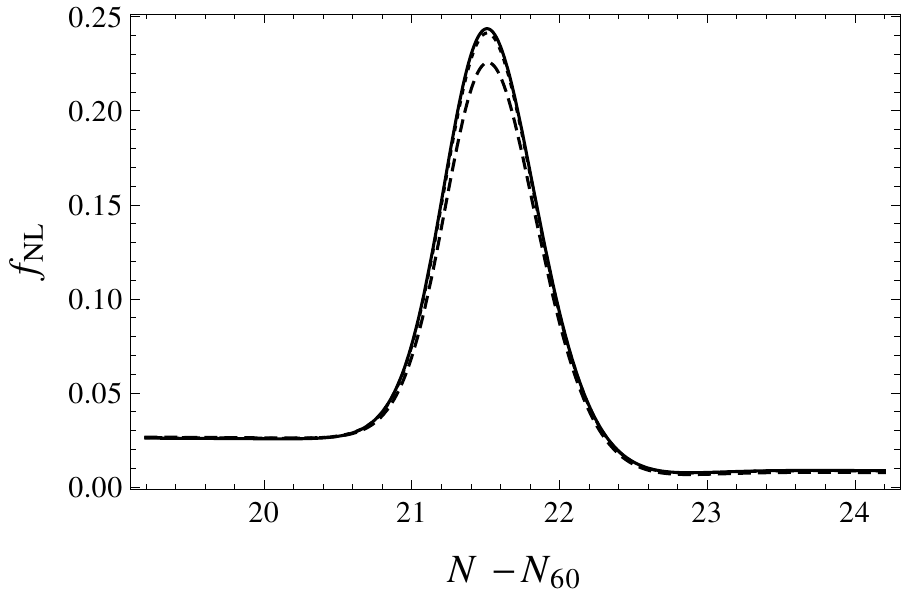}
\end{center}
\caption{The time evolution of $\fnl$ as a function of the number of e-foldings 
$N-N_{60}$ around the time of the turning of the fields,
for a triangle with $\omega=1$ (solid curve), $\omega=5/2$ (dotted curve) and $\omega=1000$ 
(dashed curve), all with fixed perimeter $K=3/2k_{60}$,
for the model (\ref{qua}) with mass ratio $m_{\phi}/m_{\gs}=9$.}
\label{fig55}
\end{figure}

After studying triangles with the same shape but varying size in the previous
section, we now turn to the scale dependence of $\fnl$ for triangles of the same perimeter but 
different shape, see the right-hand side of figure~\ref{fig01}. In figure \ref{fig55} we plot the time evolution of $\fnl$ during 
inflation for an equilateral $\omega=1$ (solid curve), an isosceles  
$\omega=5/2$ (dotted curve) and a squeezed $\omega=1000$ (dashed curve) triangle, 
all of perimeter $K=(3/2)k_{60}$, as a function of the number of e-foldings. The 
profile of the time evolution of $\fnl$ was discussed in the previous section. Here we 
are interested in the shape dependence of $\fnl$. 

\begin{figure}
\begin{tabular}{cc}
\includegraphics[width=0.46\textwidth]{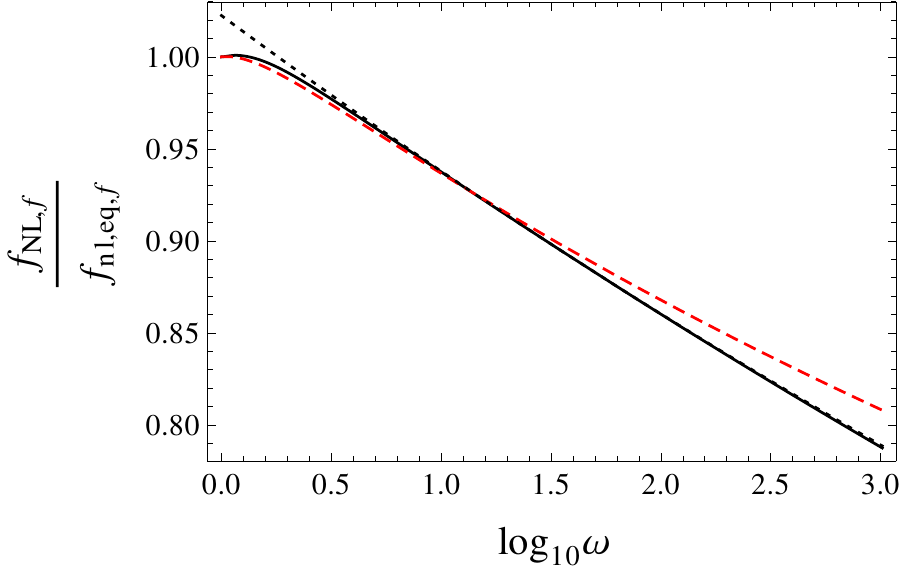}
&\includegraphics[width=0.46\textwidth]{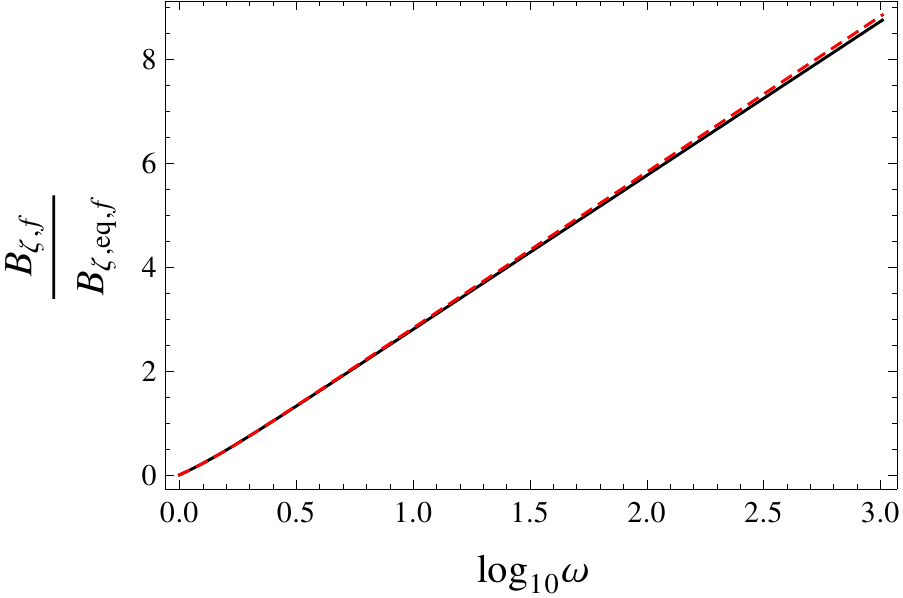}
\end{tabular}
\caption{Left: The final value of $\fnl$ normalised by the equilateral 
$\fnl$ as a function of $\omega$ for triangles with $K=
(3/2)k_{60}$ calculated exactly (black curve), using the analytical approximation (\ref{fnlb}) (dashed red curve) 
and using the shape index (\ref{into}) (dotted black curve). 
Right: The logarithm of the final value of the exact local bispectrum 
normalised by the local bispectrum computed on an equilateral triangle as a function of 
$\omega$ for triangles with $K=(3/2)k_{60}$ (black curve) and the same quantity assuming $\fnl$ scale independent (red dashed curve).  
Both figures are for the quadratic model (\ref{qua}) with mass ratio 
$m_{\phi}/m_{\gs}=9$.}
\label{fig3}
\end{figure}

Although it is during the peak that the variation of $\fnl$ for different shapes 
is more prominent, its final value is also affected. On the left-hand side of figure 
\ref{fig3} we plot the 
value of $\fnl$ at the end of inflation for triangles of perimeter $K=(3/2)k_{60}$, 
normalised by its value for the equilateral case ($\omega=1$), as a function
of $\omega$.  
The deviation of the values is small since it is related to 
horizon-exit slow-roll suppressed quantities. 
Within the long-wavelength formalism (or the $\delta N$ formalism) slow roll
at horizon crossing is a requirement. 
Nevertheless, the important conclusion here is that  
$\fnl$ decreases when the triangle becomes more squeezed. 
This can be attributed 
to the fact that the more squeezed is the triangle, the more the fluctuation $\zeta_k$ is 
frozen and behaves as part of the background when scale $k'$ crosses the horizon.  
As a result 
the correlation between $k$ and $k'$ becomes less and the resulting non-Gaussianity 
is smaller (see also the discussion below equation (\ref{fnlb}).)

An analytical formula can be found when applying the slow-roll approximation 
to expression 
(\ref{fNLgeni}) at the end of inflation, when isocurvature modes have vanished.  
We perform an integration by parts in the integral (see (\ref{tgint});
as before $\tg_\mathrm{int}=0$). 
More precisely, assuming that we are really 
in the squeezed limit $k\ll k'$, the ratio $\gamma_k^2/\gamma_{k'}^2$ becomes
very large and we can ignore the equilateral terms that depend only on $k'$
and not also on $k$. We also assume 
that the decaying mode has vanished to simplify the expressions for the Green's functions 
(see the discussion in section \ref{sour}). $G_{12k'k}$ can be set to zero as one can see in figure 
\ref{fig1} (since it is basically equal to $\bv_{12}$ and only involves times
at the very left-hand side of the figure). 
Moreover, the same figure shows that $\bv_{12k}/\bv_{12k'}\approx 1$ (in the
formula these ratios are always multiplied by slow-roll parameters, so that
the deviation from 1 would be like a second-order effect), so that we find 
in the end
\be
f_{\mathrm{NL},sq,f}
=G_{22k'k}f_{\mathrm{NL},eq,f}(k')+\frac{1-G_{22k'k}}{\lh1+\lh\bv_{12k'}\rh^2\rh^2}
\Bigg[\ge_{k'}+\getpa_{k'}+\lh\frac{2\getpe_{k'}}{\bv_{12k'}}-\chi_{k'}\rh\lh\bv_{12k'}\rh^2
\Bigg],\label{fnlb}
\ee
where $f_{\mathrm{NL},eq,f}$ is given in equation (\ref{fnlfin}). The only quantity in the above expression that depends on 
the shape of the triangle is $G_{22k'k}$, so it must be $G_{22k'k}$ that is responsible for the decreasing behaviour of $f_{\mathrm{NL},sq,f}$. 
Indeed, increasing $\omega$ for a constant perimeter $K$ 
of the triangle means increasing the interval $t_k-t_{k'}$ and hence decreasing the 
value of $G_{22k'k}$ (see the right-hand side of figure \ref{fig1}, since in the slow-roll regime $\bv_{22}=G_{22}$). 
This means that the interaction of the two modes becomes less important.   
In the complete absence of isocurvature modes $G_{22k'k}=0$ and $f_{\mathrm{NL},sq,f}$ takes its minimal value.  
It is only the isocurvature mode that interacts with itself and the 
greater is the difference between the two momenta the less is the interaction.  Notice that 
the single-field limit of this result would correspond to $G_{22k'k}=0$ and $\bv_{12}=0$. 
 
The decrease of $\fnl$ for more squeezed triangles seems contradictory 
to the well-known fact that the local bispectrum is maximized 
for squeezed configurations. In order to clarify this subtle point, we stress that 
the left-hand side of figure \ref{fig3} is essentially the ratio of the exact 
bispectrum to the bispectrum 
assuming $\fnl$ as a constant (\ref{bisp}) and hence the products 
of the power spectrum cancel out. 
We also plot on the right-hand side of figure \ref{fig3} the final value of the 
bispectrum (\ref{bisp}), 
normalised by the value of the bispectrum for equilateral triangles 
with $K=3k_{60}/2$. 
Although $\fnl$ is maximal 
for equilateral triangles, the bispectrum has the opposite 
behaviour, since it is dominated by the contribution of the products 
of the power spectrum, which leads to an increased bispectrum 
for the more squeezed shape. At the same time though we show 
that there is a small contribution of $\fnl$ itself, leading to 
smaller values of the bispectrum when compared to a bispectrum 
where $\fnl$ is assumed to be constant.

In order to quantify the above results, we examine the shape index $\tn_{\omega}$ (\ref{inddef}), 
assuming $K=\mathrm{const}$ and $\gamma=0$,
\be
 \tn_{\omega}=\frac{\partial\ln{f}}{\partial t_{k'}}\frac{1}{1-\ge_{k'}}
 \frac{1}{1+2\omega}
 -\frac{\partial\ln{f}}{\partial t_{k}}\frac{1}{1-\ge_{k}}
 \frac{2\omega}{1+2\omega}
.\label{shap}
\ee
In terms of $\tn_{\omega}$, $\fnl$ takes the form
\be
-\frac{6}{5}\fnl=\frac{1}{(2\pi^2)^2}\frac{f(k'_0,k'_0)+
2\omega^3f(k'_0,k_0)\lh\frac{\omega}{\omega_0}\rh^{\tn_{\omega}(t_{k'_0},t_{k_0})}}
{\mathcal{P}_\zeta(k'_0)^2\lh\frac{k'}{k'_0}\rh^{2(n_s(t_{k'_0})-1)}
+2\omega^3\mathcal{P}_\zeta(k'_0)\mathcal{P}_\zeta(k_0)\lh\frac{k'}{k'_0}\rh^{n_s(t_{k'_0})-1}\lh\frac{k}{k_0}\rh^{n_s(t_{k_0})-1}},
\label{nbf}
\ee
where $k'/k'_0=(1+\beta)/(1+\beta_0) \propto 2\omega/(\omega+\frac{1}{2})$ and 
$k/k_0=(1-\beta)/(1-\beta_0) \propto 1/(\omega+\frac{1}{2})$. 
This can be further simplified in the squeezed region $\omega\mg1$ to find
\ba
-\frac{6}{5}\fnl&=&\frac{1}{(2\pi^2)^2}\frac{f(k'_0,k_0)\lh\frac{\omega}{\omega_0}\rh^{\tn_{\omega}(t_{k'_0},t_{k_0})}}
{\mathcal{P}_\zeta(k'_0)\mathcal{P}_\zeta(k_0)
\lh\frac{\omega(\omega_0+\frac{1}{2})}{\omega_0(\omega+\frac{1}{2})}\rh^{n_s(t_{k'_0})-1}
\lh\frac{\omega_0+\frac{1}{2}}{\omega+\frac{1}{2}}\rh^{n_s(t_{k_0})-1}}\nn\\
&\equiv&
-\frac{6}{5}f_{\mathrm{NL},0}
\lh\frac{\omega}{\omega_0}\rh^{\tn_{\omega}(t_{k'_0},t_{k_0})+n_s(t_{k_0})-1}
\equiv
-\frac{6}{5}f_{\mathrm{NL},0}
\lh\frac{\omega}{\omega_0}\rh^{n_{\omega}(t_{k'_0},t_{k_0})}
\label{into}
\ea
with
\be
n_{\omega}\equiv\frac{\d\ln\fnl}{\d\ln\omega}=\frac{\partial\ln{\fnl}}{\partial t_{k'}}\frac{1}{1-\ge_{k'}}
\frac{1}{1+2\omega}
-\frac{\partial\ln{\fnl}}{\partial t_{k}}\frac{1}{1-\ge_{k}}
\frac{2\omega}{1+2\omega}
.\label{intodef}
\ee
The shape index $n_\omega$ describes the change of $\fnl$ due to the relative size 
of the two scales, namely due to how squeezed the triangle is, while 
keeping $K$ constant (see the right-hand side of figure 
\ref{fig01}). 

\begin{figure}
\begin{tabular}{ll}
\includegraphics[width=0.46\textwidth]{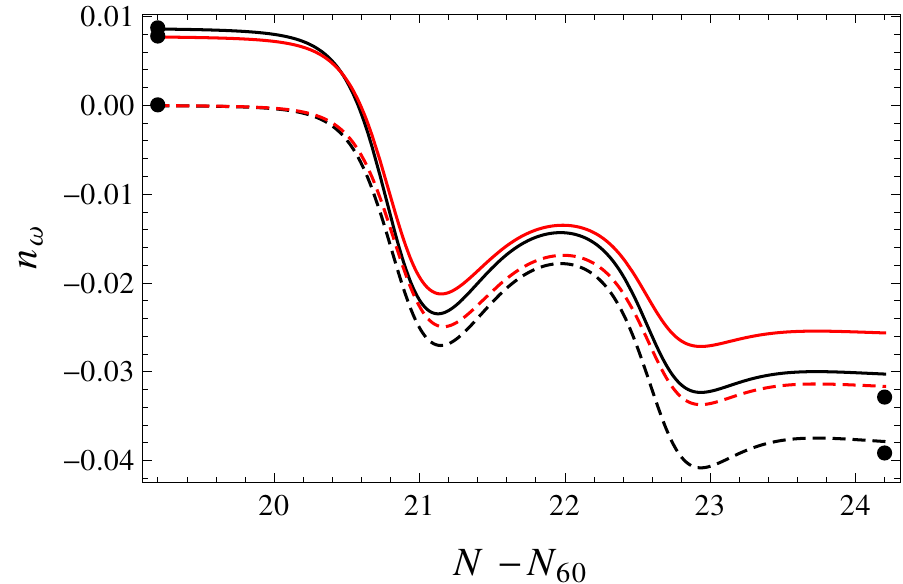}
&\includegraphics[width=0.46\textwidth]{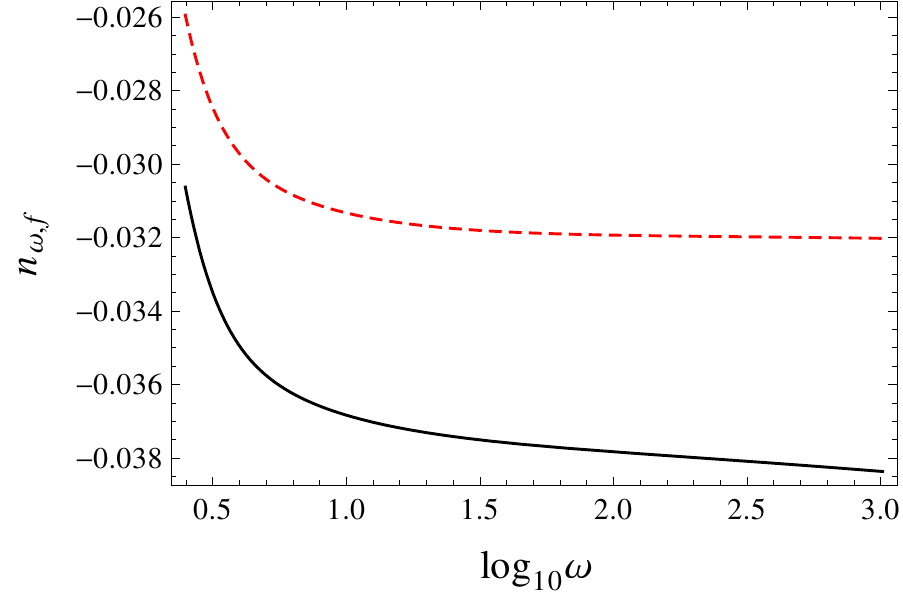}
\end{tabular}
\caption{Left: The time evolution of the shape index 
$n_\omega$ (\ref{intodef}) around the time of the turning of the fields,
for constant $K=(3/2)k_{60}$ (black curves) 
and $K=(3/2)k_{60}/10$ (red curves, above the $K=(3/2)k_{60}$ curves at the end),
and for the shapes $\omega=5/2$ (solid curve) 
and $\omega=1000$ (dashed curve). 
The points on the left correspond to the analytical values 
of the index as calculated from (\ref{noin}), while the points on the right correspond to the values of the index as calculated from (\ref{A2}) for $\omega\mg 1$.
Right: The final value of the shape index $n_\omega$ 
(\ref{intodef}) for constant $K=(3/2)k_{60}$ (black) and $K=(3/2)k_{60}/10$ (dashed red) 
as a function of $\omega$. Both figures are for the model (\ref{qua}) with mass 
ratio $m_{\phi}/m_{\gs}=9$.}\label{fig41}
\end{figure}

We studied different squeezed  
triangle configurations with constant $K$, varying $\omega$ from $\omega=5/2$ to $\omega=1000$. 
On the left-hand side of figure~\ref{fig41} we plot the time evolution of the shape index. 
The negative values of the index signify the decrease of $\fnl$ as expected. 
As one can see from the figure, for the more squeezed triangle ($\omega=1000$)
the initial value of $n_\omega$ seems to depend solely on the shape of the 
triangle and not on its magnitude, and even for the less squeezed triangle
($\omega=5/2$) the initial dependence on $K$ is negligible.  
We can find the analytical initial value of $n_\omega$ by differentiating (\ref{fnlin}):
\be
n_{\omega,in}=
\frac{1}{1+2\omega}n_{K,in}+\frac{4\omega}{1+2\omega}G_{22k'k}\frac{(\getpe_{k'})^2}{\ge_{k'}+\getpa_{k'}}.\label{noin}
\ee
For $G_{22k'k}=0$, which corresponds to the squeezed limit, $n_{\omega,in}$ is proportional to the initial shape index for equilateral 
triangles times a factor depending on the shape, which also becomes very small in the squeezed limit.

Super-horizon effects, and especially the turning of the fields, 
result in a separation of the curves of $n_\omega$ of the same shape for 
different values of $K$, due to the dependence of the evolution of $\bv_{12k'}$ 
on the scale $k'$. 
The turning of the fields increases the absolute value of $n_\omega$, which is
the opposite of the behaviour of the conformal index $n_K$. 
The shape index depends on the transfer function 
$\bv_{12k'}$ (see (\ref{A2}) for an analytical approximation). 
The smaller $K$, the less does the final value of $\bv_{12k'}$ change with respect to its initial value (see figure 
\ref{fig1}) and hence the less the shape index is affected.  
Notice that the slow-roll parameters at horizon-crossing 
have the opposite behaviour: the smaller $K$, the smaller they are. Even though $n_\omega$ also depends on the slow-roll parameters, 
it is $\bv_{12k'}$ that most affects its evolution.

On the right-hand side of figure \ref{fig41} we plot 
the value of the shape index at the end of inflation.  
It exhibits a running of about $20\%$ within the 
range of scales studied, somewhat larger than the conformal index. 
We have analytically computed the shape spectral index for models with $\tg_\mathrm{int}=0$ and with final $g_\mathrm{iso,f}=0$ 
and give the result in (\ref{A2}).
 
The dotted curve in the plot on the left-hand side of figure~\ref{fig3} shows
the final value of $\fnl$ approximated as a simple power law according to 
(\ref{into}).
Within the range of validity of 
our approximation $\omega^3\mg1$ it describes the exact result very well. 
We have also studied the shape spectral index for the potential 
(\ref{pot}). Similarly to $n_K$, its value is two orders of magnitude smaller than the value for the quadratic potential, 
but the parametrization of 
$\fnl$ in terms of the shape index is in good agreement with the exact result for a larger range of $\omega\simg 3/2$.

We repeat the calculation that we did at the end of the previous section for
$n_K$ here for the shape index $n_\omega$ (\ref{intodef}), differentiating the
squeezed $\fnl$ (\ref{fnlb}). Where needed we use the slow-roll 
approximation $G_{32k'k}=-\chi_{k'}G_{22k'k}$ and $G_{23k'k}=G_{22k'k}/3$.
The result is:
\begin{multline}
n_{\omega,sq,f}=\\
\frac{1}{f_{\mathrm{NL},sq,f}(1+(\bv_{12})^2)^2}\frac{1}{1+2\omega}\Bigg\{
 \frac{2G_{22k'k}\bv_{12k'}}{1+(\bv_{12})^2}\lh4-\lh\omega
 +(2+\omega)\bv_{12k'}\rh\frac{\chi_{k'}}{\getpe_{k'}}\rh\\
\times\Bigg[\getpe_{k'}+\bv_{12k'}\lh3(\ge_{k'}+\getpa_{k'})-2\chi_{k'}\rh+(\bv_{12k'})^2\lh \getpe_{k'}-\frac{(\ge_{k'}+\getpa_{k'}-\chi_{k'})
 \chi_{k'}}{\getpe_{k'}}\rh\Bigg]\\
-G_{22k'k}\Bigg[2(\getpe_{k'})^2-\bv_{12k'}\lh\xi^\perp_{k'}-\getpe_{k'}(11\ge_{k'}+14\getpa_{k'}-8\chi_{k'})\rh
 +\frac{(\bv_{12k'})^3}{\getpe_{k'}}\Bigg(
 \chi_{k'}\Big(3\ge_{k'}^2-2\tilde{W}_{221k'}\\
+4\ge_{k'}\getpa_{k'}+3(\getpa_{k'})^2-7(\getpe_{k'})^2
-2\chi_{k'}^2-\ge_{k'}\chi_{k'}\Big)+\xi^\parallel(\ge_{k'}+\getpa_{k'}-\chi_{k'})+\getpa_{k'}(\ge_{k'}^2-(\getpa_{k'})^2)\\
+\tilde{W}_{221k'}(\ge_{k'}+\getpa_{k'})
 +(\getpe_{k'})^2(2\ge_{k'}+5\getpa_{k'})
 -\xi^\perp_{k'}\lh\getpe_{k'}+\frac{(\ge_{k'}+\getpa_{k'}+\chi_{k'})\chi_{k'}}{\getpe_{k'}}\rh \Bigg)\\
+(\bv_{12k'})^2\lh2\tilde{W}_{221k'}-6\ge_{k'}^2+(\getpa_{k'})^2+9(\getpe_{k'})^2-\xi_{k'}^\parallel-9\getpa_{k'}\chi_{k'}+8\chi_{k'}^2-\ge_{k'}(7\getpa_{k'}+5\chi_{k'})\rh
 \Bigg]\\
-\frac{1}{1\!+\!(\bv_{12k'})^2}\Bigg[\!(\getpa_{k'})^2+3(\getpe_{k'})^2-2\ge_{k'}^2-3\ge_{k'}\getpa_{k'}-\xi_{k'}^\parallel
 -2(\bv_{12k'})^3\!\lh\xi_{k'}^\perp+\getpe_{k'}(\ge_{k'}\!-\!2\getpa_{k'}\!-\!5\chi_{k'})\rh\\
+(\bv_{12k'})^4\lh \tilde{W}_{221k'} +(\ge_{k'}-\getpa_{k'})\getpa_{k'}+3(\getpe_{k'})^2+\xi_{k'}^\parallel+2\chi_{k'}(\ge_{k'}+\chi_{k'})\rh
 -2\bv_{12k'}\big( \xi_{k'}^\perp+\getpe_{k'}(5\ge_{k'}\\
+2\getpa_{k'}+3\chi_{k'})\big)
 +(\bv_{12k'})^2\!\lh\! \tilde{W}_{221k'} \!-\!2\lh\!\ge_{k'}(\ge_{k'}+\getpa_{k'})+5(\getpe_{k'})^2-(3\ge_{k'}+3\getpa_{k'}+\chi_{k'})\chi_{k'}\!\rh\!\rh\!
 \Bigg]\!\Bigg\}
\label{A2}
\end{multline}
We have checked this approximation and we find good agreement with the exact shape index for $\omega\simg 3$.

\end{appendices}

\chapter{Non-Gaussianity in the CMB}
\label{NGCMBsec}

After having discussed in chapter~\ref{NGinflsec} how non-Gaussianity is
produced during multiple-field inflation and how we can calculate it, we will
describe in this chapter how we can observe this and other primordial
non-Gaussianity in
the cosmic microwave background radiation (CMB). These are my papers in this
subject area:\footnote{An additional paper
  \citep{Montandon:2020kuk} was published recently in collaboration with
  Guillaume Patanchon and Thomas Montandon. In that paper we perform a
  joint analysis of the power spectrum and the bispectrum of the CMB to improve
  the constraints on isocurvature modes. We analyze both the existing Planck
  data and make forecasts for future experiments. As that work was not yet
  completed when this thesis was written, it is not included.}

\begin{itemize}
\item \citep{BvTC} --- The main idea and first implementation of the binned
  bispectrum estimator to determine $\fnl$ from CMB data are explained
  (temperature-only and for an ideal sky). The local bispectrum template is
  studied in detail and two other estimators are introduced, for detecting
  the acoustic oscillations in the bispectrum.
\item \citep{BRvT} --- Many refinements (e.g.\ linear correction, inpainting,
  polarization, more templates, joint analysis) of the binned bispectrum estimator
  are treated, revealing it in its final form as used for the Planck analyses.
  The smoothed binned bispectrum as a tool for non-parametric non-Gaussianity
  searches is introduced.\footnote{This paper (from 2015) was long delayed
    because of work for Planck: most of these refinements (except the
    generalization to polarization) were already in place when work for the
    first Planck analysis started at the end of 2011.}
\item \citep{LvT1, LvT2} --- Two papers (the first short paper highlighting the
  main ideas, the second longer paper giving all the details) developing
  and studying the bispectrum templates for isocurvature non-Gaussianity,
  in temperature and polarization. Forecasts for Planck are also given.
  Constraints in the context of an explicit class of multiple-field inflation
  models are discussed.
\item \citep{JRvT} --- The bispectra of various galactic foregrounds are
  determined and studied, and used to verify that no significant
  foreground contamination remains in the cleaned Planck maps.
\item \citepalias{planck2013-24, planck2015-17, planck2018-09} --- The official
  Planck primordial non-Gaussianity papers, accompanying the three data releases
  in 2013, 2015, and 2018. A very complete analysis of the Planck CMB data
  regarding bispectral non-Gaussianity, as well as the consequences for
  inflation models.
\item \citep{Finelli:2016cyd} --- Detailed forecasts for what the proposed
  CORE satellite (Cosmic ORigins Explorer; finally not accepted by ESA) could
  do regarding constraints on inflation from both the power spectrum and the
  bispectrum of the CMB.\footnote{Despite the fact that CORE was not accepted, these
    analyses remain useful in the context of other future missions.
    However, in order not to make this thesis even longer, this work is not
    included.}
\end{itemize}  
My work on the binned bispectrum estimator \citep{BvTC, BRvT, JRvT} is mainly
a collaboration with Martin Bucher (first two) and Benjamin
Racine (last two). Gabriel Jung was a major collaborator on the last one.
Carla Carvalho was involved in the initial stages of the work for \citep{BvTC}.
The work on isocurvature non-Gaussianity is a collaboration with David Langlois.
The Planck primordial non-Gaussianity papers are of course
credited to the whole Planck collaboration, but I am one of the main authors,
having co-written with Michele Liguori at least one third of each paper.
The CORE paper was
co-written by many members of the CORE collaboration, with Michele Liguori
and myself in charge of the chapter on non-Gaussianity.

In section~\ref{theorbispecsec}, after a general introduction, we discuss
the theoretical bispectrum templates that we searched for in
our analyses, both primordial and foreground bispectra. In section~\ref{binnedbispecestsec}
all aspects of the binned bispectrum estimator are treated. Both of these sections
are mostly based on \citep{BRvT}, with some input from \citep{BvTC}.
Further extensions and applications of the estimator (including the very
important Planck results on primordial non-Gaussianity), described in the
self-contained papers \citetext{\citealp{LvT2, JRvT}; \citetalias{planck2018-09}}, are briefly
summarized in section~\ref{otherbispecsec} and included in full in the
appendices~\ref{LvT2app}--\ref{Planck2018NGapp}.

\section{Introduction and theoretical bispectrum templates}
\label{theorbispecsec}

\subsection{Introduction}

A fundamental question of observational cosmology is whether the
primordial cosmological perturbations were precisely Gaussian, or
whether small departures from exact Gaussianity can be detected at a
statistically significant level and then characterized. Here the
qualification `primordial' is essential because our goal is to probe
the new physics at play in the very early universe.
However, it is also important to study
the non-Gaussianity that was subsequently imprinted at
late times through known processes, in particular the nonlinear
dynamics of gravitational clustering,
in order to `decontaminate' the primordial non-Gaussianity.
Observations of the cosmic
microwave background anisotropies in temperature and
polarization are particularly well-suited to addressing this
fundamental question, as they provide a clean probe of the initial
conditions because most of the CMB anisotropy was
imprinted well before nonlinear effects became important.

Non-Gaussianity manifests itself in odd $n$-point 
correlation functions or in the connected 
even $n$-point correlation 
functions, from which the trivial part expressible
as combinations of two-point correlation functions
has been subtracted away.  
The extent of departures from Gaussianity can be 
characterized by ratios of higher-order
correlation functions and the appropriate 
combination of two-point correlation functions \citep{Bernardeau:2001qr}. 
The evolution of 
the primordial fluctuations of the inflaton field, 
involving both gravity and hydrodynamics,
leads to CMB anisotropies whose statistical properties 
are related to those of the primordial fluctuations. Consequently, by 
studying higher-order correlation functions of the 
CMB anisotropies, we can detect and characterize
any primordial non-Gaussianity.
The lowest order such statistic is the bispectrum, or 
three-point correlation function in Fourier space. 
The bispectrum has been shown to be an optimal statistic 
for measuring non-Gaussianity in the sense that
the signal-to-noise squared of the non-Gaussianity estimator 
based on the three-point correlation function dominates over all 
higher-order estimators \citep{Babich:2005en}. 
Consequently it would also be significantly easier to constrain. 

As discussed in detail in the previous chapters, while Gaussianity is a good
first approximation for inflationary fluctuations,
inflation does not predict exact Gaussianity no matter what
model of inflation is assumed. It cannot be modeled by a free
field theory because at a minimum the gravitational sector is
nonlinear. Additional nonlinearity will of course also arise from
other sources, such as for example from the nonlinearity of the
inflationary potential.
The departures from Gaussianity predicted within the framework of standard
single-field slow-roll inflation were calculated by
Maldacena~\citep{Maldacena:2002vr} and by
Acquaviva et al.~\citep{Acquaviva:2002ud} and are indeed unobservably small.
The bispectrum of standard single-field inflation can be viewed as a 
superposition of the local shape and the equilateral shape (see next subsection
for definitions), both terms, however, being slow-roll suppressed
\citep{Maldacena:2002vr,Fergusson:2008ra,Senatore:2009gt}. 
In particular, for the squeezed triangle configurations in single-field 
inflation the non-Gaussian signal would be proportional to the tilt of the
power spectrum and thus a detection would imply a strong 
deviation from scale invariance \citep{Creminelli:2004yq,Cheung:2007sv}.
However, many other inflation models have been introduced in the literature
that can produce non-negligible non-Gaussianity. For example, models
where multiple scalar fields play a role during inflation (as was discussed
in detail in chapter~\ref{NGinflsec}) or where inflation starts 
in an excited vacuum state (see e.g.~\citep{Holman:2007na}). In some
string-based models, as well as in 
some modified gravity or effective-field theories, the kinetic part 
of the inflaton Lagrangian can be non-standard, leading to novel bispectral
signatures (see e.g.~\citep{Alishahiha:2004eh,ArkaniHamed:2003uz,Green:2009ds}).
Deviations from the pure slow-roll phase in the inflaton
potential can also produce oscillations in the bispectrum,
see e.g.~\citep{Chen:2006xjb,Chen:2008wn}.
For more details and references, see e.g.~\citep{Liguori:2010hx,
Chen:2010xka} or \citepalias{planck2013-24,planck2015-17, planck2018-09} for a review,
the latter included in appendix~\ref{Planck2018NGapp}. Many references were also
given in the previous chapters.

After horizon crossing, 
non-linearities in both the gravitational and hydrodynamical 
evolution of the baryon-photon fluid prior to recombination 
as well as higher orders in the gravitational potential during 
recombination, can generate non-Gaussianity
\citep{Pyne:1995bs,Bartolo:2004if,Creminelli:2004pv,Nitta:2009jp,Bartolo:2008sg,
Pitrou:2010sn,Su:2012gt,Huang:2012ub}.
Other sources of non-primordial non-Gaussianity include 
secondary anisotropies such as 
weak lensing via the cross-correlation with the unlensed CMB 
arising from the integrated Sachs-Wolfe effect
\citep{Smith:2006ud, Hanson:2009kg, Mangilli:2009dr, Lewis:2011fk} 
or the Sunyaev-Zel'dovich effect \citep{Goldberg:1999xm}, as well as 
foregrounds such as dust, galactic synchrotron radiation and 
unresolved point sources.
Finally there are also instrumental effects, see e.g.~\citep{Donzelli:2009ya}.
These effects contribute spurious non-Gaussian signals, thus 
biasing the measurement of the primordial signal. It is therefore
important to develop tools to isolate the primordial signal 
from the contaminants.

Under the assumption of statistical isotropy, the bispectrum of the
map of a scalar quantity reduces to a function of three multipole
numbers $B_{\ell_1 \ell_2 \ell_3}$, where the bispectrum is
symmetric under permutations and vanishes unless the $\ell$-triplet satisfies
the triangle inequality \citep{Luo:1993xx}. If we include polarization, which 
in turn can be decomposed into $E$ and $B$ components, then the bispectrum needs
to be generalized to $B^{p_1 p_2 p_3}_{\ell_1 \ell_2 \ell_3}$, where
$p_1,p_2,p_3=T,E$ (we will not consider $B$-polarization in this thesis).
If we insist on exploiting the highest possible
spectral resolution of the CMB maps (not necessarily the best idea),
then the number of reduced bispectral coefficients that can be
measured is huge, scaling with $\ell_\mathrm{max}$ as $\ell_\mathrm{max}^3$, and
the individual coefficients are too contaminated by noise to be useful
in detecting bispectral non-Gaussianity. A major and unavoidable
contribution to this noise arises from cosmic variance --- that is, from
the departures from zero of $B^{p_1 p_2 p_3}_{\ell_1 \ell_2 \ell_3}$
that would occur even if the underlying stochastic process were exactly
Gaussian. While Gaussianity requires that the expectation value of the
bispectral coefficients, calculable only in the limit of an infinite
number of sky realizations, vanishes, the value calculated for any single sky
realization will include fluctuations about this expectation value.
For this reason, in order to make any meaningful detection of
bispectral non-Gaussianity in the data, it is necessary to combine, in
one way or another, many measured bispectral coefficients in order to
make the signal stand out over the noise.

There are basically two situations to be considered. If we have a
simple parametric model for the expected pattern of bispectral
non-Gaussianity (generally parameterized by an amplitude called
$f_\mathrm{NL}$), then an optimal estimator can be constructed by
summing the observed bispectral coefficients over
$\ell_1,\ell_2,\ell_3$ using inverse variance weighting. Another
situation to be considered involves non-parametric reconstruction of the
bispectrum, where we do not have a specific template in mind, but want
to smooth the bispectrum in order to reduce the noise and see whether
there is a broad signal that stands out over the noise at a statistically
significant level. This latter approach is particularly relevant for
studying the bispectral properties of foregrounds, for which a theory
of the expected shape of the bispectral non-Gaussianity is lacking.

Combining the bispectral coefficients is not only required from the
physical point of view (to obtain statistically significant results), but
also computationally: computing $\mathcal{O}(\ell_\mathrm{max}^3)$ bispectral
coefficients for each map is not feasible in practice. A natural
solution, motivated by the second case mentioned above as well as the
observation that many of the templates of the first case are very smooth,
is to bin the bispectrum in harmonic space. This is the basis of the
binned bispectrum estimator \citep{BvTC,BRvT} that is the subject of this
chapter (see also \citep{Casaponsa:2013mja} for
an independent investigation of the binned bispectrum estimator, and
\citep{Santos:2002df} for a first rudimentary flat-sky estimator based on a 
binned bispectrum applied to the MAXIMA data).
The binned bispectrum estimator has established itself as one of the
three main bispectrum estimators used successfully for the official
analysis of the Planck data in all three releases \citepalias{planck2013-24,
planck2015-17,planck2018-09}. The other two are the KSW 
estimator~\citep{Komatsu:2003iq,Yadav:2007rk,Yadav:2007ny} and the modal
estimator~\citep{Fergusson:2009nv,Fergusson:2010dm,Fergusson:2014gea},
and we will now briefly describe the main methodological differences between
these three estimators. In addition, other bispectrum estimators exist,
based on wavelets (e.g.~\citep{Curto:2010si}), 
needlets (e.g.~\citep{2008EJSta...2..332L}), 
and Minkowski functionals (e.g.~\citep{Ducout:2012it})
(see \citepalias{planck2013-24} for more complete references).

The KSW estimator (separable template fitting; named after its authors Komatsu,
Spergel and Wandelt) is based
on the observation that if the primordial bispectrum template is
separable as a function of $k_1,k_2,k_3$ (or alternatively the CMB bispectrum
template is separable as a function of $\ell_1,\ell_2,\ell_3$ modulo a possible
overall integral over $r$, the radial distance towards the surface of last
scattering), then the terms in the optimal estimator for
$f_\mathrm{NL}$ can be reordered as a product of terms depending only on
$k_1,\ell_1$, terms depending only on $k_2,\ell_2$, and terms depending only
on $k_3,\ell_3$ (within an overall integral over $r$). This significantly
reduces the computational cost (by effectively replacing a three-dimensional
integral and sum by the product of three one-dimensional integrals and sums),
at the cost of losing the ability for full bispectrum reconstruction. The KSW
estimator is fast, but only works for separable templates and can only be used
for the first case mentioned above (template fitting).\footnote{The
skew-$C_\ell$ extension~\citep{Munshi:2009ik} of the KSW estimator allows the
determination of a so-called bispectrum-related power spectrum, which contains
the contribution to $f_\mathrm{NL}$ (for a given shape) of all triangles with
one side equal to $\ell$.}

The modal estimator builds on the idea of the KSW estimator by first
expanding the theoretical bispectrum templates and the bispectrum of the
map in a basis of separable templates, the so-called modes. (For the Planck
2015 and 2018 analyses two pipelines were used, one with a basis of 600
polynomials, and the other with 2000, both augmented with a few special
modes to speed up convergence in the squeezed limit or for the standard
shapes.) The coefficients of the individual modes are then
computed using the KSW technique. In this way one can in principle treat
any bispectrum template, separable or not, as well as reconstruct the full
bispectrum of the map. These advantages come at the cost of often needing
a large number of modes for sufficient convergence, which can become
computationally heavy.

The binned bispectrum estimator does not use the KSW technique and keeps
the full three-dimensional sum. The required computational reduction comes
from reducing the number of terms in the sum by binning the bispectrum in
harmonic space, as will be discussed in detail in this chapter.
In this way one can do both template fitting (with templates that do not need
to be separable) and full bispectrum reconstruction as mentioned above.
Moreover, the estimator is very fast when applied to a map, has a convenient
modular structure (which means for example that one can analyze an additional
template without having to rerun the map), and gives the dependence of
$f_\mathrm{NL}$ on $\ell$ as a free bonus. The possible drawback is that the 
method works only for bispectra that are relatively smooth (or have rapid 
oscillations only in a limited $\ell$-range) in order for a limited number 
of bins (about 50--60 in practice) to suffice.

The basic output of the binned bispectrum estimator is a binned, or
coarse-grained, pseudo-bispectrum (see section~\ref{binnedbispecsubsec},
and section~\ref{implementation}
for the numerical implementation). Here `pseudo' indicates that
full-sky spherical harmonic transforms have been applied to a masked
sky, so that the recovered $a_{\ell m}$ coefficients are in fact a
convolution of the real CMB multipole coefficients with the multipole
coefficients of the mask. How one corrects for the artefacts of the
mask will be discussed in detail in section~\ref{realsky}. Below we shall
almost always assume the presence of a mask but will omit the
qualification `pseudo'. The coarse-grained pseudo-bispectrum can be
combined with a library of theoretical templates by means of an inner
product that generates optimally matched filters, see section~\ref{fNLestsec}.
It turns out that with
a modest number of bins, the loss of information compared to an
unbinned analysis is negligible. One can thus determine the
$f_\mathrm{NL}$ parameter for various templates, but one can also construct
other estimators, for example to look for the acoustic peaks in the
bispectrum (see~\citep{BvTC}).

The binned bispectrum can also be used to carry out a non-parametric,
model-independent, analysis, see section~\ref{smoothingsec}.
In such an analysis the binned
bispectrum can be smoothed to search for a serendipitous statistically
significant signal of bispectral non-Gaussianity in the CMB for which
templates have not yet been proposed, or to characterize the
bispectral properties of foregrounds without a well-motivated
theoretical template. The smoothing
complicates the statistical analysis of the significance of any 
non-Gaussian features because it introduces correlations between
neighbouring bins, but we developed a method to address this complication.
Another application of the smoothed bispectrum method is presented in
appendix~\ref{Planck2018NGapp}.

Before discussing the binned bispectrum estimator in detail in
section~\ref{binnedbispecestsec}, we first discuss in the rest of this
section what types of bispectra could be
present in the CMB. We will consider both primordial bispectra predicted by
inflation, which we want to detect, and foreground bispectra, which we have
to remove in order to measure the primordial ones.

\subsection{Standard primordial and foreground templates}
\label{templatesec}

Since the temperature and $E$-polarization fluctuations in the CMB are assumed
to originate in density fluctuations produced in the early universe, likely
during inflation, the predicted values of the 
bispectrum of the maps can be expressed in terms of the primordial
bispectrum $B_\gz(k_1,k_2,k_3)$ (defined in (\ref{defBzeta})) of the (adiabatic)
curvature  perturbation $\gz$ and the radiation transfer functions $g_\ell^p(k)$
introduced in section~\ref{CMBpowspecsec}. One finds
(see e.g.~\citep{Komatsu:2001rj})
\begin{align}
B_{\ell_1 \ell_2 \ell_3}^{p_1 p_2 p_3, \mathrm{th}} = 
h_{\ell_1 \ell_2 \ell_3}^2
\left( \frac{2}{\pi} \right)^3 
\int _0^\infty \!\!\! k_1^2 dk_1 \int _0^\infty \!\!\! k_2^2 dk_2
\int _0^\infty \!\!\! k_3^2 dk_3 & \Bigl[
g_{\ell _1}^{p_1}(k_1) g_{\ell _2}^{p_2}(k_2) g_{\ell_3}^{p_3}(k_3)
B_\gz(k_1,k_2,k_3)
\nonumber\\
& \times \int _0^\infty \!\!\! r^2dr \, j_{\ell _1}(k_1r) j_{\ell _2}(k_2r) 
j_{\ell _3}(k_3r) \Bigr]
\label{Bth}
\end{align}
where the $j_\ell$ are spherical Bessel functions.
The quantity $h_{\ell_1\ell_2\ell_3}$ is defined as (see also (\ref{Gaunt_defh})):
\be\label{real_defh}
h_{\ell_1 \ell_2 \ell_3} = 
\sqrt{ \frac{ (2\ell _1+1) (2\ell _2+1) (2\ell _3+1) }{ 4\pi } } 
\begin{pmatrix}
\ell _1 &\ell _2 &\ell _3\cr
0&0 &0\cr
\end{pmatrix}.
\ee
Expression (\ref{Bth}) includes
the non-Gaussianity amplitude parameter $f_\mathrm{NL}$ inside $B_\gz$, but later
on we will mostly consider $f_\mathrm{NL}$ an unknown parameter 
to be determined from the data and use in our estimator the theoretical
bispectrum  template $B^\mathrm{th}$ with $f_\mathrm{NL}$ set equal to
1.\footnote{\label{ftnt2}Both the
  bispectrum (\ref{Bth}) and the power spectrum (\ref{Cl_gl}) can be
  defined either in terms of the gravitational potential $\Phi$ or
  in terms of the curvature perturbation $\gz$, related by
  $\gz=-\frac{5}{3}\Phi$ during recombination. For the power spectrum this
  is invisible, as the factor $25/9$ in the power spectrum (of $\gz$ compared
  to $\Phi$) is canceled by two factors $-3/5$ in the radiation transfer
  functions of $\gz$ compared to $\Phi$. For the bispectrum, however, it
  is not invisible, as $B_\gz$ is proportional to $P_\gz^2$ as shown
  below, while there are only three radiation transfer functions in (\ref{Bth}).
  This is why the non-Gaussianity amplitude parameter $\fnl$ has to be
  multiplied by a factor $-3/5$ when considering bispectra in
  terms of $\gz$, see also footnote~\ref{ftnt1} in chapter~\ref{NGinflsec}.
  Here we use the definition
  in terms of $\gz$ preferred by theorists, for consistency with the previous
  chapters. However, in our paper \citep{BRvT} as well as in the Planck papers,
  the definition in terms of $\Phi$ is used.}

Many inflation or other early universe models predict a primordial bispectrum
that can be approximated by one (or a combination) of only a few distinct 
shapes in momentum space (see e.g.~\citep{Babich:2004gb, Fergusson:2008ra}). 
Hence it makes sense to search for these canonical
shapes. However, it should be kept in mind that these shapes are only 
approximations, and with sufficient 
sensitivity and resolution the difference between slightly 
different templates that all fall within the same approximate category can 
be resolved. Inflation models can also produce shapes that are very different 
from the canonical shapes, for example with localized features or oscillations.
See \citepalias{planck2018-09} for an overview of all the different shapes
that were tested using the Planck 2018 data, as well as more complete
references. 
The purpose of this chapter is not to give an exhaustive list of templates,
but to describe the methodology of the binned bispectrum estimator, providing 
only the most important templates as examples. 

The most well-known primordial bispectrum type is the so-called local 
bispectrum \citep{Gangui:1993tt}, 
\be
B_\gz^\mathrm{loc}(k_1,k_2,k_3) = 
-\frac{6}{5} \fnl^\mathrm{loc} \left [ \frac{2\pi^2}{k_1^3}P_\gz(k_1)
  \frac{2\pi^2}{k_2^3}P_\gz(k_2)
  + (\mathrm{2\;perms}) \right],
\label{localNG}
\ee
with the primordial (adiabatic) power spectrum $P_\gz$ defined in
(\ref{defPzeta}).
It is called local because in real space it corresponds to
the local relation
$
\Phi(\vc{x}) = \Phi_G(\vc{x}) + f_\mathrm{NL}^\mathrm{loc} 
( \Phi_G^2(\vc{x}) - \langle \Phi_G \rangle^2 )
$
\citep{Verde:1999ij,Komatsu:2001rj}
where the subscript $G$ denotes the linear (Gaussian) part.
Squeezed configurations where one $k$ (or $\ell$) is much
smaller than the other two contribute the most to the local bispectrum.
The local bispectrum shape is typically produced in multiple-field
inflation models on superhorizon scales (see (\ref{fNL_start})), or by
other mechanisms that act on superhorizon scales,
such as curvaton models (see e.g.~\citep{Bartolo:2003jx}).

The two other canonical primordial shapes are the equilateral and orthogonal 
templates.
The equilateral bispectrum is dominated by equilateral configurations
where all $k$'s (or $\ell$'s) are approximately equal, and is typically
produced at horizon crossing in inflation models with higher-derivative
or other non-standard kinetic terms (or rather, the equilateral bispectrum 
is a separable approximation to the bispectrum produced in such models, see
\citep{Creminelli:2005hu}). It is given by
\begin{align}
B_\gz^\mathrm{equ}(k_1,k_2,k_3) = &
\frac{18}{5} \fnl^\mathrm{equ} \Bigl\{ [ P(k_1)P(k_2) + (\mathrm{2\ perms}) ] 
+2 \, P^{2/3}(k_1) P^{2/3}(k_2) P^{2/3}(k_3)
\nonumber\\
&\qquad\quad - [ P(k_1) P^{2/3}(k_2) P^{1/3}(k_3) + (\mathrm{5\ perms}) ]\Bigr\},
\end{align}
where we have defined $P(k) \equiv \frac{2\pi^2}{k^3} P_\gz(k)$ for notational
simplicity.
The orthogonal bispectrum \citep{Senatore:2009gt} has 
been constructed to be orthogonal to the equilateral shape in such
a way that the bispectrum predicted by generic single-field inflation 
models can be written as a linear combination of the equilateral and
orthogonal shapes. It gets its main contribution from configurations
that are peaked both on equilateral and on flattened triangles
(where two $k$'s are approximately equal and the third is approximately
equal to their sum), with opposite sign, and is given by
\begin{align}
B_\gz^\mathrm{ort}(k_1,k_2,k_3) = &
\frac{18}{5} \fnl^\mathrm{ort} \Bigl\{ 3 [ P(k_1)P(k_2) + (\mathrm{2\ perms}) ] 
+8 \, P^{2/3}(k_1) P^{2/3}(k_2) P^{2/3}(k_3)
\nonumber\\
& \qquad\quad -3 [ P(k_1) P^{2/3}(k_2) P^{1/3}(k_3) + (\mathrm{5\ perms}) ]
\Bigr\}. 
\end{align}
It should be noted that the orthogonal shape is not at all orthogonal to
the local shape (as sometimes incorrectly stated in older literature). It
has a large correlation (about 40--50\%) with the local shape at Planck
resolution (see section~\ref{fNLestsec} and table~\ref{tab_corr_coeff}).

In addition to these three shapes, it is also interesting to look for
non-primordial contaminant bispectra, either to study these foregrounds
or to remove them. Regarding extra-galactic foregrounds, in the first place
a bispectrum will be produced by 
diffuse extra-galactic point sources. These can generally be divided into
two populations: unclustered and clustered sources. The former are radio and 
late-type infrared galaxies, while the latter are dusty star-forming galaxies 
constituting the cosmic infrared background (CIB). 
Secondly, gravitational lensing of the CMB will produce a bispectrum
that mimics the local shape, because there is a correlation between the 
lenses that produce modifications to the CMB power spectrum on small scales
and the integrated Sachs-Wolfe effect on large scales (both are due to the 
same mass distribution at low redshift).

The unclustered sources
can be assumed to be distributed according to a Poissonian distribution, 
and hence have a white noise power spectrum (i.e., with an amplitude 
independent of $\ell$). Then their bispectrum has a very simple theoretical 
shape~\citep{Komatsu:2001rj}:
\be
B_{\ell_1 \ell_2 \ell_3}^\mathrm{unclust} = 
h_{\ell_1 \ell_2 \ell_3}^2 \, b_\mathrm{ps}
\label{Bunclust}
\ee
where $b_\mathrm{ps}$, the amplitude of the unclustered point source bispectrum,
is the parameter that can be determined in the same way as the 
$f_\mathrm{NL}$ parameters for the primordial templates.
Like most foregrounds,
but unlike primordial signals, the amplitude depends on the frequency channel,
which allows a multi-frequency experiment like Planck to (partially) clean
these contaminants from its maps.
The above relation is valid both in temperature and in polarization. However,
since not all point sources are polarized, the amplitude $b_\mathrm{ps}$ is not
the same in temperature and polarization, with the difference depending on the
mean polarization fraction of the point sources. Without taking into account
that fraction, it would not make sense to look at the mixed $TTE$ and $TEE$
components of its bispectrum, nor to try to determine $b_\mathrm{ps}$ jointly 
from temperature and polarization maps. In practice for Planck the contribution
from polarized point sources is negligible (see \citepalias{planck2015-17}), so that 
we might as well consider it a temperature-only template.

The clustered point sources (CIB) have a more complicated bispectrum. A simple
template that fits the data well was established in \citep{Lacasa:2013yya}
(see also \citepalias{planck2015-17}):
\be
B_{\ell_1 \ell_2 \ell_3}^\mathrm{CIB} = 
h_{\ell_1 \ell_2 \ell_3}^2 \, b_\mathrm{CIB}
\left[ \frac{(1+\ell_1/\ell_\mathrm{break}) (1+\ell_2/\ell_\mathrm{break}) 
(1+\ell_3/\ell_\mathrm{break})}{(1+\ell_0/\ell_\mathrm{break})^3}\right]^q,
\ee
where the index is $q=0.85$, the break is located at $\ell_\mathrm{break}=70$, 
and $\ell_0=320$  is the pivot scale for normalization.
In addition, $b_\mathrm{CIB}$ is the amplitude parameter to be determined.
As for the unclustered point sources, it depends on the frequency.
The CIB is found to be negligibly polarized, so that the above template is
only used in temperature.

The theoretical shape for the lensing-ISW bispectrum was worked out in
\citep{Goldberg:1999xm, Smith:2006ud, Lewis:2011fk} and is given by
\begin{align}
B_{\ell_1 \ell_2 \ell_3}^{p_1 p_2 p_3, \mathrm{lensISW}} = 
h_{\ell_1 \ell_2 \ell_3}^2 \Bigl [ &
C_{\ell_2}^{p_2\phi} C_{\ell_3}^{p_1p_3} f_{\ell_1 \ell_2 \ell_3}^{p_1}
+ C_{\ell_3}^{p_3\phi} C_{\ell_2}^{p_1p_2} f_{\ell_1 \ell_3 \ell_2}^{p_1}
+ C_{\ell_1}^{p_1\phi} C_{\ell_3}^{p_2p_3} f_{\ell_2 \ell_1 \ell_3}^{p_2}
\nonumber\\
& + C_{\ell_3}^{p_3\phi} C_{\ell_1}^{p_1p_2} f_{\ell_2 \ell_3 \ell_1}^{p_2} + C_{\ell_1}^{p_1\phi} C_{\ell_2}^{p_2p_3} f_{\ell_3 \ell_1 \ell_2}^{p_3}
+ C_{\ell_2}^{p_2\phi} C_{\ell_1}^{p_1p_3} f_{\ell_3 \ell_2 \ell_1}^{p_3} \Bigr ].
\end{align}
Here $C_\ell^{T\phi}$ and $C_\ell^{E\phi}$ are the temperature/polarization-lensing 
potential cross power spectra, while the CMB power spectra $C_\ell^{TT}$,
$C_\ell^{TE}$, $C_\ell^{EE}$ should be taken to be the {\em lensed}
$TT$, $TE$, $EE$ power spectra. The functions $f_{\ell_1 \ell_2 \ell_3}^{p}$ are 
defined by
\begin{align}
f_{\ell_1 \ell_2 \ell_3}^T & =
\frac{1}{2} \left[ \ell_2 (\ell_2 + 1) + \ell_3 (\ell_3 + 1) - \ell_1 (\ell_1+1)
\right ], \nonumber\\
f_{\ell_1 \ell_2 \ell_3}^E & =
\frac{1}{2} \left[ \ell_2 (\ell_2 + 1) + \ell_3 (\ell_3 + 1) - \ell_1 (\ell_1+1)
\right ]  
\left(\begin{array}{ccc} \ell_1 & \ell_2 & \ell_3 \\ 2 & 0 & -2 \end{array}\right)
\left(\begin{array}{ccc} \ell_1 & \ell_2 & \ell_3 \\ 0 & 0 & 0 \end{array}\right)^{-1},
\end{align}
if $\ell_1+\ell_2+\ell_3$ is even and $\ell_1,\ell_2,\ell_3$ satisfy the
triangle inequality, and zero otherwise. Using some mathematical properties of
the Wigner 3j-symbols we find that, under the same conditions as above,
the ratio of the two Wigner 3j-symbols can be computed explicitly as
\begin{align}
& \left(\begin{array}{ccc} 
\ell_1 & \ell_2 & \ell_3 \\ 2 & 0 & -2 \end{array}\right)
\left(\begin{array}{ccc} 
\ell_1 & \ell_2 & \ell_3 \\ 0 & 0 & 0 \end{array}\right)^{-1} = \nonumber\\
& \Bigl\{ [\ell_2 (\ell_2 + 1) - \ell_1 (\ell_1 + 1) - \ell_3(\ell_3+1)]
[\ell_2 (\ell_2 + 1) - \ell_1 (\ell_1 + 1) - \ell_3(\ell_3+1) + 2] \nonumber\\ 
& - 2\ell_1 (\ell_1 + 1) \ell_3(\ell_3+1) \Bigr\} \:
\Bigl[ 4(\ell_1-1)\ell_1(\ell_1+1)(\ell_1+2)(\ell_3-1)\ell_3(\ell_3+1)(\ell_3+2)
\Bigr]^{-\frac12}.
\end{align}
Note that there is no unknown amplitude parameter in front of this template:
its $f_\mathrm{NL}$ parameter should be unity.

Apart from the extra-galactic templates provided here, one also has to take
into account galactic contaminants, although these should in principle be
absent in the cleaned maps due to the combination of component separation and
masking. Unfortunately no theorerical templates exist for those shapes, but
the binned bispectrum estimator also allows for determining and using numerical
templates. The study of galactic bispectrum contaminants is the subject of our
paper \citep{JRvT}, included in appendix~\ref{JRvTapp}.

\subsection{Isocurvature non-Gaussianity}
\label{sec_isocurv}

The generalization to the case where non-Gaussian isocurvature components
are present in
addition to the standard adiabatic component was treated in
\citep{LvT1,LvT2}. The second of those papers, which
contains the complete treatment, is included in appendix~\ref{LvT2app}.
For convenience we summarize the resulting template here. In fact this boils
down to the joint analysis of a number of additional templates.

We make two simplifying assumptions: we consider only the local shape (because
that is the shape typically produced by multiple-field inflation, and we
require multiple fields in order to produce isocurvature modes) and
assume the same spectral index for the primordial isocurvature power spectrum
and the isocurvature-adiabatic cross power spectrum as for the adiabatic
power spectrum (to limit the number of free parameters).
In that case the primordial bispectrum can be written as
\be
B^{IJK}(k_1, k_2, k_3) = 
\tilde{f}_{\rm NL}^{I, JK}  P(k_2) P(k_3) 
+ \tilde{f}_{\rm NL}^{J, KI}  P(k_1) P(k_3)
+ \tilde{f}_{\rm NL}^{K, IJ}  P(k_1)P(k_2), 
\label{primbispiso}
\ee
where $I,J,K$ label the different modes (adiabatic and isocurvature).
As a reminder, $P(k) = \frac{2\pi^2}{k^3} P_\gz(k)$. The meaning of and the
reason for the tilde on $\tilde{f}_\mathrm{NL}$ is explained below.
The invariance of this expression under the simultaneous interchange of
two of these indices and the corresponding momenta means that 
$\tilde{f}_{\rm NL}^{I, JK} = \tilde{f}_{\rm NL}^{I, KJ}$, explaining the presence
of the comma, and reducing the number of independent $\tilde{f}_\mathrm{NL}$
parameters (from 8 to 6 in the case of two modes).
Inserting this expression into (\ref{Bth}), where 
$g_{\ell _1}^{p_1}(k_1) g_{\ell _2}^{p_2}(k_2) g_{\ell_3}^{p_3}(k_3)$
should be replaced by
$\sum_{I,J,K} g_{\ell _1}^{p_1\, I}(k_1) g_{\ell _2}^{p_2\, J}(k_2) g_{\ell_3}^{p_3\, K}(k_3)$,
finally leads to the result
\be
B_{\ell_1 \ell_2 \ell_3}^{p_1 p_2 p_3} = 
\sum_{I,J,K} \tilde{f}_{\rm NL}^{I, JK} B_{\ell_1 \ell_2 \ell_3}^{p_1 p_2 p_3\, I,JK},
\ee
where
\be
B_{\ell_1 \ell_2 \ell_3}^{p_1 p_2 p_3\, I,JK}= 3 h_{\ell_1 \ell_2 \ell_3}^2
\int_0^\infty r^2 dr \, 
\alpha^{p_1\, I}_{(\ell_1}(r)\beta^{p_2\, J}_{\ell_2}(r)\beta^{p_3\, K}_{\ell_3)}(r),
\label{Bth_isocurv}
\ee
with   
\be
\alpha^{p\, I}_{\ell}(r) \equiv \frac{2}{\pi} \int k^2 dk\,  j_\ell(kr) \, 
g^{p\, I}_{\ell}(k),
\qquad\qquad
\beta^{p\, I}_{\ell}(r) \equiv \frac{2}{\pi}  \int k^2 dk \,  j_\ell(kr) 
\, g^{p\, I}_{\ell}(k)\,  P(k).
\label{iso_alpha_beta}
\ee
Here we use the notation
$(\ell_1 \ell_2 \ell_3)\equiv [\ell_1\ell_2\ell_3+ 5\,  {\rm perms}]/3!$
and it should be kept in mind that the $\ell_i$ and $p_i$ are always kept
together (so the $p_i$ are also permuted in the same way).

The tilde on $\tilde{f}_\mathrm{NL}$ indicates that we have explicitly defined
$\fnl$ parameters
here in terms of the adiabatic curvature perturbation $\zeta = e_{1A}\gz^A$ and
the total isocurvature perturbation $S$ instead of the gravitational potential
$\Phi$ (in the following we will always assume
the presence of just a single isocurvature mode in addition to the adiabatic
one, so that the total isocurvature perturbation is simply equal to the
only isocurvature perturbation). 
The relation between $S$ and the isocurvature
mode $e_{2A}\gz^A$ is given in (\ref{Szeta2rel}). As explained in
footnote~\ref{ftnt2} in this chapter and footnote~\ref{ftnt1} in
chapter~\ref{NGinflsec}, there is a factor between the $\tilde{f}_\mathrm{NL}$ and the usual
$\fnl$ defined in terms of $\Phi$, which is equal to $-6/5$ for the purely
adiabatic mode (see also (\ref{localNG})). This comes from the fact that
$\gz=-5\Phi_\mathrm{adi}/3$.
To compute the factors for the other modes, one has to use the fact that
$S=-5\Phi_\mathrm{iso}$. The final factors for the six modes $(\gz,\gz\gz)$,
$(\gz,\gz S)$, $(\gz,SS)$, $(S,\gz\gz)$, $(S,\gz S)$, and $(S,SS)$ are then
$-6/5$, $-2/5$, $-2/15$, $-18/5$, $-6/5$, and $-2/5$,
respectively.\footnote{There is a
  sign mistake in the relation between $\tilde{f}_\mathrm{NL}$ and $\fnl$
  given in \citep{LvT2} (corrected in appendix~\ref{LvT2app}), which led to a
  sign mistake in these factors given in \citepalias{planck2015-17}. This mistake
  was corrected in the revised version of \citepalias{planck2018-09}. However, as
  \citep{LvT2} only studies quadratic quantities (error bars, Fisher matrices)
  and \citepalias{planck2015-17, planck2018-09} do not use the
  $\tilde{f}_\mathrm{NL}$ but only the $\fnl$, this mistake has no consequences
  for those papers.}
\footnote{To fully understand these factors, one particularity that is not
  well documented must be noted. When writing the isocurvature
  $\beta^S_\ell$ (omitting polarization indices) from (\ref{iso_alpha_beta})
  in terms of the gravitational potential $\Phi$, it would seem logical to
  convert the $g^S_\ell P_\gz$ into $g^{\Phi_\mathrm{iso}}_\ell P_{\Phi_\mathrm{adi}}$.
  However, for historical reasons it is actually converted into
  $g^{\Phi_\mathrm{iso}}_\ell P_{\Phi_\mathrm{iso}}$, where
  $P_{\Phi_\mathrm{iso}} = \frac{1}{9} P_{\Phi_\mathrm{adi}}$ when $P_\gz=P_S$, and
  the $\fnl$ without tilde (as given in e.g.~\citepalias{planck2015-17,planck2018-09})
  are defined with respect to that $\beta^\mathrm{iso}_\ell$.}
As this factor is different for each $\tilde{f}_\mathrm{NL}$, it is not
feasible to write (\ref{primbispiso}), with the power spectra in terms of $\gz$,
in terms of the $\fnl$ without a tilde, as we could do in (\ref{localNG}) for
example. Hence the required introduction of the tilded quantities, which are
also used in \citep{LvT2} included in appendix~\ref{LvT2app}.

We can conclude that including the possibility of isocurvature non-Gaussianity
in our investigations means that we have to replace the single local 
adiabatic bispectrum template by the family of templates (\ref{Bth_isocurv}),
each with their individual $\tf_\mathrm{NL}$ parameter. In particular, if we 
assume the presence of only a single isocurvature mode in addition to the
adiabatic one (i.e.\ one of cold dark matter, neutrino density, or neutrino
velocity), we have six local $\tf_\mathrm{NL}$ parameters to determine instead
of just one, and these should always be estimated jointly (see 
section~\ref{fNLestsec}). For more details, see appendix~\ref{LvT2app}.

\section{The binned bispectrum estimator}
\label{binnedbispecestsec}

This section contains a detailed presentation of all aspects of the
binned bispectrum estimator that we developed. It is based mostly on
\citep{BRvT}, with some input from \citep{BvTC}.

\subsection{Binned bispectrum}
\label{binnedbispecsubsec}

As discussed in section~\ref{CMBpowspecsec}, a map $M^p$ of the CMB
temperature or $E$-polarization fluctuations
can be decomposed into spherical harmonics according to 
\be
M^p(\Omega) = \sum_{\ell, m} a_{\ell m}^p Y_{\ell m}(\Omega).
\ee
Here $\Omega$ is the solid angle on the sky and the $p$ label refers to either
temperature ($T$) or $E$-polarization ($E$), as we will not consider
$B$-polarization in this thesis. In the following paragraph we will omit
the explicit polarization indices, in order to lighten the notation.

The full bispectrum on the celestial sphere consists of cubic combinations
of the spherical harmonic coefficients of the form
\be 
B_{\ell_1 \ell_2 \ell_3}^{m_1 m_2 m_3}
= a_{\ell_1 m_1} a_{\ell_2 m_2} a_{\ell_3 m_3},
\ee
the expectation values of which may be calculated for a given theory.
However, under the assumption of statistical isotropy, these
expectation values are not independent and can be reduced to
quantities depending only on 
$\ell _1$, $\ell _2$, and $\ell _3$.
We may define a manifestly rotationally-invariant
reduced bispectrum, called the angle-averaged bispectrum,
in terms of integrals of triple 
products of maximally filtered maps so that
\be
B_{\ell _1 \ell _2 \ell _3} =
\int d\Omega ~
M_{\ell _1}(\Omega ) ~
M_{\ell _2}(\Omega ) ~
M_{\ell _3}(\Omega ) ,
\label{BispecFromMap}
\ee
where the maximally filtered map is defined as
\be 
M_\ell (\Omega )=\sum _{m=-\ell }^{+\ell }
a_{\ell m} Y_{\ell m}(\Omega ).
\label{TempMap}
\ee
Using the expression for the Gaunt integral\footnote{In our papers
  \citep{BvTC, BRvT} we defined a quantity $N_\triangle$ instead of $h$, equal
  to the square of $h$. However, the quantity $h$ given in
  (\ref{real_defh}) is by now more commonly used in the literature.}
\ba
\label{Gaunt_defh}
{\cal G}_{\ell_1 \ell_2 \ell_3}^{m_1 m_2 m_3} & = & \int d\Omega ~
Y_{\ell _1m_1} (\Omega )~ Y_{\ell _2m_2} (\Omega )~ Y_{\ell _3m_3} (\Omega )\\
& = & 
\sqrt{ \frac{ (2\ell _1+1) (2\ell _2+1) (2\ell _3+1) }{ 4\pi } } 
\begin{pmatrix}
\ell _1 &\ell _2 &\ell _3\cr
0&0 &0\cr
\end{pmatrix}
\begin{pmatrix}
\ell _1 &\ell _2 &\ell _3\cr
m_1 &m_2 &m_3\cr
\end{pmatrix}
\nonumber\\
& \equiv & h_{\ell_1 \ell_2 \ell_3}
\begin{pmatrix}
\ell _1 &\ell _2 &\ell _3\cr
m_1 &m_2 &m_3\cr
\end{pmatrix}
\nonumber
\ea
we obtain
\be
B_{\ell _1 \ell _2 \ell _3} = h_{\ell _1 \ell _2 \ell _3}
\sum _{m_1,m_2,m_3}
\begin{pmatrix}
\ell _1 &\ell _2 &\ell _3\cr
m_1 &m_2 &m_3\cr
\end{pmatrix}
B_{\ell_1 \ell_2 \ell_3}^{m_1 m_2 m_3}.
\ee
As a consequence of the Wigner-Eckart theorem,
$B_{\ell_1 \ell_2 \ell_3}^{m_1 m_2 m_3}$
is proportional to
$
\begin{pmatrix}
\ell _1 &\ell _2 &\ell _3\cr
m_1 &m_2 &m_3\cr
\end{pmatrix}.
$
Using this fact combined with the
Wigner-$3j$-symbol identity
\be
\sum _{m_1,m_2,m_3}
\begin{pmatrix}
\ell _1 &\ell _2 &\ell _3\cr
m_1 &m_2 &m_3\cr
\end{pmatrix}
\begin{pmatrix}
\ell _1 &\ell _2 &\ell _3\cr
m_1 &m_2 &m_3\cr
\end{pmatrix}
=1,
\label{3j_identity}
\ee
which holds whenever $\ell _1,$ $\ell _2,$ $\ell _3$
satisfy the triangle inequality ($|\ell_1-\ell_2| \leq \ell_3 \leq \ell_1+\ell_2$
and permutations) and the parity condition ($\ell _1+\ell _2+\ell _3=$ even),  
we find that
\be 
B_{\ell_1 \ell_2 \ell_3}^{m_1 m_2 m_3}
= h_{\ell_1 \ell_2 \ell_3}^{-1}
\begin{pmatrix}
\ell _1 &\ell _2 &\ell _3\cr
m_1 &m_2 &m_3\cr
\end{pmatrix}
B_{\ell _1 \ell _2 \ell _3 }.
\ee 
Again, this equality is only valid when the triangle inequality and parity
condition are respected, otherwise both $B_{\ell _1 \ell _2 \ell _3 }$ and
$h_{\ell _1 \ell _2 \ell _3 }$ are zero.\footnote{In the literature one often
  encounters
the reduced bispectrum $b_{\ell _1 \ell _2 \ell _3 }$ instead of the angle-averaged
bispectrum $B_{\ell _1 \ell _2 \ell _3 }$, defined as $b_{\ell _1 \ell _2 \ell _3 }
= B_{\ell _1 \ell _2 \ell _3 }/h_{\ell _1 \ell _2 \ell _3 }^2$, but we will not consider
it in this thesis.}
Because the angle-averaged bispectrum $B_{\ell_1 \ell_2 \ell_3}^{p_1 p_2 p_3}$
(restoring its polarization indices, and which we will call simply
``bispectrum'' in the rest of the thesis) is
symmetric under the simultaneous interchange of its 
three multipole numbers $\ell_1,\ell_2,\ell_3$ and its three polarization
indices $p_1,p_2,p_3$, it is sufficient 
to consider only the subspace $\ell _1\le \ell _2\le \ell _3$. It should be 
noted, however, that once we have both temperature and polarization, imposing 
this condition means that we no longer have the freedom to rearrange the
polarization indices, so that for example the $TTE$, $TET$, and $ETT$ 
combinations correspond to three distinct bispectra.

To compute the observed bispectrum with the 
maximum possible resolution, we would evaluate
the integral over the sky of triple products of 
maximally filtered observed sky maps, as in (\ref{BispecFromMap}).
(In practice this integral is evaluated as 
a sum over pixels.)
The total number of triplets would 
be $\mathcal{O}(10^{7})$ for a WMAP or $\mathcal{O}(10^{9})$ 
for a Planck temperature map. 
But we can also use broader filters for the integral in 
(\ref{BispecFromMap}), with very little 
loss of information because a modest resolution in $\ell$ suffices for 
many physically motivated templates for which the predicted 
$B_{\ell_1 \ell_2 \ell_3}^{p_1 p_2 p_3}$ varies slowly with its $\ell$ arguments.
We end up having to compute only $\mathcal{O}(10^{4})$ 
bin triplets, leading to an enormous reduction in the 
computational resources required. We divide the $\ell$-range 
$[\ell_\mathrm{min}, \ell_\mathrm{max}]$ into subintervals denoted by 
$\Delta_i= [\ell_i,\ell_{i+1}-1]$ where $i=0,\ldots ,(N_\mathrm{bins}-1)$
and $\ell_{N_\mathrm{bins}} = \ell_\mathrm{max}+1$, so that 
the filtered maps are
\be
M_i^p(\Omega) = 
\sum_{\ell\in\Delta_i} \sum_{m=-\ell}^{+\ell}
a_{\ell m}^p Y_{\ell m}(\Omega),
\label{Tmapbinned}
\ee
and we use these instead of $M_\ell^p$ in the expression for the bispectrum
(\ref{BispecFromMap}). The observed binned bispectrum is 
\be
B_{i_1 i_2 i_3}^{p_1 p_2 p_3, \mathrm{obs}}= \frac{1}{\Xi_{i_1 i_2 i_3}}
\int d\Omega \,
M_{i_1}^{p_1, \mathrm{obs}}(\Omega) M_{i_2}^{p_2, \mathrm{obs}}(\Omega) 
M_{i_3}^{p_3, \mathrm{obs}}(\Omega)
\label{Bobsbinned}
\ee
where $\Xi_{i_1 i_2 i_3}$ is the number of $\ell$ triplets within 
the $({i_1, i_2, i_3})$ bin triplet
satisfying the triangle inequality and parity condition selection rules. 
Because of this normalization factor,
$B_{i_1 i_2 i_3}^{p_1 p_2 p_3}$ may be considered an average over 
all valid $B_{\ell_1 \ell_2 \ell_3}^{p_1 p_2 p_3}$ inside the bin triplet.

As for the power spectrum, there will be a fundamental statistical uncertainty
in the bispectrum,
called cosmic variance, due to the fact that we want to determine an ensemble
average (to compare with inflationary predictions for example) but we can only
measure one sky. The only averaging we can do is over the $m$ indices. This
cosmic variance is in addition to other sources of uncertainty, e.g.\  due to
the finite resolution and noise of the experiment under consideration, due to
foreground residuals, etc. To compute the variance
we start by considering only the temperature bispectrum.
The covariance of the 
bispectra $B_{\ell_1 \ell_2 \ell_3}$ and $B_{\ell_4 \ell_5 \ell_6}$ equals
the average of the product minus the product of the 
averages. Under the assumption of weak non-Gaussianity 
the calculation simplifies significantly. In that case one can neglect
the average value of the bispectra, and the average of the product,
\begin{align}
\langle B_{\ell_1 \ell_2 \ell_3} B_{\ell_4 \ell_5 \ell_6} \rangle
= & \: h_{\ell_1 \ell_2 \ell_3} h_{\ell_4 \ell_5 \ell_6}
 \nonumber\\
& \times\!\!\!\!\!\!\!\!
 \sum_{\tiny\begin{array}{c@{}c@{}c} m_1, & m_2, & m_3, \\ m_4, & m_5, & m_6\\ 
\end{array}} \!\!\!\! 
\begin{pmatrix}
\ell _1 &\ell _2 &\ell _3\cr
m_1 & m_2 & m_3\cr
\end{pmatrix} 
\begin{pmatrix}
\ell _4 &\ell _5 &\ell _6\cr
m_4 & m_5 & m_6\cr
\end{pmatrix} 
\langle a_{\ell_1 m_1} a_{\ell_2 m_2} a_{\ell_3 m_3}
a_{\ell_4 m_4}^* a_{\ell_5 m_5}^* a_{\ell_6 m_6}^*  \rangle
\end{align}
(using the fact that $B$ is real so that $B=B^*$),
can be rewritten as the product of three power spectra 
$C_\ell \equiv \langle a_{\ell m} a_{\ell m}^* \rangle$ using Wick's
theorem:
\begin{align}
\langle a_{\ell_1 m_1} & a_{\ell_2 m_2} a_{\ell_3 m_3}
a_{\ell_4 m_4}^*  a_{\ell_5 m_5}^* a_{\ell_6 m_6}^*  \rangle
= C_{\ell_1} C_{\ell_2} C_{\ell_3} [  \delta_{\ell_1 \ell_4} \delta_{\ell_2 \ell_5} 
\delta_{\ell_3 \ell_6} \delta_{m_1 m_4} \delta_{m_2 m_5} \delta_{m_3 m_6}
 \nonumber\\
& + (14)(26)(35) + (15)(24)(36) + (15)(26)(34) + (16)(24)(35) + (16)(25)(34) ],
\label{6ptfunc}
\end{align}
using obvious shorthand to denote the other permutations of 
$\delta$-functions. Due to the $\delta$-functions, the covariance matrix is
diagonal, so we need to consider only the (diagonal) variance of 
$B_{\ell_1 \ell_2 \ell_3}$. We use the identity (\ref{3j_identity})
and the fact that for even parity of $\ell_1+\ell_2+\ell_3$ the columns of the
Wigner $3j$-symbol can be permuted to obtain 
\be
\Var(B_{\ell_1 \ell_2 \ell_3}) = 
g_{\ell_1 \ell_2 \ell_3} h_{\ell_1 \ell_2 \ell_3}^2
C_{\ell_1} C_{\ell_2} C_{\ell_3}
\equiv V_{\ell_1 \ell_2 \ell_3}
\ee
with $g_{\ell_1 \ell_2 \ell_3}$ equal to 6, 2, or 1, depending on whether 3,
2, or no $\ell$'s are equal, respectively, and $h$ defined
in (\ref{real_defh}). 
Similarly the variance of the binned bispectrum 
$B_{i_1 i_2 i_3} = ({\Xi_{i_1 i_2 i_3}})^{-1}
\sum_{\ell_1\in\Delta_1} \sum_{\ell_2\in\Delta_2} \sum_{\ell_3\in\Delta_3}
B_{\ell_1 \ell_2 \ell_3}$ is given by
\be
\Var(B_{i_1 i_2 i_3}) = 
\frac{g_{i_1 i_2 i_3}}{(\Xi_{i_1 i_2 i_3})^2} 
\sum_{\ell_1\in\Delta_1} \sum_{\ell_2\in\Delta_2} \sum_{\ell_3\in\Delta_3}
h_{\ell_1 \ell_2 \ell_3}^2
C_{\ell_1} C_{\ell_2} C_{\ell_3}
\equiv V_{i_1 i_2 i_3}
\label{idealVariance}
\ee
with $g_{i_1 i_2 i_3}$ equal to 6, 2, or 1, depending on whether 3,
2, or no $i$'s are equal, respectively. The $\delta$-functions in 
(\ref{6ptfunc}) lead here to conditions of equality on the bins, since due
to the sum over all $\ell$'s inside a bin, $\delta_{\ell_a \ell_b}$ will always
give 1 if $\ell_a$ and $\ell_b$ are in the same bin, and 0 if not.

With the noise and beam smoothing present in a real experiment, 
(\ref{idealVariance}) becomes
\be
V_{i_1 i_2 i_3} =
\frac{g_{i_1 i_2 i_3}}{(\Xi_{i_1 i_2 i_3})^2} 
 \sum_{\ell_1\in\Delta_1} \sum_{\ell_2\in\Delta_2} \sum_{\ell_3\in\Delta_3}
h_{\ell_1 \ell_2 \ell_3}^2 
(b_{\ell_1}^2 C_{\ell_1} + N_{\ell_1})
(b_{\ell_2}^2 C_{\ell_2} + N_{\ell_2}) 
(b_{\ell_3}^2 C_{\ell_3} + N_{\ell_3})
\label{binnedVarreal}
\ee
where $b_\ell$ is the beam transfer function and $N_\ell$ the instrument 
noise power spectrum. This expression is exact only for 
an axisymmetric beam and isotropic noise; otherwise it is an approximation
(because the beam and noise properties would include off-diagonal matrix 
elements). For a Gaussian beam, the beam transfer function is typically 
specified by the full width at half maximum $\theta_\mathrm{FWHM}$ (in radians),
so that
$
b_\ell = \exp \left[ - \frac{1}{2} \ell(\ell+1) \theta^2_\mathrm{FWHM} 
/ (8 \ln 2) \right]
$.
A pixel window function $w_\ell$ to account for pixelization effects
is combined with the beam transfer function according to 
$b_\ell \rightarrow w_\ell b_\ell$.

For bispectral elements including both $T$ and $E$, the
variance is replaced by the covariance matrix in polarization space, whose
expression without binning is
\be
\mathrm{Covar}(B_{\ell_1 \ell_2 \ell_3}^{p_1 p_2 p_3}, B_{\ell_1 \ell_2 \ell_3}^{p_4 p_5 p_6}) 
= g_{\ell_1 \ell_2 \ell_3} h_{\ell_1 \ell_2 \ell_3}^2
(\tilde{C}_{\ell_1})^{p_1 p_4} (\tilde{C}_{\ell_2})^{p_2 p_5}
(\tilde{C}_{\ell_3})^{p_3 p_6} \equiv V_{\ell_1 \ell_2 \ell_3}^{p_1 p_2 p_3 p_4 p_5 p_6},
\label{covarexact_polar}
\ee
where
\be
\tilde{C}_\ell = \left( \begin{array}{cc}
(b_\ell^T)^2 C_\ell^{TT} + N_\ell^T & b_\ell^T b_\ell^E C_\ell^{TE}\\
b_\ell^T b_\ell^E C_\ell^{TE} & (b_\ell^E)^2 C_\ell^{EE} + N_\ell^E
\end{array} \right).
\label{Cmatrix_polar}
\ee
Here noise uncorrelated in $T$ and $E$ has been assumed.
Similarly, for the binned case 
\begin{align}
\mathrm{Covar}(B_{i_1 i_2 i_3}^{p_1 p_2 p_3}, B_{i_1 i_2 i_3}^{p_4 p_5 p_6}) 
& =  \frac{g_{i_1 i_2 i_3}}{(\Xi_{i_1 i_2 i_3})^2} 
\sum_{\ell_1\in\Delta_1} \sum_{\ell_2\in\Delta_2} \sum_{\ell_3\in\Delta_3}
h_{\ell_1 \ell_2 \ell_3}^2
(\tilde{C}_{\ell_1})^{p_1 p_4} (\tilde{C}_{\ell_2})^{p_2 p_5}
(\tilde{C}_{\ell_3})^{p_3 p_6} \nonumber\\ & \equiv V_{i_1 i_2 i_3}^{p_1 p_2 p_3 p_4 p_5 p_6}.
\label{covarbinned_polar}
\end{align}

Some subtleties arise in the derivation of equation 
(\ref{covarexact_polar}). The covariance matrix is in principle
an $8 \times 8$ matrix, given that there are 8 independent polarized bispectra
$TTT$, $TTE$, $TET$, $TEE$, $ETT$, $ETE$, $EET$, and $EEE$. As mentioned 
before, note that for
example $TTE$ and $TET$ are not the same: each polarization index $p_i$ is 
coupled to a multipole index $\ell_i$, and cannot be exchanged due to the 
restriction $\ell_1 \leq \ell_2 \leq \ell_3$ that we will always impose in 
order to reduce computation time. A naive 
calculation of this $8 \times 8$ matrix appears to lead to a more complicated
expression in the case of equal $\ell$'s that is not proportional to 
$g_{\ell_1 \ell_2 \ell_3}$. However, one should treat the cases
where two or three $\ell$'s are equal separately. For example,
when $\ell_2 = \ell_3$, one {\em can} exchange the last two polarization 
indices and one finds that $TTE = TET$ and $ETE = EET$. Hence in that case there
are only 6 independent bispectra, and the covariance matrix is $6\times 6$.
Similarly, when all three $\ell$'s are equal, 
the covariance matrix is $4\times 4$.

However, it turns out that as far as computing $f_\mathrm{NL}$ is concerned,
when evaluating the sum in (\ref{innerprodexact_polar}),
properly treating the special cases where $\ell$'s are equal
by reducing the dimension of the covariance matrix and bispectrum vector,
the final result is identical to the following calculation: taking the
covariance matrix to be the $8\times 8$ matrix as computed in the case
of all $\ell$'s unequal, multiplying it by $g_{\ell_1 \ell_2 \ell_3}$, and 
then computing the sum in (\ref{innerprodexact_polar})
directly without treating the cases of equal $\ell$'s separately.
This second computation is much more convenient from a practical point
of view. Finally it can be shown that the latter expression of the 
covariance matrix can be rewritten as the separable product involving only
$2\times 2$ matrices in (\ref{covarexact_polar}).

Similarly it can be shown that the variances of the combinations
$B^{T2E} \equiv TTE+TET+ETT$ and $B^{TE2} \equiv TEE+ETE+EET$ used for the 
smoothed bispectrum (see section~\ref{smoothingsec})
are also recovered correctly when using (\ref{covarexact_polar}) or
(\ref{covarbinned_polar}). Here one should use of course that Var($B^{T2E}$) = 
Var($TTE$) + Var($TET$) + Var($ETT$) + 2 Covar($TTE,TET$) + 2 Covar($TTE,ETT$)
+ 2 Covar($TET,ETT$), and similarly for Var($B^{TE2}$).
So in the end, while one should remember the caveats regarding 
(\ref{covarexact_polar}) and (\ref{covarbinned_polar}) in the case of equal
$\ell$'s or $i$'s, for the practical purposes of this thesis they can be used
without any problem.

\subsection{$\fnl$ estimation on an ideal sky}
\label{fNLestsec}

We start by considering the case where we have only temperature.
In order to estimate $f_\mathrm{NL}$ using a template 
$B_{\ell_1 \ell_2 \ell_3}^\mathrm{th}$, the estimator 
\be
{\hat f}_\mathrm{NL}=
\frac{
\left\langle
B^\mathrm{th,exp},
B^\mathrm{obs}
\right\rangle
}{
\left\langle
B^\mathrm{th,exp},
B^\mathrm{th,exp}
\right\rangle
}
\label{fNL_estimator}
\ee
is constructed using the inner product 
\be 
\langle B^A, B^B \rangle^\mathrm{no\ binning} =
\sum_{\ell _1 \leq \ell _2 \leq \ell_3}
\frac{
B^A_{\ell _1 \ell _2 \ell _3}
B^B_{\ell _1 \ell _2 \ell _3}
}{
V_{\ell _1 \ell _2 \ell _3}
}.
\label{innerprodexact}
\ee
This definition satisfies the mathematical axioms 
of an inner product as long as bin triplets with infinite
variance are excluded from the sum.
The theoretical bispectrum for the experiment is related to the 
theoretically predicted infinite angular resolution
bispectrum by the relation 
$B^\mathrm{th,exp}_{\ell_1 \ell_2 \ell_3}= b_{\ell _1} b_{\ell _2} b_{\ell _3}
B^\mathrm{th,f_\mathrm{NL}=1}_{\ell_1 \ell_2 \ell_3}$.
For the binned estimator the template is first binned as
$
B^\mathrm{th,exp}_{i_1 i_2 i_3} = 
(\sum_{\ell_1\in\Delta_1} \sum_{\ell_2\in\Delta_2} \sum_{\ell_3\in\Delta_3}
B^\mathrm{th,exp}_{\ell_1 \ell_2 \ell_3})/\Xi_{i_1 i_2 i_3}
$
and then the above estimator can be used with the binned version of the 
inner product:
\be
\langle B^A, B^B \rangle^\mathrm{binned} =
\sum_{i_1 \leq i_2 \leq i_3}
\frac{B_{i_1 i_2 i_3}^A B_{i_1 i_2 i_3}^B}{V_{i_1 i_2 i_3}}.
\label{innerprod}
\ee

One sees that the above estimator is of the form 
$
{\hat f}_\mathrm{NL} \propto \sum [(B^\mathrm{th})^2/V] 
[B^\mathrm{obs}/B^\mathrm{th}]
$
(where from now on we drop the explicit ``exp'' label).
Since $V_{i_1 i_2 i_3}$ is the theoretical estimate of the variance of 
$B_{i_1 i_2 i_3}^\mathrm{obs}$ in the approximation of weak non-Gaussianity,
the estimator is inverse variance weighted: $B^\mathrm{obs}/B^\mathrm{th}$
is an estimate of $f_\mathrm{NL}$ based on a single bin triplet, and all these 
estimates are combined, weighted by the inverse of their variance, 
$V/(B^\mathrm{th})^2$.
The proportionality factor 
$1/\langle B^\mathrm{th},B^\mathrm{th}\rangle$ is the
normalization of the weights and gives the theoretical (Gaussian) estimate 
for the variance\footnote{If we have independent quantities
$y_i$ with variances $v_i$ and define the inverse-variance weights as
$w_i = (1/v_i)/(\sum_j 1/v_j)$, then the variance of the weighted mean
$\sum_i w_i y_i$ is $\sum_i w_i^2 v_i = (\sum_i v_i/v_i^2)/(\sum_j 1/v_j)^2
= 1/(\sum_j 1/v_j)$.}
of the total estimator $\hat{f}_\mathrm{NL}$. This is the same 
as saying that $\langle B^\mathrm{th},B^\mathrm{th}\rangle$ is the 
$\chi^2$ or $(S/N)^2$ of the estimator in the case $f_\mathrm{NL}=1$.

The generalization of the $f_\mathrm{NL}$ estimator to include polarization
in the case without binning was worked out in \citep{Yadav:2007rk}.
In that case the inner product (\ref{innerprodexact})
should be replaced by
\be
\langle B^A, B^B \rangle^\mathrm{no\ binning} =
\sum_{\ell_1 \leq \ell_2 \leq \ell_3}
\!\sum_{\tiny\begin{array}{c@{}c@{}c} p_1, & p_2, & p_3, \\ p_4, & p_5, & p_6\\ 
\end{array}} \!\!\!\!\!\!
B_{\ell_1 \ell_2 \ell_3}^{p_1 p_2 p_3, A}
(V^{-1})_{\ell_1 \ell_2 \ell_3}^{p_1 p_2 p_3 p_4 p_5 p_6}
 B_{\ell_1 \ell_2 \ell_3}^{p_4 p_5 p_6, B},
\label{innerprodexact_polar}
\ee
which involves the inverse of the covariance matrix given in 
(\ref{covarexact_polar}).
Computing this inverse simply implies inverting the three 
$2 \times 2$ matrices $\tilde{C}_\ell$ given in (\ref{Cmatrix_polar}).

Deriving an equivalent expression for the binned estimator is 
straightforward, as long as one keeps in mind that one should first bin
the elements of the covariance matrix (since that corresponds to the
covariance matrix of the binned bispectrum) and only afterwards compute 
the inverse. Trying to bin directly the elements of the inverse covariance
matrix (or one divided by these elements) is incorrect and leads to
wrong results (in particular for bins where $C_\ell^{TE}$ crosses zero).
So in the end the generalization of the binned bispectrum
estimator to include polarization is given by the prescription that the
inner product (\ref{innerprod}) should be replaced by
\be
\langle B^A, B^B \rangle^\mathrm{binned} =
\sum_{i_1 \leq i_2 \leq i_3}
\!
\sum_{\tiny\begin{array}{c@{}c@{}c} p_1, & p_2, & p_3, \\ p_4, & p_5, & p_6\\ 
\end{array}} \!\!\!\!\!\!
B_{i_1 i_2 i_3}^{p_1 p_2 p_3, A} 
(V^{-1})_{i_1 i_2 i_3}^{p_1 p_2 p_3 p_4 p_5 p_6}
B_{i_1 i_2 i_3}^{p_4 p_5 p_6, B},
\label{innerprod_polar}
\ee
involving the inverse of the binned covariance matrix given in 
(\ref{covarbinned_polar}).
However, since the multiplication with $h_{\ell_1 \ell_2 \ell_3}^2$ in
combination with the binning couples the three $\tilde{C}_\ell$ matrices 
in (\ref{covarbinned_polar}) together,
the covariance matrix can only be inverted as a full $8\times 8$ matrix
that is no longer separable in $\ell$. Fortunately this non-separability
is irrelevant for the binned bispectrum estimator.

We can quantify how much the estimator variance increases due to binning, 
compared with an ideal estimator without binning:
\begin{equation}
R \equiv \frac{\Var ( \hat{f}_\mathrm{NL}^\mathrm{ideal})}
{\Var ( \hat{f}_\mathrm{NL}^\mathrm{binned})}
=
\frac{\langle B^\mathrm{th}, B^\mathrm{th} \rangle^\mathrm{binned}}
{\langle B^\mathrm{th}, B^\mathrm{th} \rangle^\mathrm{no\ binning}} .
\label{binning_error}
\end{equation}
$R$ is a number between 0 and 1. The closer $R$ is to 1, the better the binned
approximation for the template under consideration.
To show that $0 \leq R \leq 1$ we need to rewrite (\ref{binning_error}) in terms
of a single inner product definition. It can be checked straightforwardly that 
the binned inner product of the theoretical bispectrum can be rewritten as the 
exact inner product (no binning) of the bispectrum template defined below:
\begin{equation}
\langle B^\mathrm{th}, B^\mathrm{th} \rangle^\mathrm{binned}
= \langle B^\mathrm{bin}, B^\mathrm{bin} \rangle^\mathrm{no\ binning},
\end{equation}
where
\begin{equation}
B_{\ell_1 \ell_2 \ell_3}^{p_1 p_2 p_3, \mathrm{bin}} \equiv
\frac{1}{\Xi_{i_1 i_2 i_3}} \frac{g_{i_1 i_2 i_3}}{g_{\ell_1 \ell_2 \ell_3}}
\!\!\!\! \sum_{\tiny\begin{array}{c@{}c@{}c} p_4, & p_5, & p_6, \\ p_7, & p_8, & p_9\\ 
\end{array}} \!\!\!\!\!\!
V_{\ell_1 \ell_2 \ell_3}^{p_1 p_2 p_3 p_4 p_5 p_6} (V^{-1})_{i_1 i_2 i_3}^{p_4 p_5 p_6 p_7 p_8 p_9} 
B_{i_1 i_2 i_3}^{p_7 p_8 p_9, \mathrm{th}}
\end{equation}
with $(i_1,i_2,i_3)$ the bin triplet that contains the $\ell$-triplet
$(\ell_1,\ell_2,\ell_3)$.\footnote{This result follows from the identity 
(for any function $u_{\ell_1 \ell_2 \ell_3}$)
\begin{align}
\sum_{\ell_1 \leq \ell_2 \leq \ell_3}
\frac{1}{g_{\ell_1 \ell_2 \ell_3}} \, u_{\ell_1 \ell_2 \ell_3}
& = \frac{1}{6} \sum_{\ell_1,\ell_2,\ell_3} u_{\ell_1 \ell_2 \ell_3}
= \frac{1}{6} \sum_{i_1,i_2,i_3} 
\sum_{\ell_1\in\Delta_1} \sum_{\ell_2\in\Delta_2} \sum_{\ell_3\in\Delta_3} u_{\ell_1 \ell_2 \ell_3}
\nonumber\\ & = \sum_{i_1 \leq i_2 \leq i_3} \frac{1}{g_{i_1 i_2 i_3}}
\sum_{\ell_1\in\Delta_1} \sum_{\ell_2\in\Delta_2} \sum_{\ell_3\in\Delta_3} u_{\ell_1 \ell_2 \ell_3}.
\end{align}}
In addition it is simple to show that
\begin{equation}
\langle B^\mathrm{bin}, B^\mathrm{bin} \rangle^\mathrm{no\ binning}
= \langle B^\mathrm{bin}, B^\mathrm{th} \rangle^\mathrm{no\ binning}.
\end{equation}
Now we can rewrite $R$ as
\begin{equation}
R = \frac{\langle B^\mathrm{bin}, B^\mathrm{bin} \rangle^\mathrm{no\ binning}}
{\langle B^\mathrm{th}, B^\mathrm{th} \rangle^\mathrm{no\ binning}}
= \frac{\langle B^\mathrm{bin}, B^\mathrm{th} \rangle^\mathrm{no\ binning}}
{\langle B^\mathrm{th}, B^\mathrm{th} \rangle^\mathrm{no\ binning}}
= \frac{(\langle B^\mathrm{bin}, B^\mathrm{th} \rangle^\mathrm{no\ binning})^2}
{\langle B^\mathrm{th}, B^\mathrm{th} \rangle^\mathrm{no\ binning}
\langle B^\mathrm{bin}, B^\mathrm{bin} \rangle^\mathrm{no\ binning}}.
\end{equation}
From the first expression, given that $\langle x,x \rangle \geq 0$ for an
inner product, we see that $R \geq 0$. And the last expression implies
that $R \leq 1$ using the Cauchy-Schwarz inequality.

If more than one of the above bispectrum shapes are expected to be
present in the data, then a joint estimation of the different $f_\mathrm{NL}$
parameters is required. For this the Fisher matrix 
\begin{equation}
F_{IJ} = \langle B^I, B^J \rangle,
\label{Fisher}
\end{equation}
where $I,J $ label the theoretical shapes (for example local and equilateral),
is a crucial quantity. 
The optimal estimation of the
$f_\mathrm{NL}$ parameter for shape $I$ is given by
\begin{equation}
\hat{f}_\mathrm{NL}^I = \sum_J (F^{-1})_{IJ} \langle B^J, B^\mathrm{obs} \rangle.
\label{fNL_joint}
\end{equation}
The estimate of the variance of $\hat{f}_\mathrm{NL}^I$
is $(F^{-1})_{II}$. If, on the other hand, the $\hat{f}_\mathrm{NL}^I$ parameters
would have been estimated independently using (\ref{fNL_estimator}) (as if
there is only one bispectrum shape present, but it is unknown which), then
their variance is given by $1/F_{II}$.

Another useful quantity to define is the symmetric correlation matrix 
\begin{equation}
C_{IJ} \equiv \frac{F_{IJ}}{\sqrt{F_{II} F_{JJ}}}
\label{corrmatrix}
\end{equation}
giving the correlation coefficients between any two bispectrum shapes.
By construction $-1 \leq C_{IJ} \leq +1$, with $C_{IJ}=-1,0,+1$ meaning that
the two shapes are fully anti-correlated, uncorrelated, or fully correlated, 
respectively.
Note that one could also define a correlation matrix using the inverse of
the Fisher matrix instead of the Fisher matrix itself in (\ref{corrmatrix}).
That would give us the correlation of the {\em $f_\mathrm{NL}$ parameters},
while (\ref{corrmatrix}) represents the correlation of the {\em templates}.
As an example we show the correlation coefficients between the templates of 
section~\ref{templatesec} in table~\ref{tab_corr_coeff}.

\begin{table}
\begin{center}
\begin{tabular}{l|cccccc}
\hline
& Local & Equil & Ortho & LensISW & UnclustPS & CIB \\
\hline
Local & 1 & 0.21 & -0.44 & 0.28 & 0.002 & 0.006\\
Equilateral && 1 & -0.05 & 0.003 & 0.008 & 0.03\\
Orthogonal &&& 1 & -0.15 & -0.003 & -0.001\\
Lensing-ISW &&&& 1 & -0.005 & -0.03\\
Unclustered point sources &&&&& 1 & 0.93\\
CIB point sources &&&&&& 1\\
\hline
\end{tabular}
\end{center}
\caption{Correlation coefficients between the theoretical templates
of section~\ref{templatesec}, as defined in (\ref{corrmatrix}). 
The numbers are computed using the characteristics of the Planck experiment
and are for temperature. We see a large
correlation between local and orthogonal and between local and lensing-ISW.
Equilateral and orthogonal are mostly uncorrelated, and the correlation
between the point source templates and the primordial ones is negligible.}
\label{tab_corr_coeff}
\end{table}

Suppose that we had only two shapes with non-zero correlation, but the
amplitude of the second was fixed by theory (as is the case for example for
the lensing-ISW template that has no unknown amplitude parameter).
If the theory was fully trusted, it would be a shame to do a joint estimation,
with the associated increase in variance. In that case the influence of
the second shape on the first is more properly treated as a known bias that
can be subtracted without increasing the variance. The size of the bias can
be found from (\ref{fNL_joint}), by using the second equation ($I=2$) to 
eliminate $\langle B^{(2)}, B^\mathrm{obs} \rangle$ from the first equation 
($I=1$). After expressing
the elements of the inverse Fisher matrix in terms of the elements of the
Fisher matrix, the resulting equation for the first $f_\mathrm{NL}$ parameter
simplifies to:
\begin{equation}
\hat{f}_\mathrm{NL}^{(1)} = \frac{1}{F_{11}} \langle B^{(1)}, B^\mathrm{obs} \rangle
- \frac{F_{12}}{F_{11}} f_\mathrm{NL}^{(2)},
\label{biascorreq}
\end{equation}
the second term being the bias correction.
Here $f_\mathrm{NL}^{(2)}$ is the known $f_\mathrm{NL}$ parameter of the second 
shape, most likely equal to one if the known amplitude was included in the 
template (as is the case for example for the lensing-ISW template). The variance
of $\hat{f}_\mathrm{NL}^{(1)}$ is not influenced by the bias correction and remains
equal to $1/F_{11}$, the same as for a single shape. This result
can easily be generalized to more than two shapes.

To conclude this section, we mention that in addition to studies of $\fnl$
as presented here, or non-parametric studies as presented later in
section~\ref{smoothingsec}, one can
also construct other types of parametric estimators. For example, in
\citep{BvTC} we constructed two types of estimators to look for specific
features in the CMB bispectrum related to the acoustic oscillations. At that
time (before Planck) there were hints from WMAP that $\fnl^\mathrm{loc}$
might be quite large, 30--50. If that had been true, these estimators
would have had a high signal-to-noise on the Planck data. Unfortunately, Planck
ruled out such a high value of $\fnl^\mathrm{loc}$, which makes these particular
estimators mostly irrelevant, at least as far as the Planck data is concerned.

\subsection{Extensions for a realistic sky}
\label{realsky}

The definition of the bispectrum in (\ref{BispecFromMap}) or (\ref{Bobsbinned})
assumes a
rotationally invariant CMB sky and that the bispectral expectation
values have even parity (as a consequence of the parity invariance of
the underlying stochastic process, which we assume here).  Because 
of rotational invariance, the $m$ dependence of the
expectation can be factored out, and the reduced bispectral
coefficients depending only on the $\ell$ provide a lossless compression
of the data concerning the bispectrum. 
However, in a real experiment, as opposed to idealized observations of 
the primordial sky, two sources of anisotropy arise that break rotational
invariance and require corrections to the bispectrum estimation to avoid
spurious results.

The first is anisotropic superimposed instrument noise,
due to for example an anisotropic scanning pattern of the satellite.
The second is anisotropy introduced by a mask needed to remove the brightest 
parts of our galaxy and the strongest point sources. These two anisotropic 
`contaminants', unlike for example foreground contaminants,
cannot be removed by cleaning and must be accounted for in the analysis. 
They can mimic a primordial bispectrum signal. For 
example, due to an anisotropic scanning pattern of the experiment, certain 
(large-scale) areas of the sky may have less (small-scale) noise than other
areas. This 
correlation between large and small scales produces a contaminant bispectrum 
that peaks in the squeezed limit (bispectrum configurations with one small
$\ell$ and two large ones). That is also
where the primordial local shape has its main signal. 
Since the CMB and the noise are uncorrelated, the effect will average out
to zero in the central value of the bispectrum over a large number of maps 
(no bias), but it will increase the variance. And while an unbiased estimator 
will find the correct central value when averaged over a large number of maps, 
a larger variance does mean that there is more chance to find a value far from 
the true one when applied to a single map.

These contaminants can be mitigated by subtracting from the cubic 
expression of the observed bispectrum given in (\ref{BispecFromMap}) or 
(\ref{Bobsbinned}) a linear correction term, as shown in 
\citep{Creminelli:2005hu,Yadav:2007ny}, that is,
\be
B_{i_1 i_2 i_3}^{p_1 p_2 p_3, \mathrm{obs}} \rightarrow \Bigl(
B_{i_1 i_2 i_3}^{p_1 p_2 p_3, \mathrm{obs}} - B_{i_1 i_2 i_3}^{p_1 p_2 p_3, \mathrm{lin}}\Bigr) .
\ee
`Cubic' and `linear' here mean 
cubic and linear in the observed map, respectively.
The linear correction term is
\begin{align}
B_{i_1 i_2 i_3}^{p_1 p_2 p_3, \mathrm{lin}} = \int d\Omega & \left [
M_{i_1}^{p_1, \mathrm{obs}}
\left\langle M_{i_2}^{p_2, G} M_{i_3}^{p_3, G} \right\rangle \right .
 \nonumber\\ & \left. + M_{i_2}^{p_2, \mathrm{obs}} \left\langle M_{i_1}^{p_1, G} M_{i_3}^{p_3, G} \right\rangle 
+ M_{i_3}^{p_3, \mathrm{obs}} \left\langle M_{i_1}^{p_1, G} M_{i_2}^{p_2, G} \right\rangle 
\right ],
\label{Bisp_lincorr}
\end{align}
where the average is over Gaussian CMB maps with the same beam,
(anisotropic) noise, and mask as the observed map.
A detailed derivation of the linear correction term can be found in the first
appendix of \citep{JRvT}, which paper is included in appendix~\ref{JRvTapp} of
this thesis, see section~\ref{ap:appendices}.
The linear correction is hugely significant for the local shape as explained
above, very significant for orthogonal (due to the large correlation with
local), and insignificant for equilateral.
While adding the linear correction term completely solves the
issue related to anisotropic noise, it turns out that for the issue related
to the mask we need an additional ingredient to make our bispectrum estimator
optimal again.

The region near the galactic plane and around extragalactic point
sources, where reliable subtraction of contaminants is not possible,
must be masked to prevent contamination of the primordial
bispectrum. Masking introduces a number of problems for estimating the
bispectrum because the process of filtering maps is nonlocal. If we
naively analyze a masked map in which the masked pixels are set to
zero --- or better yet, set equal to the average value of the unmasked
part of the map --- by filtering it, say with a high-pass filter, we
would observe a deficit of small scale power around the edges of the
mask. A filter in frequency space moves around the small scale power
in real space. The power is smeared, so that if there is no small
scale power in the masked region, power from the unmasked region
escapes into the masked region without there being a compensating flux
returning from the masked region. Another edge effect tending to
increase the small scale power around the border of the unmasked
region results if there is a jump discontinuity. Such a discontinuity
contains spurious small scale power that bleeds into the unmasked
region after filtering. It is therefore important to introduce
artificially the right amount of small scale power into the masked
region and to avoid spurious jumps in the maps so that the two fluxes
cancel after filtering.\footnote{Large-scale modes are much less
affected by the mask. Since these modes extend out over large parts of
the sky, they can be reconstructed reasonably accurately even when
some parts of the sky are missing. Furthermore, edge effects are also
less important for a mask with larger holes. Consequently for a
high-resolution experiment like Planck the use of inpainting algorithms
has turned out to be absolutely crucial, while for the lower resolution WMAP
experiment, which moreover had larger error bars, less care was
required.} This process of filling in the masked regions is also known
as `inpainting'.

Before showing quantitatively how masking affects the determination of
$f_\mathrm{NL}$, we first have to determine what the effect on the
error bars would be if we had none of these problems, but only less
data due to the reduced fraction of the sky. When the bispectrum is
determined according to (\ref{Bobsbinned}), it should be multiplied
with a factor $1/f_\mathrm{sky}$ to correct for the partial sky
coverage~\citep{Komatsu:2003iq}, where $f_\mathrm{sky}$ is the
fraction of the sky that is left unmasked. In practice this is done
automatically when the integral is replaced by a sum over the pixels:
the product of maps is summed over all unmasked pixels, divided by the
number of unmasked pixels, and multiplied by $4\pi$. In addition, the
partial sky coverage increases the variance of the estimator, the
theoretical estimate of which becomes $1/(f_\mathrm{sky} \langle
B^\mathrm{th,exp},B^\mathrm{th,exp}\rangle)$. The factor of $1/f_\mathrm{sky}$
can easily be understood given that the variance of a quantity determined from
$N$ data points scales as $1/N$ and here the number of data points roughly
corresponds to the number of observed pixels on the sky. If the mask is not too
large, this simple prescription for the variance works quite well.

\begin{table}
\begin{center}
\begin{tabular}{lcccccc}
\hline
& \multicolumn{3}{c}{No linear correction} &
\multicolumn{3}{c}{With linear correction} \\
& Local & Equil & Ortho & Local & Equil & Ortho \\
\hline
\multicolumn{4}{l}{No mask, isotropic noise} &&& \\
\multicolumn{1}{@{\hspace{1.3cm}}c@{\hspace{0.8cm}}}{$TTT$} & -0.1 $\pm$ 4.1 & 2 $\pm$ 58 & 5 $\pm$ 24
& -0.1 $\pm$ 4.1 & 3 $\pm$ 57 & 4 $\pm$ 25 \\
\multicolumn{1}{@{\hspace{0.5cm}}c@{}}{$EEE$} & 0.4 $\pm$ 24 & -11 $\pm$ 170 & 6 $\pm$ 92
& 0.4 $\pm$ 24 & -11 $\pm$ 171 & 7 $\pm$ 94 \\
\multicolumn{4}{l}{No mask, anisotropic noise} &&& \\
\multicolumn{1}{@{\hspace{0.5cm}}c@{}}{$TTT$} & 5.7 $\pm$ 84 & 2 $\pm$ 58 & 2 $\pm$ 35
& -0.2 $\pm$ 4.2 & 3 $\pm$ 57 & 4 $\pm$ 24 \\
\multicolumn{1}{@{\hspace{0.5cm}}c@{}}{$EEE$} & -23 $\pm$ 736 & -22 $\pm$ 193 & 15 $\pm$ 197
& 0.4 $\pm$ 24 & -20 $\pm$ 195 & 7 $\pm$ 94 \\
\multicolumn{4}{l}{Galactic mask, isotropic noise} &&& \\
\multicolumn{4}{l}{-- No filling in} &&& \\
\multicolumn{1}{@{\hspace{0.5cm}}c@{}}{$TTT$} & -0.2 $\pm$ 5.5 & 11 $\pm$ 78 & -1 $\pm$ 58
& 0.3 $\pm$ 5.1 & 6 $\pm$ 70 & 6 $\pm$ 32 \\
\multicolumn{1}{@{\hspace{0.5cm}}c@{}}{$EEE$} & 5 $\pm$ 32 & -5 $\pm$ 199 & 1 $\pm$ 108
& 2 $\pm$ 28 & -9 $\pm$ 202 & 3 $\pm$ 109 \\
\multicolumn{4}{l}{-- Diffusive filling in}&&& \\
\multicolumn{1}{@{\hspace{0.5cm}}c@{}}{$TTT$} & 0.8 $\pm$ 6.2 & 6 $\pm$ 70 & 4 $\pm$ 28
& 0.3 $\pm$ 4.6 & 7 $\pm$ 69 & 4 $\pm$ 29 \\
\multicolumn{1}{@{\hspace{0.5cm}}c@{}}{$EEE$} & 5 $\pm$ 31 & -8 $\pm$ 196 & 1 $\pm$ 109
& 2 $\pm$ 28 & -8 $\pm$ 198 & 2 $\pm$ 110 \\
\multicolumn{4}{l}{Point source mask, isotropic noise} &&& \\
\multicolumn{4}{l}{-- No filling in} &&& \\
\multicolumn{1}{@{\hspace{0.5cm}}c@{}}{$TTT$} & -0.7 $\pm$ 9.2 & 3 $\pm$ 73 & 6 $\pm$ 51
& -0.4 $\pm$ 8.4 & 3 $\pm$ 65 & 7 $\pm$ 36 \\
\multicolumn{1}{@{\hspace{0.5cm}}c@{}}{$EEE$} & 1 $\pm$ 27 & -7 $\pm$ 170 & 10 $\pm$ 92
& 0.1 $\pm$ 23 & -7 $\pm$ 170 & 9 $\pm$ 89 \\
\multicolumn{4}{l}{-- Diffusive filling in} &&& \\
\multicolumn{1}{@{\hspace{0.5cm}}c@{}}{$TTT$} & 0.2 $\pm$ 6.3 & 2 $\pm$ 59 & 5 $\pm$ 25
& -0.3 $\pm$ 4.3 & 3 $\pm$ 58 & 4 $\pm$ 24 \\
\multicolumn{1}{@{\hspace{0.5cm}}c@{}}{$EEE$} & -0.1 $\pm$ 26 & -0.1 $\pm$ 172 & 13 $\pm$ 98
& -0.5 $\pm$ 24 & -3 $\pm$ 173 & 12 $\pm$ 97 \\
\multicolumn{4}{l}{Gal + ps mask, anisotropic noise} &&& \\
\multicolumn{4}{l}{-- No filling in} &&& \\
\multicolumn{1}{@{\hspace{0.5cm}}c@{}}{$TTT$} & 0.3 $\pm$ 77 & 10 $\pm$ 93 & 3 $\pm$ 87
& -0.7 $\pm$ 9.4 & 5 $\pm$ 76 & 10 $\pm$ 39 \\
\multicolumn{1}{@{\hspace{0.5cm}}c@{}}{$EEE$} & -27 $\pm$ 719 & -11 $\pm$ 214 & 17 $\pm$ 247
& 2 $\pm$ 30 & -14 $\pm$ 207 & 4 $\pm$ 101 \\
\multicolumn{4}{l}{-- Diffusive filling in of ps mask only} &&& \\
\multicolumn{1}{@{\hspace{0.5cm}}c@{}}{$TTT$} & 1.6 $\pm$ 85 & 10 $\pm$ 78 & -2 $\pm$ 70
& 0.02 $\pm$ 5.4 & 5 $\pm$ 71 & 7 $\pm$ 32 \\
\multicolumn{1}{@{\hspace{0.5cm}}c@{}}{$EEE$} & -27 $\pm$ 752 & -5 $\pm$ 213 & 16 $\pm$ 243
& 2 $\pm$ 31 & -13 $\pm$ 210 & 2 $\pm$ 109 \\
\multicolumn{4}{l}{-- Diffusive filling in of both masks} &&& \\
\multicolumn{1}{@{\hspace{0.5cm}}c@{}}{$TTT$} & 2.7 $\pm$ 87 & 6 $\pm$ 72 & 3 $\pm$ 44
& -0.04 $\pm$ 5.0 & 6 $\pm$ 69 & 4 $\pm$ 29 \\
\multicolumn{1}{@{\hspace{0.5cm}}c@{}}{$EEE$} & -26 $\pm$ 756 & -9 $\pm$ 210 & 16 $\pm$ 242
& 2 $\pm$ 31 & -13 $\pm$ 208 & 1 $\pm$ 110 \\
\hline
\end{tabular}
\end{center}
\caption{Importance of filling in and of the linear
correction term for determining $f_\mathrm{NL}$ in the presence of
a mask and anisotropic noise. The results are based on a set of 100 Gaussian
CMB simulations at Healpix resolution $n_\mathrm{side}=2048$ with power 
spectrum according to the Planck 2013 release values. The simulations
include smoothing by a 5 arcmin FWHM Gaussian beam and noise
based on a white noise power spectrum with amplitude $1.5\times 10^{-17}$
for temperature and $6\times 10^{-17}$ for E polarization (in units made 
dimensionless by dividing by the CMB mean temperature $T_0=2.7255$~K).
Where relevant the noise has been made anisotropic by modulating it using
the hit count map of the Planck 143~GHz channel of the 2013 release.
The maps are analyzed with the binned bispectrum estimator using 54 bins 
and $\ell_\mathrm{max}=2500$, and 100 maps were used for the linear correction 
term. The masks used are shown in Fig.~\ref{fig_masks}.}
\label{table_filling_in}
\end{table}

\begin{figure}
\includegraphics[width=0.5\columnwidth]{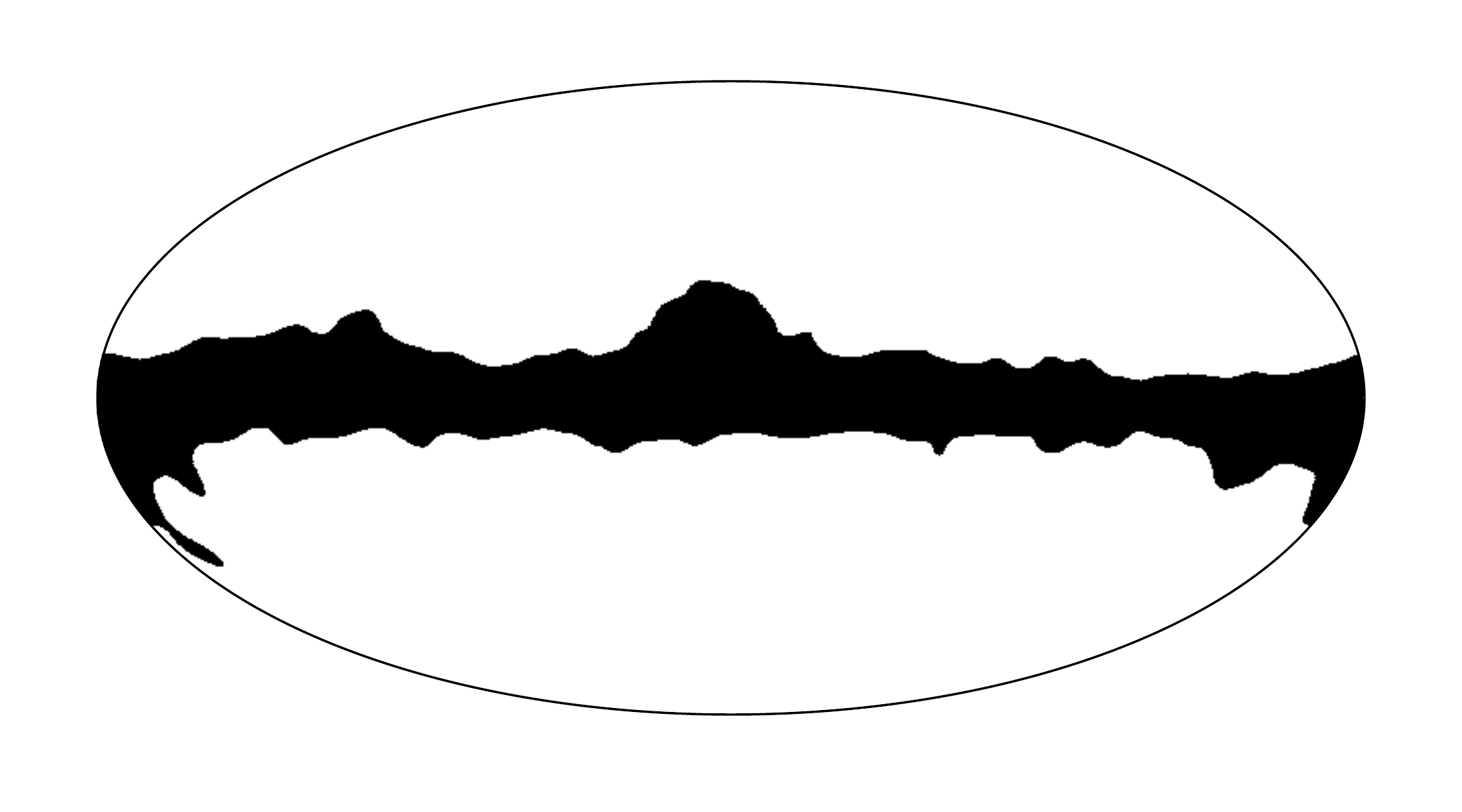}
\includegraphics[width=0.5\columnwidth]{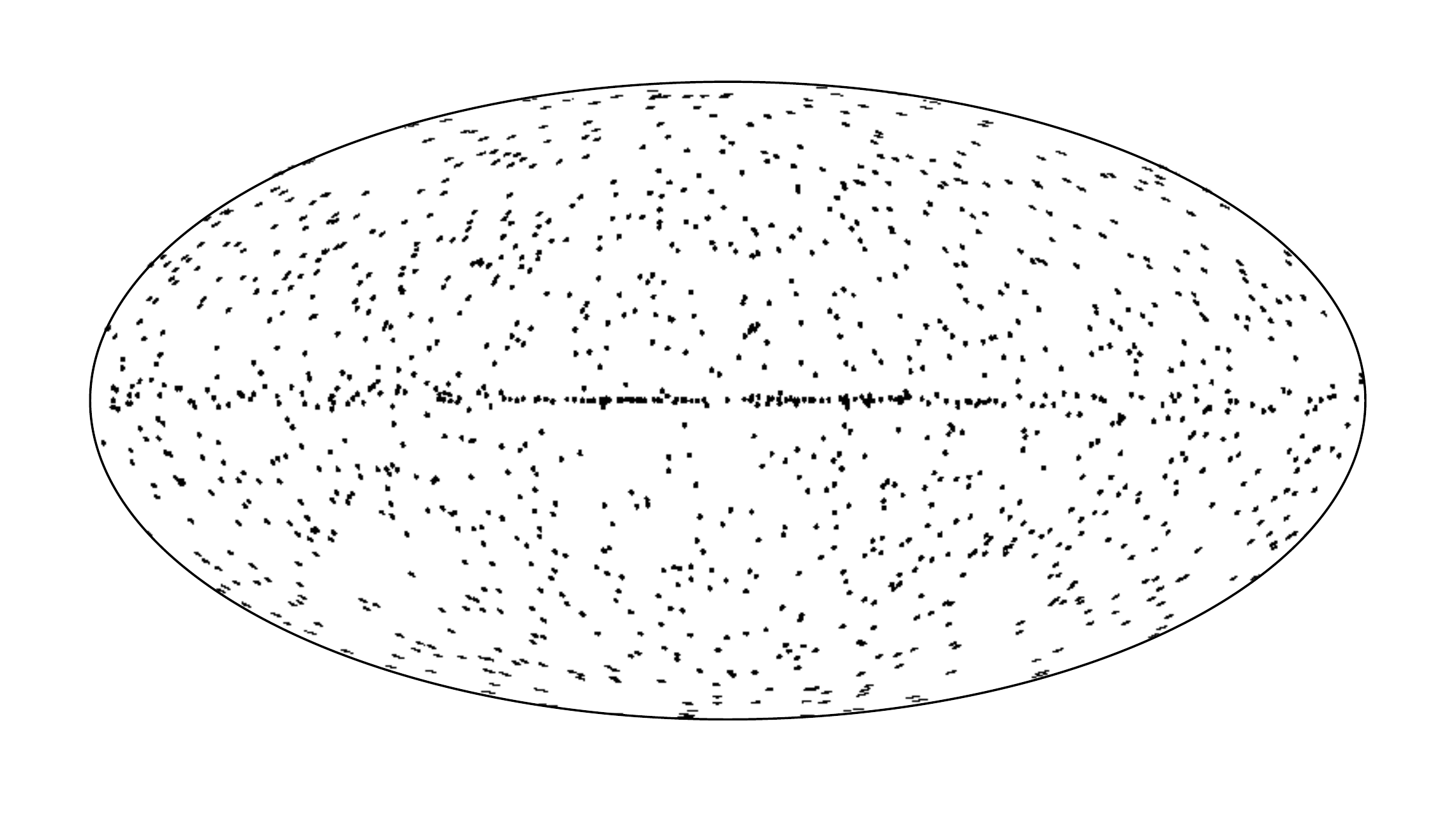}
\caption{The galactic (left) and point source mask (right) used in 
Table~\ref{table_filling_in}. The galactic mask is the $f_\mathrm{sky}=0.80$ 
galactic mask included in the Planck 2013 release, and the point source mask 
is the one based on the Planck LFI 30~GHz channel from the 2013 release 
(with a $4\sigma$ threshold level), which has $f_\mathrm{sky}=0.96$. 
The combined mask has $f_\mathrm{sky}=0.77$.}
\label{fig_masks}
\end{figure}

To illustrate quantitatively the problems encountered in determining
$f_\mathrm{NL}$ with a mask, we applied a series of tests to 
simulated CMB maps as described in Table~\ref{table_filling_in}. 
The masks used are shown in Fig.~\ref{fig_masks} while 
the details of the simulations are described in the caption of the Table. 
We find that when missing data in the masked regions are naively replaced
with the average
of the unmasked part of the map (the ``no filling in'' lines in the Table), 
the estimates of $f_\mathrm{NL}$ are unbiased but have much larger variance 
than expected, at least in temperature. The expected increase
in the standard deviation is only a factor $1/\sqrt{f_\mathrm{sky}}$ 
(i.e., $1.12$ for the galactic mask and $1.02$ for the point sources)
and in particular for the point source mask we observe
wider error bars in temperature for all three shapes.
Including the linear correction term (\ref{Bisp_lincorr}) in 
the estimator reduces this effect to some extent, but in temperature
this is clearly not enough.
The effect of the point source mask on the
local shape is exacerbated 
when the holes are smaller. For example, replacing the 
2013 Planck LFI 30~GHz point source mask with the 2013 Planck 
HFI 100~GHz 
channel mask (with a $5\sigma$ threshold level), which has a much smaller beam
and hence smaller holes ($f_\mathrm{sky}=0.99$), increases the 
``no filling in, no linear correction'' error bars for the local shape from 9.2 
to 29.5 (while the error bars for equilateral and orthogonal become smaller).
These results demonstrate the need for a suitable  filling in of the 
missing data in the masked regions of the temperature map, in particular 
for the point source mask.

\begin{figure}
\includegraphics[width=0.5\columnwidth]{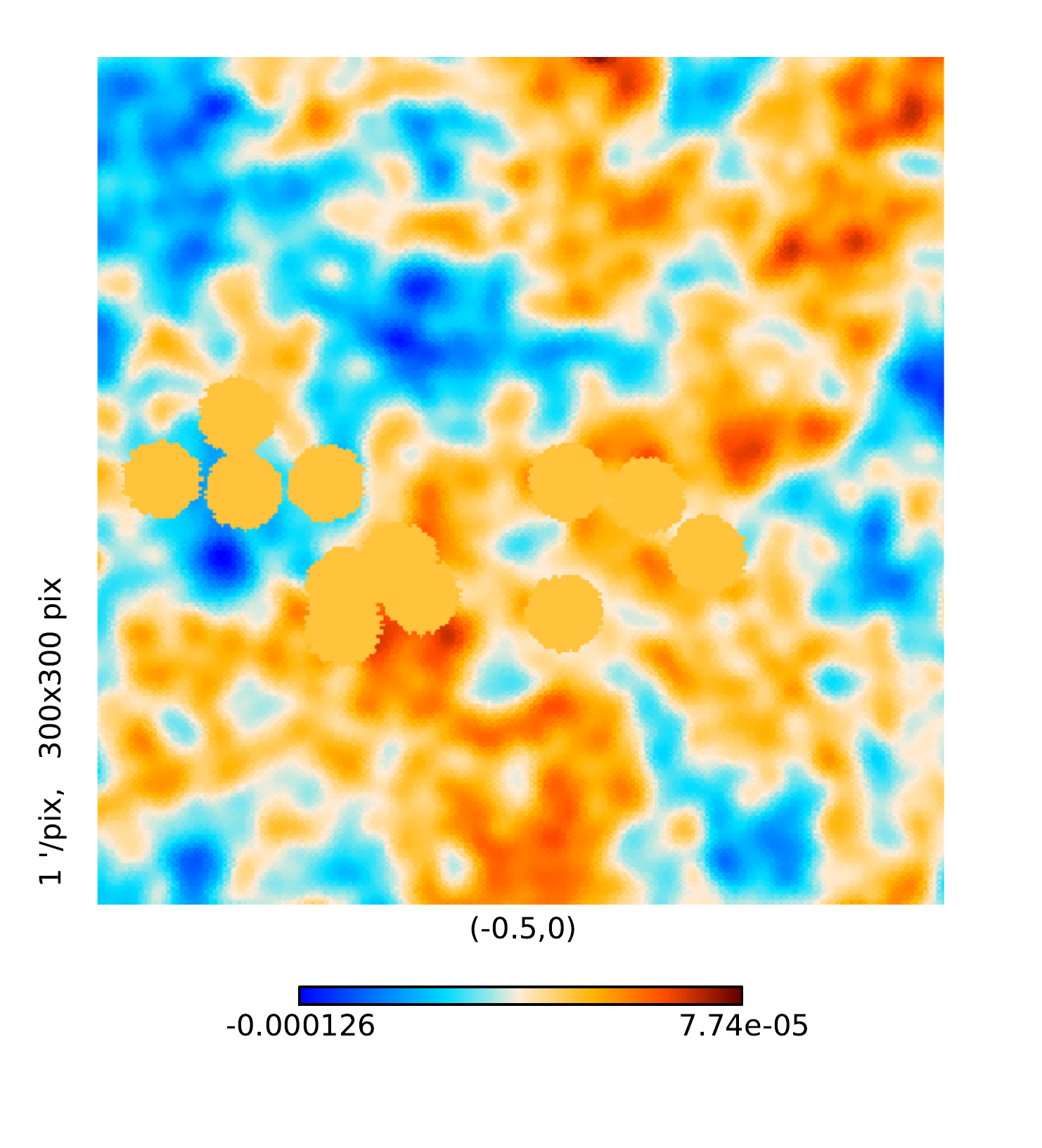}
\includegraphics[width=0.5\columnwidth]{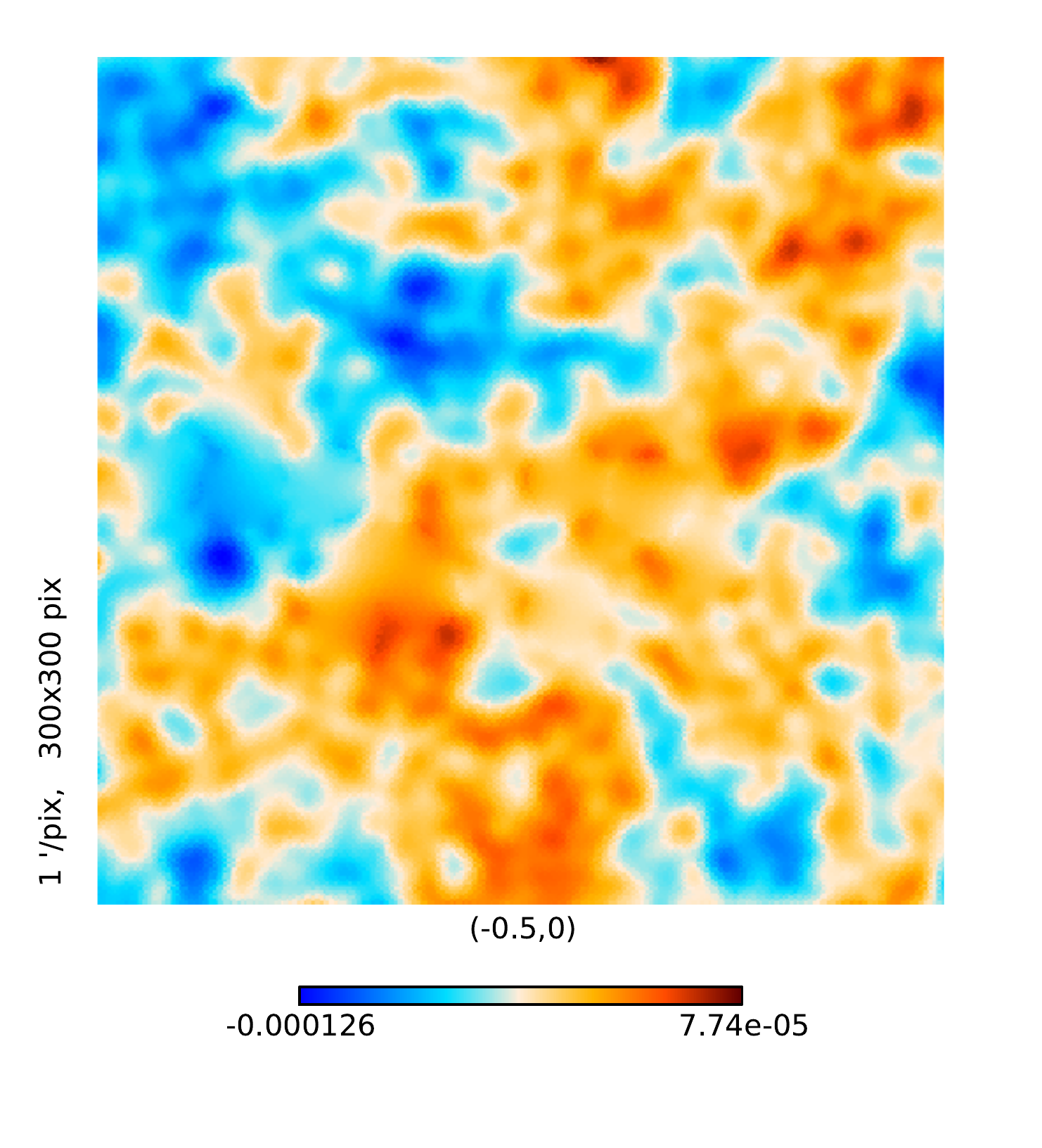}
\includegraphics[width=0.5\columnwidth]{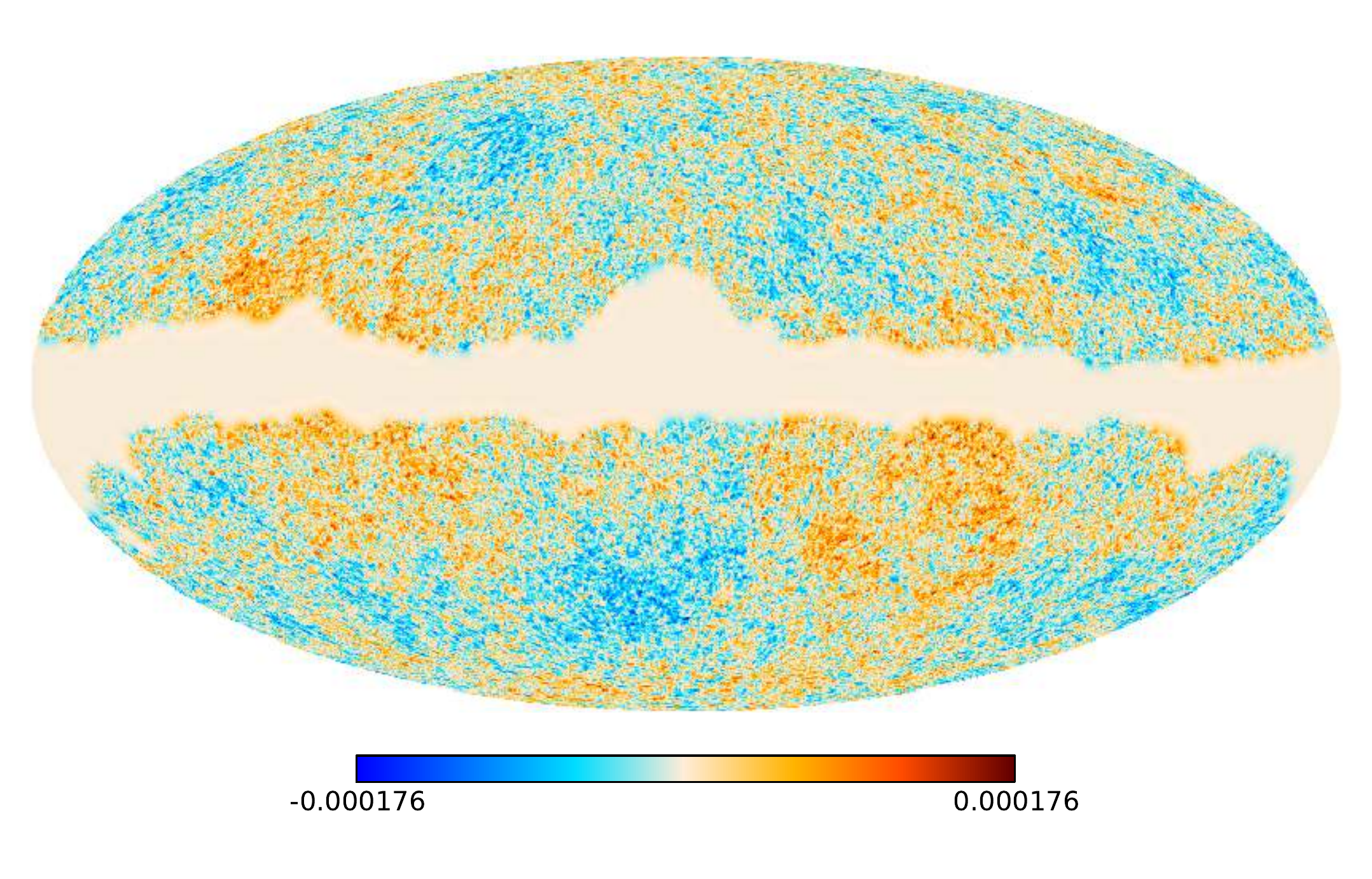}
\includegraphics[width=0.5\columnwidth]{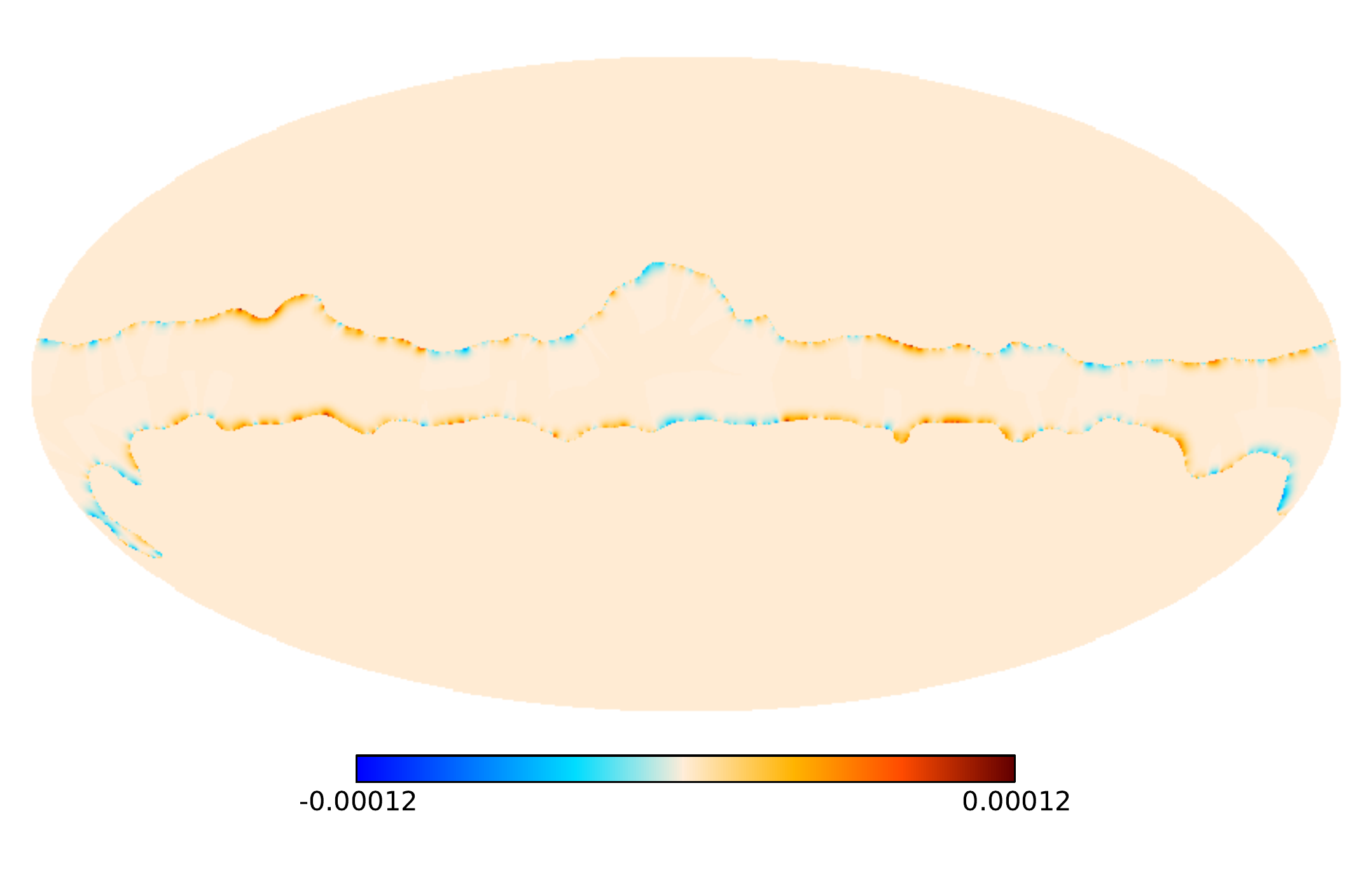}
\caption{The top two panels show a zoom of a Gaussian CMB map with holes 
from the 2013 Planck HFI 100~GHz point source mask, before (left) and after 
diffusive filling in (right) with 2000 iterations. The bottom left panel 
shows what filling in the galactic mask looks like, with the bottom right 
panel showing the difference between the maps with and without filling in.
The units are dimensionless ($\Delta T/T_0$).}
\label{fig_inpainted_maps}
\end{figure}

The simplest inpainting method is diffusive inpainting, which despite
its simplicity worked extremely well and was subsequently adopted by the 
other Planck bispectrum estimators (KSW and modal) as well. It became
the common method in all Planck releases.
After filling the masked regions with the average of the unmasked part of the
map as above, we fill each masked pixel with the average value of its
neighbouring pixels and this procedure is iterated.
We found that 2000 iterations
sufficed for the Planck maps. One can implement the iterative procedure in
two different ways: compute the average of the neighbours on the current
iteration (Gauss-Seidel method, where some of the neighbour pixels will already
have been updated and others not) or on the previous iteration kept in a buffer
(Jacobi method, where all neighbour pixels will be on the previous iteration).
While the Gauss-Seidel implementation is anisotropic, we found 
that this has no impact on the results, while on the contrary the faster 
convergence of that implementation is an advantage.
This scheme solves a discretized version of
Laplace's equation for the pixels where there is no data
with the boundary of the unmasked region providing Dirichlet boundary
data. (See~\citep{Bucher:2011nf} for a discussion of how this scheme
is related to constrained random Gaussian realizations for filling in
the missing data.) While this sort of `harmonic averaging' is simple
to implement and dulls the sharp edges, it appears at first glance not to
remedy the problem of missing small-scale power described above, as the 
resulting maps have clearly visible bald spots, see 
Fig.~\ref{fig_inpainted_maps}. However, unlike
apodization which only dulls the edges, the diffusive filling-in scheme 
does create small-scale structure inside close to the boundary of the mask. 
Given that during harmonic transforms it is the wavelength of the modes
that determines how far they propagate, this is exactly what we need:
the short wavelengths can only propagate small distances and hence need
only be reconstructed close to the edges.

After masking, filling in, and filtering the maps, we mask them once
again before integrating over products of maps.  The masked region is
never directly used in the calculation of the bispectrum, but the
filling in is crucial to avoid the influence of the masked region
spreading out over the sky when filtering the maps, as explained
above. In addition, the average of the filtered maps outside the mask 
is subtracted to remove any monopole. If this is not done, small-scale power
(whose origin is from the two-point function) will combine with this
monopole to masquerade as (local) bispectral power, and this `aliasing' 
can be a large effect.

Other more sophisticated inpainting techniques include nonlinear methods based
on sparsity (see~\citep{Abrial2007, Abrial:2008mz, Perotto:2009tv}) or
constrained Gaussian realizations~\citep{Bucher:2011nf}. Alternatively,
and even better for bispectrum determination, one can perform a full
inverse covariance weighting (Wiener filtering) of the maps 
(see e.g.,~\citep{Smith:2009jr,Elsner:2012fe}).
However, these methods
do not appear necessary, as a combination of diffusive inpainting and
the linear correction term leads to results that are effectively optimal for 
the temperature maps (meaning they cannot be distinguished from the optimal
results within the error bars).
See also \citep{Gruetjen:2015sta} for an investigation of the impact of
inpainting on masked CMB temperature maps.
For $E$ polarization the situation is even 
simpler, at least at the Planck resolution and sensitivity. Not even diffusive 
inpainting is required. Just applying the linear correction term appears 
sufficient. However, as a precaution we also applied diffusive inpainting 
to the $Q$ and $U$ maps for the Planck analysis.

\begin{figure}
\includegraphics[width=0.5\columnwidth]{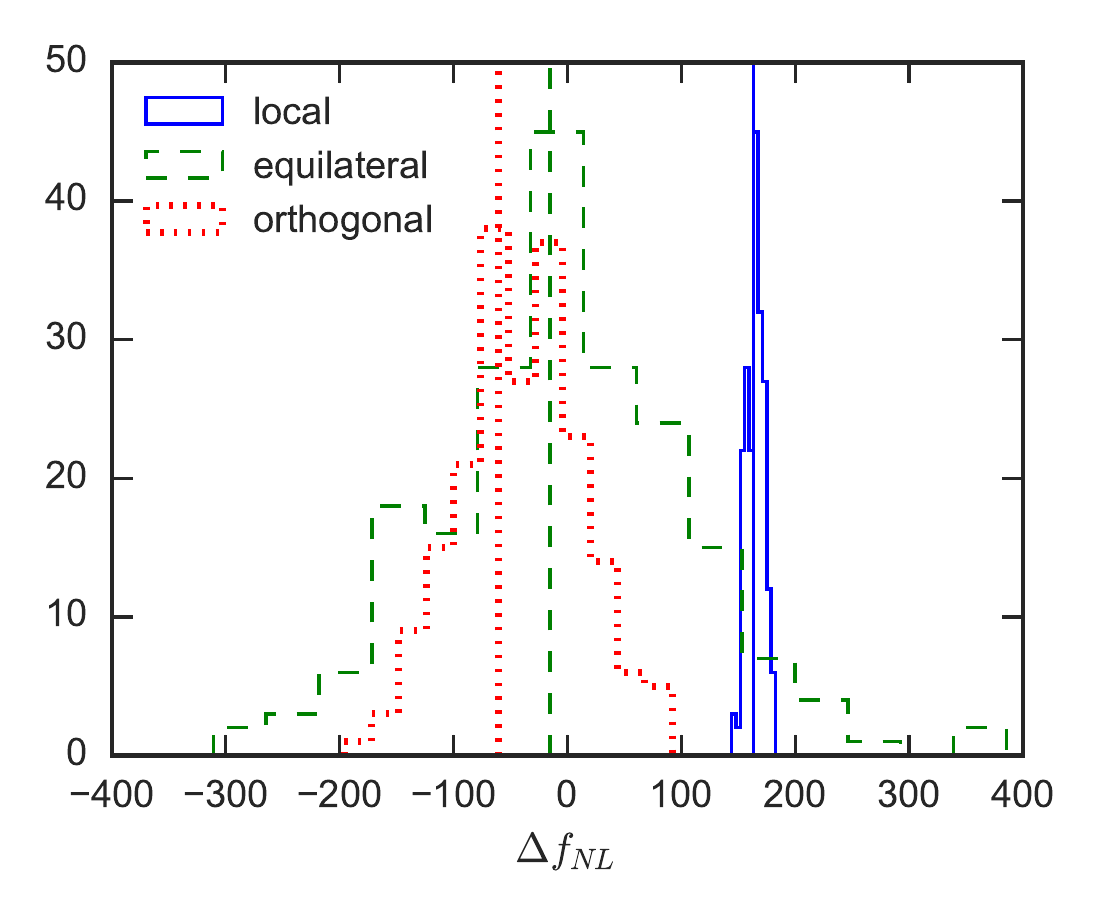}
\includegraphics[width=0.5\columnwidth]{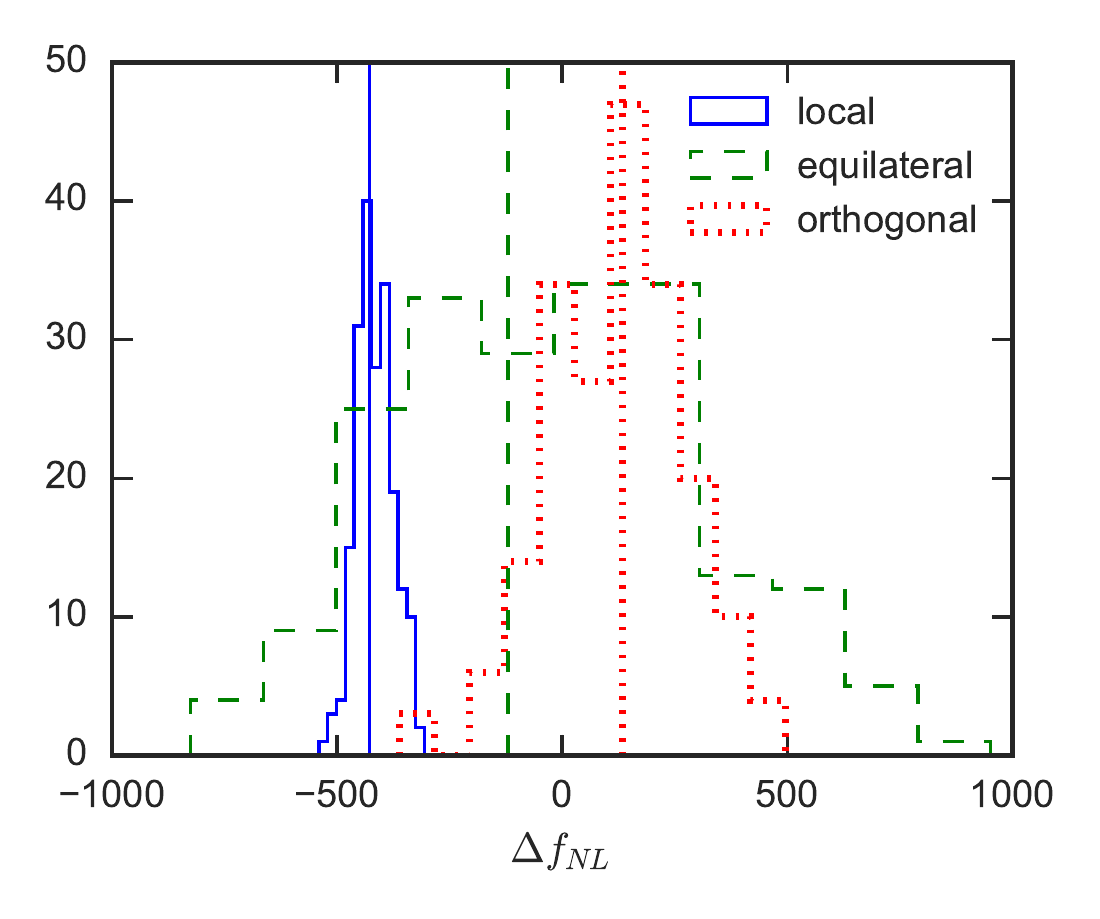}
\caption{Histogram of the contribution of 199 individual Gaussian maps
to the linear correction part of $f_\mathrm{NL}$ for one of the maps from the 
``no mask, anisotropic noise'' case of Table~\ref{table_filling_in},
for both temperature (left) and $E$ polarization (right).
Results are shown for the local (blue), equilateral (green), and orthogonal 
(red) shapes. The vertical lines correspond to the cubic (uncorrected)
part of $f_\mathrm{NL}$ for that map.}
\label{fig_hist_lincorr}
\end{figure}

Table~\ref{table_filling_in} also highlights the importance of the linear
correction term if there is anisotropic noise. While there is hardly any
impact for the equilateral shape and no bias for any shape, for the local 
shape the error bars simply explode when we add anisotropic noise to the 
map, both for temperature and for the $E$-polarization mode.
However, including the linear correction term suffices to recover the same 
error bars as in the ideal case. 
As can be seen from (\ref{Bisp_lincorr}), the linear correction to the
bispectrum of a given map, and hence to the $f_\mathrm{NL}$ parameters 
via (\ref{fNL_estimator}), involves the average over a large number of 
Gaussian maps. In Fig.~\ref{fig_hist_lincorr} we show the histogram of
the individual contributions of 199 Gaussian maps to the linear correction
part of $f_\mathrm{NL}$ for one of the maps from the 
``no mask, anisotropic noise'' case of Table~\ref{table_filling_in}.
The corresponding mean values are for $T$: local $165 \pm 0.5$, equilateral
$-4 \pm 8$, orthogonal $-40 \pm 4$, and for $E$:
local $-415 \pm 3$, equilateral $-20 \pm 24$, orthogonal $119 \pm 11$.
As expected we see a hugely significant linear correction for local, a very
significant correction for orthogonal (due to the large correlation with
local), and no significant correction for equilateral. The error bars
on the linear correction term for a single map are much smaller (in this 
case of 199 maps about a factor 7) than the error bars on the values of the 
different $f_\mathrm{NL}$ parameters determined from 100 maps in
Table~\ref{table_filling_in}, indicating that 
we have used enough maps to determine the linear correction.

\subsection{Implementation of the estimator}
\label{implementation}

A significant advantage of the binned bispectrum estimator is that it 
divides the bispectral analysis and determination of $f_\mathrm{NL}$ into 
three separate parts, the first two of which are completely independent. 
The first, slow, part is the
computation of the raw binned bispectrum of the map under consideration, 
including its linear correction. The second, much faster, part involves the
computation and subsequent binning of the theoretical bispectrum templates
one wants to test and of the expected bispectrum covariance.
Finally, the third, extremely fast, part (that runs in less than about a 
minute) is where the different analyses (for example for different templates) 
are carried out using the raw binned bispectrum from part 1 and the
quantities from part 2 as an input. In the case of $f_\mathrm{NL}$ determination,
this last part corresponds to the evaluation of the sum over the bins and
polarization indices in the inner product (\ref{innerprod_polar})
used in (\ref{fNL_estimator}).

This approach has several advantages. Firstly, the full
(binned) bispectrum is a natural output of the code 
and can be studied on its own without a particular template in mind. 
Such an analysis is the subject of the next subsection. 
Secondly, there is no need for the bispectrum template to be separable, 
since nowhere in the method does the need arise to split up the template 
into a separable form.
Thirdly, once the bispectrum of a map has been computed, modifications to 
the theoretical analysis (like for example testing additional templates) is 
fast, since there is no need to rerun the observational part (which consumes 
by far the most time). This is in contrast with competing estimators such as 
the KSW estimator, where the theoretical and observational
steps are mixed together (a separation is instead made in terms of 
$\ell_1,\ell_2,\ell_3$), so that the full code 
has to be rerun for any new template.
Fourthly, with the binned bispectrum estimator the dependence
of $f_\mathrm{NL}$ on $\ell$ is obtained almost for free, simply by leaving 
out bins from the sum when computing the final inner product. In particular
this has been used to study the dependence on $\ell_\mathrm{min}$ and 
$\ell_\mathrm{max}$ in the Planck analysis.
Finally, the binned bispectrum estimator compares
favourably to the other estimators in terms of speed: it is very fast
on a single map.

The only disadvantage of this method is that the templates
that can be studied accurately have to be reasonably smooth, or if not then
any rapid changes should be limited to a small part of $\ell$-space, in order 
for the template to be well approximated by a binned template with a not 
too large number of bins.\footnote{There are indications that the binned
bispectrum estimator might even perform well for oscillating templates that
do not satisfy these criteria. For the so-called constant feature 
model~\citepalias{planck2015-17} with a primordial bispectrum proportional to
$\sin(\omega(k_1+k_2+k_3)+\varphi)/(k_1 k_2 k_3)^2$, taking $\varphi=0$
and $\omega=100$, we find an overlap of 94\% for $T$-only with the
standard Planck binning (i.e.\ not optimized for this template). This will be 
investigated in more detail in the future.} 
For most primordial and foreground templates studied
so far, this is not a problem. Moreover, even for
templates that do not satisfy this criterion, the binned bispectrum estimator
could still perform quite well. For example, among the templates discussed in
sections~\ref{templatesec} and~\ref{sec_isocurv}, only the lensing-ISW 
template cannot be easily
binned. For a typical Planck binning the overlap is of the order of 60--70\%
(as opposed to 95\% or higher for all the other templates considered). 
Nevertheless the binned bispectrum estimator gives unbiased results even 
for this template, with error bars that are only slightly widened.

The code has been written mainly in Python, using some routines 
written in C. 
It is run on the computers of the Centre de Calcul de l'Institut 
National de Physique Nucl\'eaire et de Physique des Particules (CC-IN2P3)
in Lyon, France\footnote{\url{http://cc.in2p3.fr}} and any explicit remarks about
computing time refer to that system.

\subsubsection{Theoretical part}

The theoretical part of the code consists of two steps: first determining 
the unbinned theoretical bispectrum and power spectrum, and second, computing 
from these spectra the binned bispectrum templates $B^\mathrm{th}_{i_1 i_2 i_3}$
and the inverse of the binned covariance matrix 
$(V^{-1})_{i_1 i_2 i_3}^{p_1 p_2 p_3 p_4 p_5 p_6}$, see (\ref{fNL_estimator}) and 
(\ref{innerprod_polar}). This also requires experimental inputs in 
the form of the beam transfer function $b_\ell$ and the noise power spectrum 
$N_\ell$.

The first step is in some sense not really part of the estimator code. 
We have a code to compute all the bispectra discussed in 
sections~\ref{templatesec} and~\ref{sec_isocurv}, but in principle an 
explicitly computed theoretical bispectrum from any source could be used here. 
In our code we use the radiation transfer functions $g_\ell^{p\, I}$
(with $p$ the polarization index and $I$ the isocurvature index) computed
with CAMB (slightly modified to write them to file, since these are not a 
normal output of CAMB) to compute the primordial templates (\ref{Bth}).
For separable templates, this is a fast calculation, since the triple integral
over $k_1,k_2,k_3$ becomes a product of single integrals. For non-separable
templates a brute force calculation is much slower, but while one
might look for smarter ways to compute such bispectrum templates, it should
not be forgotten that (for a given cosmology) for use in the binned bispectrum
estimator, a template has to be computed only once. Hence even
a slow calculation might be acceptable.
While this code can also compute the power spectra from the radiation transfer
functions according to (\ref{Cl_gl}), in practice we use the power 
spectra computed by CAMB. These power spectra are used in the covariance 
matrix and some foreground bispectrum templates. 
The primordial bispectra are precomputed only on a grid (with $\Delta\ell$ 
increasing to about 10 at high $\ell$). This is denser than the binning, and 
thus accurate enough for the smooth local, equilateral, and orthogonal 
templates.

The second step involves the binning of the bispectrum templates and the
covariance matrix. The calculation of the covariance
matrix from the power spectra as well as the calculation of the foreground
templates is done directly in this step. As was seen in 
section~\ref{templatesec}, the foreground templates are simpler to
compute than the primordial templates, since there are no integrals, so there
is no need to precompute them, the required values can be computed in real 
time while binning. As for the precomputed primordial templates, since these 
have been precomputed only on a grid, other values are computed by 
three-dimensional linear interpolation.
While we developed a tetrahedral integration scheme to speed up the
calculation of all binned quantities, as described in \citep{BvTC},
we later moved away from using it. Given that the theoretical 
computation is much faster than the observational computation, there is no
point in making additional approximations to speed it up. Performing an
exact calculation of the binned quantities (where the quantities are
explicitly computed for each value of $\ell_1,\ell_2,\ell_3$ and then
summed over the bin) is fast enough. We can thus also
directly compute the overlap between the binned and the exact template using 
(\ref{binning_error}).

The final output of this step consists of two files: one containing the binned
theoretical bispectrum for all requested shapes, polarizations and 
isocurvature components; and another containing 
the inverse of the binned covariance matrix for all
polarization components. In addition the exact Fisher matrix (\ref{Fisher})
(without binning) is produced to allow for the estimation of the accuracy 
of the binning approximation using (\ref{binning_error}).

\subsubsection{Choice of binning}

The choice of binning is an important
part of the implementation. In theory the idea is very simple: one chooses
the binning that makes the overlap parameter $R$ defined in 
(\ref{binning_error}) as close to one as possible.
In practice this is not so simple, since both the number of bins
and all the bin boundaries are free parameters. Fortunately $R$ does not
depend strongly on the exact binning choice. Moreover, one does not need 
$R=1$ to obtain results statistically indistinguishable from the
exactly optimal results. For example, even with $R=0.95$, which is about the
lowest overlap for any of the templates considered in the Planck analysis 
(except for lensing-ISW), the increase in the standard
deviation is only 2.6\%. This should be compared to the 5\% uncertainty
in the standard deviation due to its determination from 200 maps.
Note that the code allows the use of separate binnings for the $T$-only,
the $E$-only, and the full $T+E$ analyses, although for reasons related
to time a single binning was used for the Planck analysis.

We developed three optimization tools: one that checks which bin boundary can be
removed with the smallest decrease of $R$ (reducing the number of bins by
one), one that checks where a bin boundary can be added with the largest
increase in $R$ (increasing the number of bins by one; the bin boundary
is added in the exact centre of an existing bin), and one that tries
moving all the bin boundaries by a given amount (relative to the size of
the bin) and tells for which bin this increases $R$ the most (leaving the number
of bins unchanged). For all of them one can indicate which shapes
and polarizations (meaning $T$ and/or $E$) should be taken into account.
These three tools are then used iteratively to optimize the
binning used as starting point, until no more significant improvements are 
obtained (as defined by a certain threshold in the change of $R$).
The starting point is arbitrary. 
For example a simple log-linear binning (with bin sizes increasing
logarithmically at low $\ell$, up to a certain value of $\ell$, after
which the binning becomes linear) or a binning that has already been
partially optimized in another way can be used. The latter could for example 
be done using the method described in \citep{BvTC}, which can 
provide a good starting point. (That method produces suboptimal binnings 
and can benefit significantly from the procedure
described here.)
While this method can likely be optimized further, for the 
Planck analysis the binning obtained in this way produces 
effectively optimal results.

\subsubsection{Observational part}

The observational component of the code consists of two parts: one to compute
the cubic part of the bispectrum of the map according to (\ref{Bobsbinned}),
and the other to compute the linear correction according to 
(\ref{Bisp_lincorr}). First the map is fully prepared, which can be as simple
as reading an existing map and doing the masking and filling in, or involve the
creation of CMB and noise realizations. It is then saved in the form 
of $a_{\ell m}$'s for later use with the linear correction term, or for 
reproducibility in the case of generated random realizations.

The maps are then filtered according to (\ref{Tmapbinned}). This leads
to some practical issues that had to be resolved, since in principle
we need to hold twice ($T$ and $E$) 50--60 maps (one for each bin) of
Planck resolution (Healpix resolution parameter $n_\mathrm{side}=2048$) in
memory for this
calculation.  However, our computer system had a limit of
16~GB per processor (after the 2015 Planck analysis this was
even reduced to only 10~GB), which makes this impossible. We managed to save
space in two ways. In the first place, while all the preprocessing of
the maps is done in double precision, the final filtered maps are only
kept in single precision, which saves a factor two in
memory. Tests have shown that this has no significant impact on the
final results for $f_\mathrm{NL}$.
Secondly, it is unnecessary to use $n_\mathrm{side}=2048$
precision for the maps that contain only low-$\ell$ bins. Hence the
filtered maps of bins up to about $\ell=400$ are produced at
$n_\mathrm{side}=512$, and those between about $400$ and $800$ at
$n_\mathrm{side}=1024$, which saves a factor of sixteen and four, respectively,
in memory for those maps (the number of pixels in the map is
$12 \, n_\mathrm{side}^2$), as well as speeding up the final computation where
three maps have to be multiplied and summed (see (\ref{Bobsbinned})).
Polarization maps are never higher resolution than $n_\mathrm{side}=1024$.
Using the nested Healpix\footnote{\url{http://healpix.sourceforge.net}} 
format, it is easy to multiply maps of different $n_\mathrm{side}$ together.

We have developed two different ways of computing the linear correction term 
of a map. In the first method, which was used for the Planck analyses in 2013
and 2015 and in the analyses presented in the previous subsection
(from \citep{BRvT}),
each job treats one of the {\em Gaussian} maps 
(see (\ref{Bisp_lincorr})), which is preprocessed and filtered as above, 
and the filtered maps are held in memory. Then a filtered map 
of only the first bin of the {\em observed} map is created and all required sums
of products involving that map are computed. Next this process is repeated
for the second bin of the observed map, etc.\footnote{In an earlier version 
of the code these filtered maps of the observed map, which are also produced 
during the cubic calculation, were saved to disk at that time, and then read in
here. However, the required I/O turned out to make this actually slower than 
when these filtered maps are recreated on the fly, which also has the advantage
of using much less disk space.} The final result of this job
is a temporary file with a linear correction term computed with
just one Gaussian map. Once all jobs have finished (with the results for the
other Gaussian maps), the results are summed and averaged to obtain the
final linear correction term for the map. 
This whole process (preprocessing the map and computing the cubic and linear 
terms) for a single map at Planck resolution for all $T+E$ (including
mixed) components takes a few hours, which is quite fast compared to other
bispectrum estimators. (Computing the theoretical part is much faster and 
requires only a single job, so can easily be done on the side.) 
With this method one can simply add more Gaussian maps to the linear 
correction term at a later stage if required, and investigate its convergence 
as a function of the number of Gaussian maps. 
However, this first method of computing the linear correction term scales
very badly with the number of observed maps. Since the object
$\langle M_{i_1}^{p_1, G} M_{i_2}^{p_2, G} \rangle$ in (\ref{Bisp_lincorr}) is too 
large to compute directly
and save to file, if one has a set of similar maps (for example to
compute error bars), the linear correction term has to be recomputed 
for each map in the same way as above, making this a very slow process. 

For this reason we later developed another way to compute the linear
correction term, which was used for the Planck analysis in 2018.
This second method is based on the observation that while
the object $\langle M_{i_1}^{p_1, G} M_{i_2}^{p_2, G} \rangle$ (consisting of 6612
maps for a full $T+E$ calculation in the case of 57 bins) is too large to
handle, saving it in the form of $a_{\ell m}$'s is doable. Moreover, we make
use of the fact that when multiplying several masked maps together (all with
the same mask), it is enough if only one of the maps is masked. Hence if
the observed map in (\ref{Bisp_lincorr}) is properly masked, the Gaussian maps
can be left unmasked (since the Gaussian maps are based on simulations, they
are full-sky maps). This has the advantage that no filling-in needs to
be performed on these maps, which would otherwise be required before conversion
to $a_{\ell m}$'s, as explained in section~\ref{realsky}. By limiting the
number of considered bins per job in such a way that both the filtered maps 
for those bins and all the product maps involving those bins can be kept in 
memory at the same time, one job can compute the full average for
the considered bins by treating one Gaussian map after the other. Only
at the end are the final maps converted to $a_{\ell m}$ format and written to 
disk. This precomputation for the linear correction term can be run with a 
modest number of jobs (about 100) in a reasonable amount of time (less than
a day for 200 maps). Once the precomputation has finished, the linear
correction for any map can be quickly computed using (\ref{Bisp_lincorr}).
Each job reads in a number of product maps (i.e.\ for certain values of
$i_1$ and $i_2$; the number being determined by memory considerations), 
and converts them back to pixel space. They are then multiplied with the
filtered observed maps as explained above for the first method. The main 
difference is that the results are now for the full average of all the 
Gaussian maps, instead of for a single one. Another (small) advantage of this
second method is that at this step we only need to multiply two maps together
and not three. Once all jobs are finished, the temporary files containing
results for different $i_1$-$i_2$ bins are combined to get the full linear
correction for the observed maps.
While this second method with precomputation is slower if one is only 
interested in a single map, its much better scaling with the number of 
maps makes it by far the preferred method when dealing with a set of maps,
for example to compute error bars.

The final result of this part are two files for each map, one containing the 
binned cubic-only bispectrum of the map and the second its linear correction,
both containing all requested polarization components. These can then be
combined with the results from the theoretical part to compute $f_\mathrm{NL}$
according to (\ref{fNL_estimator}), which takes less than a minute even when
producing convergence plots and dependence on $\ell$ as well, or
be studied directly without the assumption of a theoretical template, as
discussed in the next subsection.

\subsection{Non-parametric bispectrum studies}
\label{smoothingsec}

The previous subsections described how the binned bispectrum
of a map can be analysed parametrically by computing the
$f_\mathrm{NL}$ parameters corresponding to a selection of theoretically
motivated templates. But one advantage of the binned bispectrum
estimator is that the full (binned) three-dimensional bispectrum is a
direct output of the code, which can be studied 
non-parametrically, thus searching for any deviations from
Gaussianity even when no suitable template is available.  
Here we start by describing the smoothing procedure that must first be applied
to the binned bispectrum in order to enhance the 
signal-to-noise of any possible non-Gaussian features, which 
otherwise would remain hidden in the noise. 
After that we will discuss the statistical analysis subsequently 
applied to this smoothed binned bispectrum to assess the statistical 
significance of any non-Gaussian features appearing as extreme values. 

We first normalize the binned data by dividing by the square root of the 
expected
bin variance, so that each bin triplet in the absence of a bispectral signal 
would have noise obeying a normalized Gaussian distribution. Thus for the 
bin triplets for which there is data, we define
\begin{align}
&{\cal B}_{i_1 i_2 i_3}^{TTT} =
\frac{B_{i_1 i_2 i_3}^{TTT} }{\sqrt{V_{i_1 i_2 i_3}^{TTTTTT}}},
\qquad
{\cal B}_{i_1 i_2 i_3}^{EEE} =
\frac{B_{i_1 i_2 i_3}^{EEE} }{\sqrt{V_{i_1 i_2 i_3}^{EEEEEE}}},
\nonumber\\
&{\cal B}_{i_1 i_2 i_3}^{T2E} =
\frac{B_{i_1 i_2 i_3}^{T2E} }{\sqrt{\mathrm{Var}(B^{T2E})_{i_1 i_2 i_3}}},
\qquad
{\cal B}_{i_1 i_2 i_3}^{TE2} =
\frac{B_{i_1 i_2 i_3}^{TE2} }{\sqrt{\mathrm{Var}(B^{TE2})_{i_1 i_2 i_3}}}.
\end{align}
For the mixed $T$ and $E$ components we analyzed only the combinations
$B^{T2E} \equiv TTE+TET+ETT$ and $B^{TE2} \equiv TEE+ETE+EET$, with 
corresponding variance Var($B^{T2E}$) = 
Var($TTE$) + Var($TET$) + Var($ETT$) + 2 Cov($TTE,TET$) + 2 Cov($TTE,ETT$)
+ 2 Cov($TET,ETT$), and similarly for Var($B^{TE2}$). 
This projection entails a loss of information but allows the same analysis to
be used as for $TTT$, as described below.

Only bin triplets 
containing $\ell$'s that satisfy both the parity condition and the
triangle inequality contain data. However, among the bin triplets containing 
data, we noticed that some triplets systematically produced outliers. It 
turned out that
these bin triplets contained very few valid $\ell$-triplets (for example,
the hypothetical bin triplet $([50,100],[50,100],[200,300])$ would contain only 
one valid $\ell$-triplet (100,100,200), since the triangle inequality imposes
that $\ell_3 \leq \ell_1+\ell_2$). While the theoretical variance calculation is exact, 
the computation of the observed bispectrum using Healpix spherical harmonic 
transforms contains some numerical inaccuracies, so that the 
bispectrum in points outside the triangle inequality is not zero but
contains leakage.\footnote{This results because the pixelization breaks the 
spherical symmetry as must be the case with any pixelization of the sphere.}
For bin triplets like the above example with many $\ell$-triplets violating 
the triangle inequality, a significant mismatch between
the theoretical and the actual standard deviation of the bispectrum in that bin
is observed. 
The obvious solution is to remove such bin triplets from the data. Moreover,
the statistical analysis described below
assumes that bin triplets contain many valid $\ell$-triplets
in order for Gaussian statistics to apply to the noise from cosmic variance,
which constitutes another reason to exclude such triplets. 
After some experimentation, we adopted the
selection criterion that the ratio of valid $\ell$-triplets to the ones 
satisfying only the parity condition (but not the triangle inequality) in a
bin triplet should be at least 1\%, finding this a good threshold for 
rejecting systematic outliers. The results are insensitive
to the precise threshold used. For the Planck binning with 57 bins (which
is used for the results in this section), this 
criterion excluded 293 out of 13020 bin triplets.

If we were looking for a sharp bispectral feature of a linewidth
narrow compared to the binwidth, there would be no
motivation to smooth. We would simply examine the statistical
significance of the extreme values of the renormalized binned
bispectrum described above, taking into account the look-elsewhere
effect. However, for broad features, as are likely to arise from
galactic foregrounds, smoothing increases statistical significance 
by averaging over and thus diminishing the noise.
One approach would be to use binning with a range of bin widths, but
this approach has the disadvantage that the statistical significance
for detecting a feature depends on how it is situated relative to
the neighbouring bin boundaries. Instead we rather smooth using a
Gaussian kernel and renormalize so that in the absence of a signal the
single pixel distribution function is again a unit Gaussian. For a
Gaussian kernel $K_{\sigma_\mathrm{bin}}$ of width $\sigma_\mathrm{bin}$, 
we have
\begin{equation}
\mathcal{B}^{p_1 p_2 p_3, \text{smoothed}}_{i_1i_2i_3} = 
\sum_{i'_1} \sum_{i'_2} \sum_{i'_3} 
K_{\sigma_{\text{bin}}}(i_1-i'_1, i_2-i'_2, i_3-i'_3)  
\mathcal{B}^{p_1 p_2 p_3} _{i'_1i'_2i'_3} 
\end{equation}
where the Gaussian smoothing kernel 
\begin{equation}
K_{\sigma_{\text{bin}}}(
\Delta i_1, 
\Delta i_2, 
\Delta i_3) 
=
\left( 2\pi {\sigma _\mathrm{bin}}^2\right) ^{-3/2} ~
\exp \left[ 
-\frac{1}{2}\frac{
{\Delta i_1}^2
+{\Delta i_2}^2 
+{\Delta i_3}^2 
}{
{\sigma_\mathrm{bin} }^2}
\right]
\end{equation}
is used. Numerically the kernel is applied in the Fourier domain.

Without boundaries this smoothing and renormalization procedure would
be straightforward.  However, near the boundary the Gaussian 
smoothing kernel would extend
into the region where there is no data.  To minimize 
boundary effects, we first extend the fundamental domain
(where $i_1 \leq i_2 \leq i_3$) to the five identical domains
obtained by permuting $(i_1, i_2, i_3)$ and pad with zero data 
beyond the boundaries of this extended domain as well
as for triplets inside the domain for which there is no data.  
The smoothing causes power to leak out into 
the zero padded regions, and to correct for this leakage, we construct a mask 
consisting of ones in the domain of definition and zeros outside. After 
smoothing the signal-to-noise bispectrum $\mathcal{B}$, we renormalize by 
dividing by the mask that has undergone the same smoothing procedure. 
For the bin triplet statistic to be a Gaussian of unit variance, we generate 
1000 Monte Carlo realizations going through the same procedure and compute 
the variance, with which we divide our smoothed renormalized bispectrum.

\begin{figure}
\centering
\includegraphics[trim = 3mm 0mm 12mm 0mm,clip, width=0.31\columnwidth]{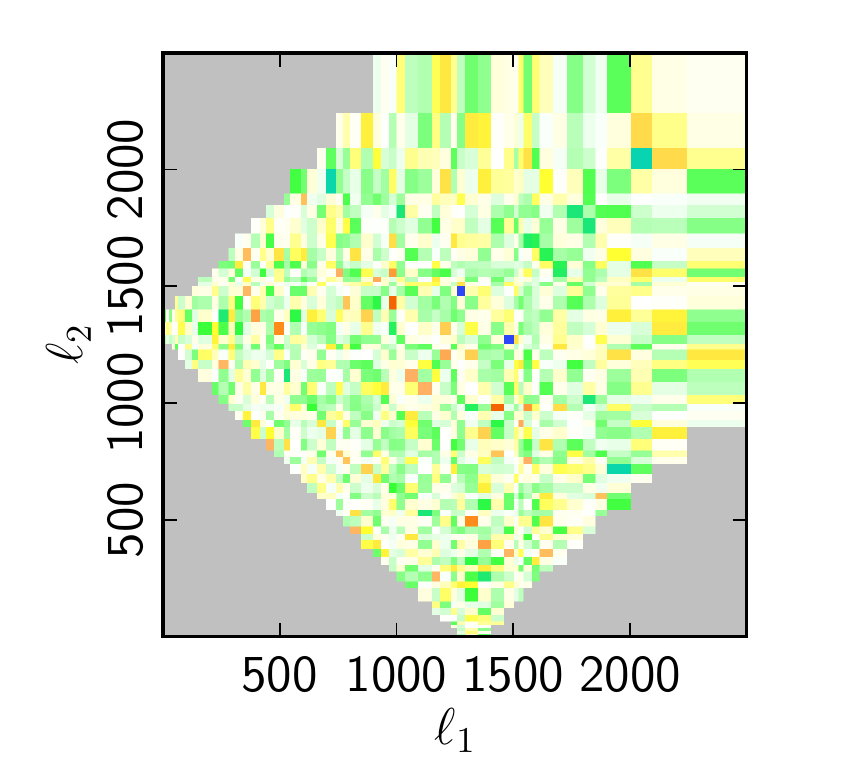}
\includegraphics[trim = 3mm 0mm 12mm 0mm,clip, width=0.31\columnwidth]{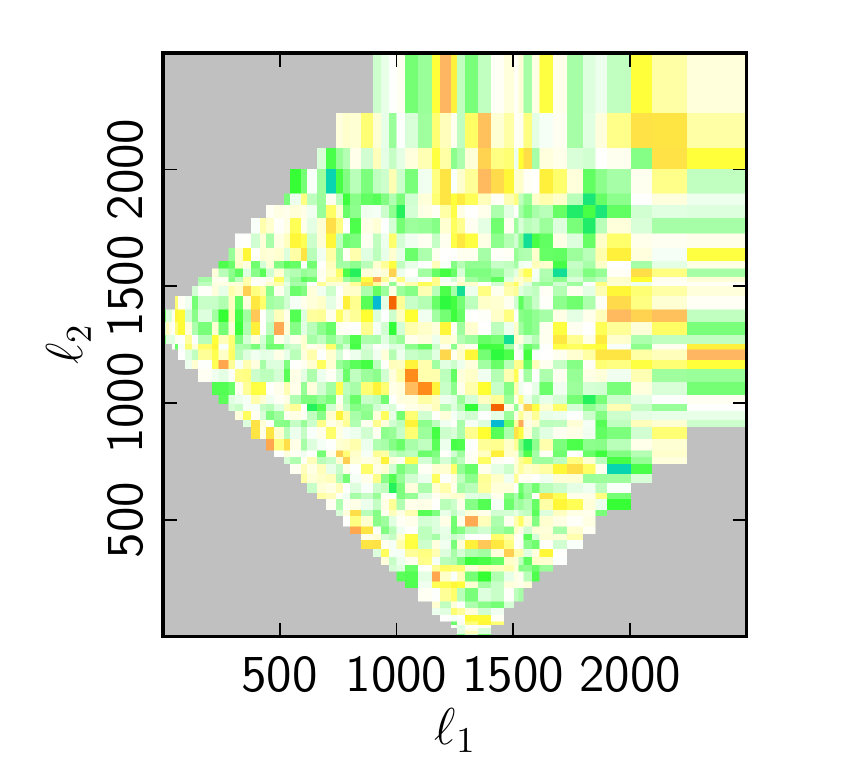}
\includegraphics[trim = 3mm 0mm 12mm 0mm,clip, width=0.31\columnwidth]{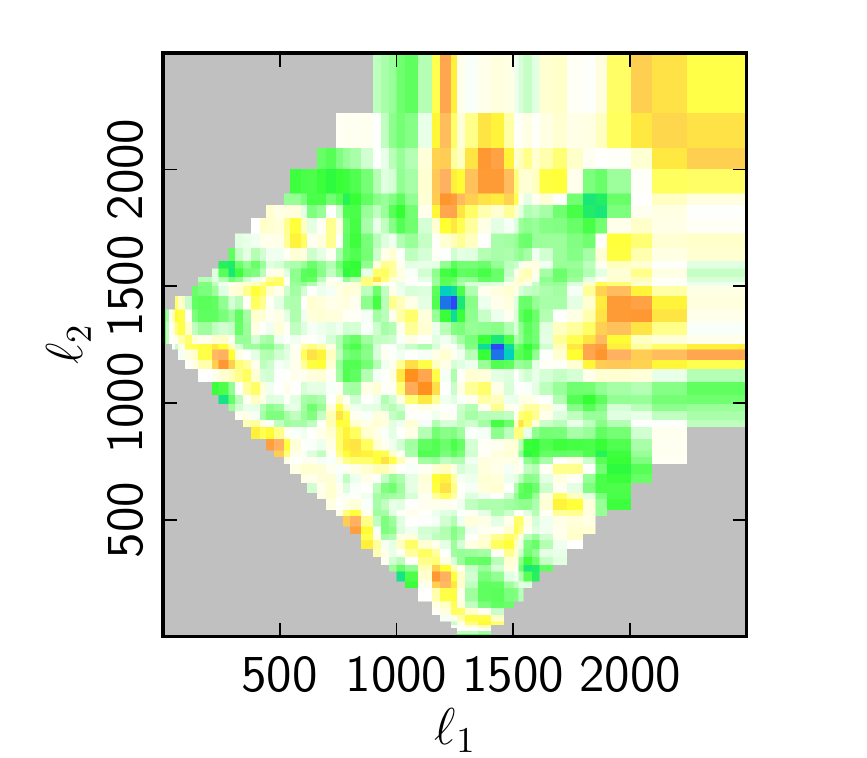}
\includegraphics[trim = 3mm 0mm 12mm 0mm,clip, width=0.31\columnwidth]{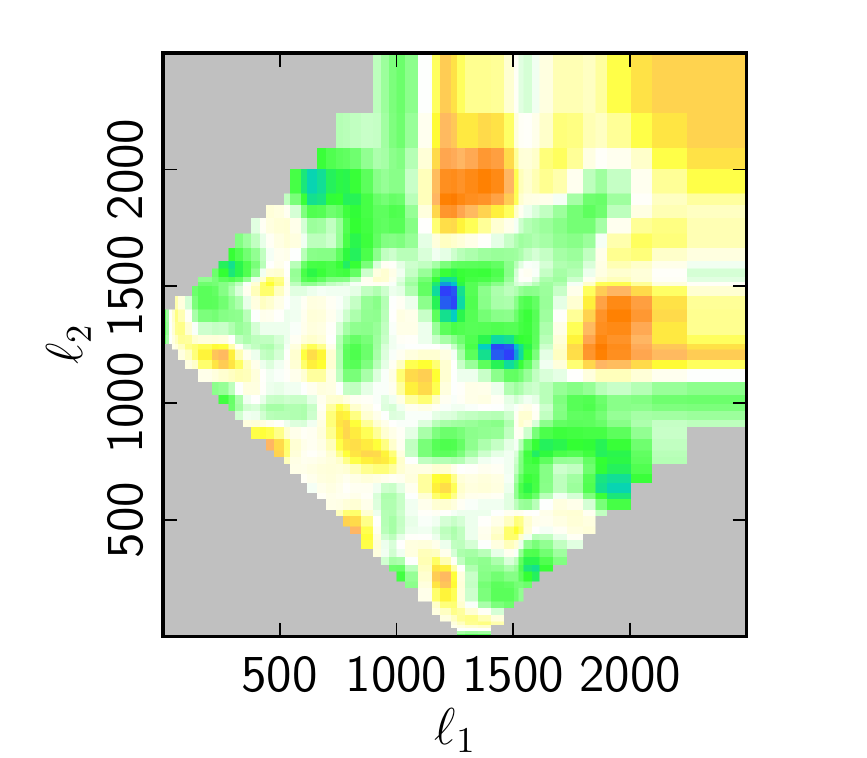}
\includegraphics[trim = 3mm 0mm 12mm 0mm,clip, width=0.31\columnwidth]{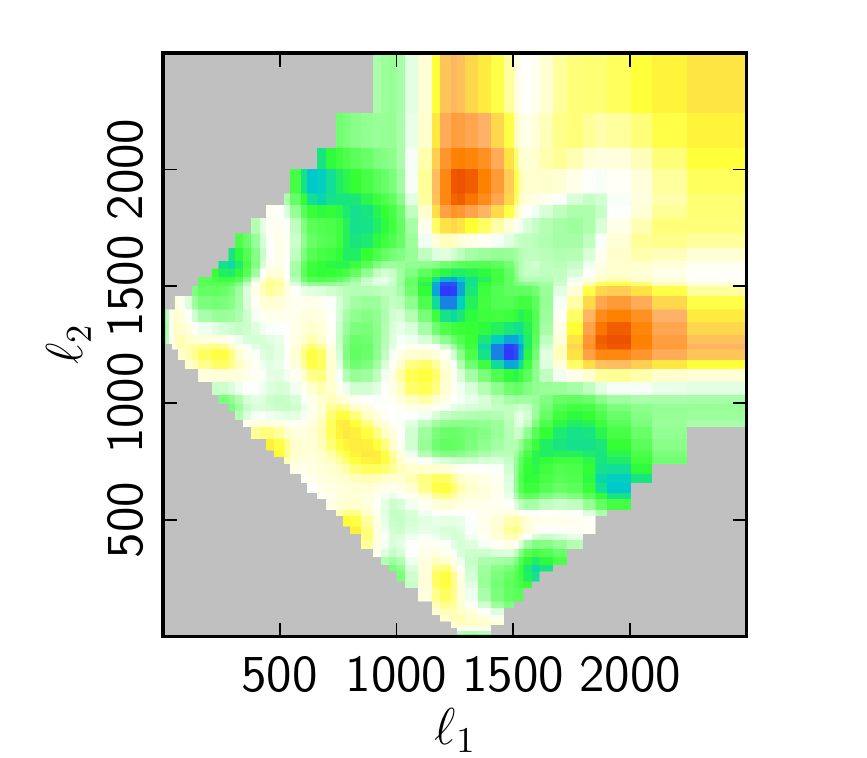}
\includegraphics[trim = 3mm 0mm 12mm 0mm,clip, width=0.31\columnwidth]{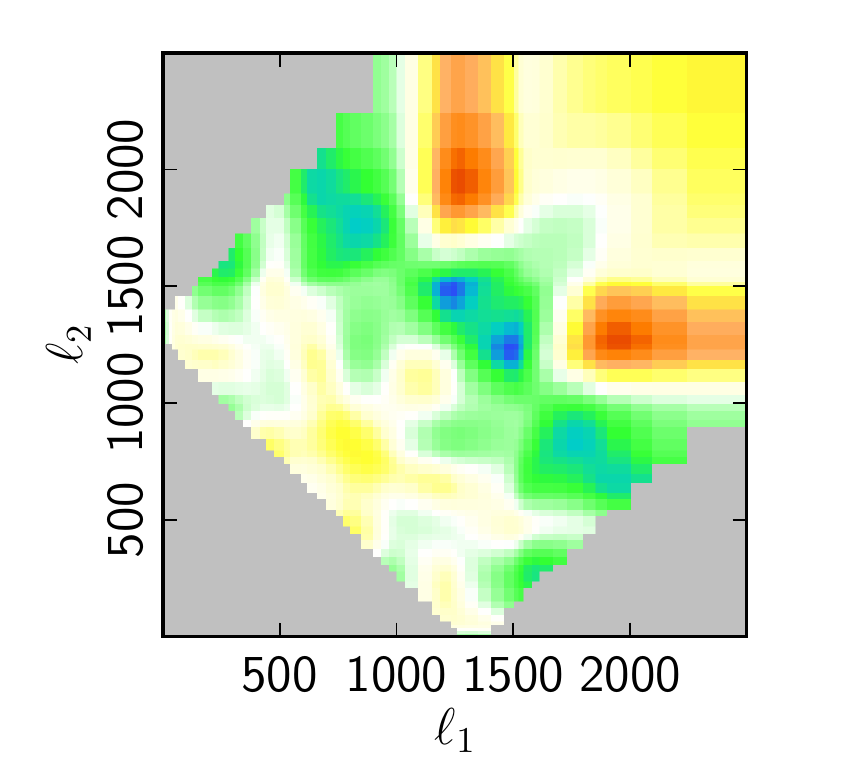}
\includegraphics[trim = 3mm 0mm 12mm 0mm,clip, width=0.31\columnwidth]{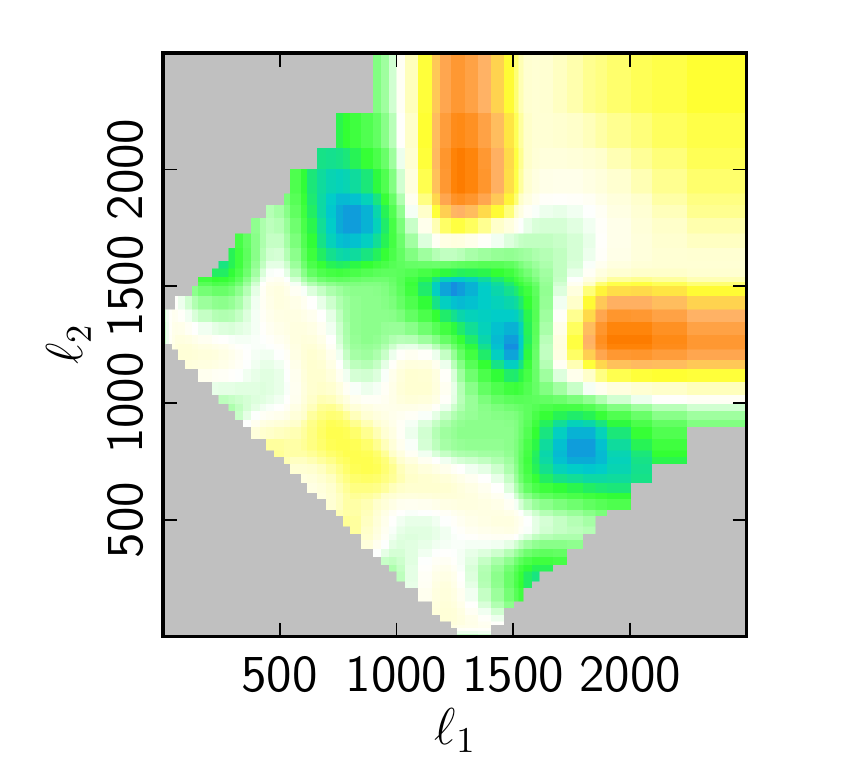}
\includegraphics[trim = 0mm 0mm 25mm 10mm,clip,width=0.75\columnwidth]{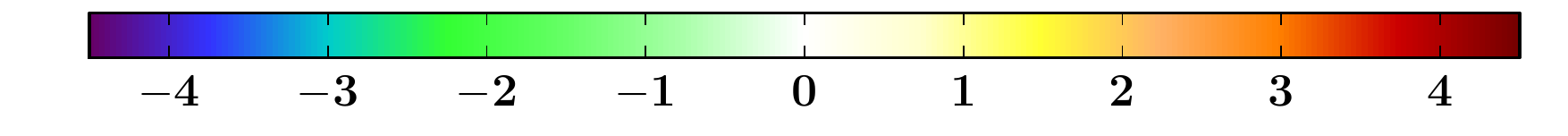}
\caption{Effect of different smoothing lengths on the bispectrum of one of 
the Gaussian maps from Table~\ref{table_filling_in} with 
galactic and point source masks and anisotropic noise. From left to right 
and top to bottom are shown: no smoothing, smoothing using $\sigma_{\mathrm{bin}} 
= $ 0.5, 1, 1.5, 2, 2.5 and 3. 
The slices correspond to $\ell_3 \in [1291, 1345]$. }
\label{figure_smolen}
\end{figure}

The result using different smoothing lengths is
illustrated in Fig.~\ref{figure_smolen} as two-dimensional
slices showing $\mathcal{B}$ as a function of $\ell_1$ and $\ell_2$ for a
fixed bin in $\ell_3$. 
With the colour scale used in Fig.~\ref{figure_smolen}, both dark red and dark 
blue represent 
extreme values with small $p$ values if Gaussianity is assumed, and thus 
suggest the presence of statistically significant bispectral non-Gaussianity.
A correct analysis of the significance would also take into account the
look-elsewhere effect --- that is, that the small probability to exceed,
calculated for a fixed bin, is too small because it does not reflect that an
improbable value could have occurred in any of a number of bins. 
The analysis of this issue is complicated by the correlations between the 
bins that result from the smoothing, an issue analyzed next.

In the absence of smoothing, we face the following statistical
problem. We have a binned bispectrum that has been rescaled so that we
have $N$ bins and the bispectrum value in each bin $x_i,$ where
$i=1,\ldots , N$, has a probability distribution function well
approximated by a normalized Gaussian distribution. Moreover, values in
different bins are almost statistically independent. The quadratic
correlation vanishes, but some of the higher-order joint correlations
do not precisely vanish, a feature that we shall neglect here. The
corrections to Gaussianity and to statistical independence are
suppressed when $\ell $ is large and when there are many $\ell$-triplets 
containing data in a bin.  Thus we have the distribution function
\begin{eqnarray}
p(x_1,\ldots ,x_N)=(2\pi )^{-N/2}~\exp \left[ -\frac{1}{2}\sum _{i=1}^N {x_i}^2
\right] ,
\label{pdfMV}
\end{eqnarray}
and since we are interested in extreme values, we define two new derived 
statistics
\begin{eqnarray}
X_\mathrm{min}=\min (x_1,\ldots , x_N); \qquad
X_\mathrm{max}=\max (x_1,\ldots , x_N),
\end{eqnarray}
and accordingly define the $p$-values
\begin{eqnarray}
p_\mathrm{min}(X)=P(X_\mathrm{min}<X); \qquad 
p_\mathrm{max}(X)=P(X_\mathrm{max}>X)
\end{eqnarray}
where $X_\mathrm{min}$ and $X_\mathrm{max}$ are the derived random variables 
defined above. If either of these $p$-values are extremely small, then we 
have evidence of bispectral non-Gaussianity directly in the unsmoothed 
binned bispectrum, and this $p$-value can be converted into a $\sigma $ 
for the normal distribution using the inverse error function as is customary.

For this simple unsmoothed case it is not hard to give the probability 
distribution function for the extreme value statistics $X_\mathrm{min}$ and 
$X_\mathrm{max}$. 
Given the (complementary) cumulative distribution function for the normal 
distribution (integrating from right to left)
\begin{eqnarray}
\Phi (x)=\frac{1}{\sqrt{2\pi }}\int _x^{+\infty }dt~\exp \left[ -\frac{1}{2}t^2
\right] ,
\label{ccdf}
\end{eqnarray}
the analogous distribution for the maximum extreme value for $N$ variates is 
given by 
\begin{eqnarray}
\Phi_\mathrm{max}(X_\mathrm{max}; N)=1-\Bigl( 1- \Phi (X_\mathrm{max})\Bigr) ^N
\label{evCDF}
\end{eqnarray}
and we may straightforwardly obtain an analogous expression for the case
of the minimum value. (Below we shall only give results for the case of
the maximum.) For $X\gg 1$ we obtain an approximation to 
$\Phi_\mathrm{max}(X_\mathrm{max}; N)$ by inserting the following expression
\citep{Abramowitz}
\begin{eqnarray}
\ln \Bigl[ \Phi (X)\Bigr] \approx -\left[ 
\frac{X^2}{2}+ \ln (X)+ \frac{1}{2}\ln (2\pi ) 
\right] 
\label{evCDFapprox}
\end{eqnarray}
into (\ref{evCDF}). 

When we consider extreme values of multivariate Gaussian distributions
with correlations, there is, as far as we know, no way of obtaining an
analytic result for the extreme value distribution for 
$\Phi_\mathrm{max}(X_\mathrm{max})$. After the smoothing described above is
applied, the probability distribution defined in
(\ref{pdfMV}) must be replaced with
\begin{eqnarray}
p({\bf x})=(2\pi )^{-N/2}~\exp \left[ -\frac{1}{2} {\bf x}^T {\bf C}~{\bf x} 
\right] 
\end{eqnarray}
where the correlation matrix ${\bf C}$ has all ones on the diagonal,
but also a lot of positive off-diagonal elements as the result of the
smoothing process, rather than all zeros away from the diagonal. It is
these off-diagonal elements that prevent us from solving analytically for the
extreme value statistic probability distribution function.

Instead we postulate an Ansatz to approximate the cumulative
distribution function (CDF) of the extreme value statistic $\Phi_\mathrm{max}$, 
which has one adjustable parameter $N_\mathrm{eff}$, the effective
number of independent bins, which will be smaller than the actual
number of bins $N$ as the result of the smoothing. The Ansatz states
that the CDF given in (\ref{evCDF}) (and approximated using
(\ref{evCDFapprox})) holds where $N$ has been replaced with
$N_\mathrm{eff}$. For a given level of smoothing, we fit $N_\mathrm{eff}$ to the
tail of the CDF, which has been determined empirically by Monte Carlo
simulations. We then assess the quality of the approximation, in
particular in the tail of the distribution where $X$ is very large,
which is the range of values of particular interest here.  It should
be stressed that we do not need a good approximation to the entire
CDF. It suffices to have an approximation that works well
asymptotically, in the extreme tail of the distribution where
$p$-values cannot feasibly be obtained by Monte Carlo methods. Thus
the Ansatz serves as an asymptotic approximation for the tail of the
distribution.

\begin{figure}
\begin{center}
\includegraphics[width=0.49\textwidth]{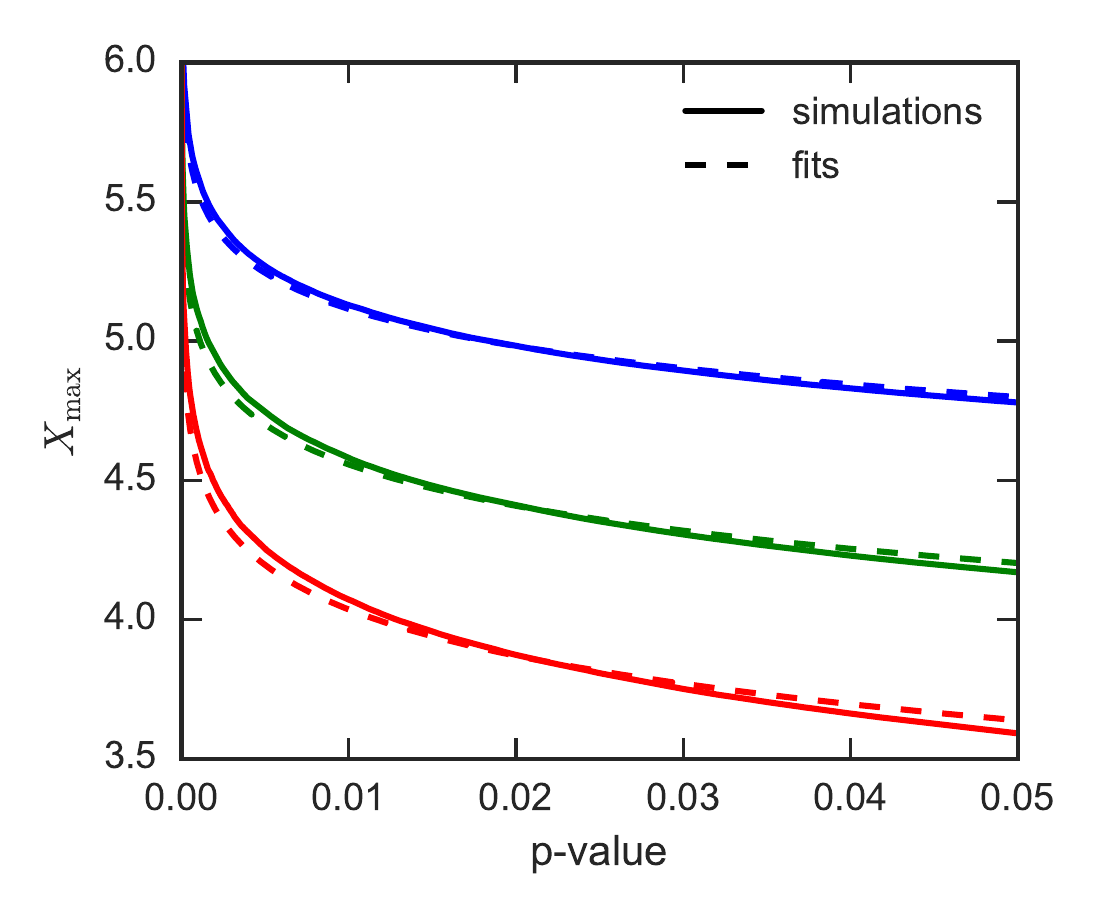}
\includegraphics[width=0.49\textwidth]{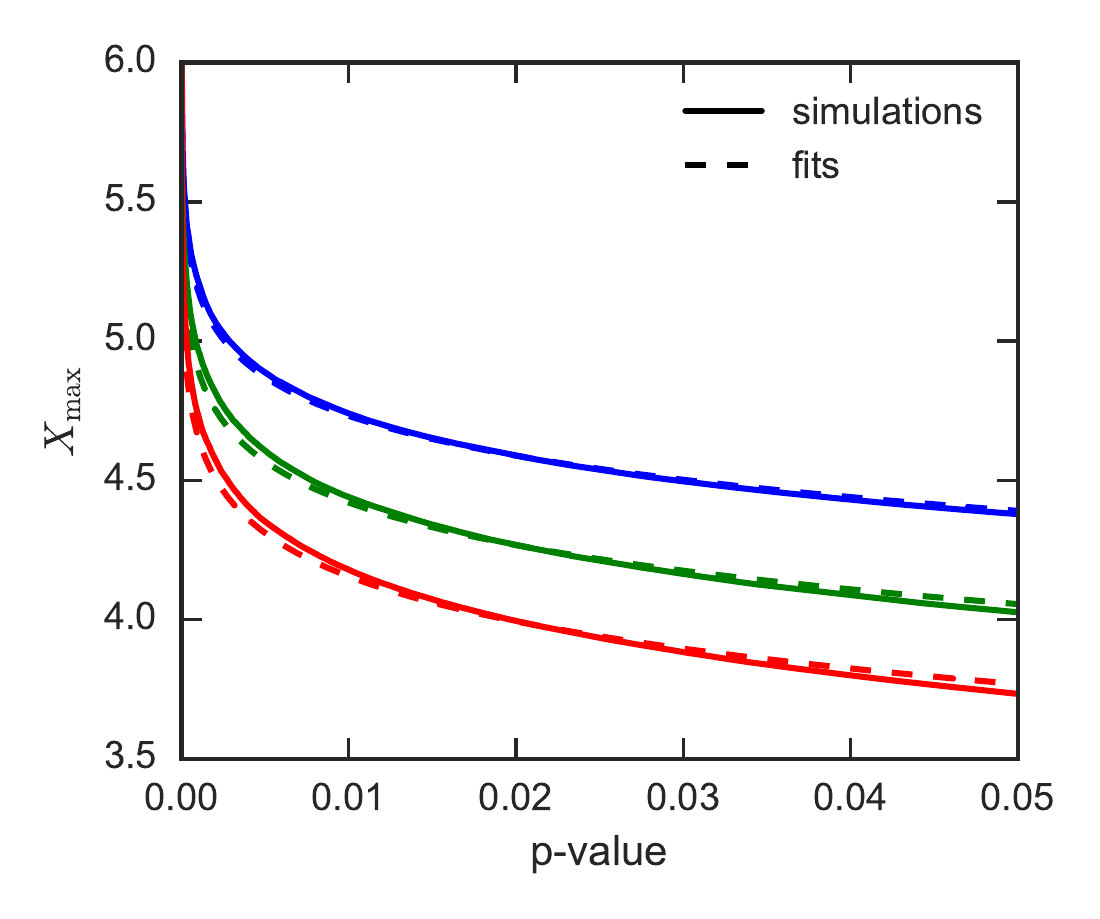}
\end{center}
\caption{Fitting the extreme values of the smoothed bispectrum.
The horizontal axis gives the $p$-value and the vertical axis
the maximum bin deviation in number of $\sigma$.
The solid curves show the CDFs for the smoothed $64^3$ cubes 
(figure on the left) for $\sigma_\mathrm{smoothing} =2, 5,$ and $10$ 
(respectively from top to bottom blue, green and red)
and for the more realistic situation (figure on the right) for 
$\sigma_\mathrm{smoothing} = 1, 2,$ and $3$ as described in the main text.
We show only the 5\% most extreme values, since we are
interested in the extreme tail of the distribution. The dashed curves show 
non-linear least square fits to these curves using the Ansatz where a 
distribution with a smaller number $N_\mathrm{eff}$ degrees of freedom is used 
to approximate the distribution with correlations between bins. Here (from 
top to bottom in the figure on the left) $N_\mathrm{eff} = 61900, 3709,$ and 
$351$ are used to represent the $64^3 = 2.6\times 10^5$ degrees of freedom
with various degrees of correlation as the result of the smoothing.
For the figure on the right the numbers are (from top to bottom)
$N_\mathrm{eff}=8664, 1948,$ and $588$ to represent the original $12727$
bin-triplets. The fit
provides a good approximation, especially at the small $p$-values that
are of the greatest interest for this application.
}
\label{extremeDist}
\end{figure}

To demonstrate the validity of our Ansatz
in a simplified context very similar to the case of interest,
we generate a three-dimensional periodic cubic lattice filled with $64^3$ 
independent realizations of a normal Gaussian random variable. This cube
is then smoothed using a Gaussian smoothing kernel with 
widths $\sigma _\mathrm{smoothing}=2, 5,$ and $10$. The smoothed cube is rescaled
so that the variable at each lattice point has unit variance.
For each smoothing width, the extreme value statistic (maximum) is taken 
for $10^6$ realizations and only the greatest $5\% $ of the extreme values 
are retained. Fig.~\ref{extremeDist} (left) shows the empirical CDF for the 
extreme values, which are compared to the functional form of the Ansatz for the 
best-fit values of $N_\mathrm{eff}$ according to
the approximation given in~(\ref{evCDFapprox}).

The above discussion demonstrates that an extreme value distribution
for $N_\mathrm{eff}$ independent Gaussian variates can be used to approximate
the distribution for $N_\mathrm{bin}$ variates with correlations due to
smoothing. However, the geometry of the allowed bins is complicated
and must be taken into account. Thus simulations must be carried out
to determine $N_\mathrm{eff}$ for a given smoothing scale in the more realistic
case of the actual bins used for the bispectrum analysis. We show that the 
Ansatz is still a good approximation also in this case in 
Fig.~\ref{extremeDist} (right). We generate random
numbers in the domain of definition of the binned bispectrum, and
smooth it as for the real data. Restricting to those bin triplets that
contain enough valid data (see the discussion above), these simulations
are a good approximation to a Gaussian CMB map's bispectrum.

We now illustrate this method by applying it to a realistic situation. 
We add a point source map to a simulated Gaussian CMB map with anisotropic
noise, generated as described in the caption of
Table~\ref{table_filling_in}. The point source simulation was created
with the Planck Sky Model, at 143~GHz, with a beam with a FWHM of 5
arcmin, and contains faint infrared sources, as described
in~\citep{Delabrouille:2012ye}, and faint radio sources with the
improved parameters described in~\citepalias{planck2015-12}. The galactic and
point source masks were applied as described in Section~\ref{realsky}.

The binned bispectrum of this map was evaluated applying the linear
correction and the filling-in procedure, and the $f_{\rm NL}$'s were
determined individually for each of the templates described in
Section~\ref{templatesec}.  The unclustered point
source contribution was detected with high significance in this
contaminated map: $b_{\rm ps} = (64.8\pm 0.8)\times 10^{-29}$. This
signal is much stronger than the one detected in the {\em cleaned} Planck maps, 
but of the same order of magnitude as the forecast at
217~GHz (see \citepalias{planck2015-17}).  No statistically significant
detection of a nonzero $f_{\rm NL}$ was obtained for the other
templates, with the exception of the CIB template.  But the CIB
bispectrum has significant overlap with the unclustered point
source bispectrum (see Table~\ref{tab_corr_coeff}), so this result is
not surprising.  The nonzero result for $b_\mathrm{CIB}$ disappears in a joint 
analysis of the unclustered point source and CIB templates.
Finally we smooth the bispectrum with a few different values of the smoothing
length, namely $\sigma_{\rm smoothing}=1,2,3$. 

Apart from studying the contaminated bispectrum, we can also try to remove the 
estimated point source contribution from the measured binned bispectrum, 
simply by subtracting the corresponding smoothed template (\ref{Bunclust}) 
with the measured amplitude. We can then check if there are remaining 
non-Gaussian features in this cleaned bispectrum using the method described 
above.
Table~\ref{ataTable} gives the maximum bin values for the
bispectrum before and after the template cleaning is applied. The
minima are not given because the inclusion of the point sources tends
to gives a positive bispectral contribution. The $p$-values were calculated 
using simulations of $10^6$ Gaussian realizations and fitting the CDF for 
$N_\mathrm{eff}$ to the empirical distribution.  We found that the smoothing 
lengths $\sigma_\mathrm{smoothing} =1, 2,$ and $3$ correspond to 
$N_\mathrm{eff}=8664, 1948,$
and $588$ respectively.
We see that a highly statistically significant detection is found using the 
above procedure on the uncleaned bispectrum. We also observe that the 
template cleaning
procedure is successful; however, some detectable unsubtracted residual remains.
This residual, however, has little overlap with the known 
theoretically motivated primordial templates.

\begin{table}
\begin{center}
\begin{tabular}{c|llll}
\hline
Smoothing  & \multicolumn{2}{c}{Before} & \multicolumn{2}{c}{After}\\
length     & \multicolumn{2}{c}{template cleaning} 
& \multicolumn{2}{c}{template cleaning}\\
($\sigma_\mathrm{smoothing}$) &   $X_{max}$ & $p$-value  &  $X_{max}$ & $p$-value \\
\hline
1 & 33.4 & $e^{-553}$ & 5.1 & $1.5 \times 10^{-3}$ \\
2 & 51.2 & $e^{-1308}$ & 3.6 & 0.15 \\
3 & 59.2 & $e^{-1751}$ & 4.4 & $3.3 \times 10^{-3}$  \\
\hline
\end{tabular}
\end{center}
\caption{Maximum bin values and associated $p$-values for the bispectrum of
a simulated Gaussian CMB + point sources map before and after cleaning of the
point source contribution, for various values of the smoothing 
length. See main text for a more detailed explanation.}
\label{ataTable}
\end{table}

\section{Summary of applications and results}
\label{otherbispecsec}

Having discussed the binned bispectrum estimator in detail, we can now start
looking at further extensions, and in particular at applications and results.
As in chapter~\ref{NGinflsec}, these self-contained papers have been added
verbatim to the appendices of this chapter, removing only their conclusions.
Those conclusions are used to provide a summary of the papers in this section.
The papers in question are \citep{LvT2} (section~\ref{summLvT2} and
apppendix~\ref{LvT2app}) containing the extension to isocurvature
non-Gaussianity, \citep{JRvT} (section~\ref{summJRvT} and appendix~\ref{JRvTapp})
with an application to galactic foregrounds, and \citepalias{planck2018-09}
(section~\ref{summPlanck2018NG} and appendix~\ref{Planck2018NGapp}) with
the Planck 2018 primordial non-Gaussianity analysis.

\subsection{Isocurvature non-Gaussianity}
\label{summLvT2}

In the paper \citep{LvT2}, reproduced in appendix~\ref{LvT2app},
we systematically investigated the angular bispectra generated by initial conditions that combine the usual adiabatic mode with an isocurvature mode, assuming local non-Gaussianity. We studied successively the four types of isocurvature modes, namely cold dark matter (CDM), 
baryon, neutrino density and neutrino velocity isocurvature modes. In each case, the total bispectrum can be decomposed into six elementary bispectra and we estimated the expected uncertainties on the corresponding coefficients, which are extensions of the usual purely adiabatic $f_\mathrm{NL}$ parameter, in the context of the Planck data of the cosmic microwave background radiation (CMB). As we showed, the results
for baryon isocurvature can be obtained from a simple rescaling of the
CDM isocurvature results, but the others are distinct.
At the time of this paper (2012), Planck results were not yet available, and in
the 2013 Planck release, due to lack of time, isocurvature non-Gaussianity was
not studied. Because this release did not include polarization, and large
improvements for the isocurvature non-Gaussianity results were expected with
polarization as explained in this paper, this was a logical decision. However, starting
from the 2015 Planck release, results for isocurvature non-Gaussianity have
been included. The 2018 results can be found in appendix~\ref{Planck2018NGapp}.

In the squeezed limit, where one multipole $\ell$ is much smaller than the other
two (which then have to be almost equal due to the triangle inequality), we showed that  the six elementary bispectra  factorize as a function of the small $\ell$ times the power
spectrum as a function of the large $\ell$.  Since the  squeezed limit components dominate the bispectrum for {\it local} non-Gaussianity, we were able,  
using this factorization,  to give simple explanations for the various interesting results that we observed.

By enlarging the space of initial conditions, one obviously expects a larger uncertainty on the purely adiabatic $f_\mathrm{NL}$ coefficient. Interestingly, this uncertainty is increased only by a factor $2$ for CDM and baryon isocurvature modes, whereas it increases by a factor $6$ in the neutrino density isocurvature case and by a factor $5$ in the neutrino velocity isocurvature case.
This can be explained by the fact that the CDM isocurvature power spectrum 
decreases much faster with $\ell$ than the adiabatic and neutrino isocurvature
ones. As we showed, this means that the uncertainties on $\tf_\mathrm{NL}^{\zeta,\zeta\zeta}$ 
and $\tf_\mathrm{NL}^{\zeta, \zeta S}$ in the case of CDM isocurvature continue to improve
as one increases the number of available multipoles, while the other four
saturate at a much lower $\ell$. As a consequence the first two can be
determined much more accurately than the other four, and are only weakly 
correlated with them. This small correlation also means that it was 
important to look at the data with the full $f_\mathrm{NL}$ estimator, and
not just the adiabatic one, as a large CDM isocurvature non-Gaussianity could
have been hiding behind a small adiabatic signal.

We showed that the E-polarization often plays a crucial role in reducing the uncertainties. In the CDM isocurvature case, polarization improves slightly the precision on the coefficients $\tf_\mathrm{NL}^{\zeta,\zeta\zeta}$ and $\tf_\mathrm{NL}^{\zeta, \zeta S}$, but the precision of the other four coefficients improves by a factor of order five.
Polarization is also very important for some of the parameters in the
neutrino velocity case; the uncertainty on the purely 
isocurvature $\tf_\mathrm{NL}^{S,SS}$, for example, improves by a factor 8 when
polarization is included. Again we were able to explain these results
from the behaviour of the power spectra, using the factorization of the
squeezed bispectrum.

The decomposition of the bispectrum into six elementary bispectra and
the CMB constraints on the six $\tf_\mathrm{NL}^{(i)}$ parameters do not depend on any
assumptions about the specific model of the early universe. We only assumed
that (possibly correlated) primordial adiabatic and isocurvature modes are 
produced, with a primordial bispectrum of local type and with power spectra 
that all have the same shape. Note that neither of these assumptions appears
to be essential; they were only made for simplicity.

If, however, one does consider an explicit early universe model, there are
often relations between the six parameters, and an observational detection
or constraint could then be used to check for such a relation and put
constraints on the parameters of the model. 
We discussed a general class
of models with two scalar fields, where only one of the fields generates
both the isocurvature perturbations and all non-linearities. 
We also
considered a specific implementation of this general model, where the
two fields are an inflaton and a curvaton. In this model, a  CDM isocurvature mode
is produced and the six $\tf_\mathrm{NL}^{(i)}$ parameters only depend on two model parameters. In some ranges of the model parameters,
the isocurvature mode is subdominant in the power spectrum but provides observable  non-Gaussianity that can dominate the usual adiabatic non-Gaussianity. Looking for these new angular shapes in  the CMB data would thus provide interesting information on the very early universe.

\subsection{Galactic foregrounds}
\label{summJRvT}

In the paper \citep{JRvT}, included in appendix~\ref{JRvTapp},
we used the binned bispectrum estimator to determine
the bispectra of different galactic foreground maps, as produced by
the \texttt{Commander} component separation method from Planck 2015 data
(rescaled to amplitudes representative for the 143~GHz Planck channel).
These galactic foreground
bispectra were then used as templates for other runs of the binned
bispectrum estimator applied to various types of maps: simulations,
raw sky maps, and cleaned CMB maps.

This paper serves different purposes. In the first place it is a proof
of concept. The possibility to determine the (binned) bispectrum of any
map is a clear advantage of the binned bispectrum estimator, and was
used in the official Planck releases to
present the bispectrum of the observed CMB. The fact that any provided
bispectrum, not only if an analytical template is known but also simply
any numerical bispectrum, can be used as template in the binned bispectrum
analysis pipeline, had also long been presented as an advantage of the
method. In fact the
possibility of combining these two advantages to do an analysis as
presented here was already mentioned in the original
paper of the binned bispectrum estimator~\citep{BvTC} and was
one of the motivations for developing it in the first place, but
had until this paper never been worked out explicitly. This paper proved that
this idea also works in practice.

Secondly, this paper shows and discusses the bispectra of the various
galactic foregrounds, which is an interesting result in itself, even if
for the purposes of the paper it was only an intermediate step.
We found that the dust, the free-free and the anomalous microwave emissions
have very squeezed bispectra (similar to the local shape, but with an
opposite sign). The small-scale fluctuations of the dust radiation are
stronger in the large-scale dust clouds, so small-scale and large-scale
fluctuations are correlated (and a similar explanation is valid for
the other foregrounds). The synchrotron map as provided is different, as
its bispectrum is more similar to the equilateral shape, but we were able
to show that at least a large part of this effect is due to a residual
contamination by unresolved extra-galactic point sources. At 143 GHz (the
most important Planck frequency for CMB analysis) only the dust really 
contaminates the CMB signal, the other foregrounds giving contributions 
that are orders of magnitude smaller.

An issue with the numerical
templates we determined is that they also depend on the mask applied
to the foregrounds and contain the characteristics of the experiment
like the beam and the noise. We showed that the choice of the mask is
very important because the foregrounds are localized in the galactic
plane close to the galactic mask, so a small change of mask
could give a large difference of bispectrum. This means the same mask
should be used for determining the template as for the final analysis.
It should be pointed out that for the purpose of studying the non-Gaussianity
of galactic foregrounds as goal in itself, the bispectrum is likely not the
best tool: a pixel-space based statistic to take into account the localized
nature of these foregrounds would seem more logical. However, our main
purpose was to investigate the impact these galactic foregrounds have on the
determination of primordial $f_\mathrm{NL}$ parameters in a bispectrum
analysis.

The third and final result of this paper is the $f_\mathrm{NL}$ analysis of real
sky maps, both raw and cleaned CMB, with these galactic bispectrum templates,
where we investigated in particular if any observable galactic residuals remained
in the cleaned CMB map and if a joint analysis of primordial and galactic
templates improved the determination of the primordial $f_\mathrm{NL}$.
But before we did that analysis we obviously
first tested and validated our methodology and our new analysis pipeline
on simulations. These tests were based on Gaussian realizations of the CMB
to which we added noise simulations and a known amount of dust. We showed
that both with isotropic and with anisotropic noise we managed to detect
the expected amount of dust in our maps. However, we also showed that, to do a joint analysis
with the primordial and the dust shapes, the usual choice of bins, while
acceptable, can be improved. With more bins at low $\ell$ one can better
discriminate between the templates that peak in the squeezed
configuration (local and dust especially). We also discussed the
effects of the (small) breakdown of the weak non-Gaussianity
approximation that occurs when we add the full dust map to the CMB
simulations (i.e.\ the expected amount of dust in raw-sky
observations). The main consequence is that the real error bars become
larger than the Fisher forecasts.

The testing and validation having been successful, we then used the
numerical galactic templates on the cleaned \texttt{SMICA} CMB map of the 2015 Planck
release. Fortunately, we did not detect any residual of the dust. The error bars for the dust and
local shapes increased in the joint analysis with the usual
binning, again because of the choice of bins that was not optimal to
differentiate them. Finally, we applied the foreground templates to
the raw sky map at 143~GHz and the binned bispectrum estimator
succeeded in detecting the dust in it at the expected level (the
intensities of the other foregrounds at 143 GHz being too small 
to detect even if they were present in the map).

The work presented in this paper can be extended in several ways.
The additional functionality built into the binned bispectrum estimator
code to use numerical bispectra as templates opens new possibilities, and 
allows us to include the template of any component of which a map exists in
our bispectrum and $f_\mathrm{NL}$ analyses.
It would also be interesting to further study the galactic bispectra,
or their non-Gaussianity in general, together with an expert on galactic
astrophysics, to see if they can be physically understood. This could maybe
lead to building an analytical template for these bispectra that can also be
used by other bispectrum estimator codes.
Finally, the analysis of this paper obviously had to be repeated on
the final 2018 Planck data. This was indeed done, although for temperature
only, as presented in section~6.3.1 of \citepalias{planck2018-09}, reproduced
in appendix~\ref{Planck2018NGapp}. The hope is that the improved treatment of
the polarization maps in that release will make an extension to E-polarization
of this analysis viable as well.

\subsection{Planck 2018 results}
\label{summPlanck2018NG}

In the paper \citepalias{planck2018-09}, included in
appendix~\ref{Planck2018NGapp},
we presented constraints on primordial non-Gaussianity (NG), using
the Planck full-mission CMB
temperature and E-mode polarization maps. 
Compared to the Planck 2015 release, the low-$\ell$
($4 \leq \ell < 40$) polarization multipole range was this time also included.

Our analysis produced the following final results
(68\,\%~CL, statistical): $f_{\rm NL}^{\rm local} = -0.9 \pm 5.1$;
$f_{\rm NL}^{\rm equil} = -26 \pm 47$; and $f_{\rm NL}^{\rm ortho} = - 38 \pm 24$.
These results are overall stable with respect to our
constraints from the 2015 Planck data. They show no real improvement in
errors, despite the additional polarization modes. This is due to a
combination of two factors. Firstly, the local shape, which is most sensitive to
low-$\ell$ modes and where one would naively expect an improvement, is
actually less sensitive to polarization than the equilateral and orthogonal
shapes. This means that in the end none of the three shapes are very sensitive
to low-$\ell$ polarization modes. Secondly, the temperature and polarization
simulations used to determine the errors had a more realistic but slightly
higher noise level than in the previous release.

On the other hand, the quality of polarization data shows a clear improvement
with respect to our previous analysis. This was confirmed by a large 
battery of tests on our data set, including comparisons between different
estimator implementations (KSW, 
Binned, and two Modal estimators) and foreground-cleaning methods
(\texttt{SMICA}, \texttt{SEVEM}, \texttt{NILC}, and \texttt{Commander}), 
studies of robustness under changes in sky coverage and multipole range, and
an analysis of the impact of noise-related systematics.
While in our previous release we had cautioned the reader to take polarization
bispectra and related constraints 
as preliminary, in light of these tests we consider our results presented here
based on the combined temperature and polarization data set to be fully
reliable.
This also implied that polarization-only, EEE bispectra could now be used for
independent tests, which led to primordial NG constraints at a sensitivity level
comparable to that of WMAP from temperature bispectra, and 
yielding statistical agreement.

As in the previous analyses, we went beyond the local, equilateral, and orthogonal
$f_{\rm NL}$ constraints by considering a large number of additional
cases, such as scale-dependent feature and resonance bispectra, running
$f_{\rm NL}$ models, 
isocurvature primordial NG, and parity-breaking models. We set tight
constraints on all these scenarios, but did not detect any significant signals. 

On the other hand, the non-primordial lensing bispectrum
was detected with an improved significance compared to 2015,
excluding the null hypothesis at $3.5\,\sigma$. The amplitude of the signal is
consistent with the expectation from the Planck best-fit cosmological
parameters, further indicating the absence of significant foreground
contamination or spurious systematic effects.
We also explicitly checked for the presence of various non-primordial
contaminants, like unclustered extragalactic point sources, CIB, galactic
thermal dust, and the thermal SZ effect, but apart from the first, none of these
were detected. The small amount of remaining point-source signal in the
cleaned maps had no impact on our other constraints because of its
negligible correlations.

We updated our trispectrum constraints, finding
$g_{\rm NL}^{\rm local} = (-5.8 \pm 6.5) \times 10^4$ (68\,\%~CL, statistical),
while also constraining additional shapes, generated by different operators in
an effective field-theory approach to inflation.

In addition to estimates of bispectrum and trispectrum amplitudes, we produced
model-independent reconstructions and analyses of the Planck CMB
bispectrum. Finally, we used our measurements to obtain constraints on
early-universe scenarios 
that can generate primordial NG. We considered, for example, general single-field
models of inflation, curvaton models,
models with axion fields producing parity-violating tensor bispectra, and
inflationary scenarios generating directionally-dependent bispectra 
(such as those involving vector fields). 

In our data analysis efforts, which started with the 2013 release, we achieved
a number of crucial scientific goals. In particular we reached an unprecedented
level of sensitivity in the determination of the bispectrum and trispectrum
amplitude parameters ($f_{\rm NL}$, $g_{\rm NL}$) and significantly extended the
standard local, equilateral, and orthogonal analysis, encompassing a large
number of additional shapes motivated by a variety of inflationary models.
Moreover, we produced the first polarization-based CMB bispectrum constraints
and the first detection of the (non-primordial) bispectrum induced by
correlations between CMB lensing and secondary anisotropies.  
Our stringent tests of many types of non-Gaussianity are fully consistent with
expectations from the standard single-field slow-roll paradigm and provide
strong constraints on alternative scenarios. 
Nevertheless, the current level of sensitivity does not allow us to rule out
or confirm most alternative scenarios. It is natural at this stage to ask
ourselves what should be the $f_{\rm NL}$ sensitivity goal for future
cosmological experiments. A number of studies has identified $f_{\rm NL} \sim 1$
as a target.  Achieving such sensitivity for local-type NG would enable us to
either confirm or rule out a large class of multi-field models. 
A similar target for equilateral, orthogonal, and scale-dependent shapes would
allow us to distinguish standard slow-roll from more complex single-field
scenarios, such as those characterized by higher-derivative kinetic terms or
slow-roll-breaking features in the inflaton potential
(see e.g.~\citep{Alvarez:2014vva,Finelli:2016cyd} and references therein).
With this aim in mind, the challenge for future cosmological observations will
therefore be that of reducing the $f_{\rm NL}$ errors from this paper
by at least one order of magnitude.

\begin{appendices}
\renewcommand{\thechapter}{\ref*{NGCMBsec}\Alph{chapter}}
\setcounter{chapter}{0}  

\chapter{Isocurvature modes in the CMB bispectrum}
\label{LvT2app}

This appendix contains the full paper \citep{LvT2}, except for the conclusions
that were used as a summary in section~\ref{summLvT2}. It was written
in collaboration with David Langlois.

We study the angular bispectrum of local type arising from the
(possibly correlated) combination of a primordial adiabatic mode with
an isocurvature one. Generically, this bispectrum can be decomposed
into six elementary bispectra. We estimate how precisely CMB data,
including polarization, can enable us to measure or constrain the six
corresponding amplitudes, considering separately the four types of
isocurvature modes (CDM, baryon, neutrino density, neutrino velocity).
Finally, we discuss how the model-independent constraints on the
bispectrum can be combined to get constraints on the parameters of
multiple-field inflation models.

\section{Introduction}

Inflation is currently  the best candidate to explain the generation of primordial perturbations, but many of its realizations remain compatible with the present data. One can hope that future data will enable us to find additional 
information in the primordial perturbations that could help to discriminate between  the
various mechanisms that can have taken place in the very early universe. 

In this respect, it is important 
to test the adiabatic nature of the primordial perturbations. 
Since single-field inflation predicts only adiabatic perturbations, the detection of a fraction of an isocurvature mode in the cosmological data would rule out the simplest models of inflation. By contrast, multiple-field inflation could easily account for the presence of isocurvature modes~\citep{Linde:1985yf}, which can even be correlated with the adiabatic component~\citep{Langlois:1999dw,Langlois:2000ar}. 

As shown in \citep{Bucher:1999re}, the most  general primordial perturbation is a priori a linear combination of the usual adiabatic mode with four types of isocurvature modes, respectively the Cold Dark Matter (CDM), baryon, neutrino density  and neutrino velocity isocurvature modes. The existence, and amplitude, of  isocurvature modes depends on the details of the thermal history of the universe. Various scenarios that  can lead to observable isocurvature modes have been discussed in the literature (double inflation \citep{Silk:1986vc,Polarski:1994rz,Langlois:1999dw}, axions \citep{Seckel:1985tj}, curvatons 
\citep{Linde:1996gt,Lyth:2001nq,Moroi:2001ct,Moroi:2002rd,Lyth:2003ip}).

In parallel to the possible presence of isocurvature modes, another property that could distinguish multiple-field models from single-field models is  a detectable primordial non-Gaussianity of the local type. 
 So far\footnote{Once more, I have decided to keep the original text and references from the paper, which dates from before Planck and even from before the final WMAP release.}, the   WMAP measurements  of the CMB anisotropies~\citep{Komatsu:2010fb}  have set the present limit $f_{\rm
  NL}^{\rm local} =32\pm 21$ (68\,\% CL) [and $-10 < f_{\rm NL}^{\rm
  local} < 74$ (95\,\% CL)] on the parameter $f_{\rm
  NL}^{\rm local} $ that characterizes the amplitude of the simplest type of non-Gaussianity, namely the local shape.   
Similarly to isocurvature modes,  a detection of local primordial non-Gaussianity   would rule out all inflation models based on a single scalar field, since they generate only 
unobservably small local non-Gaussianities~\citep{Creminelli:2004yq}. 
Scenarios with additional scalar fields, 
 such as another inflaton (see e.g.\ \citep{Byrnes:2010em,TvT1}),  a curvaton~\citep{Lyth:2002my} or a modulaton~\citep{Dvali:2003em,Kofman:2003nx,Langlois:2009jp}, which can 
 produce  detectable local non-Gaussianity, would then move to the front stage.

Isocurvature modes are usually investigated by constraining the  power spectrum of primordial perturbations with CMB or large-scale structure data (see e.g.\ \citep{Bean:2006qz,Sollom:2009vd,Mangilli:2010ut,Li:2010yb,Kawasaki:2011ze,Kasanda:2011np,DiValentino:2011sv,Valiviita:2012ub}).
However, isocurvature modes could also contribute to non-Gaussianities as discussed in several works~\citep{Bartolo:2001cw,Kawasaki:2008sn,Langlois:2008vk,Kawasaki:2008pa,Hikage:2008sk,Kawakami:2009iu, Langlois:2011zz,Langlois:2010fe,LvT1}. Moreover, there exist models 
\citep{Langlois:2011zz}  where isocurvature modes, while remaining a small fraction at the linear level,  would dominate the non-Gaussianity. As shown in \citep{LvT1}, these CDM isocurvature modes would be potentially detectable via their non-Gaussianity in the CMB data such as collected by Planck. Non-Gaussianity can thus be considered as a complementary probe of isocurvature modes. 

In the present work, we refine and extend our  previous analysis~\citep{LvT1} by considering all types of isocurvature modes, not only the CDM isocurvature mode. We analyse the bispectrum generated by the  adiabatic mode together with   one of the four isocurvature modes. The total  angular bispectrum can be decomposed  into six distinct components: the usual purely adiabatic bispectrum, a purely isocurvature bispectrum, and four other bispectra that arise from the possible 
 correlations between the  adiabatic and isocurvature mode. 
 Because these six bispectra have  different shapes in $\ell$-space, their amplitude can  in principle be measured  in the CMB data and  we have computed, for each type of isocurvature mode,  the associated $6\times 6$  Fisher matrix to estimate what precision on these six parameters could be reached with the Planck data. 
 We also show that the inclusion of polarization measurements improves the predicted precision of some
isocurvature non-Gaussianity parameters significantly. 
 
The elementary bispectra discussed above depend only on the adiabatic and isocurvature transfer functions  and are thus independent of the details of the generation mechanism. Now, by assuming a specific class of inflationary models, one obtains particular relations between the six bispectra, which can be used as consistency relations for the model or to constrain  the model parameters. 
We illustrate this in the context of curvaton-type models, generating adiabatic and CDM  isocurvature perturbations. 

The outline of the paper is the following. In the next section, we present the various isocurvature perturbations and discuss their impact on the CMB angular power spectrum. The following section is devoted to the angular bispectrum and its decomposition into six elementary bispectra. We then discuss the observational prospects to detect these elementary local bispectra in the future data, distinguishing the various isocurvature modes. Finally, we consider  models where primordial perturbations are generated by an inflaton and a curvaton, 
and show that the amplitudes of all six bispectra depend on only two coefficients, which can be constrained from the data.

\section{Isocurvature perturbations}
In this section, we recall  the definition of isocurvature modes in the context of {\it linear} cosmological perturbations. At the time of last scattering, the main components in the universe are the CDM (c), the baryons (b), the photons ($\gamma$) and the neutrinos ($\nu$). All these components are characterized by their individual energy density perturbation $\delta\rho_i$ and their velocities ${\cal V}_i$ (as well as higher momenta of their phase space distribution functions, which we do not  discuss here; see \citep{Ma:1995ey} for details). The ``primordial'' perturbations for each Fourier mode $k$ are usually defined on super-Hubble scales, i.e.\ when $k\ll aH$, deep in the radiation dominated era. 

The most common type of perturbation is the adiabatic mode, 
characterized by
\be
\frac{\delta n_c}{n_c}=\frac{\delta n_b}{n_b}=\frac{\delta n_\nu}{n_\nu}=\frac{\delta n_\gamma}{n_\gamma}\,,
\ee
which means that the number of photons (or neutrinos, or CDM particles) per baryon is not fluctuating.
 In terms of the energy density contrasts ($\delta\equiv \delta\rho/\rho$), the above condition is expressed as
\be
\label{adiabatic}
\delta_c=\delta_b= \frac34\delta_\nu=\frac34\delta_\gamma\,,
\ee
where the $3/4$ factor, for photons and neutrinos, comes from the relation $\rho\propto n^{4/3}$ for relativistic species. 

Assuming adiabatic initial conditions is  natural if all particles have been created by the decay of a single degree of freedom, such as a single inflaton, and,  so far, the CMB data are fully compatible with purely adiabatic perturbations.
However, other types of perturbations can be included in a more general framework. In addition to the adiabatic mode, one can consider  four distinct types of so-called isocurvature modes~\citep{Bucher:1999re}: the CDM isocurvature mode $S_c$, the baryon isocurvature mode $S_b$, the neutrino density  isocurvature mode $S_{\nu d}$ and the neutrino velocity  isocurvature mode $S_{\nu v}$. 
At zeroth order in $k\tau$, where $\tau$ is the conformal time, the  first three are  characterized, respectively,  by 
\begin{eqnarray}
\delta_c&=&S_c+\frac34 \gd_\gamma, \qquad 
\delta_b= \frac34\delta_\nu=\frac34\delta_\gamma \qquad {\rm (CDM\ isocurvature)}
\\
\delta_b &=& S_b +\frac34 \gd_\gamma, \qquad 
\delta_c= \frac34\delta_\nu=\frac34\delta_\gamma \qquad {\rm (baryon \ isocurvature)}
\\
\frac34\delta_\nu &=& S_{\nu d} +\frac34 \gd_\gamma, \qquad 
\gd_b=\delta_c=\frac34\delta_\gamma \qquad {\rm (neutrino \ density \ isocurvature)},
\end{eqnarray}
while  the corresponding velocities tend to zero. 
As for the neutrino velocity isocurvature mode, it  is characterized by non vanishing ``initial velocities'',
\be
{\cal V}_\nu= S_{\nu v}, \qquad  {\cal V}_{\gamma b}=-\frac78 N_\nu \left(\frac{4}{11}\right)^{4/3} S_{\nu v}
\qquad  {\rm (neutrino \ velocity \ isocurvature)},
\ee
where ${\cal V}_{\gamma b}$ is the common velocity of the photon-baryon plasma (the photons and baryons are initially tightly coupled via the Thomson scattering off free electons) and $N_\nu$ is the number of species of massless neutrinos. The above relation between the two velocities ensures that they exactly cancel in the momentum density, 
while the energy densities satisfy the adiabatic condition (\ref{adiabatic}). 

In the following, each mode will be characterized by its amplitude: the curvature perturbation on constant energy  hypersurfaces,  $\zeta$,  for the adiabatic mode, and $S_c$, $S_b$, $S_{\nu d}$ and $S_{\nu v}$ for the four isocurvature modes. These variables will be  denoted collectively as $X^I$. 
In the context of inflation, a necessary condition for these isocurvature modes to be created is that several light degrees of freedom exist during inflation. Moreover, since the adiabatic and isocurvature modes can be related in various ways to these degrees of freedom during inflation, one can envisage the existence of correlations between these modes.

These various modes lead to {\it different} predictions for the CMB temperature and polarization.  
Let us consider for instance the temperature anisotropies, which can be decomposed into spherical harmonics:
\be
\frac{\Delta T}{T}=\sum_{\ell m} a_{\ell m} Y_{\ell m}\,.
\ee
The multipole coefficients $a_{\ell m}$ can be related linearly to  any of the primordial modes. The precise correspondance can be computed numerically and written in the form
\be
a_{\ell m}^{I}=4\pi (-i)^\ell \int \frac{d^3\mathbf{k}}{(2\pi)^3} X^I(\mathbf{k})\,  g^I_\ell(k)\, Y^*_{\ell m}(\hat{\mathbf{k}})\,,
\ee
where  $g^I_\ell(k)$ is the transfer function associated with the corresponding primordial perturbation ($g^I_\ell(k)$ depends also on the various cosmological parameters). 

For each type of perturbation, the angular power spectrum  is thus given by
\be
C_\ell^{I}=\langle a_{\ell m}^{I}a_{\ell m}^{I*}\rangle=\frac{2}{\pi}\int_0^\infty dk\, k^2 \left[g_\ell^I(k)\right]^2  P_I(k),
\ee
where we have introduced the primordial power spectrum $P_I(k)$ defined by 
\be
\langle X^{I}(\mathbf{k}_1) X^I(\mathbf{k}_2)  \rangle \equiv  (2 \pi)^3 \delta ( \mathbf{k}_1+\mathbf{k}_2) P_{I}(k_1)\,.
 \ee
For our purposes, the crucial point is that the transfer functions associated with isocurvature perturbations are very different from the adiabatic transfer function. 
Moreover, each isocurvature mode leads to a specific signature that enables one to distinguish it from the other isocurvature modes. The only exception are  the CDM and baryon isocurvature modes  which give  exactly  the same pattern, up to the rescaling:
\be
\label{omega_bc}
S_b=\omega_{bc} \, S_c\,,\qquad \omega_{bc}\equiv\frac{\Omega_b}{\Omega_c}\,,
\ee
where the parameters  $\Omega_b$ and $\Omega_c$ denote, as usual, the present energy density fractions, respectively for baryons and CDM (note however that these two modes can in principle be discriminated via other effects, see e.g.\ \citep{
Holder:2009gd,Gordon:2009wx, Kawasaki:2011ze,Grin:2011tf}).

More generally, when we also allow for possible correlations between the modes
and include E-polarization, the angular power spectra are given by
\be
C_\ell^{IJ;\, \alpha \alpha'}=\frac{2}{\pi}\int_0^\infty dk\, k^2 g_\ell^{I; \alpha}(k)  
g_\ell^{J;\alpha'}(k) P_{IJ}(k),
\ee
where $I$ and $J$ label the isocurvature mode and $\alpha$ and $\alpha'$ the polarization
(i.e.\ either T (temperature) or E (polarization)). The primordial power spectrum
$P_{IJ}(k)$ is now defined by
\be
\langle X^{I}(\mathbf{k}_1) X^J(\mathbf{k}_2)  \rangle \equiv  (2 \pi)^3 \delta ( \mathbf{k}_1+\mathbf{k}_2) P_{IJ}(k_1)\,,
 \ee
which generalizes the previous definition to include the presence of 
correlations between the modes, which corresponds to the situation
where $P_{IJ}$ with different $I$ and $J$ does not vanish. 

\begin{figure}
\centering
\includegraphics[width=0.49\textwidth, clip=true]{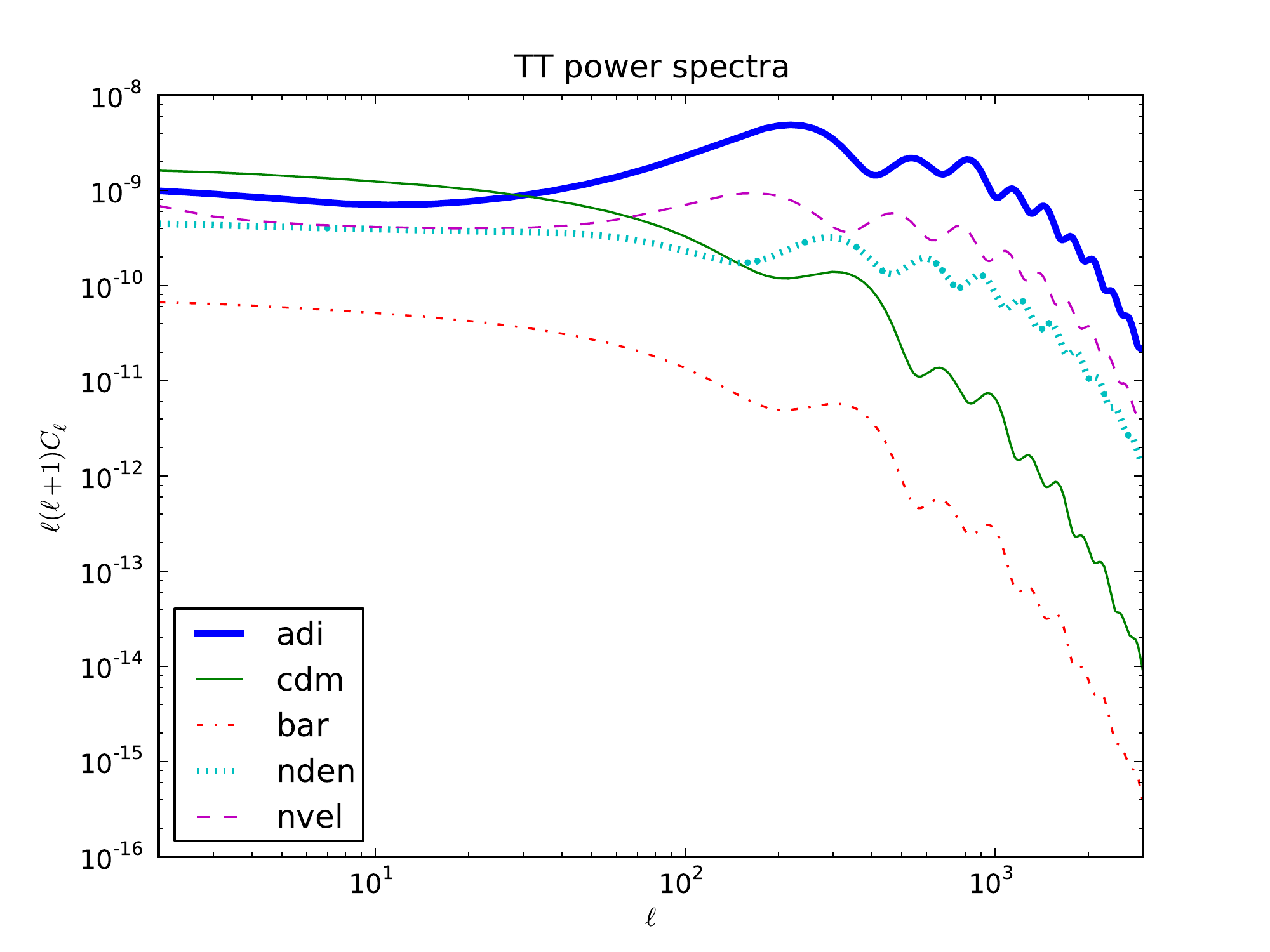}
\includegraphics[width=0.49\textwidth, clip=true]{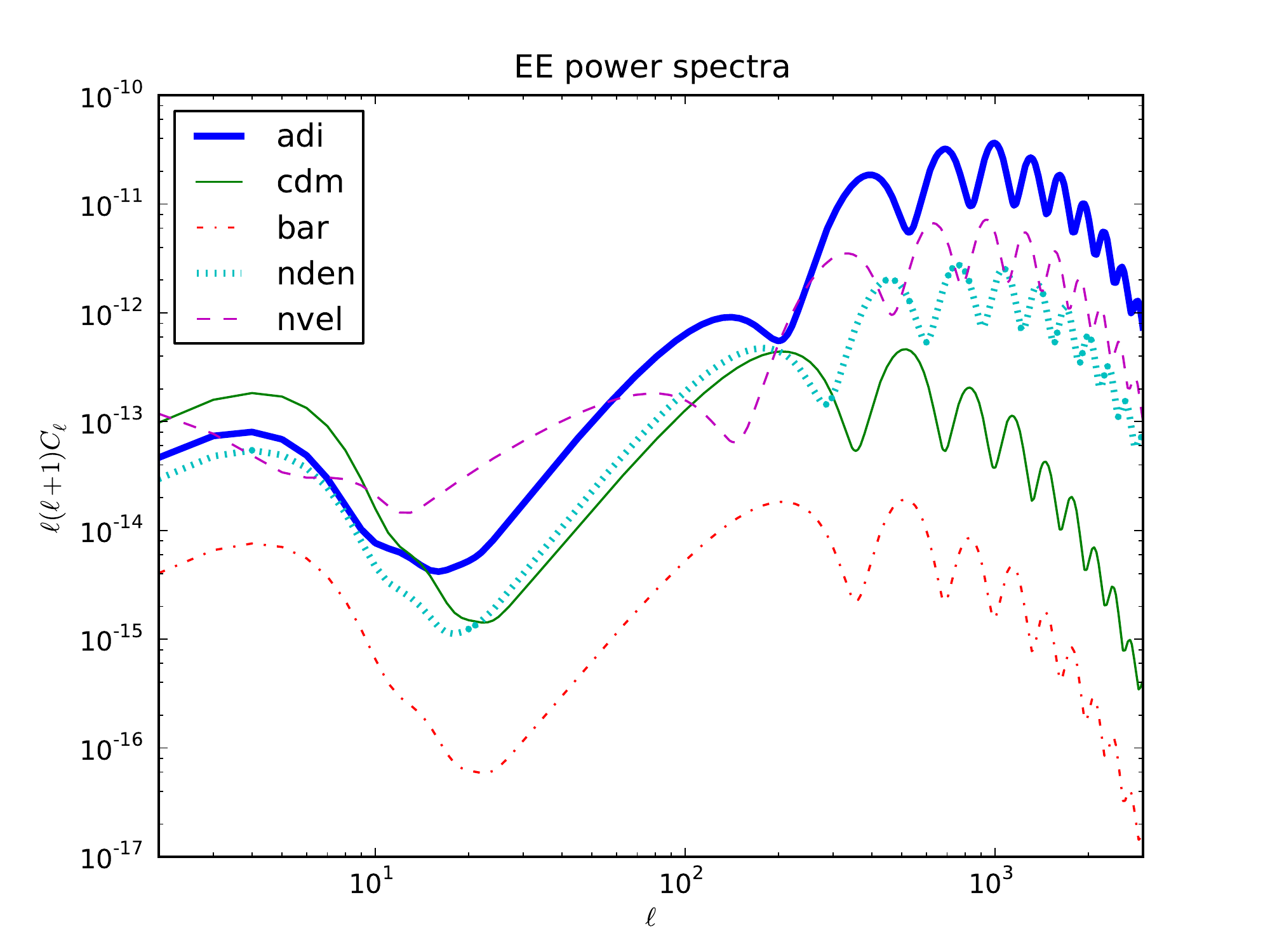}
\caption{Angular power spectra (multiplied by $\ell(\ell+1)$) for the temperature 
(left) and polarization (right) obtained from
purely adiabatic  or purely isocurvature initial conditions.  The amplitude 
and spectral index of the primordial power spectrum, as well as the 
cosmological parameters, on which the transfer functions depend, correspond 
to the WMAP7-only best-fit parameters 
(WMAP7-only best-fit parameters are used for all the figures and explicit
computations in this paper).}
\label{spectra_fig}
\end{figure}

\begin{figure}
\centering
\includegraphics[width=0.49\textwidth, clip=true]{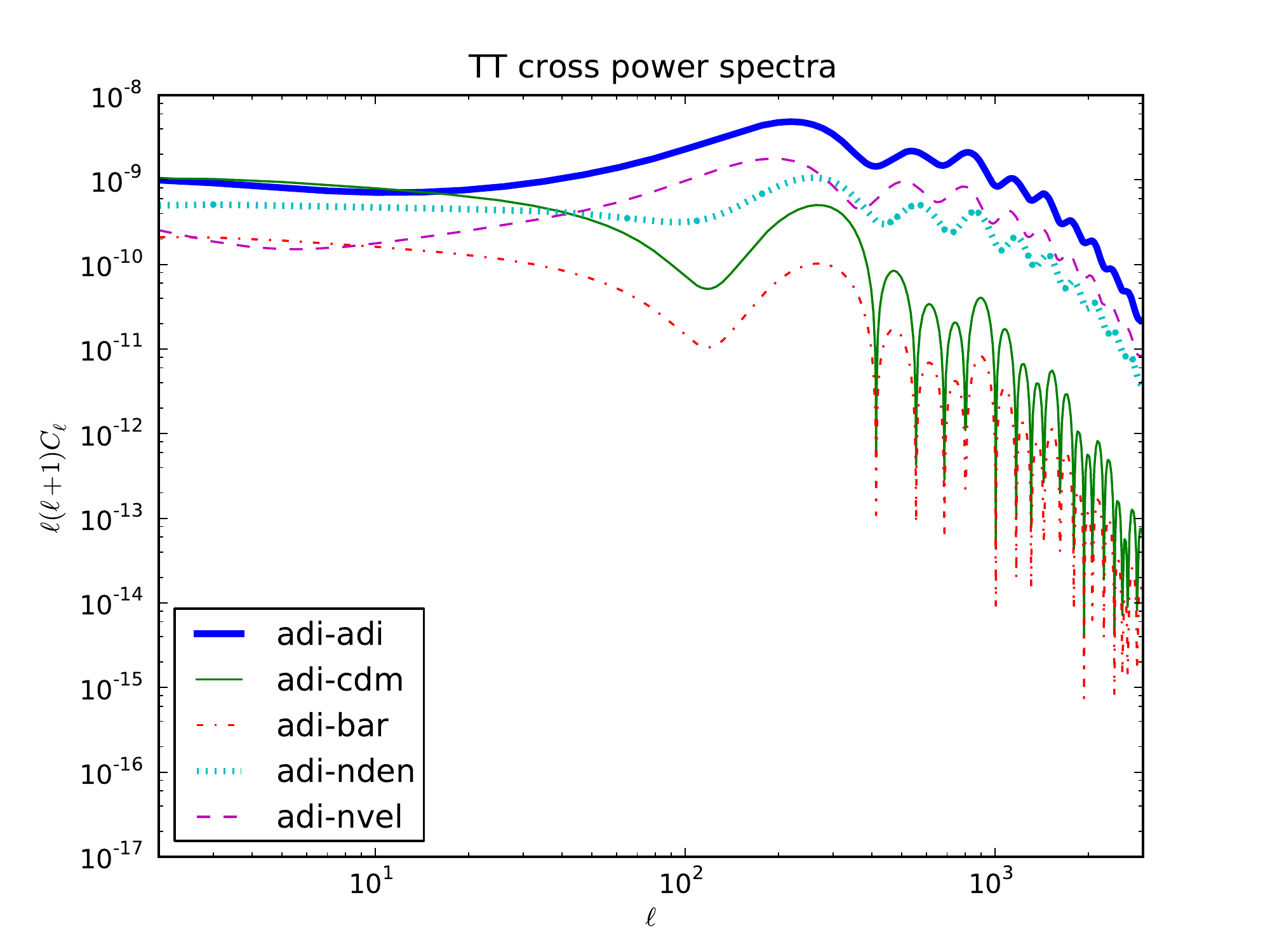}
\includegraphics[width=0.49\textwidth, clip=true]{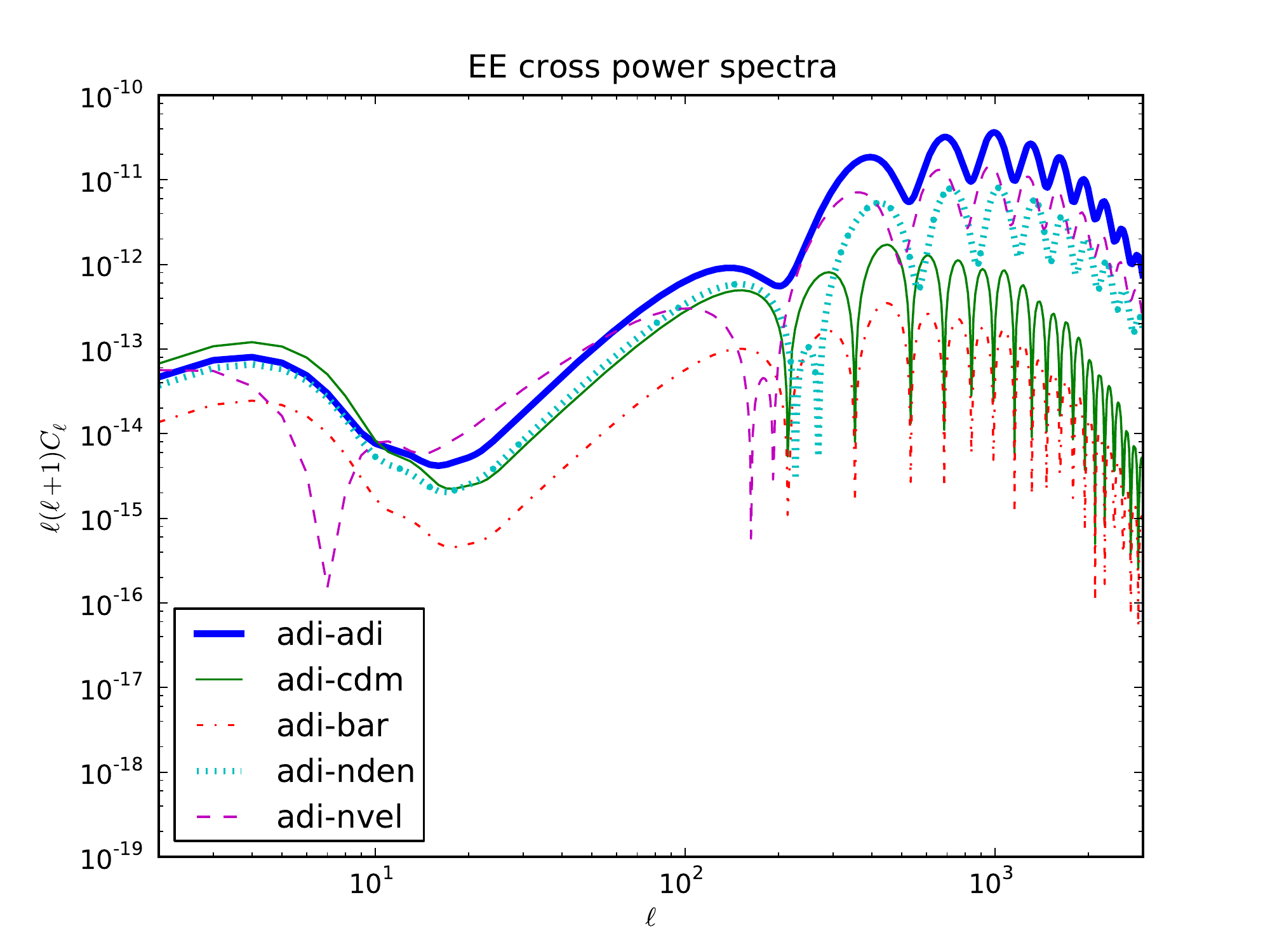}
\caption{The isocurvature cross power spectra (multiplied by $\ell(\ell+1)$ and
in absolute value) with one component fixed to be the adiabatic one.
The figure on the left shows temperature (TT), the one on the right 
polarization (EE).}
\label{cross_spectra_fig}
\end{figure}

\begin{figure}
\centering
\includegraphics[width=0.49\textwidth, clip=true]{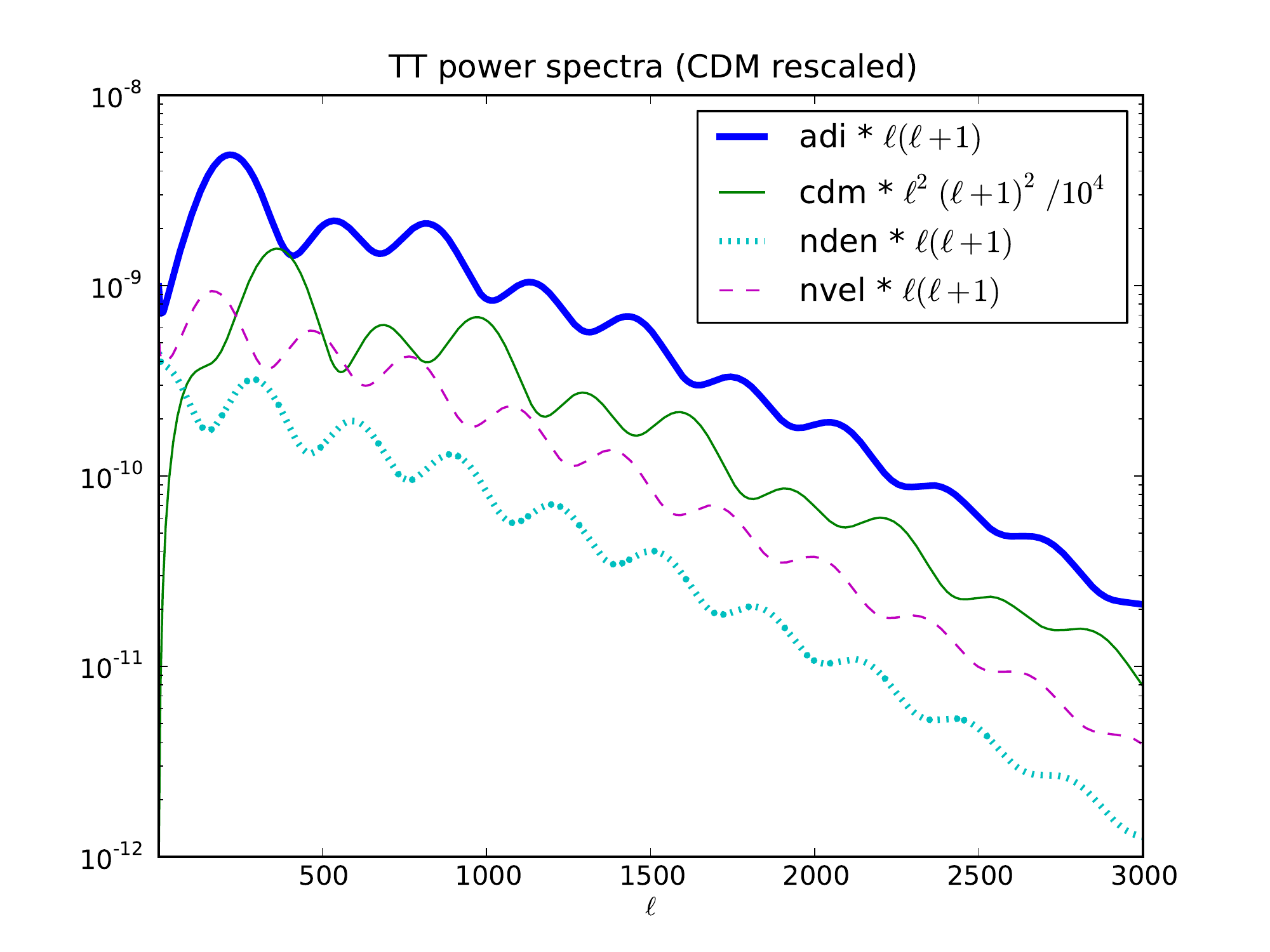}
\includegraphics[width=0.49\textwidth, clip=true]{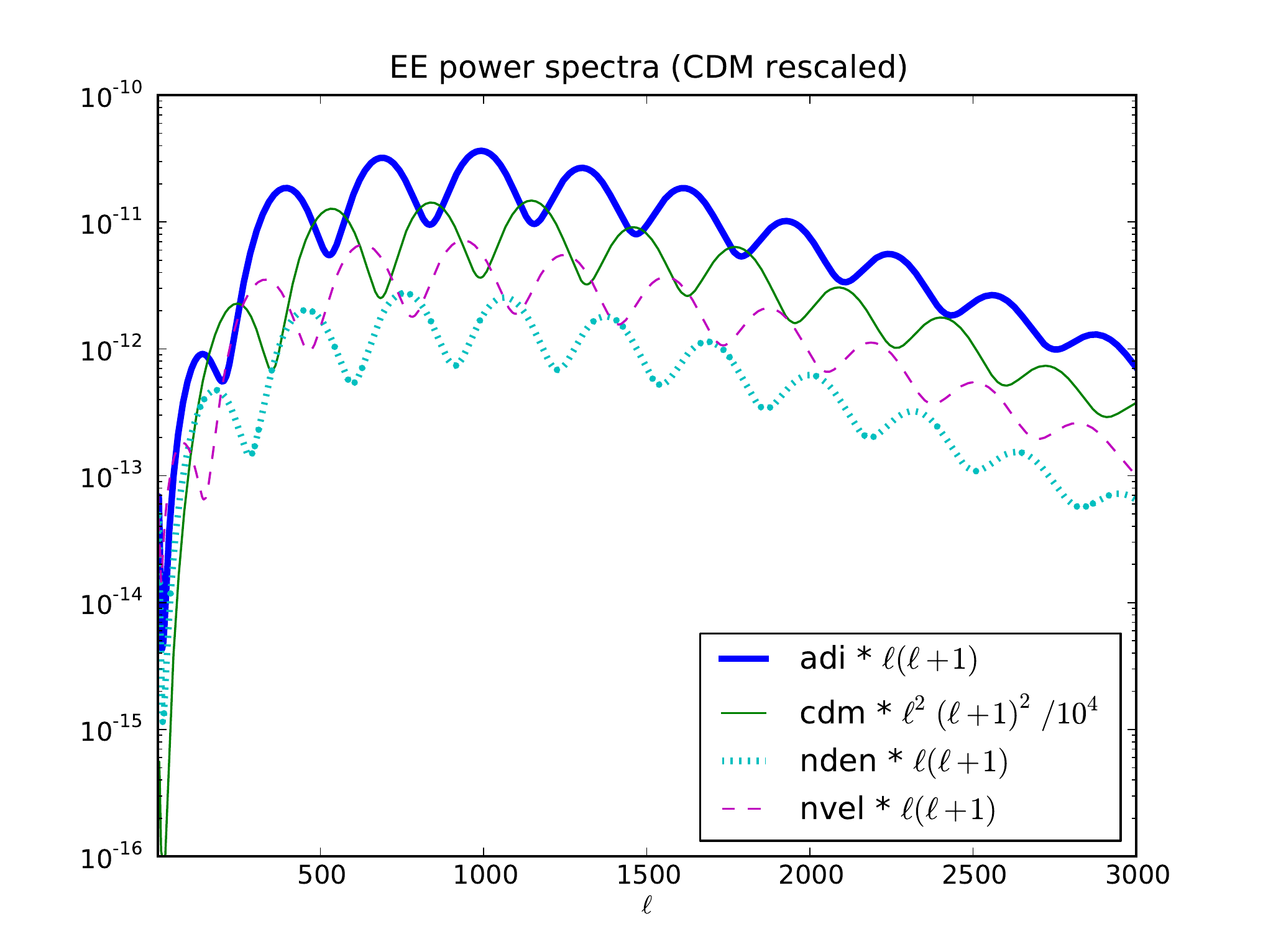}
\caption{Same angular power spectra for the temperature and the polarization as in Fig.~\ref{spectra_fig}, except that the horizontal axis is now linear, and that the CDM isocurvature spectrum is multiplied by $10^{-4} \ell^2(\ell+1)^2$ instead of $\ell(\ell+1)$.}
\label{spectra_rescaled}
\end{figure}

All this is illustrated in Fig.~\ref{spectra_fig} and \ref{cross_spectra_fig}, 
where we have plotted the angular power spectra for all the various modes 
(Fig.~\ref{spectra_fig}) and  the isocurvature cross power spectra where one of the 
components is adiabatic (Fig.~\ref{cross_spectra_fig}), assuming the same primordial power spectrum for all. 
As one can see from these figures, the CDM (and baryon) isocurvature mode
decreases much faster with $\ell$ than the other modes. In fact it turns out that
if one multiplies the CDM isocurvature power spectrum by $\ell^2(\ell+1)^2$ 
instead of $\ell(\ell+1)$, it falls off roughly in the same way as the other modes 
at large $\ell$, as illustrated in Fig.~\ref{spectra_rescaled}. This figure also 
nicely shows the relative phases of the acoustic peaks for the different modes.

When the ``primordial'' perturbation is a superposition of several modes, 
the multipole coefficients depend on a linear combination of the ``primordial'' modes, 
\be
\label{a_lm}
a_{\ell m}=4\pi (-i)^\ell \int \frac{d^3\mathbf{k}}{(2\pi)^3} \left(\sum_I X^I(\mathbf{k}) \, g^I_\ell(k)\right) Y^*_{\ell m}(\hat{\mathbf{k}}).
\ee
(Here we have once again omitted the polarization indices, as we will do 
in most of the equations of the paper, in order to improve readability.)
As a result,  the total angular power spectrum is now given by
\be
\label{C_l}
C_\ell=\langle a_{\ell m}a_{\ell m}^*\rangle=\sum_{I,J}\frac{2}{\pi}\int_0^\infty dk\, k^2 g_\ell^I(k) g_\ell^J(k) P_{IJ}(k).
\ee
We infer from CMB observations that the ``primordial'' perturbation is mainly of the adiabatic type. However, this does not preclude the presence, in addition to the adiabatic mode, of an isocurvature component, with a smaller amplitude. Precise measurement of the CMB fluctuations could lead to a detection
of such an extra component, or at least put constraints on its amplitude. 
For example, constraints on the CDM isocurvature to adiabatic ratio,
\begin{equation}
\label{eq:alpha}
\alpha = \frac{{\cal P}_{S_c}}{{\cal P}_{\zeta}} ,
\end{equation}
based on the WMAP7+BAO+SN data, have been published for the uncorrelated and fully correlated cases (the impact of isocurvature perturbations on the observable power spectrum  indeed depends  on the correlation between adiabatic and isocurvature perturbations, as illustrated in~\citep{Langlois:2000ar}). 
In terms of  the parameter $a\equiv \alpha/(1+\alpha)$, the limits 
given in  \citep{Komatsu:2010fb} 
are\footnote{Our notation differs from that of \citep{Komatsu:2010fb}: our $a$ corresponds to their $\alpha$ and our fully {\sl correlated} limit corresponds to their fully {\sl anti-correlated} limit, because their definition of the correlation has the opposite sign 
(see also \citep{Komatsu:2008hk} for a more detailed discussion).}
\be
a_0<0.064\quad (95 \% {\rm CL}), \qquad a_{1}< 0.0037 \quad (95 \% {\rm CL})\,,
\ee
respectively for the uncorrelated case  and for the fully correlated case.

\section{Generalized angular bispectra}
In this section, we turn to non-Gaussianities, including both adiabatic and isocurvature modes.
 
\subsection{Reduced and angular-averaged bispectra}
The angular bispectrum corresponds to the three-point function of the multipole coefficients:
\be
B_{\ell_1 \ell_2 \ell_3}^{m_1m_2m_3} \equiv \langle a_{\ell_1 m_1} a_{\ell_2 m_2} a_{\ell_3 m_3}\rangle\,.
\ee
Substituting the  expression (\ref{a_lm}) into the angular bispectrum, one can write it in the form
\be
\label{B_lm}
B_{\ell_1 \ell_2 \ell_3}^{m_1m_2m_3} ={\cal G}^{m_1m_2m_3}_{\ell_1\ell_2\ell_3}b_{\ell_1\ell_2\ell_3}\,,
\ee
where the first, purely geometrical,  factor is the  Gaunt integral
\be
{\cal G}^{m_1m_2m_3}_{\ell_1\ell_2\ell_3}\equiv \int d\Omega\,  Y_{\ell_1m_1}(\Omega)\,  Y_{\ell_2m_2}(\Omega)\,  Y_{\ell_3m_3}(\Omega)\,.
\ee
The second factor, usually called the {\it reduced} bispectrum, is given by 
\begin{eqnarray}
\label{reduced_bispectrum}
b_{\ell_1\ell_2 \ell_3} & = & \sum_{I,J,K}  \left(\frac{2}{\pi}\right)^3\int \left(\prod_{i=1}^3k_i^2 dk_i\right)  \ g^I_{\ell_1}(k_1) g^J_{\ell_2}(k_2) g^K_{\ell_3}(k_3) 
B^{IJK}(k_1,k_2, k_3) \nonumber\\
&& \times
\int_0^\infty r^2 dr j_{\ell_1}(k_1r) j_{\ell_2}(k_2 r) j_{\ell_3}(k_3 r)\,, 
\end{eqnarray}
which depends on  the bispectra of the primordial $X^I$:
\be
\label{B_IJK}
\langle X^{I}(\mathbf{k}_1) X^J(\mathbf{k}_2) X^{K}(\mathbf{k}_3) \rangle \equiv  (2 \pi)^3 \delta (\Sigma_i \mathbf{k}_i) B^{IJK}(k_1, k_2, k_3)\,.
 \ee
 The reduced bispectrum (\ref{reduced_bispectrum}) is  the sum of several contributions, corresponding to different values of  the indices $I$, $J$ and $K$ that vary over the range of  modes included in the primordial perturbations. This expression thus generalizes the purely adiabatic expression given in \citep{Komatsu:2001rj}. 
 
It is also useful to define the angle-averaged  bispectrum
 \ba
 B_{\ell_1 \ell_2 \ell_3} & \equiv & \sum_{m_1, m_2, m_3} 
  \left(
\begin{array}{ccc}
\ell_1 & \ell_2 & \ell_3 \cr
m_1 & m_2 & m_3
\end{array}
\right)
B_{\ell_1 \ell_2 \ell_3}^{m_1 m_2 m_3} \nonumber\\
& = & \sqrt{\frac{(2\ell_1+1)(2\ell_2+1)(2\ell_3+1)}{4\pi}}  \left(
\begin{array}{ccc}
\ell_1 & \ell_2 & \ell_3 \cr
0 & 0 & 0
\end{array}
\right)b_{\ell_1 \ell_2 \ell_3}\,,
 \ea
 where the second relation is obtained by substituting (\ref{B_lm}) and by using the identity
 \be
 \sum_{m_1, m_2, m_3} 
  \left(
\begin{array}{ccc}
\ell_1 & \ell_2 & \ell_3 \cr
m_1 & m_2 & m_3
\end{array}
\right)
{\cal G}^{m_1m_2m_3}_{\ell_1\ell_2\ell_3}=\sqrt{\frac{(2\ell_1+1)(2\ell_2+1)(2\ell_3+1)}{4\pi}}  \left(
\begin{array}{ccc}
\ell_1 & \ell_2 & \ell_3 \cr
0 & 0 & 0
\end{array}
\right)
\,.
\ee

\subsection{Non-Gaussianities of local type}
 
To proceed further, one must make some assumption about the functional dependence of the 
bispectra $B^{IJK}(k_1, k_2, k_3)$ in Fourier space. This corresponds to the so-called 
``shape'' of the bispectrum~\citep{Babich:2004gb}, 
which has been discussed at length 
 in the literature
in the  purely adiabatic case where the $B^{IJK}$ reduce to the single bispectrum $B^{\zeta\zeta\zeta}$. 
In the present work, we consider the simplest form of non-Gaussianity, namely  the local shape. 
In the purely adiabatic case, it is defined by 
\be
\zeta(\mathbf{x})=\zeta_L(\mathbf{x})-\frac{3}{5}f_{\rm NL}
\left(\zeta_L(\mathbf{x})^2-\langle \zeta_L\rangle^2\right)\,,
\ee
in physical space, where the factor $-3/5$ appears because $f_{\rm NL}$ was originally defined with respect to the gravitational potential $\Phi=-(3/5)\zeta$, instead of $\zeta$. 
The subscript $L$ here denotes the linear part of the perturbation, which is assumed to be Gaussian.

In Fourier space, this leads to a bispectrum that depends quadratically on the power spectrum:
\be
\label{B_adiab}
B^{\zeta\zeta\zeta}= \tf_{\rm NL} \left[ P_\zeta(k_2)P_\zeta(k_3)+P_\zeta(k_3)P_\zeta(k_1)+P_\zeta(k_1)P_\zeta(k_2)\right], \qquad \tf_{\rm NL}\equiv-\frac65 f_{\rm NL}\,.
\ee
In the present context where we assume the presence of an isocurvature mode in addition to the dominant adiabatic mode, the simplest extension of (\ref{B_adiab}) is to assume that all the 
generalized  bispectra $B^{IJK}(k_1, k_2, k_3)$ can be written  as the sum of terms  quadratic in the adiabatic power spectrum (note that this implicitly assumes that the power spectrum of the isocurvature mode 
and the isocurvature cross power spectrum,
 if non-vanishing, 
 have the same spectral dependence as the adiabatic one). However, in contrast with (\ref{B_adiab}) where all terms share the same coefficient, as a consequence of the invariance of the bispectrum under 
the exchange of momenta, this is no longer the case for the generalized bispectra when the  indices $I$, $J$ and $K$ are not identical. What  the definition (\ref{B_IJK}) implies  is simply that the bispectra are left unchanged under  the {\it simultaneous} change of two indices and the corresponding momenta (e.g.\ $I$ and $J$, $k_1$ and $k_2$). 
This leads to the decomposition
\begin{eqnarray}
\label{bispectrum_local}
 B^{IJK}(k_1, k_2, k_3)=
 \tf_{\rm NL}^{I, JK}  P_\zeta(k_2) P_\zeta(k_3) 
 +\tf_{\rm NL}^{J, KI}  P_\zeta(k_1) P_\zeta(k_3)
+\tf_{\rm NL}^{K, IJ}   P_\zeta(k_1)P_\zeta(k_2)\,, 
  \end{eqnarray}
  where the coefficients $\tf_{\rm NL}^{I, JK}$ must satisfy the condition 
\be
\label{f_sym}
\tf_{\rm NL}^{I, JK} =\tf_{\rm NL}^{I, KJ} \,.
\ee
To keep track of  this symmetry, we  separate the first index from the last two indices with a comma.

\subsection{Link with multiple-field inflation}

It is instructive to show that our definition of generalized local non-Gaussianity is the natural outcome of a generic model of multiple-field inflation. 
Indeed, allowing for several light degrees of freedom during inflation, one can relate, in a very generic way, 
the ``primordial'' perturbations $X^I$ (defined during the standard radiation era)  to the fluctuations of light primordial fields $\phi^a$, generated  at Hubble crossing during inflation, so that one can write, up to second order, 
\be
\label{X_I}
X^I= N^I_a\,  \delta\phi^a+\frac12 N^{I}_{ab}\,  \delta\phi^a \delta\phi^b + \dots
\ee
where the $\delta\phi^a$ can usually be treated as  independent quasi-Gaussian fluctuations, i.e. 
\be
\langle \delta\phi^a (\mathbf{k}) \, \delta\phi^b (\mathbf{k}')\rangle=
 (2\pi)^3 \, \delta^{ab}P_{\delta\phi}(k) \, \delta(\mathbf{k} + \mathbf{k}')\,, \qquad P_{\delta\phi}(k)=2\pi^2k^{-3}\left(\frac{H_*}{2\pi}\right)^2\,,
 \ee
 where  a star denotes Hubble crossing time. The relation (\ref{X_I}) is very general, and all the details of the inflationary model are embodied by the coefficients $N_a^I$ and $N_{ab}^I$.

 Substituting (\ref{X_I}) into (\ref{B_IJK}) and using Wick's theorem, one finds that the bispectra $B_{IJK}$ can be expressed  in  the form
\be
\label{bispectrum_fields}
 B^{IJK}(k_1, k_2, k_3)=
 \lambda^{I, JK}  P_{\delta\phi}(k_2) P_{\delta\phi}(k_3) 
 +\lambda^{J, KI}  P_{\delta\phi}(k_1) P_{\delta\phi}(k_3)
+\lambda^{K, IJ}   P_{\delta\phi}(k_1)P_{\delta\phi}(k_2)\,, 
\ee
with the coefficients 
\be
\label{lambda}
\lambda^{I, JK} \equiv \delta^{ac}\delta^{bd}N^I_{ab} N^J_{c} N^K_{d}
\ee
 (the summation over scalar field indices $a$, $b$, $c$ and $d$  is implicit), which are  symmetric under the interchange  of the last two indices, by construction. Since the 
 adiabatic power spectrum is given by
 \be
 P_\zeta=(\delta^{ab} N_a^\zeta N_b^\zeta) P_{\delta\phi} \equiv A P_{\delta\phi},
 \ee
 one obtains finally (\ref{bispectrum_local}) with
 \be
 \label{f_NL}
\tf_{\rm NL}^{I,JK}= \lambda_{NL}^{I, JK} /A^2\,,
\ee
where it is implicitly assumed that the coefficients $N^I_a$ are weakly time dependent so that the scale dependence of $A^2$ can be neglected.  One can notice that the first index is related to the second-order terms in the decomposition (\ref{X_I}), while  the last two indices come  from   the first-order terms. 

Except in the last section devoted to a specific class of early universe models, all our considerations will simply follow from our assumption (\ref{bispectrum_local}) and thus apply to any model leading to this local form, whether based on inflation or not.

\subsection{Decomposition of the angular bispectrum}
After substitution of (\ref{bispectrum_local}) into (\ref{reduced_bispectrum}), the reduced bispectrum can 
finally 
be written as 
\be
\label{b_I}
b_{\ell_1\ell_2 \ell_3}=   \sum_{I,J,K}\tf_{\rm NL}^{I,JK}b_{\ell_1\ell_2 \ell_3}^{I,JK},
\ee
where each contribution is of the form\footnote{We use the standard notation: $(\ell_1 \ell_2 \ell_3)\equiv [\ell_1\ell_2\ell_3+ 5\,  {\rm perms}]/3!$.}
\begin{eqnarray}
\label{b_IJK}
b_{\ell_1\ell_2 \ell_3}^{I,JK}= 3   \int_0^\infty r^2 dr \, \alpha^I_{(\ell_1}(r)\beta^{J}_{\ell_2}(r)\beta^{K}_{\ell_3)}(r),
\end{eqnarray}
with   
\begin{eqnarray}
\label{alpha}
\alpha^I_{\ell}(r)&\equiv& \frac{2}{\pi} \int k^2 dk\,   j_\ell(kr) \, g^I_{\ell}(k),
\\
\label{beta}
\beta^{I}_{\ell}(r)&\equiv& \frac{2}{\pi}  \int k^2 dk \,  j_\ell(kr) \, g^I_{\ell}(k)\,  P_\zeta(k)\,.
\end{eqnarray}
While we have omitted the polarization indices, the reader should keep in mind
that each transfer function carries, in addition to the isocurvature index,
a polarization index, and hence the same is true for $\alpha_\ell$ and $\beta_\ell$.
As a consequence, the bispectrum has three polarization indices that we do not
show.

\begin{figure}
\centering
\includegraphics[width=0.49\textwidth, clip=true]{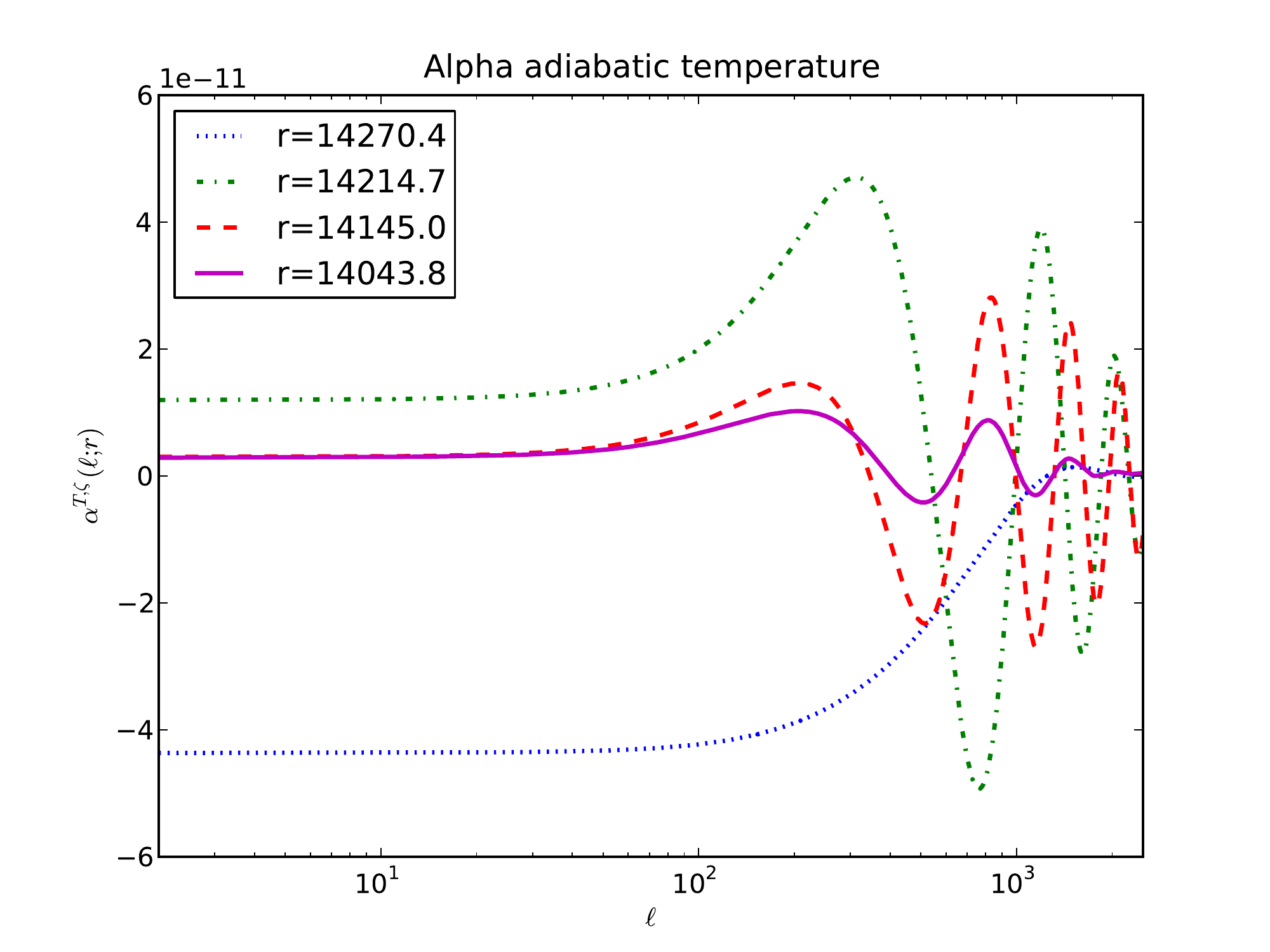}
\includegraphics[width=0.49\textwidth, clip=true]{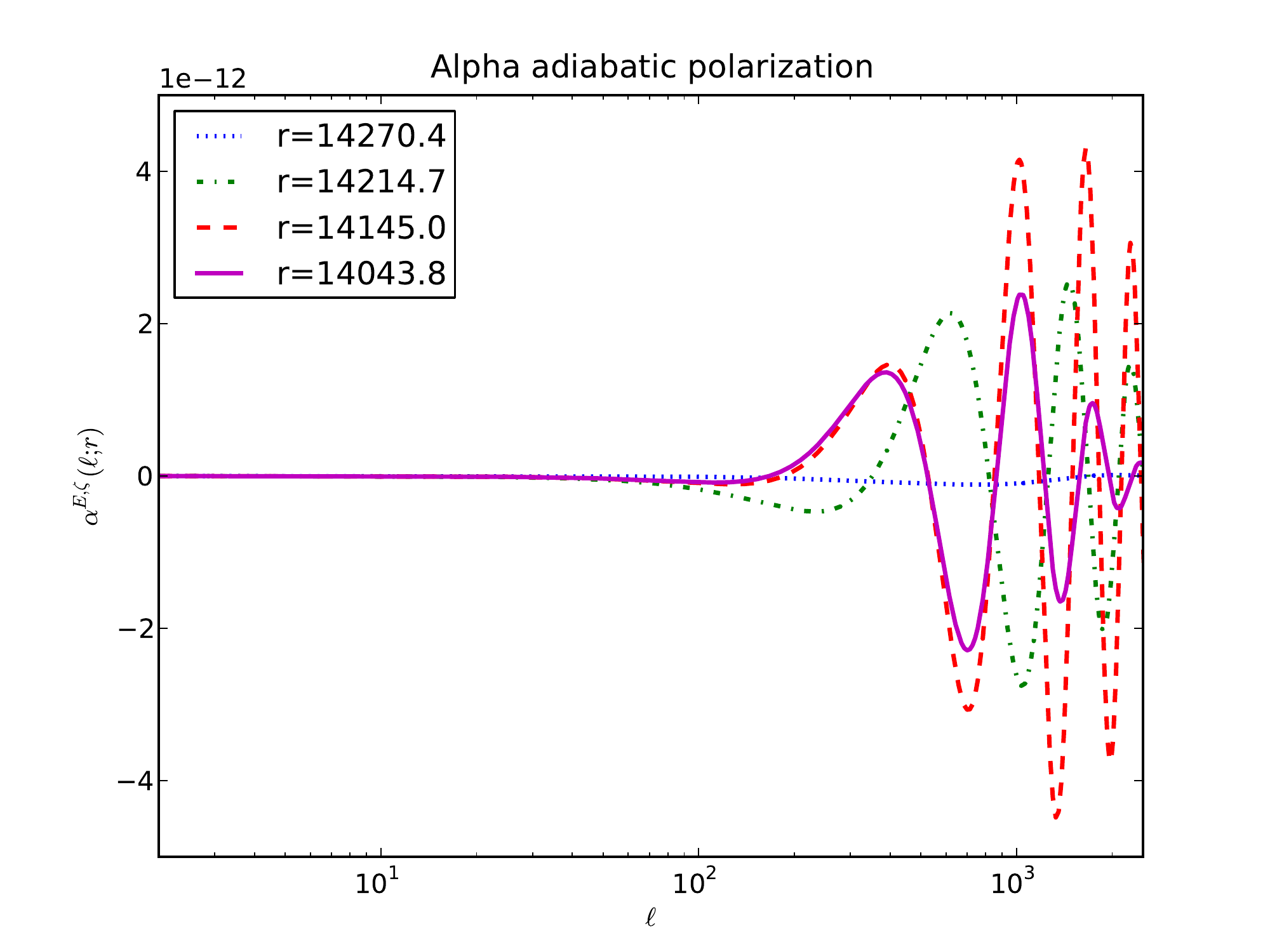}
\caption{The adiabatic $\alpha^{\zeta}_\ell(r)$ as a function of $\ell$ for 
temperature (left) and polarization (right). It has been evaluated at
four different values of $r$: 14043.8, 14145.0, 14214.7, and 14270.4.}
\label{alpha_a}
\end{figure}

\begin{figure}
\centering
\includegraphics[width=0.49\textwidth, clip=true]{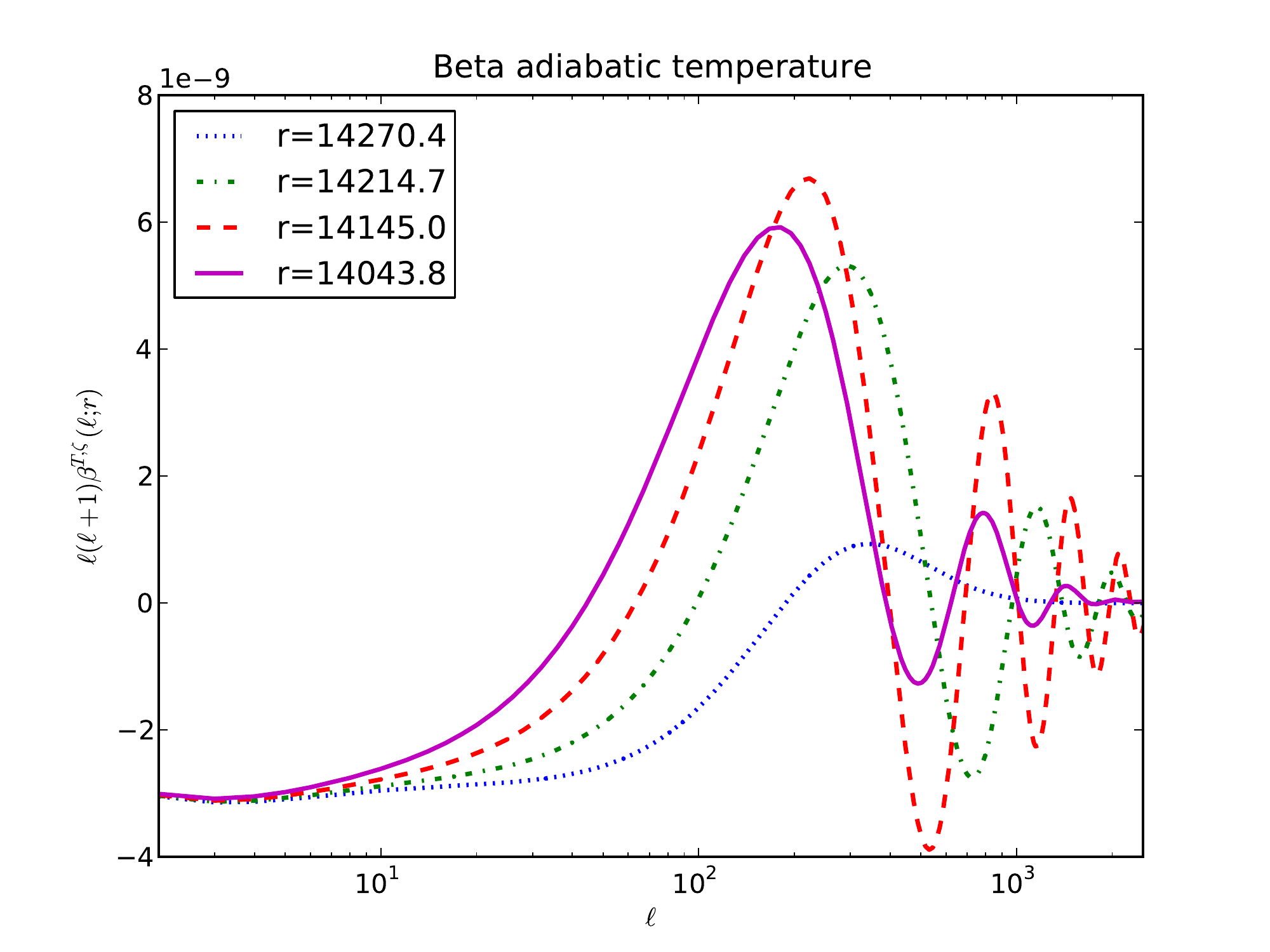}
\includegraphics[width=0.49\textwidth, clip=true]{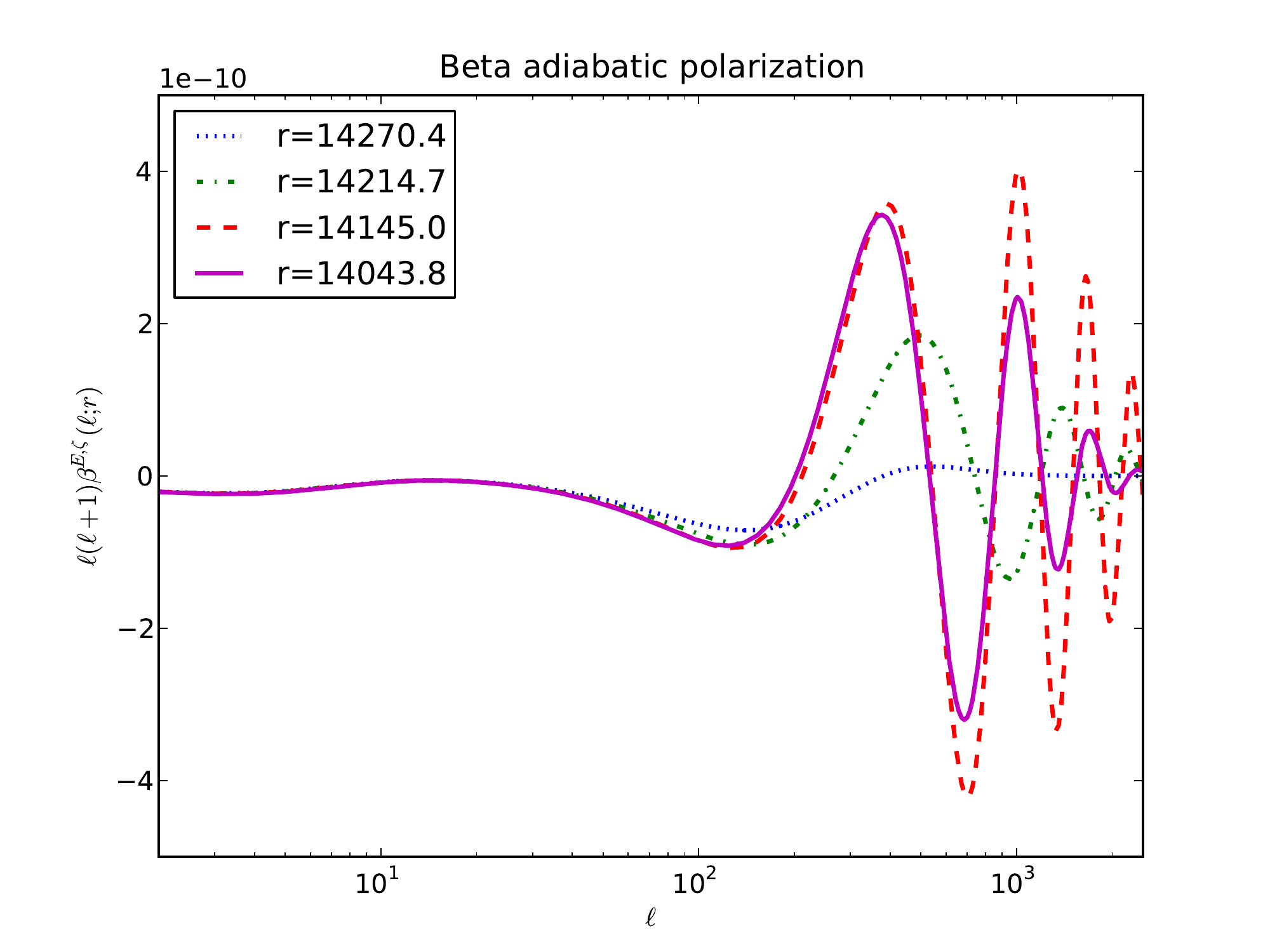}
\caption{The adiabatic $\ell(\ell+1)\beta^\zeta_\ell(r)$ as a function of $\ell$ for 
temperature (left) and polarization (right) (for the same values of  
$r$ as in Fig.~\ref{alpha_a}).}
\label{beta_a}
\end{figure}

The purely adiabatic bispectrum, usually the only one considered, can be expressed as 
\begin{eqnarray}
\!\!\!\!b^{\zeta,\zeta\zeta}_{\ell_1\ell_2 \ell_3}&\!\!\!\!=& \!\!\!\! 3 \int_0^\infty r^2 dr \alpha^\zeta_{(\ell_1}(r)\beta^{\zeta}_{\ell_2}(r)\beta^{\zeta}_{\ell_3)}(r)
\cr
&\!\!\!\!=& \!\!\!\!  \int_0^\infty r^2 dr \left[\alpha^\zeta_{\ell_1}(r)\beta^{\zeta}_{\ell_2}(r)\beta^\zeta_{\ell_3}(r)+\alpha^\zeta_{\ell_2}(r)\beta^\zeta_{\ell_3}(r)\beta^\zeta_{\ell_1}(r)+\alpha^\zeta_{\ell_3}(r)\beta^\zeta_{\ell_1}(r)\beta^\zeta_{\ell_2}(r)\right]\!\!.
\label{b_adiab}
\end{eqnarray}
The functions $\alpha^\zeta_{\ell}(r)$ and $\beta^\zeta_\ell(r)$ are plotted respectively in Fig.~\ref{alpha_a} and Fig.~\ref{beta_a}. We have considered both the temperature and polarization transfer functions. 
The radial distance $r$ can
be expressed as the speed of light times the difference in conformal time
between now and the time in the past we consider. In the figures we have chosen 
some sample values around the time of last scattering, which corresponds to
$r \approx 14100$~Mpc using the WMAP7-only best-fit parameters.

Since we consider local non-Gaussianity, the main contribution to the
bispectrum comes from the squeezed limit, i.e.\ one of the multipole
numbers is much smaller than the other two. To simplify the analysis,
let us assume that $\ell_1 \ll \ell_2 = \ell_3 \equiv L$. One finds that, in
this limit, the integrand in (\ref{b_adiab}) is dominated by (twice)
$\alpha^\zeta_{L}(r)\beta^\zeta_{L}(r)\beta^\zeta_{\ell_1}(r)$, whereas
the first term is negligible: 
one can see from Fig.~\ref{alpha_a} and \ref{beta_a} that $\beta^\zeta_{\ell_1}(r)$
is much larger (in absolute value) than $\alpha^\zeta_{\ell_1}(r)$.
This is true both for temperature and polarization (one should also keep
in mind that the $\beta$'s have been multiplied by $\ell(\ell+1)$ in the plots,
which makes them look much larger at large $\ell$). 

\begin{figure}
\centering
\includegraphics[width=0.49\textwidth, clip=true]{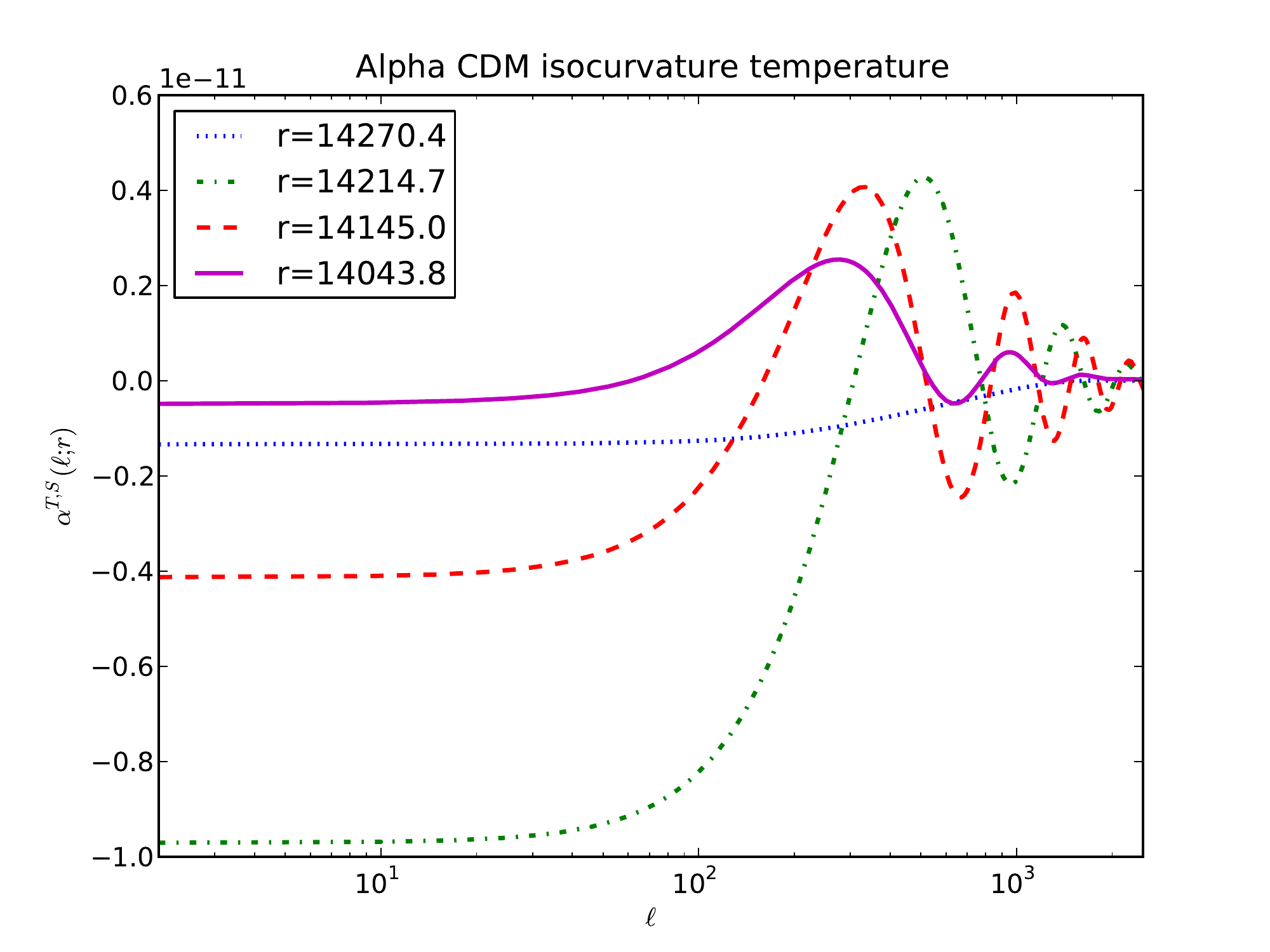}
\includegraphics[width=0.49\textwidth, clip=true]{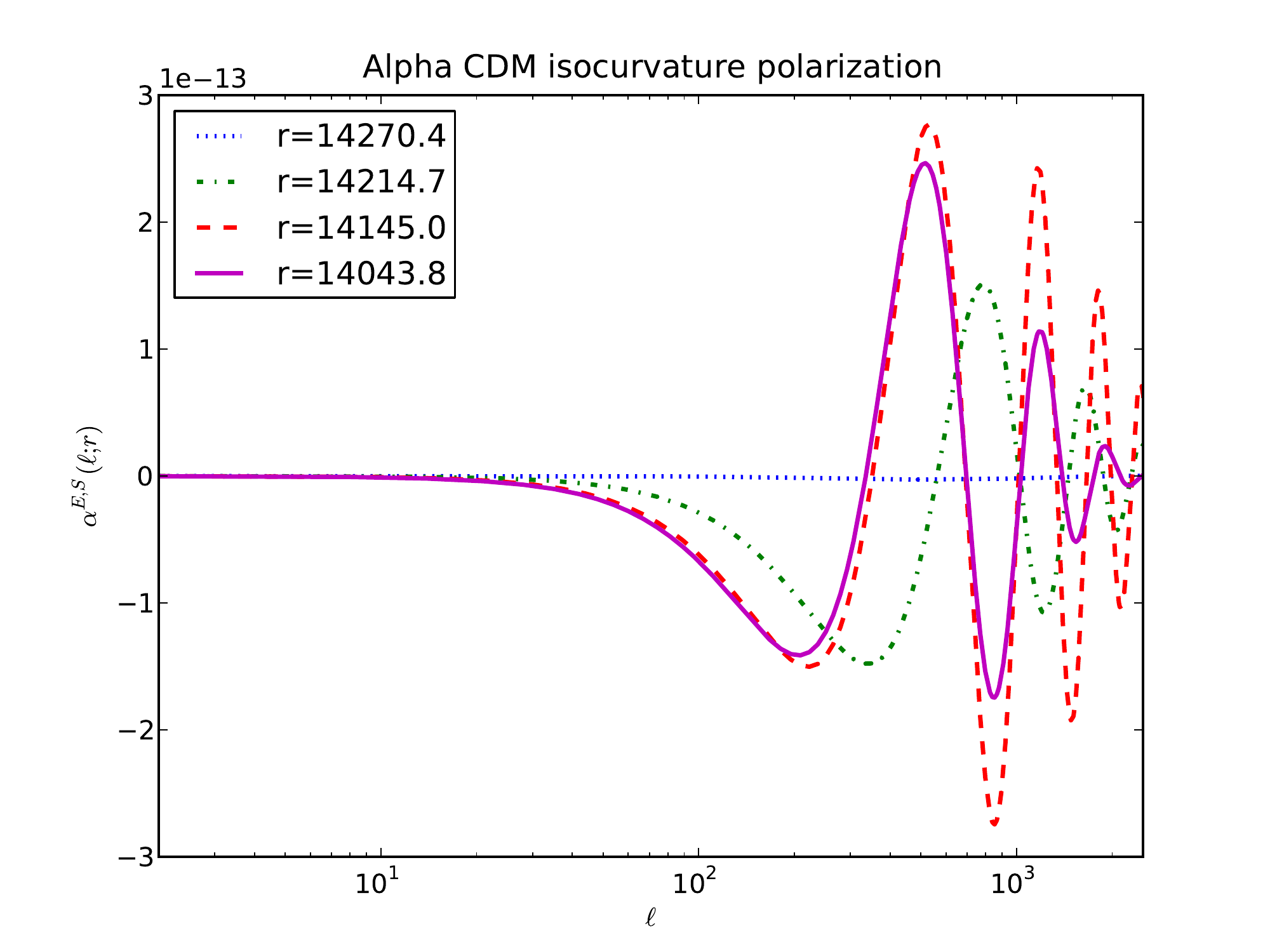}
\caption{The CDM isocurvature $\alpha^{S_c}_\ell(r)$ as a function of $\ell$ 
for temperature (left) and polarization (right) (for the same values of  
$r$ as in Fig.~\ref{alpha_a}).}
\label{alpha_cdm}
\end{figure}

\begin{figure}
\centering
\includegraphics[width=0.49\textwidth, clip=true]{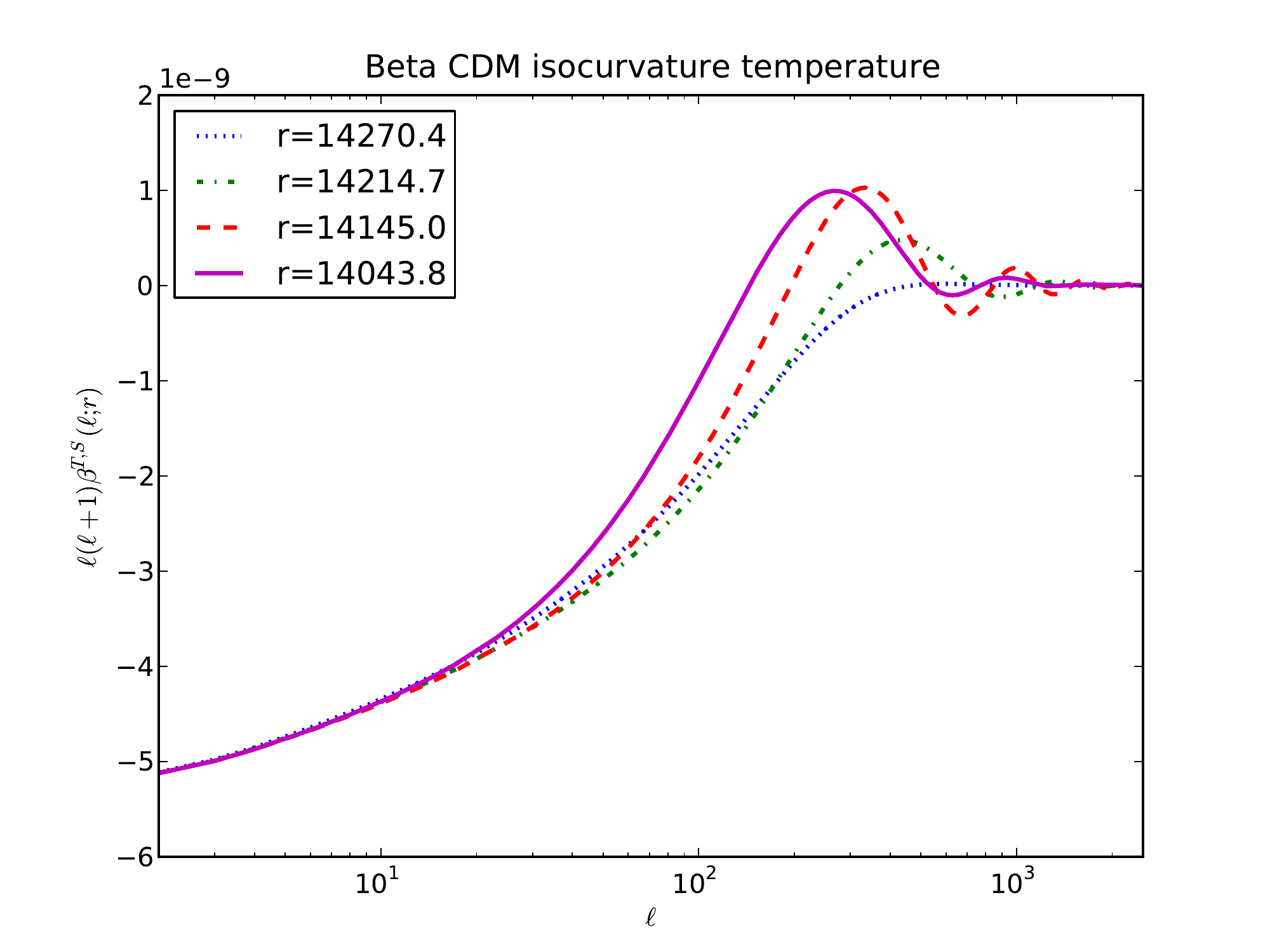}
\includegraphics[width=0.49\textwidth, clip=true]{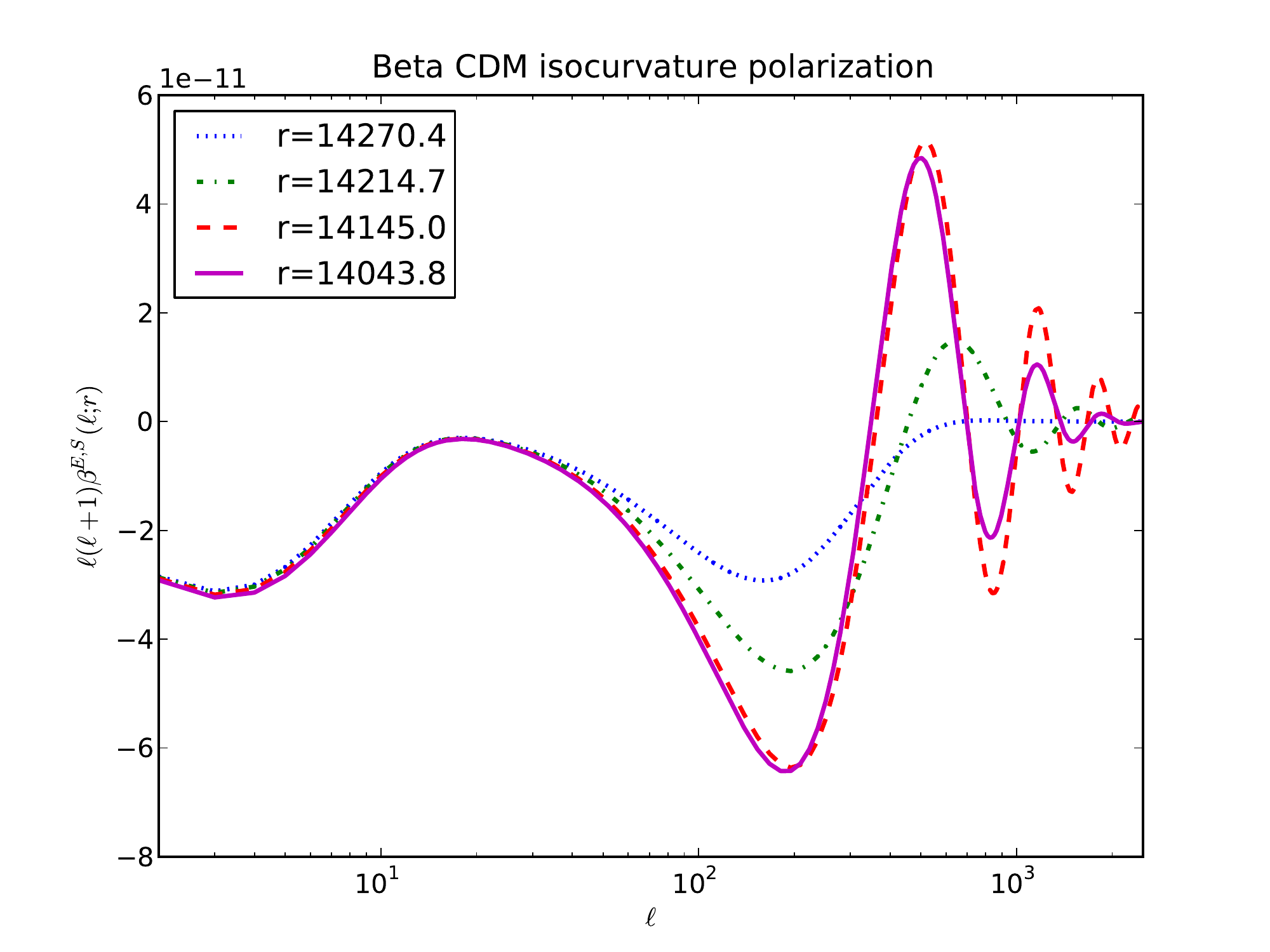}
\caption{The CDM isocurvature $\ell(\ell+1)\beta^{S_c}_\ell(r)$ as a function of $\ell$ 
for temperature (left) and polarization (right) (for the same values of  
$r$ as in Fig.~\ref{alpha_a}).}
\label{beta_cdm}
\end{figure}

\begin{figure}
\centering
\includegraphics[width=0.49\textwidth, clip=true]{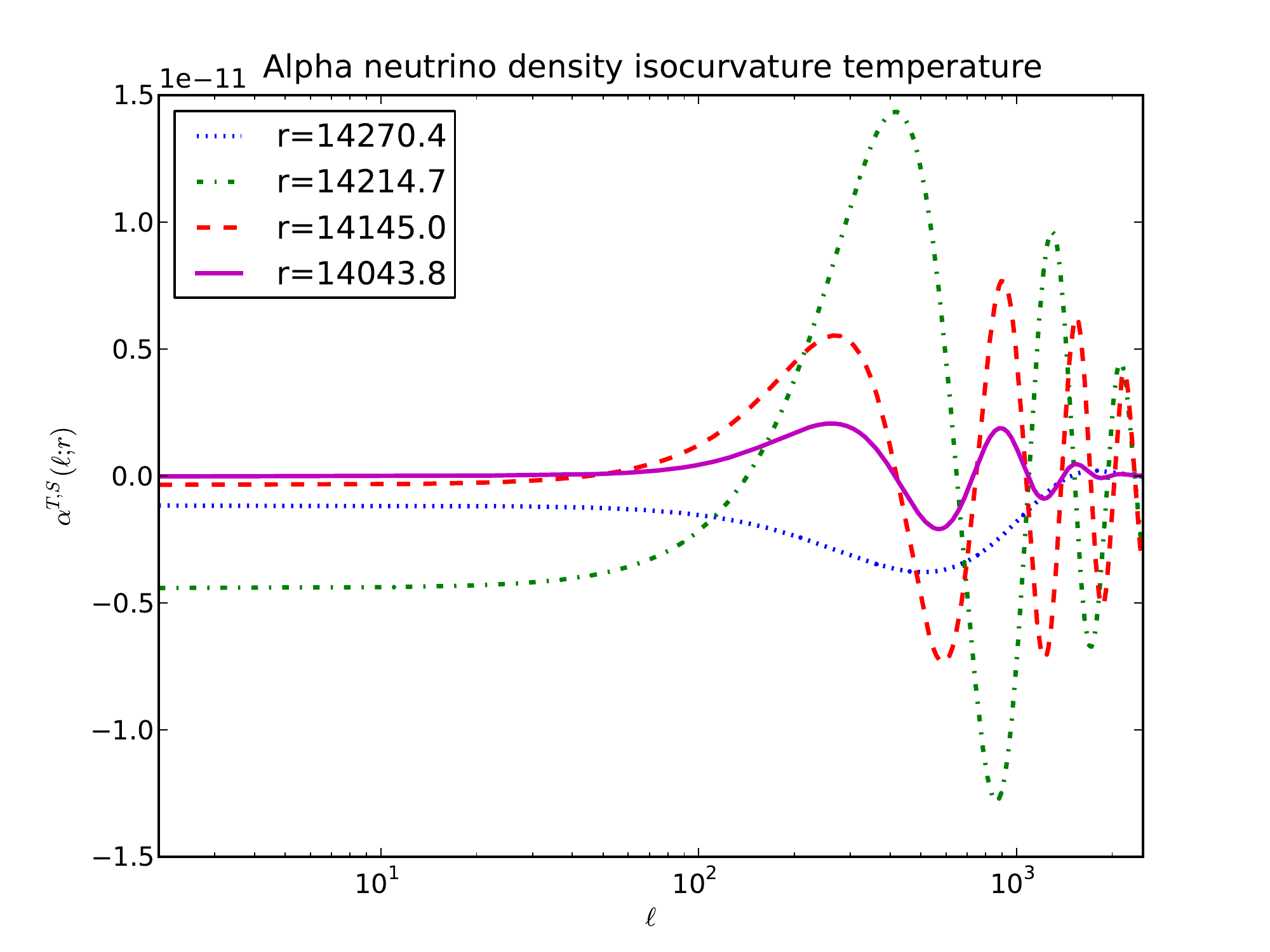}
\includegraphics[width=0.49\textwidth, clip=true]{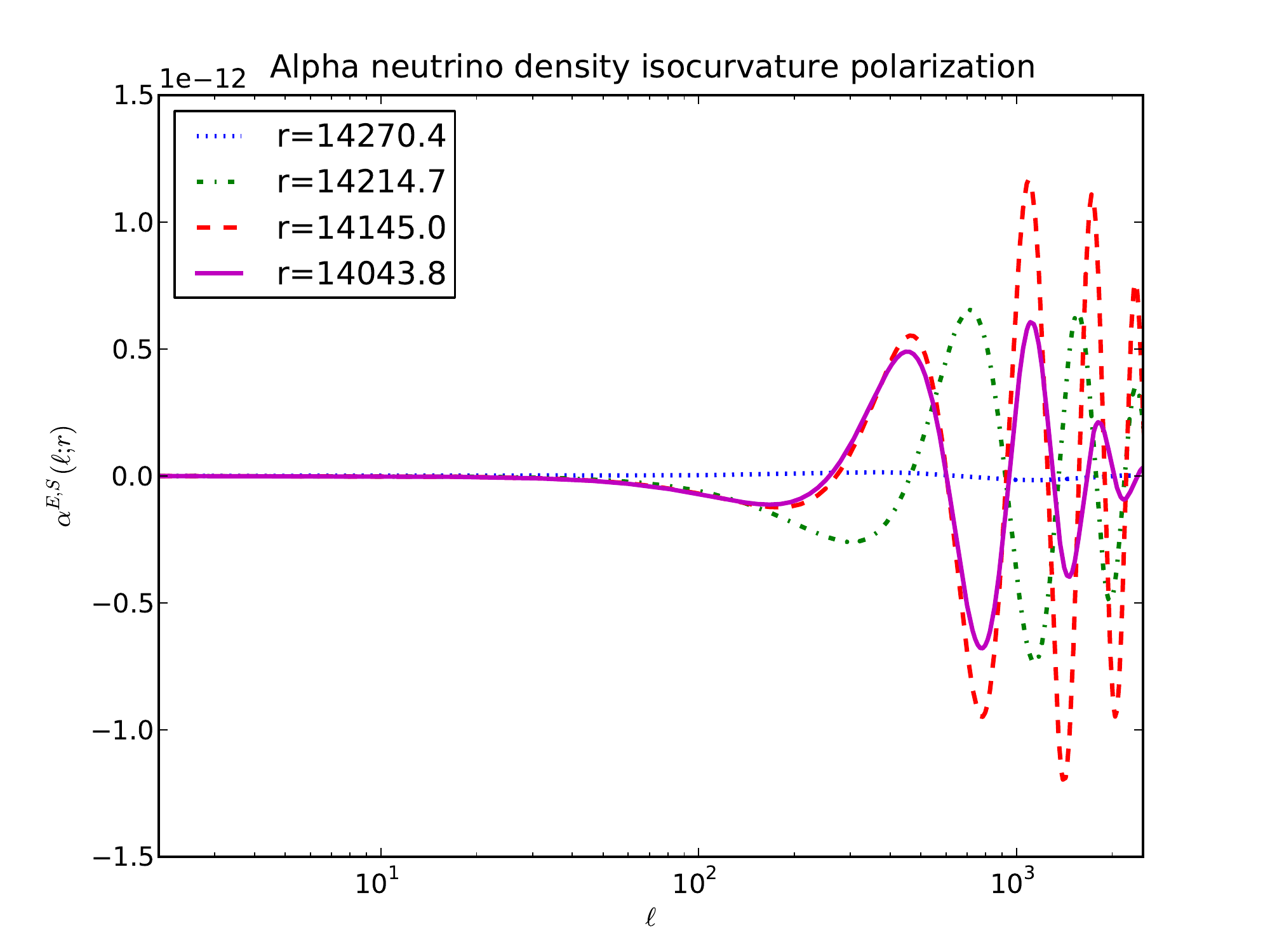}
\caption{The neutrino density isocurvature $\alpha^{S_{\nu d}}_\ell(r)$ as a 
function of $\ell$ for temperature (left) and polarization (right) (for the 
same values of $r$ as in Fig.~\ref{alpha_a}).}
\label{alpha_nd}
\end{figure}

\begin{figure}
\centering
\includegraphics[width=0.49\textwidth, clip=true]{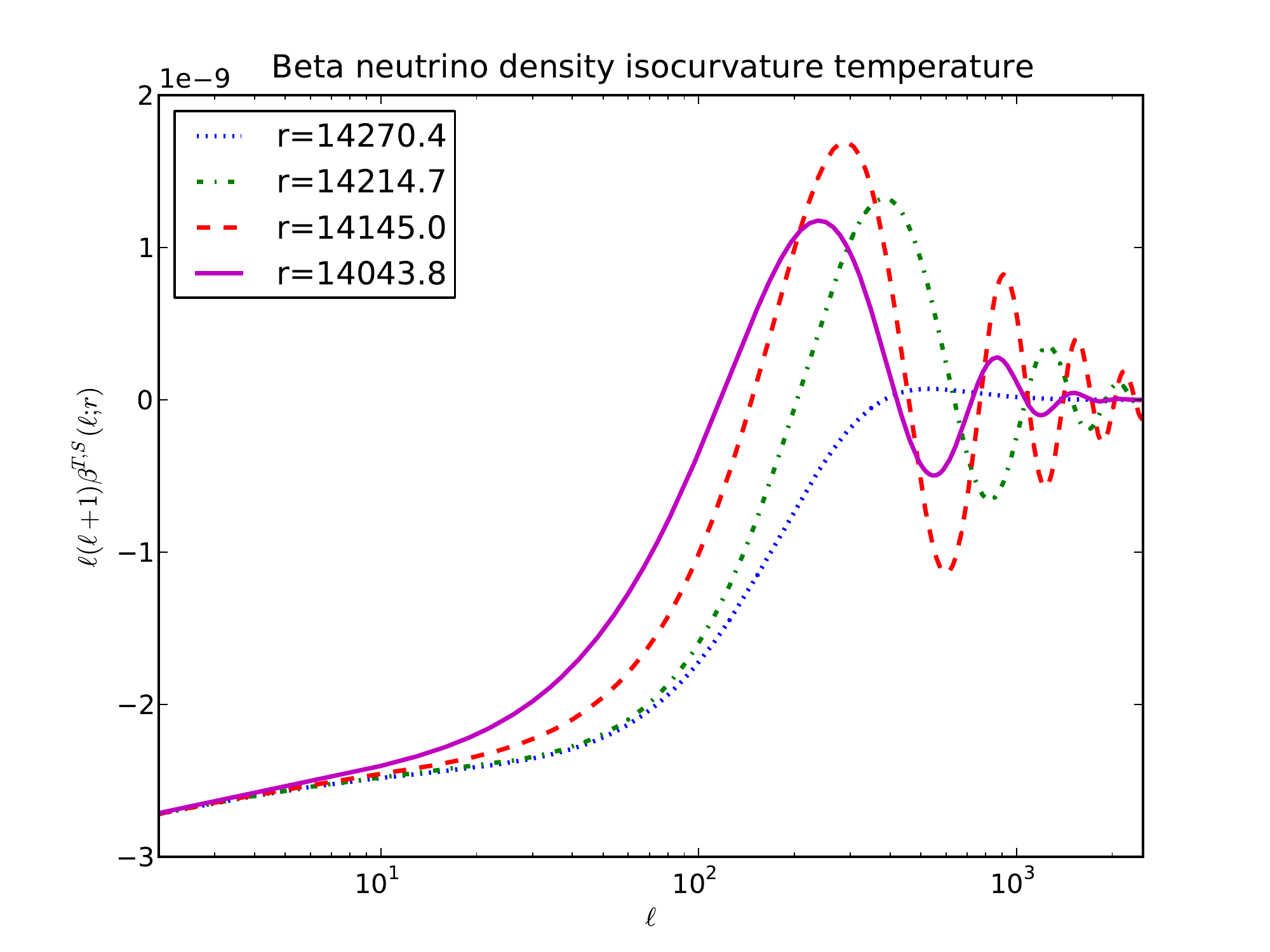}
\includegraphics[width=0.49\textwidth, clip=true]{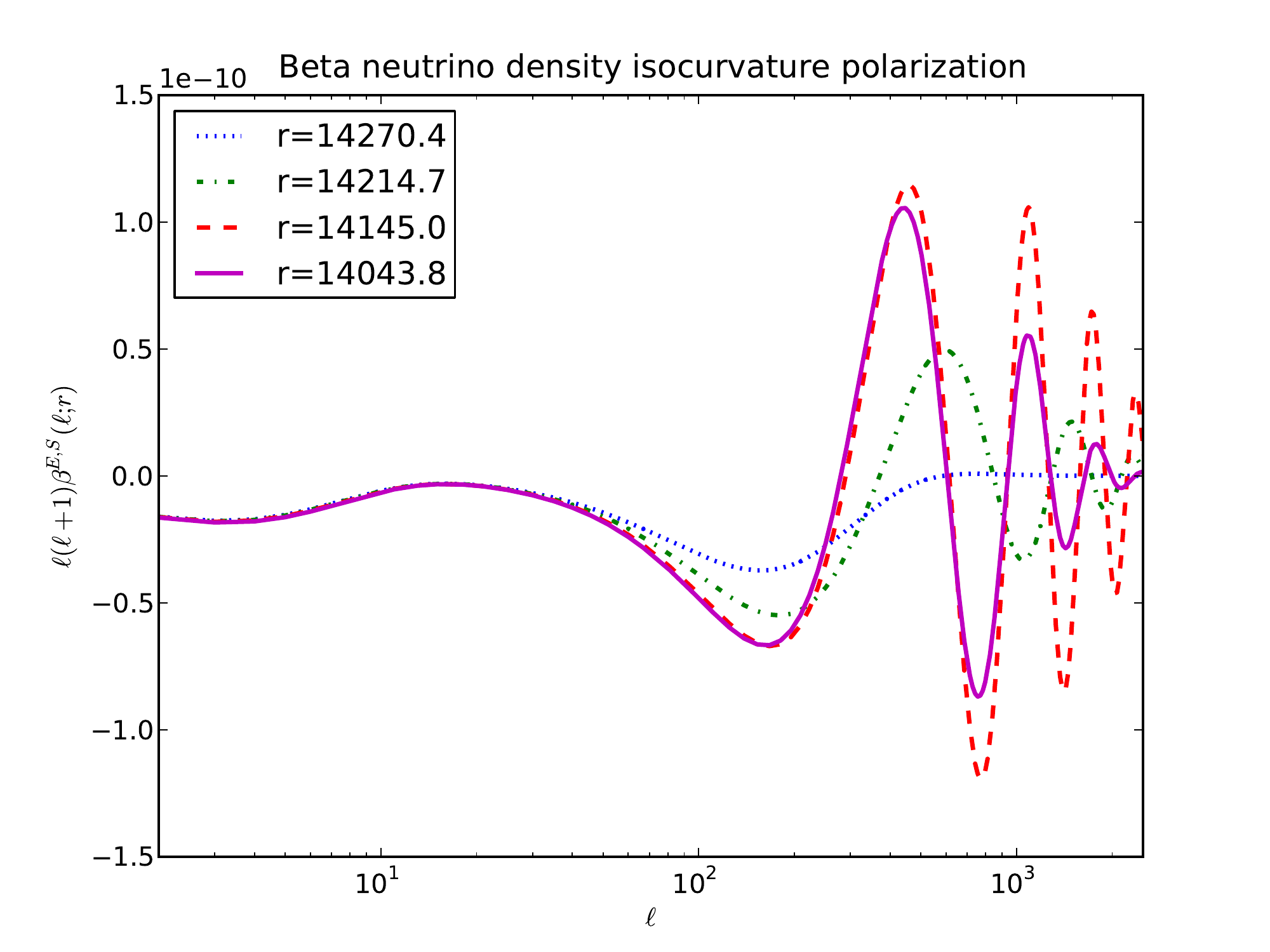}
\caption{The neutrino density isocurvature $\ell(\ell+1)\beta^{S_{\nu d}}_\ell(r)$ 
as a function of $\ell$ for temperature (left) and polarization (right) 
(for the same values of $r$ as in Fig.~\ref{alpha_a}).}
\label{beta_nd}
\end{figure}

 \begin{figure}
\centering
\includegraphics[width=0.49\textwidth, clip=true]{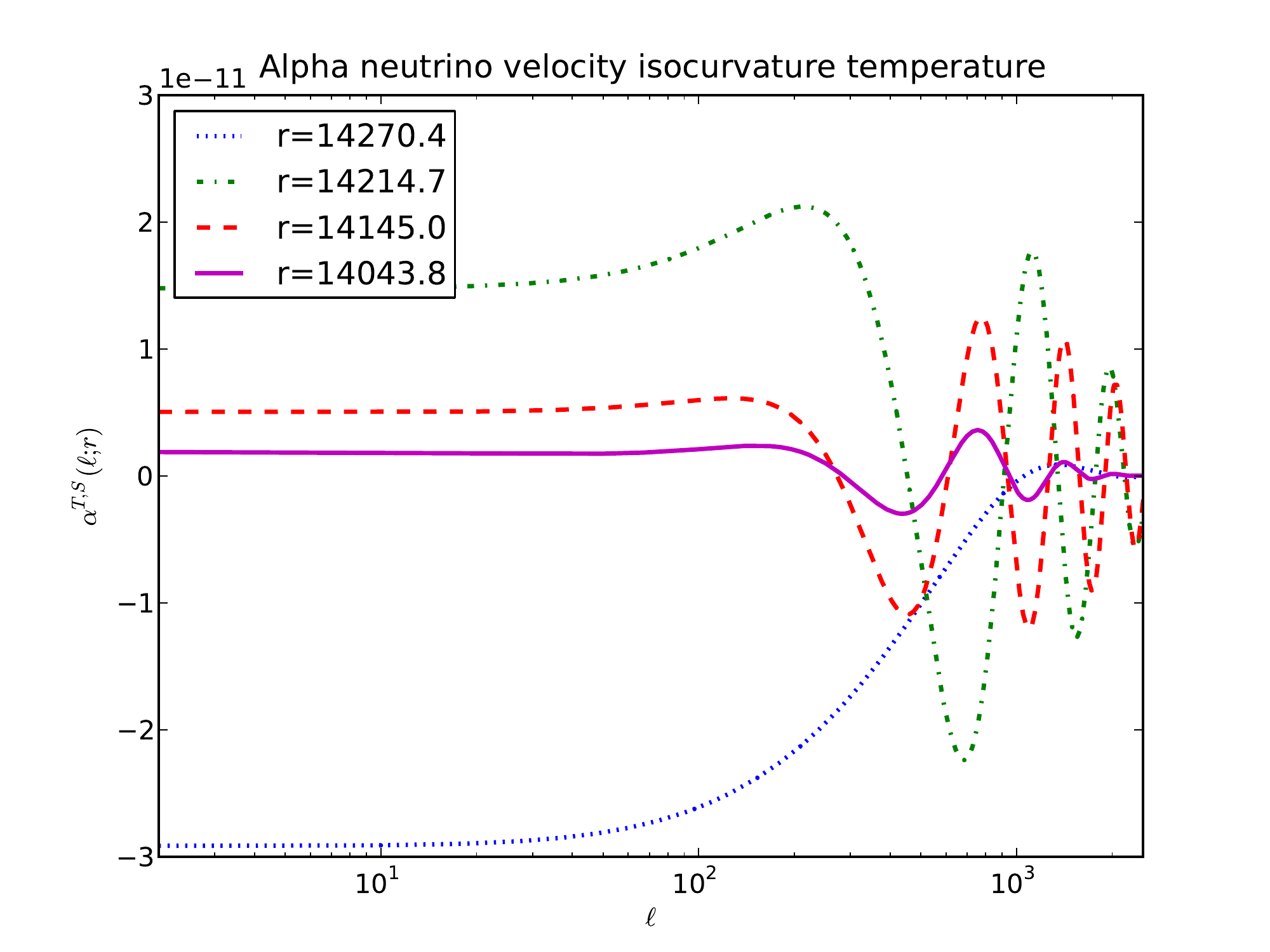}
\includegraphics[width=0.49\textwidth, clip=true]{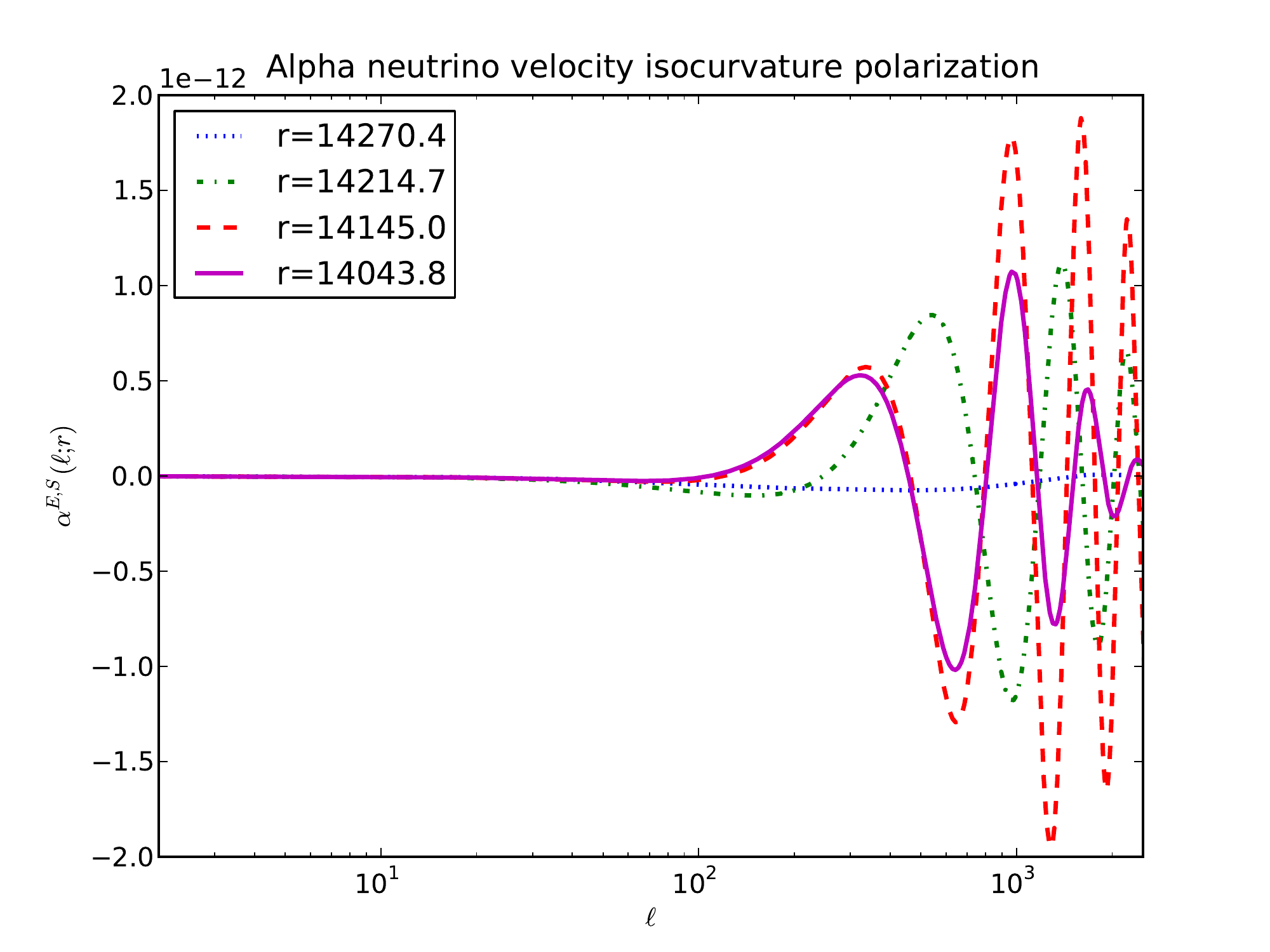}
\caption{The neutrino velocity isocurvature $\alpha^{S_{\nu v}}_\ell(r)$ as a 
function of $\ell$ for temperature (left) and polarization (right) (for the 
same values of $r$ as in Fig.~\ref{alpha_a}).}
\label{alpha_nv}
\end{figure}

\begin{figure}
\centering
\includegraphics[width=0.49\textwidth, clip=true]{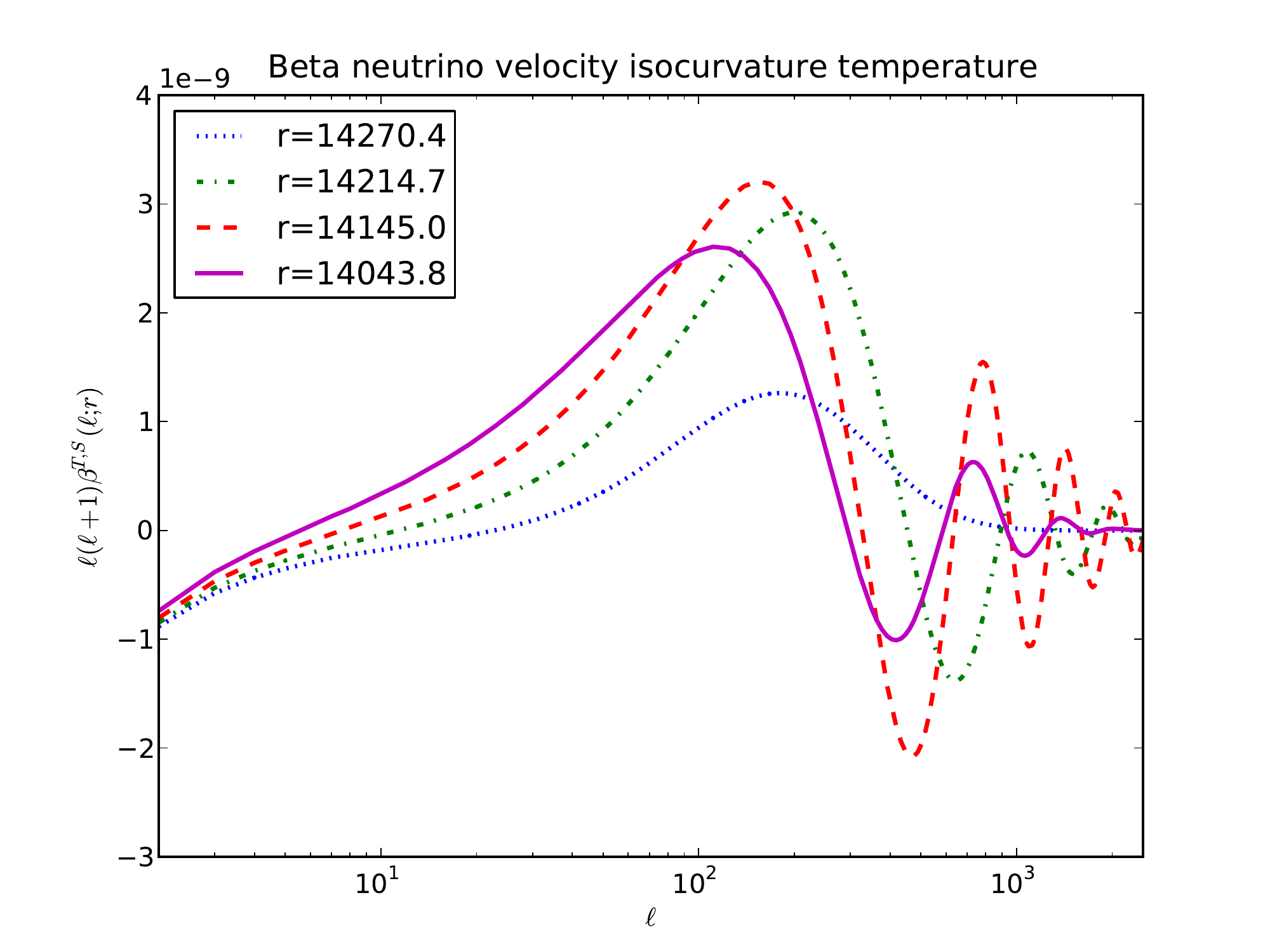}
\includegraphics[width=0.49\textwidth, clip=true]{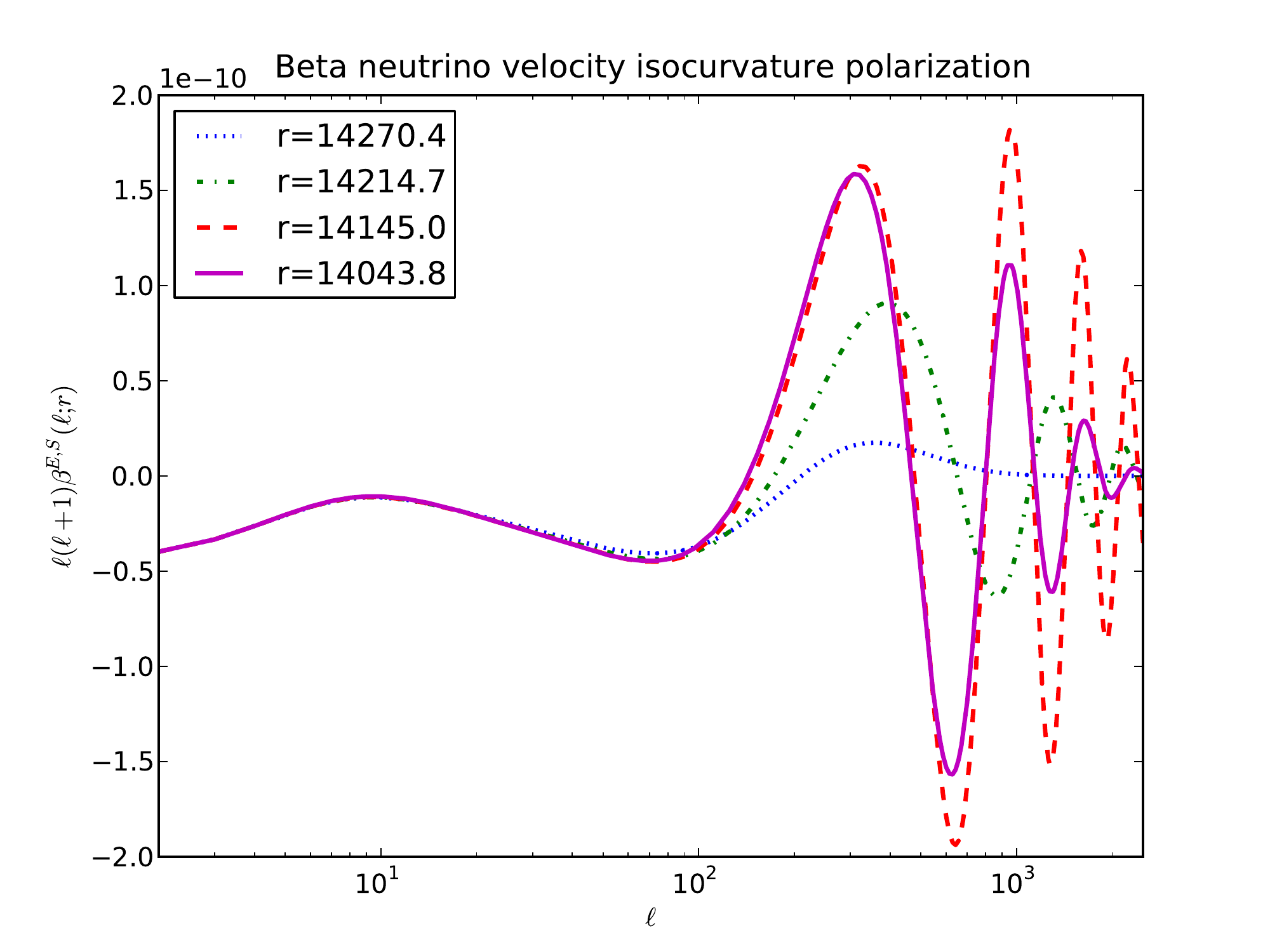}
\caption{The neutrino velocity isocurvature $\ell(\ell+1)\beta^{S_{\nu v}}_\ell(r)$ 
as a function of $\ell$ for temperature (left) and polarization (right) 
(for the same values of $r$ as in Fig.~\ref{alpha_a}).}
\label{beta_nv}
\end{figure}

The purely isocurvature bispectrum has exactly the same structure as (\ref{b_adiab}), but 
with the functions $\alpha^S_{\ell}(r)$ and $\beta^S_\ell(r)$ replacing $\alpha^\zeta_{\ell}(r)$ and $\beta^\zeta_\ell(r)$. Moreover, the shapes of $\alpha^S_{\ell}(r)$ and $\beta^S_\ell(r)$ depend on the type of isocurvature mode:
Fig.~\ref{alpha_cdm} and Fig.~\ref{beta_cdm} correspond to the CDM isocurvature mode, Fig.~\ref{alpha_nd} and Fig.~\ref{beta_nd}  to the neutrino density isocurvature mode and Fig.~\ref{alpha_nv} and Fig.~\ref{beta_nv}  to the neutrino velocity isocurvature mode.
The functions $\alpha^S_{\ell}(r)$ and $\beta^S_\ell(r)$ for  the baryon 
isocurvature mode can be deduced from the CDM isocurvature functions by a simple  rescaling, according to (\ref{omega_bc}):
\be 
\alpha^{S_b}_{\ell}(r)=\omega_{bc} \, \alpha^{S_c}_{\ell}(r), \qquad 
\beta^{S_b}_{\ell}(r)=\omega_{bc}\,  \beta^{S_c}_{\ell}(r)\,.
\ee

The other bispectra depend on a mixing of the adiabatic and isocurvature functions. For example, one finds 
\begin{eqnarray}
b^{\zeta, \zeta S}_{\ell_1\ell_2 \ell_3}+  b^{\zeta, S \zeta}_{\ell_1\ell_2 \ell_3}&\!\!\!\!=& \!\!\!\! 6 \int_0^\infty r^2 dr \alpha^\zeta_{(\ell_1}(r)\beta^{\zeta}_{\ell_2}(r)\beta^{S}_{\ell_3)}(r)
\cr
&\!\!\!\!=& \!\!\!\! \int_0^\infty r^2 dr \left\{\alpha^\zeta_{\ell_1}(r)
\left[\beta^{\zeta}_{\ell_2}(r)\beta^{S}_{\ell_3}(r)+\beta^{\zeta}_{\ell_3}(r)\beta^{S}_{\ell_2}(r)\right]
\right.
\\
&&\!\!\!\!
\left.
+
\alpha^\zeta_{\ell_2}(r)\left[\beta^{\zeta}_{\ell_3}(r)\beta^{S}_{\ell_1}(r)+\beta^{\zeta}_{\ell_1}(r)\beta^{S}_{\ell_3}(r)\right]
+\alpha^\zeta_{\ell_3}(r)\left[ \beta^{\zeta}_{\ell_1}(r)\beta^{S}_{\ell_2}(r)+ \beta^{\zeta}_{\ell_2}(r)\beta^{S}_{\ell_1}(r)\right]
\right\}.
\nonumber
\end{eqnarray}
Since $b^{\zeta, \zeta S}_{\ell_1\ell_2 \ell_3}$ and $b^{\zeta, S \zeta}_{\ell_1\ell_2 \ell_3}$ cannot be distinguished, we will always consider the sum of the two, and similarly for $b^{S, \zeta S}_{\ell_1\ell_2 \ell_3}$ and $b^{S, S \zeta}_{\ell_1\ell_2 \ell_3}$.

\begin{figure}
\centering
\includegraphics[width=0.49\textwidth, clip=true]{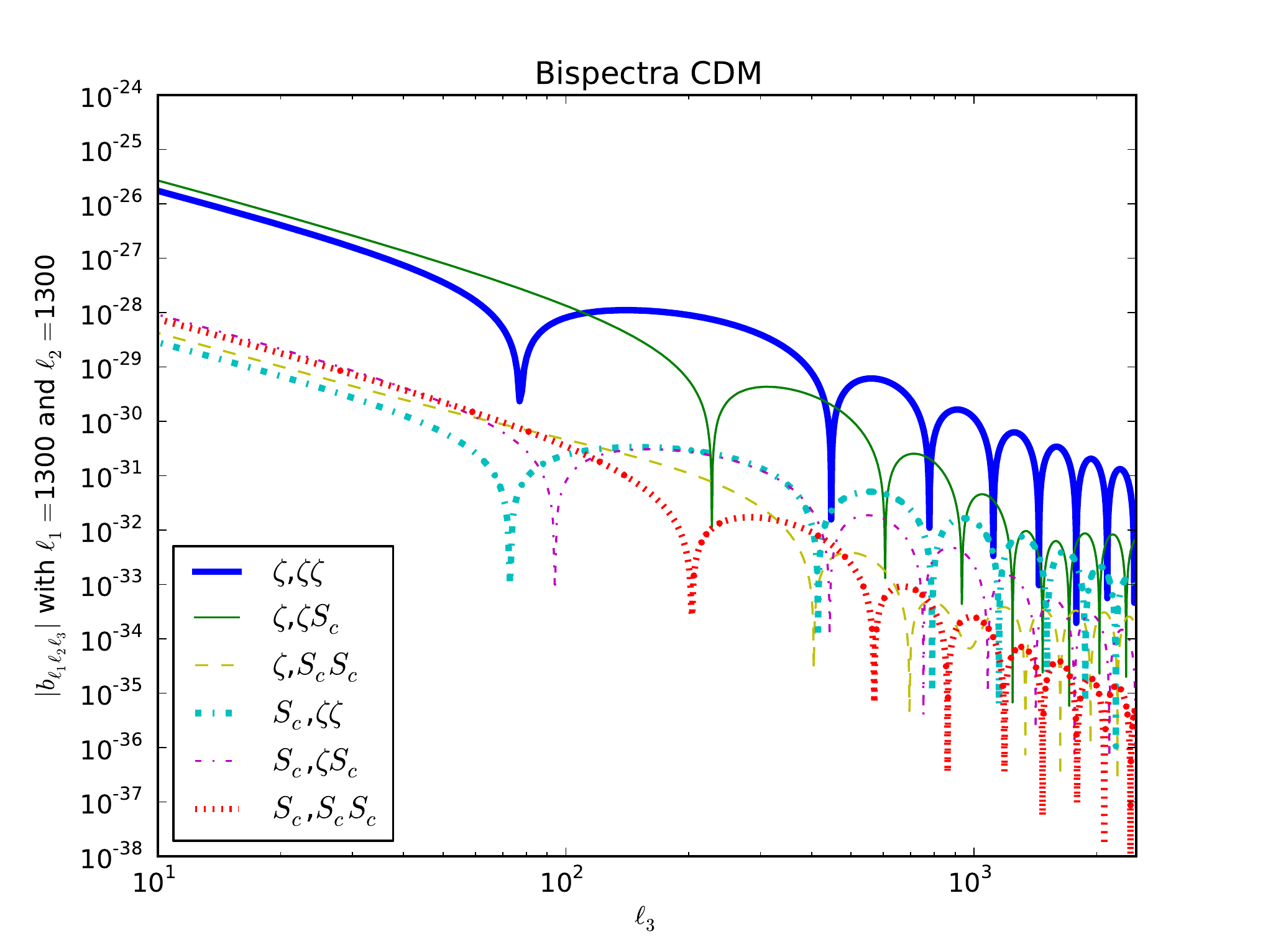}
\includegraphics[width=0.49\textwidth, clip=true]{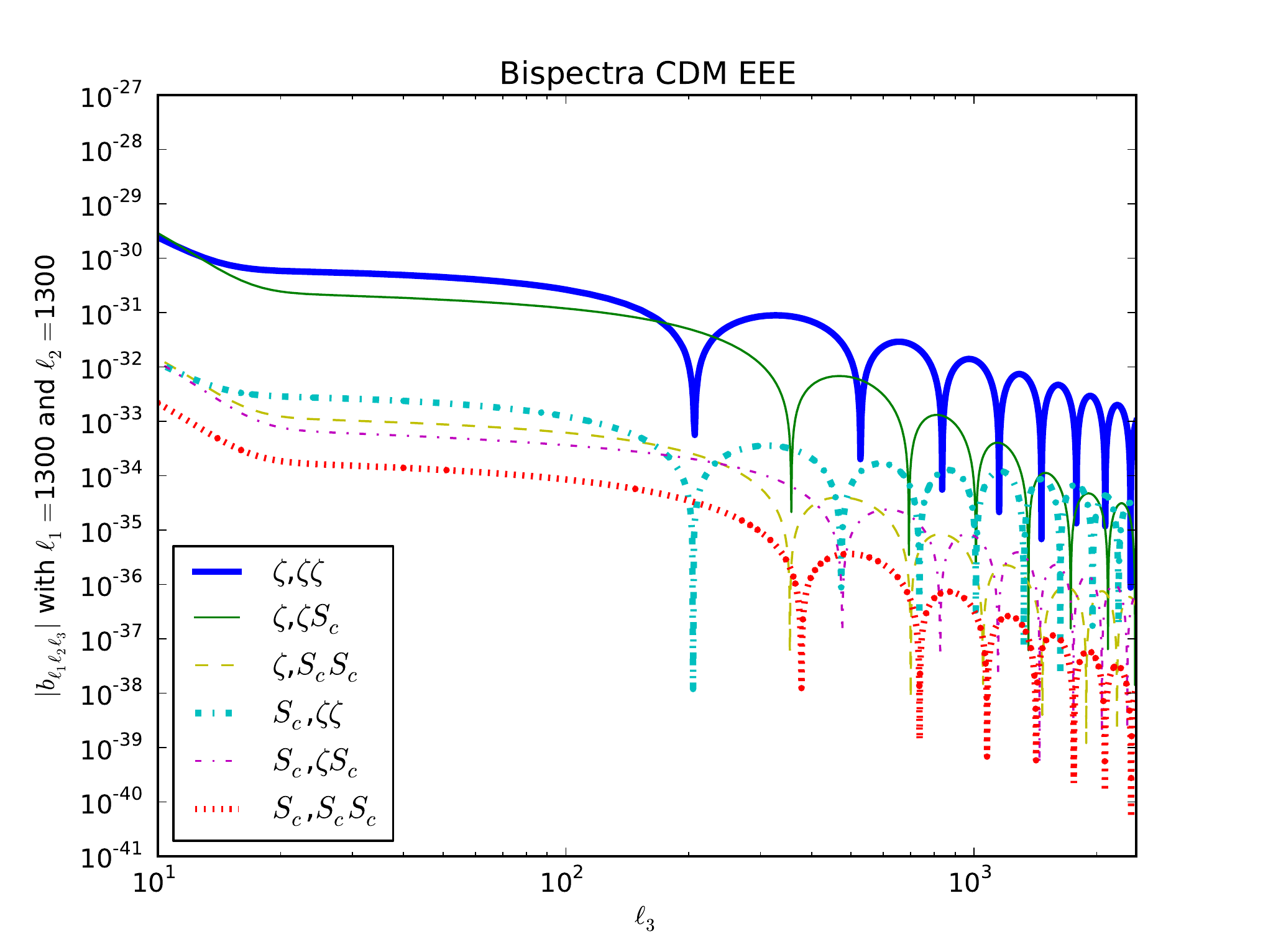}
\caption{Plot of $|b^{I,JK}_{\ell_1\ell_2\ell_3}|$ in the CDM isocurvature case, as functions of $\ell_3$, for $\ell_1=\ell_2=1300$. The figure on the left shows the temperature-only TTT bispectra, the one on the right the pure polarization EEE bispectra.}
\label{fig_bispectra_cdm}
\end{figure}

\begin{figure}
\centering
\includegraphics[width=0.49\textwidth, clip=true]{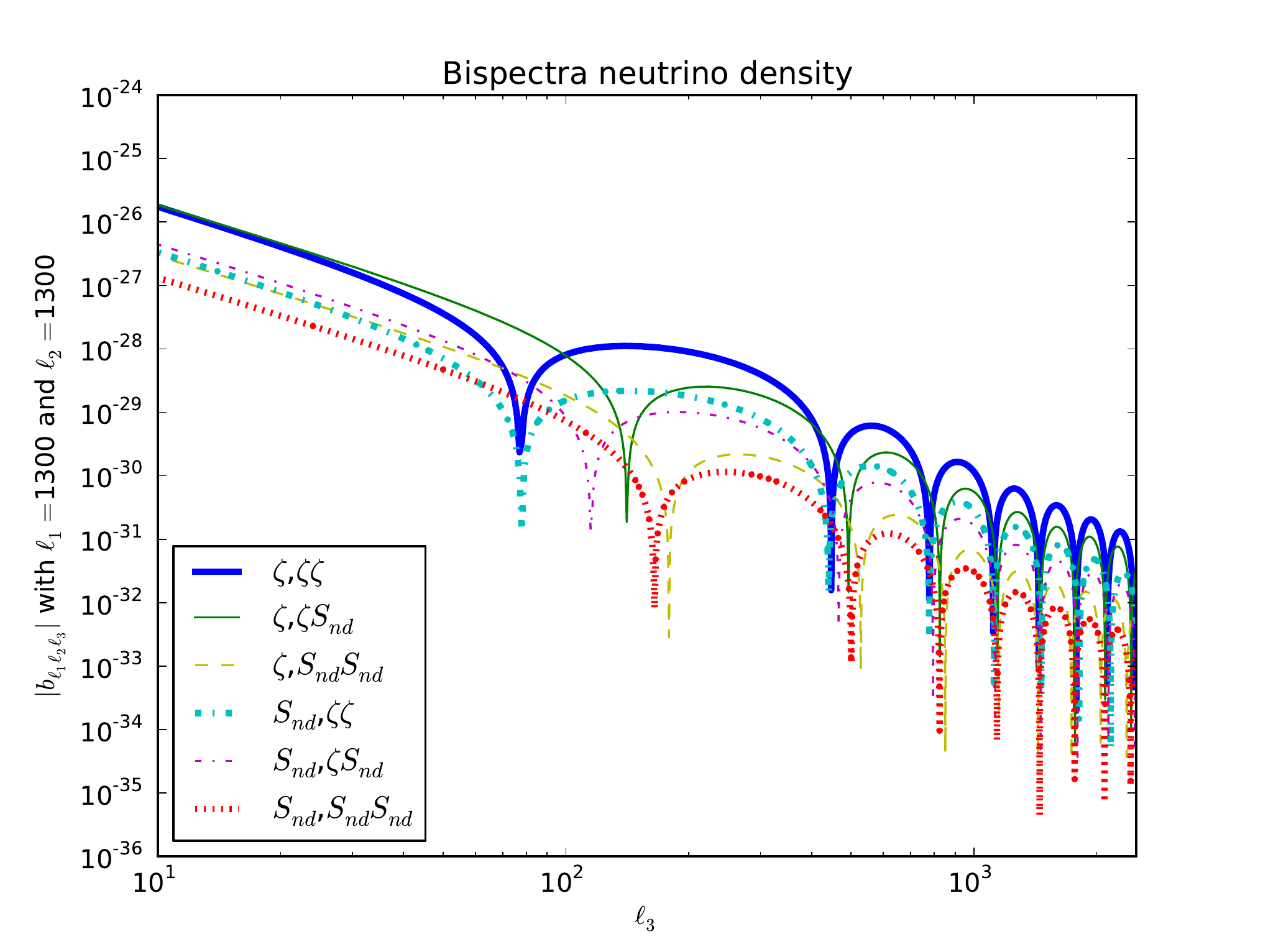}
\includegraphics[width=0.49\textwidth, clip=true]{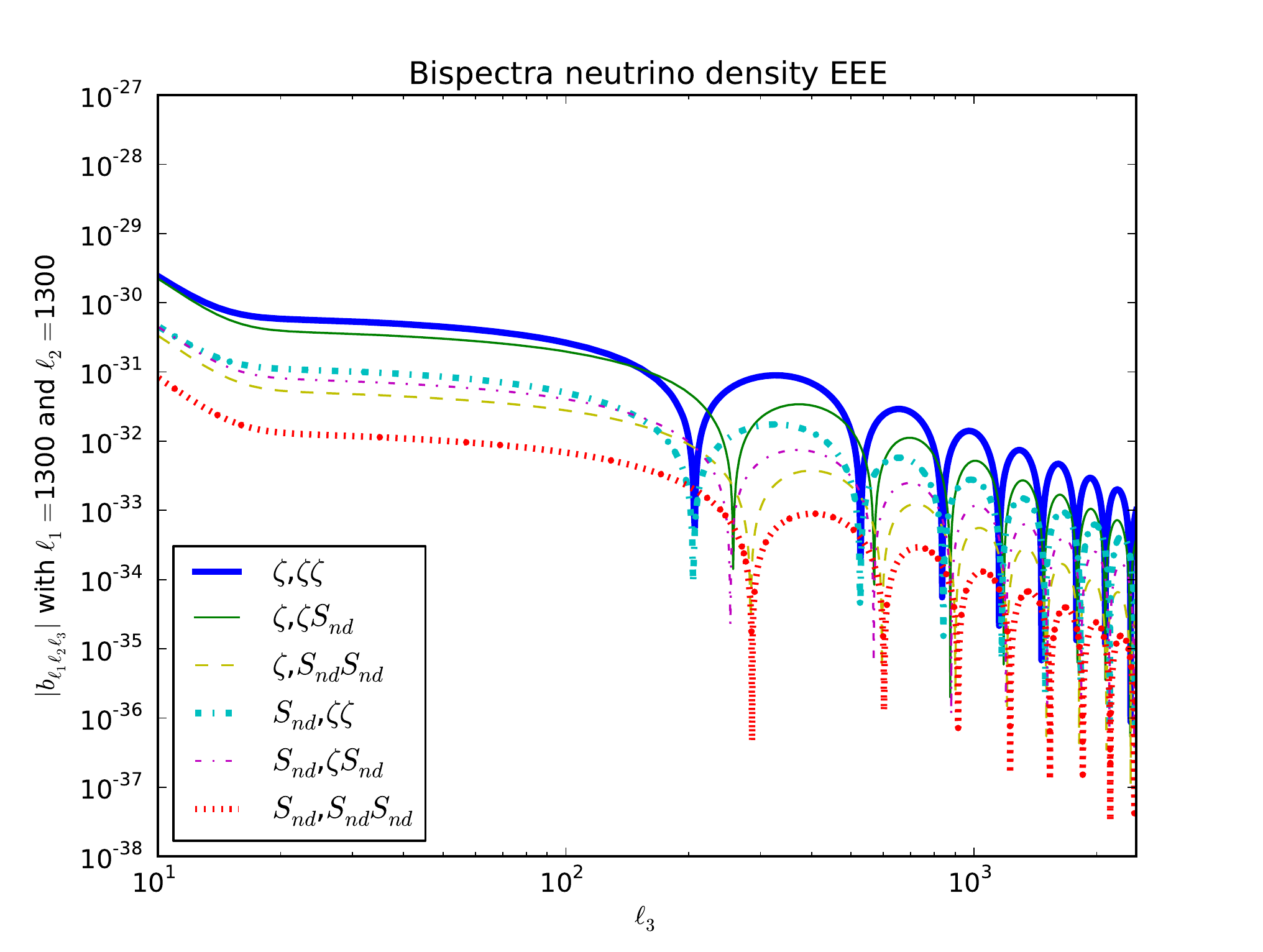}
\caption{Plot of $|b^{I,JK}_{\ell_1\ell_2\ell_3}|$ in the neutrino density isocurvature case, as functions of $\ell_3$, for $\ell_1=\ell_2=1300$. The figure on the left shows the temperature-only TTT bispectra, the one on the right the pure polarization EEE bispectra.}
\label{fig_bispectra_nd}
\end{figure}

\begin{figure}
\centering
\includegraphics[width=0.49\textwidth, clip=true]{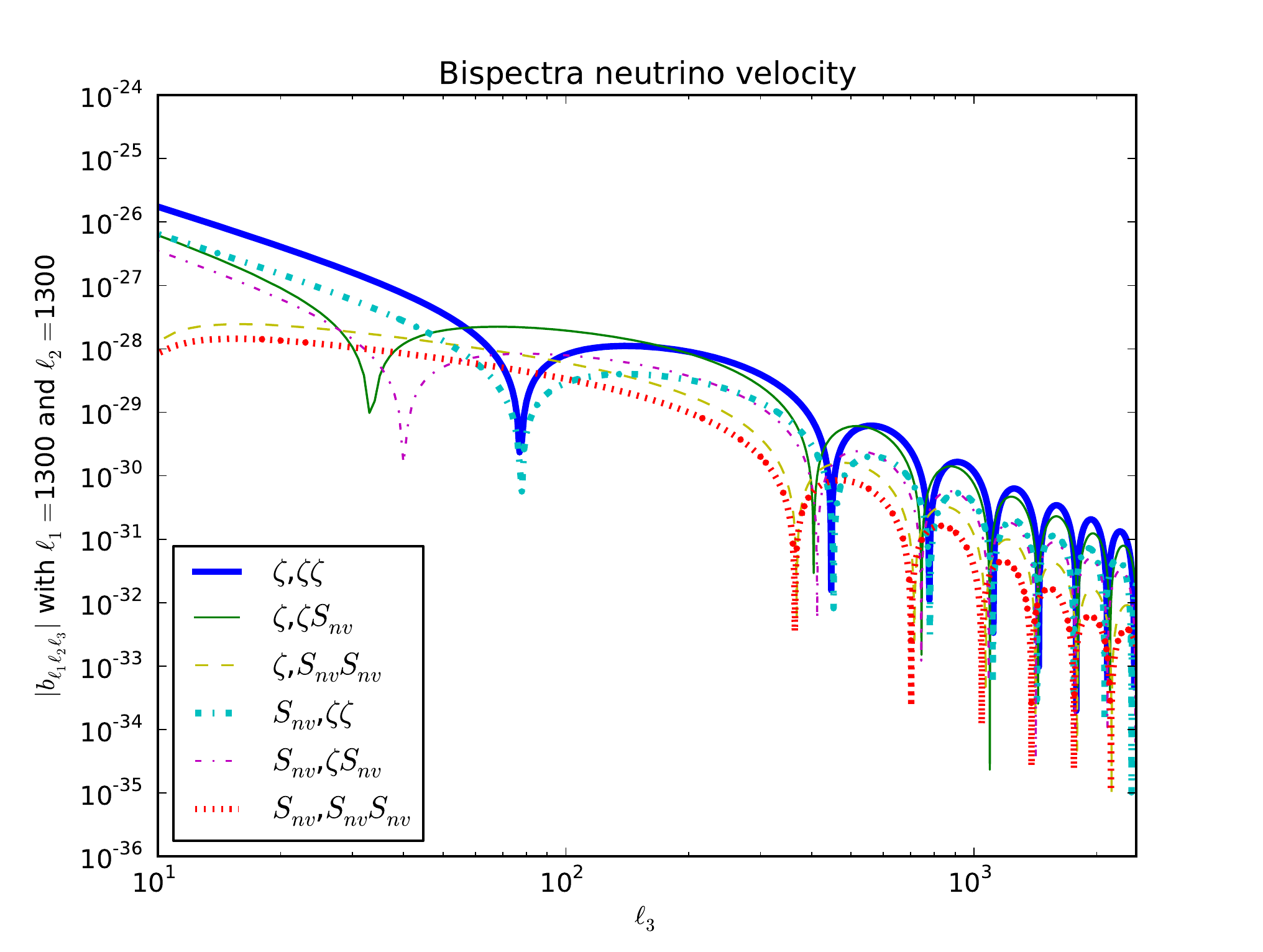}
\includegraphics[width=0.49\textwidth, clip=true]{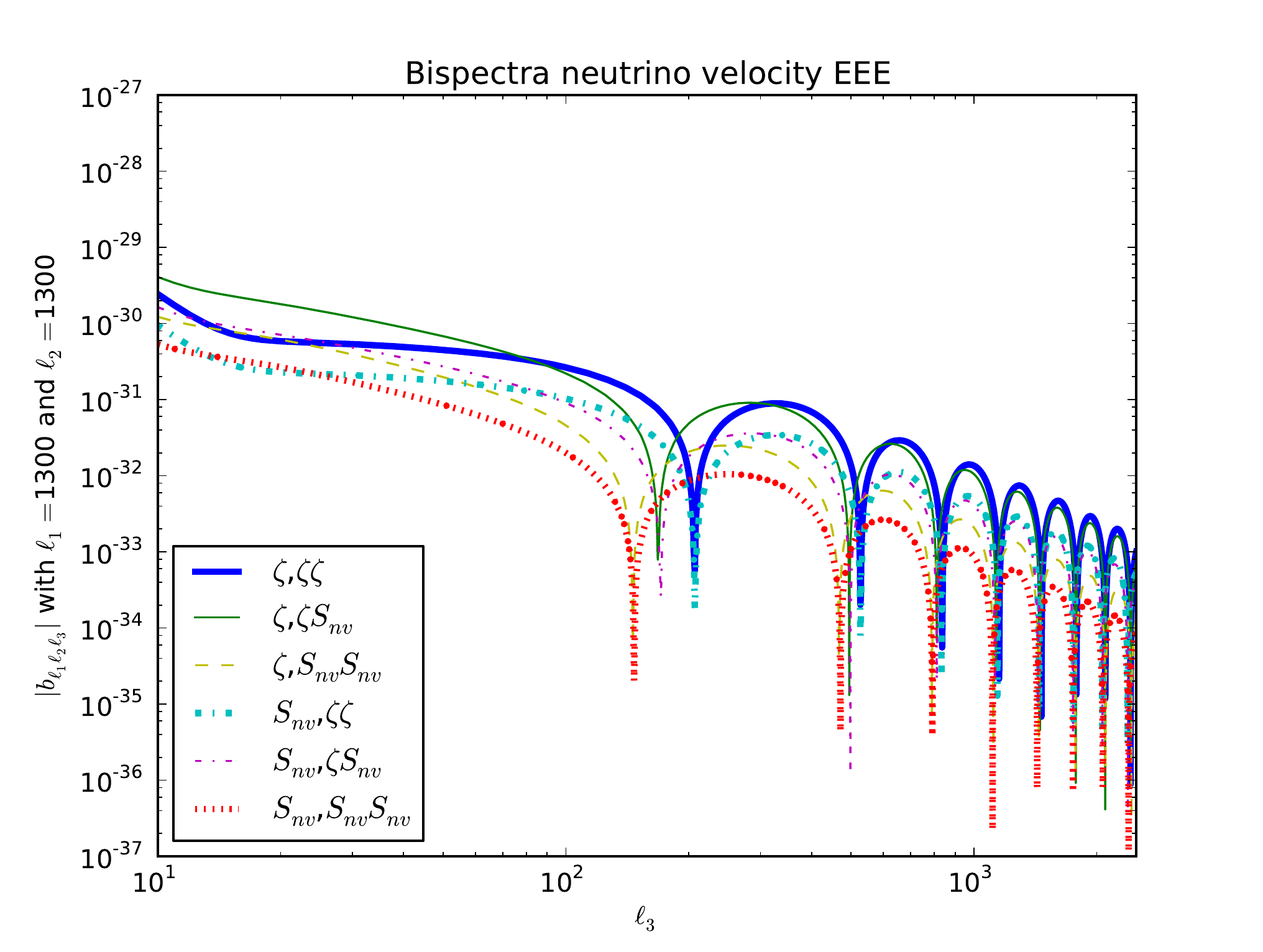}
\caption{Plot of $|b^{I,JK}_{\ell_1\ell_2\ell_3}|$ in the neutrino velocity isocurvature case, as functions of $\ell_3$, for $\ell_1=\ell_2=1300$. The figure on the left shows the temperature-only TTT bispectra, the one on the right the pure polarization EEE bispectra.}
\label{fig_bispectra_nv}
\end{figure}

In summary, after integration over $r$ of these various combinations of $\alpha$ and $\beta$ functions,  we  obtain six independent bispectra, for each type of isocurvature mode. To illustrate the typical angular dependence of these bispectra, we have plotted them as  functions of $\ell_3$, for fixed values of $\ell_1$ and $\ell_2$, respectively in the CDM isocurvature case 
(Fig.~\ref{fig_bispectra_cdm}), in the neutrino density isocurvature case (Fig.~\ref{fig_bispectra_nd}) and in the neutrino velocity case (Fig.~\ref{fig_bispectra_nv}). 
We plot only the pure temperature (TTT) and pure polarization (EEE) bispectra,
but of course one also has all the polarization cross bispectra.
As mentioned before, the $(\zeta, \zeta S)$ curve corresponds to twice $b^{\zeta, \zeta S}_{\ell_1\ell_2 \ell_3}$ since we consider the sum of
$b^{\zeta, \zeta S}_{\ell_1\ell_2 \ell_3}$ and $b^{\zeta, S \zeta}_{\ell_1\ell_2 \ell_3}$,  which cannot be distinguished. The same applies to the $(S, \zeta S)$ curve.
As usual, the bispectra for the baryon isocurvature mode are deduced from the CDM bispectra by the appropriate rescalings. 

\section{Observational prospects}

As we will see in the next section, one can envisage early universe scenarios that generate significant isocurvature non-Gaussianity, which could dominate the purely adiabatic component, even if the adiabatic mode is dominant in the power spectrum as required by observations. 
 This is why it is important to assess how precisely one can hope to measure and to discriminate the various isocurvature bispectra in the future.
  
  The most general analysis would  require to consider simultaneously all five possible modes, which corresponds to a total of $n^2(n+1)/2=75$ coefficients, taking into account the symmetry 
  (\ref{f_sym}).
  In order to simplify our analysis, we will consider separately the various isocurvature perturbations. In other words, we will assume that the primordial perturbation is the superposition of a dominant  adiabatic mode and of a {\it single} isocurvature mode. In this case, the total bispectrum is characterized by six parameters, which we now denote  $\tf^{(i)}$,  
  \begin{eqnarray}
b_{\ell_1\ell_2 \ell_3}&\!\!\!\!=&\!\!\!\!\tf^{\zeta,\zeta\zeta}\,b_{\ell_1\ell_2 \ell_3}^{\zeta,\zeta\zeta}+2\tf^{\zeta,\zeta S}\, b_{\ell_1\ell_2 \ell_3}^{\zeta,\zeta S}+
 \tf^{\zeta,SS}\, b_{\ell_1\ell_2 \ell_3}^{\zeta,SS}+\tf^{S,\zeta\zeta}\, b_{\ell_1\ell_2 \ell_3}^{S,\zeta\zeta}+2\tf^{S,\zeta S}\, b_{\ell_1\ell_2 \ell_3}^{S,\zeta S}+\tf^{S,SS}\, b_{\ell_1\ell_2 \ell_3}^{S,SS}
 \nonumber
 \\
 &\!\!\!\!=&\!\!\!\! \sum_{(i)}\tf^{(i)} b_{\ell_1\ell_2 \ell_3}^{(i)}\,,
\end{eqnarray}
where the index $i$ varies between $1$ to $6$, following the order indicated in the upper line. Note that, because of  the factor $2$ in front of  $\tf^{\zeta,\zeta S}$ and 
$\tf^{S,\zeta S}$, we define $b_{\ell_1\ell_2 \ell_3}^{(2)}\equiv 2b_{\ell_1\ell_2 \ell_3}^{\zeta,\zeta  S}$ and $b_{\ell_1\ell_2 \ell_3}^{(5)}\equiv2  b_{\ell_1\ell_2 \ell_3}^{S,\zeta S}$ whereas there is no such factor $2$ for  the other terms.

\subsection{The Fisher matrix} 
 
 To estimate these six parameters, given some data set, the usual 
procedure is to minimize 
\be
\chi^2=\left\langle (B^{obs}-\sum_i  \tf^{(i)} B^{(i)}), 
(B^{obs}-\sum_i  \tf^{(i)} B^{(i)})\right\rangle.
\ee
For an ideal experiment (no noise and no effects due to the beam size) without
polarization, the scalar product is defined  by
\be
\langle B, B' \rangle\equiv \sum_{\ell_1 \leq \ell_2 \leq \ell_3}
\frac{B_{\ell_1\ell_2\ell_3}B'_{\ell_1\ell_2\ell_3}}{\sigma^2_{\ell_1\ell_2\ell_3}}.
\ee
The bispectrum variance in that case is given by
\be
\sigma^2_{\ell_1\ell_2\ell_3}\equiv \langle B^2_{\ell_1\ell_2\ell_3}\rangle
-\langle B_{\ell_1\ell_2\ell_3} \rangle^2\approx
\Delta_{\ell_1\ell_2\ell_3} C_{\ell_1}C_{\ell_2} C_{\ell_3}
\ee
in the approximation of weak non-Gaussianity,
where  
\be
\Delta_{\ell_1\ell_2\ell_3} \equiv  1+\delta_{\ell_1\ell_2}+\delta_{\ell_2\ell_3}+\delta_{\ell_3\ell_1}+2\, \delta_{\ell_1\ell_2}\delta_{\ell_2\ell_3}\,.
\ee 
The best estimates for the parameters are thus obtained by solving
\be
\sum_j \langle B^{(i)}, B^{(j)}\rangle \tf^{(j)}
=\langle B^{(i)}, B^{obs}\rangle\, ,
\ee
while the statistical error on the parameters is deduced from the second-order 
derivatives of $\chi^2$,   which define the Fisher matrix, given in our case by
\be
F_{ij}\equiv \langle B^{(i)}, B^{(j)}\rangle.
\ee 
The Fisher matrix is a symmetric matrix, which can be determined by computing
the 21 different scalar products between the six elementary bispectra.

For a real experiment, and if E-polarization is included as well, the above
equations remain valid, except that the definition of the scalar product
has to be replaced by a more complicated expression (see e.g.\ 
\citep{Yadav:2007rk}):
\be
\langle B^{(i)}, B^{(j)} \rangle \equiv 
\sum_{\alpha\beta\gamma\alpha'\beta'\gamma'}\ 
\sum_{\ell_1 \leq \ell_2 \leq \ell_3} \frac{1}{\Delta_{\ell_1 \ell_2 \ell_3}}
B_{\ell_1 \ell_2 \ell_3}^{(i) \, \alpha\beta\gamma}
({\cal C}_{\ell_1}^{-1})^{\alpha\alpha'}
({\cal C}_{\ell_2}^{-1})^{\beta\beta'}
({\cal C}_{\ell_3}^{-1})^{\gamma\gamma'}
B_{\ell_1 \ell_2 \ell_3}^{(j) \, \alpha'\beta'\gamma'},
\ee
where $\alpha, \beta, \gamma, \alpha', \beta', \gamma'$ are polarization
indices taking the two values $T$ and $E$.
The covariance matrix ${\cal C}_\ell$ (a matrix in polarization space) is given by
\be
{\cal C}_\ell = \left(\begin{array}{cc}
b_\ell^2 C_\ell^{TT} + N_\ell^T & b_\ell^2 C_\ell^{TE}\\
b_\ell^2 C_\ell^{TE} & b_\ell^2 C_\ell^{EE} + N_\ell^E
\end{array} \right),
\ee
where $b_\ell$ is the beam function and $N_\ell$ the noise power spectrum.
We assumed the same beam function for temperature and polarization detectors,
as well as no correlated noise, but the generalization  is straightforward. In the calculation of the covariance
matrix we only take the adiabatic power spectrum, since from observations
we know that the isocurvature contribution to the power spectrum must be
very small.

For each type of isocurvature mode, we have computed the corresponding Fisher
matrix by extending the numerical code
described in \citep{BvTC} to include isocurvature modes and
E-polarization, according to the expressions presented above. 
We have taken into account the noise
characteristics of the Planck satellite~\citepalias{Planck:2006aa}, using only
the 100, 143, and 217 GHz channels, combined in quadrature. Our
computation goes up to $\ell_\mathrm{max} = 2500$ and uses the WMAP-only
7-year best-fit cosmological parameters~\citep{Komatsu:2010fb}.

From the Fisher matrix, one can compute the statistical uncertainty on each of the parameters:
\be
\label{errors}
\Delta \tf^i=\sqrt{(F^{-1})_{ii}}\,.
\ee
This takes into account the correlations between the various bispectra. By contrast, if one assumes that the data contain only a single elementary bispectrum, for example the purely adiabatic one, then the corresponding statistical error is 
\be
\Delta \tf^i=\frac{1}{\sqrt{F_{ii}}} \qquad ({\rm single\    parameter})\,.
\ee
One can also determine the correlations between any two bispectra:
\be
\label{correlation_matrix}
{\cal C}_{ij}=\frac{(F^{-1})_{ij}}{\sqrt{(F^{-1})_{ii}(F^{-1})_{jj}}}\,.
\ee

\subsection{CDM isocurvature mode}

\def\text{}
\begin{table}
\footnotesize
\begin{center}
\begin{tabular}{|cccccc|}
\hline
$(\zeta, \zeta\zeta)$ & $(\zeta, \zeta S)$ & $(\zeta, S S)$ &  $(S,\zeta\zeta)$  & $(S,\zeta S)$ &  $(S,SS)$\\
\hline
$3.9 \,(2.5) \times 10^{\text{-2}}$ & $4.5 \,(3.6) \times 10^{\text{-2}}$ & $2.3 \,(2.1) \times 10^{\text{-4}}$ &
   $2.4\,(1.6) \times 10^{\text{-4}}$ & $6.9 \,(4.3)\times 10^{\text{-4}}$ & $5.3 \,(3.1) \times 10^{\text{-4}}$
\\ 
- & $7.1 \,(6.0) \times 10^{\text{-2}}$ & $5.3 \,(3.8)\times 10^{\text{-4}}$ &
   $3.8\,(2.1)\times 10^{\text{-4}}$ & $11 \,(7.4)\times 10^{\text{-4}}$ & $8.8 \,(5.5)\times 10^{\text{-4}}$
\\ 
- & - & $28 \,(6.4) \times 10^{\text{-5}}$ &
   $16 \,(3.7)\times 10^{\text{-5}}$ & $33 \,(9.5)\times 10^{\text{-5}}$ & $11 \,(5.0)\times 10^{\text{-5}}$
\\ - & - & - & $15 \,(3.0)\times 10^{\text{-5}}$ & $22 \,(5.8)\times 10^{\text{-5}}$ & $7.5 \,(3.2)\times 10^{\text{-5}}$
\\- & - & - & - & $5.1 \,(1.6)\times 10^{\text{-4}}$ & $2.4 \,(1.0)\times 10^{\text{-4}}$
\\- & - & - & - & - & $21 \,(8.3)\times 10^{\text{-5}}$
\\ \hline
\end{tabular}
\caption{Fisher matrix for the CDM isocurvature mode. Only the upper half coefficients are indicated,  since the matrix is symmetric. The value between parentheses corresponds to the Fisher matrix components obtained  without including the polarization.}
\label{table_cdm}
\end{center}
\end{table}

Our results for this mode have already been presented elsewhere \citep{LvT1}, but we discuss here  in more detail the peculiarities of the corresponding Fisher matrix, which is given in Table~\ref{table_cdm}.
One can immediately notice the intriguing fact that the coefficients of the upper left \mbox{$2\times 2$} submatrix, corresponding to the purely adiabatic component and the correlated $(\zeta, \zeta S)$ component, are typically  two orders of magnitude larger than all the other coefficients. 
The correlation matrix,  defined in (\ref{correlation_matrix}) and given in Table \ref{correlation_cdm} shows that  the first two bispectra are strongly (anti-)correlated while their correlation with the four other bispectra is weak.

\begin{table}
\begin{center}
\begin{tabular}{|cccccc|}
\hline
$(\zeta, \zeta\zeta)$ & $(\zeta, \zeta S)$ & $(\zeta, S S)$ &  $(S,\zeta\zeta)$  & $(S,\zeta S)$ &  $(S,SS)$\\
\hline

 $1.$ & $-0.84 \,(0.92)$ & $0.18 \,(0.21)$ & $0.003 \,(-0.12)$& $-0.15 \,(0.17)$ & $0.13 \,(0.18)$ \\
 - & $1.$ & $-0.15 \,(0.18)$ & $-0.007 \,(+0.13)$ & $0.13 \,(0.14)$& $-0.16 \,(0.16)$ \\
 - & - & 1. & $-0.069 \,(+0.24)$ & $-0.80 \,(0.97)$ & $0.58 \,(0.92)$ \\
 - & - & - & 1. & $-0.42 \,(0.43)$& $0.29 \,(0.39)$ \\
 - & - & - & - & $1.$ & $-0.82\,(0.98)$ \\
 - & - & - & - & - & 1.
\\ \hline
\end{tabular}
\caption{Correlation matrix for the CDM isocurvature mode. Only the upper half coefficients are indicated,  since the matrix is symmetric. The value between parentheses corresponds to the correlations obtained  without including the polarization (the absence of  sign in the parentheses means that it is unchanged with respect to the value taking into account the polarization).}
\label{correlation_cdm}
\end{center}
\end{table}

From this Fisher matrix, one finds that the  $68$ \% error on the parameters $\tf^i$ is given 
by\footnote{The tiny differences in the 3rd, 5th, and 6th value compared to 
\citep{LvT1} are due to small improvements in the computer code.}
\be
\Delta \tf^i=\sqrt{(F^{-1})_{ii}}=\{9.6, 7.1, 160, 150, 180, 140\}\,.
\label{errors_CDMpol}
\ee
For ease of readability, we have written $160$ instead of $1.6\cdot 10^2$, etc.,
but we are not claiming more than two digits of significance.
We also remind the reader that in the purely adiabatic case, our 
$\tf^1$, i.e.\ the $(\zeta,\zeta\zeta)$ component, is $-6/5$ times the
standard $f_\mathrm{NL}$. 
One sees that the first two uncertainties are typically one order of magnitude  smaller than  the last four. 

It is also interesting to estimate how much the inclusion of the polarization data in the analysis improves the precision of the non-linear parameters. The components of the Fisher matrix when the polarization is not taken into account can be read between the parentheses in Table~\ref{table_cdm}. One notices that whereas the coefficients of the first two lines are reduced by a factor inferior to two, the other coefficients are significantly suppressed when one removes the polarization data. As a consequence, one finds that  
the uncertainties on the parameters without polarization, given by
\be
\Delta \tf^i=\sqrt{(F^{-1})_{ii}}=\{17, 11, 980, 390, 1060, 700\}\qquad {\rm (no \ polarization)},
\ee
increase by less than a factor two for the first two parameters, whereas the 
increase is much bigger for the four other ones.

\begin{figure}
\centering
\includegraphics[width=0.6\textwidth, clip=true]{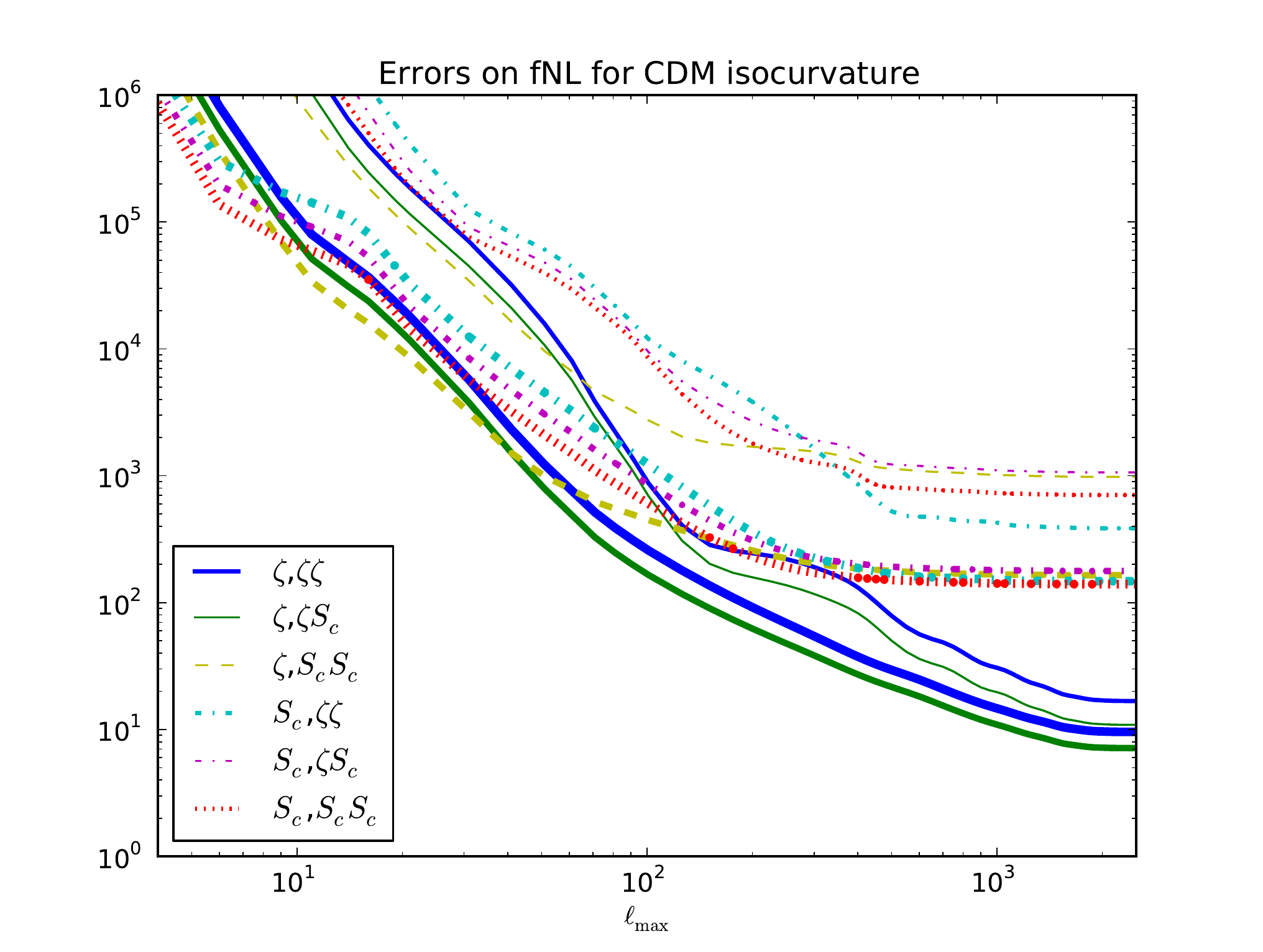}
\caption{Evolution of the $f_{\rm NL}$ parameter uncertainties as one increases 
the cut-off $\ell_{\rm max}$, for the CDM isocurvature mode. The six thinner 
curves describe the situation if only temperature data is used, while for the 
six thicker curves both temperature and E-polarization data are included.}
\label{V}
\end{figure}

The evolution of these 
uncertainties as a function of the cut-off $\ell_\mathrm{max}$ is shown
in Fig.~\ref{V}, both for the case where temperature and E-polarization data 
are used and for the case where only temperature data is included.
One can see that the curves for temperature-only typically look 
bumpier than the curves that include polarization as well.
This can be explained as follows. First,  unlike the power spectrum, the bispectrum is an 
alternating function, so that for certain regions in $\ell_1 \ell_2 \ell_3$ space it is 
zero or close to zero, and the  contribution  to the determination of  $f_\mathrm{NL}$, which is quadratic in the bispectrum, is then negligible in these regions.  Second, as one can
see for example in Fig.~\ref{fig_bispectra_cdm}, the acoustic peaks of the
polarization bispectrum are out of phase with the ones of the temperature
one, so that including polarization neatly fills in the holes 
in $\ell_1 \ell_2 \ell_3$ space and leads to a smoother determination of $f_\mathrm{NL}$,
as first pointed out by \citep{Komatsu:2003iq}.

\begin{figure}
\centering
\includegraphics[width=0.49\textwidth, clip=true]{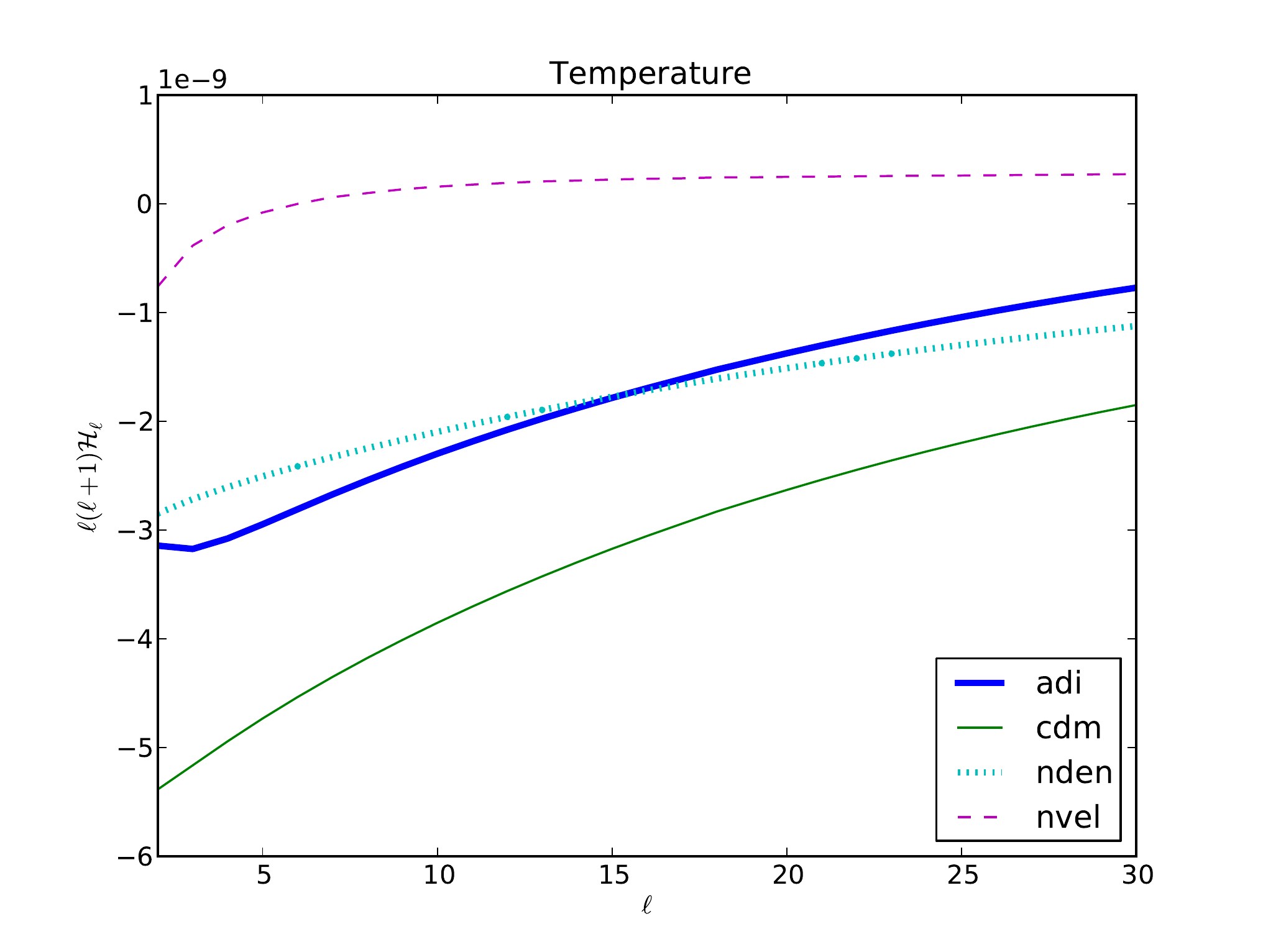}
\includegraphics[width=0.49\textwidth, clip=true]{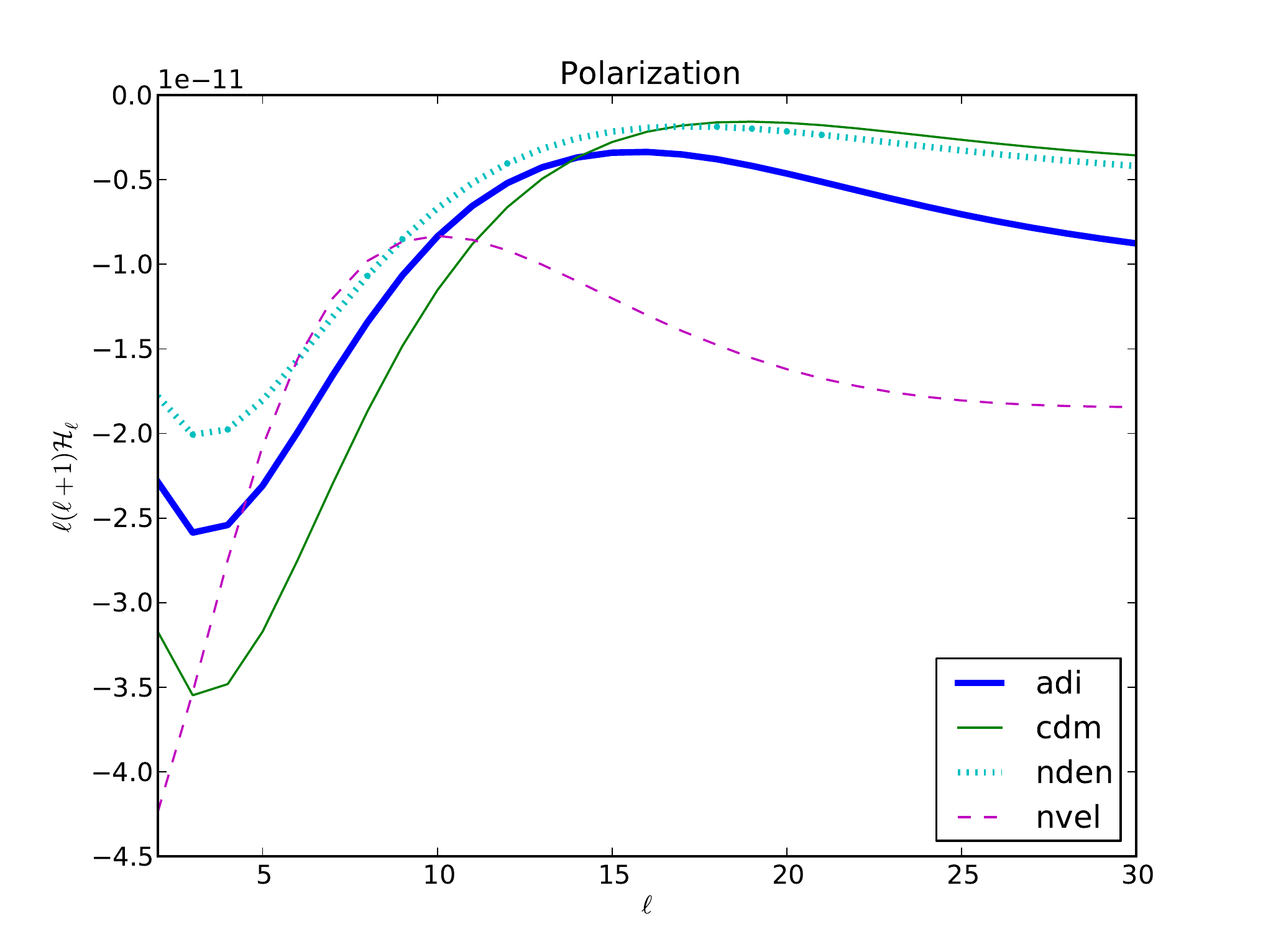}
\caption{The function $\ell(\ell+1){\cal H}_K$ plotted as a function of $\ell \leq 30$ 
for different values of $K$. The figure on the left
shows temperature (T), the one on the right polarization (E).}
\label{HK_fig}
\end{figure}

\begin{figure}
\centering
\includegraphics[width=0.49\textwidth, clip=true]{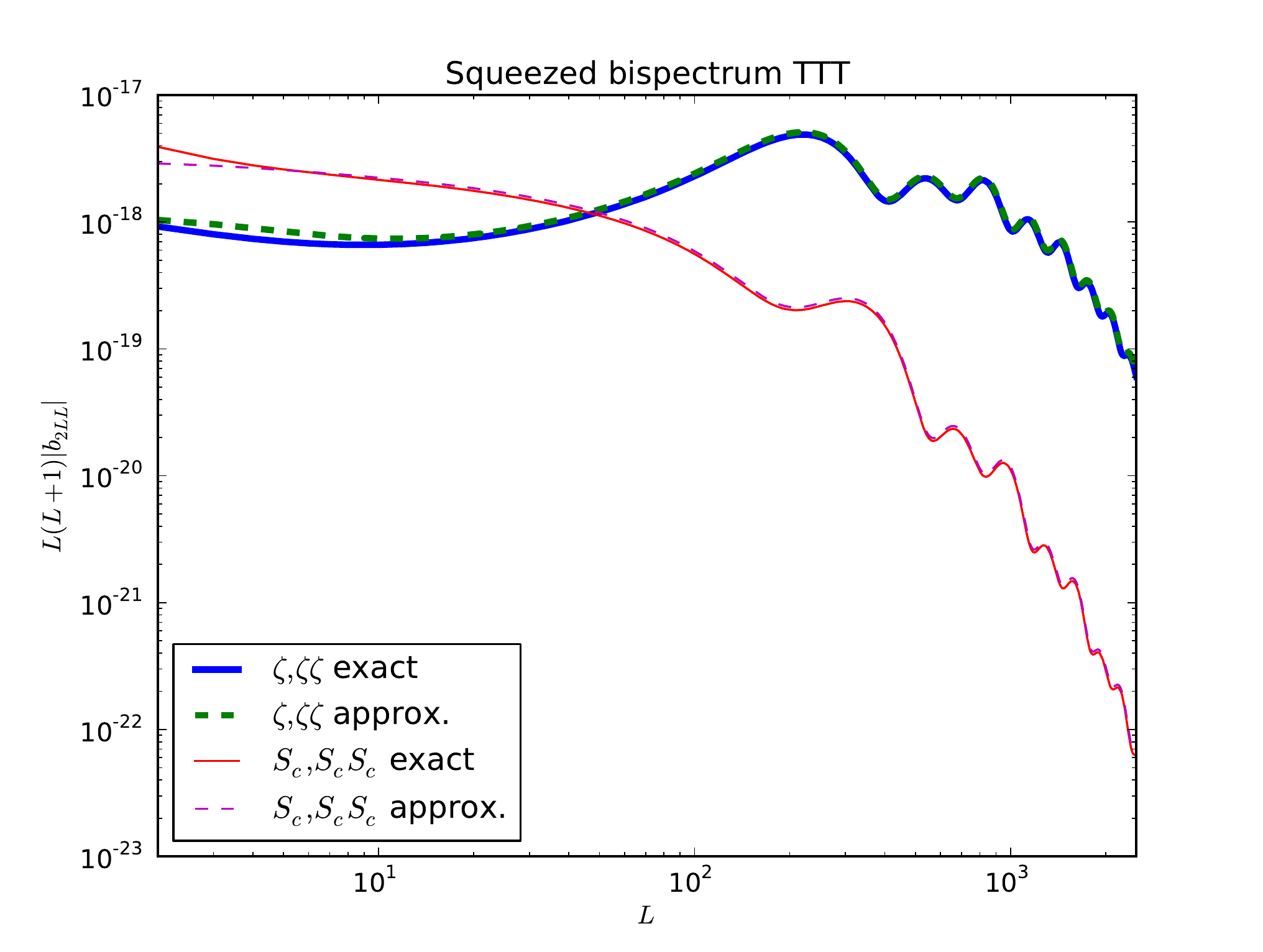}
\includegraphics[width=0.49\textwidth, clip=true]{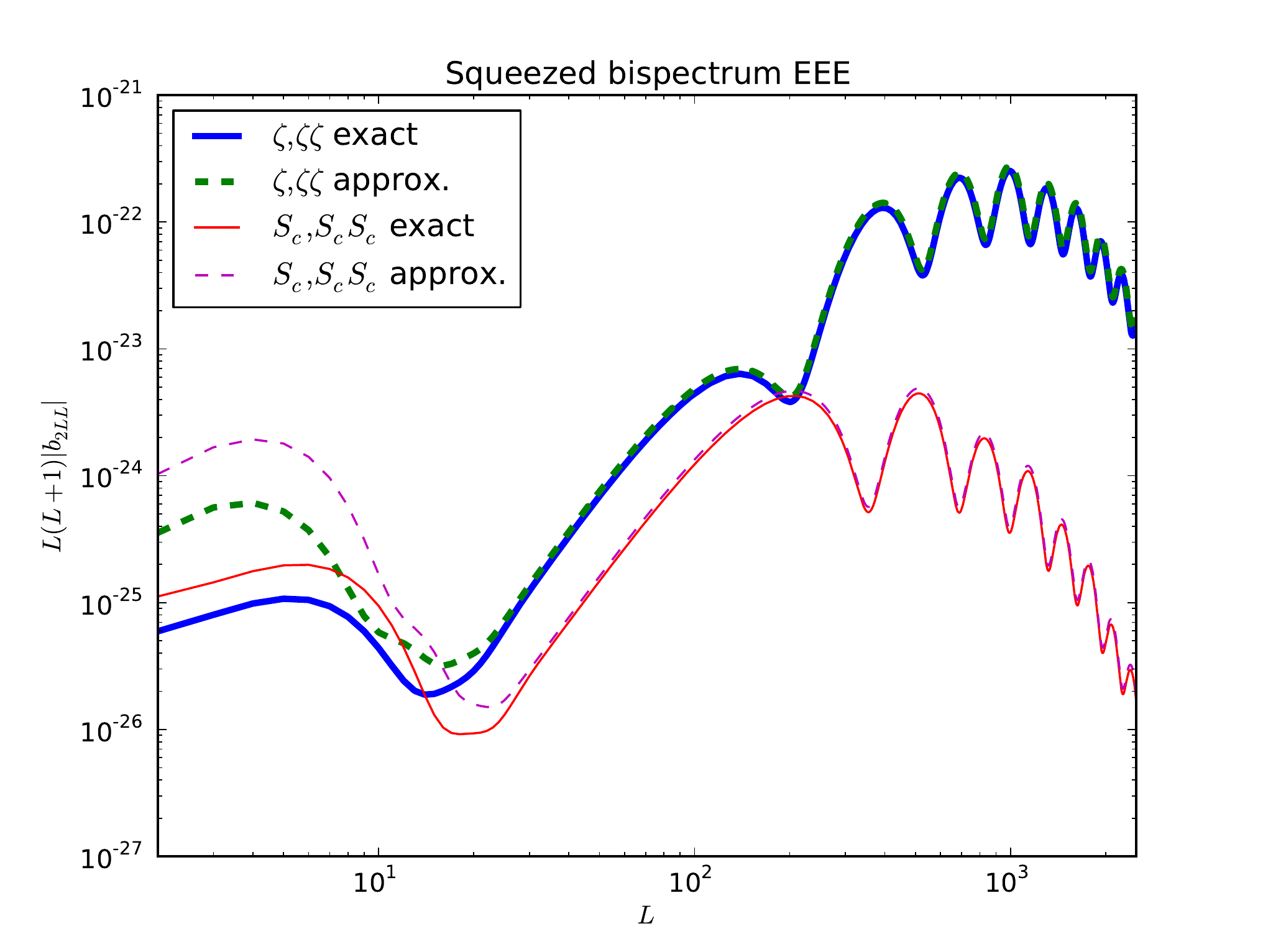}
\caption{The absolute value of the squeezed bispectrum (multiplied by
  $L(L+1)$) with $\ell_1=2$ and $\ell_2=\ell_3\equiv L$ plotted as a function
  of $L$. Both the exact bispectrum and (twice) the approximation
  given in (\ref{G_and_H}) are plotted, for the purely adiabatic mode
  and the purely CDM isocurvature mode. The figure on the left shows
  the temperature-only TTT bispectra, the one on the right the pure
  polarization EEE bispectra. Note that the configuration is only
  squeezed in the case of large $L$; for small $L$ the approximation
  should not be used.}
\label{bispec_approx_fig}
\end{figure}

Our results can be understood by the following analysis in the squeezed limit
(based on \citep{BvTC}), assuming that 
$\ell_1\equiv \ell\ll \ell_2\approx \ell_3\equiv L$. In this limit, 
the bispectra defined in (\ref{b_IJK}) can be decomposed as
\be
b_{\ell L L}^{I,JK}=    \int_0^\infty r^2 dr \left[\alpha^I_{\ell}(r)\beta^{J}_{L}(r)\beta^{K}_{L}(r)+\alpha^I_{L}(r)\beta^{J}_{\ell}(r)\beta^{K}_{L}(r)+\alpha^I_{L}(r)\beta^{J}_{L}(r)\beta^{K}_{\ell}(r)\right].
\label{squeezed_bispec}
\ee
The first term is subdominant, since, like the power spectrum, $|\beta_L|$ 
decreases as $1/(L(L+1))$ (or even faster as $1/(L^2(L+1)^2)$ for large $L$), 
as can be seen for example in Fig.~\ref{beta_cdm}. 
The last term, for instance, is explicitly given by
\be
 \left(\frac2\pi\right)^3 \int k_1^2 dk_1  \, k_2^2 dk_2 \, k_3^2 dk_3 \, g^I_{L}(k_1)g^J_{L}(k_2)g^K_{\ell}(k_3)\,  P_\zeta(k_2)P_\zeta(k_3)
\int_0^\infty r^2 dr   j_{L}(k_1r)  j_{L}(k_2r)   j_{\ell}(k_3 r) \,,
\ee
where the last Bessel function  oscillates slowly while the first  two oscillate very rapidly. This leads to a cancellation of the radial integral unless $k_1$ is very close to $k_2$. 
We find that the above expression can thus be approximated by
\ba
&\displaystyle\left(\frac2\pi\right)^3 & \!\!\!\! \int k_1^2 dk_1 \int k_2^2 dk_2 \int k_3^2 dk_3 \, g^I_{L}(k_1)g^J_{L}(k_2)g^K_{\ell}(k_3) P_\zeta(k_2)P_\zeta(k_3) \frac{5\delta(k_1-k_2)}{k_1 k_2}\int_0^\infty \frac{dr}{r} j_{\ell}(k_3r) 
\cr
&\approx& \!\!\!\! \left\{\left(\frac2\pi\right)^2\int k^2 dk \, g^I_{L}(k)g^J_{L}(k) P_\zeta(k) \right\} \left\{ \frac{10}{\pi}\int k_3^2 dk_3 \, g^K_{\ell}(k_3) P_\zeta(k_3) \int_0^\infty \frac{dr}{r} j_{\ell}(k_3r) \right\}
\cr
&\equiv& \!\!\!\! {\cal G}_{IJ}(L)\, {\cal H}_K(\ell) .
\label{G_and_H}
\ea
The $\delta(k_1-k_2)$ is explained above, but together with the $1/(k_1 k_2)$
also motivated by the closure relation for spherical Bessel functions,
$\int_0^\infty r^2 j_L(k_1 r) j_L (k_2 r) dr = \pi \delta(k_1-k_2)/(2k_1k_2)$.
The $1/r$ follows from a dimensional analysis, and the $5$ has been 
determined heuristically by comparing with the exact bispectrum: the ratio
is approximately $5$ and only weakly dependent on the small $\ell$.
The full squeezed bispectrum (\ref{squeezed_bispec}) is thus approximated by 
\be
 b_{\ell L L}^{I,JK}\approx {\cal G}_{IJ}(L)\, {\cal H}_K(\ell)+{\cal G}_{IK}(L)\, {\cal H}_J(\ell)\,.
 \ee
Assuming all primodial power spectra to be equal, the functions ${\cal G}_{IJ}$
are the angular (cross) power spectra plotted in Fig.~\ref{spectra_fig} and
\ref{cross_spectra_fig}. 
The function ${\cal H}_K(\ell)$ is simply an integral over $\beta_\ell^K(r)$:
\be
{\cal H}_K(\ell) = 5 \int_0^\infty \frac{dr}{r} \, \beta_\ell^K(r)
\ee
and is shown for small $\ell$ in Fig.~\ref{HK_fig}. The good agreement
of our approximation with the exact squeezed bispectrum is shown in
Fig.~\ref{bispec_approx_fig}.


In the squeezed limit, one thus finds that only the first two elementary bispectra, $(\zeta, \zeta\zeta)$ and $(\zeta, \zeta S)$, depend on ${\cal G}_{\zeta\zeta}(L)$. The four others depend on ${\cal G}_{\zeta S}(L)$ and/or ${\cal G}_{SS}(L)$.
The large $L$ limit of ${\cal G}_{\zeta S}(L)$ and ${\cal G}_{SS}(L)$ are strongly suppressed with respect to ${\cal G}_{\zeta\zeta}(L)$, which explains why the uncertainty on the first two non-Gaussianity parameters can be reduced by probing high multipoles (the bispectrum there is still sufficiently large compared to the noise) 
while the uncertainty on the four other ones saturates as shown in Fig.~\ref{V}.
One can even understand why the curve for the $(\zeta,\zeta S)$ mode is below
the one for $(\zeta,\zeta\zeta)$: their dominant terms both depend on the same 
${\cal G}_{\zeta\zeta}(L)$, but different ${\cal H}(\ell)$, and 
$|{\cal H}_S(\ell)| > |{\cal H}_\zeta(\ell)|$.
Finally, one can understand why including polarization helps much more for
the uncertainty on e.g.\ the $(S,SS)$ mode than for the $(\zeta,\zeta\zeta)$
mode. As one can see from Fig.~\ref{V}, it is in particular in the region
$50 \leq \ell \leq 200$ that the distance between the two $(S,SS)$ curves increases
compared to the distance between the two $(\zeta,\zeta\zeta)$ curves. A quick
look at Fig.~\ref{spectra_fig} shows that in that region of multipole space
the TT CDM isocurvature power spectrum (i.e.\ ${\cal G}_{SS}(L)$) becomes
very small compared to the TT adiabatic spectrum (i.e.\ 
${\cal G}_{\zeta\zeta}(L)$), but the EE CDM isocurvature spectrum still remains
comparable to the EE adiabatic one.

It is also instructive to  compare (\ref{errors_CDMpol}) with the uncertainties
\be
\Delta\tf^i= 1/\sqrt{F_{ii}}=\{5.0, 3.7, 60, 82, 44, 69\} \qquad {\mathrm{(single \ parameter)}}
\ee
  obtained by ignoring the correlations, or, equivalently, by assuming that only one parameter  is nonzero. In particular, the contamination of the purely adiabatic signal by the other shapes induces an increase of the uncertainty, but only by a factor 2, which is rather moderate. 
  
Assuming that the adiabatic and isocurvature modes are {\it uncorrelated} implies that only the purely adiabatic and isocurvature bispectra are relevant. The corresponding  
two-parameter
Fisher matrix, which is the submatrix of $F_{(ij)}$ with entries  $F_{11}$, $F_{66}$ and $F_{16}$,   leads to uncertainties on  $\tf^{(1)}$ and $\tf^{(6)}$ that are almost identical to the corresponding single-parameter errors. 

Finally, let us note that if the observed bispectrum is mainly purely isocurvature with amplitude $\tf^{(6)}$, a naive analysis using 
only the purely adiabatic estimator would lead to an apparent adiabatic coefficient 
\be
\tf^{(1)}=  \frac{F_{16}}{F_{11}}\, \tf^{(6)}\simeq 10^{-2} \, \tf^{(6)},
\ee
thus hiding the isocurvature signal with larger amplitude.

\subsection{Baryon isocurvature mode}

The Fisher matrix for the baryon isocurvature mode can be easily deduced from the CDM Fisher matrix. Indeed, as discussed earlier, the CDM and baryon isocurvature transfer functions are identical up to a rescaling by $\omega_{bc}$ introduced in (\ref{omega_bc}).
Consequently, the $\alpha$ and $\beta$ functions are simply rescaled:
\be
\alpha^{S_b}_{\ell}(r)=\omega_{bc}\, \alpha^{S_c}_{\ell}(r),\qquad \beta^{S_b}_{\ell}(r)=\omega_{bc}\, \beta^{S_c}_{\ell}(r)\,.
\ee
The rescaling of the various bispectra will thus depend on the number of $S$ indices, i.e.
\be
b^{I,JK}_{\ell_1\ell_2 \ell_3}(S_b)=(\omega_{bc})^p \, b^{I,JK}_{\ell_1\ell_2 \ell_3}(S_c),
\ee
where 
the power $p$ is the number of $S$ among the indices $\{IJK\}$. In summary, 
all coefficients of the baryon isocurvature Fisher matrix can be deduced from 
Table \ref{table_cdm} by using the rescaling 
\be
F_{ij}^{S_b}=\mathcal{N}_i \, \mathcal{N}_j F_{ij}^{S_c} \quad ({\rm no \, summation}), \qquad \mathcal{N}_i=\{1, \omega_{bc}, \omega_{bc}^2,\omega_{bc},\omega_{bc}^2,\omega_{bc}^3\},
\ee
where, in our computation, $\omega_{bc}=0.2036$.

The parameter uncertainties can also be deduced from the CDM results via the rescalings $\mathcal{N}_i$: $\Delta \tilde f^i(S_b)=\Delta \tilde f^i(S_c)/\mathcal{N}_i$. One thus obtains:
\be
\Delta \tilde f^i=
\left\{9.6, 35, 4000, 720, 4300, 16600\right\}\,.
\ee
Except for the purely adiabatic coefficient, we thus find that the uncertainties on all the other coefficients are significantly larger  than the uncertainties obtained in (\ref{errors_CDMpol}) in the  CDM  case, simply because the elementary bispectra have a smaller amplitude than their CDM counterparts. By contrast, the correlation matrix, which is independent of the normalization of the bispectra, is exactly the same as in the CDM  case. 

\subsection{Neutrino density isocurvature mode}

\def\text{}
\begin{table}
\footnotesize
\begin{center}
\begin{tabular}{|cccccc|}
\hline
$(\zeta, \zeta\zeta)$ & $(\zeta, \zeta S)$ & $(\zeta, S S)$ &  $(S,\zeta\zeta)$  & $(S,\zeta S)$ &  $(S,SS)$\\
\hline
$3.9 \,(2.5) \times 10^{\text{-2}}$ & $3.6 \,(2.6) \times 10^{\text{-2}}$ & $5.6 \,(4.1) \times 10^{\text{-3}}$ &
   $7.9\,(5.1) \times 10^{\text{-3}}$ & $8.8 \,(6.1)\times 10^{\text{-3}}$ & $2.2 \,(1.6) \times 10^{\text{-3}}$
\\ 
- & $3.8 \,(2.9) \times 10^{\text{-2}}$ & $6.3 \,(4.8)\times 10^{\text{-3}}$ &
   $7.4 \,(5.2)\times 10^{\text{-3}}$ & $9.2 \,(6.7)\times 10^{\text{-3}}$ & $2.5 \,(1.8)\times 10^{\text{-3}}$
\\ 
- & - & $11 \,(8.1) \times 10^{\text{-4}}$ &
   $12 \,(8.5)\times 10^{\text{-4}}$ & $1.6 \,(1.1)\times 10^{\text{-3}}$ & $4.4 \,(3.1)\times 10^{\text{-4}}$
\\ - & - & - & $1.8 \,(1.1)\times 10^{\text{-3}}$ & $2.0 \,(1.3)\times 10^{\text{-3}}$ & $5.0 \,(3.2)\times 10^{\text{-4}}$
\\- & - & - & - & $2.5 \,(1.6)\times 10^{\text{-3}}$ & $6.8 \,(4.4)\times 10^{\text{-4}}$
\\- & - & - & - & - & $2.1 \,(1.2)\times 10^{\text{-4}}$
\\ \hline
\end{tabular}
\caption{Fisher matrix for the neutrino density isocurvature mode. Only the upper half coefficients are indicated,  since the matrix is symmetric. The value between parentheses corresponds to the Fisher matrix components obtained  without including the polarization.}
\label{table_nd}
\end{center}
\end{table}
For a neutrino density isocurvature mode, we have obtained the Fisher matrix in Table~\ref{table_nd}. 
Unlike the case of CDM isocurvature, here the difference between the different
entries in the Fisher matrix is smaller, although the coefficients in the 
upper left $2\times 2$ submatrix are still about one order of magnitude 
larger than the others. Also in contrast to the CDM isocurvature case, we see 
that all coefficients increase about equally when polarization is included.

\begin{figure}
\centering
\includegraphics[width=0.6\textwidth, clip=true]{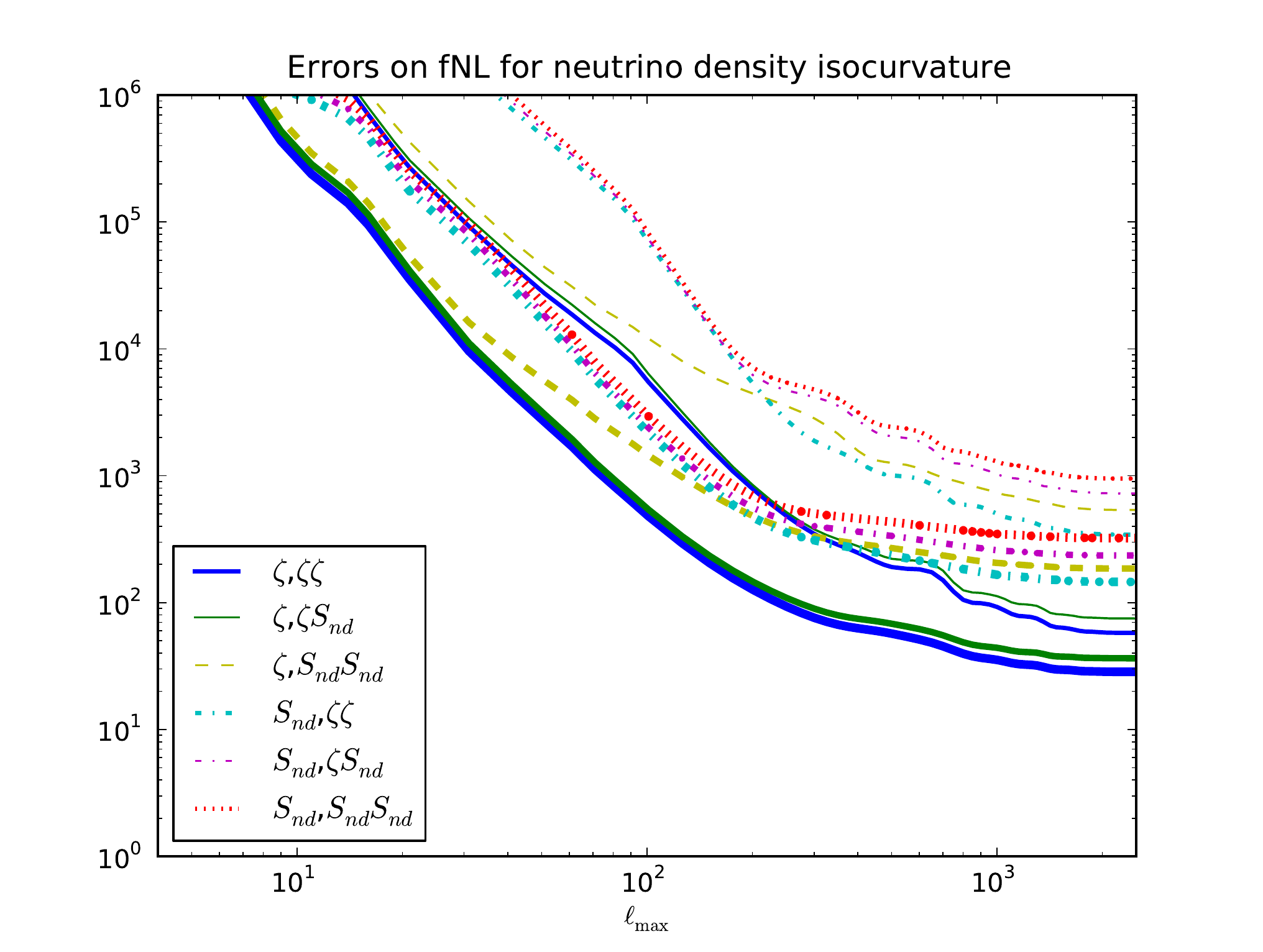}
\caption{Evolution of the $f_{\rm NL}$ parameter uncertainties as one increases 
the cut-off $\ell_{\rm max}$, for the neutrino density isocurvature mode. 
The six thinner curves describe the situation if only temperature data is 
used, while for the six thicker curves both temperature and E-polarization 
data are included.}
\label{Va}
\end{figure}

The corresponding uncertainties on the six non-Gaussianity parameters are 
(taking into
account the correlations)\footnote{While we were finalizing our manuscript, 
we became aware of the work \citep{Kawakami:2012ke}, where the authors 
also investigate neutrino density isocurvature non-Gaussianity. Their 
numbers for the uncertainties are very similar to ours 
(note that they use the six non-Gaussianity parameters that we introduced in 
\citep{LvT1} but in a different ordering), although they use a different
selection of Planck channels.}
\be
\Delta \tilde f^i=\{28, 36, 190, 150, 240, 320\}.
\ee  
When using temperature only, the uncertainties increase to
\be
\Delta \tf^i=\{58, 75, 540, 340, 720, 950\}\qquad {\rm (no \ polarization)}.
\ee
The evolution of the uncertainties as a function of $\ell_\mathrm{max}$ is shown 
in Fig.~\ref{Va}.
As in the CDM case, the $(\zeta,\zeta\zeta)$ and $(\zeta,\zeta S)$ 
non-Gaussianity parameters can be determined more accurately than the other
four, although the difference is not as big as for CDM. Unlike for CDM, all
parameters gain about the same from the inclusion of polarization, and all
uncertainties continue to decrease when higher multipoles are probed,
since the neutrino density isocurvature power spectrum does not decrease
as steeply as the CDM isocurvature one.

The correlation matrix is given in Table~\ref{correlation_nd}.
If one assumes the parameters to be independent, one finds
\be
\Delta\tf^i= \{5.0, 5.1, 30, 24, 20, 69\} \qquad {\mathrm{(single \ parameter)}}.
\ee
One sees that the increase of the uncertainties due to the correlations is much
more important here than for CDM, due to the larger correlations between
the various modes.

\begin{table}
\begin{center}
\begin{tabular}{|cccccc|}
\hline
$(\zeta, \zeta\zeta)$ & $(\zeta, \zeta S)$ & $(\zeta, S S)$ &  $(S,\zeta\zeta)$  & $(S,\zeta S)$ &  $(S,SS)$\\
\hline
 $1.$ & $-0.83\,(0.82) $ & $0.42\,(0.31) $ & $-0.73\,(0.67)$& $0.26\,(0.17) $ & $-0.16\,(0.09) $ \\
 - & $1.$ & $-0.58\,(0.45)$ & $0.58\,(0.55)$ & $-0.31\,(0.22)$& $0.25\,(0.20)$  \\
 - & - &  1. & $0.03\,(0.35)$ & $-0.45\,(0.69)$ & $0.19\,(0.41)$ \\
 - & - & - & 1. & $-0.76\,(0.81)$ & $0.58\,(0.64)$ \\
 - & - & - & - & 1. & $-0.84\,(0.86)$ \\
 - & - & - & - & - & 1.
\\ \hline
\end{tabular}
\caption{Correlation matrix for the neutrino density isocurvature mode. Only the upper half coefficients are indicated,  since the matrix is symmetric. The value between parentheses corresponds to the correlations obtained  without including the polarization (the absence of  sign in the parentheses means that it is unchanged with respect to the value taking into account the polarization).}
\label{correlation_nd}
\end{center}
\end{table}

\subsection{Neutrino velocity isocurvature mode}

\def\text{}
\begin{table}
\footnotesize
\begin{center}
\begin{tabular}{|cccccc|}
\hline
$(\zeta, \zeta\zeta)$ & $(\zeta, \zeta S)$ & $(\zeta, S S)$ &  $(S,\zeta\zeta)$  & $(S,\zeta S)$ &  $(S,SS)$\\
\hline
$3.9 \,(2.5) \times 10^{\text{-2}}$ & $3.5 \,(1.1) \times 10^{\text{-2}}$ & $74 \,(5.1) \times 10^{\text{-4}}$ &
   $15\,(9.5) \times 10^{\text{-3}}$ & $16 \,(5.6)\times 10^{\text{-3}}$ & $40 \,(2.6) \times 10^{\text{-4}}$
\\ 
- & $73 \,(7.6) \times 10^{\text{-3}}$ & $23 \,(1.3)\times 10^{\text{-3}}$ &
   $13 \,(4.0)\times 10^{\text{-3}}$ & $30 \,(3.4)\times 10^{\text{-3}}$ & $12 \,(0.65)\times 10^{\text{-3}}$
\\ 
- & - & $80 \,(4.3) \times 10^{\text{-4}}$ &
   $29 \,(1.6)\times 10^{\text{-4}}$ & $96 \,(4.8)\times 10^{\text{-4}}$ & $45 \,(2.1)\times 10^{\text{-4}}$
\\ - & - & - & $5.9 \,(3.6)\times 10^{\text{-3}}$ & $6.3 \,(2.1)\times 10^{\text{-3}}$ & $16 \,(0.88)\times 10^{\text{-4}}$
\\- & - & - & - & $13 \,(1.6)\times 10^{\text{-3}}$ & $54 \,(2.5)\times 10^{\text{-4}}$
\\- & - & - & - & - & $28 \,(1.1)\times 10^{\text{-4}}$
\\ \hline
\end{tabular}
\caption{Fisher matrix for the neutrino velocity isocurvature mode.
Only the upper half coefficients are indicated,  since the matrix is symmetric. The value between parentheses corresponds to the Fisher matrix components obtained  without including the polarization.}
\label{table_nv}
\end{center}
\end{table}

\begin{figure}
\centering
\includegraphics[width=0.6\textwidth, clip=true]{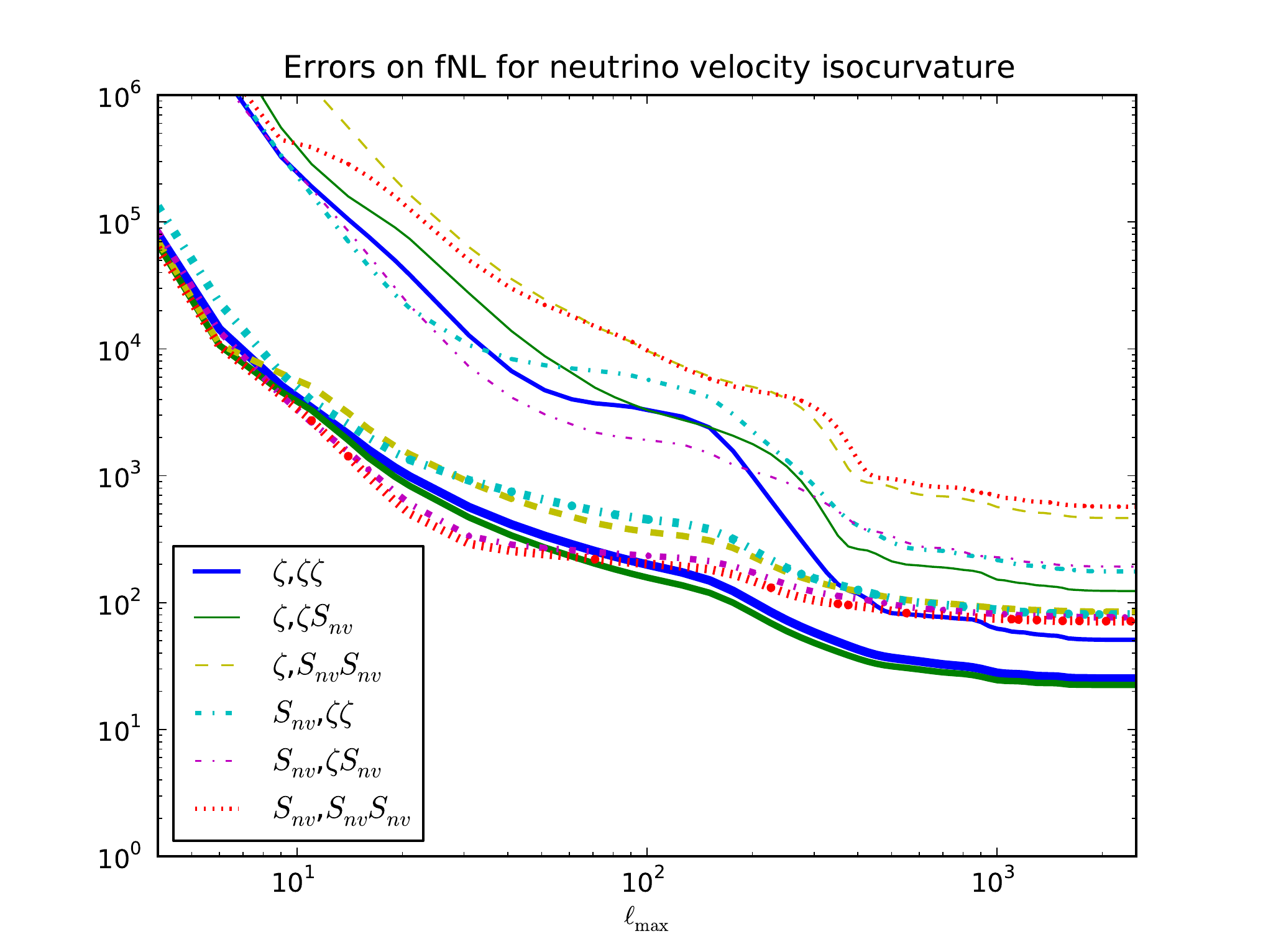}
\caption{Evolution of the $f_{\rm NL}$ parameter uncertainties as one increases 
the cut-off $\ell_{\rm max}$, for the neutrino velocity isocurvature mode. 
The six thinner curves describe the situation if only temperature data is 
used, while for the six thicker curves both temperature and E-polarization 
data are included.}
\label{Vb}
\end{figure}

For a neutrino velocity isocurvature mode, we have obtained the Fisher matrix in Table~\ref{table_nv}. One notices that, including polarization, all entries
are roughly of the same order of magnitude, but without polarization, they
vary a lot. 
The corresponding uncertainties on the six non-Gaussianity parameters are 
(taking into account the correlations) 
\be
\Delta \tilde f^i=\{25, 22, 85, 81, 77, 71\}.
\ee  
When using temperature only, the uncertainties increase to
\be
\Delta \tf^i=\{51, 120, 460, 180, 190, 570\}\qquad {\rm (no \ polarization)}.
\ee
The evolution of the uncertainties as a function of $\ell_\mathrm{max}$ is shown 
in Fig.~\ref{Vb}.

One sees that in this case the difference between the first two and
the other four uncertainties (when including polarization) is much smaller 
than for CDM or neutrino
density isocurvature (a factor of about 3 compared to a factor of
about 15 in the CDM case). In particular, the latter four can be determined
more accurately than in the case of CDM or neutrino density isocurvature. 
However, the improvement due to
polarization is much more important than in the neutrino density case,
in particular for the $(\zeta,\zeta S)$, the $(\zeta, S S)$ and the
$(S,S S)$ parameters. One can understand why the $(\zeta,\zeta S)$
mode, for example, gains much more from polarization than the
$(\zeta,\zeta\zeta)$ mode with a similar argument as the one presented
for CDM isocurvature. The dominant contributions to these modes both
depend on ${\cal G}_{\zeta\zeta}(L)$ (defined in (\ref{G_and_H})), but
for $(\zeta,\zeta\zeta)$ this is multiplied by ${\cal H}_\zeta(\ell)$
and for $(\zeta,\zeta S)$ by ${\cal H}_S(\ell)$. And as one can see from
Fig.~\ref{HK_fig}, the ratio ${\cal H}_S/{\cal H}_\zeta$ for neutrino
velocity isocurvature increases enormously when one passes from
temperature to polarization (remember that it is the lowest values of
$\ell$ that contribute most to the squeezed configuration).

The correlation matrix is given in Table~\ref{correlation_nv}.
If one assumes the parameters to be independent, one finds
\be
\Delta\tf^i= \{5.0, 3.7, 11, 13, 8.7, 19\} \qquad {\mathrm{(single \ parameter)}}.
\ee
Hence one sees that the correlations in the case of neutrino velocity 
isocurvature are more important than for CDM, but less than for neutrino 
density.

\begin{table}
\begin{center}
\begin{tabular}{|cccccc|}
\hline
$(\zeta, \zeta\zeta)$ & $(\zeta, \zeta S)$ & $(\zeta, S S)$ &  $(S,\zeta\zeta)$  & $(S,\zeta S)$ &  $(S,SS)$\\
\hline
 1. & $-0.50\,(0.42)$ & $0.08\,(0.11)$ & $-0.82\,(0.66)$ & $0.34\,(0.34)$ & $-0.13\,(+0.05)$ \\
 - & 1. & $-0.60\,(0.80)$ & $0.25\,(-0.18)$ & $-0.26\,(0.35)$ & $0.36\,(0.39)$ \\
 - & - &   1. & $0.37\,(0.44)$ & $-0.51\,(+0.07)$ & $-0.29\,(0.75)$ \\
 - & - & - & 1. &  $-0.74\,(0.61)$ & $0.25\,(-0.10)$ \\
 - & - & - & - & 1. & $-0.46\,(0.36)$ \\
 - & - & - & - & - & 1.
\\ \hline
\end{tabular}
\caption{Correlation matrix for the neutrino velocity isocurvature mode.
Only the upper half coefficients are indicated,  since the matrix is symmetric. The value between parentheses corresponds to the correlations obtained  without including the polarization (the absence of  sign in the parentheses means that it is unchanged with respect to the value taking into account the polarization).}
\label{correlation_nv}
\end{center}
\end{table}

\section{Constraints on early universe models}

In the previous section, we have studied how to obtain constraints on the six non-linearity coefficient $\tf_{NL}^{(i)}$ without assuming any particular relation between them.
In the context of an early universe model, or in a class of models, one can go further and use the results of the previous section to  obtain some constraints on the parameters of the model itself.

\subsection{General analysis}

As we have seen earlier, isocurvature perturbations require the existence of at least two degrees of freedom in the early universe. So, for simplicity, let us focus on models with two scalar fields, $\phi$ and $\sigma$, such that isocurvature perturbations are generated only by the fluctuations 
of $\sigma$ and all non-linearities are also dominated by their contribution from $\sigma$. This means that we have
\be
\label{ZS}
\zeta=N_\phi^\zeta\delta\phi+N_\sigma^\zeta \delta\sigma+\frac12 N_{\sigma\sigma}^\zeta \delta\sigma^2\,, \qquad 
S=N_\sigma^S \delta\sigma+\frac12 N_{\sigma\sigma}^S \delta\sigma^2\,.
\ee

Using the general expressions (\ref{lambda}--\ref{f_NL}), one easily finds that the coefficients are interdependent and can be written in the form
\begin{eqnarray}
\tf_{NL}^{I, \zeta\zeta}&=& \mu_I \, \Xi^2,\\
\tf_{NL}^{I, \zeta S}&=& \varepsilon_{_{\zeta S}}\, \mu_I \, \alpha^{1/2} \,  \Xi^{3/2},\\
\tf_{NL}^{I, S S}&=& \mu_I \,\alpha \, \Xi,
\end{eqnarray}
where we have introduced the contribution of $\sigma$ in the adiabatic power spectrum,
\be
\Xi\equiv \frac{(N^{\zeta}_\gs)^2}{(N^{\zeta}_\phi)^2+(N^{\zeta}_\gs)^2}\,,
\ee
and the isocurvature to adiabatic ratio,
\be
\alpha\equiv \frac{(N^{S}_\gs)^2}{(N^{\zeta}_\phi)^2+(N^{\zeta}_\gs)^2}\,.
\ee
Note that the extraction of an isocurvature component in the power spectrum would fix $\alpha$ while,  in principle, $\Xi$ could be determined from observations by measuring both the bispectrum and the trispectrum coefficients, since 
they satisfy consistency relations~\citep{Langlois:2010fe} similar to the purely adiabatic relation $\tau_{NL}=\tf_{NL}^2/\Xi$.

The two coefficients 
\be
\mu_I=\frac{N^I_{\gs\gs}}{(N^{\zeta}_\gs)^2}\qquad (I=\zeta, S)
\ee
fully characterize the non-Gaussianity of the adiabatic and isocurvature perturbations, respectively, while $\varepsilon_{_{\zeta S}}$ denotes the relative sign of $\zeta$ and $S$: $\varepsilon_{_{\zeta S}}=1$ if they have the same sign, $\varepsilon_{_{\zeta S}}=-1$ otherwise.
Interestingly, $\tf_{NL}^{I, \zeta\zeta}$ and $\tf_{NL}^{I, SS}$ share the same sign as $\mu_I$, whereas $\tf_{NL}^{I, \zeta S}$ can have a different sign. The hierarchy between the coefficients $\tf_{NL}^{I, JK}$, $I$ being fixed, depends on the relative amplitude of $\alpha$ and $\Xi$: $\tf_{NL}^{I, \zeta\zeta}$ dominates if $\Xi\gg \alpha$, whereas $\tf_{NL}^{I, SS}$ dominates if $\Xi\ll \alpha$. 

For given values of $\alpha$ and $\Xi$, the uncertainties on the two parameters $\mu_\zeta$ and $\mu_S$ are determined from the ``projected'' Fisher matrix
\be
{\cal F}_{IJ}=\sum_{i,j}{\cal \pi}_{(i)I} \, F_{ij} \, {\cal \pi}_{(j)J}
\ee
with
\be
{\cal \pi}_{(i)\zeta}=\{\Xi^2, \varepsilon_{_{\zeta S}} \alpha^{1/2}\Xi^{3/2}, \alpha\Xi, 0,0,0\},\quad 
{\cal \pi}_{(i)S}=\{0,0,0, \Xi^2, \varepsilon_{_{\zeta S}}\alpha^{1/2}\Xi^{3/2}, \alpha\Xi\}\,.
\ee
From this $2\times 2$ Fisher matrix, one can easily deduce the expected uncertainties on the two parameters $\mu_\zeta$ and $\mu_S$, by using the analog of (\ref{errors}).

\subsection{Illustrative example}

In  \citep{Langlois:2011zz,LvT1} we have studied a class of models which produces perturbations of the above type. In these models, $\sigma$ is a curvaton which decays into radiation and CDM. Since part of the CDM can have been produced before the decay of the curvaton, one can introduce as a parameter the fraction of CDM created by the decay as
\be
f\equiv\frac{\gamma_c\,  \Omega_\gs}{\Omega_c+\gamma_c\Omega_\gs},
\ee 
where the $\Omega$'s represent the relative abundances just before the decay
and $\gamma_c$ is the fraction of the curvaton energy density transferred 
into CDM.
The second relevant parameter,
\be
r\equiv \frac{3 (1-\gamma_c)\,  \Omega_\gs}{(1-\gamma_c\, \Omega_\gs)(4-\Omega_\gs)}\,,
\ee
quantifies the transfer between the pre-decay and post-decay perturbations~\citep{Langlois:2011zz} (one finds $\zeta_\gamma^\mathrm{after}=(1-r)\zeta_\gamma^\mathrm{before}$ at the linear level).

As shown in \citep{Langlois:2011zz,LvT1}, one can derive the ``primordial'' perturbations $\zeta$ and $S$ as expansions, up to second order, in terms of $\delta\gs$, which yield the coefficients $N^I_{\gs}$ and $N^I_{\gs\gs}$. Using these results and assuming $r\ll1$, one obtains 
\be
\mu_\zeta=\frac{3}{2r}, \qquad 
\mu_S=\frac{9}{2 r^2}\left(f(1-2f)-r\right)\,,
\ee
while $\varepsilon_{_{\zeta S}}= {\rm sgn}(f-r)$ and
\be
\alpha= 9\left(1-\frac{f}{ r}\right)^2 \,\Xi \, .
\ee
In the regime $f\ll r \ll 1$, one finds $\mu_\zeta=3/(2r)$, 
$\mu_S=-9/(2r)$ and $\alpha=9\Xi$, with 
$\varepsilon_{_{\zeta S}}=-1$ and the amplitudes of non-Gaussianities depend only on the parameter $r$. 
By contrast, in the opposite regime $r \ll f_c  \ll 1$, $\mu_\zeta=3/(2r)$, 
$\mu_S=9f/(2 r^2)$ 
and $\alpha= 9\, \Xi (f/r)^2$. The coefficients $\tf_{NL}^{S, IJ}$, which depend on $\mu_S$, are thus enhanced with respect to the coefficients $\tf_{NL}^{\zeta, IJ}$ in the latter case.
As discussed in more detail in \citep{LvT1}, the above results show that a
small isocurvature fraction in the power spectrum can, for certain parameter
values, be compatible with a dominantly isocurvature bispectrum 
detectable by Planck (e.g.\ $\alpha\simeq 10^{-2}$ and $r\ll f_c\simeq 10^{-8}$
yields $\tf_{\rm NL}^{S,SS}\simeq 5\times 10^3$).

\chapter{The bispectra of galactic CMB foregrounds and their impact on
  primordial non-Gaussianity estimation}
\label{JRvTapp}

This appendix contains the paper \citep{JRvT}, with the exception of its
section~2 (which only contains a summary of the binned bispectrum estimator
as presented earlier in this thesis) and of its conclusions, used in the summary
in section~\ref{summJRvT}. This paper was written in collaboration with
Gabriel Jung and Benjamin Racine.

We use the binned bispectrum estimator to determine the bispectra of
the dust, free-free, synchrotron, and AME galactic foregrounds using
maps produced by the \texttt{Commander} component separation method from Planck
2015 data. We find that all of these peak in the squeezed configuration,
allowing for potential confusion with in particular the local primordial
shape. Applying an additional functionality implemented in the binned
bispectrum estimator code, we then use these galactic bispectra as
templates in an $f_\mathrm{NL}$ analysis of other maps. After testing
and validating the method and code with simulations, we show that
we detect the dust in the raw 143~GHz map with the expected amplitude
(the other galactic foregrounds are too weak at 143~GHz to be detected)
and that no galactic residuals are detected in the cleaned CMB map.
We also investigate the effect of the mask on the templates and the effect
of the choice of binning on a joint dust-primordial $f_\mathrm{NL}$
analysis.

\section{Introduction}
\label{sec:introduction}

The exploration of the CMB as a source of high-precision information on
cosmology and as the best (if somewhat opaque) window on
the primordial universe started in earnest with the first WMAP release in 2003 \citep{Spergel:2003cb}.  
The Planck satellite with its three releases in 2013, 2015, and 2018 \citepalias{planck2013-01, planck2015-01, planck2018-01} raised
the game to unprecedented levels of precision. Still, the amount of information
we have about the primordial universe, and in particular on the period of
inflation, remains very limited. Apart from looking for possible new
observables, it is also very important to work as much as possible on the
observables that we do have, from both ends: from the observational side to
improve estimators and data cleaning to get as precise a value as we can, and
from the theoretical side to improve (inflationary) predictions so that we can
draw the theoretical consequences from the observations.
Microwave observations are contaminated by astrophysical
foregrounds, which can be extra-galactic or galactic in origin. In order to
improve the quality of the data used for the CMB analyses, these foregrounds
are first removed as much as possible by component separation methods, which
produce so-called cleaned CMB maps, although these still contain foreground
residuals at some level.

Some of the most important inflationary observables are the non-Gaussianity
parameters $f_\mathrm{NL}$. Non-Gaussianity means that not all information
about the CMB is contained in its two-point correlation function / power 
spectrum, as would be the case for a Gaussian distribution. The lowest-order 
deviation from Gaussianity will lead to a non-zero bispectrum, the Fourier 
transform (or spherical harmonic transform on the celestial sphere) of the 
three-point correlation function. Standard single-field slow-roll
inflation produces an unobservably small non-Gaussian signal \citep{Maldacena:2002vr, Acquaviva:2002ud}, but other
inflation models predict larger amounts. Moreover, different types of models
predict differently shaped bispectra,
and we can look for the presence of any of them. The amplitude of each is
parametrized by its own $f_\mathrm{NL}$ parameter. Some of the most important
bispectrum templates are the local shape \citep{Gangui:1993tt} (typically produced by multiple-field
inflation models) and the equilateral \citep{Creminelli:2005hu} and orthogonal \citep{Senatore:2009gt} shapes (typically produced
by single-field models with non-standard kinetic terms), see \citep{Chen:2010xka} for a review. So far there is no
detection of a primordial $f_\mathrm{NL}$ value inconsistent with zero, but this
null detection with precise error bars has led to the exclusion of inflation
models that predict too much non-Gaussianity~\citepalias{planck2013-24, planck2015-17}.

In order to extract any information about primordial bispectral non-Gaussianity
from the CMB data, given that this information is
primarily parametrized in the form of the $f_\mathrm{NL}$ amplitude
parameters of the different bispectrum shapes, we need an estimator for
$f_\mathrm{NL}$. This estimator should be unbiased as well as optimal (or
effectively optimal given the accuracy of the experiment), which
means it has the smallest variance theoretically possible, to extract the
primordial $f_\mathrm{NL}$ from real data contaminated by astrophysical
foreground residuals and experimental effects like noise. In addition the
estimator implementation should be fast enough to make data analysis
possible in practice. Three such estimators were developed and used for the
official Planck analysis: the KSW estimator \citep{Komatsu:2003iq, Yadav:2007rk, Yadav:2007ny} (this was the only one used for
the official WMAP analysis as well, as the other ones did not yet exist at
that time), the binned estimator~\citep{BvTC, BRvT}, and
the modal estimator \citep{Fergusson:2009nv, Fergusson:2010dm, Fergusson:2014gea}. All three are based
on the same theoretical exact $f_\mathrm{NL}$ estimator, but differ in the
approximations made in their implementations to make them fast enough for
practical use. In addition to being $f_\mathrm{NL}$ estimators, the binned
and modal estimators also allow for the determination of the full bispectrum
of the data. 

In this paper we will use the binned bispectrum estimator to determine the
bispectra of various galactic foregrounds (dust, free-free, anomalous microwave
emission (AME), synchrotron), and then use those bispectra as templates to 
determine the corresponding $f_\mathrm{NL}$ parameters in other
maps. The aim of the paper is threefold. In the first place it is a proof
of concept. In fact, the ability to determine bispectra from maps and then
use them as templates was one of the original motivations for developing the
binned bispectrum estimator, but so far this potential ability had not yet
been put to the test in practice. Secondly, it is interesting to study the
bispectra of these galactic foregrounds as an aim in itself, and see how they
correlate with the primordial templates.\footnote{It should be noted that for 
the purpose of studying the non-Gaussianity of a galactic foreground in itself, 
the bispectrum would probably not be the best tool. Due to their localized 
(non-isotropic) nature, an approach in pixel space instead of harmonic space 
would seem more logical. Minkowski functionals, for example, have been used to study galactic synchrotron radiation in the context of 21-cm line studies \citep{Rana:2018oft}. However, in this paper we are primarily interested in 
seeing how much these foregrounds contaminate a determination of primordial 
non-Gaussianity, which is in general isotropic and for which the bispectrum is 
then an optimal tool.} 
Finally, in the third place we want to test if any detectable 
galactic non-Gaussianity remains in the cleaned Planck CMB maps.
The quality of these maps has been tested in many different ways, mostly
using the power spectrum, but also by seeing if primordial $f_\mathrm{NL}$
measurements remained optimal, and they passed these tests. Still, it is good 
to also test for the presence of non-Gaussian galactic residuals directly.

The fact that we restrict ourselves to galactic foregrounds is because for
the most important extra-galactic foregrounds templates already exist in
analytic form \citep{Komatsu:2001rj, Lacasa:2013yya} (theoretically or heuristically determined), and can for example be used to compute biases \citep{Hill:2018ypf}. 
But no such templates exist for galactic foregrounds. One of the advantages of the
binned bispectrum estimator is that it does not necessarily require templates
in analytic form, but can also deal with a numerical binned template.
We restrict ourselves in this paper to temperature maps only. A preliminary
exploration of the polarization maps and simulations of the 2015 Planck
release showed that these were not yet sufficiently accurate to make
a similar analysis in polarization meaningful.

The paper is organized as follows.
In sections~\ref{sec:foregrounds} and~\ref{sec:analyses}, our data
analysis results using the binned bispectrum estimator on data from
the 2015 Planck release are presented. In
section~\ref{sec:foregrounds}, several galactic foregrounds are
studied at the bispectral level, with some additional results in the second
appendix in section~\ref{ap:appendices}. The newly determined templates from
these foregrounds are then applied to several CMB maps (Gaussian
simulations and real data) in section~\ref{sec:analyses}.
One of the early questions we had, was if we should include
a linear correction term in the foreground bispectrum templates. To answer
that, we had to look at the theoretical derivation of the linear term and
its assumptions, and found that it makes no sense to add a linear correction
for highly non-Gaussian bispectra. We include
a full derivation in the first appendix in section~\ref{ap:appendices}.

\section{Galactic foregrounds}
\label{sec:foregrounds}

In this section, we study several galactic foregrounds with the binned bispectrum estimator. As discussed previously, when studying non-Gaussianity in CMB data maps, the usual method is to compare the observed bispectrum to different theoretical shapes using the inner product \eqref{innerprod}. The determined parameters $\fnl$ simply indicate to what extent these shapes are present in the data. Usually this method is applied to several shapes which have analytical expressions and it includes primordial non-Gaussianity (generated during inflation) but also late-time bispectra (generated after recombination) like extra-galactic foregrounds.
However, when observing the CMB, the main source of contamination is our own galaxy and there is no equivalent theoretical expression to describe the non-Gaussianity of galactic foregrounds yet. There are many techniques to clean the maps from the presence of different galactic foregrounds (see \citepalias{planck2015-09, planck2015-10, planck2015-25} for a review) and CMB analyses at the bispectral level are generally performed on these clean maps. In this section, we use the fact that an analytical formulation of theoretical shapes is not mandatory for use with the binned bispectrum estimator, allowing us to examine these foregrounds too. Indeed, to use the inner product \eqref{innerprod}, one only needs the numerical binned theoretical bispectrum. This means that in principle, the binned bispectrum of any map determined numerically could be used as theoretical template for the analysis of another map under the condition that the binning is the same. In this way we determine templates using the maps of different galactic foregrounds from the 2015 Planck release obtained by the \texttt{Commander} component separation technique \citep{Eriksen:2004ss, Eriksen:2007mx}.\footnote{\url{https://pla.esac.esa.int}}

In section \ref{sec:analyses} we will use these new numerical templates on the CMB cleaned maps studied in \citepalias{planck2015-17}. To be more precise, we will use the \texttt{SMICA} \citep{Cardoso:2008qt} CMB map from the 2015 Planck release. We will also study the raw 143 GHz map, which is the dominant frequency channel in the \texttt{SMICA} map (see figure D.1 of \citepalias{planck2015-09}). While it is the best channel to observe the CMB (best combination of a low noise level and a good resolution), that is not the case for the different foregrounds (at least if the goal was to study the physics of these foregrounds). Nevertheless, here we only need to estimate their eventual contamination to the CMB signal. At that frequency, the CMB dominates the sky after masking the brightest parts (galactic plane and strong point sources). For this, we use the temperature common mask of the Planck 2015 release, which is a combination of the masks of the different component separation methods \citepalias{planck2015-09}. In section~\ref{sec:noise-masks}, we will discuss the influence of the mask on the different foregrounds by using a smaller one (\texttt{Commander} mask). Finally, another important choice is the binning which was determined using the ratio $R$ defined in \eqref{binning_error} to be optimal for the primordial shapes. It is true that this criterion has nothing to do with the galactic foregrounds, but our ultimate goal is to determine the primordial shapes optimally, not the galactic ones. To illustrate the method, we start by studying the case of thermal dust.

\subsection{Thermal dust}
\label{sec:dust}

Above 100 GHz, the strongest contamination from galactic foregrounds is due to small dust grains ($\sim$ 1 $\mu$m or smaller) present in the interstellar medium. This dust plays an important role in galactic evolution (chemistry of interstellar gas, etc., see the textbook \citep{draine2010physics} for example), but it also has a large influence on astrophysical observations. Indeed dust grains are heated by the UV starlight they absorb, so they emit a thermal radiation (infrared) in the frequency range of CMB experiments. This emission is  well described by a modified blackbody model also called greybody (see \citepalias{planck2013-12, planckint-17, planck2015-10})
\begin{equation}
  \label{eq:greybody-dust}
  I(\nu) = A \nu^{\beta_d}B_\nu(T_d), 
\end{equation}
where $B_\nu$ describes Planck's law, $T_d \sim 20 K$ is the mean temperature and $\beta_d \sim 1.5$ is the free emissivity spectral index.

\begin{figure}
  \centering
   \includegraphics[width=0.49\linewidth]{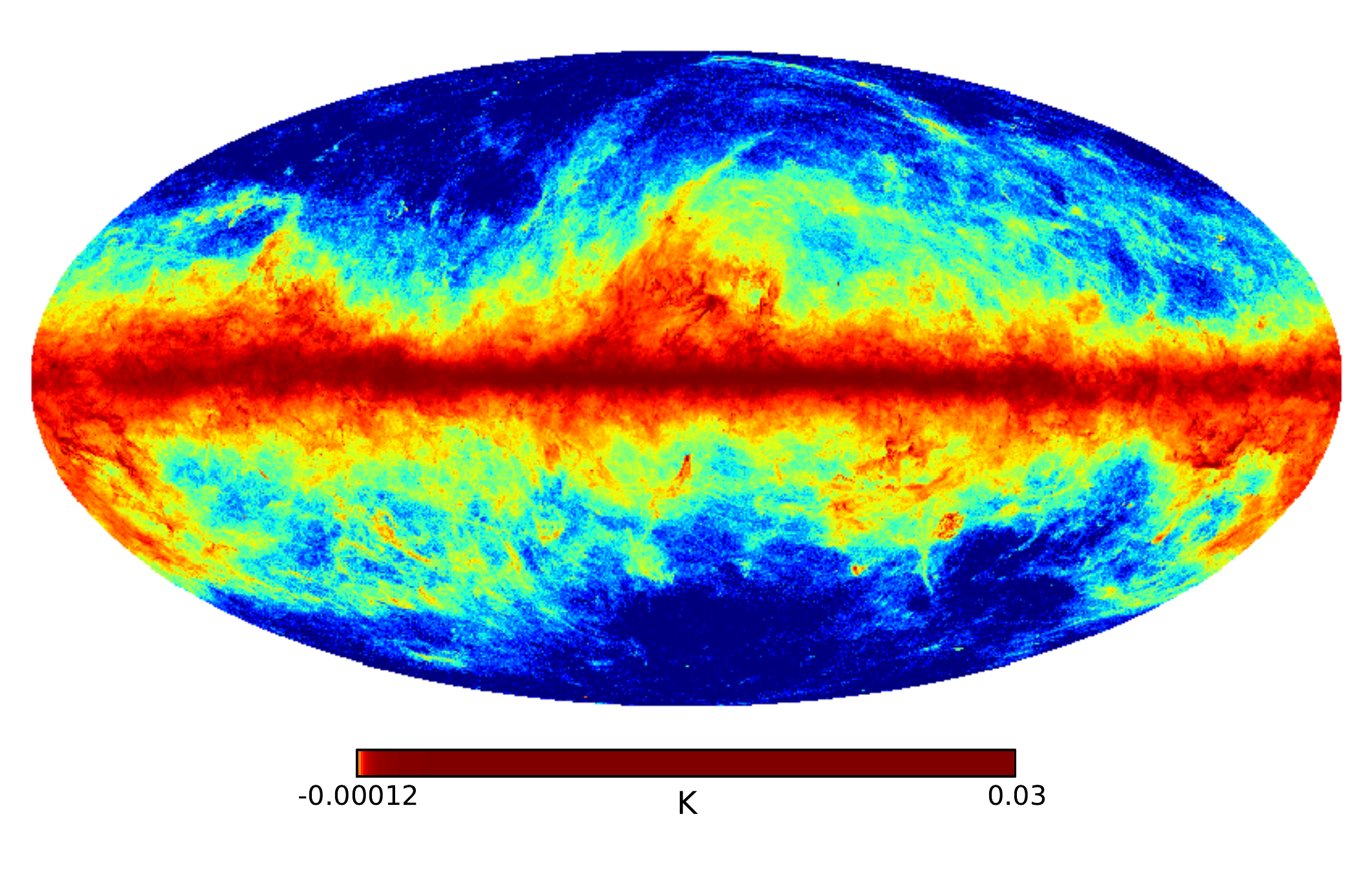}
 \includegraphics[width=0.49\linewidth]{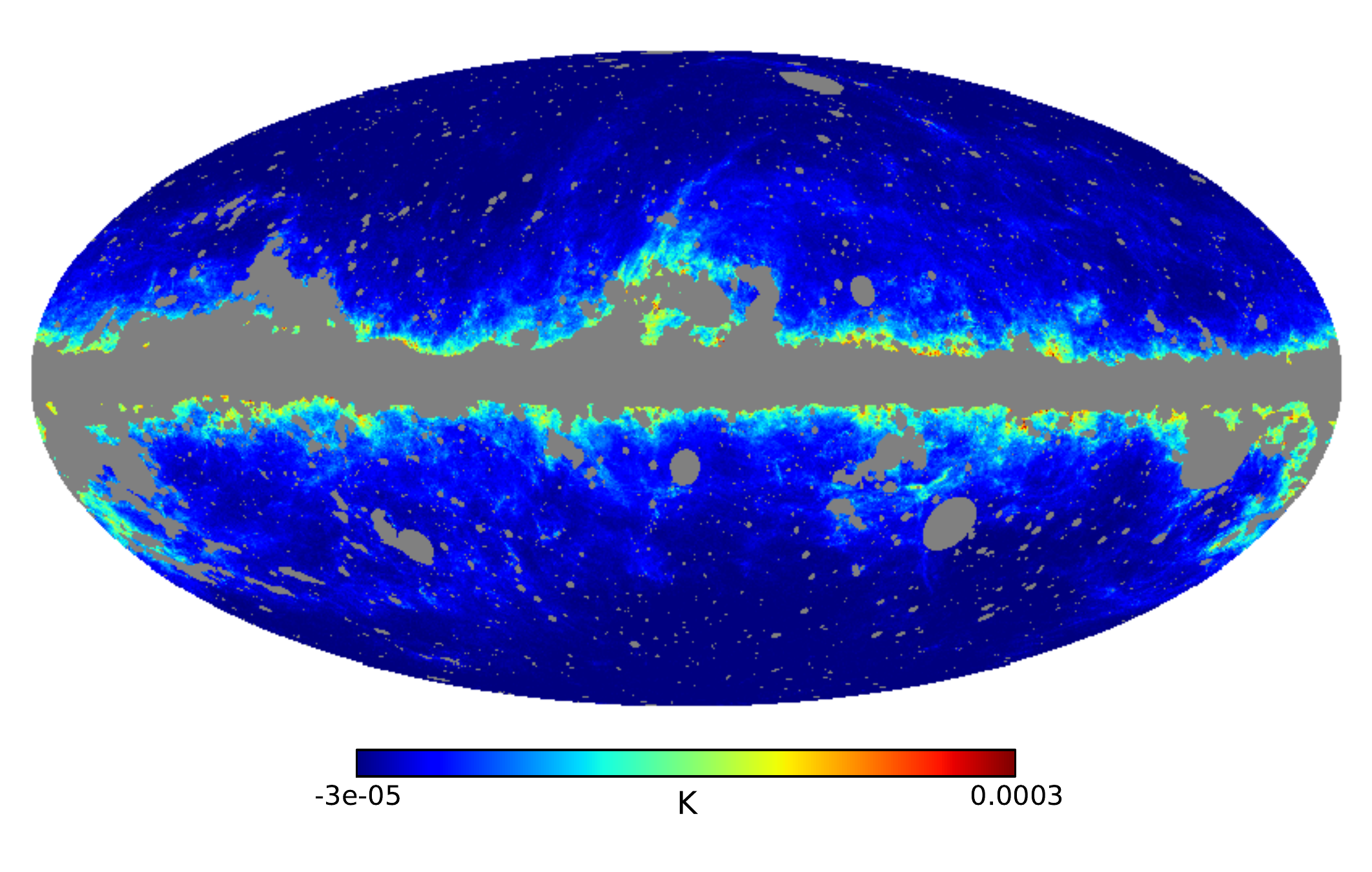}
  \caption{Unmasked (left) and masked (right) maps of thermal dust at 143 GHz from \texttt{Commander} (Planck 2015), using the common mask. On the left, the map is normalized using histogram equalization to highlight all the regions containing dust. On the right, the scale is linear, this shows that most of the signal is coming from the galactic plane near the mask. The range of the scale is also different.}
  \label{fig:dust-map}  
\end{figure}

Figure \ref{fig:dust-map} shows the map of the galactic thermal dust at 143 GHz, before and after applying the common mask. As mentioned before, we are interested in the contribution of the foregrounds in a CMB analysis (where a mask is always used to hide the galactic plane). Hence, the map on the right is the most important here because it is the actual contribution of dust that could be seen in a CMB analysis. In the following, we will be interested in the power spectrum and the bispectrum of this map. As expected, most of the signal comes from the galactic plane, and it is strongest close to the mask. Because of the dust localization, this emission is very non-Gaussian \citep{MivilleDeschenes:2007ya} and anisotropic (and this is also the case for the other galactic foregrounds studied in the next subsection). The bispectrum is not the best tool to describe such a localized non-Gaussianity (an estimator in pixel space would be better). However, we are only interested in the impact of this galactic foreground on the primordial shapes. This requires us to be careful with the different expressions of the binned bispectrum estimator, mostly derived using the weak non-Gaussianity approximation (see the first appendix in section~\ref{ap:appendices}). Concerning the observed bispectrum of the dust map, which is exactly what we need to make a dust template, it is still defined by \eqref{Bobsbinned} (divided by $f_\mathrm{sky}$ and only integrated over the non-masked part of the sky), without the linear correction terms which are not justified here.

\begin{figure}
  \centering
  \includegraphics[width=0.66\linewidth]{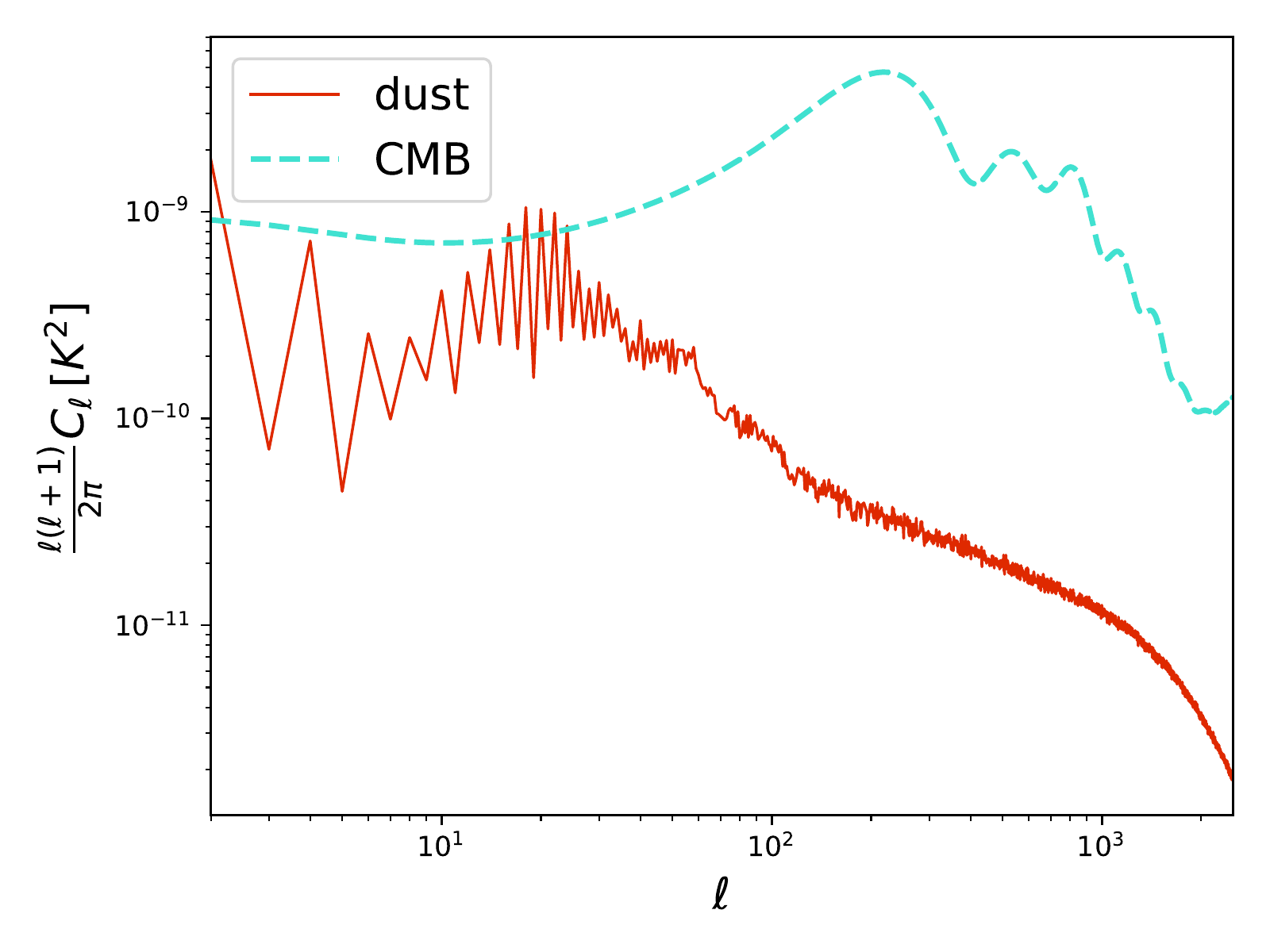}
  \caption{This figure shows the dust (at 143 GHz) and the CMB (Planck 2015 best fit) temperature power spectra (including the beam and the noise for both) as a function of the multipole $\ell$.}
  \label{fig:dust-power-spectrum}
\end{figure}

Before describing the dust bispectrum, it is interesting to examine the power spectra of the dust and CMB maps shown in figure \ref{fig:dust-power-spectrum}. It is clear that at 143 GHz, the CMB dominates except for the largest scales (smallest $\ell$) where the dust power spectrum has a sawtooth pattern. We can see that it is smaller (up to an order of magnitude) for each odd $\ell$ up to $\ell \sim \mathcal{O}(20)$. This is in fact due to the symmetry of the masked map in figure \ref{fig:dust-map} around the galactic plane when viewed on the largest scales. Because of this symmetry, the temperature is an even function of the angle $\theta$ (with the usual $\hat{\Omega}=(\theta,\varphi)$, where $\theta$ describes the latitude position), using the simple approximation that the mask can be seen as a band with all the dust signal on the border. The spherical harmonics $Y_{\ell 0}$ also have a similar symmetry around the galactic plane so they are the main contribution when decomposing in harmonic space. However, the $Y_{\ell 0}$ are even in $\theta$ only for $\ell$ even and they are odd for $\ell$ odd, so the odd terms have to be small. For the same reasons similar effects are expected in the dust bispectrum as far as large scales are concerned.

\begin{figure}
  \centering
\includegraphics[width=0.49\linewidth]{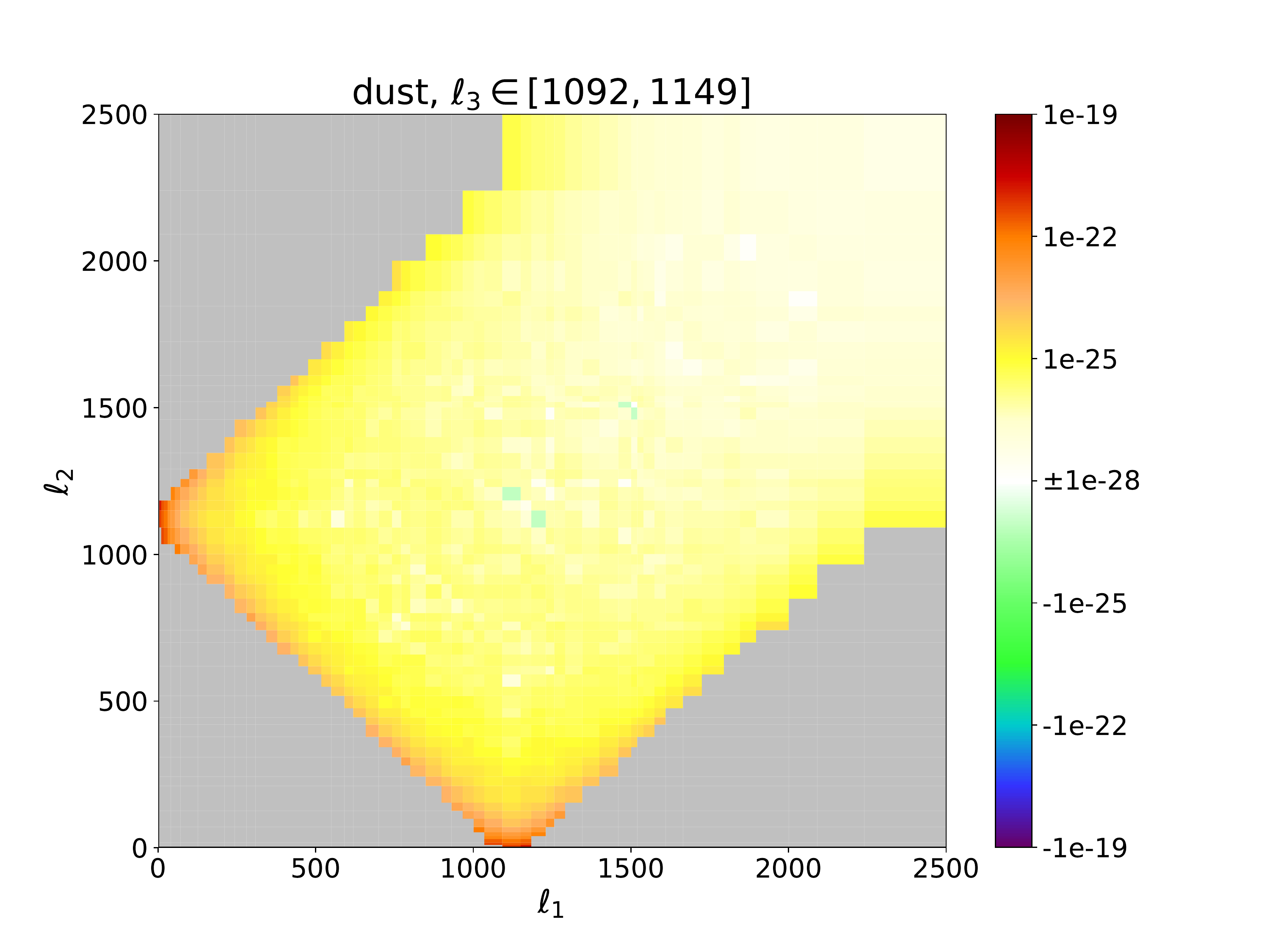}
  \includegraphics[width=0.49\linewidth]{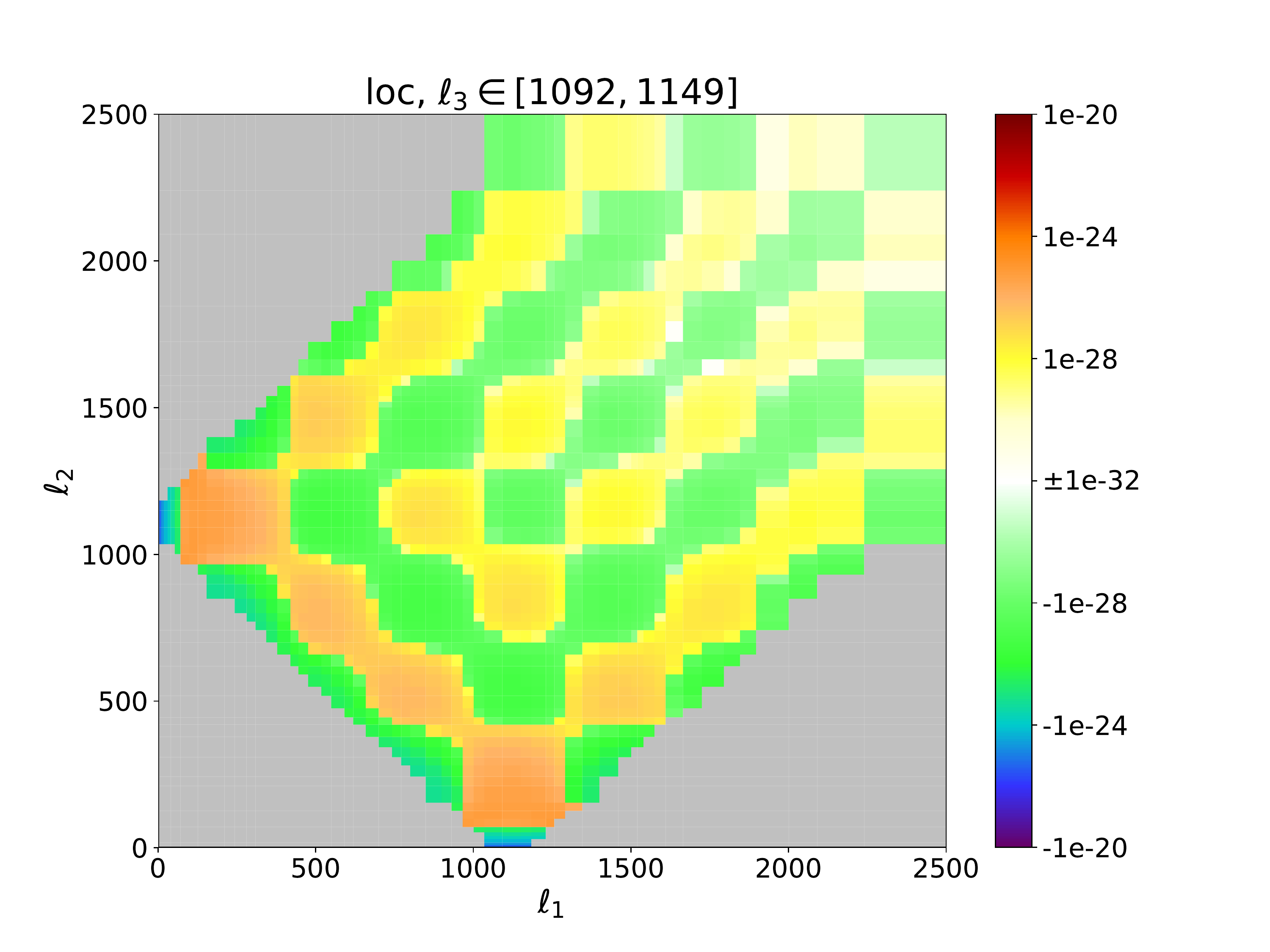}
  \caption{Left: observed binned bispectrum of the thermal dust as a function of the multipoles $\ell_1$ and $\ell_2$ for $\ell_3 \in [1092,1149]$.
  Right: theoretical bispectrum for the local shape in the case $\fnl^\mathrm{local}=1$ for the same bin of $\ell_3$. Note the difference of colour scale (both for maximal and minimal values).}
  \label{fig:dust-template}
\end{figure}

Moving on to the dust bispectrum, we use 2D-slices where the multipoles $\ell_1$ and $\ell_2$ go from 2 to 2500 but $\ell_3$ is in a chosen bin, in order to make it easy to visualize. Figure \ref{fig:dust-template} shows a slice ($\ell_3 \in [1092,1149]$) of the binned dust bispectrum compared to the local shape in the case $\fnl^\mathrm{local}=1$.\footnote{$\fnl^\mathrm{local}=1$ is still well within the observational bounds, but is very large compared to the predictions of standard slow-roll single-field inflation $\fnl^\mathrm{local}\sim\mathcal{O}(10^{-2})$.} If we compare the bispectrum amplitudes, it is clear that the dust is several orders of magnitude larger than the local shape. Moreover, as expected, acoustic oscillations present in both the CMB power spectrum and the local theoretical bispectrum are not there in the case of thermal dust.

\begin{figure}
  \centering
  \includegraphics[width=0.49\linewidth]{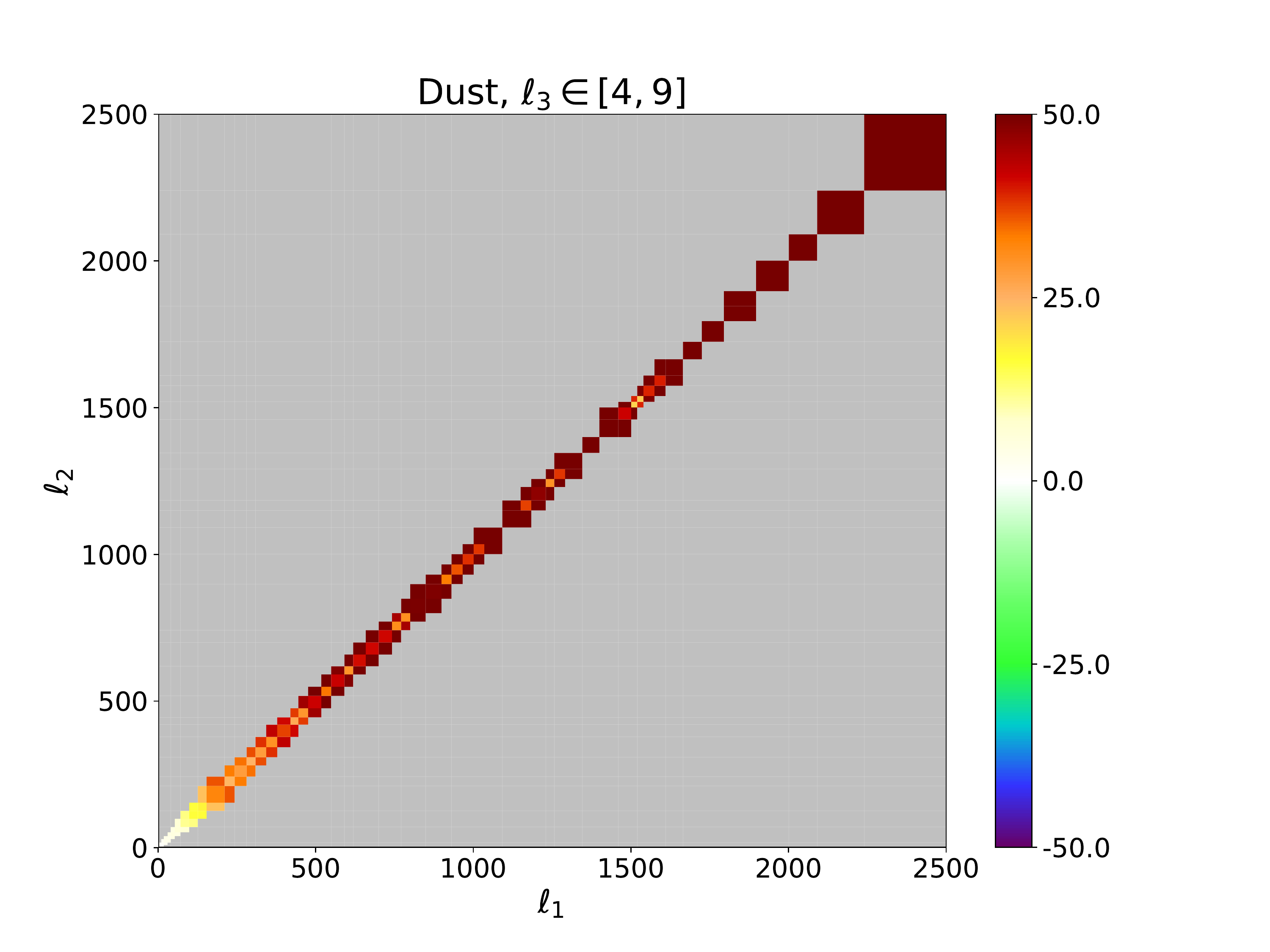}
  \includegraphics[width=0.49\linewidth]{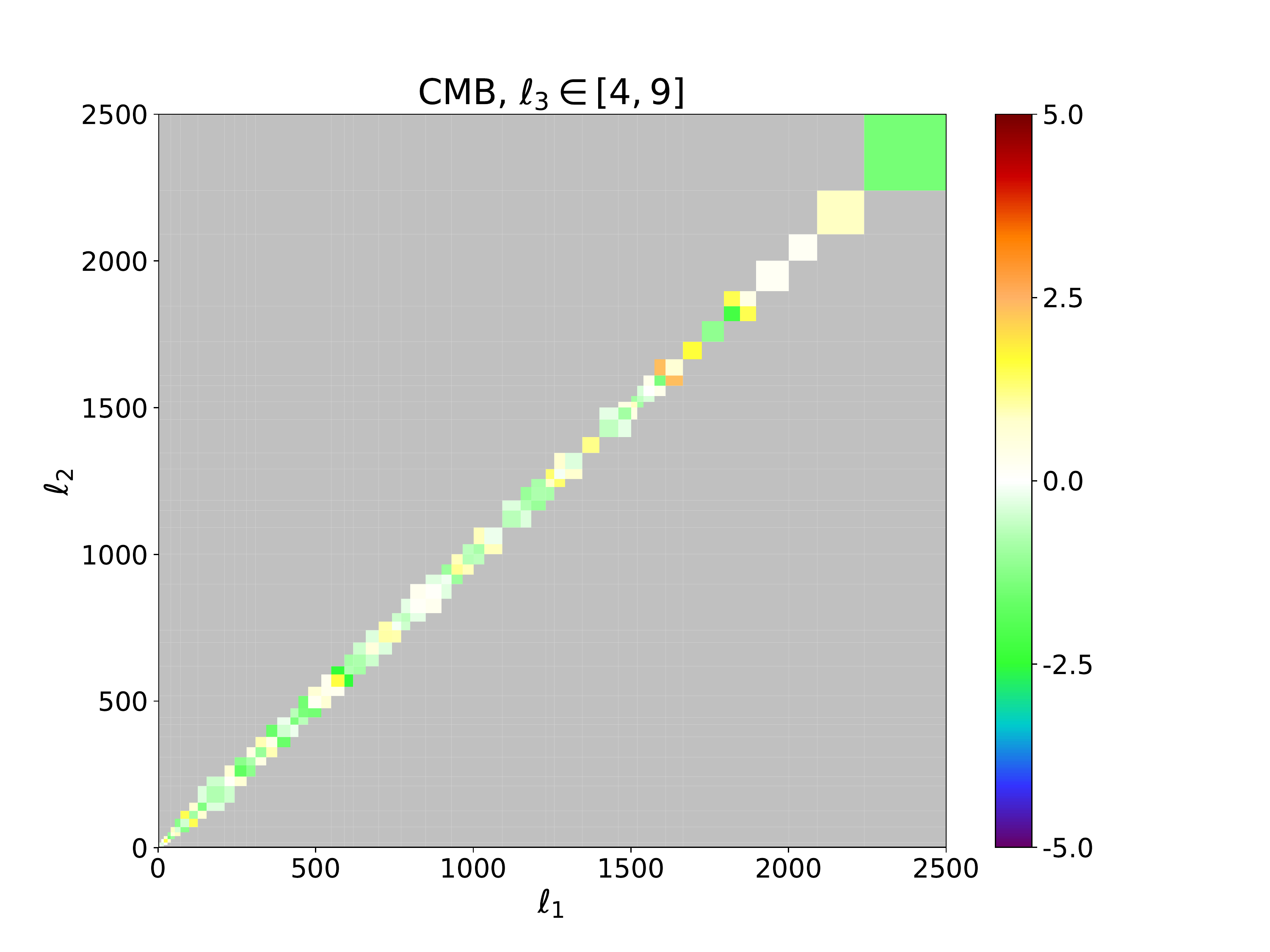}
  \includegraphics[width=0.49\linewidth]{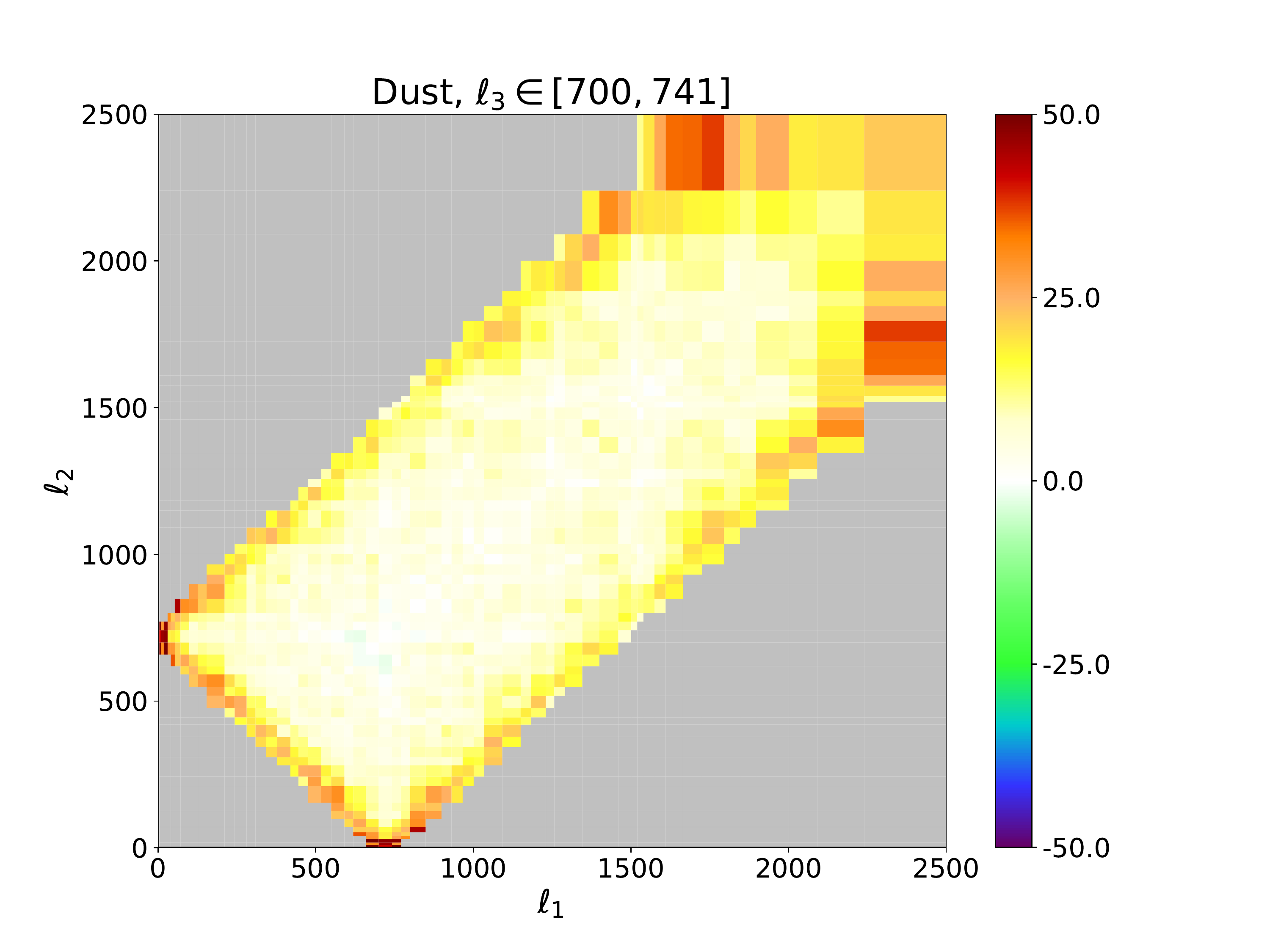}
  \includegraphics[width=0.49\linewidth]{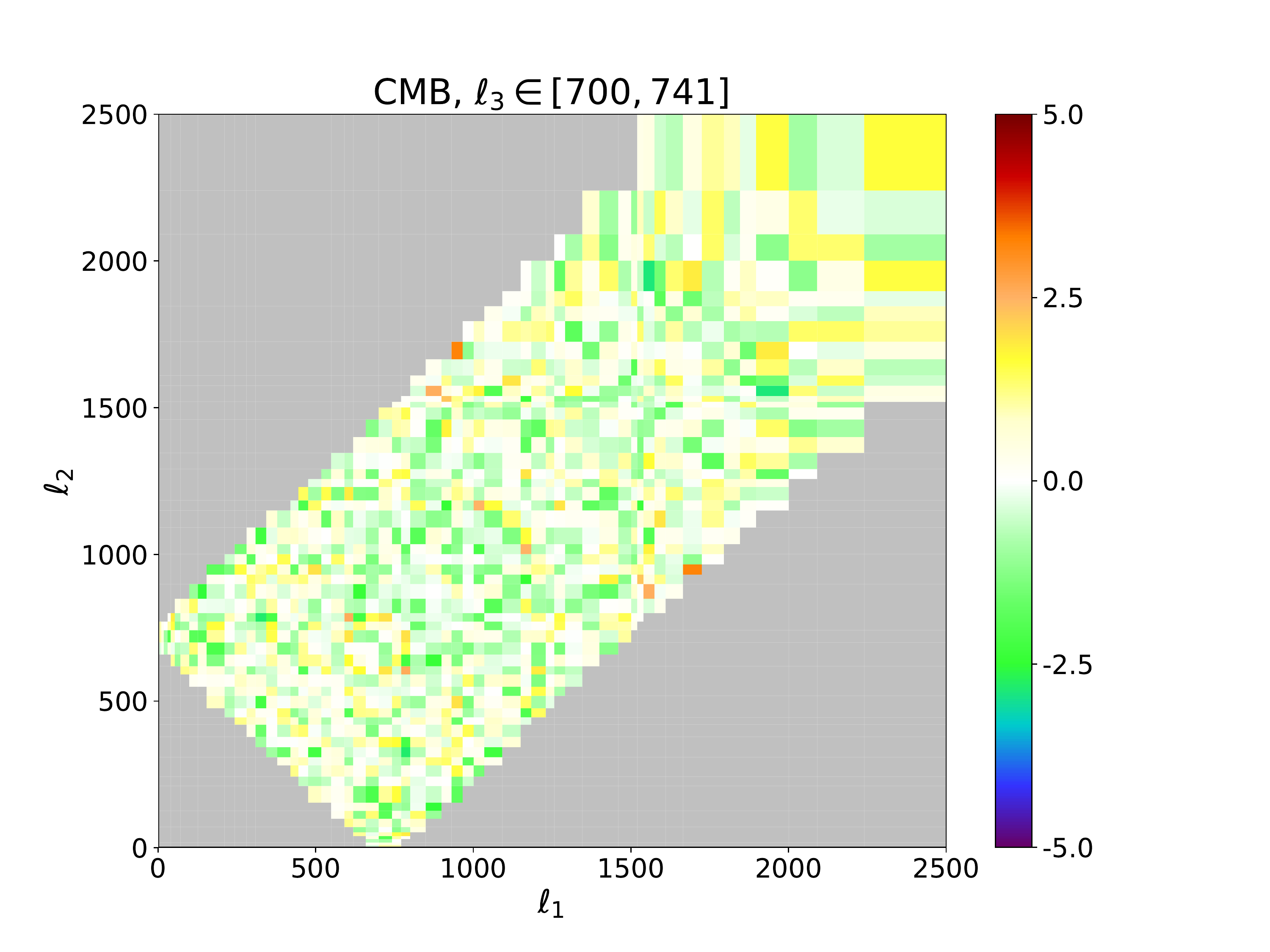}
  \includegraphics[width=0.49\linewidth]{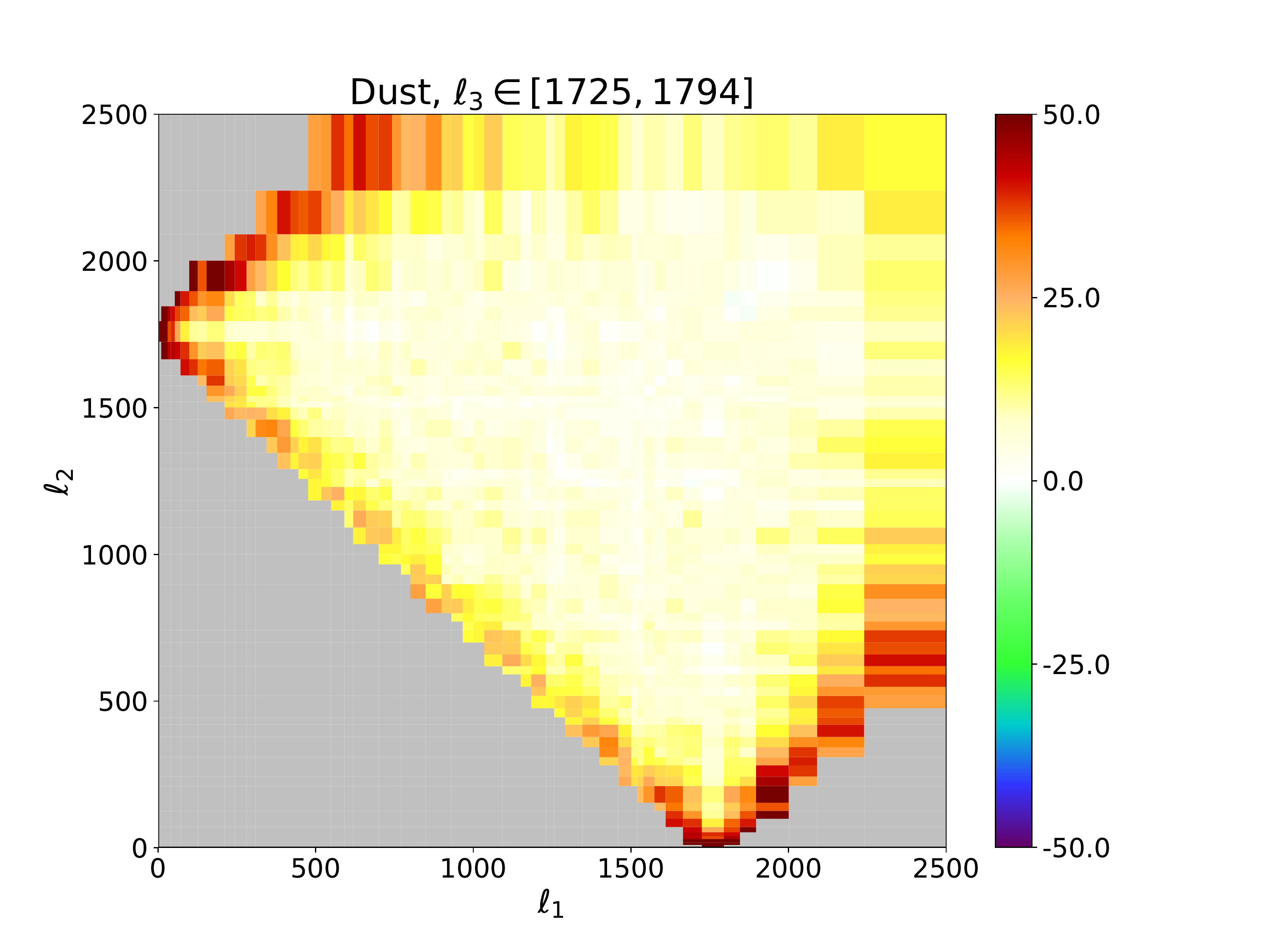}
  \includegraphics[width=0.49\linewidth]{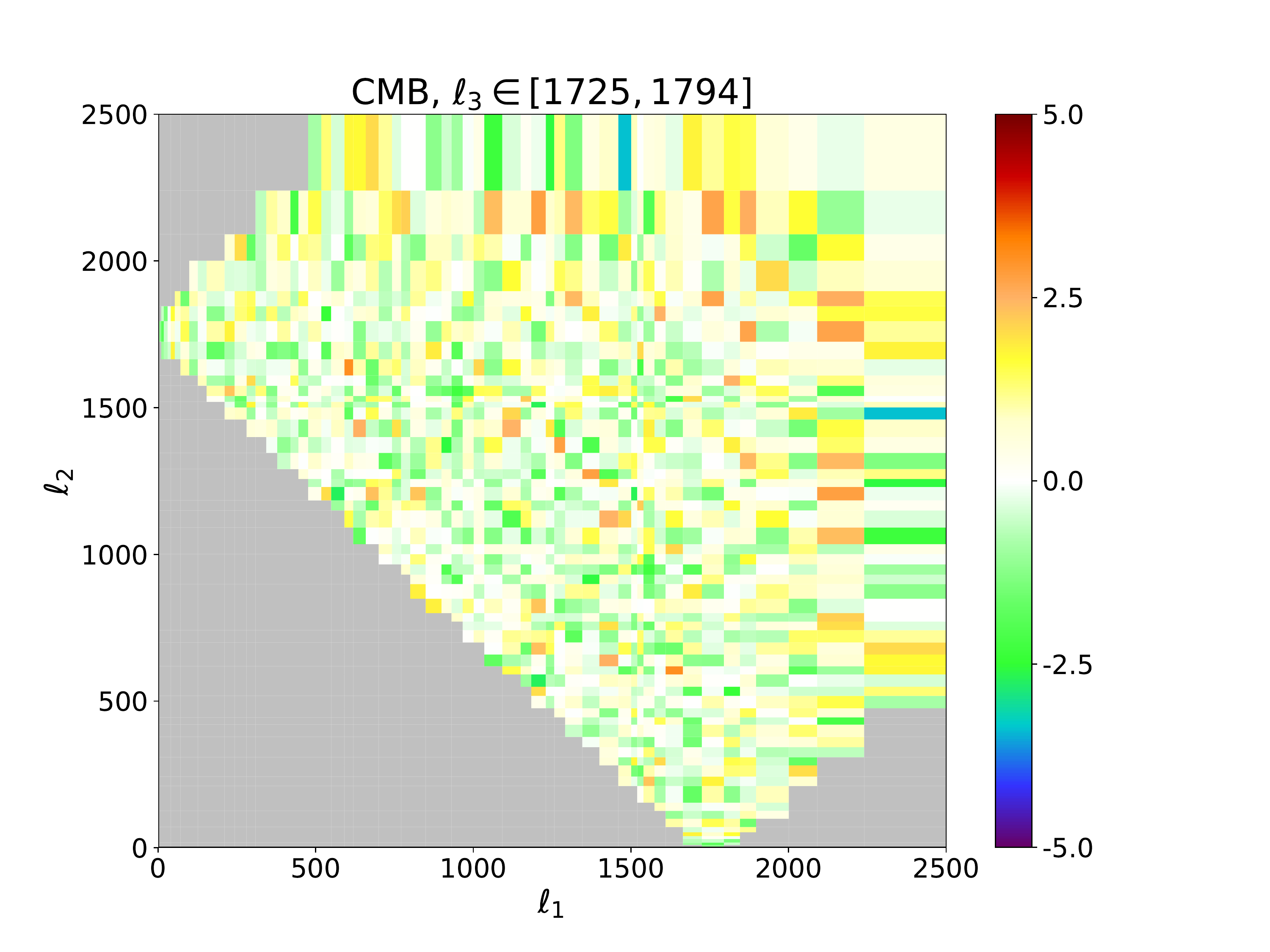}
  \caption{Left: bispectral signal-to-noise of the thermal dust as a function of the multipoles $\ell_1$ and $\ell_2$ for three different bins of $\ell_3$.
  Right: same for the CMB map studied later in section \ref{sec:analyses} for the same bins of $\ell_3$. Note the factor 10 difference in colour scale.}
  \label{fig:dust-bispectrum}
\end{figure}

However, the plots of figure \ref{fig:dust-template} are not well suited to describe quantitatively the non-Gaussian nature of these shapes. As in the case of the power spectrum, which peaks at low $\ell$ if we do not multiply by the factor $\ell(\ell+1)$, the CMB bispectrum as well as the dust bispectrum have a strong $\ell$ dependence. This means that we should use an adapted function of $\ell$ to highlight the true nature of a bispectral signal. A good choice is to use signal-to-noise plots \citep{BRvT} as shown in figure \ref{fig:dust-bispectrum}: the bispectrum is divided by the square root of the variance of the map computed using the power spectrum, see \eqref{binnedVarreal} (divided by $f_\mathrm{sky}$). It is important to note that this is different from the correlation coefficients \eqref{corrmatrix} that we discuss later in this section where the variance of the cleaned CMB map is used. In this kind of plots, non-Gaussianity is simply represented by values large compared to $\mathcal{O}(1)$.

Figure \ref{fig:dust-bispectrum} shows the bispectral signal-to-noise ratio for three different slices of the dust template (on the left), but also for a cleaned CMB map which we will study in detail in section~\ref{sec:analyses} (on the right). It is now obvious that the dust map is very non-Gaussian and that indeed its bispectrum peaks in the squeezed configuration. This effect can be seen in the top plot (low $\ell_3$) but also on the left (low $\ell_1$) and on the bottom (low $\ell_2$) of the other plots. A squeezed configuration is expected when there are correlations between small-scale and large-scale effects. There is a simple physical explanation for the origin of these correlations. The large clouds of dust (i.e.\ large-scale fluctuations) have the highest intensity where they are the thickest along the line of sight. Moreover, the brightest parts have stronger fluctuations (small-scale), see \citep{MivilleDeschenes:2007ya} for a discussion, so the small-scale fluctuations are modulated by the large-scale ones which corresponds to a squeezed bispectrum.

The squeezed signal present in both the dust and the local shapes is a good indication that they are correlated. This can be verified in table \ref{tab:corr_coeff_dust} which gives the correlation coefficients between the dust and the standard shapes computed using \eqref{corrmatrix} in the context of a CMB analysis (more details in section \ref{sec:analyses}), so the denominator of the inner product is the CMB bispectrum variance. There is an anti-correlation between the dust and local shapes (60~$\%$) because they have opposite signs (this anti-correlation was pointed out in \citep{Yadav:2007yy}). The local shape is itself correlated to the other primordial shapes (see table~\ref{tab_corr_coeff}). However, this does not mean that the dust template has to be correlated to them too. And indeed, the dust and equilateral shapes are uncorrelated because the latter does not peak in the squeezed configuration. The correlations between local and dust (squeezed) do not come from the same multipole triplets as the correlations between local and equilateral (acoustic peaks). However, the orthogonal and dust shapes are a little correlated (around 15 $\%$), because the orthogonal bispectrum in the squeezed limit is large. The dust bispectrum template is very weakly correlated to extra-galactic foreground templates like unclustered point sources and CIB, but anti-correlated to lensing-ISW (which is known to be highly correlated to the local shape). An alternative representation of the bispectra of the different shapes, which shows in which regions of multipole space they dominate, is given in the second appendix in section~\ref{ap:appendices}.

\begin{table}
  \begin{center}
    \begin{tabular}{l|cccccc}
      \hline
      & Local & Equilateral & Orthogonal & Lensing-ISW & Point sources & CIB \\
      \hline
      Dust & -0.6 & 0.004 & 0.15 & -0.34 & 0.054 & 0.083\\
      \hline
    \end{tabular}
  \end{center}
  \caption{Correlation coefficients between the standard theoretical templates and the observed dust bispectrum computed using the characteristics of the Planck experiment (temperature).}
  \label{tab:corr_coeff_dust}
\end{table}

\subsection{Other foregrounds}

Apart from dust, there exist other foregrounds which have a greater effect at low frequencies, of which we will study three here. In this section we use maps produced by the \texttt{Commander} method to separate foregrounds, but this time in addition to the Planck data, observations from WMAP between 23 and 94 GHz \citep{Bennett:2012zja} and a 408 MHz survey map \citep{Haslam:1982zz} were also used to determine them. They have a lower resolution ($n_{\mathrm{side}}=256$) and a larger beam (60' FWHM Gaussian beam). For the sake of comparison of these foregrounds with the dust we discussed in the previous section, we will also use here a dust map with the same characteristics.

In the case the dust grains rotate rapidly (in addition to their thermal vibrations), they can produce a microwave emission which probably corresponds to the anomalous microwave emission (AME) \citep{Leitch:1997dx, Draine:1998gq}, large at low frequencies.

Dust is not the only component responsible for the contamination of the CMB signal; some interactions of electrons with the interstellar medium can also generate emissions. On the one hand, ultra-relativistic electrons (cosmic rays) spiraling in the galactic magnetic fields radiate. This synchrotron emission can be described by a power law $\nu^\beta$ with $\beta \simeq -3$ indicating that indeed, this radiation is significant at low frequencies \citep{Haslam:1982zz}. On the other hand, electrons can be slowed down by scattering off ions. This generates the free-free emission \citep{Dickinson:2003vp}, also called bremsstrahlung.

The frequency dependence of the foregrounds and the CMB signal can be seen in figure 51 of \citepalias{planck2015-10}. As discussed, the synchrotron, the free-free and the spinning dust (AME) emissions dominate at low frequencies. The dust thermal emission is the main contribution at high frequencies and is of the same order as the CMB at 143 GHz (this of course depends on the choice of mask). 

\begin{figure}
  \centering  \includegraphics[width=0.49\linewidth]{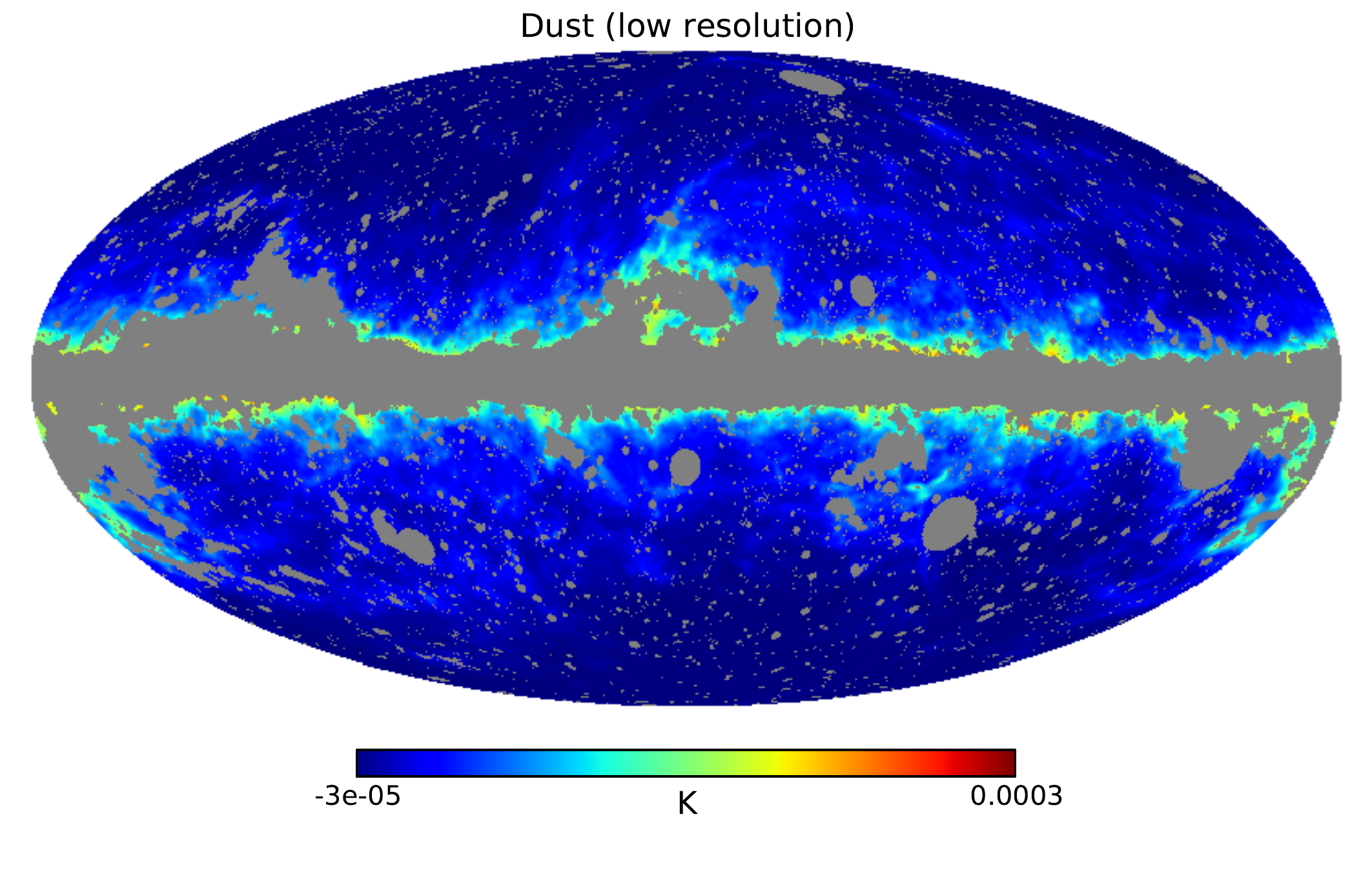}
\includegraphics[width=0.49\linewidth]{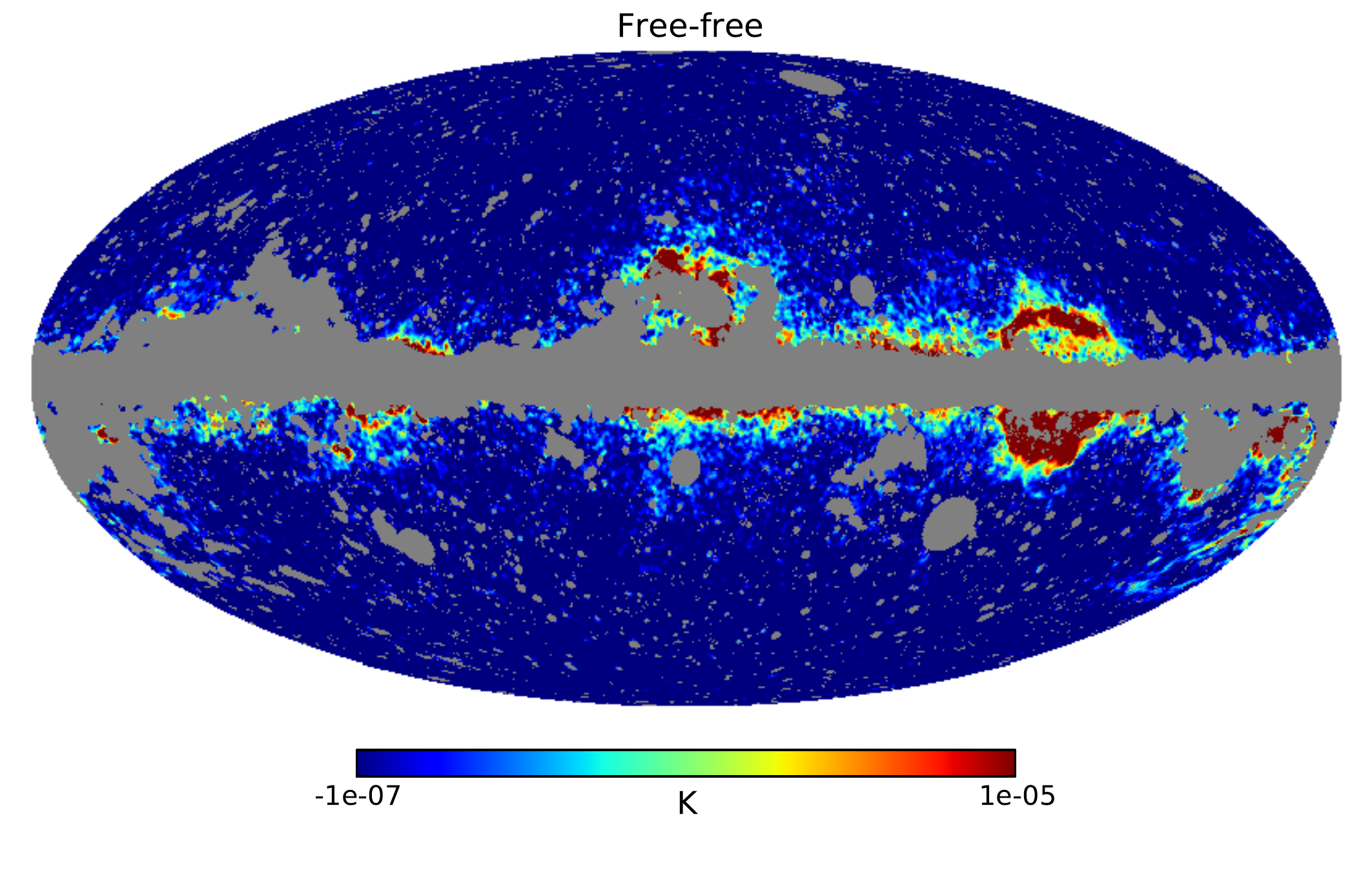}
\includegraphics[width=0.49\linewidth]{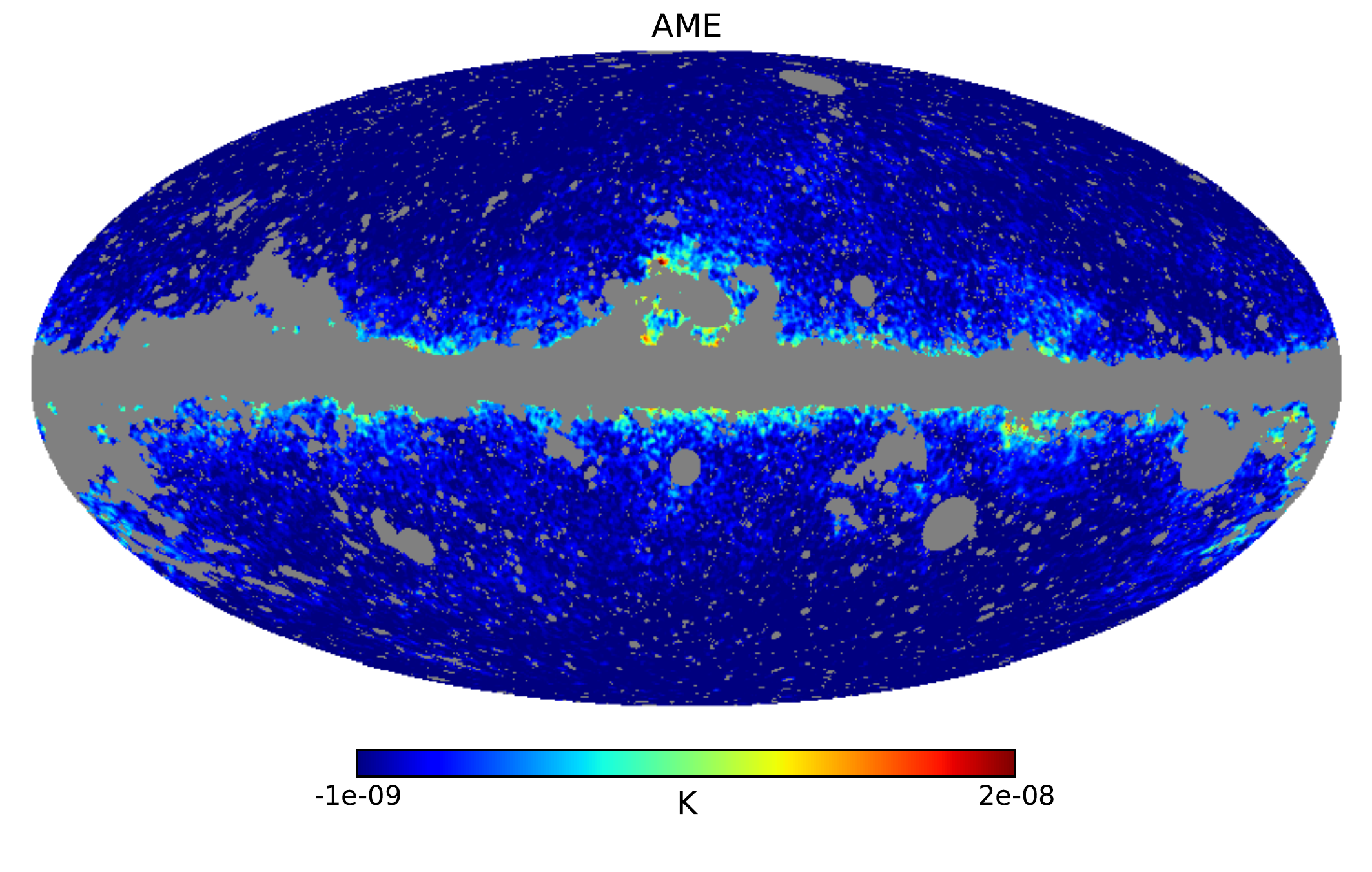}
\includegraphics[width=0.49\linewidth]{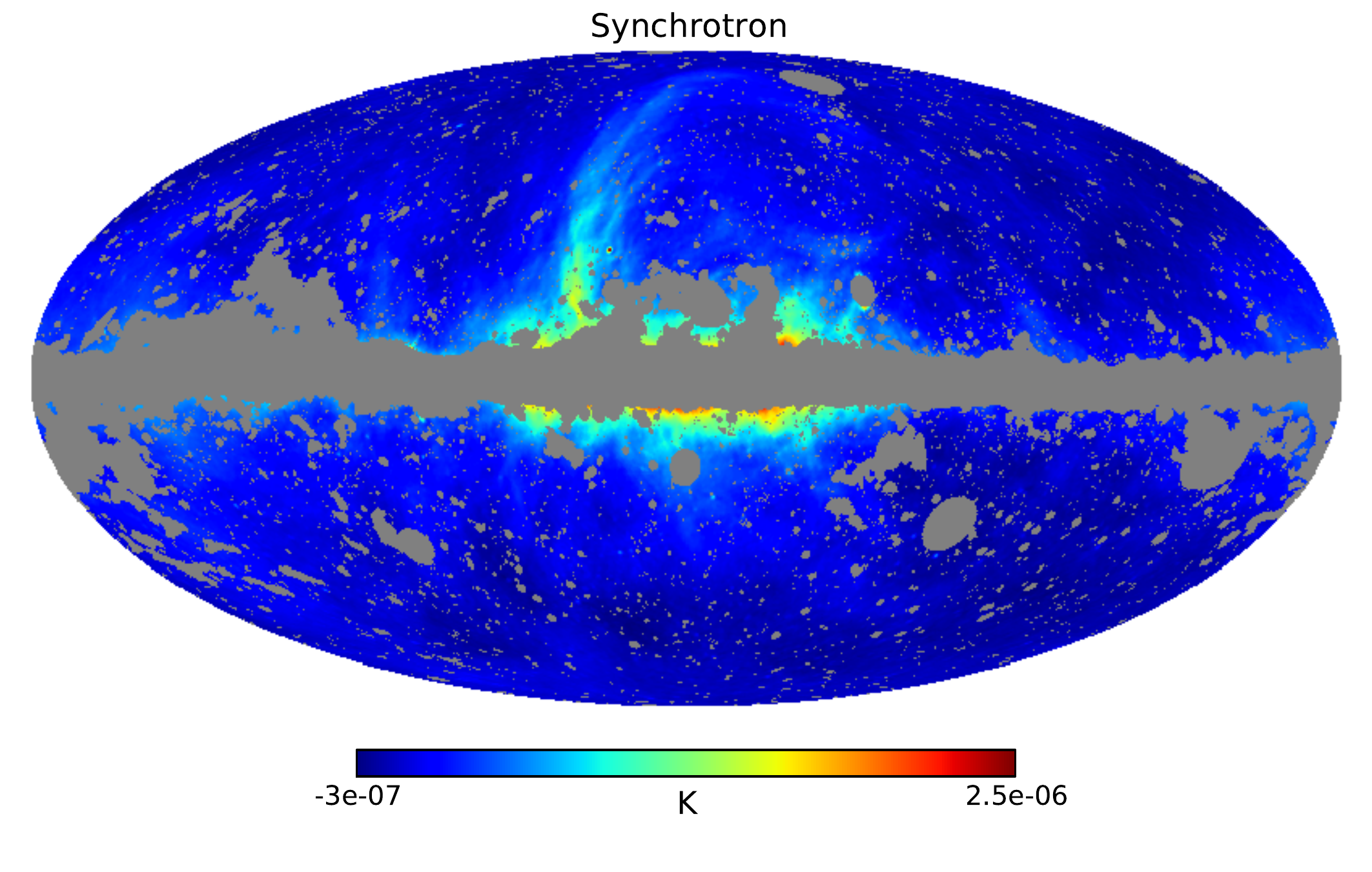}
\caption{Masked maps of the different galactic foregrounds we consider in this paper (dust, free-free, AME, synchrotron) at 143 GHz derived using the \texttt{Commander} method from Planck 2015 data. Note the different colour scales.}
  \label{fig:other-maps}
\end{figure}

\begin{figure}
  \centering
  \includegraphics[width=0.66\linewidth]{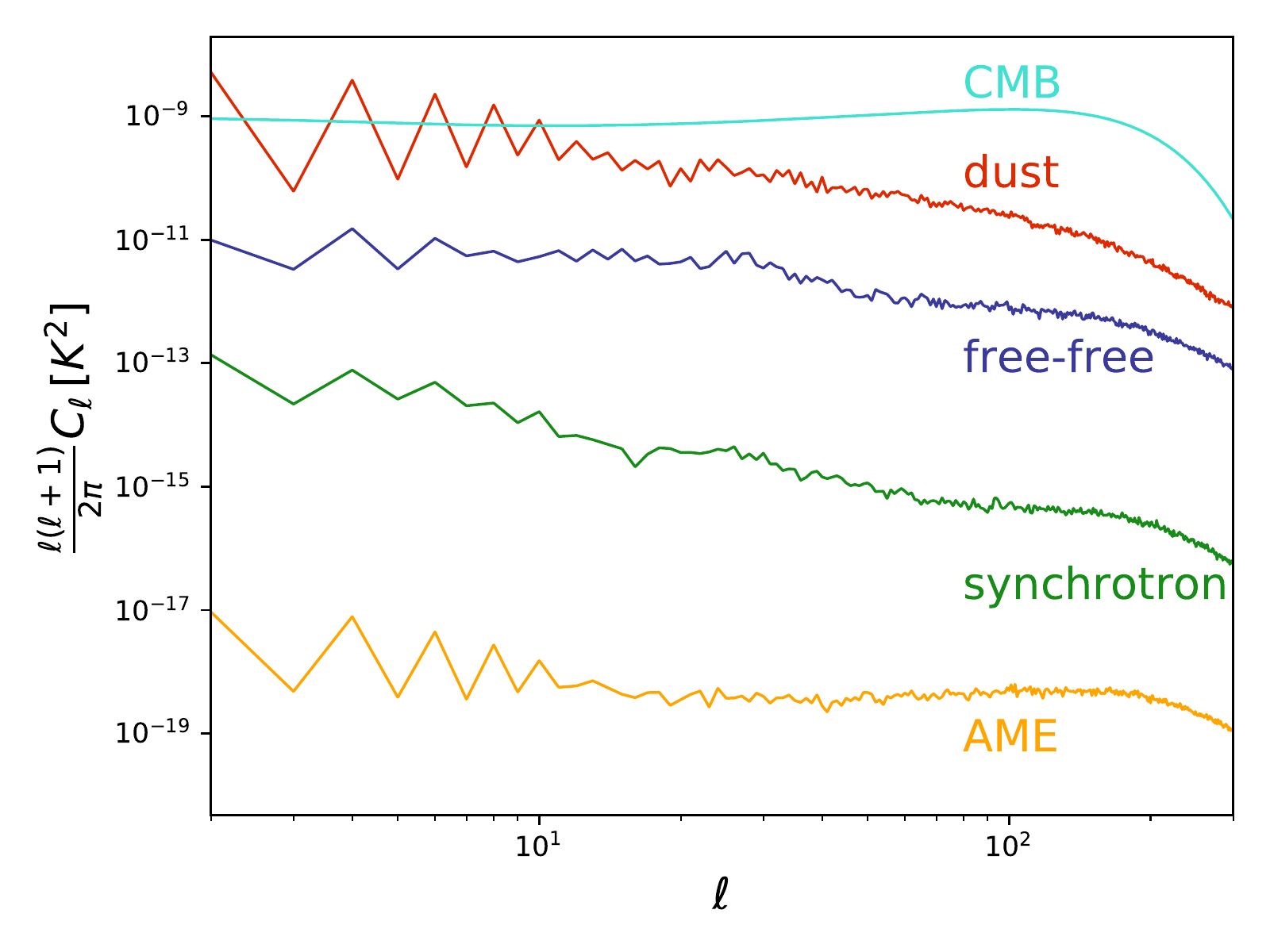}
  \caption{The power spectra of the different foregrounds (dust, free-free, synchrotron and AME) and of the CMB at 143 GHz (including the 60 arcmin beam and the noise) as a function of the multipole $\ell$.}
  \label{fig:all-power-spectra}
\end{figure}

Figure \ref{fig:other-maps} shows the contributions of all these foregrounds at 143 GHz. Similarly to the dust in the previous section, they are all localized in the galactic plane. Moreover, we can see that the dust signal has a higher intensity and therefore is the dominant foreground contribution at 143 GHz. The same hierarchy can be seen in the power spectra, as shown in figure \ref{fig:all-power-spectra}. It is clear that at 143 GHz, the contributions of AME, synchrotron and free-free are negligible compared to the CMB (remember that the brightest parts of the sky are masked). Note that because of the low resolution of the map and the 60 arcmin beam, the range of multipoles is a lot smaller than in the previous section ($\ell_\mathrm{max}=300$ here). This also means that we were able to use smaller bins for the binned bispectrum estimator. We simply took the usual binning, with each bin split into three when possible (two otherwise). 

\begin{figure}
  \centering 
\includegraphics[width=0.32\linewidth]{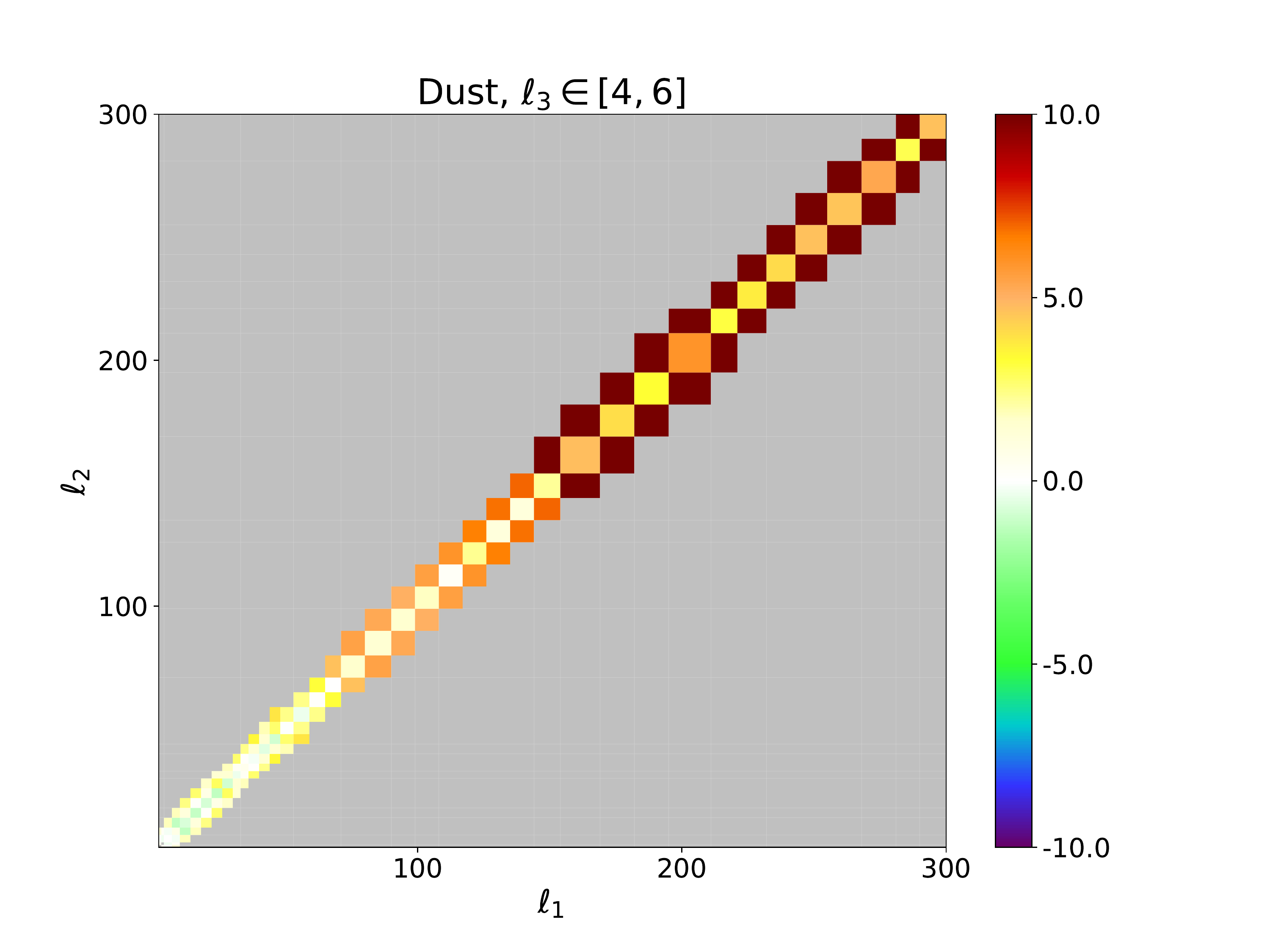}  \includegraphics[width=0.32\linewidth]{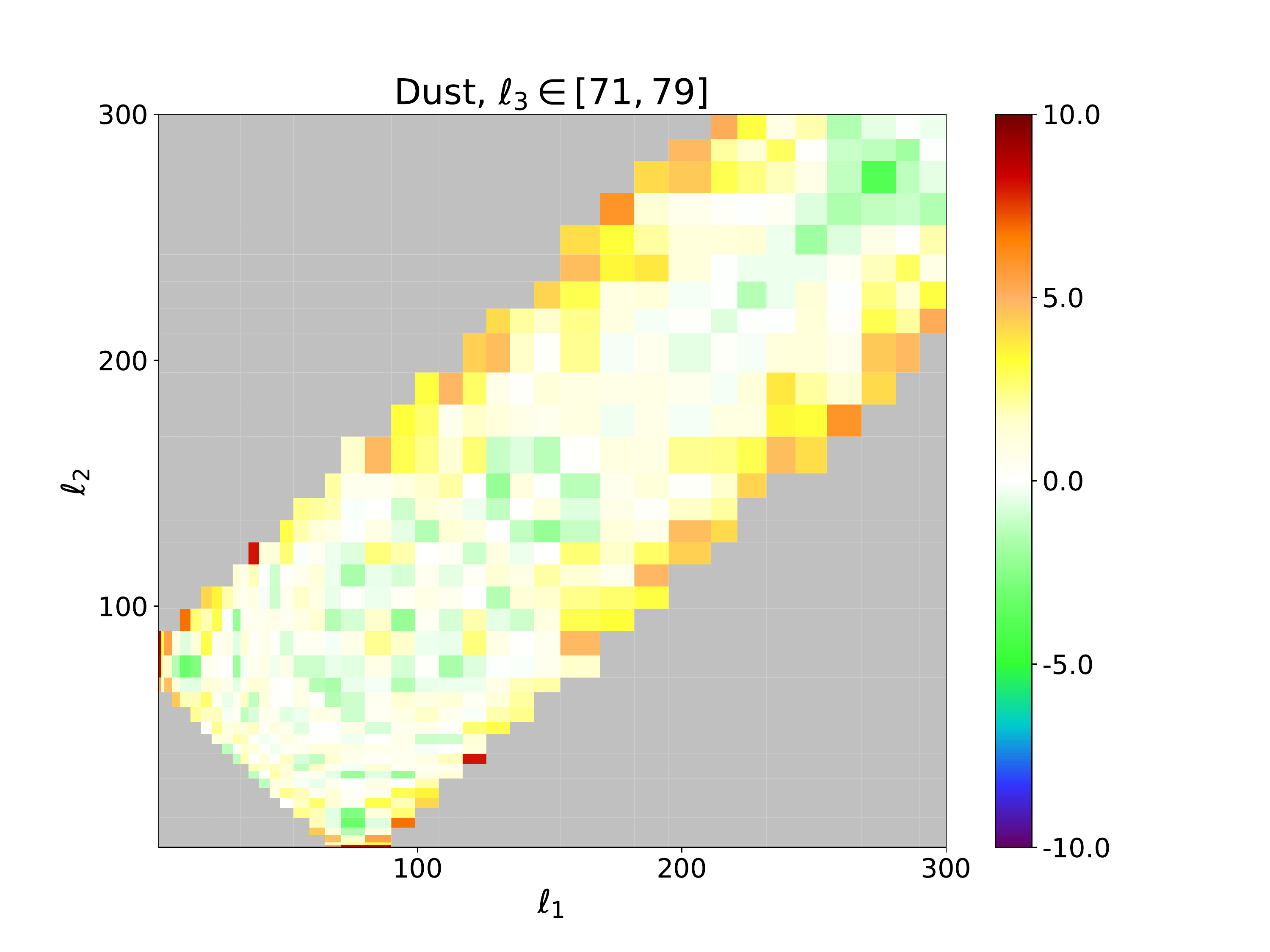} \includegraphics[width=0.32\linewidth]{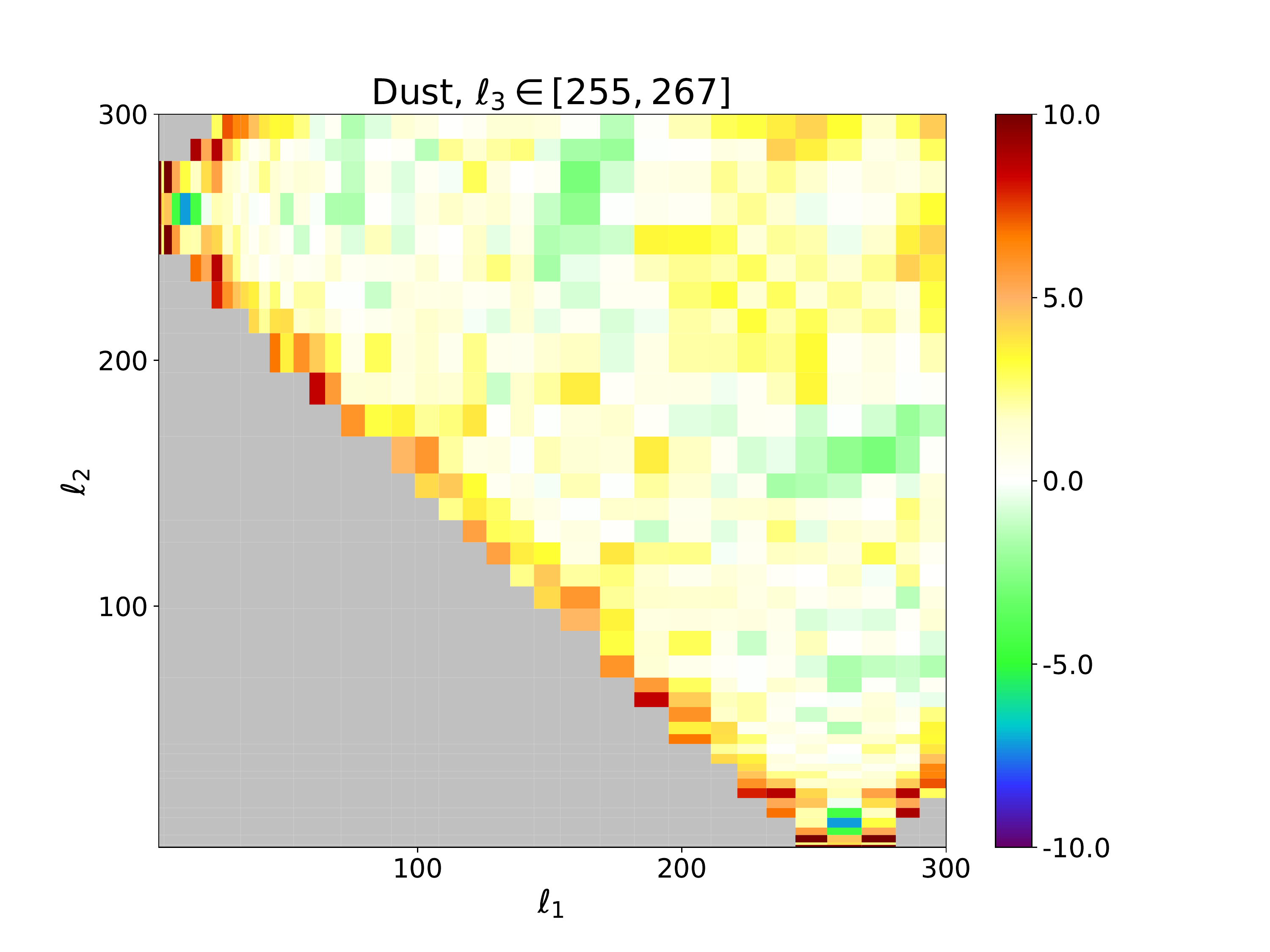}
\includegraphics[width=0.32\linewidth]{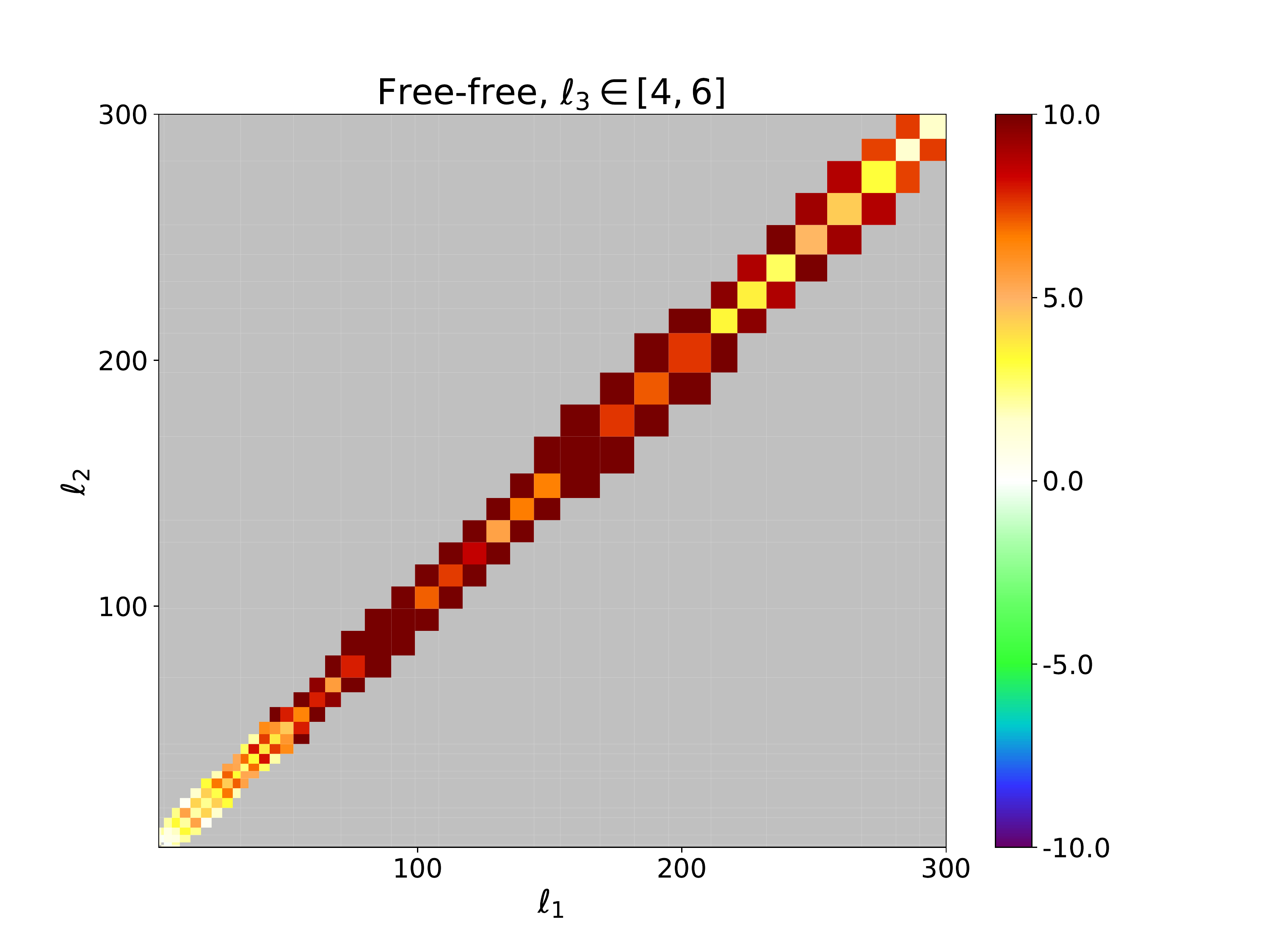}   \includegraphics[width=0.32\linewidth]{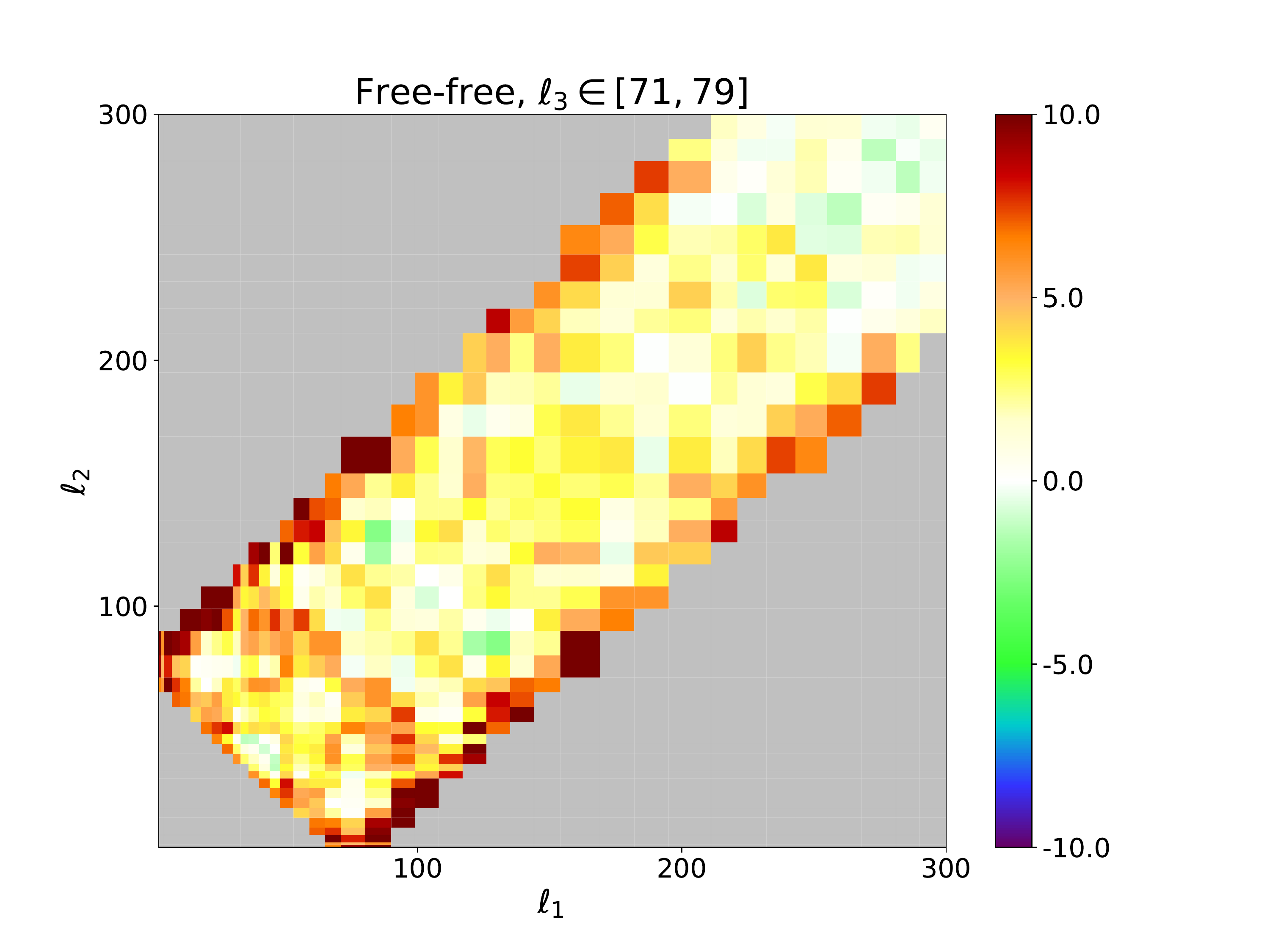} 
\includegraphics[width=0.32\linewidth]{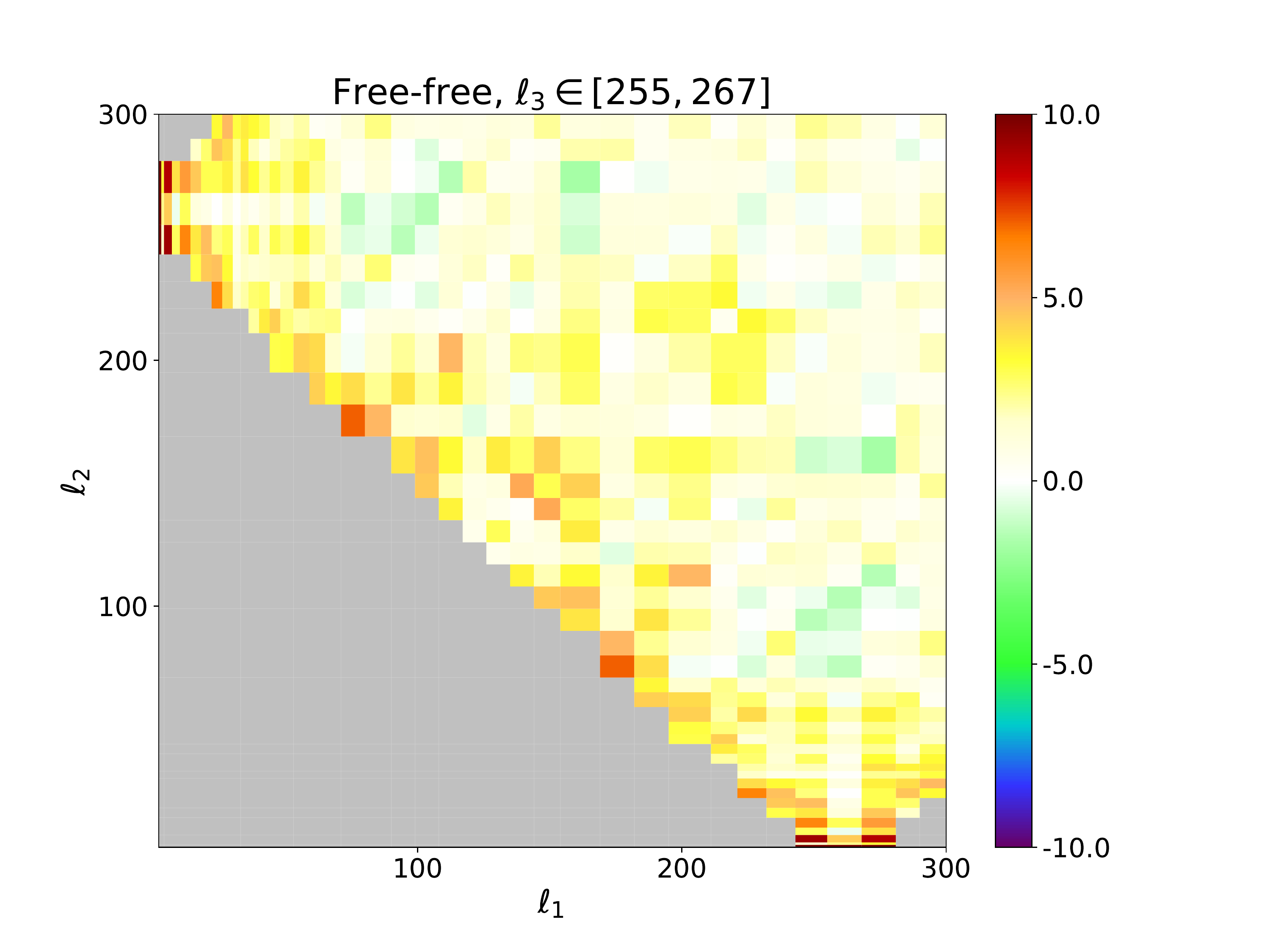}
\includegraphics[width=0.32\linewidth]{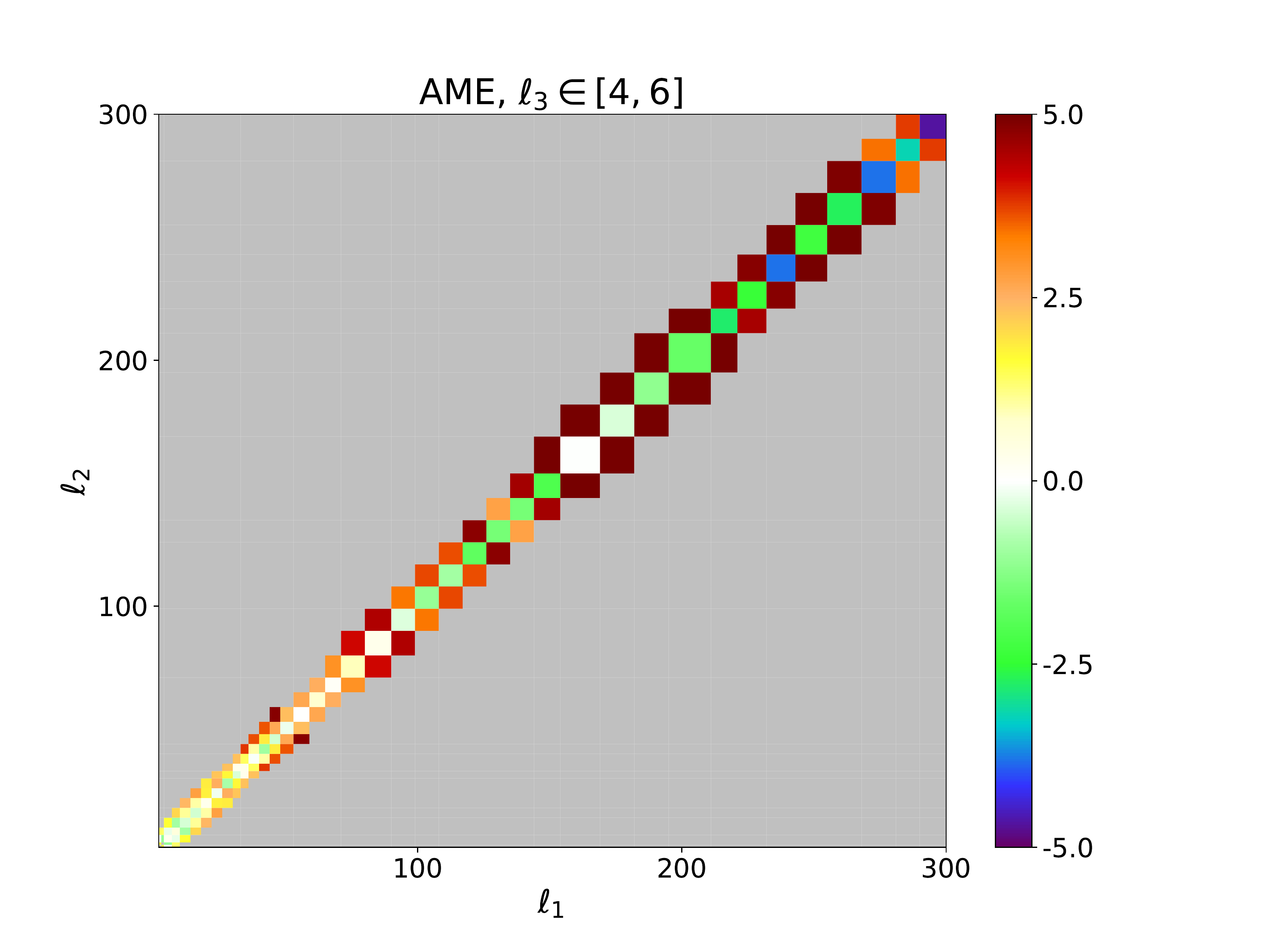} 
\includegraphics[width=0.32\linewidth]{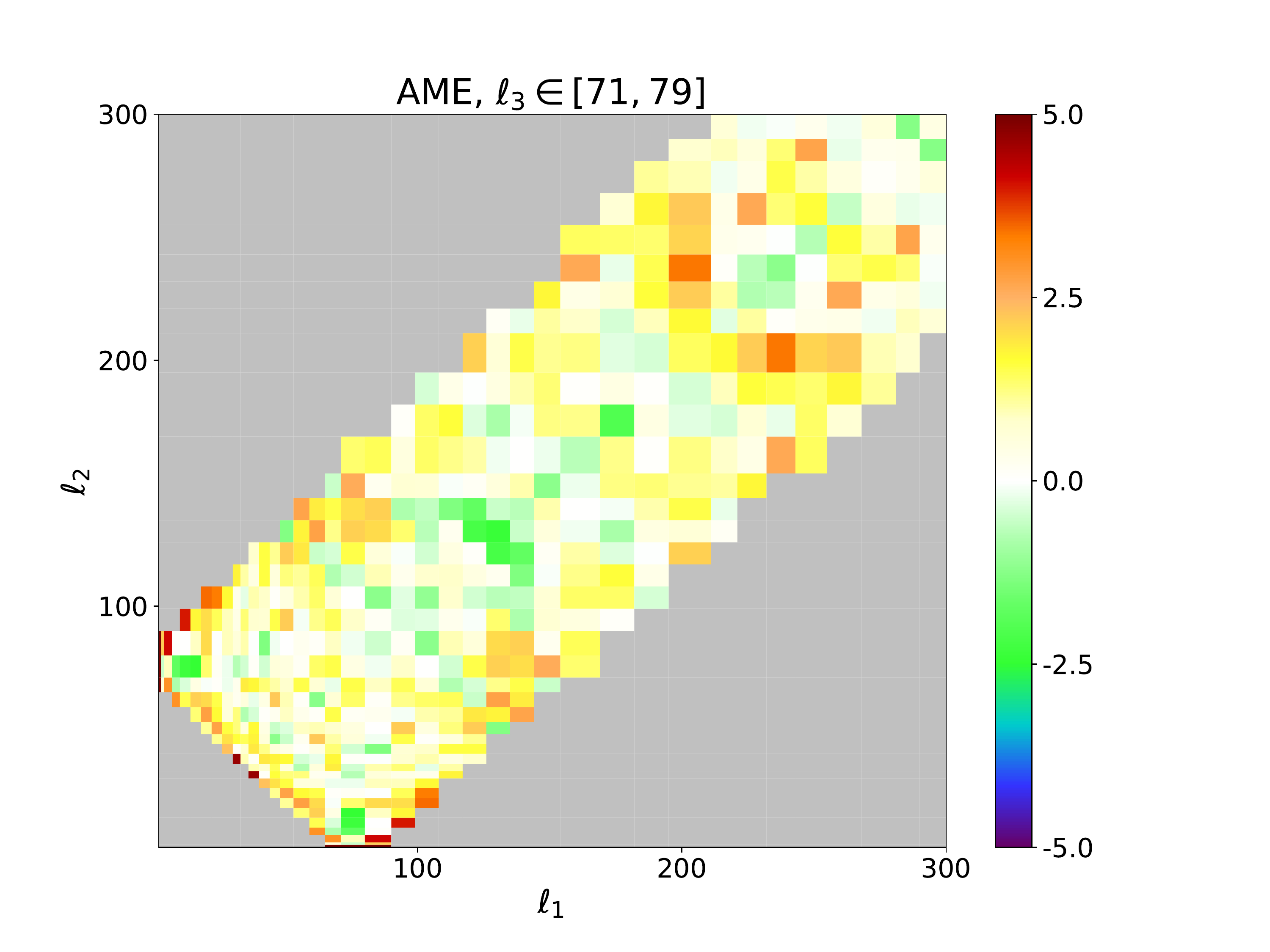}  \includegraphics[width=0.32\linewidth]{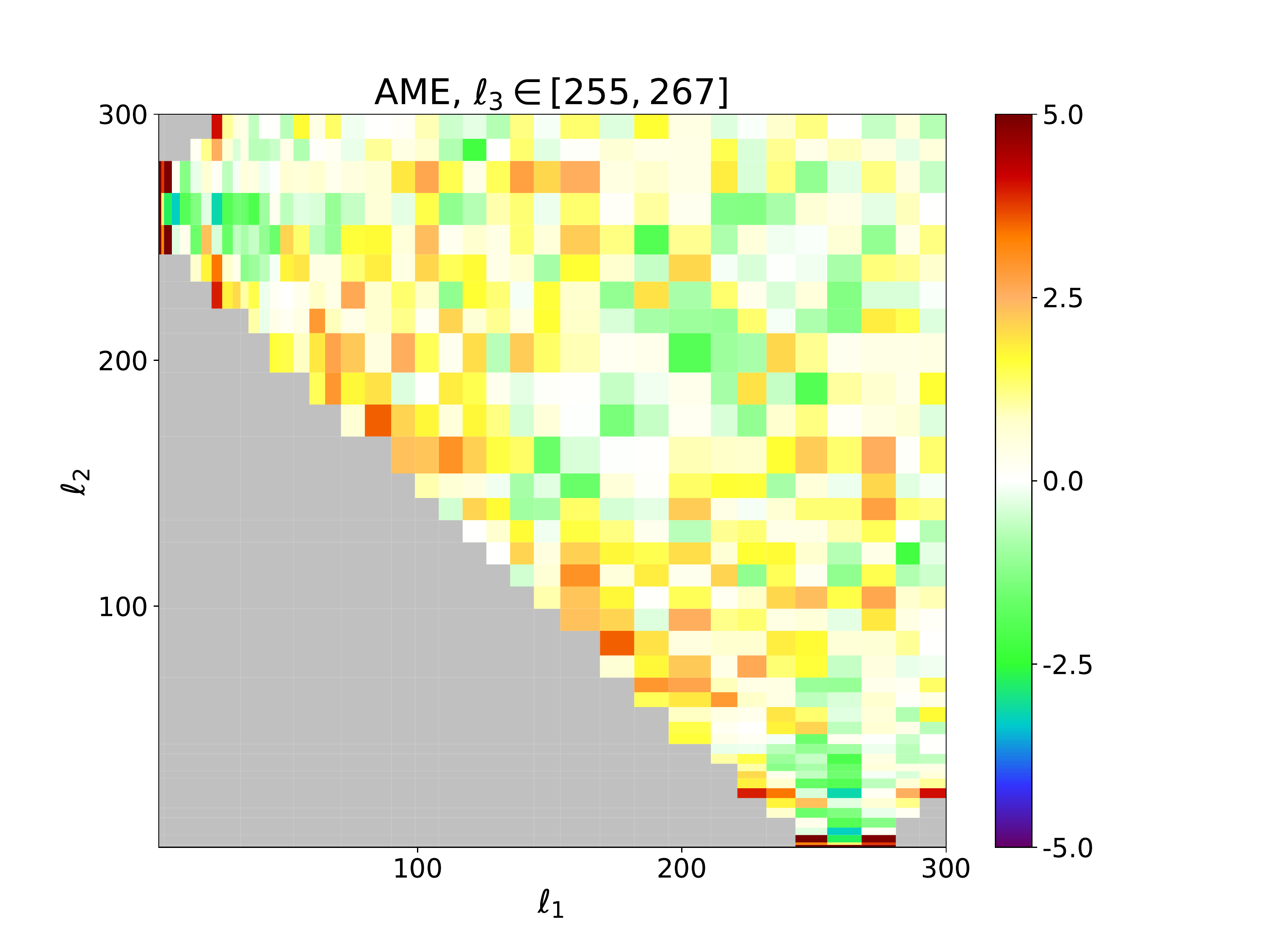}
\includegraphics[width=0.32\linewidth]{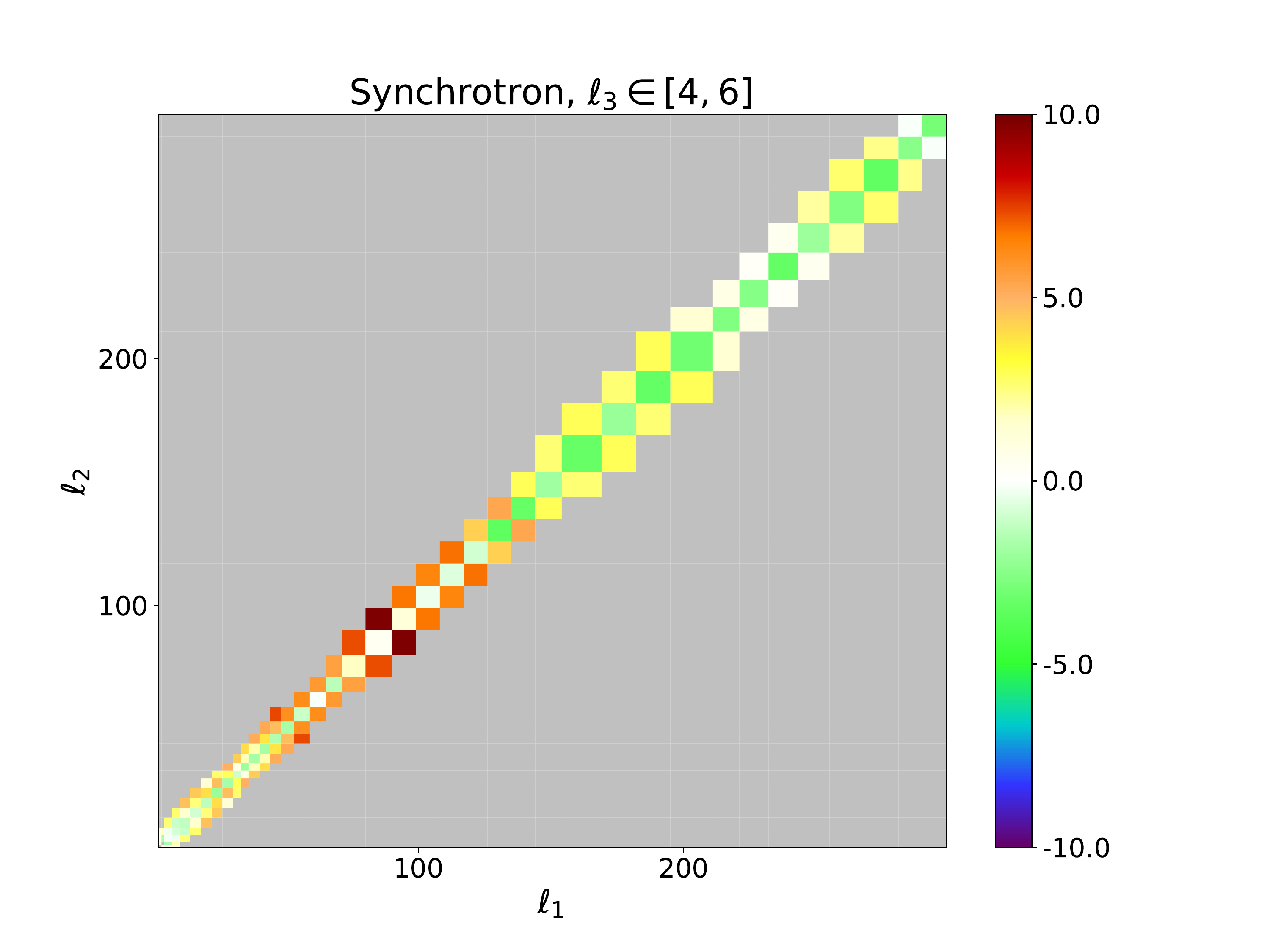} 
\includegraphics[width=0.32\linewidth]{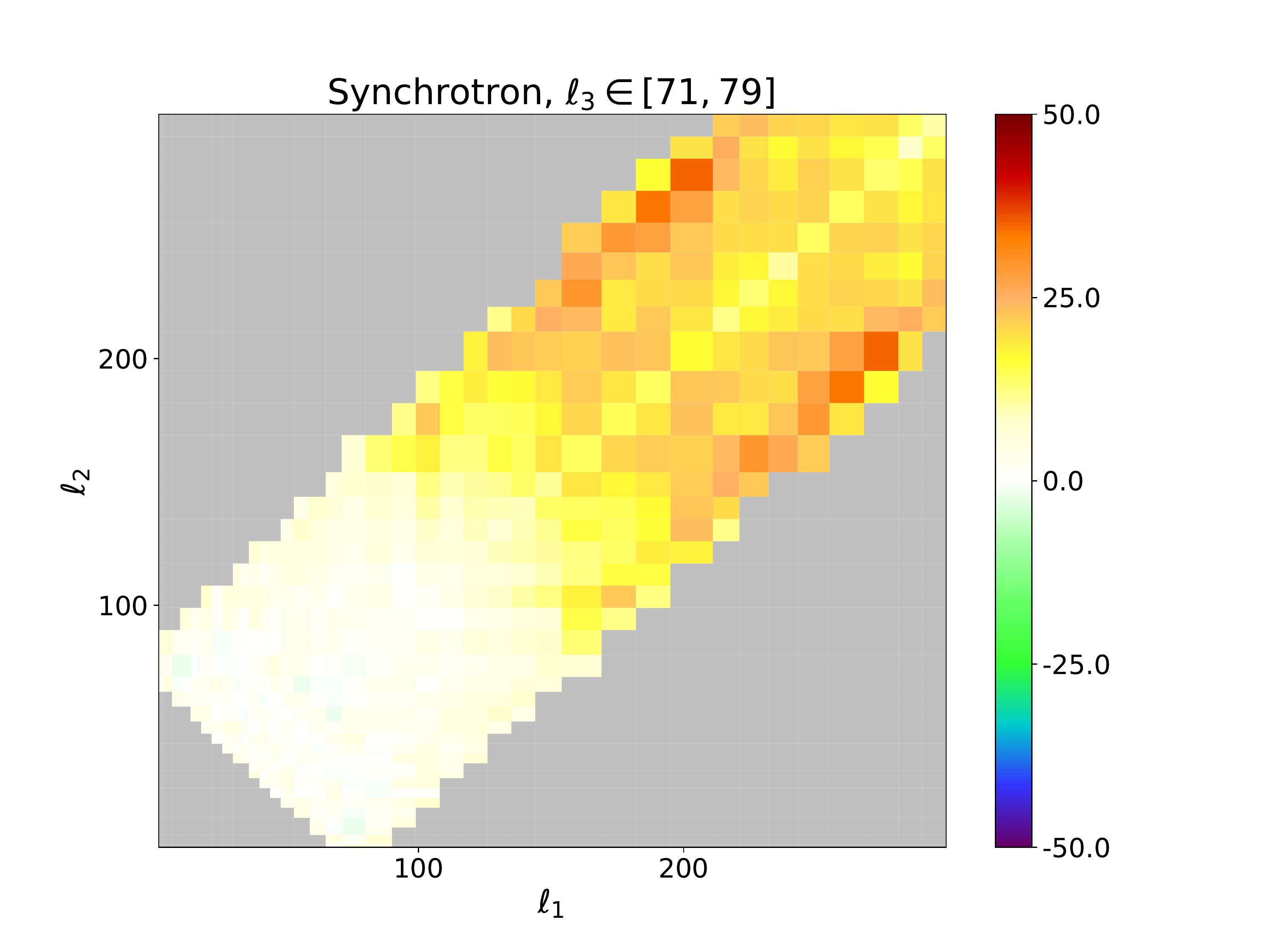}  \includegraphics[width=0.32\linewidth]{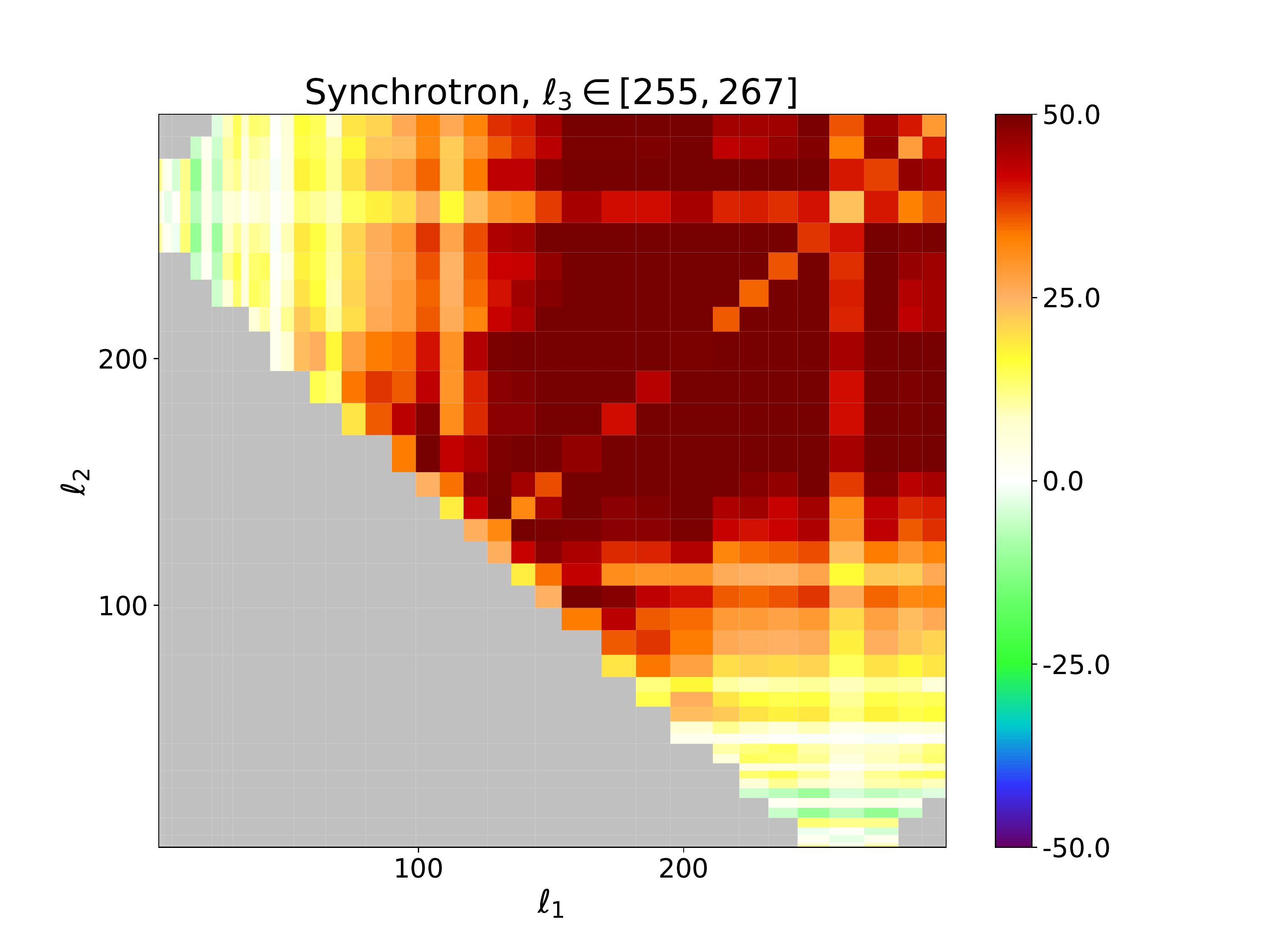}
  \caption{Bispectral signal-to-noise of the different foregrounds for $\ell_3 \in [4,6]$, $\ell_3 \in [71,79]$ and $\ell_3 \in [255,267]$. Note the different colour scales.} 
  \label{fig:other-templates}
\end{figure}

The same behaviour is of course present in the bispectra (i.e.\ the templates) where the dust dominates everything. However, as discussed in the previous section, it is more interesting to study the bispectral signal-to-noise to study the form of these bispectra. Figure \ref{fig:other-templates} shows these bispectra for three different slices of $\ell_3$. Free-free, dust and AME peak in the squeezed configuration (but for AME, the signal is so low that it could be only noise). An argument similar to the dust case described in the previous section can explain this bispectral configuration. We can also verify this in table \ref{tab:corr_coeff_others} where we have computed the correlation coefficients of these shapes with the ones previously introduced. As expected, the dust, free-free and AME bispectra are anti-correlated to the local shape (and for the other shapes see the previous section, the discussion is similar) and are correlated between themselves (they share the squeezed configuration). For a visual representation that helps to understand the correlations, see the second appendix in section~\ref{ap:appendices}.

\begin{table}
  \begin{center}
    \small
    \begin{tabular}{l|cccccc}
      \hline
      & Local & Equilateral & Orthogonal & Lensing-ISW & Point sources & CIB \\
      \hline
      Dust (low resolution) & -0.14 & 0.0097 & 0.087 & -0.036 & 0.0083 & 0.012\\
     Free-free & -0.44 & -0.045 & 0.43 & 0.043 & 0.069 & 0.11\\
     AME & -0.23 & 0.032 & 0.052 & -0.051 & 0.033 & 0.037\\
     Synchrotron & -0.057 & 0.33 & 0.29 & 0.051 & 0.44 & 0.38\\
      \hline
    \end{tabular}
    
    \vspace{0.5cm}
    
   \begin{tabular}{l|cccc}
      \hline
      & Dust (low resolution) & Free-free & AME & Synchrotron \\
      \hline
     Dust (low resolution) & 1 & 0.24 & 0.28 & 0.56 \\
     Free-free &  & 1 & 0.37 & 0.32\\
     AME &  &  & 1 & 0.32 \\
     Synchrotron &  &  &  & 1 \\
      \hline
    \end{tabular}
  \end{center}
  \caption{Correlation coefficients between the standard theoretical templates and the observed foreground templates (low resolution) computed using the characteristics of the Planck experiment (temperature).}
  \label{tab:corr_coeff_others}
\end{table}

\begin{figure}
  \centering 
\includegraphics[width=0.32\linewidth]{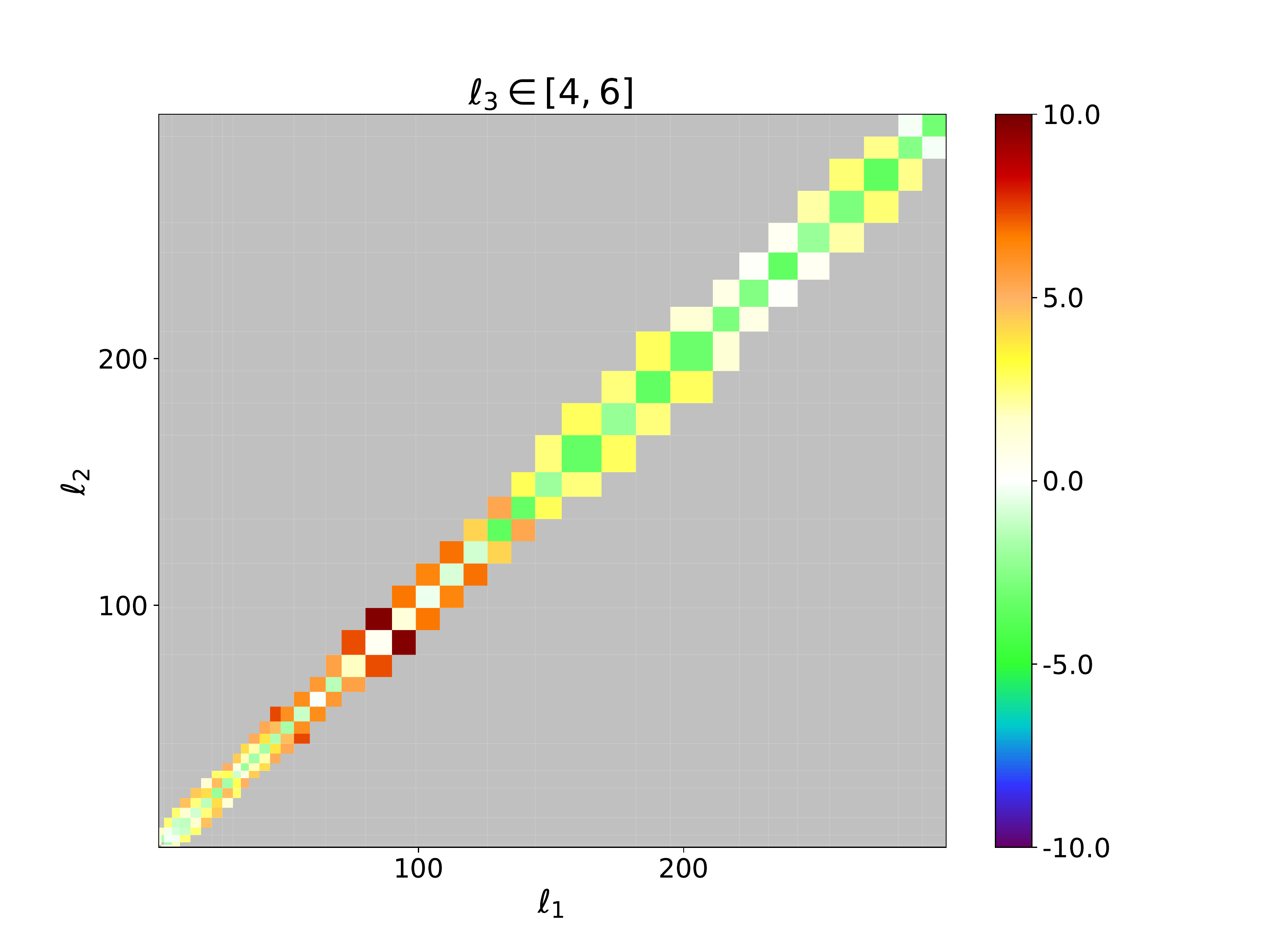}  
\includegraphics[width=0.32\linewidth]{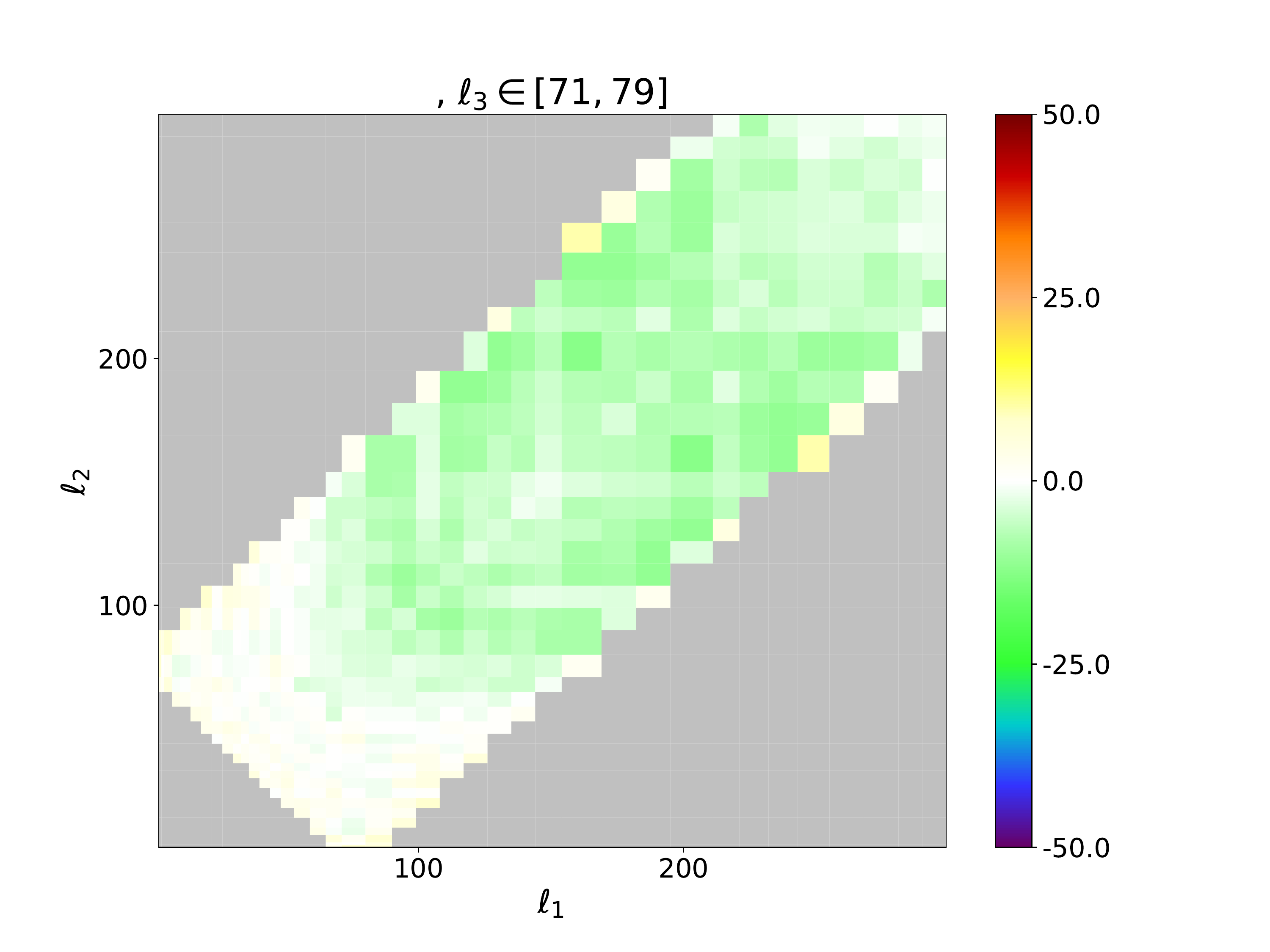} 
\includegraphics[width=0.32\linewidth]{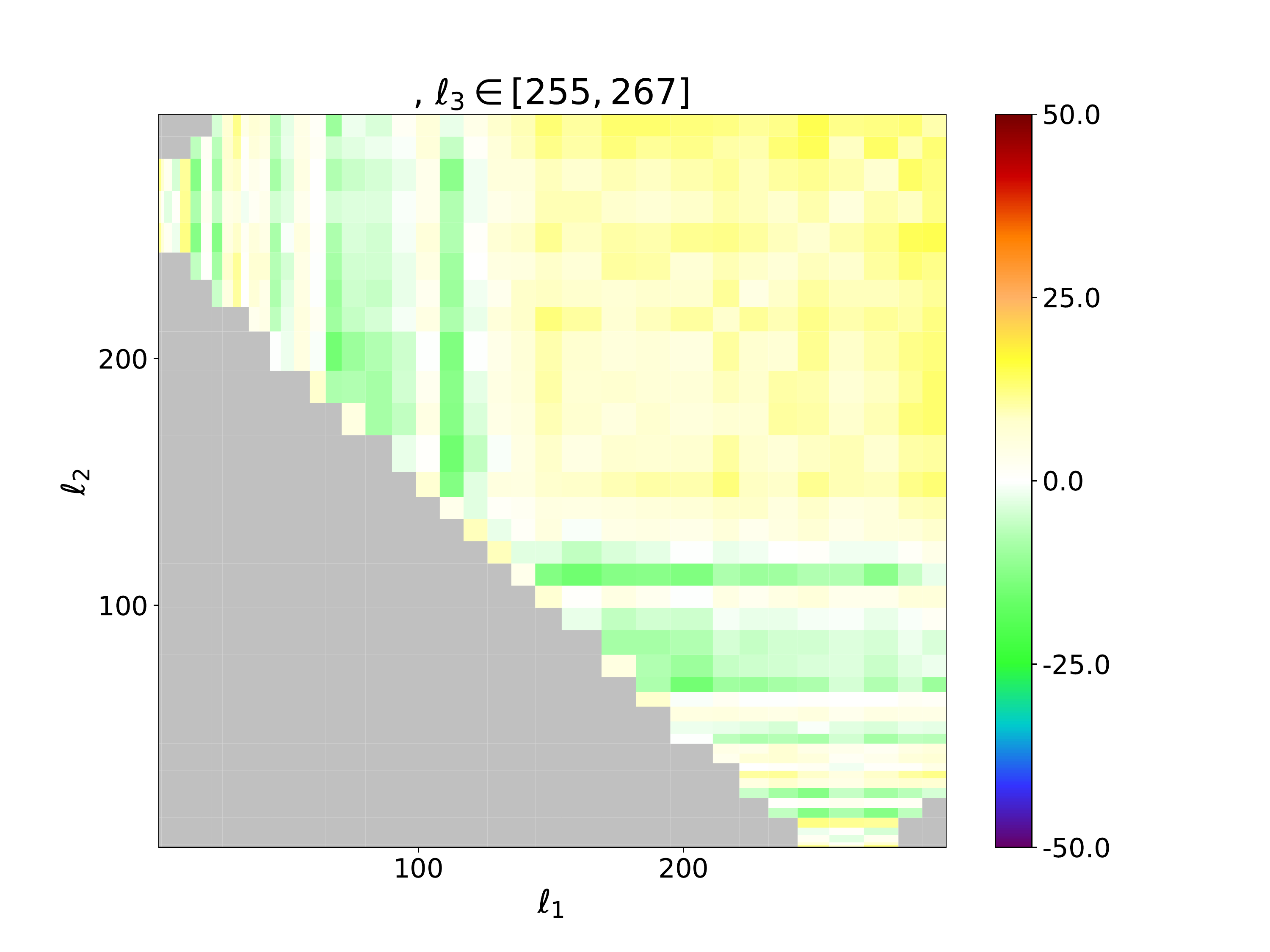}
  \caption{Bispectral signal-to-noise of the synchrotron after subtracting the unresolved point sources contamination for $\ell_3 \in [4,6]$, $\ell_3 \in [71,79]$ and $\ell_3 \in [255,267]$. Note the different colour scales.} 
  \label{fig:synch-ps-template}
\end{figure}

The case of synchrotron is different. The signal seems to be larger for three ``high'' values of $\ell$, so it is similar to the equilateral shape. This is also the typical shape produced by unresolved point sources and by the CIB. Indeed, the synchrotron is correlated (around 40~$\%$) to the point sources and CIB shapes as well as to equilateral and orthogonal (around 30~$\%$). However, it is also correlated to the other foregrounds (more than 30~$\%$), meaning that the synchrotron bispectrum also peaks in the squeezed limit, as shown in the bottom left plot of figure \ref{fig:other-templates}, even if it is not at all its dominant part. Physically that makes sense because we expect a squeezed signal for similar reasons as the other foregrounds. The simplest explanation for the equilateral shape is a contamination of the map by point sources and this possibility is mentioned in \citepalias{planck2015-10}. To verify it, we performed the simple test of subtracting the unresolved point sources bispectral template (of which the amplitude was determined using the estimator (\ref{fNL_estimator})) from the bispectrum of the synchrotron map. The cleaned bispectrum is shown in figure \ref{fig:synch-ps-template} where one can see that the left plot (showing the squeezed part of the bispectrum) has not changed from the one of figure \ref{fig:other-templates}, while the other two are much less non-Gaussian (but not perfectly cleaned either). This is also illustrated in table \ref{tab:corr_coeff_synch}, where the correlation of the synchrotron bispectrum with the local shape increases (to around 15 $\%$) and becomes of the same order as for the other foreground bispectra, while the anomalous correlation with the equilateral, point sources and CIB templates vanishes. From now on, when we mention the synchrotron bispectrum, it will be the one cleaned from the unresolved point sources contamination.

\begin{table}
  \begin{center}
    \small
    \begin{tabular}{l|cccccc}
      \hline
      & Local & Equilateral & Orthogonal & Lensing-ISW & Point sources & CIB  \\
      \hline
     Cleaned synchrotron & -0.14 & 0.025 & 0.13 & -0.022 & 0.059 & 0.033\\
      \hline
    \end{tabular}
        
    \vspace{0.5cm}
    
   \begin{tabular}{l|cccc}
      \hline
      & Dust (low resolution) & Free-free & AME & Synchrotron \\
      \hline
     Cleaned synchrotron & 0.62 & 0.32 & 0.34 & 0.92 \\
      \hline
    \end{tabular}
  \end{center}
  \caption{Correlation coefficients between the synchrotron bispectrum cleaned from the point sources contamination and the templates of table \ref{tab:corr_coeff_others} computed using the characteristics of the Planck experiment (temperature).}
  \label{tab:corr_coeff_synch}
\end{table}

\subsection{Noise and masks}
\label{sec:noise-masks}

The main source of anisotropy in the foreground maps are the foregrounds themselves as they are mostly present in the galactic plane, but we still need to examine the influence of the other sources discussed in section \ref{realsky}.

\begin{figure}
  \centering
\includegraphics[width=0.66\linewidth]{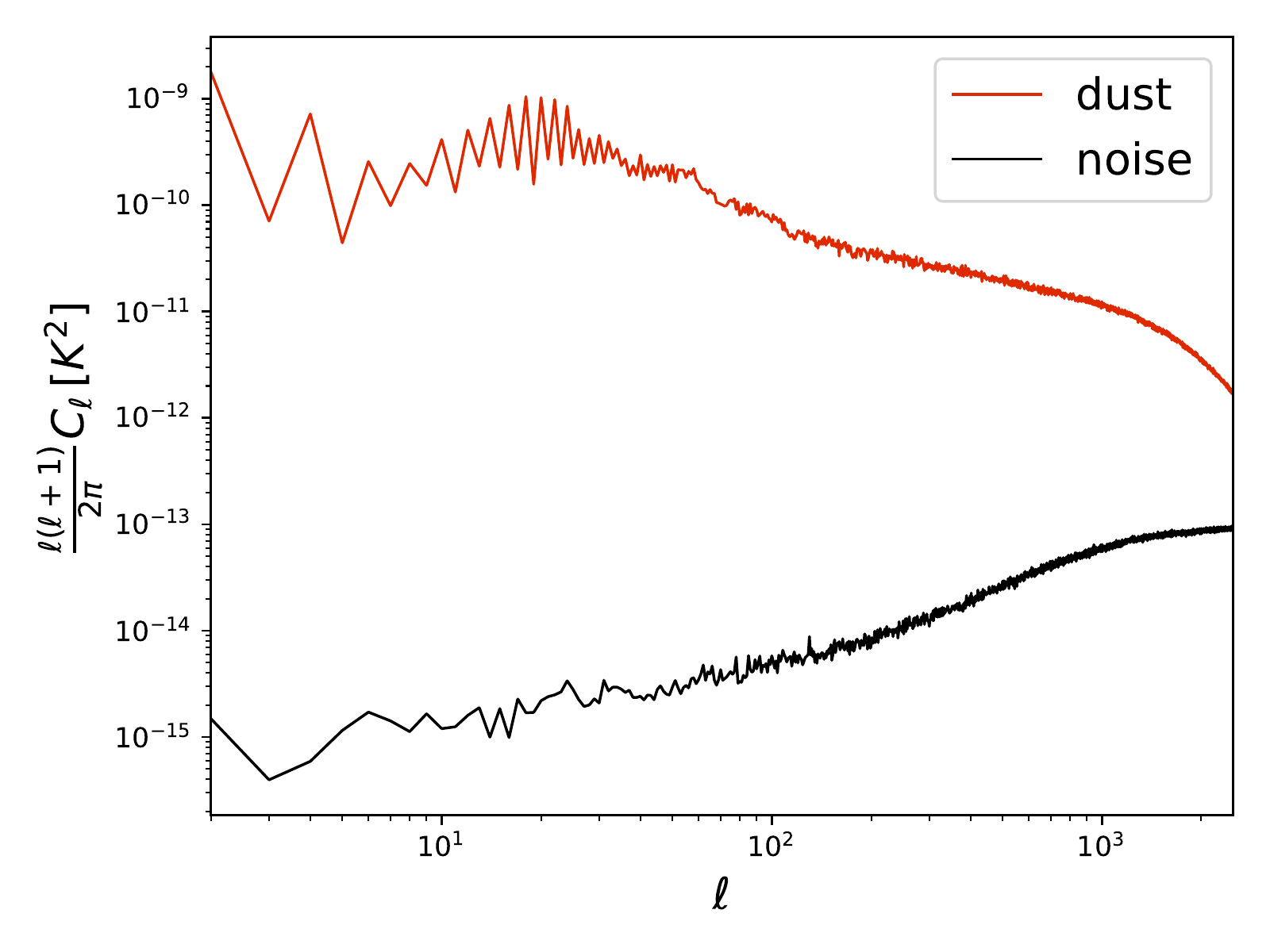}
  \caption{Dust and noise power spectra for the map studied in section~\ref{sec:dust} as a function of $\ell$.}
  \label{fig:dust-noise}
\end{figure}

We start by the noise, which for the CMB has a large effect at high $\ell$. Hence, it is sufficient to look at the best resolution dust map. Figure \ref{fig:dust-noise} shows the noise power spectrum of the dust map evaluated using half-mission maps. Even at high $\ell$, it seems that it is small compared to the signal. Hence, we will not discuss it further in this paper.

\begin{figure}
  \centering  \includegraphics[width=0.49\linewidth]{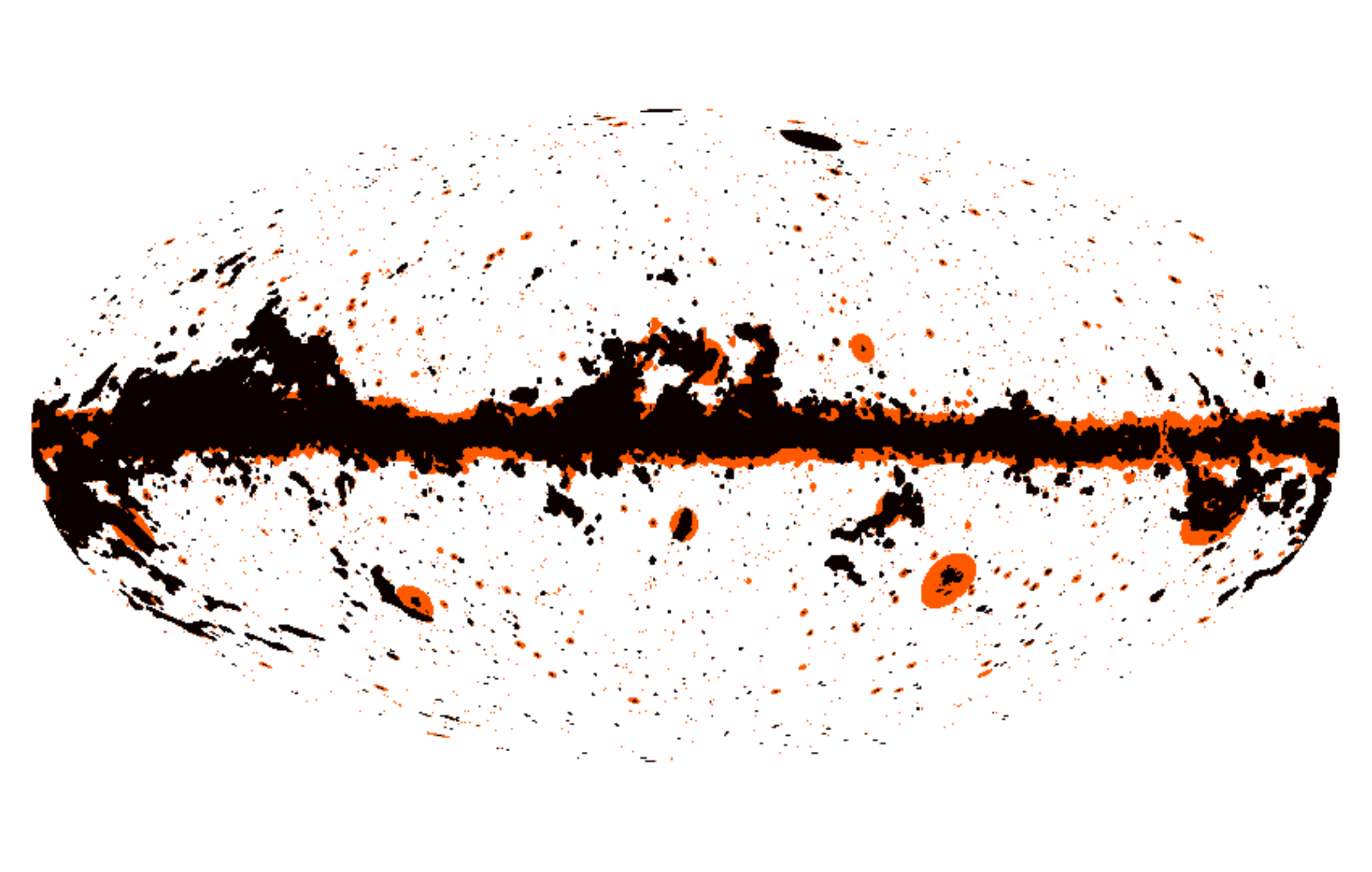} \includegraphics[width=0.49\linewidth]{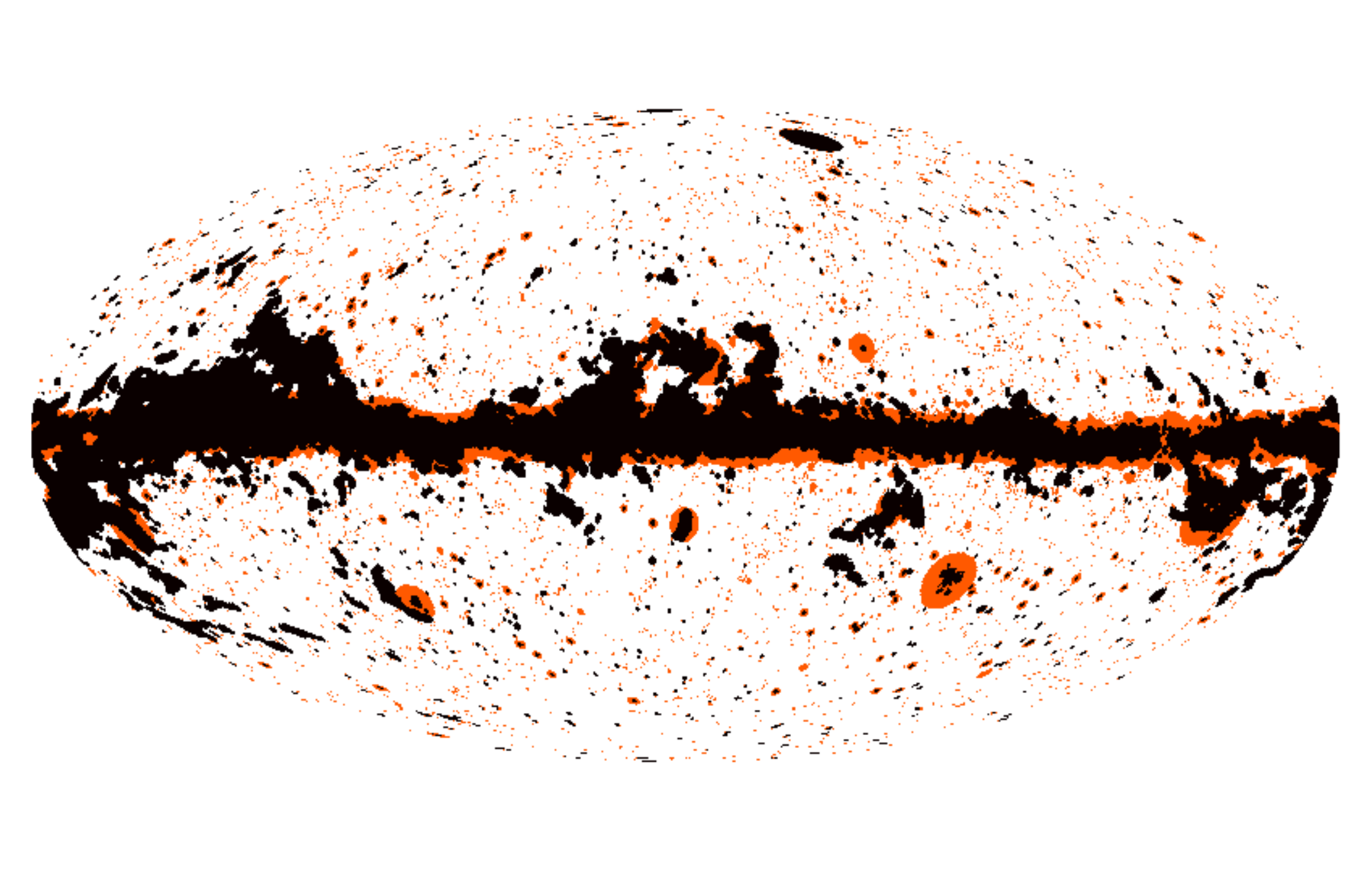}
  \caption{The different masks used in sections \ref{sec:foregrounds} and \ref{sec:analyses}. On the left, the high resolution versions ($n_\mathrm{side}=2048$) of the \texttt{Commander} mask ($f_\mathrm{sky} = 0.822$) in black only and of the common mask ($f_\mathrm{sky} = 0.776$) in black and orange. On the right, the masks have a low resolution ($n_\mathrm{side}=256$) and have been obtained by degrading the resolution of the masks on the left. In the process their size has slightly increased ($f_\mathrm{sky} = 0.804$ and $f_\mathrm{sky} = 0.745$). This effect is easily visible for the point sources.}
  \label{fig:masks}
\end{figure}

The choice of mask should also be examined more carefully. That is why here we compare our previous results obtained with the common mask ($f_\mathrm{sky} = 0.776$) to those obtained with the mask provided by the \texttt{Commander} component separation method which is slightly smaller ($f_\mathrm{sky} = 0.822$) and of course fully included in the common mask. Figure \ref{fig:masks} shows these two masks in the high and low resolution cases.

\begin{figure}
  \centering \includegraphics[width=0.49\linewidth]{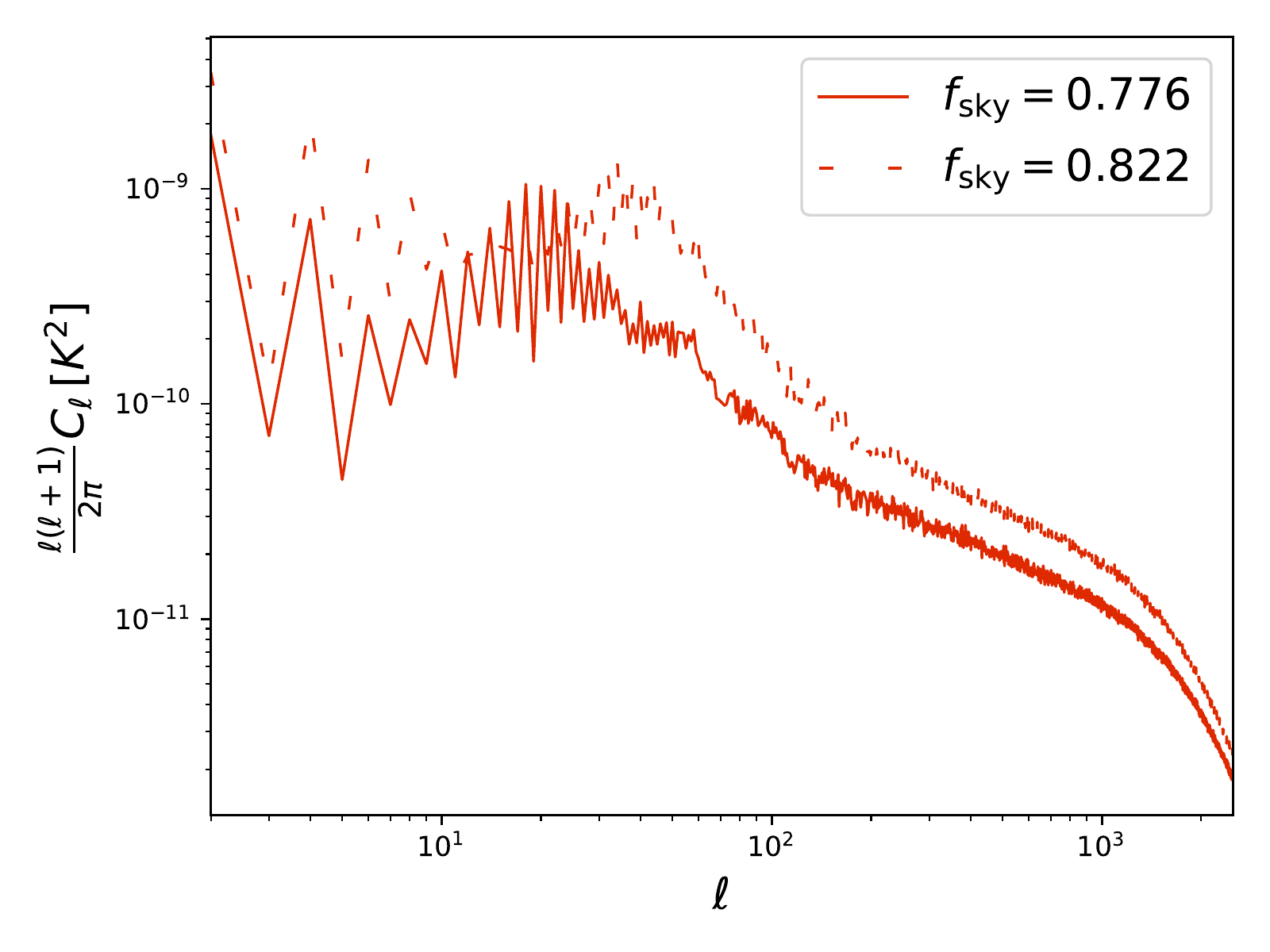}
  \includegraphics[width=0.49\linewidth]{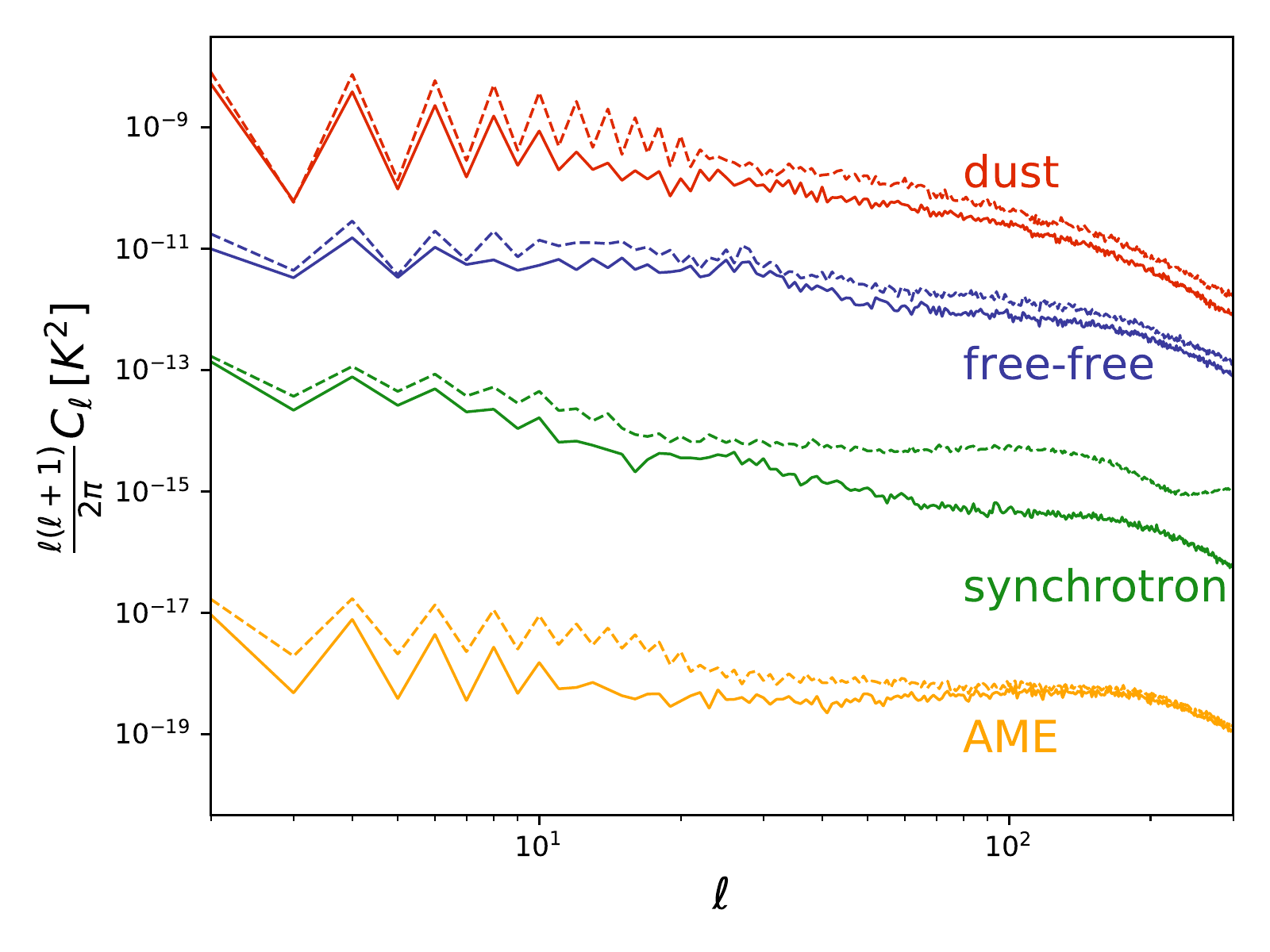}
  \caption{Power spectra of the different foregrounds using the common mask (solid line, $f_\mathrm{sky} = 0.776$) and the \texttt{Commander} mask (dashed line, $f_\mathrm{sky} = 0.822)$. On the left, the high resolution dust map ($n_{\mathrm{side}}=2048$) and on the right, the low resolution foreground maps ($n_{\mathrm{side}}=256$).}
  \label{fig:masks-power-spectra}
\end{figure}

Figure \ref{fig:masks-power-spectra} shows the power spectra of the different foregrounds with these two masks and highlights the large difference between the two cases. The reason for this difference is quite obvious because the masks have been constructed to hide most of the foregrounds, so with a smaller mask, there is a lot more of the foregrounds to detect. Moreover, as they are anisotropic, both the amplitude and the form are different depending on the mask, this is especially true for the synchrotron signal (as we will explain below).

\begin{figure}
  \centering \includegraphics[width=0.32\linewidth]{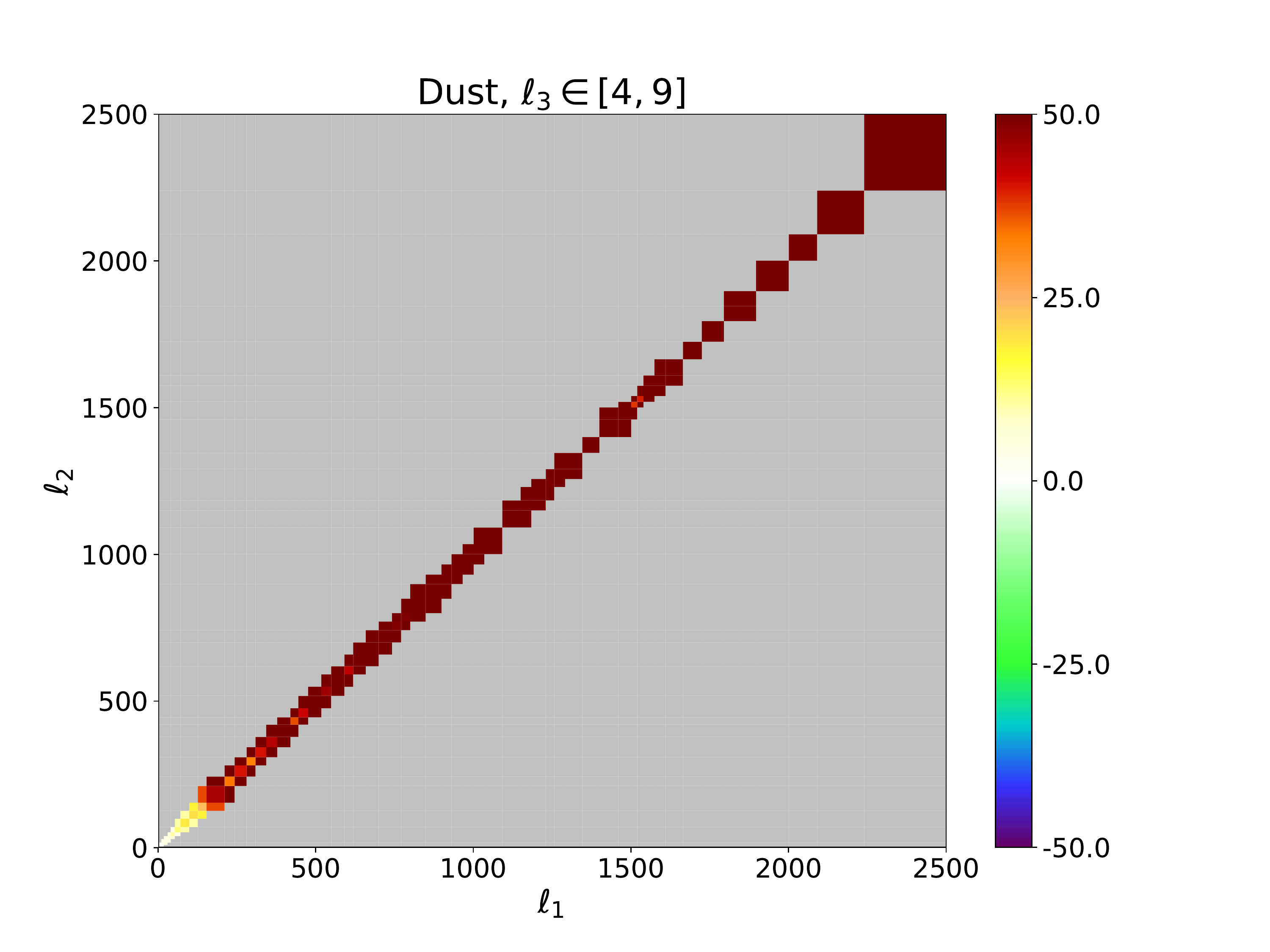}  \includegraphics[width=0.32\linewidth]{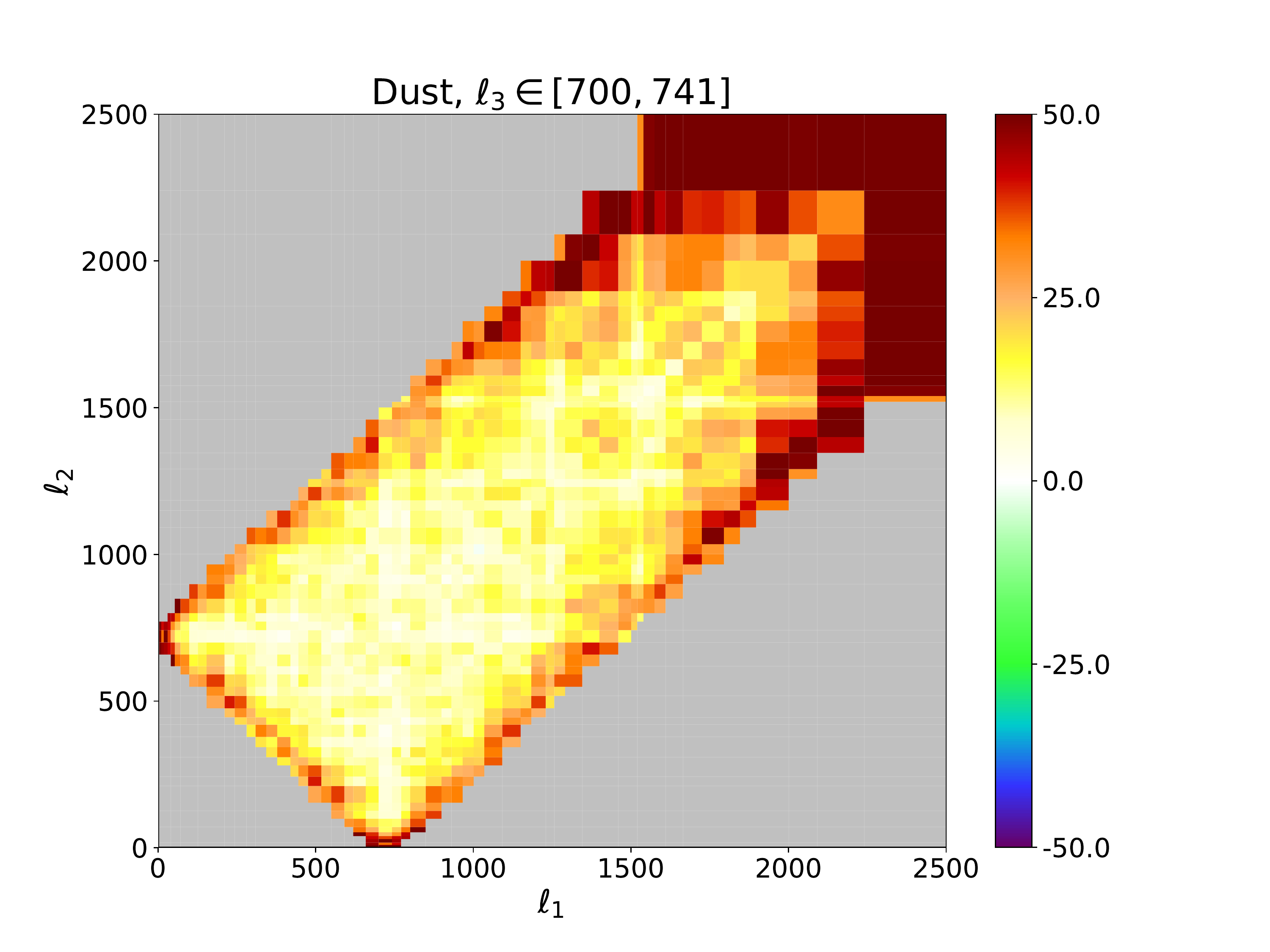} \includegraphics[width=0.32\linewidth]{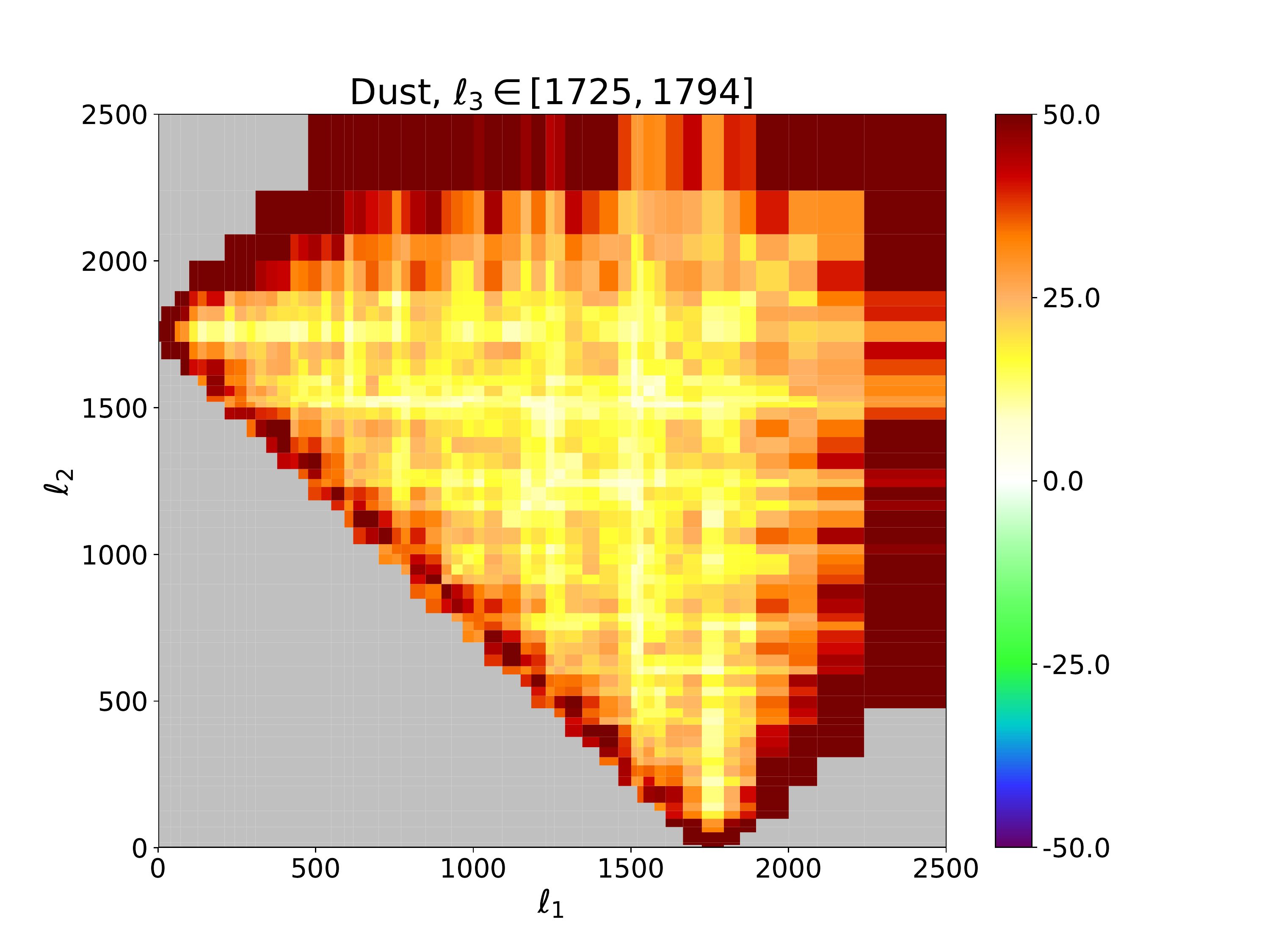}
\includegraphics[width=0.32\linewidth]{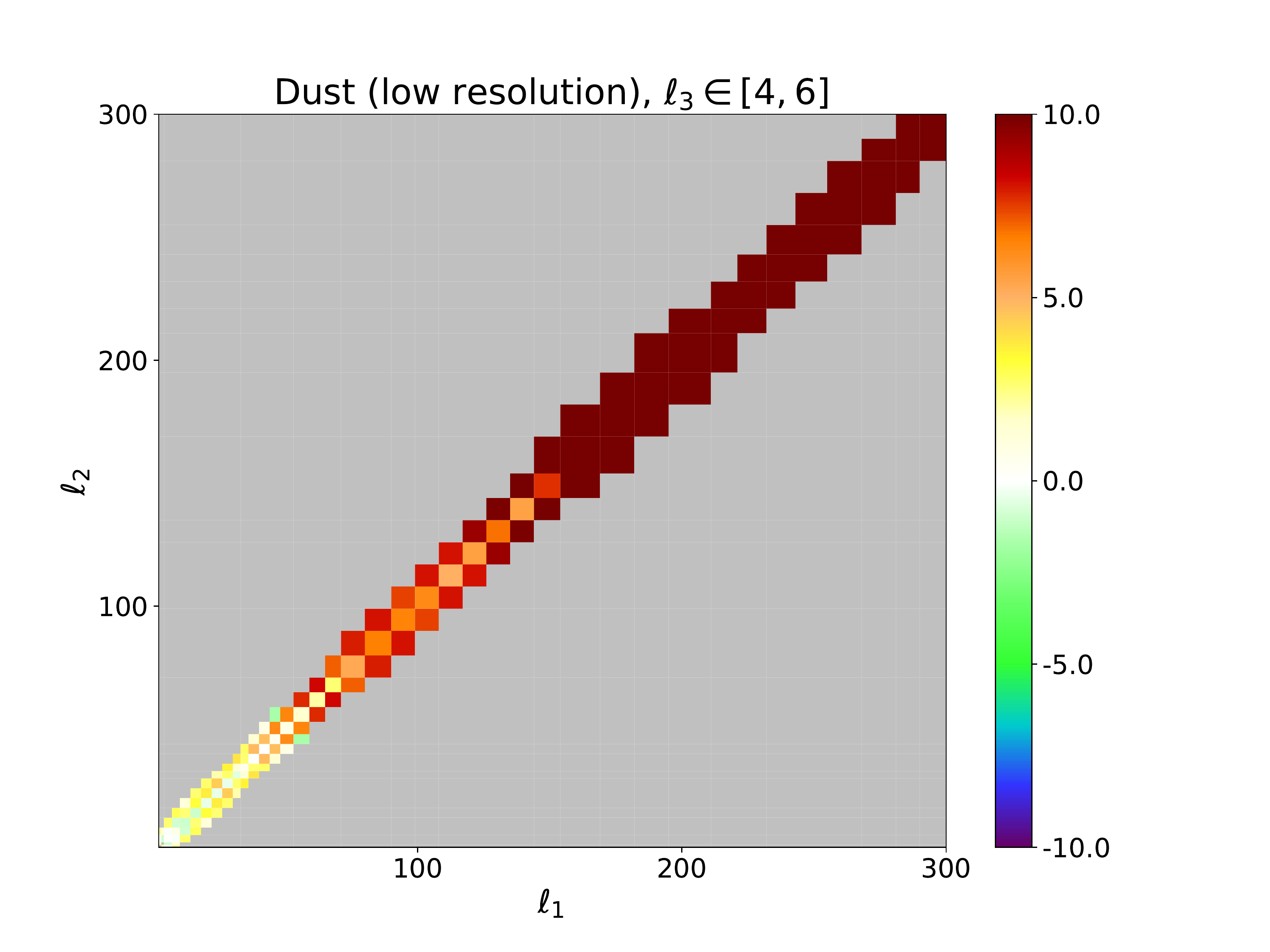}  \includegraphics[width=0.32\linewidth]{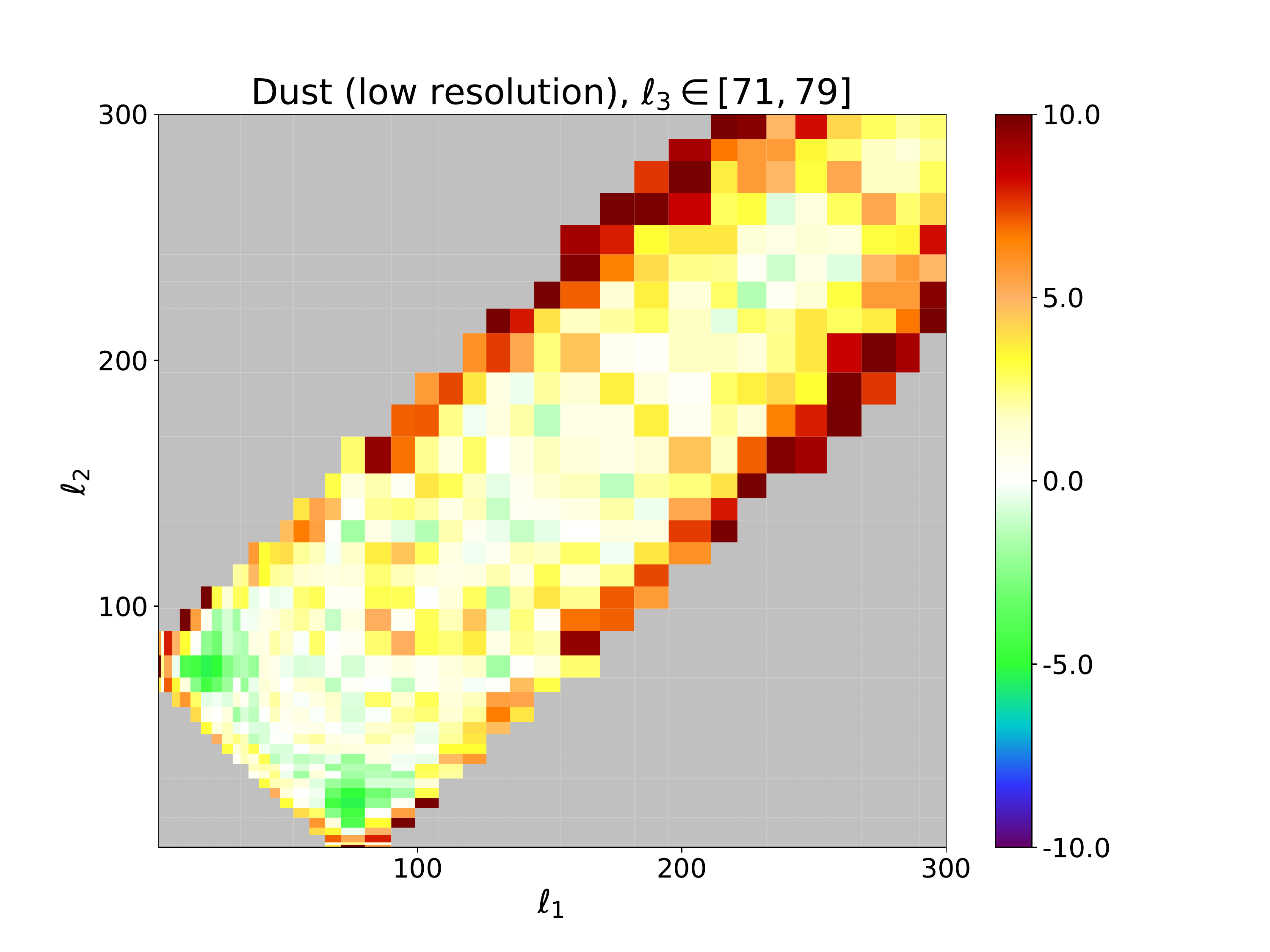} \includegraphics[width=0.32\linewidth]{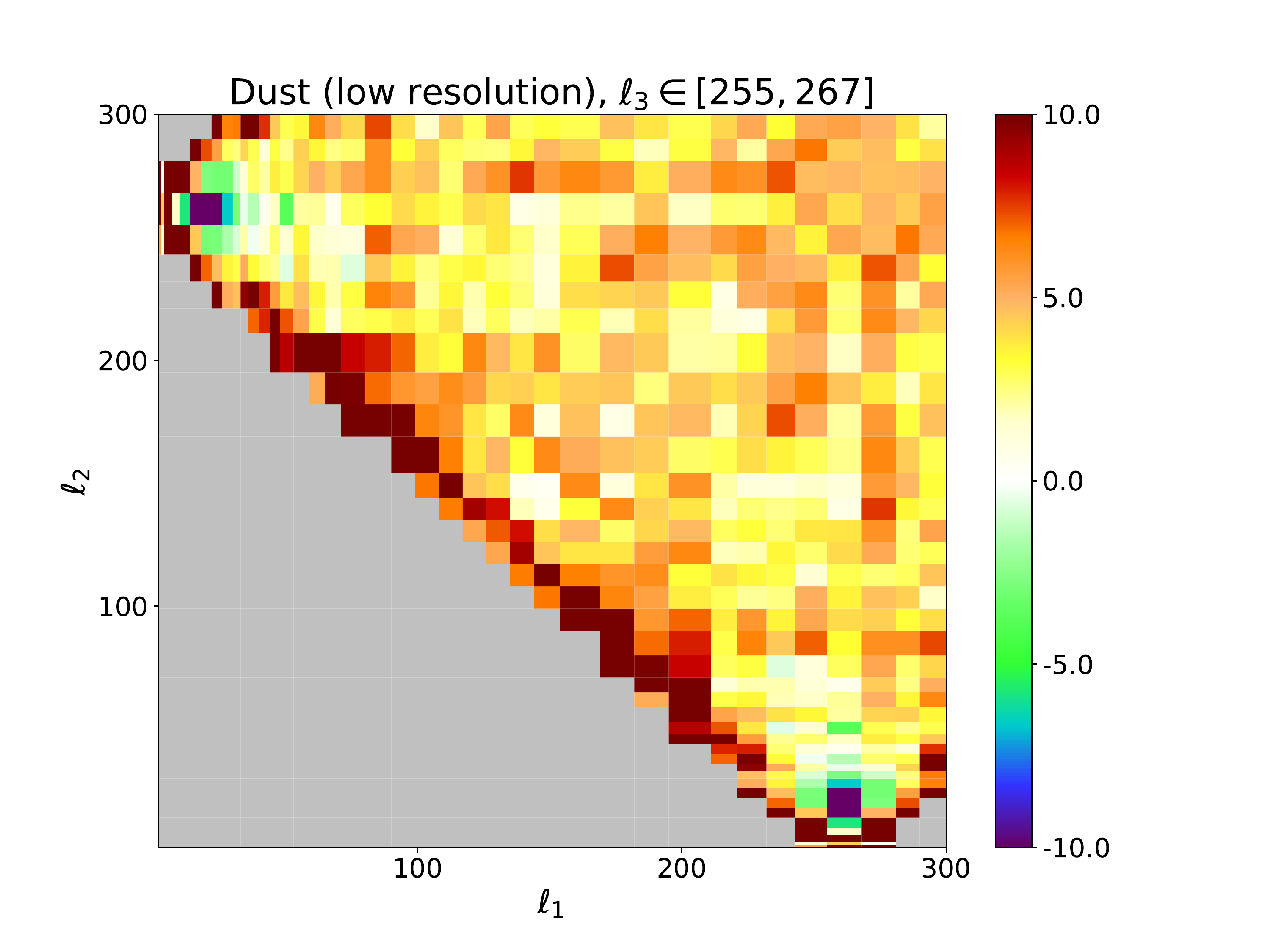}
\includegraphics[width=0.32\linewidth]{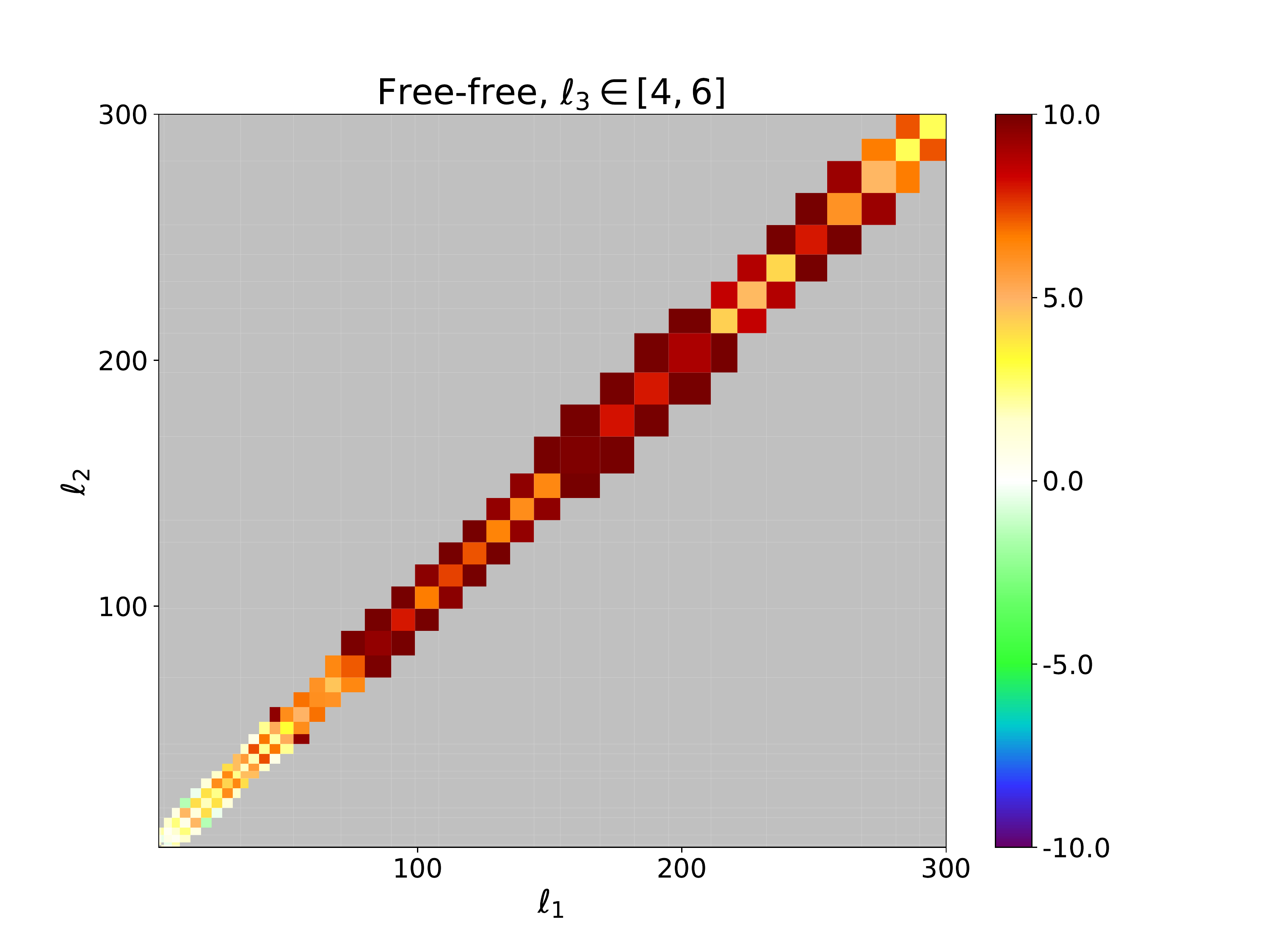}   \includegraphics[width=0.32\linewidth]{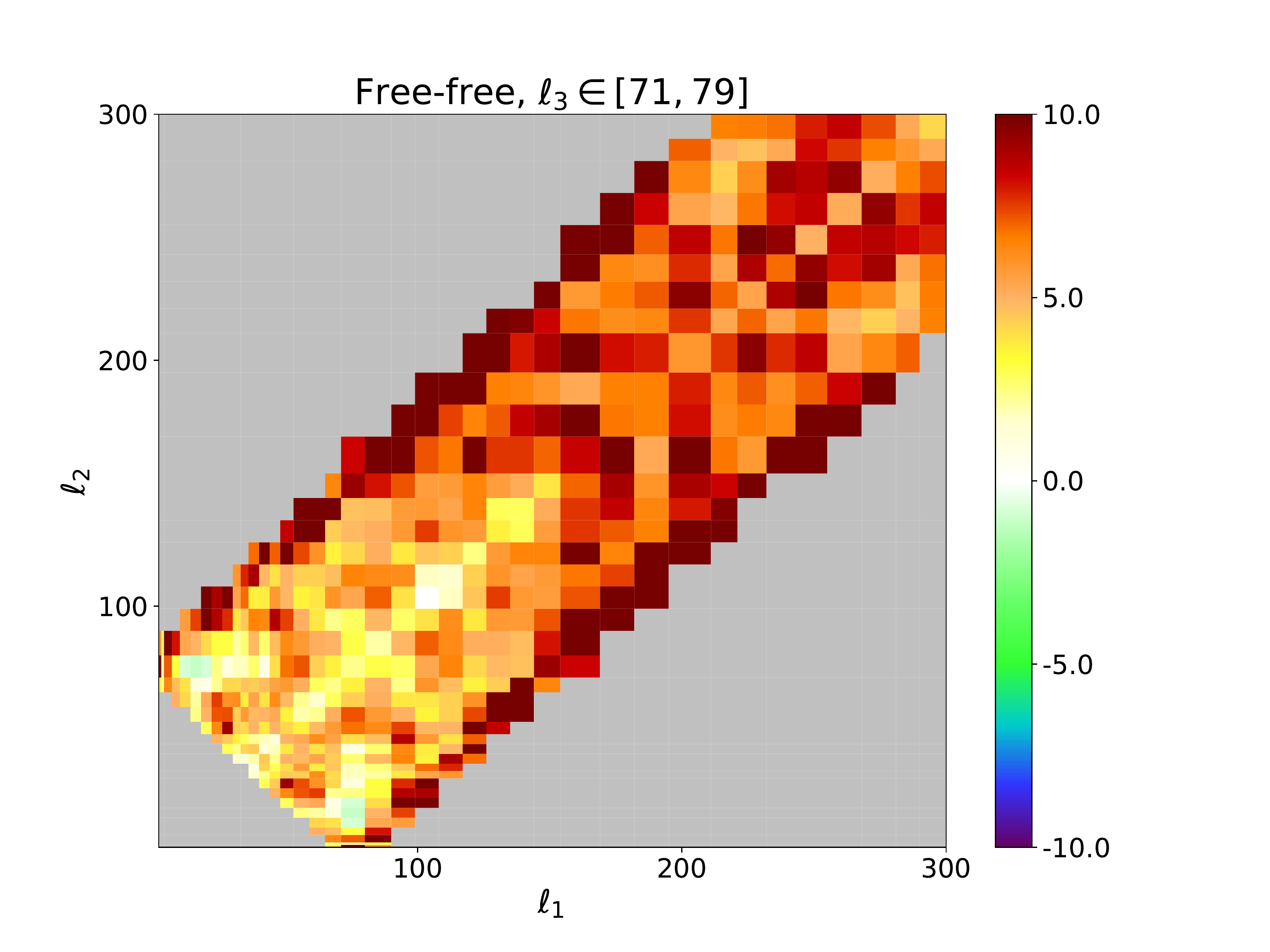} 
\includegraphics[width=0.32\linewidth]{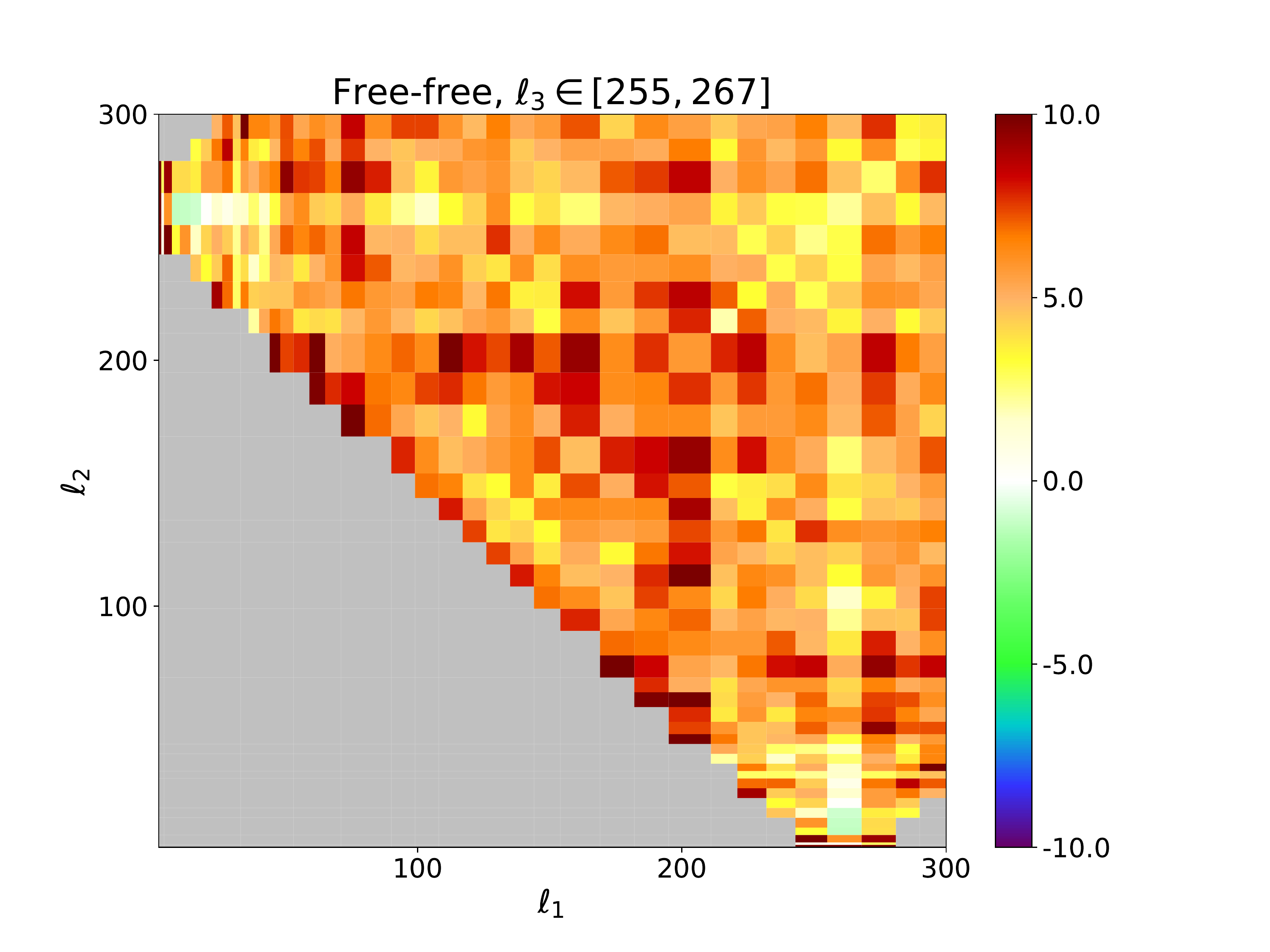}
\includegraphics[width=0.32\linewidth]{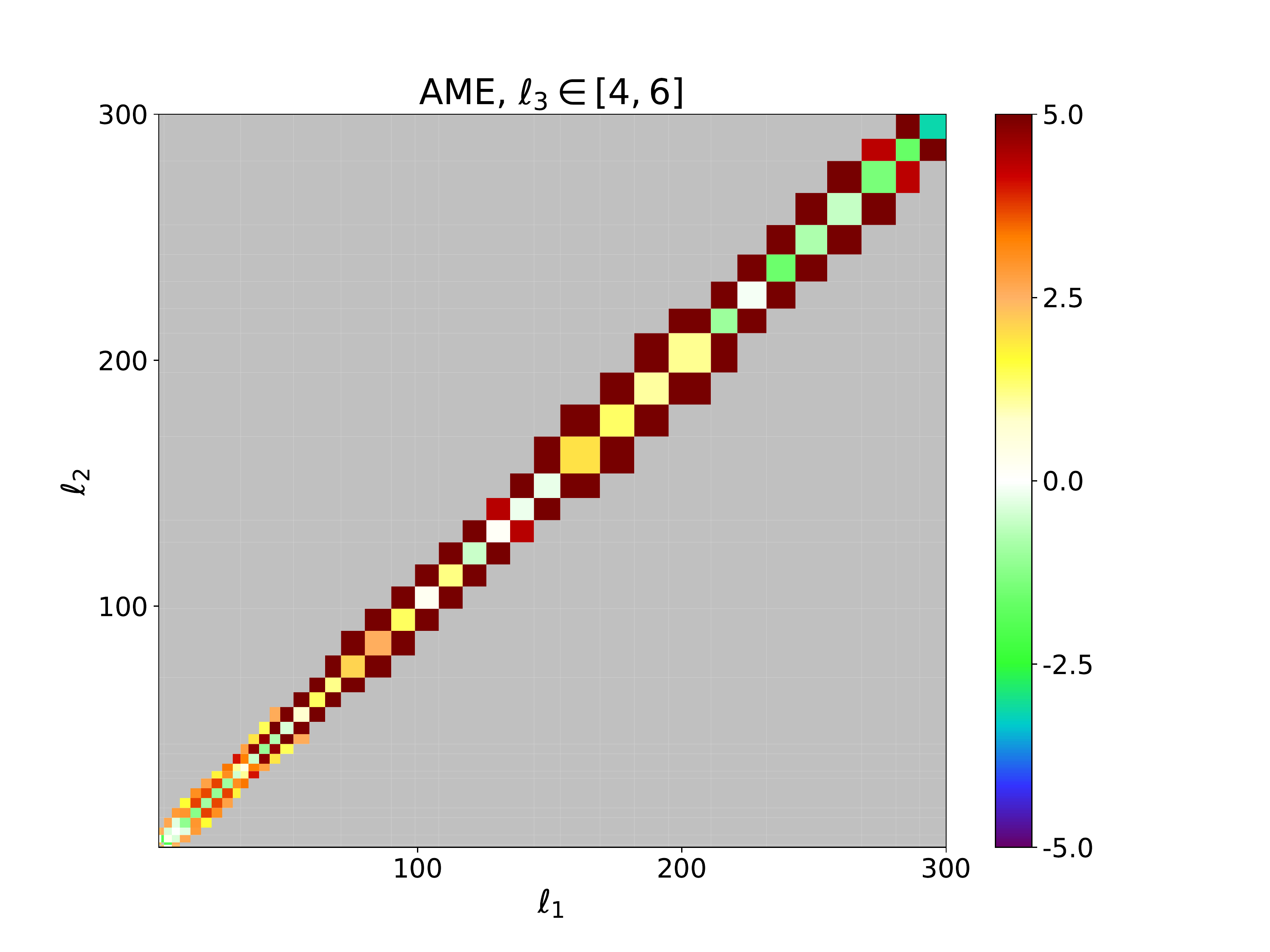} 
\includegraphics[width=0.32\linewidth]{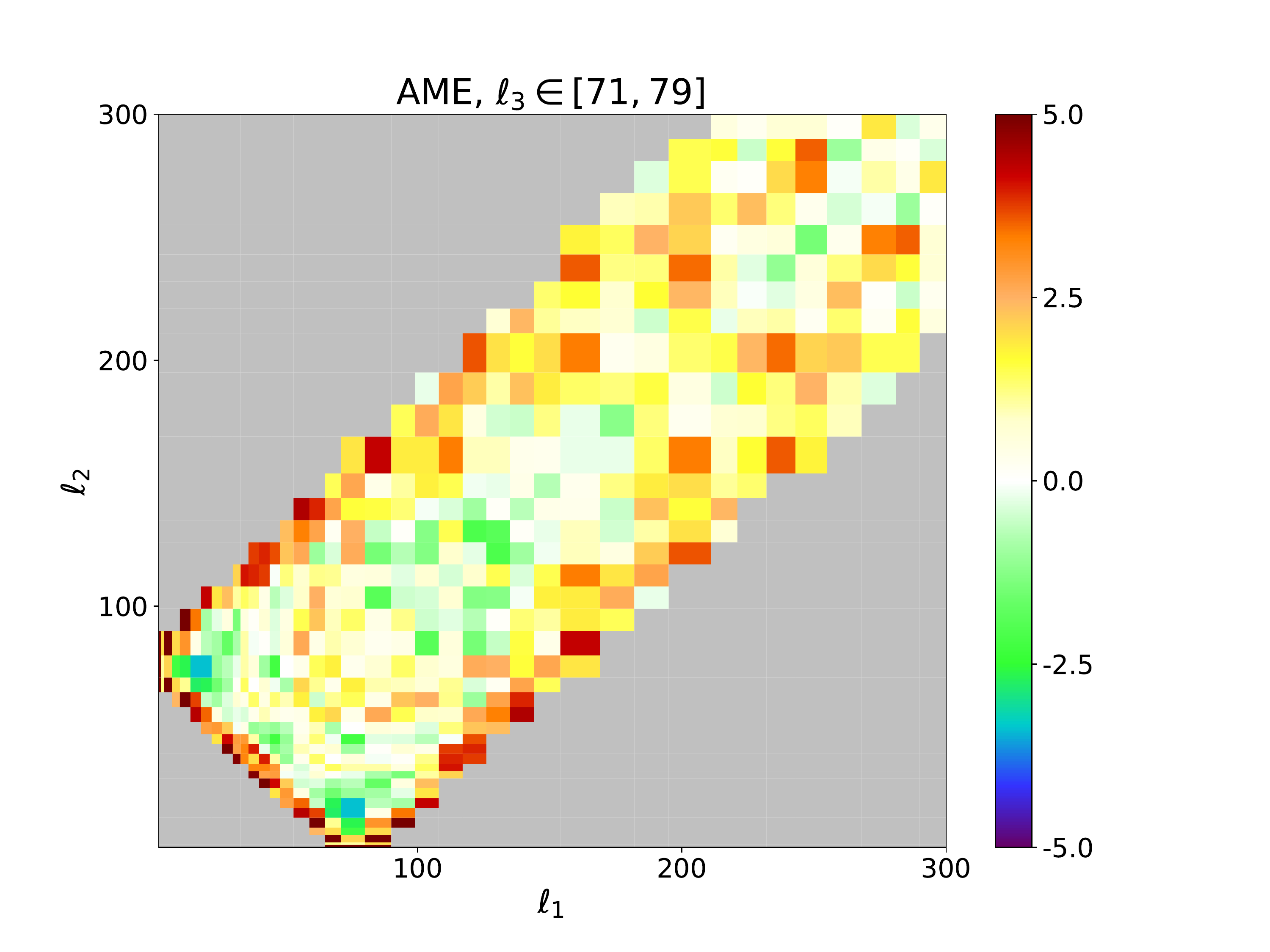}  \includegraphics[width=0.32\linewidth]{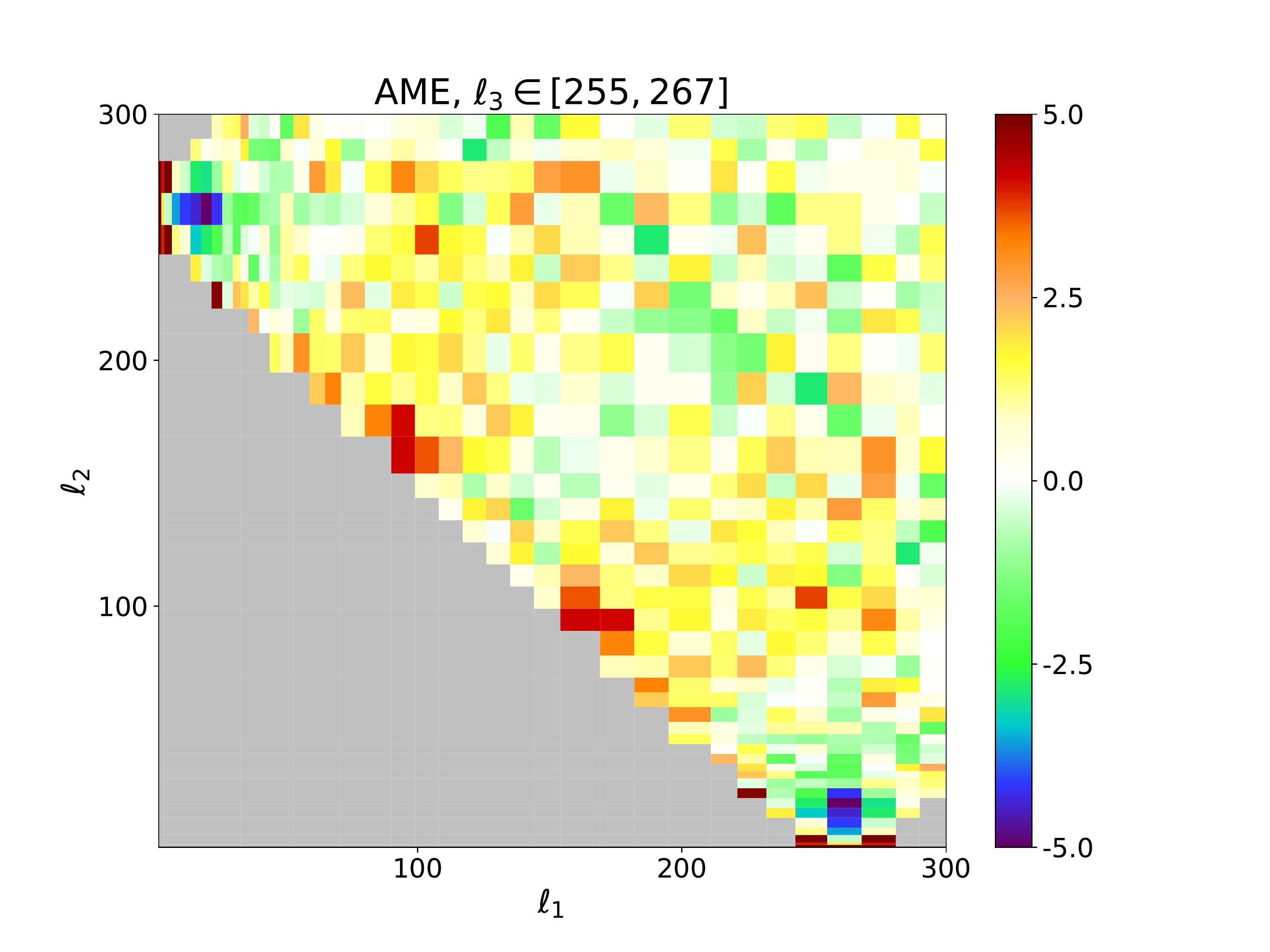}
\includegraphics[width=0.32\linewidth]{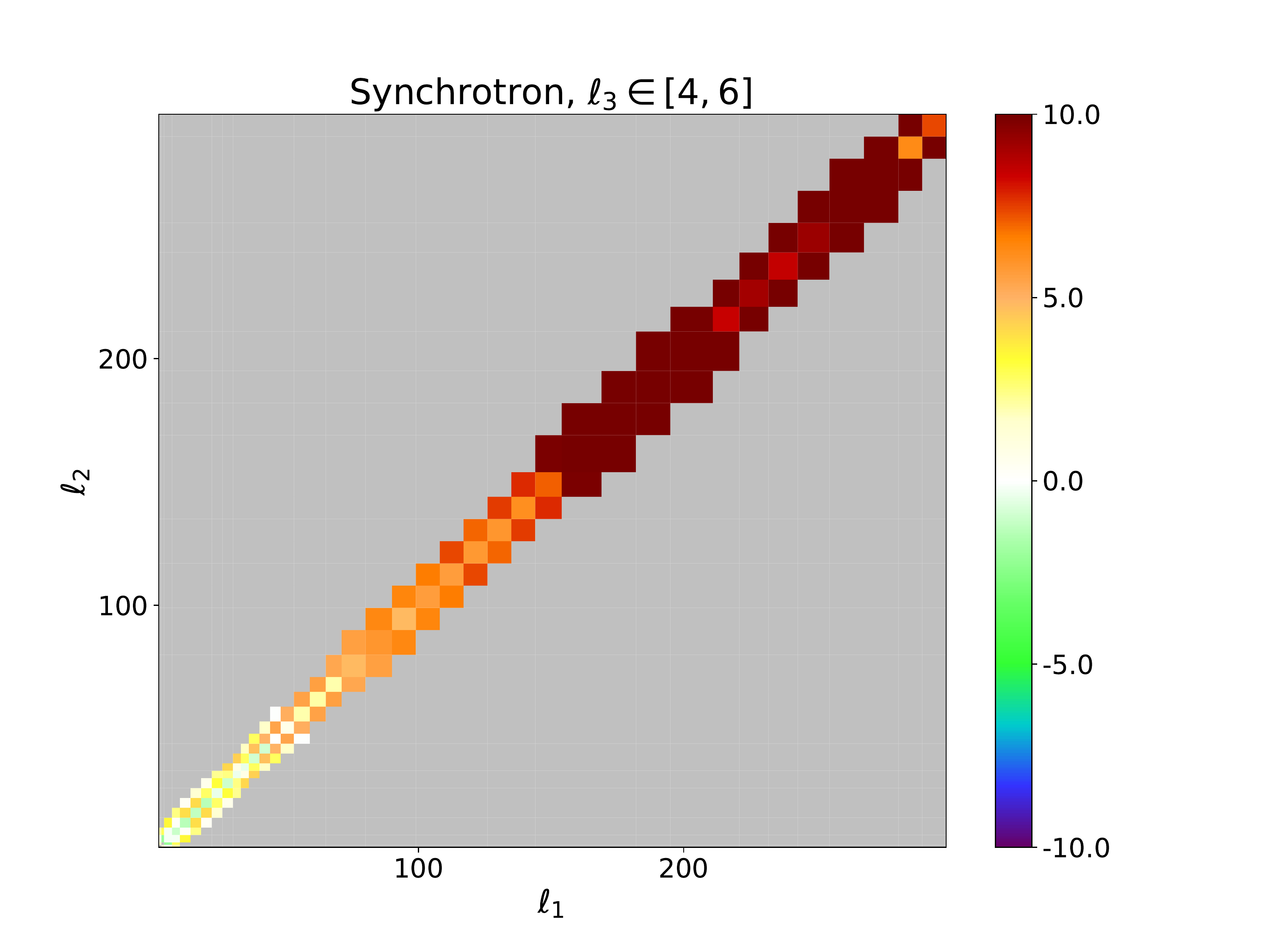} 
\includegraphics[width=0.32\linewidth]{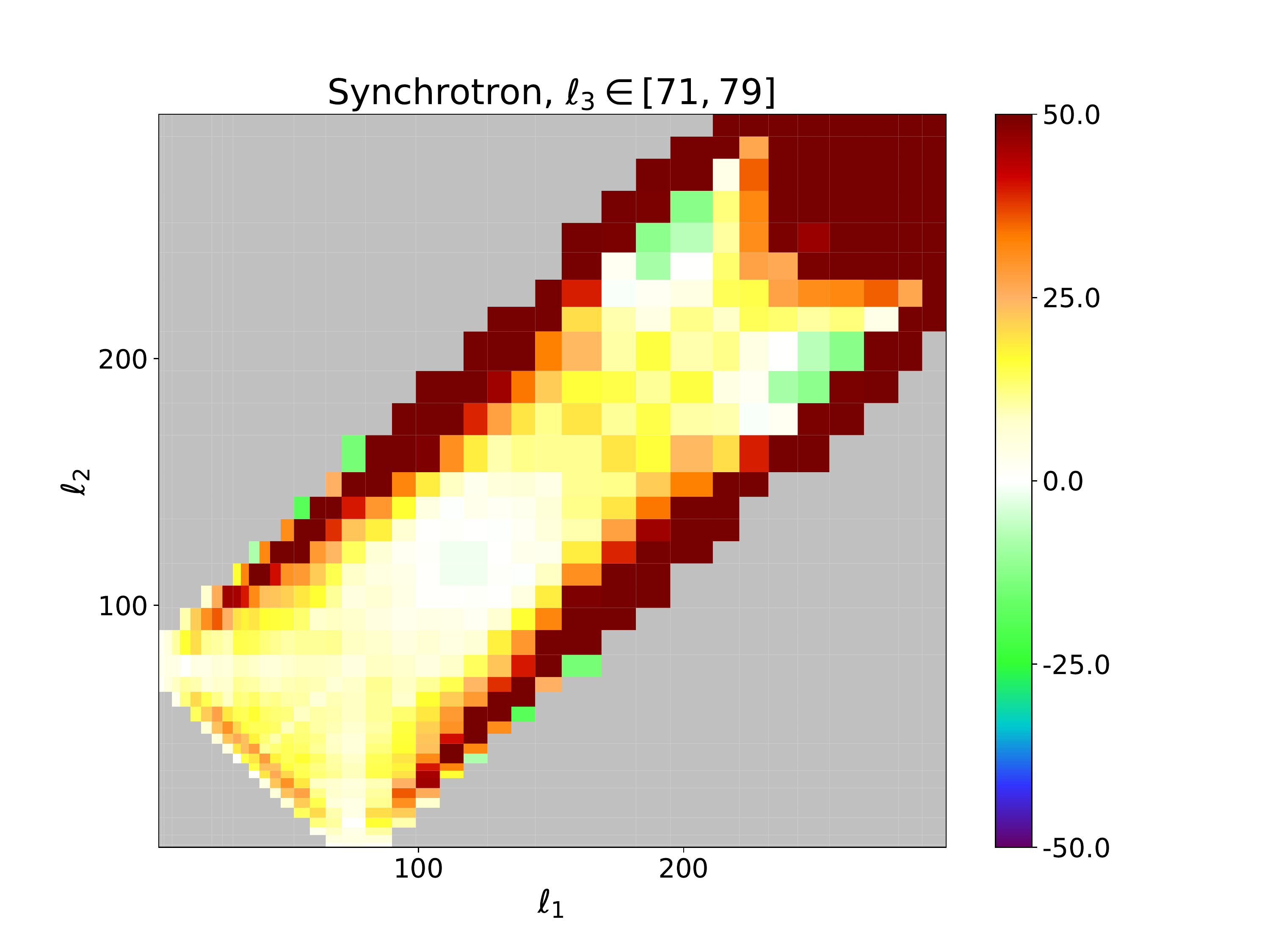}  \includegraphics[width=0.32\linewidth]{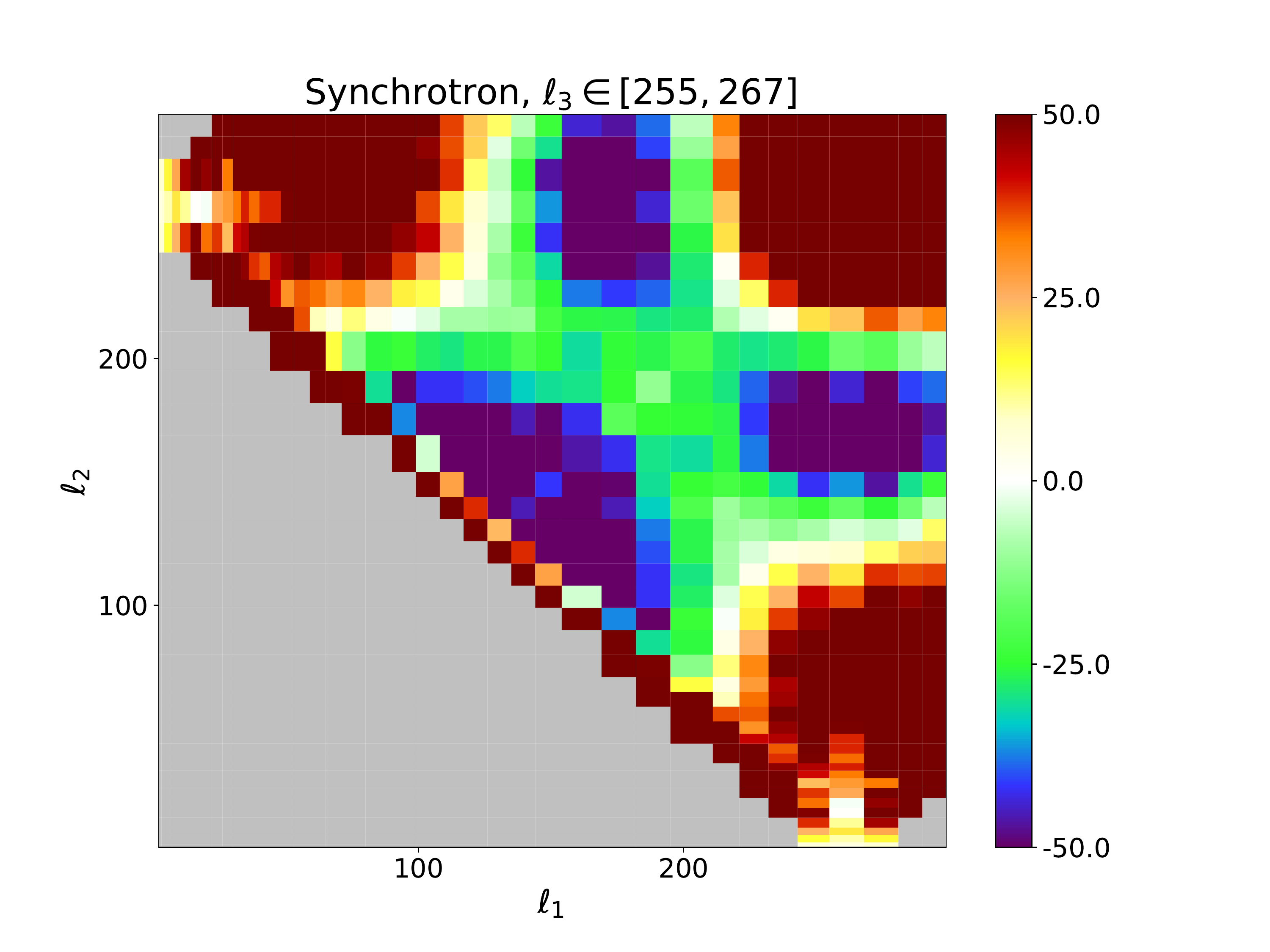}
  \caption{Bispectral signal-to-noise of the different foregrounds for the same three bins of $\ell_3$ as figures \ref{fig:dust-template} and \ref{fig:other-templates} using the \texttt{Commander} mask. Note the different colour scales.}
  \label{fig:bispectrum-masks}
\end{figure}

This is also checked for the bispectra as shown in figure \ref{fig:bispectrum-masks}. With the \texttt{Commander} mask (the smallest one), all the signals are a lot more non-Gaussian. To verify that is not only a difference of amplitude, we have at our disposal a useful tool: the correlation coefficients defined in \eqref{corrmatrix}. For each foreground, we have computed the correlation between the templates determined using the two masks. The results are given in table \ref{tab:correlation-masks}. For the dust, free-free and AME emissions, the templates are correlated (above 80 $\%$) and indeed we can see that the bispectra peak in the squeezed configuration as discussed previously. However, the fact that the correlation is not 100 $\%$ shows that the difference is not only the amplitude.

\begin{table}
  \begin{center}
    \begin{tabular}{lcccc}
      \hline
      Dust & Dust (low resolution) & Free-free & AME & Synchrotron \\
      \hline
      0.90 & 0.85 & 0.88 & 0.91 & 0.11 \\
      \hline
    \end{tabular}
  \end{center}
  \caption{Correlation coefficients between the bispectral templates determined using the common mask and the \texttt{Commander} mask for each foreground.}
  \label{tab:correlation-masks}
\end{table}

\begin{figure}
  \centering
\includegraphics[width=0.49\linewidth]{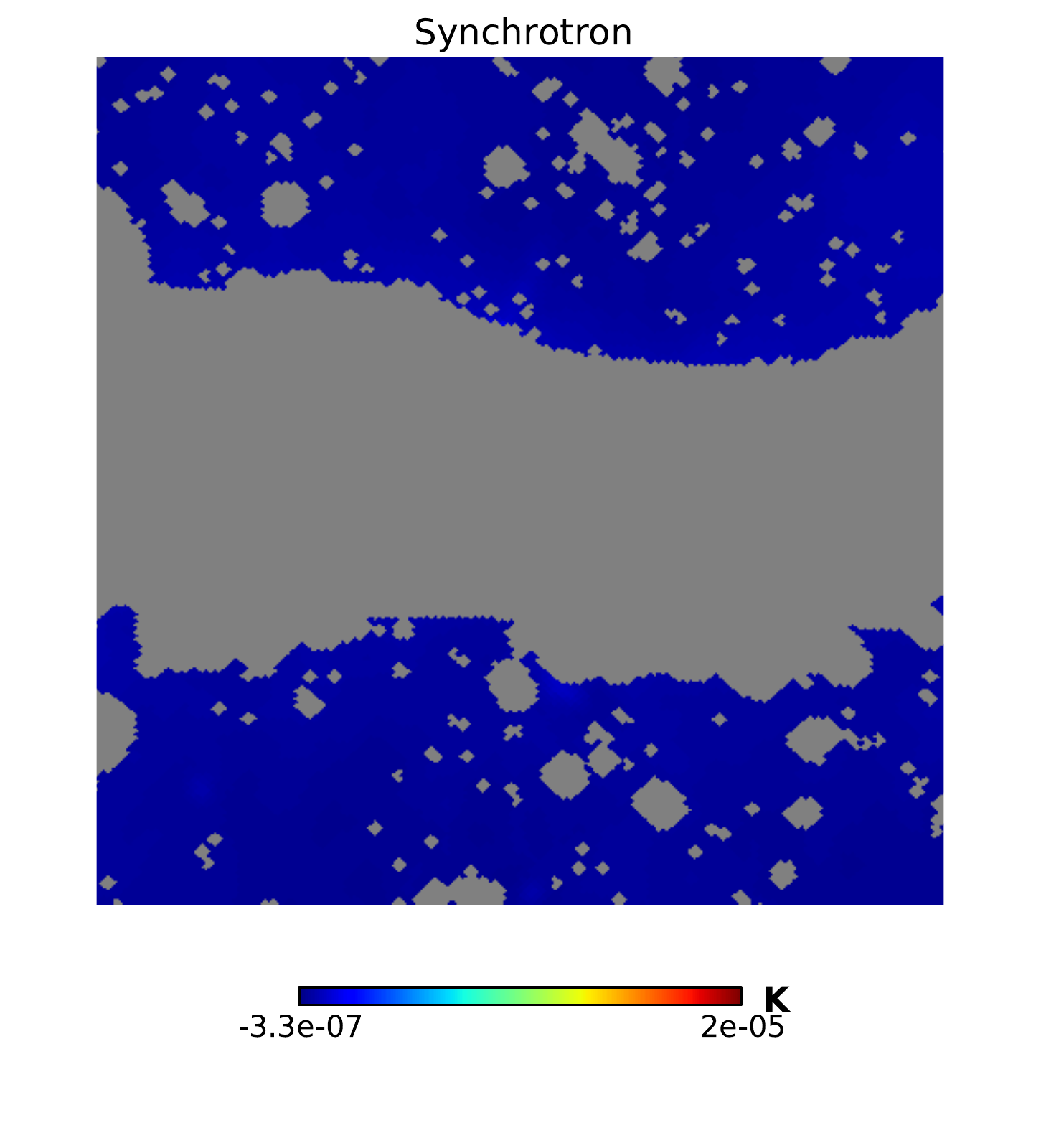}  \includegraphics[width=0.49\linewidth]{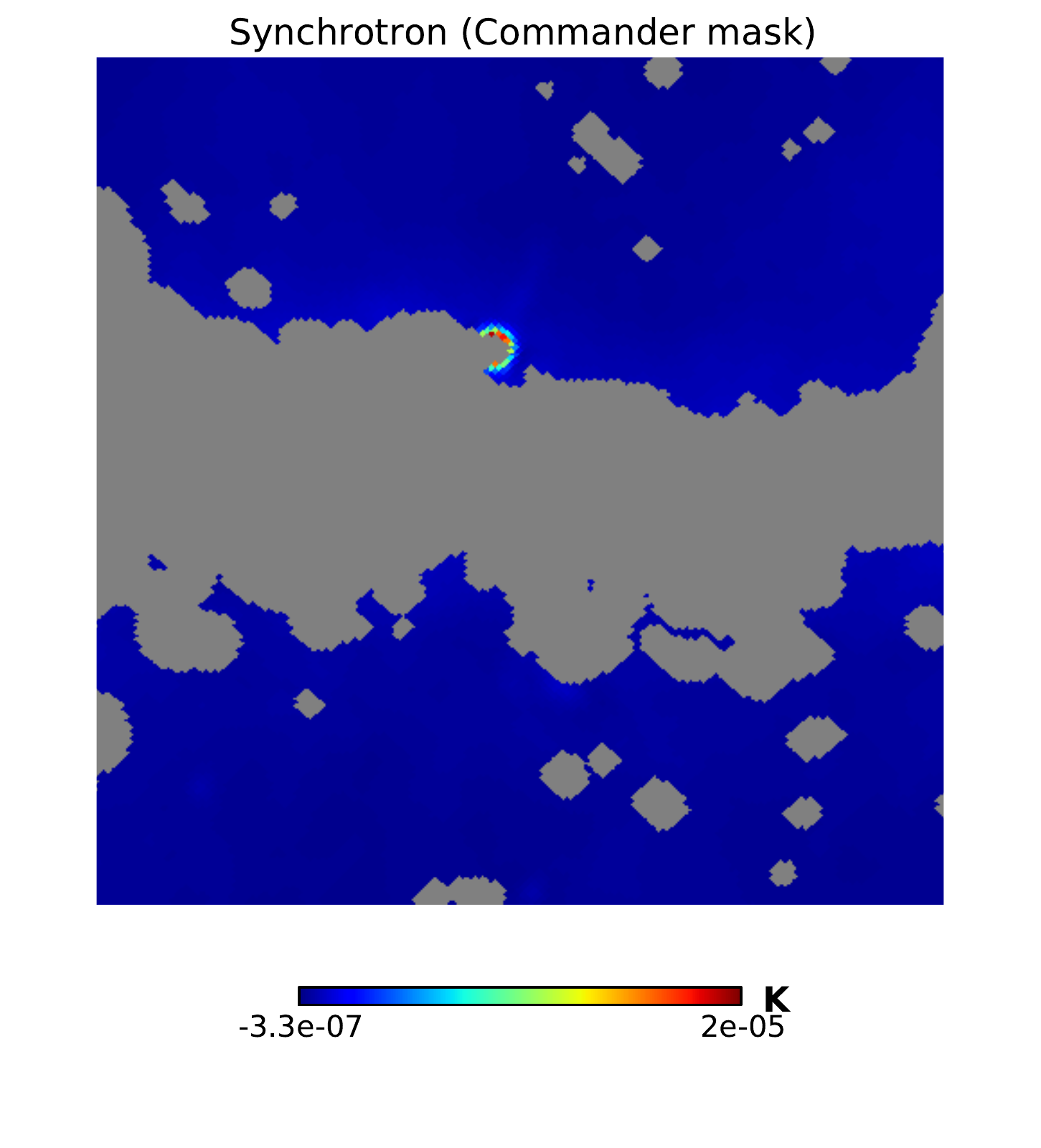}
  \caption{Zoom on the synchrotron map at 143 GHz after application of the masks (left: common mask, right: \texttt{Commander} mask).}
  \label{fig:synchrotron-commander-mask}
\end{figure}

However, for synchrotron the situation is more complicated, like in the power spectrum case. Indeed the new template is very different from the one in figure \ref{fig:synch-ps-template} and it is confirmed by the low correlation between the two synchrotron templates determined with the two different masks. To understand this result, it is interesting to examine directly the data map with the \texttt{Commander} mask in figure \ref{fig:synchrotron-commander-mask}. One can see that there are a few pixels where the intensity is ten times larger than with the common mask (where they are hidden). The influence of this very small region dominates the power spectrum and the bispectrum because the transition is so important. It could be modeled as a Heaviside step function, the Fourier transform of which is a sinc function, meaning that these two pixels have a large influence over the whole multipole space and we can see oscillations as expected in both the power spectrum (there is a minimum at $\ell \approx 240$) and the bispectrum (there are three regions of negative bispectrum with positive bispectrum around them on the plot for $\ell_3\in[255,267]$).

In conclusion, the choice of mask has a large influence on the templates we are determining because of the localization of the foregrounds in the galactic plane. This means that when we apply these templates to other maps in the next section, it is mandatory to use the same mask at every step. From now on, we will exclusively use the common mask.

\section{Data analyses}
\label{sec:analyses}

The aims of this section are twofold. First, we want to verify that the numerical templates we just determined can be used in the context of a CMB data analysis. The first basic test to check this is to apply the template to the map it comes from. The expected answer for the amplitude parameter of this specific shape is then obviously $\fnl=1$. Moreover, if we perform a correlated analysis with other shapes like the primordial ones, their own $\fnl$ has to be negligible. Indeed, that is what we observe (see table \ref{tab:dust}) and we can now discuss more interesting tests based on CMB maps.
It is important to recall that the galactic foregrounds are highly anisotropic while most other shapes have an isotropic origin (primordial, lensing-ISW or extra-galactic foregrounds). These galactic numerical templates also contain mask and noise effects, but we will show that it is not an issue. For this, we ran a series of tests with the simple idea of artificially adding dust to the maps containing CMB realizations (simulations, but also the observed data) to check that we indeed detect the right amount of dust and that it has no impact on the other shapes.

\begin{table}
  \begin{center}
    \small
  \begin{tabular}{lcccccc}
  \hline
 & Local & Equilateral & Orthogonal & P.S.$/10^{-29}$ & CIB$/10^{-27}$ & Dust\\
  \hline
  \multicolumn{1}{@{\hspace{0.5cm}}c@{\hspace{0.3cm}}}{$Indep$} & $-5.3$ & $16.5 $ & $15.4$ & $1.76$ & $1.32$ & $1.0$ \\
    \multicolumn{1}{@{\hspace{0.0cm}}c@{}}{$Joint$} & $-7 \times 10^{-11}$ & $2 \times 10^{-10}$ & $-2 \times 10^{-11}$ & $-1 \times 10^{-13}$ & $7 \times 10^{-14}$ & $1$ \\
  \hline
  \end{tabular}
  \end{center}
  \caption{Determination of $\fnl$ for the local, equilateral, orthogonal, point sources, CIB and dust shapes using the high-resolution dust map studied in section \ref{sec:dust}. The only error bars at our disposal are Fisher forecasts; they are not indicated because for every case given here, they are several (at least three) orders of magnitude smaller than the determined values for $\fnl$ in the independent case, which makes them many orders of magnitude larger than the non-dust values in the joint analysis.} 
  \label{tab:dust}
  \end{table}

Then, we will focus on the second aim which is to analyze the CMB map from the 2015 Planck data. We will apply the numerical templates to the cleaned \texttt{SMICA} CMB map \citepalias{planck2015-09}, both at low and high resolutions, for which we expect not to detect any galactic foregrounds. Finally we will perform a similar analysis on raw sky observations at 143 GHz.

\subsection{Gaussian simulations}
\label{sec:gaussian-simulations}

For the first tests, we constructed a set of 100 Gaussian simulations of the CMB obtained using the best fit of the CMB power spectrum from the 2015 Planck release \citepalias{planck2015-11} at the resolution $n_{\mathrm{side}}=2048$. There are several reasons to use these simulations instead of the observed CMB  map. First, it is important to check the validity of this new use of the binned bispectrum estimator with a large number of maps. Moreover, even the cleaned CMB map still contains contamination from extra-galactic foregrounds and the ISW-lensing. Here, these effects are not present. However, we need the Gaussian realizations of the CMB to have the characteristics of the \texttt{SMICA} CMB map. Hence, we smoothed the maps using a 5~arcmin FWHM Gaussian beam and we added noise based on the noise power spectrum of the \texttt{SMICA} CMB map (moreover, our choice of bins is optimal only if this noise is present in the maps, because it diminishes the weights of the bins at high $\ell$ following \eqref{binnedVarreal}). In this section, we will discuss two different cases for the noise. First, we will assume it has an isotropic distribution in pixel space. In the second case we will make it anisotropic by modulating it in pixel space using the hit-count map corresponding to the scanning pattern of the Planck satellite. Finally, we add some dust to these maps using the high-resolution dust map at 143 GHz discussed in section \ref{sec:dust}. Every analysis presented in this section uses the common mask introduced in section \ref{sec:foregrounds}, see figure~\ref{fig:masks}.

The determination of the amplitude parameters is performed using the binned bispectrum estimator, including a linear correction term to the bispectrum as discussed in section \ref{realsky}. In practice, the linear correction term is computed using Gaussian simulations of the analyzed maps with the same characteristics (beam, noise, mask). We use the average power spectrum of our 100 maps (CMB + dust) to generate the maps necessary for the computation of the linear correction. Here we use 80 maps for the linear correction. We have verified that this number is sufficient to detect squeezed bispectra like the local and the dust shapes to high precision. The first analysis is performed with the same choice of 57 bins as in the 2015 Planck analysis \citepalias{planck2015-17} which was shown to be optimal to determine the primordial shapes, using multipoles from $\ell_\mathrm{min}=2$ to $\ell_\mathrm{max}=2500$ (remember that our analysis is temperature only). We add the dust map to the simulations of the CMB, thus the expected value of the $\fnl$ for the dust template is $1$. We also determine the amplitude parameters $\fnl$ for the primordial shapes, the point sources and CIB bispectra in both the independent and the joint case.

\begin{table}
\begin{center}
  \small
\begin{tabular}{lcccccc}
\hline
& Local & Equilateral & Orthogonal & P.S.$/10^{-29}$ & CIB$/10^{-27}$ & Dust\\
\hline
\multicolumn{4}{l}{Dust 100$\%$, 57 bins (expected $\fnl^{\mathrm{dust}}=1$)} &&& \\
\multicolumn{1}{@{\hspace{0.7cm}}c@{\hspace{0.5cm}}}{$Indep$} & $-86 \pm 14$ & $27 \pm 67$ & $103 \pm 38$ & $1.4 \pm 0.9$ & $1.1\pm 0.5$ & $1.03 \pm 0.20 $ \\
  \multicolumn{1}{@{\hspace{0.0cm}}c@{}}{$Joint$} & $-6 \pm 14$ & $16 \pm 77$ & $-10 \pm 45$ & $0.1 \pm 2.6$ & $0.0 \pm 1.5$ & $1.00 \pm 0.24$ \\
\multicolumn{4}{l}{Dust 100$\%$, 70 bins (expected $\fnl^{\mathrm{dust}}=1$)} &&& \\
\multicolumn{1}{@{\hspace{0.7cm}}c@{\hspace{0.5cm}}}{$Indep$} &  -67 $\pm$ 11 & 20 $\pm$ 68 & 92 $\pm$ 34 & 1.4 $\pm$ 1.0 & 1.0 $\pm$ 0.5 & 1.00 $\pm$ 0.20 \\
\multicolumn{1}{@{\hspace{0.0cm}}c@{}}{$Joint$} & 0 $\pm$ 14 & -5 $\pm$ 75 & -1 $\pm$ 39 & 0.0 $\pm$ 2.6 & 0.0 $\pm$ 1.4 & 1.01 $\pm$ 0.24 \\
  \multicolumn{4}{l}{Dust 75$\%$, 70 bins (expected $\fnl^{\mathrm{dust}}=0.42$)} &&& \\
\multicolumn{1}{@{\hspace{0.7cm}}c@{\hspace{0.5cm}}}{$Indep$} & -30 $\pm$ 8 & 11 $\pm$ 66 & 41 $\pm$ 36 & 0.6 $\pm$ 0.9 & 0.4 $\pm$ 0.5 & 0.42 $\pm$ 0.12 \\
\multicolumn{1}{@{\hspace{0.0cm}}c@{}}{$Joint$} & 0 $\pm$ 9 & 1 $\pm$ 70 & -2 $\pm$ 42 & 0.0 $\pm$ 2.6 & 0.0 $\pm$ 1.4 & 0.42 $\pm$ 0.13 \\
  \multicolumn{4}{l}{Dust 0$\%$, 70 bins (expected $\fnl^{\mathrm{dust}}=0$)} &&& \\
\multicolumn{1}{@{\hspace{0.7cm}}c@{\hspace{0.5cm}}}{$Indep$} & -0.1 $\pm$ 0.5 & -1.7 $\pm$ 6.1 & -3.1 $\pm$ 3.4 & -0.03 $\pm$ 0.09 & -0.01 $\pm$ 0.05 & 0.001 $\pm$ 0.003 \\
\multicolumn{1}{@{\hspace{0.0cm}}c@{}}{$Joint$} & -0.3 $\pm$ 0.7 & -1.6 $\pm$ 6.4 & -4.2 $\pm$ 4.1 & -0.15 $\pm$ 0.26 & 0.06 $\pm$ 0.13 & 0.001 $\pm$ 0.003 \\
\hline
\end{tabular}
\end{center}
\caption{Determination of $\fnl$ for the local, equilateral, orthogonal, point sources, CIB and dust shapes using a set of 100 Gaussian simulations of the CMB with isotropic noise to which we added a known amount of dust (the high-resolution dust map of section \ref{sec:dust} multiplied by a factor 1 or 0.75, or no dust at all). The analysis is performed using 57 bins or 70 bins and the error bars are given at 1$\sigma$. For the reason behind the much smaller error bars in the 0 $\%$ dust case, see the main text.} 
\label{tab:gaussian-isotropic}
\end{table}

\begin{figure}
  \centering
 \includegraphics[width=0.49\linewidth]{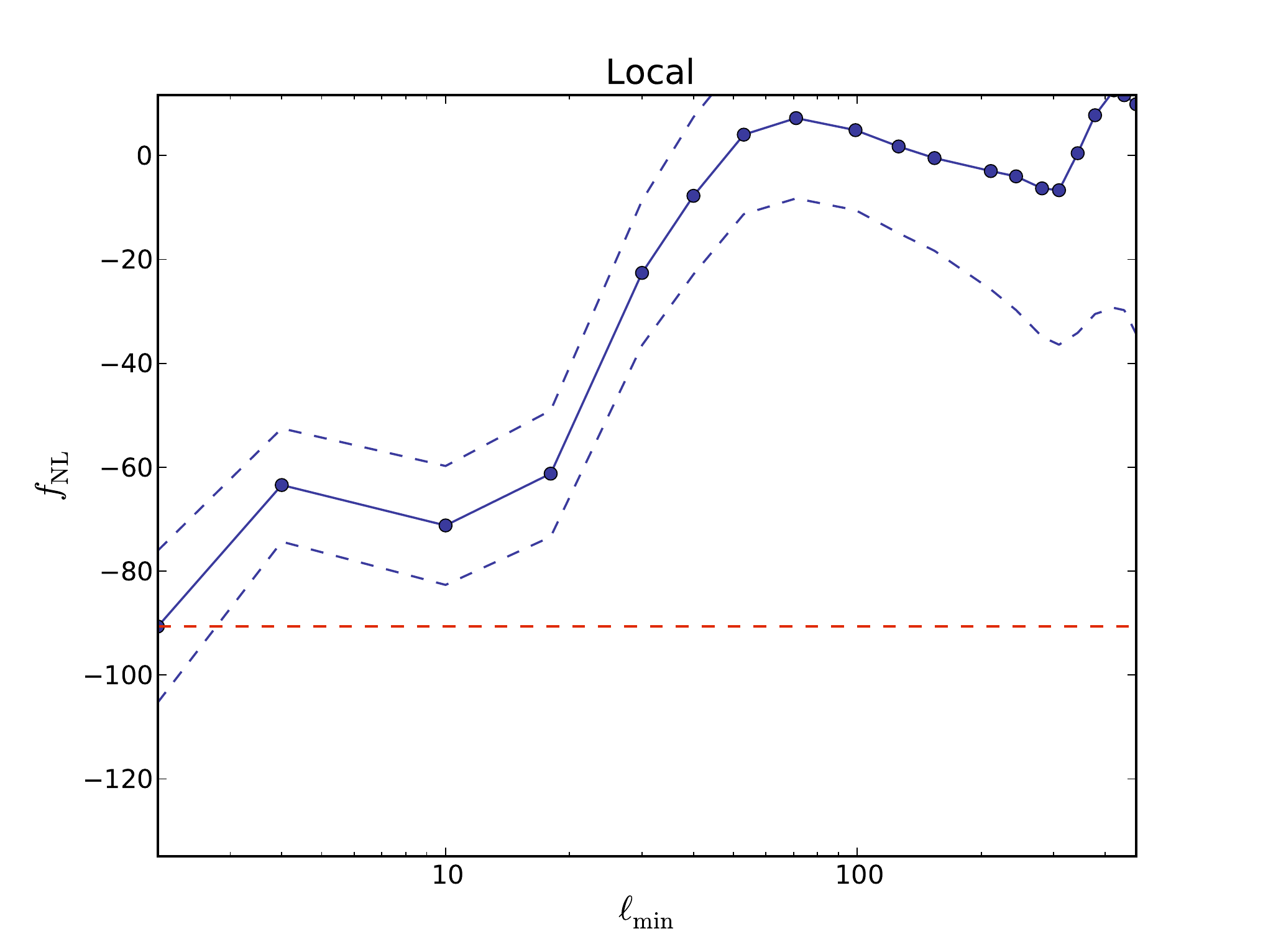}
  \includegraphics[width=0.49\linewidth]{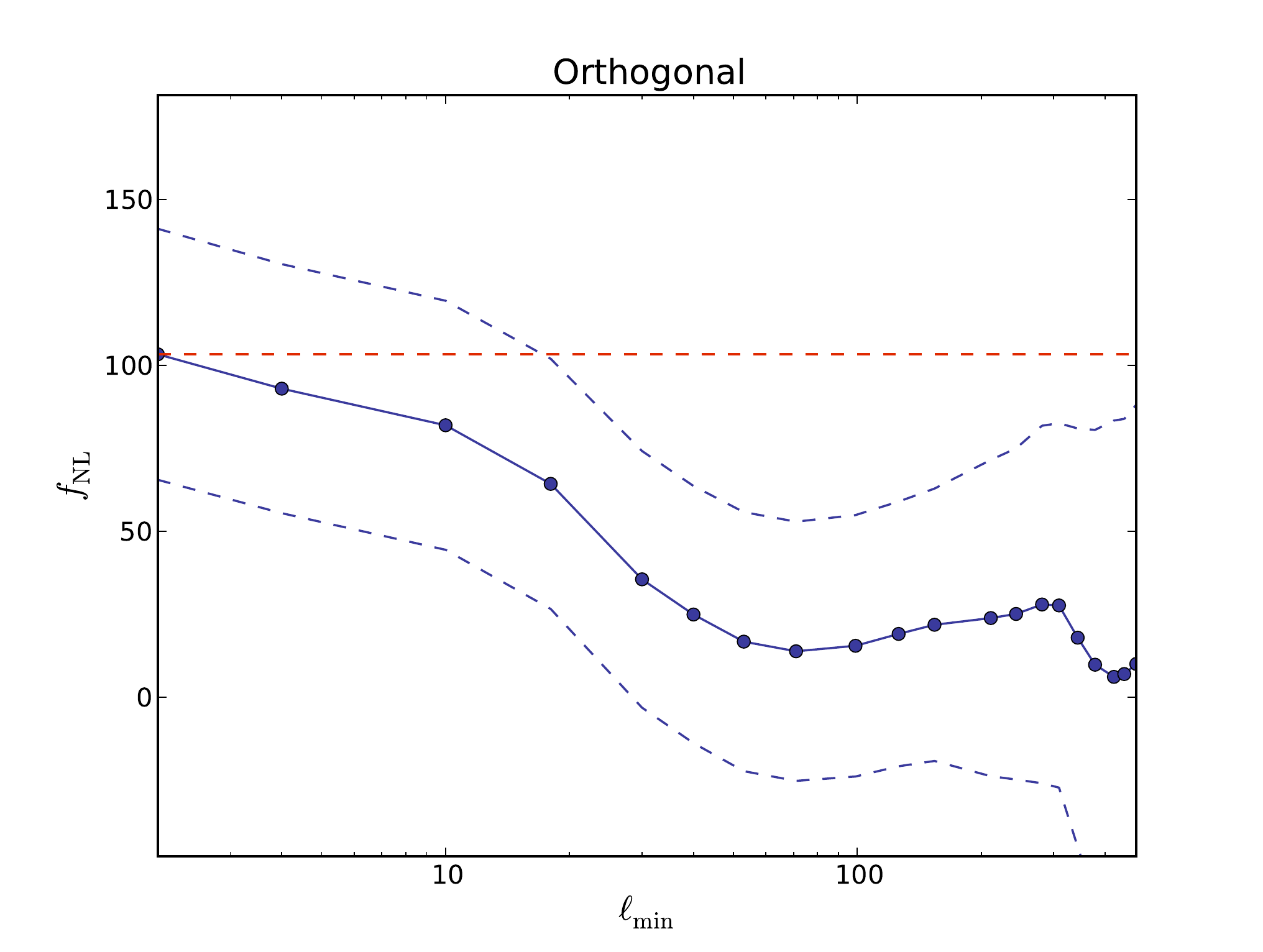}
  \includegraphics[width=0.49\linewidth]{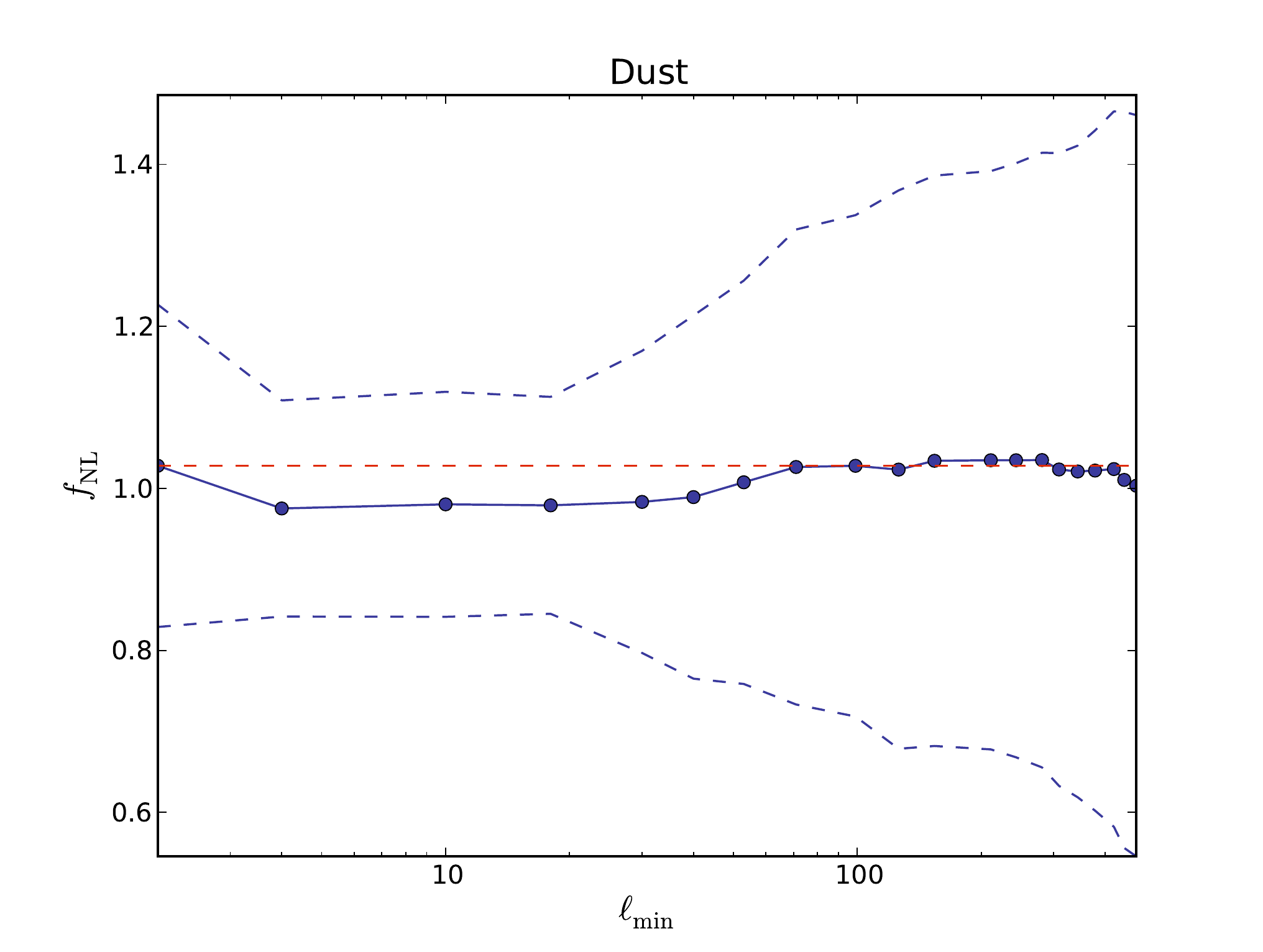}
  \includegraphics[width=0.49\linewidth]{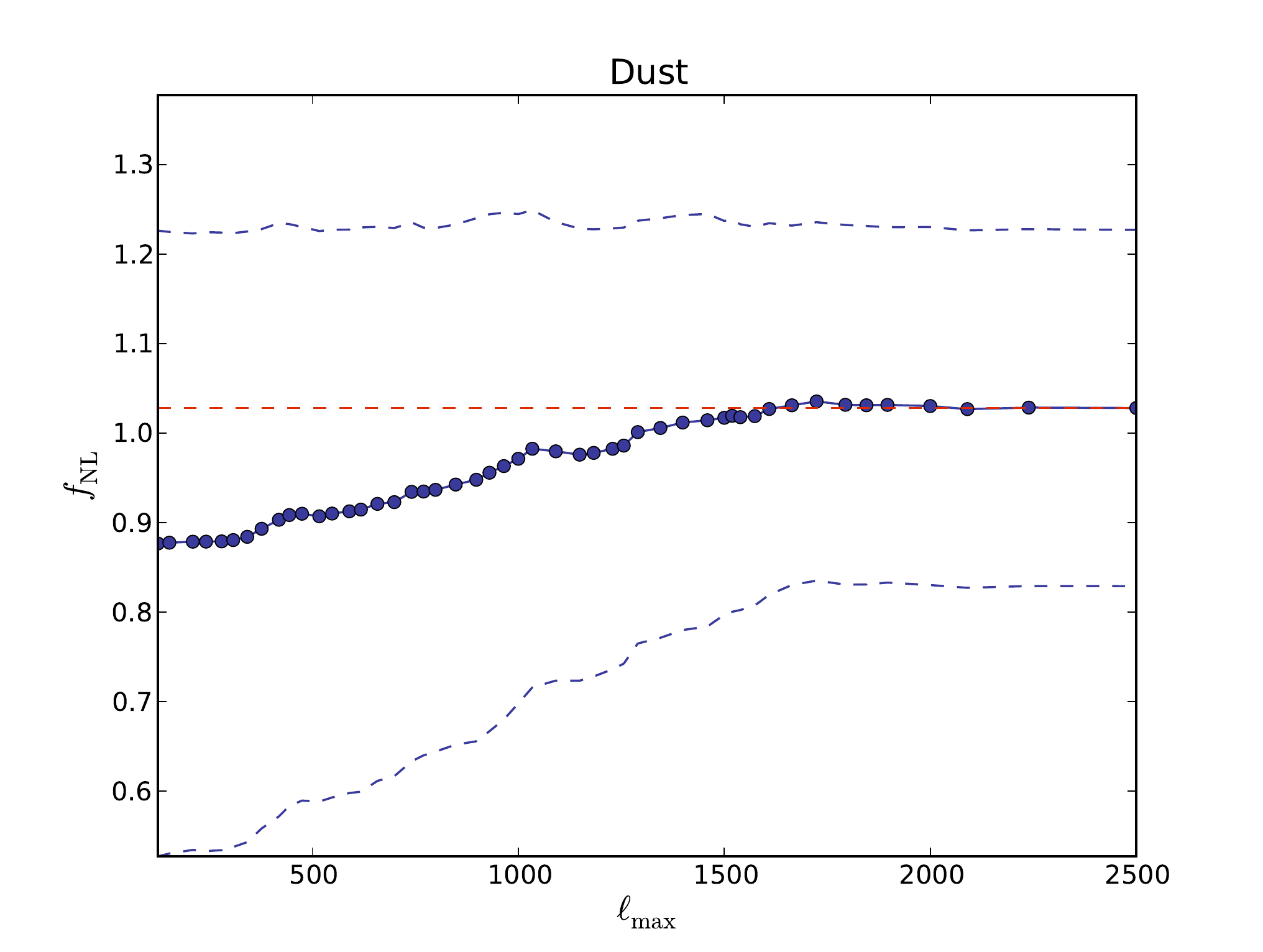}
  \caption{Convergence of $\fnl$ in the independent analysis of the 100 Gaussian CMB simulations + dust map as a function of  $\ell_\mathrm{min}$ for the local (top left), orthogonal (top right) and dust (bottom left) shapes, and as a function of $\ell_\mathrm{max}$ for the dust shape (bottom right). The blue dots correspond to the values determined by the binned bispectrum estimator when $\ell_\mathrm{min}$ (or $\ell_\mathrm{max}$ for the fourth plot) is inside the corresponding bin. The $68\%$ confidence interval is given by the blue dashed lines. The horizontal red dashed line corresponds to the determined value of $\fnl$ using the whole multipole interval (from 2 to 2500 with 57 bins).}
  \label{fig:convergence-fnl}
\end{figure}

Results are given in table \ref{tab:gaussian-isotropic}. First, we see that we detect the expected amount of dust with a good accuracy. We also observe that the shapes correlated to the dust template (see table~\ref{tab:corr_coeff_dust}), because they also peak in the squeezed configuration, are strongly detected in the independent case. However, in the joint analysis all the non-Gaussianity of the maps is attributed to the dust, with only a small impact on the error bars of the primordial shapes, meaning that this test is successful. However, this choice of bins is only optimized to detect the primordial bispectra and not the dust. Then, it is important to verify if the results can be improved by adding a few bins at very low $\ell$ (below $30$) to better measure the dust contribution. This can be seen in the second appendix of section~\ref{ap:appendices} where we observe that only the very low $\ell_1$ are important for the template (it is more squeezed than the local shape). Figure~\ref{fig:convergence-fnl} can also be used to highlight this effect. It shows the convergence of $\fnl$ when using a smaller multipole interval to determine $\fnl$. In the two top plots, we can see that if we exclude the very low $\ell$ (below 30) both the local and orthogonal $\fnl$ are consistent with 0. If we exclude the region of multipole space where the dust template is the strongest, there is no detection of the primordial shapes, even in an independent analysis. The two bottom plots are interesting as they show that the determination of $\fnl$ for the dust template is very stable when increasing $\ell_{\mathrm{min}}$ or decreasing $\ell_{\mathrm{max}}$. Note however that the error bars on $\fnl^{\mathrm{dust}}$ increase a lot if we use $\ell_{\mathrm{min}}>30$. This is visible with the dashed blue lines which correspond to the 68$\%$ confidence intervals.

There is another important effect in the dust template when a very large scale (small $\ell$) is concerned: the sawtooth pattern in the dust power spectrum (see figure \ref{fig:dust-power-spectrum}) is also expected in the dust bispectrum for the same reason (the only large harmonic coefficients describing the dust at low $\ell$ are the $a_{\ell 0}$ with $\ell$ even). In principle, it could be used to differentiate between the dust and the local shapes, but this effect is hidden if the bins are large because it is averaged over several $\ell$'s, thus providing another motivation to add some bins at low $\ell$. One issue when adding bins is that the memory constraints on the computer system we use limit us to a number of bins between 50 and 60 at most at the Planck resolution when including the polarization too. Here we can use 70 bins, because we only look at the temperature data and because we only add bins for the largest scales, where it is possible to downgrade the resolution of the filtered maps. With this new binning, the correlation coefficient between the local and dust shapes becomes $-0.48$ (instead of $-0.60$ for 57 bins, see table~\ref{tab:corr_coeff_dust}). So indeed adding a few bins at low $\ell$ helps to differentiate these squeezed shapes. The results of the same test with 70 bins are also given in table~\ref{tab:gaussian-isotropic}. In the independent analysis, the amount of local non-Gaussianity and its error bar decreases which is consistent with the fact that the dust and the local templates are easier to differentiate with the new binning. However, in the joint analysis there is no clear difference, except that the different central values are now very close to the expected values.

We also have to note that the approximation of weak non-Gaussianity, which is needed for the validity of the linear correction of the bispectrum to take into account the effects of the mask here, starts to break down when we observe a local shape at more than $6\sigma$ (independent case). This is why it is important to verify how a similar analysis works with a smaller amount of dust in the map. Hence, with the same choice of 70 bins, we perform two other tests with the 100 CMB simulations. For one we multiply the dust map by a factor 0.75 before adding it to the CMB realizations and the expected value of $\fnl$ is then $0.75^3 \approx 0.42$. For the other test, we use the the Gaussian CMB maps without adding dust, to verify that we do not detect any bispectral shape. These results are also given in table \ref{tab:gaussian-isotropic} and are exactly as expected.

Note that the error bars for the case of the CMB only are roughly one order of magnitude smaller than for the rest. The reason is that we made a distinction between the standard deviation (square root of the variance) and the standard error (standard deviation divided by the square root of the number of maps, so divided by 10 here). The standard error gives the expected error on the determination of the mean value of $\fnl$ with our sample of 100 Gaussian maps. The standard deviation gives the $1\sigma$ interval in which we would detect $\fnl$ if we study one map. It is clear that the standard error has to be used in the CMB-only analyses because we determine the mean value of each $\fnl$ from a sample of 100 maps. However, when we add dust to these maps, the situation is different because we only have one realization of the dust so the standard error cannot be used. We are however very conservative by using the standard deviation, the real error bars on the mean values of the different $\fnl$ are probably between the standard error and the standard deviation (the more dust in the map, the closer to the standard deviation it will be). However, the fact that for the two amounts of dust with 70 bins the central values in the joint analysis are so close to the expected values is an indication that the error bars are likely overestimated for these two cases (the results with dust would still be correct if we divided the standard deviation by 10 to obtain the standard error, which is not true with 57 bins).

\begin{figure}
  \centering \includegraphics[width=0.49\linewidth]{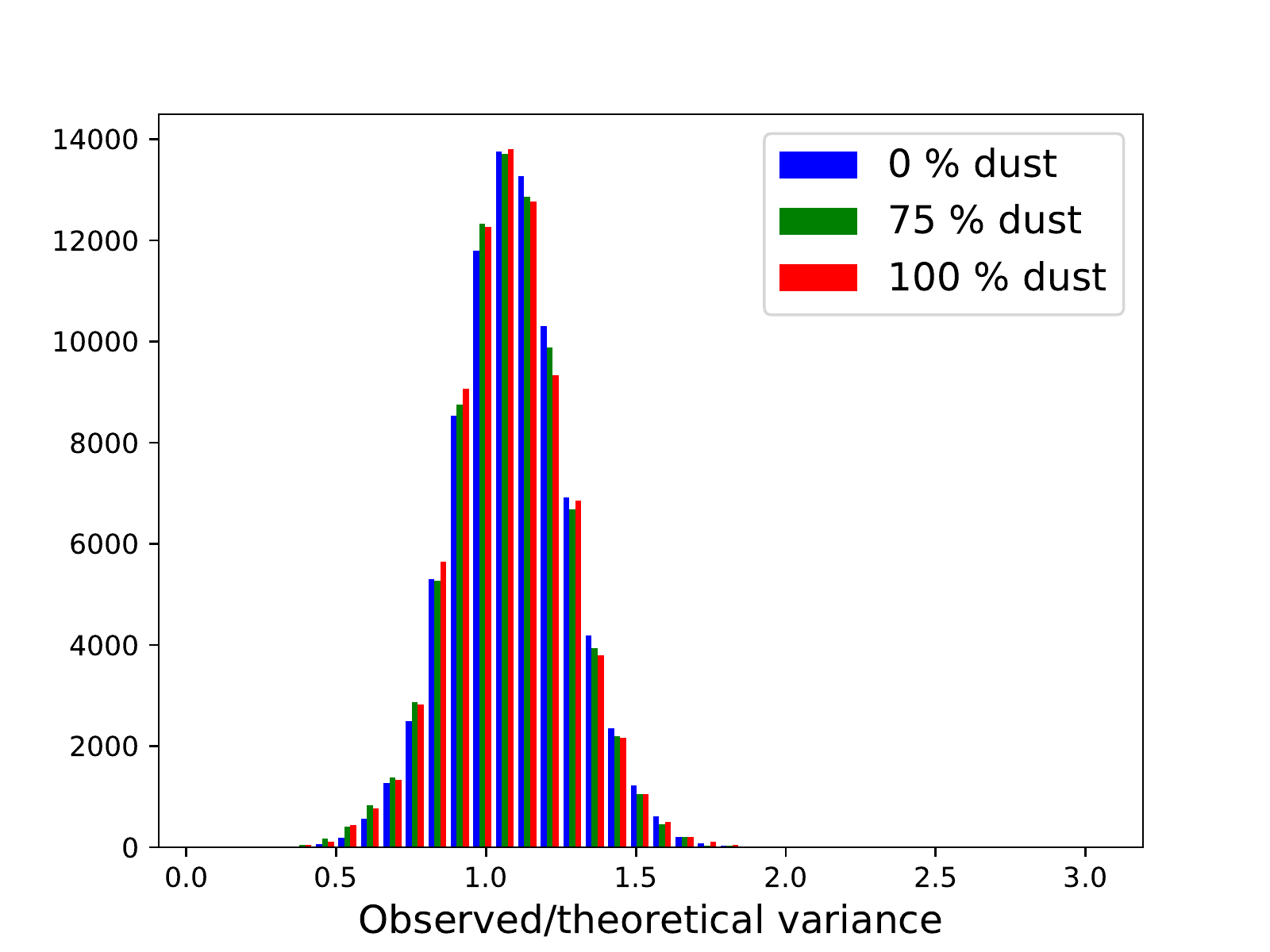}
  \includegraphics[width=0.49\linewidth]{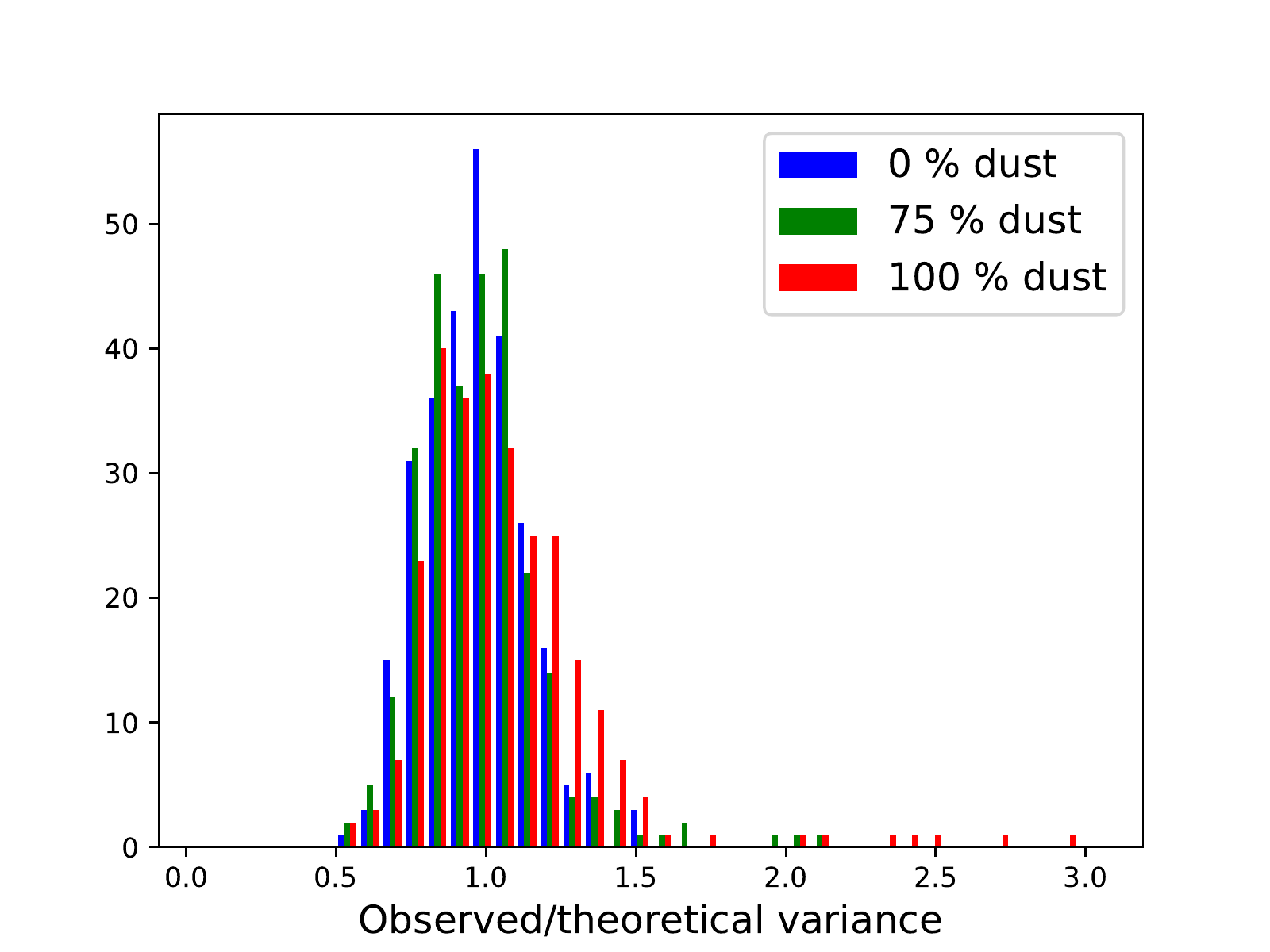}
  \caption{Distribution of the ratio of the observed variance over the theoretical prediction for the three different amounts of dust (0 $\%$ in blue, 75 $\%$ in green and 100 $\%$ in red). On the left, all the valid $\ell$-triplets are included while on the right only very squeezed $\ell$-triplets are shown (one small multipole $\ell_1 \leq 13$ and two large ones $\ell_{2,3} \geq 742$).}
  \label{fig:ratio-variance}
\end{figure}

We can illustrate the breakdown of the weak non-Gaussianity approximation using the variance of the bispectrum. Indeed, we have at our disposal a theoretical prediction for the variance, given in \eqref{binnedVarreal}, that scales as the power spectrum cubed and for which the derivation relies on the weak non-Gaussianity approximation. However, we can also directly compute the variance of the bispectrum from our 100 maps, which we call here observed variance. Figure \ref{fig:ratio-variance} shows the distribution of the ratio of the observed variance over the theoretical variance for the three different amounts of dust in two different configurations. First, we examine this ratio over the whole triplet space (on the left) where there is no difference between the three cases and the values are distributed around 1 as expected. That is logical because the non-Gaussianity of the dust is very localized in multipole space; the bispectrum is large only in the very squeezed configuration. This is why on the right we consider only the triplets where one $\ell$ is very small (in the first five bins i.e.\ $\ell \leq 13$) and the two others large (in the last 30 bins, i.e.\ $\ell \geq 742$). Adding or removing a few bin triplets here does not change the results. Here we can see that if there is more dust (in red),  there are several values which strongly deviate from one. This effect is even more obvious when we examine the mean and the standard deviation of these distributions, which are given in table \ref{tab:ratio-variance}. When considering the full space of multipole triplets, there is no significant difference between the three cases. However, when we examine only the squeezed part of the bispectrum, the standard deviation increases slightly with a small amount of dust (75 $\%$), and is three times larger for 100 $\%$ dust compared to the CMB-only case. Hence, the weak non-Gaussianity approximation stops being valid, but not enough to invalidate the results (only a few bin-triplets deviate strongly). However, if we were to add even more dust, we would have to take this effect into account.

\begin{table}
  \begin{center}
    \small
    \begin{tabular}{l|cc|cc|cc}
      \hline
      & \multicolumn{2}{c|}{0 $\%$ dust} & \multicolumn{2}{c|}{75 $\%$ dust} & \multicolumn{2}{c}{100 $\%$ dust}\\
  & \multicolumn{1}{c}{Full} & \multicolumn{1}{c|}{Squeezed} & \multicolumn{1}{c}{Full} & \multicolumn{1}{c|}{Squeezed} & \multicolumn{1}{c}{Full} & \multicolumn{1}{c}{Squeezed}   \\ 
      \hline
     Mean & 1.09 & 0.97 & 1.08 & 0.97 & 1.08 & 1.11\\
     Standard deviation & 0.19 & 0.16 & 0.19 & 0.21 & 0.20 & 0.51\\
    \end{tabular}
    \caption{Means and standard deviations of the distributions of the ratio of the observed variance over the theoretical prediction, shown in figure \ref{fig:ratio-variance}, for the three amounts of dust, including the full bispectrum or only a very squeezed part of it.}
    \label{tab:ratio-variance}
  \end{center}
\end{table}

In addition to looking at the variance of the bispectrum itself, we can also investigate the variance of the $\fnl$ parameters with regard to the validity of the weak non-Gaussianity approximation.
Every error bar given in table \ref{tab:gaussian-isotropic} was computed from the observed variance of the set of 100 maps. However, we can also compute Fisher error bars from the theoretical prediction of the variance and they are given in table~\ref{tab:error-bars-gaussian} for the three cases studied in this section for the local and dust shapes. We can see that for both, the more non-Gaussian the map is, the more important is the difference between Fisher and observed error bars. This is related to the breakdown of the weak non-Gaussianity approximation. For a local $|\fnl|$ of around 70 (corresponding to 100 $\%$ dust), the difference is a factor 2 between the two kinds of error bars. The difference is larger for the dust template, where for this case the observed error bars are four times larger than the Fisher forecasts. For both templates, when there is no dust (so purely Gaussian maps), the observed error bars agree with the Fisher forecasts up to the expected precision (the relative error in the standard deviation is $1/\sqrt{2(N-1)}$, which is $7~\%$ for 100 maps).

\begin{table}
\begin{center}
  \small
  \begin{tabular}{l|ccc}
    \hline
        & 100 $\%$ dust & 75 $\%$ dust & 0 $\%$ dust \\
        \hline
    Local &&& \\
    \;\;{\em Fisher} & 6.6 & 6.4 & 5.6 \\
    \;\;{\em Observed} & 14 & 7.7 & 5.2 \\
    Dust &&& \\
    \;\;{\em Fisher} & 0.05 & 0.04 & 0.031 \\
    \;\;{\em Observed} & 0.20 & 0.12 & 0.030 \\
    \hline
  \end{tabular}
\end{center}
\caption{Fisher and observed standard deviations on $\fnl^{\mathrm{local}}$ and $\fnl^{\mathrm{dust}}$ (independent analysis) determined from 100 Gaussian simulations of the CMB with isotropic noise to which we added a known amount of dust (100 $\%$, 75 $\%$ or 0 $\%$ of the dust map of figure  \ref{fig:dust-map}) using 70 bins.} 
\label{tab:error-bars-gaussian}
\end{table}

\begin{figure}
  \centering
   \includegraphics[width=0.49\linewidth]{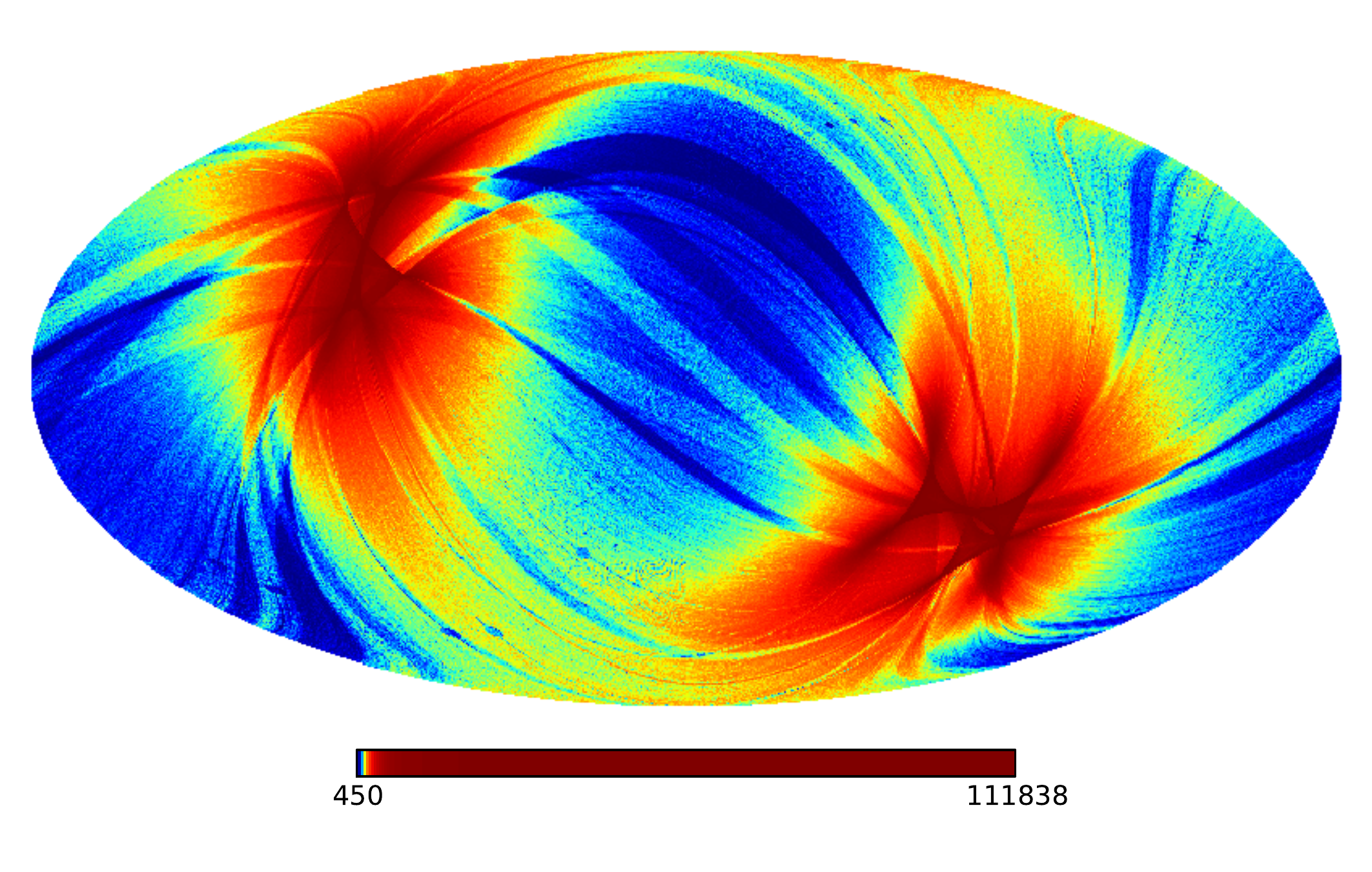}
 \includegraphics[width=0.49\linewidth]{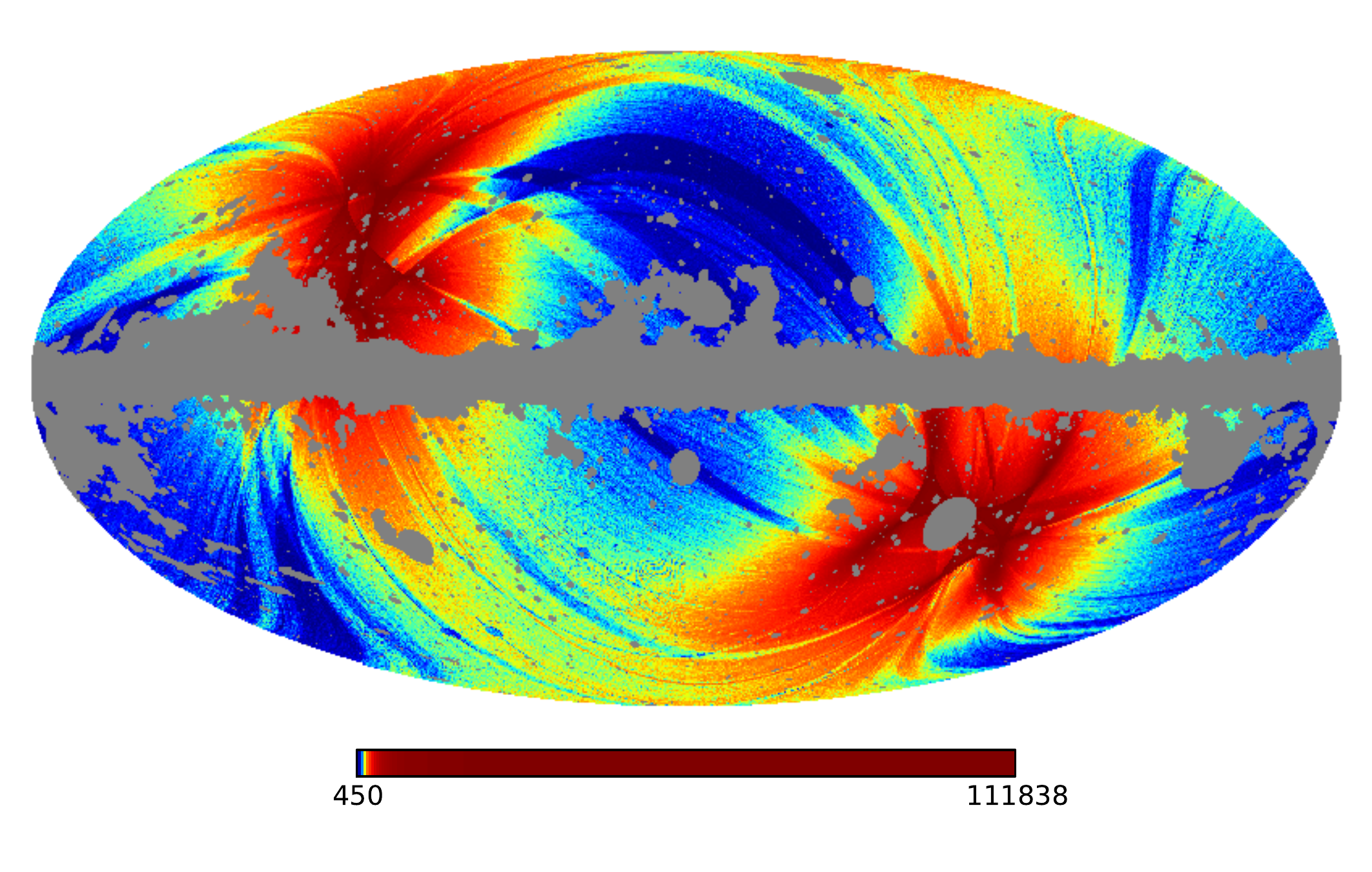}
  \caption{Unmasked (left) and masked (right) hit-count map of Planck (number of observation samples per pixel).}
  \label{fig:hitcount-map}  
\end{figure}

As explained before, adding noise realizations with the correct power spectrum to the CMB simulations is necessary for the optimization of the binning and to make the simulations more realistic. However, the real instrument noise does not have an isotropic distribution in pixel space because some parts of the sky are observed more often than others, as shown in figure \ref{fig:hitcount-map}. Without a linear correction, the anisotropic noise also gives a large squeezed contribution to the bispectrum for the usual reason: small-scale fluctuations are larger (more noise) in the large-scale regions which are less observed and vice versa.
This is why we also verify the previous results with an anisotropic distribution of the noise following the scanning pattern of the Planck satellite. The results are given in table~\ref{tab:gaussian-anisotropic}. Here we use only the best choice of bins (70 bins) and the results are given for the same three amounts of dust as in table~\ref{tab:gaussian-isotropic}. Results are very similar with isotropic and with anisotropic noise for the three cases; each time we detect successfully the amount of dust we added to the maps. 

\begin{table}
\begin{center}
  \small
\begin{tabular}{lcccccc}
\hline
& Local & Equilateral & Orthogonal & P.S.$/10^{-29}$ & CIB$/10^{-27}$ & Dust\\
\hline
\multicolumn{4}{l}{Dust 100$\%$ (expected $\fnl^{\mathrm{dust}}=1$)} &&& \\
\multicolumn{1}{@{\hspace{0.7cm}}c@{\hspace{0.5cm}}}{$Indep$} & -67 $\pm$ 11  & 24 $\pm$ 65 & 93 $\pm$ 34 & 1.5 $\pm$ 1.1 & 1.1 $\pm$ 0.5 & 1.00 $\pm$ 0.20\\
  \multicolumn{1}{@{\hspace{0.0cm}}c@{}}{$Joint$} & -1 $\pm$ 14 & 4 $\pm$ 73 & -4 $\pm$ 40 & 0.2 $\pm$ 2.8 & 0.0 $\pm$ 1.4 & 1.00 $\pm$ 0.24\\
\multicolumn{4}{l}{Dust 75$\%$ (expected $\fnl^{\mathrm{dust}}=0.42$)} &&& \\
\multicolumn{1}{@{\hspace{0.7cm}}c@{\hspace{0.5cm}}}{$Indep$} & -30 $\pm$ 7 & 4 $\pm$ 62 & 44 $\pm$ 33 & 0.5 $\pm$ 1.1 & 0.4 $\pm$ 0.5 & 0.42 $\pm$ 0.12 \\
\multicolumn{1}{@{\hspace{0.0cm}}c@{}}{$Joint$} & 0 $\pm$ 10 & -5 $\pm$ 69 & 1 $\pm$ 38 & -0.1 $\pm$ 2.8 & 0.0 $\pm$ 1.4 & 0.42 $\pm$ 0.14\\
  \multicolumn{4}{l}{Dust 0$\%$ (expected $\fnl^{\mathrm{dust}}=0$)} &&& \\
\multicolumn{1}{@{\hspace{0.7cm}}c@{\hspace{0.5cm}}}{$Indep$} & -0.15 $\pm$ 0.50 & 0.3 $\pm$ 6.6 & -1.4 $\pm$ 3.8 & -0.08 $\pm$ 0.10 & -0.05 $\pm$ 0.05 & 0.000 $\pm$ 0.003\\
\multicolumn{1}{@{\hspace{0.0cm}}c@{}}{$Joint$} & -0.30 $\pm$ 0.64 & 1.2 $\pm$ 6.8 & -2.2 $\pm$ 4.3 & 0.08 $\pm$ 0.23 & -0.08 $\pm$ 0.11 & 0.000 $\pm$ 0.004 \\
\hline
\end{tabular}
\end{center}
\caption{Determination of $\fnl$ for the local, equilateral, orthogonal, point sources, CIB and dust shapes using a set of 100 Gaussian simulations of the CMB with anisotropic noise to which we added a known amount of dust (the high-resolution dust map of section \ref{sec:dust} multiplied by a factor 1 or 0.75, or no dust at all). The analysis is performed using 70 bins and the error bars are given at 1$\sigma$.} 
\label{tab:gaussian-anisotropic}
\end{table}

With these different tests, we have proven that the binned bispectrum estimator can be used to detect a galactic foreground shape that we determined numerically. It works well with the amount of dust that is expected at 143 GHz, the dominant frequency channel in the cleaned CMB map. The next logical step is then to use the template on real data.

\subsection{CMB analyses}

The previous tests have shown that detecting the dust is possible when there is a large amount of it. We can now apply the dust template to a real CMB analysis. Here, we follow the analysis of the Planck 2015 paper~\citepalias{planck2015-17} (note that we only study the temperature bispectrum, while for Planck the polarization was also taken into account). We use a set of 160 simulation maps for the computation of the error bars and the linear correction. The power spectrum is the best-fit cosmological model from the 2015 Planck analysis. This time, we also include the ISW-lensing shape in the analysis because it is present in the data. The amplitude of this template is known, so it can be used to subtract the bias (see equation \eqref{biascorreq}) from the bispectral non-Gaussianity of the map. Results are given in table \ref{tab:cmb-high}.\footnote{The difference between the values in this table and those in the Planck paper~\citepalias{planck2015-17}, in particular for equilateral, is mainly due to our use here of a slightly different mask (the preferred temperature mask from~\citepalias{planck2015-09} instead of the slightly extended mask used in~\citepalias{planck2015-17}).} We include the results with and without taking into account the ISW-lensing bias, and we perform two different joint analyses for comparison, with and without the dust. As expected, there is no detection of the primordial shapes or the dust. However, it is important to note that the error bars of the local and dust shapes in the joint analysis increase because these shapes are correlated. Similarly to the previous section, one way to improve the situation would be to find a binning that is optimal for both shapes. 

\begin{table}
\begin{center}
  \footnotesize
\begin{tabular}{lccccccc}
\hline
& Local & Equilateral & Orthogonal & P.S.$/10^{-29}$ & CIB$/10^{-27}$ & Dust/$10^{-2}$ & Lensing-ISW\\
\hline
\multicolumn{4}{l}{No ISW-lensing bias subtraction} &&&& \\
\multicolumn{1}{@{\hspace{0.3cm}}c@{\hspace{0.2cm}}}{$Indep$} & 8.7 $\pm$ 5.5 & 8 $\pm$ 67 & -34 $\pm$ 33 & 9.6 $\pm$ 1.0 &  4.6 $\pm$ 0.5 & -0.8 $\pm$ 3.8 & 0.59 $\pm$ 0.29 \\
  \multicolumn{1}{@{\hspace{0.0cm}}c@{}}{$Joint$} & 6 $\pm$ 8 & -21 $\pm$ 69 & -3 $\pm$ 38 & 7.3 $\pm$ 2.7 & 1.2 $\pm$ 1.4 & -2.2 $\pm$ 5.2 & 0.57 $\pm$ 0.31 \\
  \multicolumn{1}{@{\hspace{0.0cm}}c@{}}{$Joint\setminus dust$} & 4.2 $\pm$ 6.7 & -15 $\pm$ 68 & -6.6 $\pm$ 37 & 7.2 $\pm$ 2.7 & 1.3 $\pm$ 1.4 & & 0.55 $\pm$ 0.31\\
\multicolumn{4}{l}{ISW-lensing bias subtracted} &&&& \\
\multicolumn{1}{@{\hspace{0.3cm}}c@{\hspace{0.2cm}}}{$Indep$} &  1.2 $\pm$ 5.5 & 6 $\pm$ 67 & -8 $\pm$ 33 &  9.6 $\pm$ 1.0 & 4.6 $\pm$ 0.5 &  -4.0 $\pm$ 3.8 &  \\
  \multicolumn{1}{@{\hspace{0.0cm}}c@{}}{$Joint$} &  -5 $\pm$ 8 & -16 $\pm$ 69 & 1 $\pm$ 38 & 7.1 $\pm$ 2.7 & 1.3 $\pm$ 1.4 & -3.5 $\pm$ 5.1 & \\
   \multicolumn{1}{@{\hspace{0.0cm}}c@{}}{$Joint\setminus dust$} & 1.0 $\pm$ 6.3 & -7 $\pm$ 68 & -5 $\pm$ 37 & 7.0 $\pm$ 2.7 & 1.4 $\pm$ 1.4 & & \\
\hline
\end{tabular}
\end{center}
\caption{Determination of $\fnl$ for the local, equilateral, orthogonal, point sources, CIB, dust and ISW-lensing shapes in the cleaned \texttt{SMICA} CMB map from the 2015 Planck release. In the three first lines, the ISW-lensing shape is considered as the others. In the last three, the ISW-lensing bias is subtracted. The joint analysis is performed with and without the dust template. The binning consists of 57 bins.} 
\label{tab:cmb-high}
\end{table}

We also performed a similar analysis on a low resolution cleaned CMB map ($n_\mathrm{side}=256$) with a 60 arcmin FWHM Gaussian beam to look for the other foreground templates with the usual choice of bins. Results are given in table \ref{tab:cmb-low}. Because of the resolution and the beam, we only analyze multipoles in the interval $[2, 300]$, which is the reason for the very large error bars. As that would leave only few bins from the original binning, we split all the bins below $\ell=300$ into three (where possible, two otherwise), which gives 39 bins in total. Moreover, we did not subtract the ISW-lensing bias as its contribution is small compared to the error bars. The point sources and the CIB are not given in the table because they were not observed here. The results are consistent with zero non-Gaussianity in the map. But the new foreground shapes (AME, free-free and synchrotron) have very large error bars and even if they were present in the map, it would not be possible to detect them.

\begin{table}
\begin{center}
  \small
\begin{tabular}{lccccccc}
\hline
& Local & Equilateral & Orthogonal & Dust & Free-free & Synch$/10^5$ & AME$/10^{10}$\\
\hline
\multicolumn{1}{@{\hspace{0.3cm}}c@{\hspace{0.2cm}}}{$Indep$} & 13 $\pm$ 30 & 49 $\pm$ 155 & 66 $\pm$ 130 & -0.01 $\pm$ 0.07 & -1 $\pm$ 32 & -0.1 $\pm$ 4.4 & -10 $\pm$ 7 \\
  \multicolumn{1}{@{\hspace{0.0cm}}c@{}}{$Joint$} & 17 $\pm$ 51 & 281 $\pm$ 406 & 50 $\pm$ 287 & -0.01 $\pm$ 0.10 & 22 $\pm$ 43 & 2 $\pm$ 6 & -12 $\pm$ 8 \\
\hline
\end{tabular}
\end{center}
\caption{Determination of $\fnl$ of some primordial and all galactic templates in the cleaned \texttt{SMICA}  CMB map at low resolution $n_{\mathrm{side}}=256$ with a 60 arcmin FWHM Gaussian beam from the 2015 Planck release. Because of the low $\ell_{\mathrm{max}}=300$, the analysis is performed using 39 bins.} 
\label{tab:cmb-low}
\end{table}

\subsection{Raw sky}

\begin{figure}
  \centering
\includegraphics[width=0.66\linewidth]{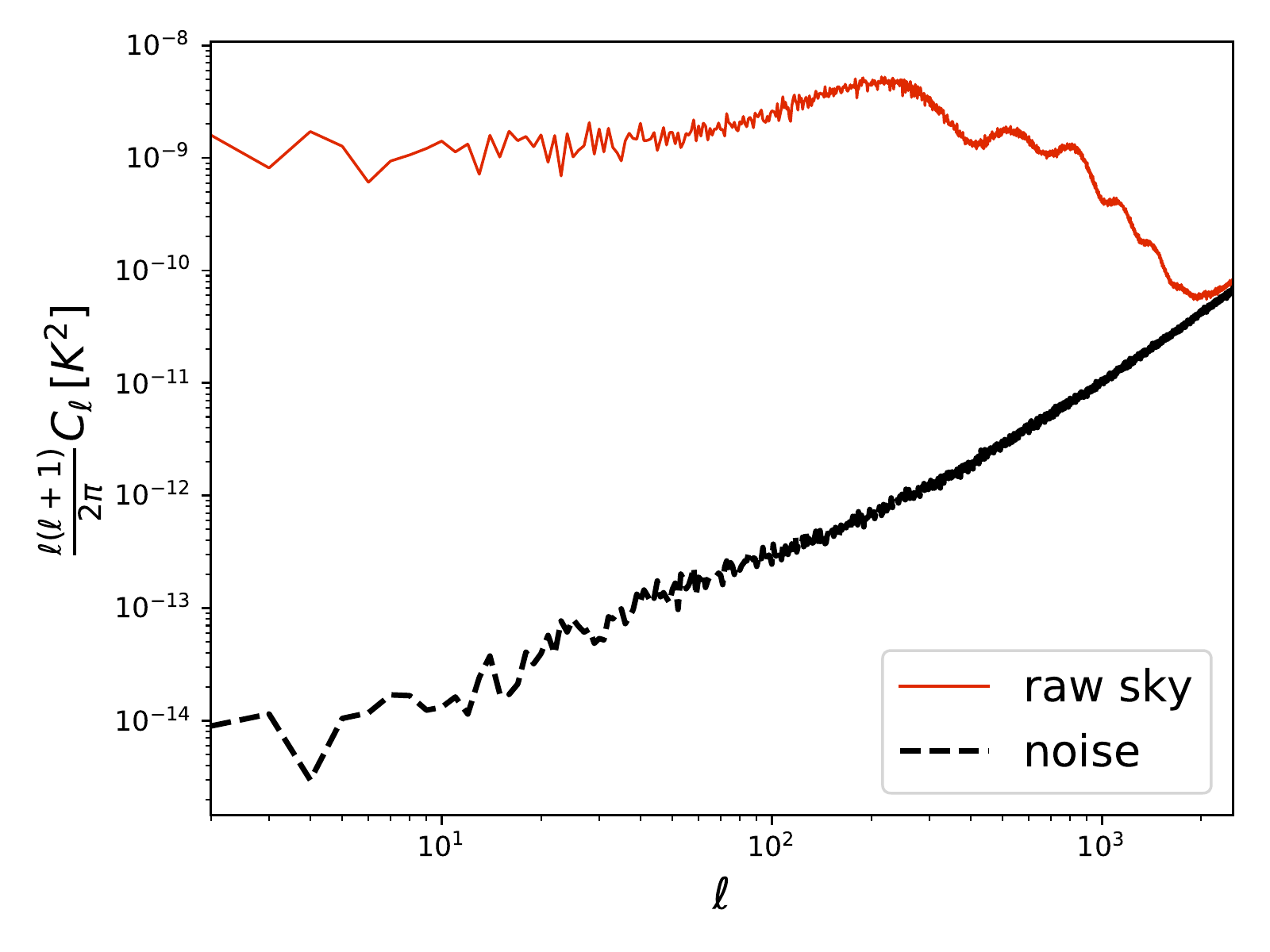}
  \caption{Power spectrum of the raw 143~GHz map as well as the estimated noise power spectrum.}
  \label{fig:raw-noise}
\end{figure}

After applying the foreground templates to the cleaned CMB map that is not supposed to contain any galactic foreground (which we confirmed), it is also interesting to study the raw 143~GHz Planck map. Again, we had to generate Gaussian simulations of this map to compute the linear correction. For this, we used the power spectrum of the map and we determined the noise power spectrum using the half-mission maps. The power spectra are shown in figure~\ref{fig:raw-noise}. The noise is modulated in pixel space using the hit-count map of figure~\ref{fig:hitcount-map} to make it anisotropic. The beam of this map can be approximated by a 7.3~arcmin FWHM Gaussian beam.

Results are given in tables \ref{tab:raw-high} and \ref{tab:raw-low}. We detected the expected amount of dust since $\fnl^{\mathrm{dust}}=1$ is within the $1\sigma$ error bars in both the independent and joint analyses. For the other foregrounds, the situation is similar to the previous section: the error bars are far too large for a detection (the synchrotron and AME shapes are not given here because the error bars are many orders of magnitude larger than the expected quantity in the map). To determine error bars, we did not have good simulations of the data map but we had of course Fisher forecasts. We made the simple but reasonable hypothesis that the factor between the real error bars and the Fisher ones due to the breakdown of the weak non-Gaussianity approximation is the same as for the anisotropic case in section~\ref{sec:gaussian-simulations}. Then it was easy to determine error bars for the primordial and the dust shapes that are a bit larger than in table~\ref{tab:gaussian-anisotropic}. However, for the CIB and the point sources, which are not the main object of study here, the situation is different because they were not present in the Gaussian simulations, so we could not apply this method. Hence we only give Fisher error bars for those two shapes, but because of the strong detection we know that they are underestimated. This is not an issue because they are uncorrelated to the local and the dust shapes. For the low-resolution case in table~\ref{tab:raw-low} we only have Fisher error bars for all shapes. In conclusion, the method also works correctly when applied to a raw sky map.

\begin{table}
\begin{center}
  \small
\begin{tabular}{lcccccc}
\hline
& Local & Equilateral & Orthogonal & P.S.$/10^{-29}$ & CIB$/10^{-27}$ & Dust \\
\hline
\multicolumn{1}{@{\hspace{0.7cm}}c@{\hspace{0.5cm}}}{$Indep$} & -61 $\pm$ 13 & 22 $\pm$ 71 & -12 $\pm$ 39 & 90 $\pm$ 4 & 28 $\pm$ 1 & 1.09 $\pm$ 0.25  \\
  \multicolumn{1}{@{\hspace{0.0cm}}c@{}}{$Joint$} & 13 $\pm$ 18 & -37 $\pm$ 81 & -81 $\pm$ 47 & 115 $\pm$ 9 & -11 $\pm$ 3 & 1.08 $\pm$ 0.32 \\
\hline
\end{tabular}
\end{center}
\caption{Determination of $\fnl$ in the raw 143~GHz map at high resolution $n_{\mathrm{side}}=2048$ from the 2015 Planck release. The analysis is performed with the usual choice of 57 bins. For the details on the error bars for the primordial and the dust shapes, see the main text. The CIB and point sources error bars are Fisher forecasts.} 
\label{tab:raw-high}
\end{table}

\begin{table}
\begin{center}
  \small
\begin{tabular}{lccccc}
\hline
& Local & Equilateral & Orthogonal & Dust & Free-free \\
\hline
\multicolumn{1}{@{\hspace{0.7cm}}c@{\hspace{0.5cm}}}{$Indep$} & -41 $\pm$ 54 & 232 $\pm$ 198 & 277 $\pm$ 177 & 0.88 $\pm$ 0.28 & 39 $\pm$ 50\\
  \multicolumn{1}{@{\hspace{0.0cm}}c@{}}{$Joint$} & -33 $\pm$ 94 & 543 $\pm$ 536 & 112 $\pm$ 388 & 0.85 $\pm$ 0.39 & -9 $\pm$ 69\\
\hline
\end{tabular}
\end{center}
\caption{Determination of $\fnl$ in the raw 143~GHz map at low resolution $n_{\mathrm{side}}=256$ from the 2015 Planck release. The analysis is performed with 39 bins. The error bars are all Fisher forecasts.} 
\label{tab:raw-low}
\end{table}

\section{Appendices}
\label{ap:appendices}

This section contains the two appendices of the paper.

\subsection{Derivation of the variance of the bispectrum and the linear correction}

In this appendix, we recall the derivation of the variance of the bispectrum to explain the role of the linear correction. By definition, the variance is given by
\begin{equation}
  \label{eq:bispectrum-variance}
  \mathrm{Var}(B^\mathrm{obs}_{\ell_1 \ell_2 \ell_3}) = \langle (B^\mathrm{obs}_{\ell_1 \ell_2 \ell_3})^2 \rangle - \langle B^\mathrm{obs}_{\ell_1 \ell_2 \ell_3} \rangle^2
  \equiv V_{\ell_1 \ell_2 \ell_3}.
\end{equation}
In the weak non-Gaussianity regime, the average value of the bispectrum is negligible. This leaves us with computing the mean value of the product of two bispectra.

\subsubsection{Isotropic case}

\begin{equation}
  \label{eq:variance-first-term}
\langle B_{\ell_1 \ell_2 \ell_3}(\hat\Omega) B_{\ell_4 \ell_5 \ell_6}(\hat\Omega') \rangle = \int_{S^2 \times S^2}\d \hat\Omega \d \hat\Omega'\, \langle M_{\ell_1}(\hat\Omega) M_{\ell_2}(\hat\Omega) M_{\ell_3}(\hat\Omega) M_{\ell_4}(\hat\Omega') M_{\ell_5}(\hat\Omega') M_{\ell_6}(\hat\Omega') \rangle.
 \end{equation}
One can use Wick's theorem for Gaussian fields to reduce the six-point correlation function to the sum of fifteen products of two-point correlation functions:
\begin{itemize}
\item 6 terms: each $\ell$ is paired with an element of the other triplet like in \\$\langle M_{\ell_1}(\hat\Omega) M_{\ell_4}(\hat\Omega')\rangle \langle M_{\ell_2}(\hat\Omega) M_{\ell_5}(\hat\Omega')\rangle \langle M_{\ell_3}(\hat\Omega) M_{\ell_6}(\hat\Omega') \rangle$;
 \item 9 terms: the rest (example: $\langle M_{\ell_1}(\hat\Omega) M_{\ell_2}(\hat\Omega)\rangle \langle M_{\ell_3}(\hat\Omega) M_{\ell_4}(\hat\Omega')\rangle \langle M_{\ell_5}(\hat\Omega') M_{\ell_6}(\hat\Omega') \rangle$).
 \end{itemize}

 We will explicitly compute the contribution of the six first terms below. But first, we show that in the isotropic case, the nine last terms are zero. For this, we use the addition theorem,
 \begin{equation}
  \label{eq:ylm-addition-theorem}
  \sum\limits_{m=-\ell}^{\ell}\, Y_{\ell m}(\hat{\Omega})Y_{\ell m}^*(\hat{\Omega}') = \frac{2\ell+1}{4\pi}P_\ell(\hat{\Omega}\cdot\hat{\Omega}'),
\end{equation}
where $P_\ell$ is a Legendre polynomial, and the fact that the maps are real, to compute the two-point correlation function
 \begin{equation}
   \label{eq:cancellation-variance}
   \begin{split}
   \langle M_{\ell}(\hat\Omega) M_{\ell'}(\hat\Omega')\rangle =
   \sum\limits_{m=-\ell}^{\ell}\sum\limits_{m'=-\ell'}^{\ell'} \langle a_{\ell m} a_{\ell' m'}^*\rangle Y_{\ell m}(\hat\Omega) Y_{\ell' m'}^*(\hat\Omega') 
   =C_\ell \delta_{\ell \ell'} \frac{2\ell+1}{4\pi}P_\ell(\hat\Omega\cdot\hat\Omega').   
   \end{split}
 \end{equation}
 Then we can perform the integration of our example term (and the eight others follow the same computation):
 \begin{equation}
   \label{eq:cancellation-variance-2}
   \begin{split}
     &\int_{S^2 \times S^2}\d \hat\Omega \d \hat\Omega'\,\langle M_{\ell_1}(\hat\Omega) M_{\ell_2}(\hat\Omega)\rangle \langle M_{\ell_3}(\hat\Omega) M_{\ell_4}(\hat\Omega')\rangle \langle M_{\ell_5}(\hat\Omega') M_{\ell_6}(\hat\Omega') \rangle\\
     &= \frac{(2\ell_1+1)(2\ell_3+1)(2\ell_5+1)}{(4\pi)^3} C_{\ell_1}C_{\ell_3}C_{\ell_5} \delta_{\ell_1 \ell_2}\delta_{\ell_3 \ell_4}\delta_{\ell_5 \ell_6}
     \int_{S^2 \times S^2}\d \hat\Omega \d \hat\Omega'\,
     P_{\ell_1}(1) P_{\ell_3}(\hat\Omega\cdot\hat\Omega') P_{\ell_5}(1)
   \end{split}
 \end{equation}
This integral can be solved using well-known properties of Legendre polynomials. First, we have $P_\ell(1) = 1$ and then we can use the integral
 \begin{equation}
   \label{eq:legendre-integral}
     \int_{S^2 \times S^2}\d \hat\Omega \d \hat\Omega'\,  P_\ell(\hat\Omega\cdot\hat\Omega')=0,
   \end{equation}
and we find the announced result that these terms vanish.   

Concerning the six first terms, we will also explicitly compute only the given example, but the correct permutations to obtain the five other terms will be in the final result. Substituting \eqref{TempMap} into the integral and using the fact that the maps are real, one obtains
 \begin{equation}
   \label{eq:remaining-variance}
   \begin{split}
     \int_{S^2 \times S^2}\d \hat\Omega &\d \hat\Omega'\,\langle M_{\ell_1}(\hat\Omega) M_{\ell_4}(\hat\Omega')\rangle \langle M_{\ell_2}(\hat\Omega) M_{\ell_5}(\hat\Omega')\rangle \langle M_{\ell_3}(\hat\Omega) M_{\ell_6}(\hat\Omega') \rangle\\
     &=
     C_{\ell_1}C_{\ell_2}C_{\ell_3} \delta_{\ell_1 \ell_4}\delta_{\ell_2 \ell_5}\delta_{\ell_3 \ell_6} \sum\limits_{m_1, m_2, m_3}\lh\int_{S^2}\d \hat\Omega  Y_{\ell_1 m_1}(\hat\Omega)\, Y_{\ell_2 m_2}(\hat\Omega) Y_{\ell_3 m_3}(\hat\Omega)\rh\\ &\quad\times \lh\int_{S^2}\d\hat\Omega'\, Y_{\ell_1 m_1}^*(\hat\Omega') Y_{\ell_2 m_2}^*(\hat\Omega') Y_{\ell_3 m_3}^*(\hat\Omega') \rh.     
   \end{split}
 \end{equation}
 where one can recognize the Gaunt integral \eqref{Gaunt_defh}. Substituting it here and using the identity relation
 \begin{equation}
  \label{eq:3-j-identity}
  \sum\limits_{m_1 m_2 m_3}\begin{pmatrix}
     \ell_1 &  \ell_2 &  \ell_3\\
    m_1 & m_2 & m_3
  \end{pmatrix}^2=1,
\end{equation}
and the fact that the columns of Wigner 3$j$-symbols can be permuted when the parity condition is respected, one can find that the 6 terms give
 \begin{equation}
   \label{eq:variance-iso-1}
   \begin{split}
   \langle B_{\ell_1 \ell_2 \ell_3} B_{\ell_4 \ell_5 \ell_6} \rangle = h_{\ell_1 \ell_2 \ell_3}^2 C_{\ell_1}C_{\ell_2}C_{\ell_3} \biggl[&\delta_{\ell_1 \ell_4}\delta_{\ell_2 \ell_5}\delta_{\ell_3 \ell_6}+ \delta_{\ell_1 \ell_4}\delta_{\ell_2 \ell_6}\delta_{\ell_3 \ell_5} + \delta_{\ell_1 \ell_5}\delta_{\ell_2 \ell_4}\delta_{\ell_3 \ell_6}\\
   &+ \delta_{\ell_1 \ell_5}\delta_{\ell_2 \ell_6}\delta_{\ell_3 \ell_4}
   + \delta_{\ell_1 \ell_6}\delta_{\ell_2 \ell_4}\delta_{\ell_3 \ell_5}
   + \delta_{\ell_1 \ell_6}\delta_{\ell_2 \ell_5}\delta_{\ell_3 \ell_4}\biggr].
   \end{split}
 \end{equation}
 Hence the variance is
 \begin{equation}
   \label{eq:bispectrum-variance-iso}
   V_{\ell_1 \ell_2 \ell_3} = g_{\ell_1 \ell_2 \ell_3} h_{\ell_1 \ell_2 \ell_3}^2 C_{\ell_1}C_{\ell_2}C_{\ell_3}
 \end{equation}
with $g_{\ell_1 \ell_2 \ell_3}$ equal to 6, 2, or 1, depending on whether 3,
2, or no $\ell$'s are equal, respectively.

\subsubsection{Anisotropic case}

As explained in section \ref{realsky}, with observational data from an actual experiment we cannot use the isotropy assumption. This would lead to a large increase of the variance \eqref{eq:bispectrum-variance-iso}, because the nine terms described above are no longer zero. However, it is possible to show that adding the simple linear correction given in \eqref{Bisp_lincorr} to the cubic term of the angle-averaged bispectrum solves this issue. To verify this, we will derive the variance similarly to the previous section, the main difference being that integrations are performed on $S^2 \setminus \mathcal{M}$ instead of $S^2$. Again, in the weak non-Gaussianity regime, we only have to compute the average of the product of two bispectra $\langle B_{\ell_1 \ell_2 \ell_3}^{\mathrm{obs}}(\hat\Omega) B_{\ell_4 \ell_5 \ell_6}^{\mathrm{obs}}(\hat\Omega')\rangle$ and there are three types of terms:
\begin{itemize}
  \item 1 term: product of the two cubic terms: $\langle M_{\ell_1} M_{\ell_2} M_{\ell_3} M_{\ell_4} M_{\ell_5} M_{\ell_6} \rangle$ (this is the only term present in the isotropic case);
  \item 6 terms: product of a linear term with a cubic term, e.g.\ $\langle - M_{\ell_1} M_{\ell_2} M_{\ell_3} M_{\ell_4}  \rangle\langle M_{\ell_5} M_{\ell_6}\rangle$;
  \item 9 terms: product of two linear terms, e.g.\ $\langle M_{\ell_1} M_{\ell_4} \rangle\langle M_{\ell_2} M_{\ell_3}\rangle\langle M_{\ell_5} M_{\ell_6}\rangle$.
\end{itemize}
We have seen how the first term gives 15 contributions if we use Wick's theorem to transform the six-point correlation function into a combination of products of three two-point correlation functions. We have also seen that in the isotropic case, only the six terms where each multipole among $(\ell_1,~\ell_2,~\ell_3)$ is coupled with an element of the other triplet $(\ell_4,~\ell_5,~\ell_6)$ are non-zero. The same can be done for the four-point correlation function and each combination of a linear with a cubic term will give three terms, hence a total of eighteen terms. Note also that each term derived from the linear correction contains necessarily two $\ell$'s of the same triplet that are coupled (it is in the definition of the linear term). Hence, it means that they cannot cancel the six terms of the isotropic case. The new terms (i.e.\ the terms that are not present in the isotropic case) are nine from the six-point correlation function, nine from the product of two linear terms, and eighteen from terms with the four-point correlation functions (with a minus sign) and it is then easy to check that they exactly cancel each other. So finally
\begin{equation}
  \begin{split}
  \langle B^{\mathrm{obs}}_{\ell_1 \ell_2 \ell_3} B^{\mathrm{obs}}_{\ell_4 \ell_5 \ell_6} \rangle =
  &\int \d \hat\Omega \d \hat\Omega'\, \left[ \langle  M_{\ell_1} M_{\ell_4}\rangle\langle  M_{\ell_2} M_{\ell_5}  \rangle\langle M_{\ell_3} M_{\ell_6}\rangle + (14)(26)(35) \right. \\
  &\qquad\left.+ (15)(24)(36) + (15)(26)(34) + (16)(24)(35) + (16)(25)(34)\right],
  \end{split}
\end{equation}
where we use an obvious shorthand notation to indicate the other permutations of filtered maps. It is important to note that we recover the same variance as in the isotropic case without a linear term (except for the integration interval). This proves that the estimator with this linear correction is optimal, with the assumption that $C_{\ell m, \ell' m'}$ is diagonal (thus equation \eqref{eq:cancellation-variance} is valid here). However, the integration interval for the two integrals is $S^2 \setminus \mathcal{M}$. In the
$f_\mathrm{sky}$ approximation, which many tests have shown to be a good approximation, we calculate the
integrals as if the interval were the full sky $S^2$, and then add appropriate factors of $f_\mathrm{sky}$
at the end to compensate for the partial sky.
Then, performing the same last steps of the calculation as before, the variance is given by
\begin{equation}
  \label{eq:variance-appendix}
  V_{\ell_1 \ell_2 \ell_3} =
  g_{\ell_1 \ell_2 \ell_3} \frac{h_{\ell_1 \ell_2 \ell_3}^2}{f_\mathrm{sky}}
  (b_{\ell_1}^2 C_{\ell_1} + N_{\ell_1}) (b_{\ell_2}^2 C_{\ell_2} + N_{\ell_2}) (b_{\ell_3}^2 C_{\ell_3} + N_{\ell_3}),
\end{equation}
when including the effect of the beam and the noise and has a form similar to the isotropic case
\eqref{eq:bispectrum-variance-iso}. The factor of $1/f_\mathrm{sky}$ can easily be understood given that
the variance of a quantity determined from $N$ data points scales as $1/N$ and here the number of data
points roughly corresponds to the number of observed pixels on the sky.

\subsection{Weights of bispectral shapes}

In this appendix, we give another representation of the different bispectra discussed in this paper (primordial shapes, ISW-lensing, extra-galactic and galactic foregrounds) well-suited to understand the correlation coefficients given in e.g.~tables~\ref{tab_corr_coeff}, \ref{tab:corr_coeff_dust} and \ref{tab:corr_coeff_others}.

\begin{figure}
  \centering 
  \includegraphics[width=0.48\linewidth]{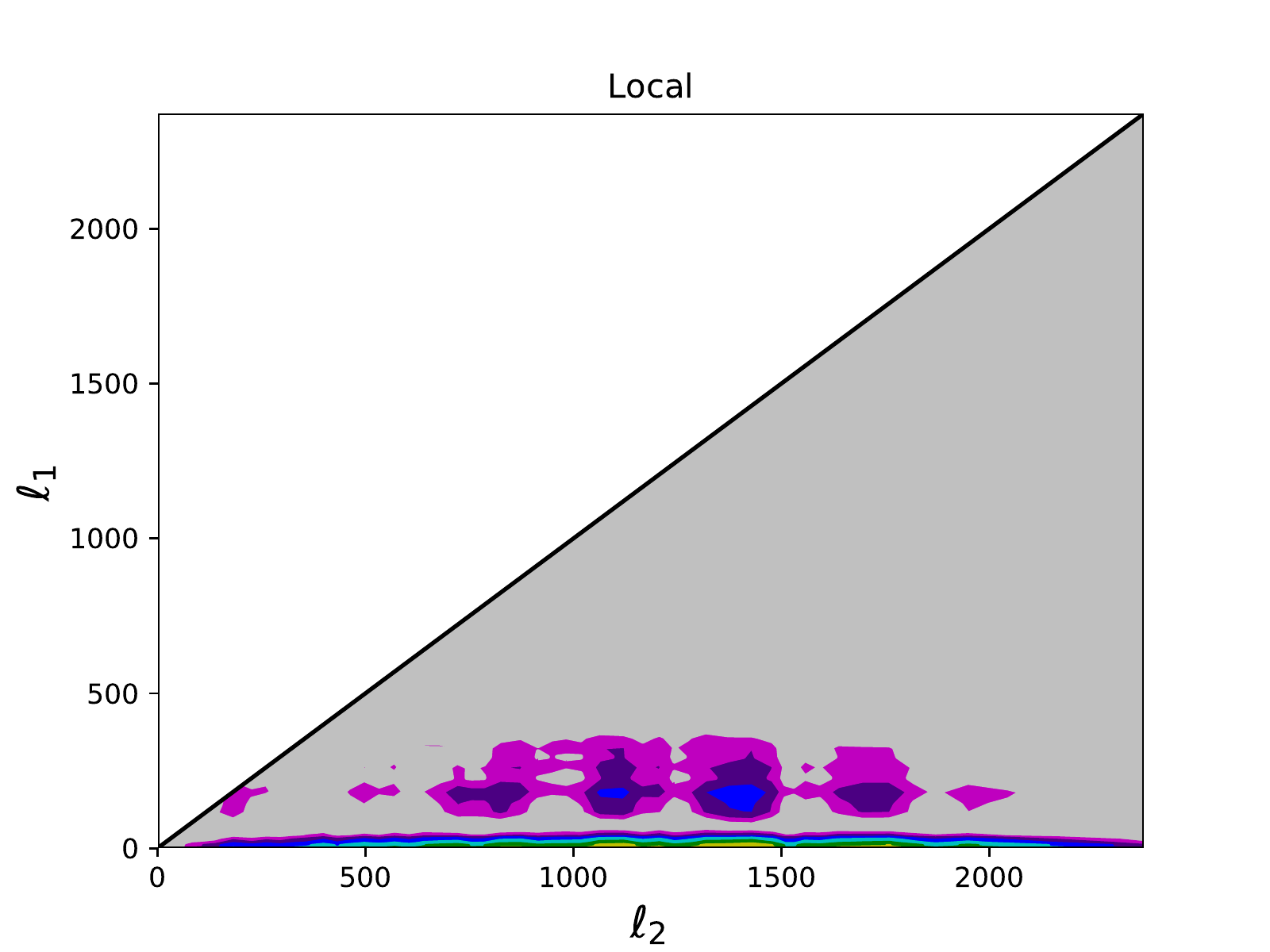} 
  \includegraphics[width=0.48\linewidth]{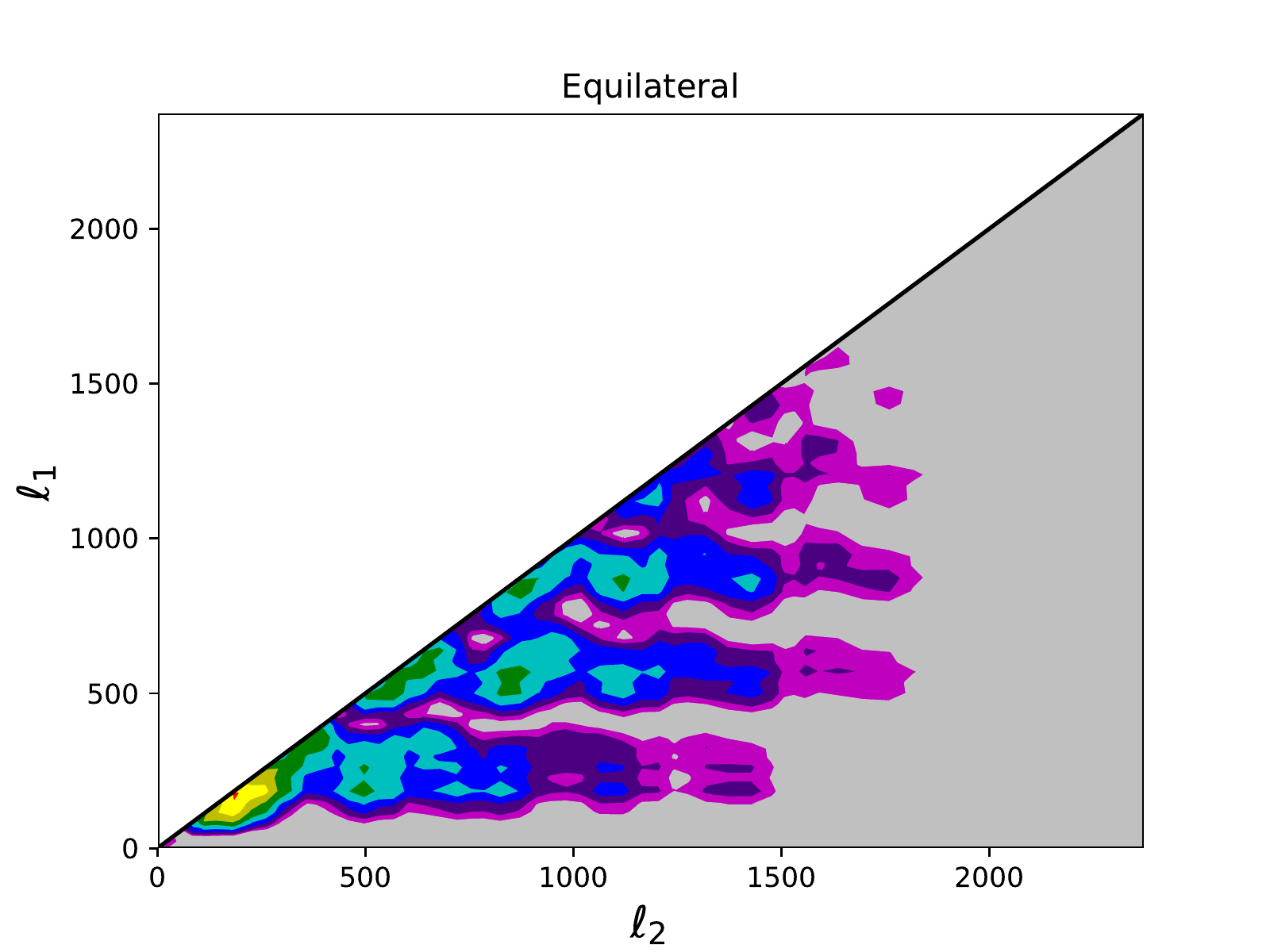}
  \includegraphics[width=0.48\linewidth]{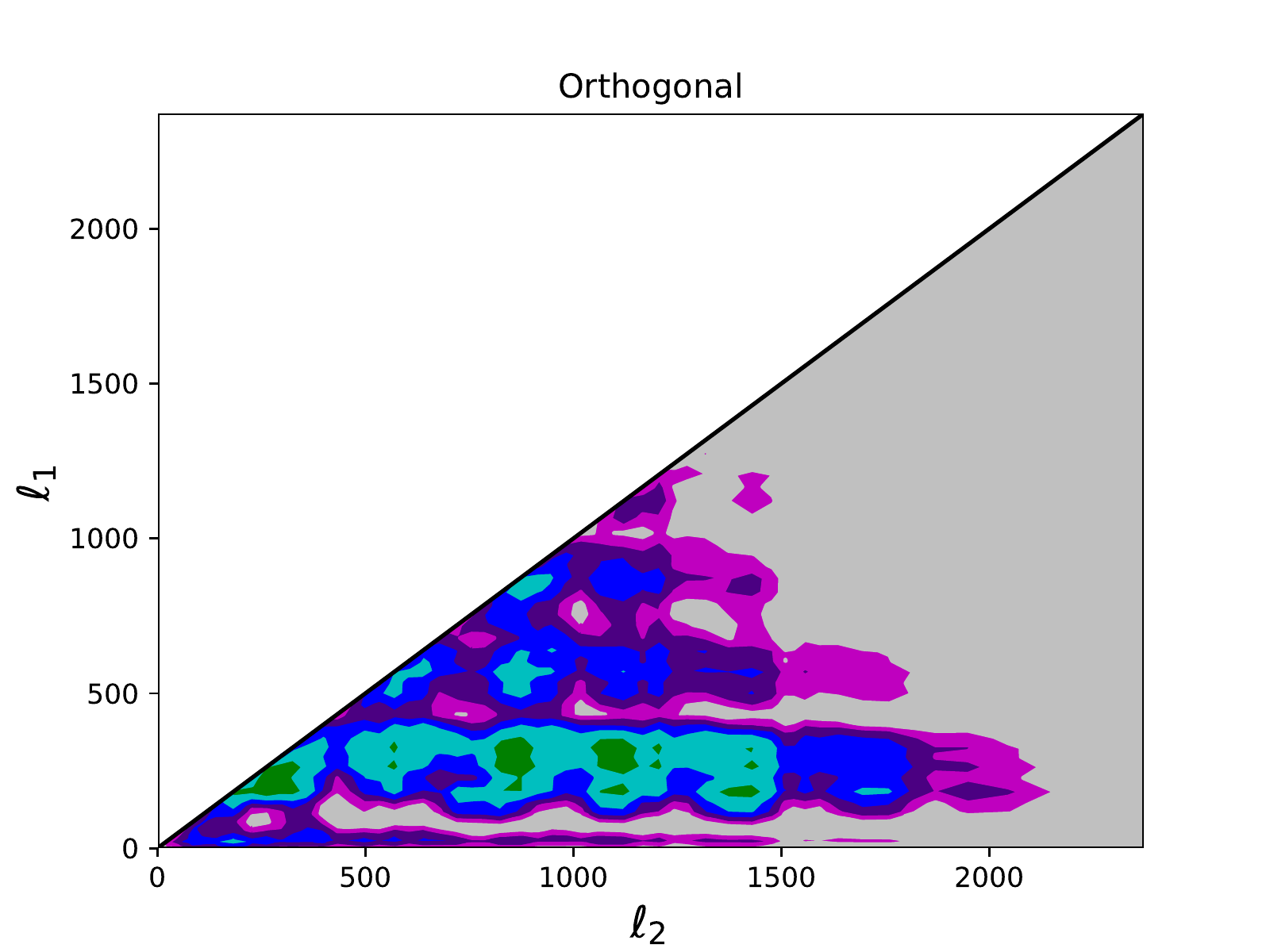}
  \includegraphics[width=0.48\linewidth]{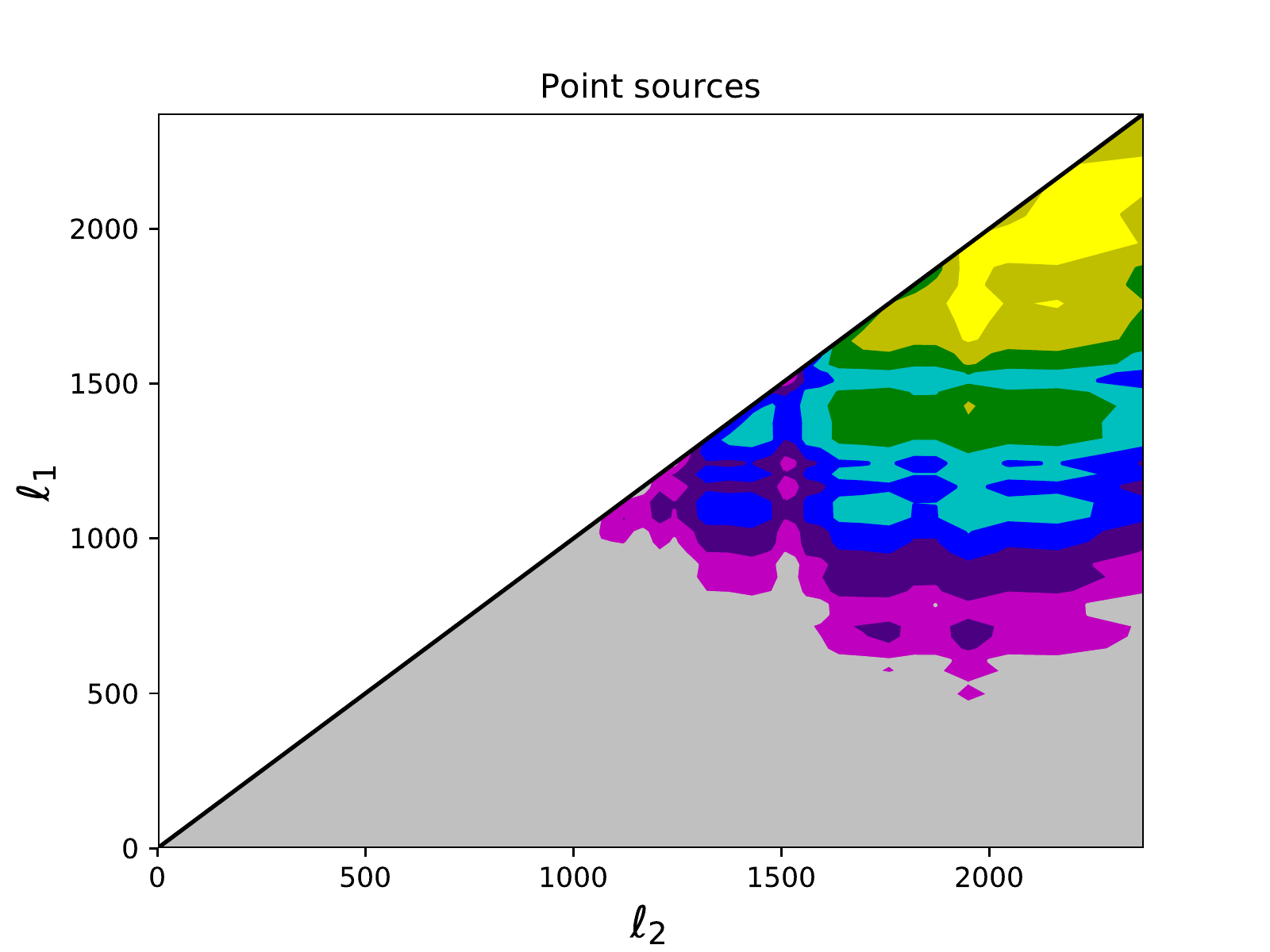}
  \includegraphics[width=0.48\linewidth]{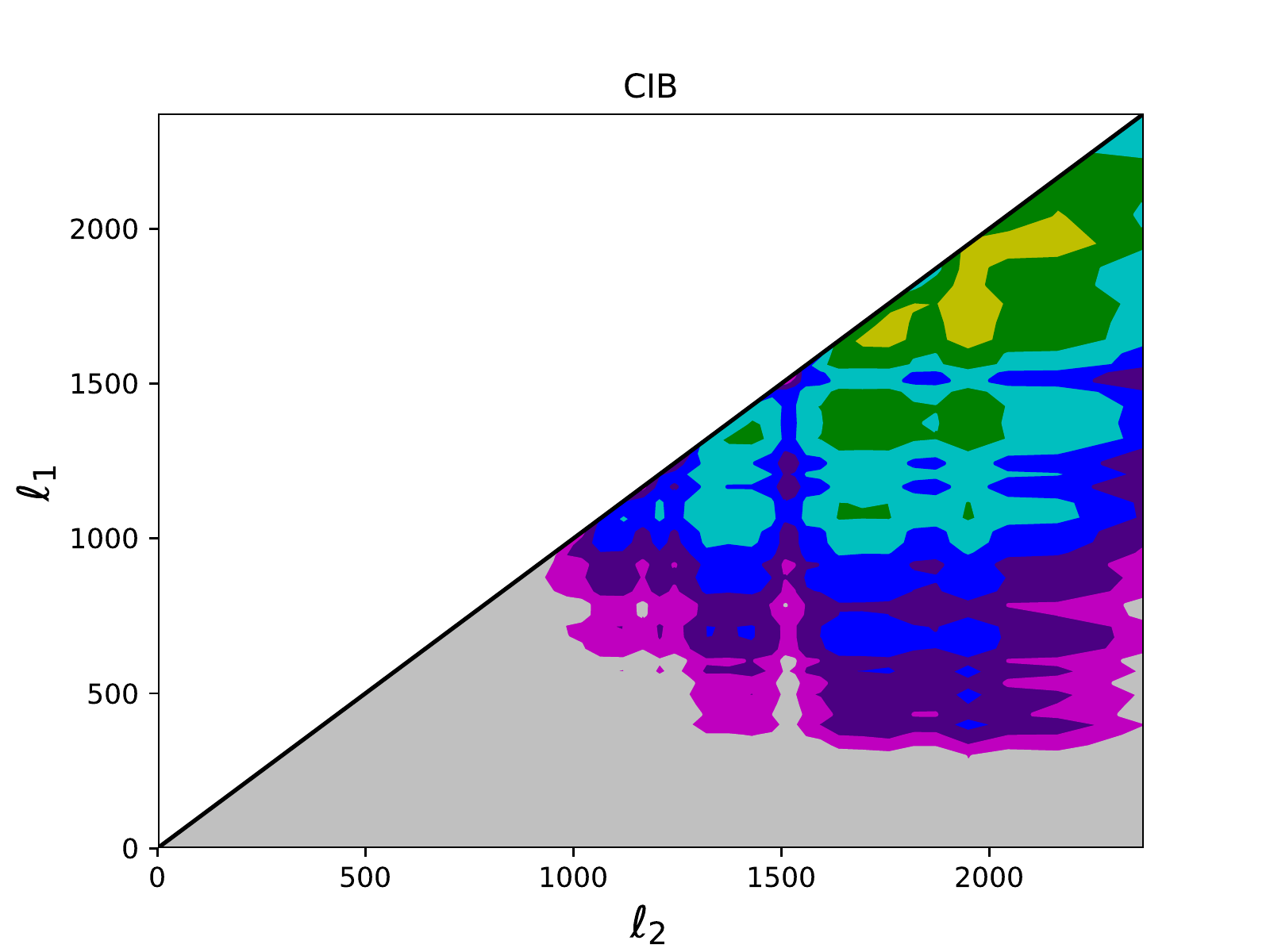} 
  \includegraphics[width=0.48\linewidth]{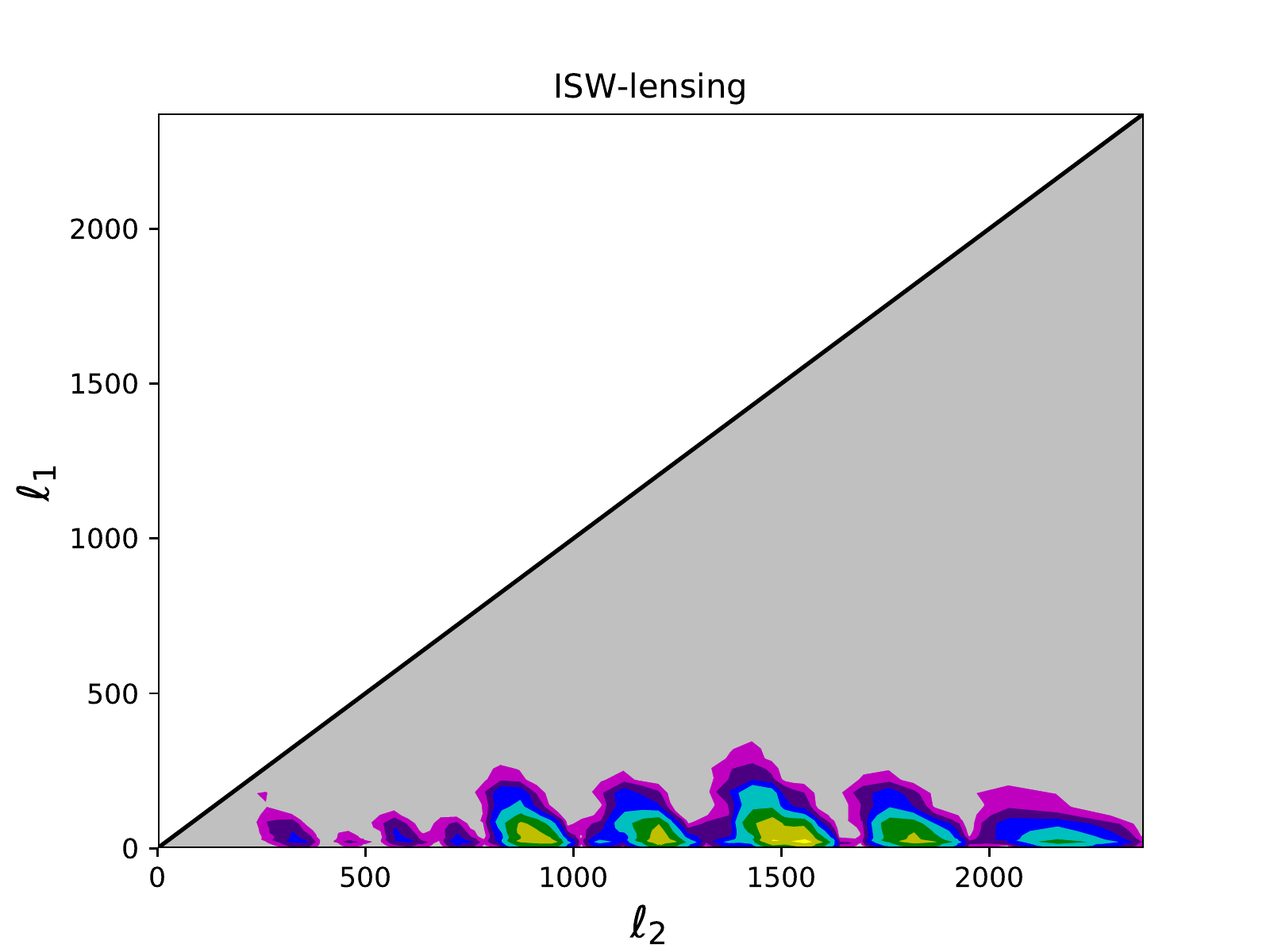}
  \includegraphics[align=c,width=0.48\linewidth]{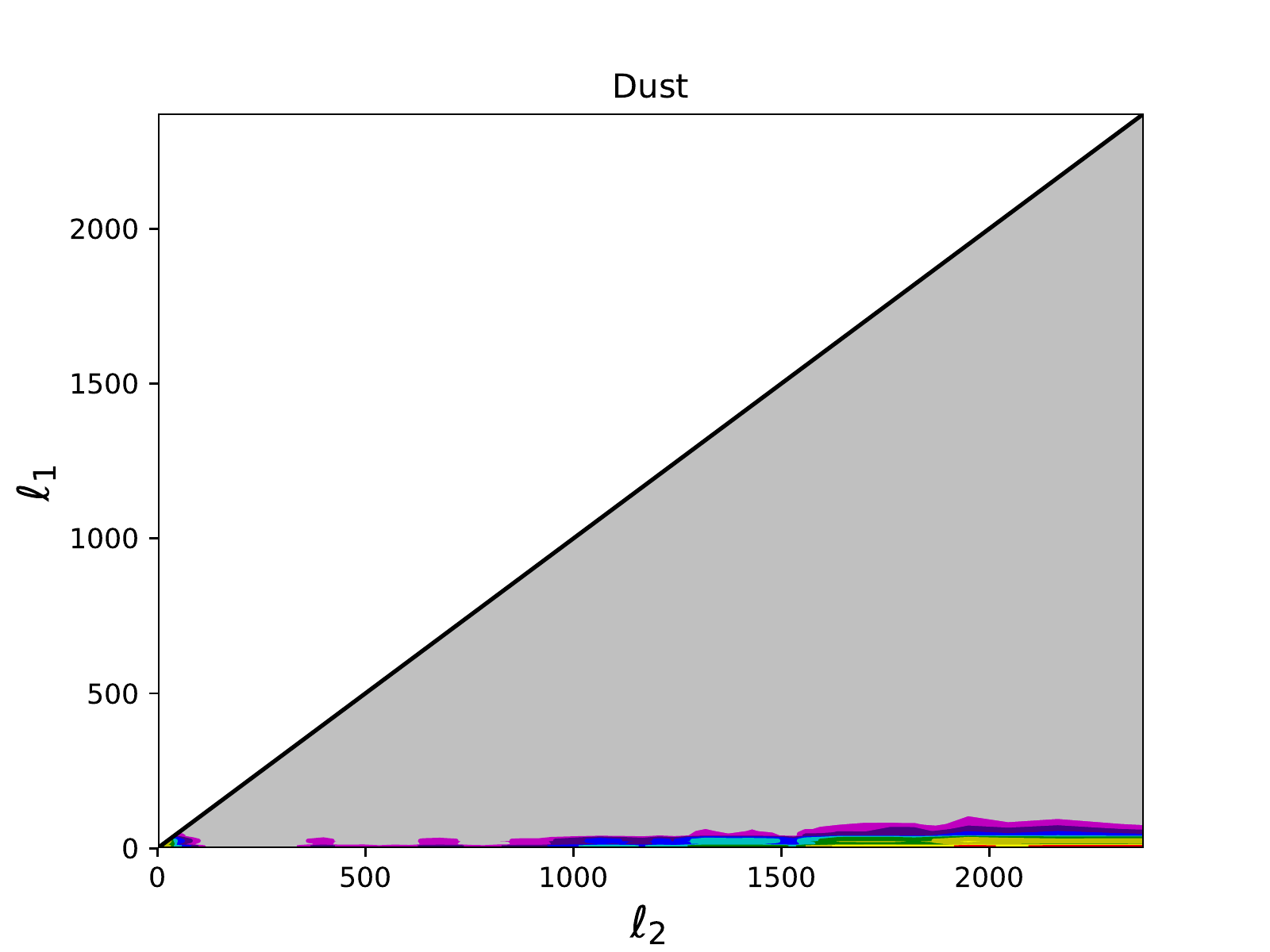}    
  \includegraphics[align=c,width=0.50\linewidth]{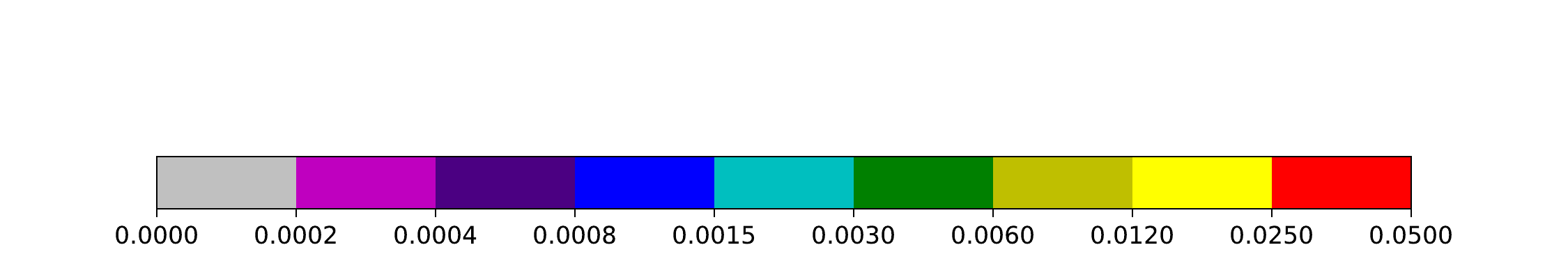}
  \caption{Weights of the bispectral shapes discussed in this paper at high resolution. Note that the colour scale is logarithmic.}
  \label{fig:weights-highres}
\end{figure}

\begin{figure}
  \centering    
  \includegraphics[width=0.48\linewidth]{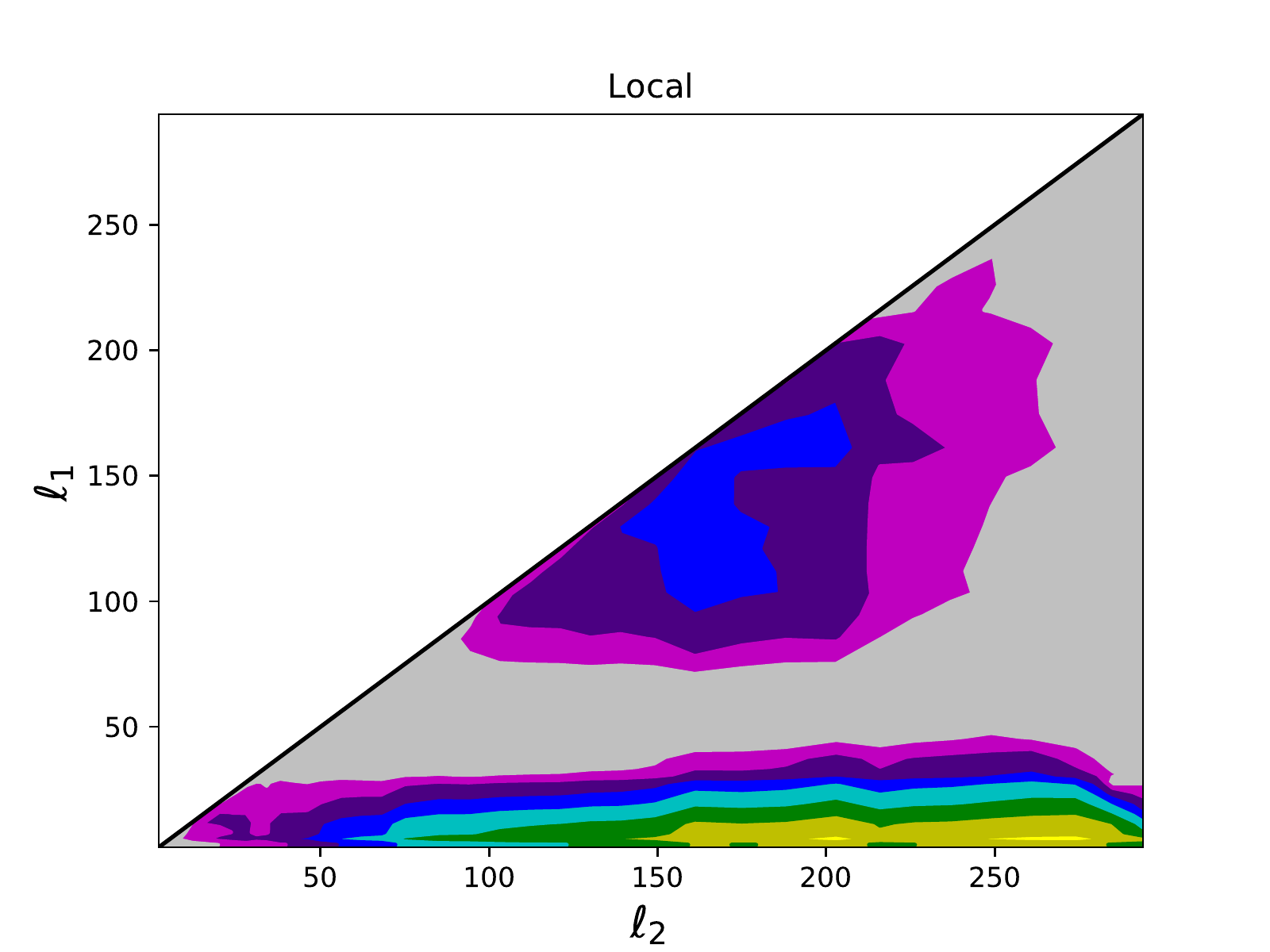}
  \includegraphics[width=0.48\linewidth]{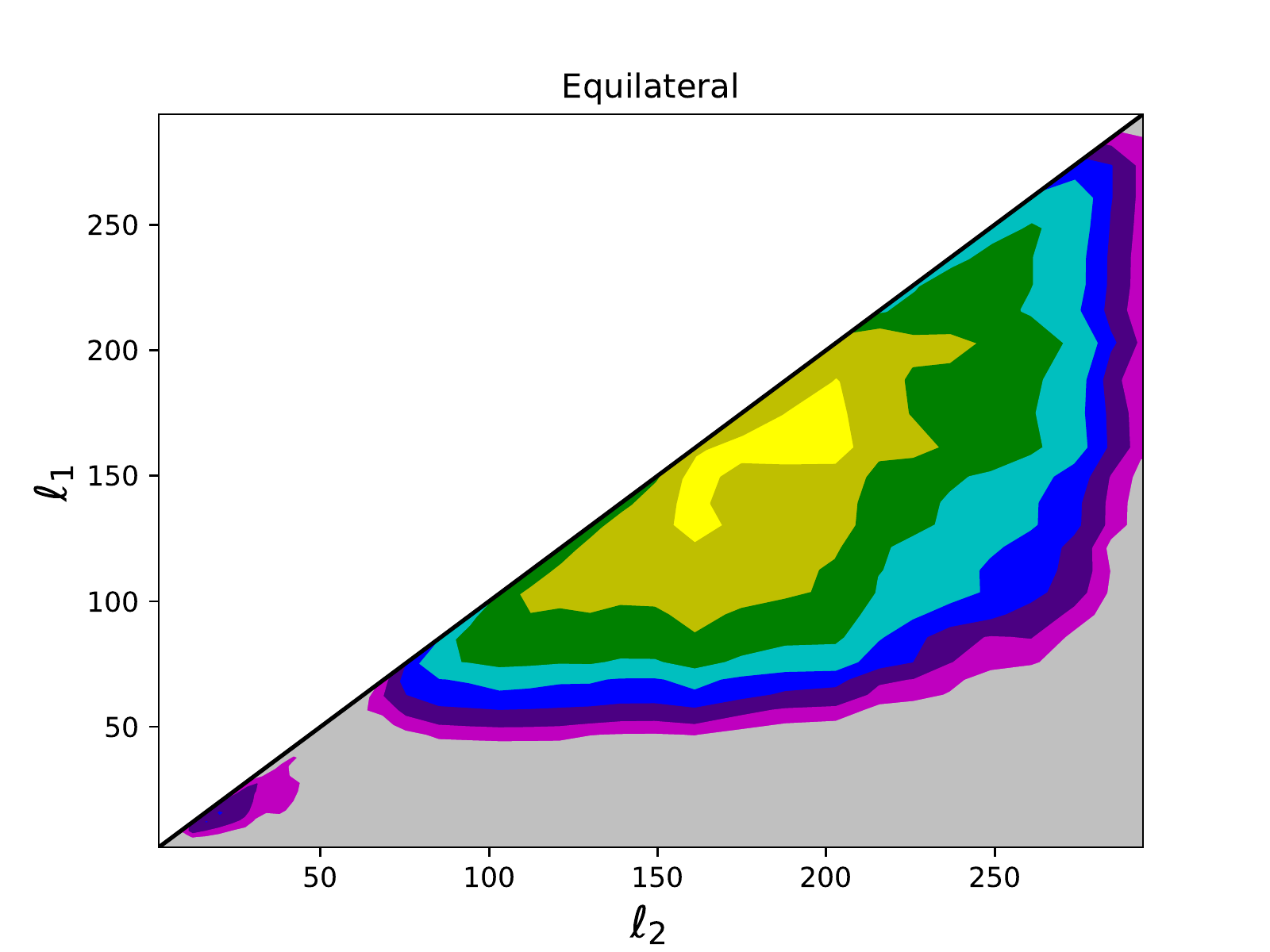}
  \includegraphics[width=0.48\linewidth]{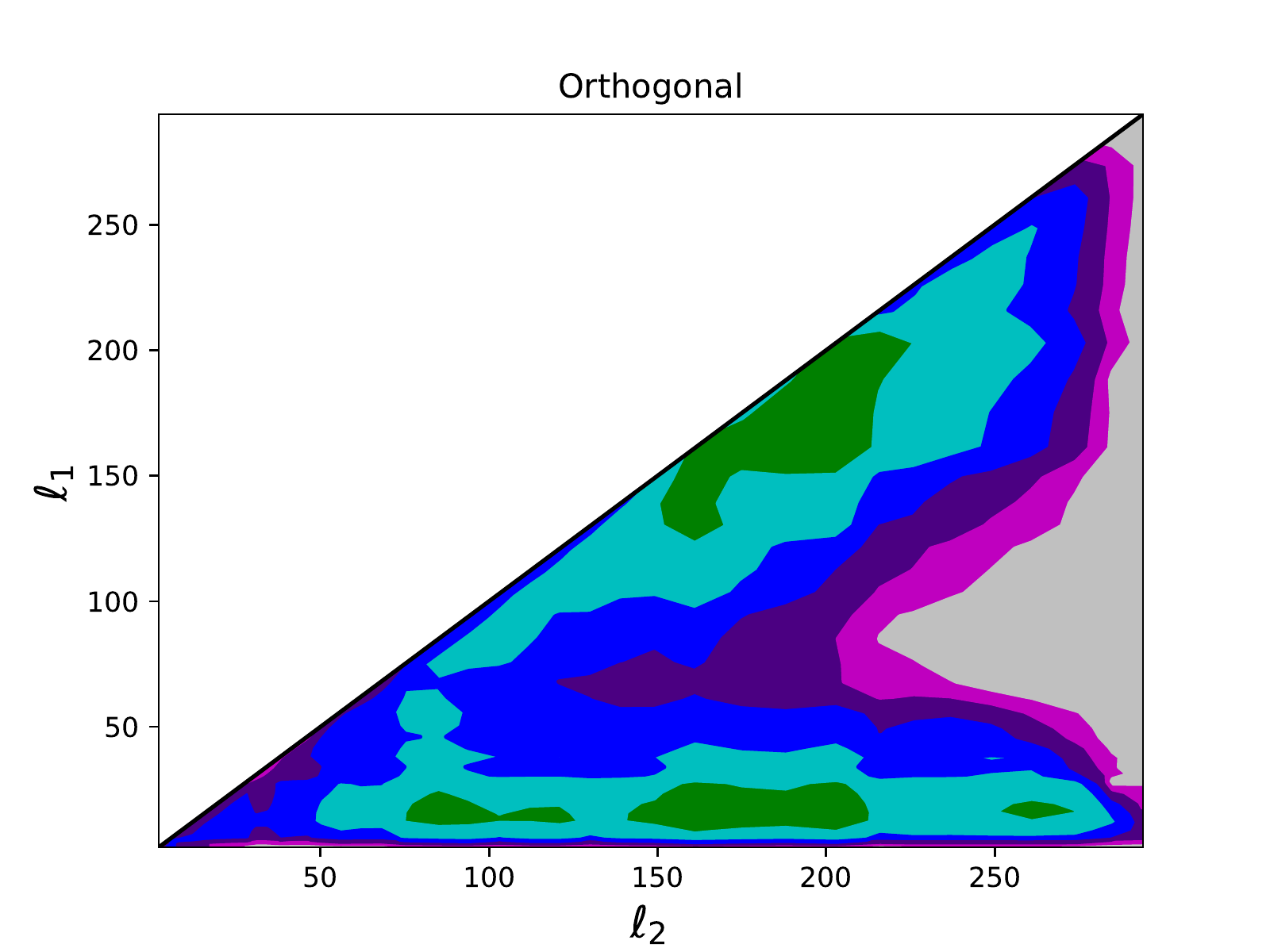} 
  \includegraphics[width=0.48\linewidth]{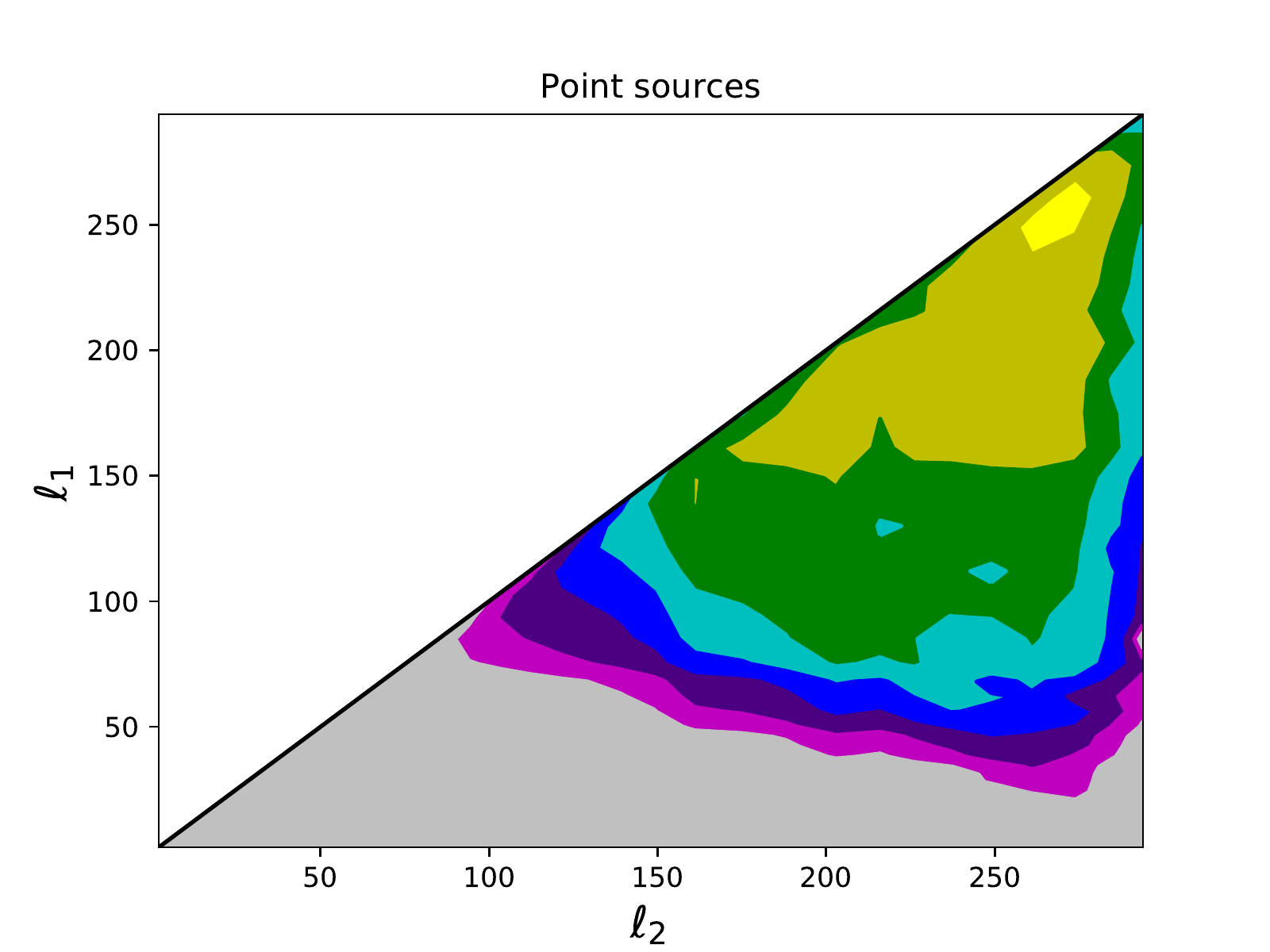} 
  \includegraphics[width=0.48\linewidth]{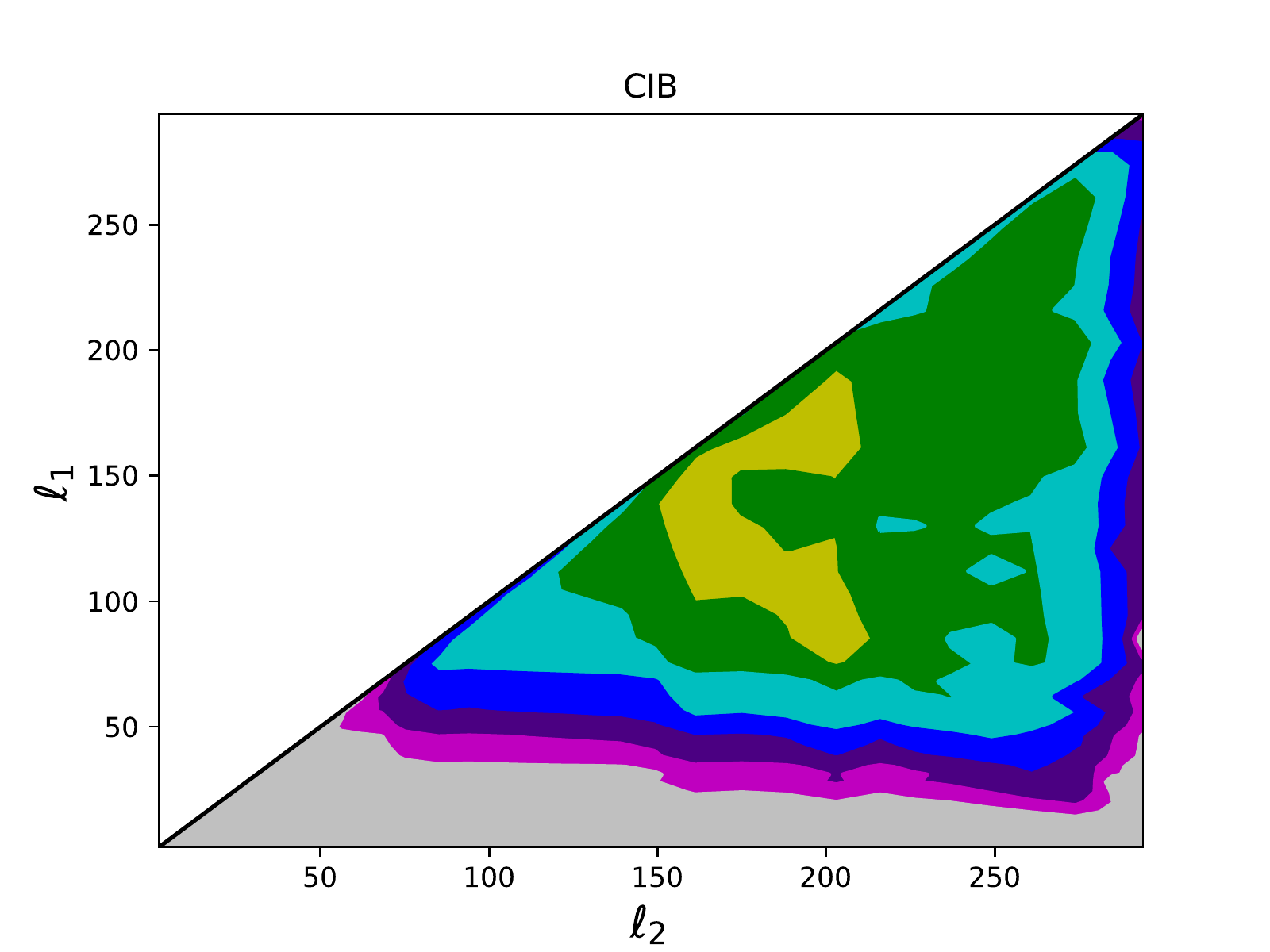} 
  \includegraphics[width=0.48\linewidth]{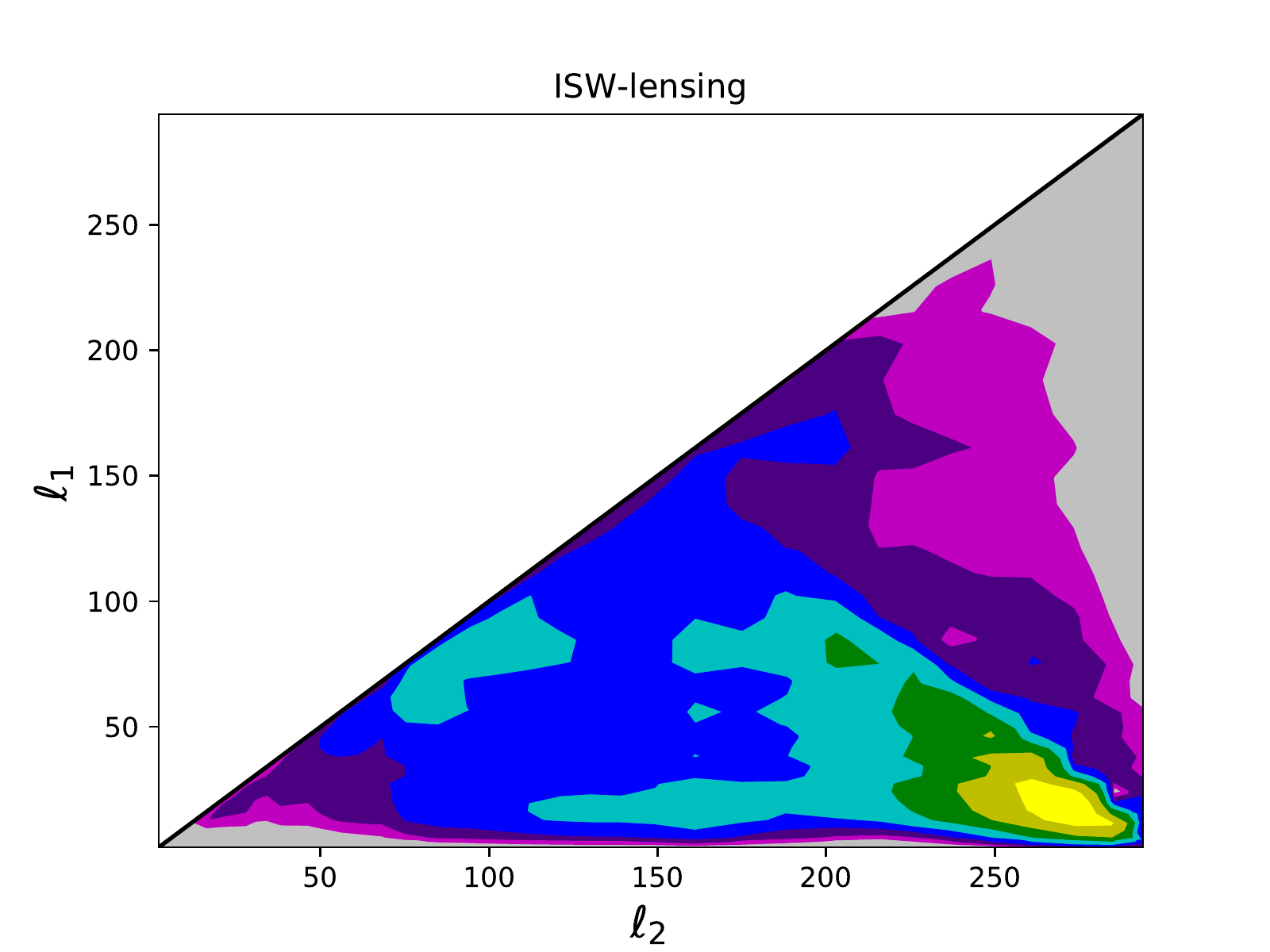}
  \includegraphics[align=c, width=0.49\linewidth]{plots_JRvT/appendices/colorbar.pdf}
  \caption{Weights of the standard bispectral shapes at low resolution. Note the difference on the axes with the previous figure. The colour scale is the same, but the weights are normalized to unity here over a much reduced region of multipole space with $\ell<300$.}
  \label{fig:weights-lowres_1}
\end{figure}

\begin{figure}
  \centering
  \includegraphics[width=0.48\linewidth]{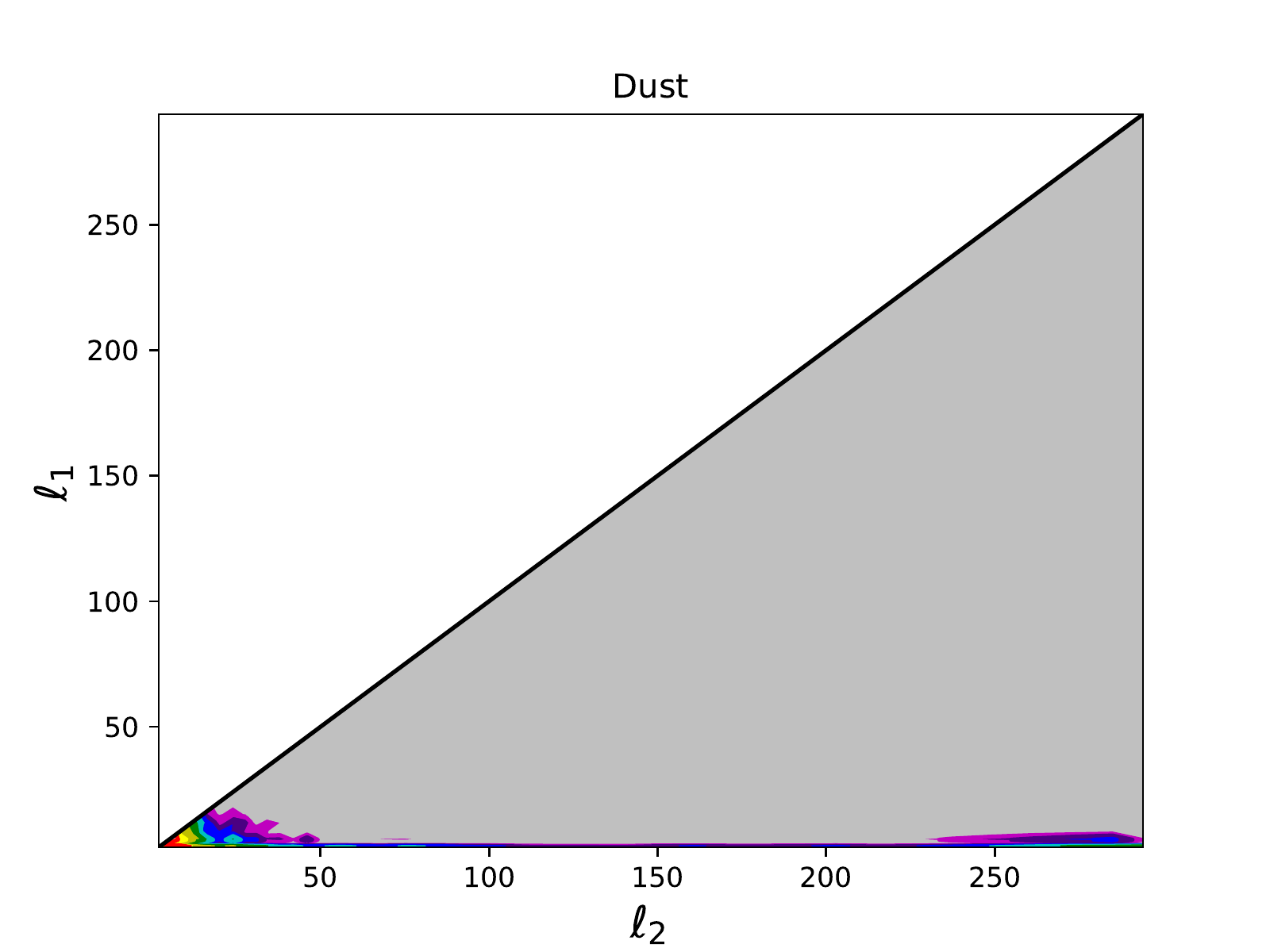}
  \includegraphics[width=0.48\linewidth]{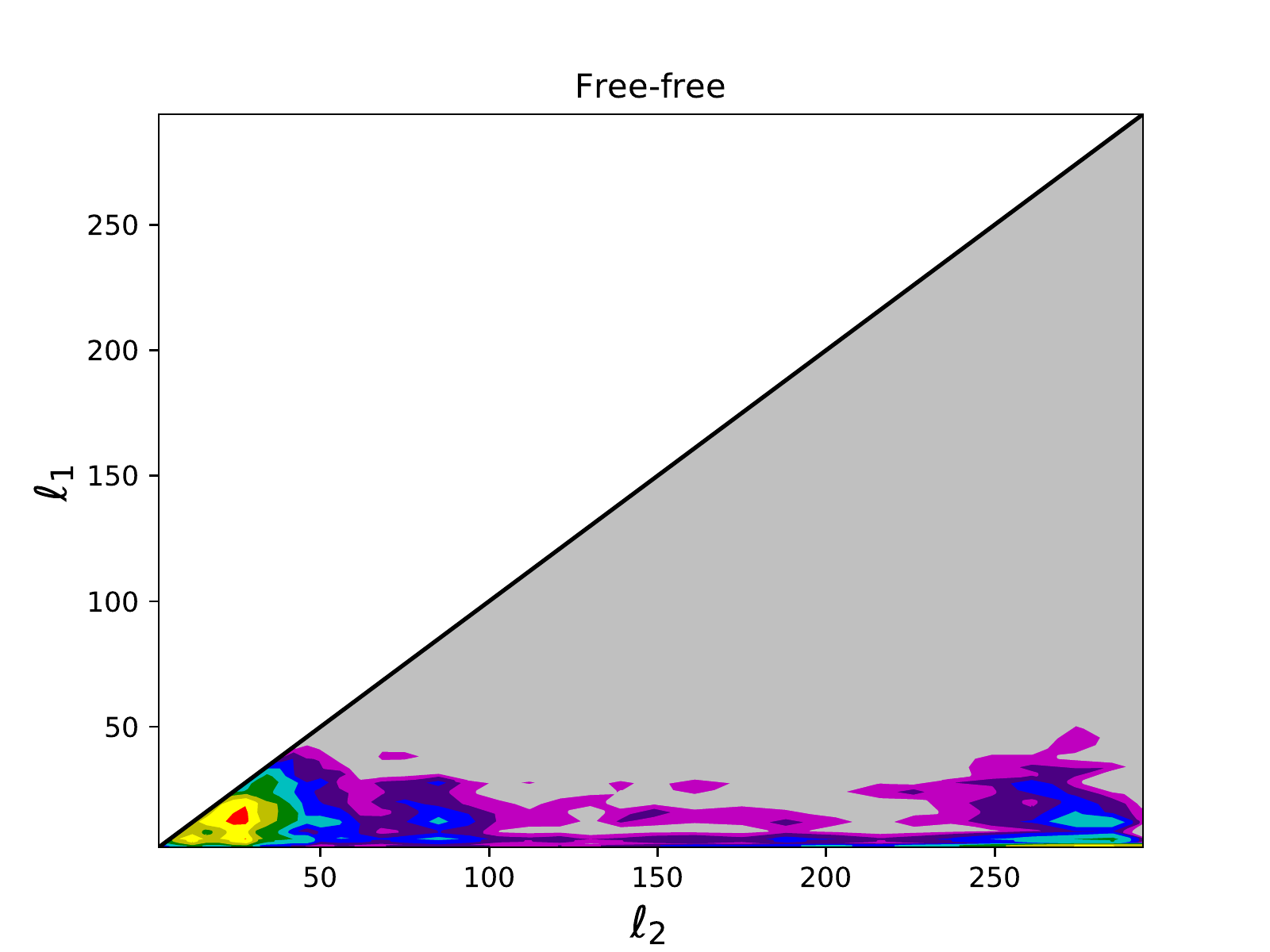}
  \includegraphics[width=0.49\linewidth]{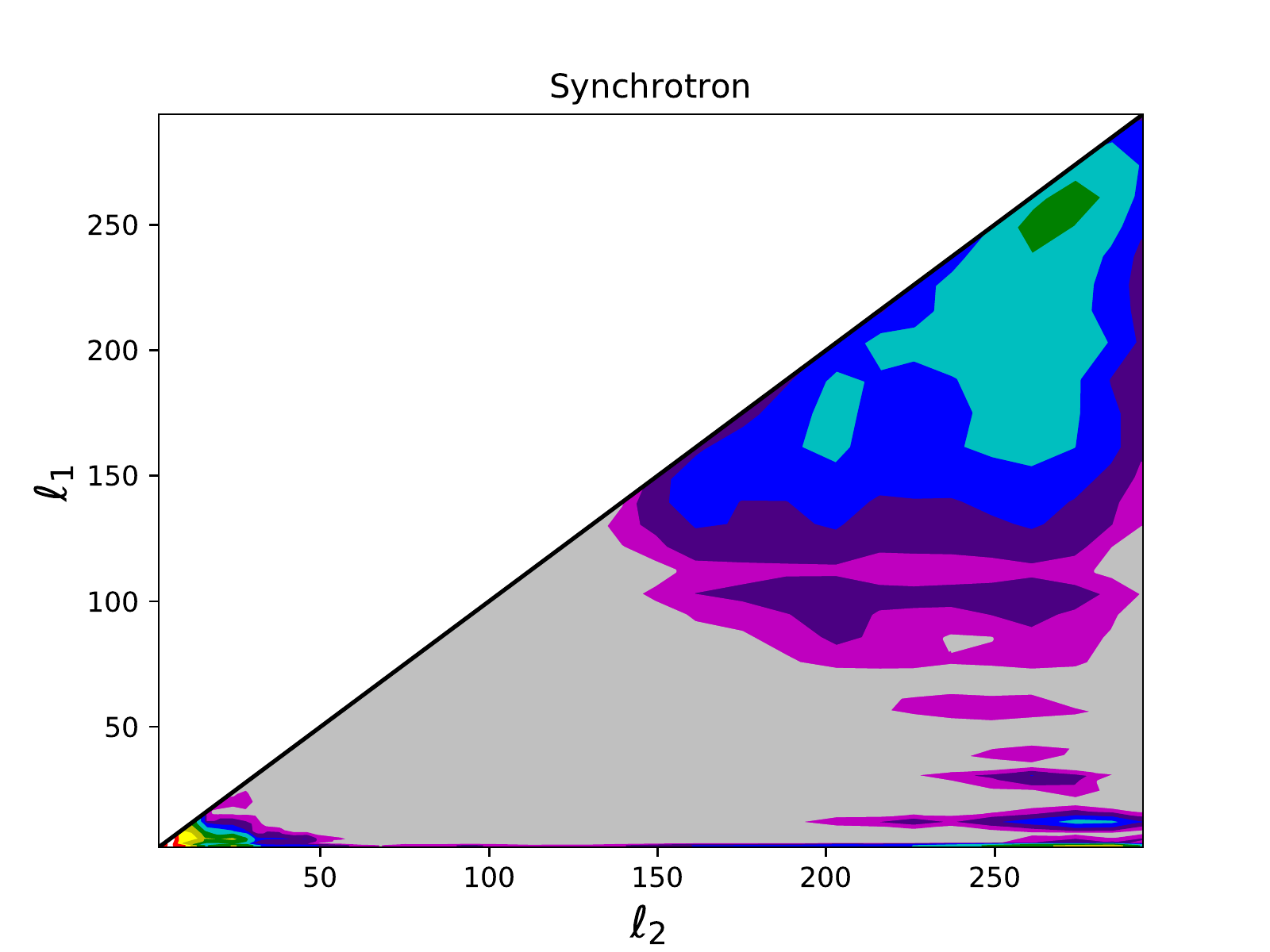}
  \includegraphics[width=0.49\linewidth]{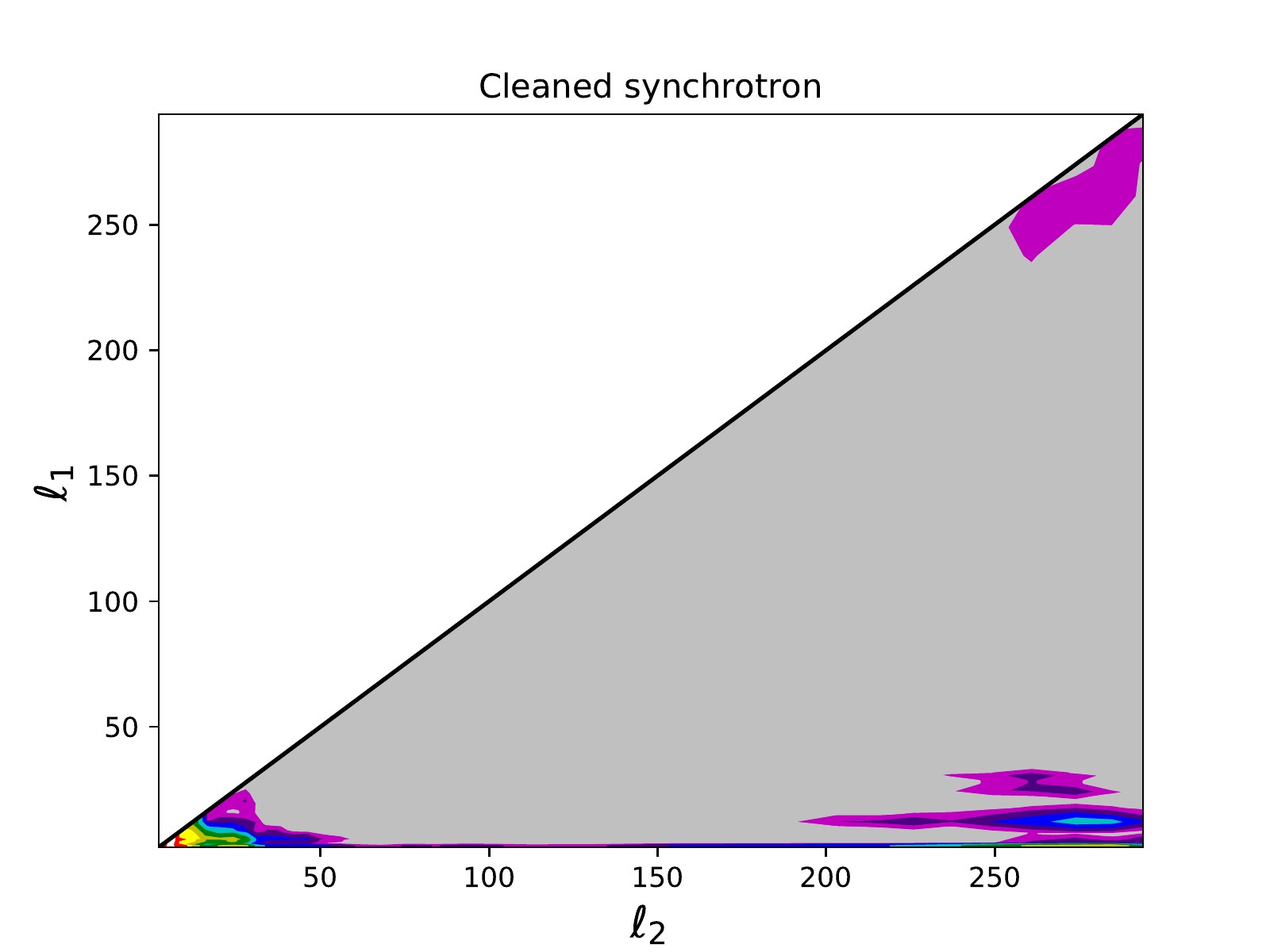}
  \includegraphics[align=c, width=0.49\linewidth]{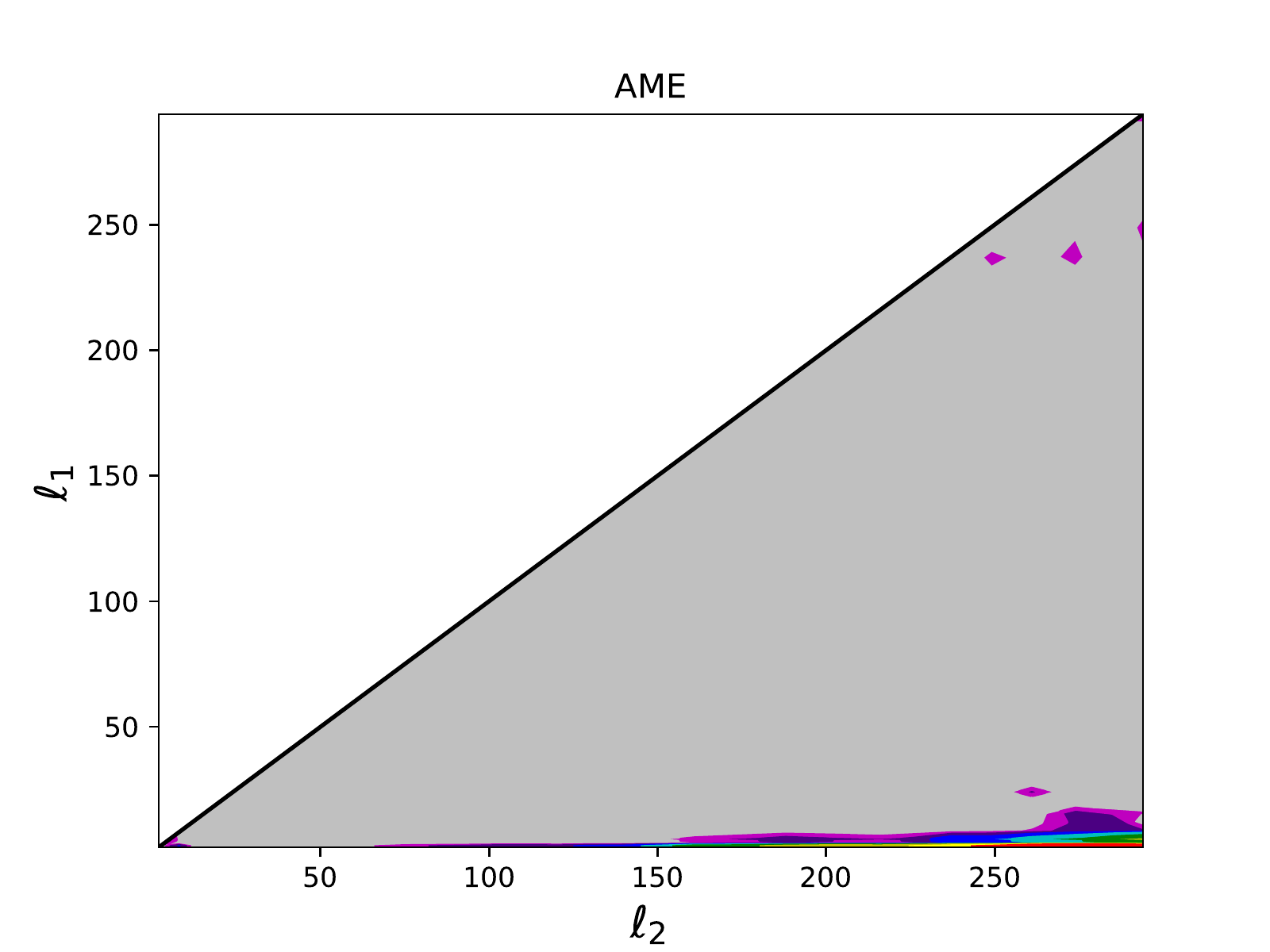}
  \includegraphics[align=c, width=0.49\linewidth]{plots_JRvT/appendices/colorbar.pdf}
  \caption{Weights of the foreground bispectral shapes discussed in this paper at low resolution.}
  \label{fig:weights-lowres}
\end{figure}

The weight of a single multipole configuration ($\ell_1, \ell_2, \ell_3$) of a bispectral shape $B_{\ell_1 \ell_2 \ell_3}$ is defined by \citep{BvTC}
\begin{equation}
  \label{eq:weight}
  w_{\ell_1 \ell_2 \ell_3}= \frac{1}{\langle B, B\rangle}\frac{(B_{\ell_1 \ell_2 \ell_3})^2}{V_{\ell_1 \ell_2 \ell_3}}.
\end{equation}
It is the inverse of the variance of the ratio of the observed and theoretical bispectra divided by $\langle B, B\rangle$ which is the denominator of the estimator for $\fnl$ and normalizes the sum of the weights to one. In other words,
\begin{equation}
  \label{eq:estimator-weight}
  \hat{f}_\mathrm{NL} = \sum\limits_{\ell_1 \ell_2 \ell_3}w_{\ell_1 \ell_2 \ell_3} \frac{B_{\ell_1 \ell_2 \ell_3}^{\mathrm{obs}}}{B_{\ell_1 \ell_2 \ell_3}},
\end{equation}
where $B^\mathrm{obs}/B$ can be viewed as an $\fnl$ estimator based on just a single $\ell$-triplet. These equations are the same for bin-triplets ($i_1, i_2, i_3$). Figures~\ref{fig:weights-highres}, \ref{fig:weights-lowres_1}, and~\ref{fig:weights-lowres} show the weights of the different theoretical and numerical shapes discussed in this paper at both high and low resolution, with the usual choice of 57 bins. Instead of using a few slices of $\ell_3$ like in section \ref{sec:foregrounds}, we summed over $\ell_3$. It has the advantage that now the whole bispectrum is used in one figure, but of course we lose the information about the variation of the bispectrum as a function of $\ell_3$. A larger weight means that the region of multipole space is more important for the template. Conversely, a large observed non-Gaussianity in that region of multipole space means that it is more likely to be that particular shape.

In this kind of plot, shapes that peak in squeezed configurations will have a colored band/line at the bottom of the figure (low $\ell_1$). As expected it is present for the different foregrounds for very low $\ell_1$ ($< \mathcal{O}(20)$), including synchrotron (which was not visible in figure~\ref{fig:other-templates}). As expected, the characteristic line of a squeezed bispectrum can be seen for the local and the ISW-lensing shapes, but also for the orthogonal shape (which explains why it is somewhat correlated to the foregrounds).

Shapes that peak in equilateral configurations have a large weight along the diagonal black line of these plots, when the three $\ell$'s are of the same order. The primordial equilateral shape is the strongest for three low $\ell$'s, while the point sources and the CIB are more non-Gaussian at higher multipoles. It is easy to see the correlation between the point sources and the synchrotron bispectra which peak when the three $\ell$'s are over 150. An additional remark is necessary about the orthogonal shape. Indeed by definition it is orthogonal to the equilateral shape (uncorrelated), which is not visible in these figures because they both have similar acoustic peaks. It is an effect of the sum over $\ell_3$ which hides the differences of these bispectra.


\chapter{Planck 2018 results. IX. Constraints on primordial non-Gaussianity}
\label{Planck2018NGapp}

This appendix, containing the full paper \citepalias{planck2018-09} with the
primordial non-Gaussianity analysis of the Planck 2018 release, has been
removed from the arXiv version of this thesis due to size restrictions.
Please check out the original paper!

\end{appendices}

\chapter{Summary and conclusion}
\label{conclusion}

In this thesis my work on non-Gaussianity in cosmology has been presented. It
has two main parts: 1) the development of the long-wavelength formalism
to compute the non-Gaussianity produced in multiple-field
inflation models and the application of this formalism to study those models;
and 2) the development of the binned bispectrum estimator to extract
information about non-Gaussianity from CMB data and its application to the
Planck data.

The development of the long-wavelength formalism was a long process that
took place over many years, first in collaboration with Gerasimos Rigopoulos
and Paul Shellard, and later with Eleftheria Tzavara. The formalism reached
its final form in \citep{TvT1}. Based on the work from all my papers on this
subject, this thesis provides a detailed description and derivation of
the long-wavelength formalism.

The formalism is based on the long-wavelength approximation valid on
super-horizon scales (which boils down to neglecting second-order spatial
gradients in the equations of motion) and a smart choice of gauge-invariant
variables constructed from spatial gradients. In this way the fully non-linear
equations of motion of the fluctuations on super-horizon scales are determined,
and can be used for numerical simulations. The set of equations is completed
with source terms describing the continuous influx of sub-horizon perturbations
into the long-wavelength system. For analytic results a perturbative
approach is required, and in this thesis the full second-order calculation
for two-field inflation of the (local) bispectrum and its amplitude parameter
$\fnl$ is given, the final result being \eqref{fNLresult}.

This expression for $\fnl$ contains
three types of terms: terms that are slow-roll suppressed, terms that are
directly proportional to the isocurvature mode, and an integrated contribution. 
The isocurvature terms can easily become very large, but have the practical
disadvantage that with an isocurvature mode still present at the end of
inflation, the adiabatic perturbation will continue to evolve on super-horizon
scales even after inflation. Computing any observational predictions
for the CMB would then require to follow their evolution during (p)reheating
and afterwards, a major challenge. Hence we always assume these
terms to have disappeared by the end of inflation, so that the only potentially
detectable non-Gaussianity comes from the integrated contribution.

Some study of the consequences of this integrated expression for non-Gaussianity
predictions of two-field inflation models (and classes of models) was
included in earlier papers, but a much more complete analysis was presented
in \citep{JvT} in collaboration with Gabriel Jung. That paper is
included in one of the appendix chapters of this thesis. After first deriving
an alternative expression for the integrated contribution to $\fnl$, we
investigate the class of two-field sum potentials in great detail. We show that
the region in parameter space where $\fnl$ is large enough to be detected in the
near future, but that is compatible with the current Planck constraint on the
spectral index, is quite limited, although we do manage to construct such a
model. We also extend our analytic results to models where slow roll
is broken during the turn of the field trajectory, and show why the slow-roll
conclusions remain valid even in that case.

The other two papers included in the appendix chapters of the first part of the
thesis are \citep{TvT2} and \citep{TvT3}. The former paper contains an
investigation of gauge issues at second order and a derivation of the full
cubic action of the adiabatic and the isocurvature perturbations. It
definitively settles a number of previously open questions regarding the
long-wavelength formalism. Although this has not yet been followed up on,
this full cubic action should make it possible to compute the non-Gaussianity
of any two-field inflation model without any approximation.\footnote{The
  generalization of
  this full cubic action to the case of non-standard kinetic terms was computed
  in \citep{TMvT}, but that work is not included in this thesis.}
The latter paper investigates the issue of the
scale-dependence of $\fnl$ and defines two spectral indices for the
bispectrum, one related to the size of the momentum triangle and the
other to its shape.

The second part of this thesis is dedicated to the binned bispectrum estimator
and its applications. It was developed in collaboration with Martin Bucher and
later also Benjamin Racine (and with Carla Carvalho being involved in the
initial stages of the project).
The final estimator and its implementation were published in \citep{BRvT}.
The detailed description and derivation of the binned bispectrum estimator in
this thesis is based mainly on that paper.

The binned bispectrum estimator is motivated by the observation that most
bispectrum templates we are interested in are relatively smooth in
harmonic space. This means that we lose very little information if we pursue
a binned data analysis with moderately sized bins in $\ell$-space. On the
other hand, the computational gains of this approximation are enormous, turning
a completely unfeasible computation into a fast estimator. 
The binned bispectrum estimator determines the full three-dimensional
bispectrum of a map, binned in harmonic space. This binned bispectrum can then
be combined with a library of theoretical bispectrum templates to determine
the $\fnl$ parameters. Alternatively, the binned bispectrum of the map can
also be used directly for non-parametric (blind) non-Gaussianity searches,
by smoothing it.
The binned bispectrum estimator is very fast, has a convenient modular
structure (which means for example that one can analyze an additional template
without having to rerun the map), and gives the dependence of $\fnl$ on $\ell$
as a free bonus. It accepts many different types of templates, even
non-separable theoretical templates or templates that have been determined
numerically from a map.

Various other aspects of the estimator are also discussed in detail in this
thesis: the standard theoretical templates, the extension to include
polarization, analyzing templates jointly or including a bias, the linear
correction to deal with anisotropic noise, and how to determine the
statistical significance of features in the smoothed binned bispectrum. We
also look at how to deal with the presence of a galactic and a point source
mask. The method of diffusive inpainting that we developed for this was
later adopted by all other bispectrum estimators as well. Finally we discuss
the choice of binning and some issues of the implementation of the
estimator as a computer code, for example related to memory limitations.

In the paper \citep{LvT2}, written in collaboration with David Langlois and
included as an appendix chapter, we investigate
the local bispectrum arising from the (possibly correlated) combination
of the adiabatic mode with an isocurvature mode (which can be CDM,
neutrino density or neutrino velocity). We show that this leads to six
different $\fnl$ parameters, and estimate how precisely CMB data can constrain
these parameters, emphasizing in particular the importance of polarization.
The actual analysis was later performed in the Planck primordial non-Gaussianity
papers, confirming our estimates.\footnote{Recently the paper
  \citep{Montandon:2020kuk} was published, where we look in even more detail at
  the Planck constraints on isocurvature modes, by performing a joint analysis
  of the power spectrum and the bispectrum, but that work is not part of
  this thesis.}
The second appendix chapter of this part contains \citep{JRvT}, written
in collaboration with Gabriel Jung and Benjamin Racine.
We numerically determine and study the bispectra of galactic foregrounds and use
these bispectra as templates with the binned bispectrum estimator to show
that the cleaned Planck maps do not have any detectable galactic foreground
residuals.

The binned bispectrum estimator is one of the three bispectrum estimators
used for the official Planck analyses of 2013, 2015 and 2018. The third and
final of these \citepalias{planck2018-09} is included in the final appendix
chapter of this thesis, as the crowning achievement of a massive amount of
work over the last 10 years. While unfortunately no primordial non-Gaussianity
has been detected, very tight constraints are put on many different bispectrum
templates, including also the isocurvature non-Gaussianity mentioned above.
On the other hand, some non-primordial non-Gaussianity templates are detected,
in particular the lensing bispectrum. A large number of validation tests are
performed, increasing our confidence in the results. The consequences for
various inflation models are worked out. It is likely that this paper will
remain the definitive reference for non-Gaussianity constraints for the next
10 years, until the results from the LiteBIRD satellite or the ground-based
CMB-S4 experiments come in (and even those will still need to be combined
with the Planck data for the best results).

\clearpage
\phantomsection
\addcontentsline{toc}{chapter}{Bibliography}

\bibliographystyle{abbrvnat_BJW}
\footnotesize
\bibliography{biblio}

\end{document}